\documentclass[twocolumn,superscriptaddress,prb]{revtex4}
\usepackage{epsfig}
\usepackage{verbatim}
\usepackage{bm}
\usepackage{mathrsfs}
\usepackage{yfonts}

\usepackage{times}
\usepackage{amsmath, amsthm}
\usepackage{amsfonts}
\usepackage{amssymb}
\usepackage{graphicx}
\usepackage{pst-all}
\usepackage{leftidx}

\newcommand{\gammabar}{\ensuremath\gamma\kern-0.53em-}
\renewcommand{\ol}[1]{\overline{#1}}
\newcommand{\ket}[1]{|#1\rangle}
\newcommand{\bra}[1]{\langle #1|}
\newcommand{\mb}[1]{\mathbf{#1}}
\newcommand{\sminus}{\scalebox{0.55}[1.0]{\( - \)}}

\newcommand{\gauged}[2]{( \mathcal{#1}_{#2}^\times )^{#2}}
\newcommand{\graded}[2]{\mathcal{#1}_{#2}}
\newcommand{\ext}[2]{\mathcal{#1}_{#2}^\times}
\newcommand{\cohosub}[1]{\scalebox{0.7}{\textswab{#1}}}
\newcommand{\coho}[1]{\textswab{#1}}
\newcommand{\openonesub}{\scalebox{0.75}{\openone}}
\newcommand{\Ref}[1]{Ref.~\onlinecite{#1}}
\newcommand{\Refs}[1]{Refs.~\onlinecite{#1}}
\renewcommand{\eqref}[1]{Eq.~(\ref{#1})}

\newcommand{\mcc}{\mathcal{C}}

\newcommand{\mcg}{\mathcal{G}}

\newcommand{\mcm}{\mathcal{M}}

\newcommand{\Pic}{\underline{\mathrm{Pic}(\mcc)}}

\DeclareMathOperator{\tr}{Tr}
\DeclareMathOperator{\Hom}{Hom}
\DeclareMathOperator{\mdim}{dim}

\newtheorem{definition}{Definition}
\newtheorem{theorem}{Theorem}[section]

\makeatletter
\def\l@subsubsection#1#2{}
\makeatother

\begin{document}

\title{Symmetry Fractionalization, Defects, and Gauging of Topological Phases}
\author{Maissam Barkeshli}
\affiliation{Station Q, Microsoft Research, Santa Barbara, California 93106-6105, USA}
\author{Parsa Bonderson}
\affiliation{Station Q, Microsoft Research, Santa Barbara, California 93106-6105, USA}
\author{Meng Cheng}
\affiliation{Station Q, Microsoft Research, Santa Barbara, California 93106-6105, USA}
\author{Zhenghan Wang}
\affiliation{Station Q, Microsoft Research, Santa Barbara, California 93106-6105, USA}
\affiliation{Department of Mathematics, University of California, Santa Barbara, California 93106, USA}
\date{\today}
\begin{abstract}
We examine the interplay of symmetry and topological order in $2+1$ dimensional topological quantum phases of matter. We present a precise definition of the \it topological symmetry \rm group $\text{Aut}(\mathcal{C})$, which characterizes the symmetry of the emergent topological quantum numbers of a topological phase $\mathcal{C}$, and we describe its relation with the microscopic symmetry of the underlying physical system. This allows us to derive a general framework to characterize and classify symmetry fractionalization in topological phases, including phases that are non-Abelian and symmetries that permute the quasiparticle types and/or are anti-unitary. We develop a theory of extrinsic defects (fluxes) associated with elements of the symmetry group, extending previous results in the literature. This provides a general classification of $2+1$ dimensional symmetry-enriched topological phases derived from a topological phase of matter $\mathcal{C}$ with on-site symmetry group $G$. We derive a set of data and consistency conditions, the solutions of which define the algebraic theory of the defects, known as a $G$-crossed braided tensor category $\mathcal{C}_{G}^{\times}$. This allows us to systematically compute many properties of these theories, such as the number of topologically distinct types of defects associated with each group element, their fusion rules, quantum dimensions, zero modes, braiding exchange transformations, a generalized Verlinde formula for the defects, and modular transformations of the $G$-crossed extensions of topological phases. We also examine the promotion of the global on-site symmetry to a local gauge invariance (``gauging the symmetry''), wherein the extrinsic $G$-defects are turned into deconfined quasiparticle excitations, which results in a different topological phase $\gauged{C}{G}$. We present systematic methods to compute the properties of $\gauged{C}{G}$ when $G$ is a finite group. The quantum phase transition between the topological phases $\gauged{C}{G}$ and $\mathcal{C}$ can be understood to be a ``gauge symmetry breaking'' transition, thus shedding light on the universality class of a wide variety of topological quantum phase transitions. A number of instructive and/or physically relevant examples are studied in detail.
\end{abstract}

\maketitle
\tableofcontents

\section{Introduction}

The last two decades of research in condensed matter physics have yielded remarkable progress
in the understanding of gapped quantum states of matter. In the absence of any symmetry,
gapped quantum systems at zero temperature may still form distinct phases of
matter that exhibit \emph{topological order}, which is a new kind of order characterized by
patterns of long range entanglements~\cite{wen04,Nayak08}.
Topologically ordered phases possess numerous remarkable properties, including quasiparticle excitations with exotic, possibly non-Abelian, exchange transformations (statistics), robust patterns of long range quantum entanglement, robust topology-dependent ground state degeneracies, and protected gapless edge modes.

Recently, a number of exciting new directions have emerged in the study of topological phases of matter, one of which is the study of extrinsic
defects~\cite{Bravyi98, bombin2010,kitaev2012, barkeshli2012a,you2012,
clarke2013,lindner2012,cheng2012, you2013,barkeshli2013genon,brown2013,barkeshli2013defect,barkeshli2013defect2,vaezi2013,teo2013,kapustin2013,petrova2013,teo2013b,khan2014}.
This includes the study of extrinsically imposed point-like defects, which are not finite-energy
quasiparticle excitations, but nevertheless have a nontrivial interplay with the topological order.
These point-like defects can themselves give rise to topologically protected degeneracies,
non-Abelian braiding exchange transformations, and exotic localized zero modes. From a practical standpoint, they might
be useful in enhancing the computational power of a topological phase used for topologically protected quantum information processing~\cite{Kitaev03,Freedman98,Preskill98,Freedman02a,Freedman03b,Freedman02b,Nayak08}. For example, one may engineer non-Abelian defects in an Abelian topological phase, or even defects that realize a computationally universal braiding gate set in a non-Abelian phase that otherwise would not have computationally universal braiding~\cite{barkeshli2013genon}. Several microscopic realizations of such defects have been proposed in the past few years, ranging from lattice
dislocations in certain microscopic models~\cite{bombin2010,barkeshli2012a,you2012, you2013,teo2013,teo2013b,brown2013,petrova2013} to
unconventional methods of coupling fractional quantum Hall (FQH) edge states~\cite{barkeshli2012a,lindner2012,clarke2013,cheng2012,barkeshli2013genon,barkeshli2013defect,barkeshli2013defect2}.
In addition to point-like extrinsic defects, topological phases also support a rich variety of extrinsic
line-like defects. These may either be gapped or gapless, and in both cases there is necessarily a nontrivial
interplay with the topological order. In particular, gapped line-like defects, such as gapped boundaries~\cite{Bais2009b,kitaev2012,beigi2011,wang2012,levin2013,barkeshli2013defect,barkeshli2013defect2,wan2014,lan2014},
have recently been proposed to be used for robust experimental signatures of certain
topologically ordered states, such as fractionalization in spin liquids and topological degeneracy in FQH states~\cite{barkeshli2013,barkeshli2014deg,barkeshli2014sledge,ladecola2014}.

A second direction that has generated intense research is the interplay of symmetry with topological
order. In the presence of symmetries, gapped quantum systems acquire a
finer classification~\cite{wen2002psg,schnyder2008,kitaev2009,fidkowski2010,hasan2010,qiRMP2011,chen2011,fidkowski2011,turner2011,
chen2013,levinPRB2012,lu2012,freed2013,essin2013,mesaros2013,lu2013, wang2014, XuNLSM,neupert2014,KapustinSPT1,metlitski2014, Freed2014, ElseSPT}. Specifically, it is possible for two phases of matter to be equivalent in the absence of the symmetry, but distinct in
the presence of symmetry. These are referred to as symmetry-protected topological (SPT) states if the gapped
phase is trivial in the absence of symmetry, and as symmetry-enriched topological (SET) states if the gapped phase is topologically
nontrivial, even when all symmetries are broken.
One-dimensional Haldane phases in spin chains~\cite{Haldane1, Haldane2}, two-dimensional quantum spin Hall insulators~\cite{Kane2005b,Kane2005a, Bernevig2006}, and three-dimensional
time-reversal-invariant topological insulators~\cite{Moore_PRB07,Roy_PRB2009,Fu_PRL07} are all well-known examples of SPT states.
In contrast, FQH states and gapped quantum spin liquids are examples of SET states, because they possess
symmetries (particle number conservation or spin rotational invariance) together with topological order.

In the presence of symmetries, quasiparticles of a topological phase of matter can acquire
fractional quantum numbers of the global symmetry. For example,
in the $\nu=\frac{1}{3}$ Laughlin FQH state~\cite{Laughlin83}, the quasiparticles carry charge in units of $e/3$;
in gapped $\mathbb{Z}_2$ quantum spin liquids \cite{balents2010}, the quasiparticles can carry unit charge and no spin (chargeons/holons), or
zero charge and spin-$\frac{1}{2}$ (spinons). With symmetry, an even larger class of extrinsic defects
is possible, as one can always consider a deformation of the Hamiltonian that forces a flux associated with the symmetry into a region of
the system, even if this flux is not associated with any deconfined
quasiparticle excitation.

When a Hamiltonian that realizes a topological phase of matter possesses a global symmetry, it is natural
to consider the topological order that is obtained when this global symmetry is promoted to a
local gauge invariance, i.e. ``gauging the symmetry.'' This is useful for a number of reasons: (1) The properties of the resulting
gauged theory can be used as a diagnostic to understand the properties of the original, ungauged
system~\cite{levin2012,swingle2014,hung2012,ChengPRL2014}. (2) Gauging the symmetry provides a relation between two different topological phases of matter, and can give insight
into the nature of the quantum phase transition between them~\cite{bais2009,barkeshli2010prl,barkeshli2011orb,burnell2012}.
(3) Understanding the relation between such phases may aid in the development of microscopic Hamiltonians for exotic topological phases (described by the gauged theory),
by starting with known models of simpler topological phases (described by the ungauged theory).

Although a remarkable amount of progress has been made on these deeply interrelated topics,
a completely general understanding is lacking, and many questions remain. For example, although there are many partial results, the
current understanding of fractionalization of quantum numbers, along with the classification and characterization of
SETs is incomplete. Moreover, while there have been many results towards understanding the
properties of extrinsic defects in topological phases, there has been no general systematic understanding and, in particular,
no concrete method of computing all the rich topological properties of the defects for an arbitrary topological phase.
The study of topological phase transitions between different topological phases is also missing a general theory.

In this paper, we develop a general systematic framework to understand these problems.
We develop a way to characterize the interplay of symmetry and topological order in $2+1$ dimensions,
thus leading us to a general understanding of how symmetries can be consistently fractionalized
in a given topological phase. Subsequently, we develop a mathematical framework to describe and compute
the properties of extrinsic point-like defects associated with symmetries of the topological phase.
Our construction utilizes results and ideas from recent mathematical literature~\cite{turaev2000,kirillov2004,ENO2009,turaev2010}. However, since our focus is on concrete applications to physics, our approach and formalism are quite different from the more abstract categorical formalism that has been presented in the mathematical literature. Our framework for understanding the topological properties
of extrinsic defects then provides us with a way to systematically classify and characterize SETs (including SPTs) in
$2+1$ dimensions. Finally, we again build on results from the mathematics literature~\cite{kirillov2004,Burciu2013} to provide a systematic
prescription for gauging the symmetry of a system in a topological phase of matter.

\subsection{Summary of Main Results}

Due to the length of this paper, we will briefly summarize the main results of our work here.
Before we proceed, we note that our starting framework to describe a topological phase without symmetry
is in terms of an anyon model $\mathcal{C}$, for which we provide a detailed review of the general theory in Sec.~\ref{sec:Anyons}.
Mathematically, $\mathcal{C}$ is referred to as a unitary modular tensor category (UMTC). Physically,
it can be thought of as the set of topological charges, which label the topologically distinct types of quasiparticles (anyons), together with
data that self-consistently specifies their fusion, associativity, and braiding exchange transformations. As this paper draws upon a number of technical mathematical concepts, we have made an effort to include precise definitions and explanations of most of these concepts, in order to make it as self-contained as possible.

\subsubsection{Symmetry and Fractionalization}

Symmetry fractionalization refers to the manner in which topologically nontrivial quasiparticles carry quantum numbers that are (in a sense) fractions of the quantum numbers of the underlying local constituents of the system, such as electrons or spins. We show that for a symmetry $G$ (continuous or discrete, unitary or anti-unitary), symmetry fractionalization is characterized by a pair of objects, $\rho$ and $\eta$, which we briefly describe here. There are non-physical redundancies, i.e. a sort of gauge freedom, associated with these objects that should be factored out, and the resulting equivalence classes provide a classification of symmetry fractionalization.

We first define the group of \textit{topological symmetries}, denoted $\text{Aut}(\mathcal{C})$, of a topological phase of matter described by $\mathcal{C}$. Roughly speaking, this corresponds to all of the different ways the theory $\mathcal{C}$ can be mapped back onto itself, including permutations of topological charges, in such a way that the topological properties are left invariant. A subset of such auto-equivalence maps called ``natural isomorphisms,'' which do not permute topological charges and leave all the basic data unchanged, provide the redundancy under which one equates the auto-equivalence maps to form the group $\text{Aut}(\mathcal{C})$. Simple examples of auto-equivalence maps include layer permutations in multi-layer systems that consist of multiple identical copies of a topological phase, or electric-magnetic duality in phases described by a $\mathbb{Z}_N$ gauge theory.

We next consider a physical system in a topological phase described by $\mathcal{C}$, which also has a global symmetry described by the group $G$ acting on the physical degrees of freedom. One must specify how $G$ acts upon the topological degrees of freedom and thus interplays with the topological symmetry. This is characterized by a group action
\begin{align}
[\rho]: G \rightarrow \text{Aut}(\mathcal{C})
.
\end{align}
The notation means that we assign an auto-equivalence map $\rho_{\bf g}$ to each group element ${\bf g} \in G$ and take the equivalence classes of these maps under natural isomorphism. (It is useful to work with a specific choice $\rho \in [\rho]$ when deriving results, and then demonstrate invariance within the equivalence class for certain quantities at the end.)

Once $[\rho]$ is specified, we examine the symmetry action in an underlying physical system described by a microscopic Hamiltonian. We show that symmetry fractionalization is possible only when a certain object $[\coho{O}] \in H^3_{[\rho]}(G, \mathcal{A})$ vanishes. This object $[\coho{O}]$ is uniquely defined by $[\rho]$, and hence called the fractionalization obstruction class of $[\rho]$. Here $\mathcal{A}$ corresponds to the group whose elements are the Abelian topological charges in $\mathcal{C}$, where group multiplication is defined by fusion. $H^3_{[\rho]}(G, \mathcal{A})$ is the 3rd cohomology group of $G$ with coefficients in the group $\mathcal{A}$, where the subscript $[\rho]$ indicates the inclusion of the symmetry action in the definition of the cohomology, which, in this context, is a potential permutation of the topological charge values in $\mathcal{A}$ (and, hence, is independent of the choice of $\rho \in [\rho]$).

When the obstruction vanishes, it is possible to consistently fractionalize the symmetry in the system, meaning one can specify a local projective symmetry action that is compatible with the symmetry action on the topological degrees of freedom. This local projective symmetry action has associated projective phases $\eta_{a}({\bf g},{\bf h})$ for each topological charge $a$. There is also non-physical redundancy in how the localized symmetry operators are defined, which transforms the corresponding projective phases. Factoring out this redundancy yields symmetry fractionalization classes corresponding to the equivalence classes $[\eta]$. The different ways (up to redundancy) in which the symmetry can be fractionalized is shown to be classified by the 2nd cohomology group $H^2_{[\rho]}(G, \mathcal{A})$, with there being a distinct fractionalization class for each element $[\coho{t}] \in H^2_{[\rho]}(G, \mathcal{A})$. More precisely, the set of symmetry fractionalization classes form an $H^2_{[\rho]}(G, \mathcal{A})$ torsor, which means the classes are not themselves elements of $H^2_{[\rho]}(G, \mathcal{A})$, but rather the distinct fractionalization classes are related to each other by an action of distinct elements $H^2_{[\rho]}(G, \mathcal{A})$. The precise definitions of these mathematical objects will appear in the main text and appendices.

\subsubsection{Extrinsic Defects}

When the physical system has a symmetry $G$, one can consider the possibility of point-like defects associated with group elements ${\bf g} \in G$, which may be thought of as fluxes. In many ways, a defect behaves like a quasiparticle. However, an important distinction is that when a quasiparticle is transported around a ${\bf g}$-defect, it is acted upon by the corresponding symmetry action $\rho_{\bf g}$, possibly permuting the quasiparticle's topological charge value. Another important distinction is that, since $G$ describes a global symmetry and not a local gauge invariance in this context, these defects do not correspond to finite-energy excitations of the system. Thus, they must be \textit{extrinsically imposed} by modifying the Hamiltonian in a manner that forces the $\mb{g}$-flux into the system. If the position of the defects are allowed to fluctuate quantum mechanically, the energy cost of separating such defects will grow either logarithmically or linearly in their separation. Therefore they may also be viewed as \textit{confined} excitations of the system.

The extrinsic defects of a topological phase have many rich topological properties, and one purpose of this paper is to develop a
concrete algebraic formalism, analogous to the algebraic theory of anyons, that can be used to characterize and
systematically compute the many topological properties of such defects. For this, we begin by generalizing the notion of topological charge to apply to defects, with distinct types of ${\bf g}$-defects carrying distinct values of topological charge. We then extend the description of the original anyon model $\mathcal{C}$, describing the topological phase, to a $G$-graded fusion theory
\begin{align}
\mathcal{C}_{G} = \bigoplus_{{\bf g} \in G} \mathcal{C}_\mb{g}
,
\end{align}
where each sector $\mathcal{C}_\mb{g}$ describes the topologically distinct types of $\mb{g}$-defects and the fusion and associativity relations respect the group multiplication of $G$, i.e. a ${\bf g}$-defect and an ${\bf h}$-defect fuse to a ${\bf gh}$-defect. In this way, the quasiparticles of the original topological phase correspond to the ${\bf 0}$-defects, i.e. $\mathcal{C}_{\bf 0} = \mathcal{C}$.

Subsequently, we introduce a generalized notion of braiding transformations that incorporates the symmetry action $\rho_{\bf g}$ on topological charges as a quasiparticle or defect passes around a ${\bf g}$-defect. This is referred to as ``$G$-crossed braiding'' and defines a $G$-crossed braided tensor category (BTC), which we denote as $\mathcal{C}_{G}^{\times}$. Additionally, the symmetry action on states and fractionalization ($\rho$ and $\eta$) are incorporated when considering fusion spaces. Similar to anyon models, we provide a diagrammatic representation of the states and operators of the theory and identify the basic data that fully characterizes the theory. We introduce consistency conditions on the basic data, which generalize the famous hexagon equations for braiding consistency to ``heptagon equations'' for $G$-crossed braiding, and impose consistency of the incorporation of the symmetry action and its fractionalization within the theory.

Given the basic data of the $G$-crossed theory, we are able to compute all properties of the defects, including their fusion rules, quantum dimensions, localized zero modes, and braiding statistics. We find that topological twists, which characterize the braiding statistics of objects, is not a gauge invariant quantity for defects, which meshes well with the notion that the defects are associated with confined objects. Another important property that we derive is that the total quantum dimension $\mathcal{D}_\mb{g}$ of the sector $\mathcal{C}_\mb{g}$ is the same for all $\mb{g}\in G$, i.e. $\mathcal{D}_\mb{g} = \mathcal{D}_{\bf 0}$ (this holds generally for a $G$-graded fusion category). We also find that the number of topologically distinct $\mb{g}$-defects, $|\mathcal{C}_\mb{g}|$, is equal to the number of $\mb{g}$-invariant topological charges [i.e. those for which $\rho_\mb{g}(a) = a$] in the original UMTC $\mathcal{C}_{\bf 0}$.

We describe the notion of $G$-crossed modular transformations when the system inhabits a torus or surfaces of arbitrary genus.
These extend the usual definition of modular transformations, generated by $\mathcal{S}$ and $\mathcal{T}$ matrices, to cases where there are defect branch lines wrapping the cycles of the torus or higher genus surface. We derive a $G$-crossed generalization of the Verlinde formula, which relate the fusion rules of defects (and quasiparticles) to the $G$-crossed $S$-matrix.

For every $2+1$ dimensional SET phase, one can construct a corresponding $G$-crossed theory $\ext{C}{G}$ describing the defects in
the topological phase, which also incorporates the symmetry fractionalization. Therefore, the $G$-crossed defect theories $\ext{C}{G}$ provide both a classification and a characterization of SET phases in $2+1$ dimensions. In this way, one can classify SETs by solving the $G$-crossed consistency relations. Topological invariants
that can distinguish different SET phases are associated with gauge-invariant quantities of the $G$-crossed theory $\ext{C}{G}$. A partial
list of such topological invariants is presented in Table~\ref{tab:inv} of Sec.~\ref{sec:Classification}.

Importantly, not every fractionalization class corresponds to a well-defined SET in $2+1$ dimensions. In some cases, there can be an additional obstruction that prevents the existence of a solution of the $G$-crossed consistency relations (such as the heptagon equations). The inability to solve these consistency conditions and, thus, to construct a consistent defect theory $\ext{C}{G}$ indicates that the symmetry fractionalization class is anomalous. Similar to the classification and possible obstruction of symmetry fractionalization for a topological phase when the symmetry action is specified, the classification and possible obstruction of defectification (i.e. the existence of a consistent defect theory) for a topological phase when the symmetry action and fractionalization are specified can be reduced to a simpler cohomological structure.
In particular, it has been proven~\cite{ENO2009} that, for a finite group $G$, which describes unitary on-site symmetries, the defectification obstruction takes values in $\mathcal{H}^4(G, U(1))$. Moreover, this obstruction only depends on $\mathcal{C}_{\bf 0}$, the symmetry action, and the symmetry fractionalization class. Using the formalism of our paper, we explicitly derive an expression in Eq.~(\ref{eq:obstruction}) for such an obstruction to solving the $G$-crossed consistency conditions for the case where the symmetry action $[\rho]$ does not permute quasiparticle types. A number of recent examples have shown that anomalous realizations of symmetry fractionalization, while they cannot exist in $2+1$ dimensions, can instead exist as a surface termination state of a $3+1$ dimensional SPT state~\cite{Walker2012,vonKeyserlingk2013,VishwanathPRX2013,wang2013,wang2013b,Bonderson13d,chen2014b,metlitski2013,chen2014,cho2014,kapustin2014}.

Furthermore, it has been proven~\cite{ENO2009} that, when there are solutions of the $G$-crossed consistency relations (i.e. when the defectification obstruction vanishes) for a specified $\mathcal{C}_{\bf 0}$, symmetry action, and symmetry fractionalization, the set of gauge inequivalent solutions, i.e. the defectification classes, form a $H^3(G, \text{U}(1))$ torsor. More specifically, this means that distinct $G$-crossed theories (with the same $\mathcal{C}_{\bf 0}$, symmetry action, and fractionalization class) are related to each other by an action of the distinct elements of $H^3(G, \text{U}(1))$. This action by $[\alpha] \in H^3(G, \text{U}(1))$ is essentially ``gluing'' a SPT state, whose symmetry group is $G$ and associativity is defined by $[\alpha]$, onto the $G$-crossed theory such that the group labels of the defects of both theories match up. Whether gluing on a SPT state in this way actually produces a distinct $G$-crossed theory can be determined in our framework by checking whether the old and new $G$-crossed theories are equivalent under relabeling the defect topological charge values. Thus, these results imply that the possible $G$-crossed extensions $\ext{C}{G}$, i.e. the possible symmetry enrichments of a $2+1$ dimensional topological phase described by $\mathcal{C}$ for finite on-site unitary $G$ symmetry, are fully classified (possibly up to relabeling topological charges) by three properties: (1) the symmetry action $[\rho]: G \rightarrow \text{Aut}(\mathcal{C})$, (2) the symmetry fractionalization class, which is an element of an $H^2_{[\rho]}(G, \mathcal{A})$ torsor, and (3) the defectification class, which is an element of an $H^3(G, \text{U}(1))$ torsor.

\subsubsection{Gauging the Symmetry}

Given a topological phase of matter $\mathcal{C}$, together with its symmetry-enriched class, i.e. its $G$-crossed defect theory
$\ext{C}{G}$, one can promote the symmetry $G$ to a local gauge invariance (``gauging the symmetry''). This results in a
different topological order, which we denote $\gauged{C}{G}$, in which the ${\bf g}$-defects become deconfined quasiparticle excitations.
Importantly, the gauged theory $\gauged{C}{G}$ depends on the particular $G$-crossed extension $\mathcal{C}_G^\times$ of $\mathcal{C}$, which thus forms
the input data necessary to construct the gauged theory. The topological properties of the gauged theory $\gauged{C}{G}$ can alternatively be viewed from a different
perspective as topological invariants of the associated SET, which is described by $\ext{C}{G}$.

We first examine the question of how one may obtain a microscopic Hamiltonian that realizes the topological phase $\gauged{C}{G}$, given a Hamiltonian that realizes a topological phase $\mathcal{C}$. Along this line, we provide a concrete model demonstrating how this may be done in the case where $G$ is an Abelian finite group.

Next, we provide a review of some known results from the mathematics literature for obtaining the properties of $\gauged{C}{G}$ from those of $\ext{C}{G}$, in particular the topological charge content, quantum dimensions, and fusion rules. It follows from these results that the total quantum dimension of the gauged theory $\gauged{C}{G}$ is always related to the total quantum dimension of the original theory $\mathcal{C}$ and its $G$-crossed extension by $\mathcal{D}_{\gauged{C}{G} }= |G|^{\frac{1}{2}} \mathcal{D}_{\mathcal{C}_G} = |G| \mathcal{D}_{\mathcal{C}}$. We further provide a formula for the topological twists of quasiparticles in $\gauged{C}{G}$, which were not previously given in the literature. We confirm the validity of this expression based on physical considerations and consistency. Using the expression for the topological twists of the gauged theory, we show that the chiral central charge (mod 8) is the same in these theories. We also derive an expression for the topological $S$-matrix of $\gauged{C}{G}$ in terms of the data of $\ext{C}{G}$. Finally, we discuss how to compute the ground state degeneracy of $\gauged{C}{G}$ on higher genus surfaces in terms of the properties of $\ext{C}{G}$, without needing to derive the full fusion rules of $\gauged{C}{G}$. This is useful for practical computations of the number of topological charge types and their quantum dimensions.

We close the discussion of gauging the symmetry by observing that, since $\gauged{C}{G}$ and $\mathcal{C}$ are related to each other by gauging $G$, the topological quantum phase transition between them can be understood as a discrete $G$ ``gauge symmetry breaking'' transition. This point of view provides insight into the universality class of the topological phase transitions between a wide variety of distinct topological phases.

\subsubsection{Examples}

After developing the general theory, summarized above, we study many concrete examples. We focus on examples that are physically relevant
and/or which illustrate different technical aspects and subtleties of using the theory and methods developed in this paper to derive the
various properties of $G$-crossed extensions and gauged theories. One large class of examples for which we have obtained all the basic data
of $\ext{C}{G}$ by solving the consistency conditions is the case where $\mathcal{C}_{\bf 0}$ is a MTC and the symmetry action $[\rho]$ does not permute quasiparticle types.
Another particularly interesting example that we examine is the ``three-fermion theory,'' also known as $\textrm{SO}(8)_1$, with the
non-Abelian symmetry group $G = S_3$ acting nontrivially. Gauging the $S_3$-symmetry of the three-fermion theory results in a rank
$12$ (weakly integral) UMTC that has not been previously described elsewhere.

\subsection{Relation to Prior Work}

The background context of our work is closely related to a large number of works spanning many different fields. Here we briefly comment on the relation to some of the most closely related works.

A framework, called the projective symmetry group (PSG),  to address the problem of classifying SETs was originally introduced in \Ref{wen2002psg}. As we discuss in Sec.~\ref{sec:PSG}, the PSG framework only captures a subset of possible types of symmetry fractionalization and, thus, misses a large class of possible SETs for a given topological phase. Our results on the general classification of symmetry fractionalization in terms of $H^2_{[\rho]}(G ,\mathcal{A})$ extends the previous result of \Ref{essin2013}, which specifically applies to Abelian topological phases where the symmetries do not permute the topological charge values. A preliminary consideration of some of these ideas can also be found at a more abstract level in the discussion in Appendix F of \Ref{Kitaev06a}.

The notion of a $G$-crossed braided tensor category (BTC) was originally introduced in the mathematics literature in \Refs{turaev2000,turaev2010}. Similarly, the classification and possible obstructions of $G$-crossed extensions, which we summarized in the previous subsection, has previously appeared in the mathematics literature~\cite{ENO2009} in the problem of extending a fusion category or a braided fusion category by a finite group $G$.

With respect to these prior mathematical results, our results can be viewed as both providing (1) a new and detailed concrete formulation of the theory of $G$-crossed BTC, and (2) providing the physical context and interpretations of the abstract mathematical results by directly linking them to their physical realizations. In particular, we provide a physical interpretation of these mathematical objects in terms of the fusion and braiding properties of extrinsic defects associated with group elements ${\bf g} \in G$. Moreover, since the mathematical constructions are highly abstract, they may obscure many of the important details that are of interest for physical applications. For example, we provide concrete definitions of the symmetry action $[\rho]$, the fractionalization obstruction $[\coho{O}]$, and local projective phases $[\eta]$ that classify fractionalization in terms of the symmetry action on the states of quasiparticles. The mathematical treatment that we utilize in this paper, working directly with the topologically distinct classes of simple objects (quasiparticles and defects), their basic data ($F$-symbols, $R$-symbols, etc.), and their consistency conditions, is referred to in mathematical parlance as a ``skeletonization'' of a category. Our work may, thus, also be viewed as a new mathematical result that introduces the skeletonization of $G$-crossed BTCs and provides a new definition of the theory of $G$-crossed BTCs.

Extrinsic defects in topological phases of matter have been increasingly studied in various examples in the condensed matter physics literature~\cite{bombin2010, kitaev2012, barkeshli2012a,you2012,lindner2012,clarke2013,cheng2012,barkeshli2013genon,you2013,barkeshli2013defect,barkeshli2013defect2, vaezi2013, teo2013,teo2013b,brown2013,petrova2013,kapustin2013,khan2014}. One purpose of our work is to provide a totally general treatment of extrinsic twist defects that captures all of their topologically nontrivial properties, provides a framework for computing them, and can be applied to arbitrary topological phases of matter. In recent years, such defects have also been studied in the mathematical physics literature, both for conformal field theory (CFT)~\cite{frohlich2007} and for topological quantum field theory (TQFT)~\cite{kapustin2010,fuchs2012}. While our work has some overlap with these, our approach is quite different. Our emphasis is on developing concrete methods that can be used to compute various topological properties of the defects and direct physical interpretations that apply in the condensed matter physics setting.

The idea of ``gauging'' a discrete symmetry of a topological phase of matter is closely related to the concept of ``orbifolding'' in rational CFT~\cite{dijkgraaf1989,difrancesco}. However, while there are often close relations between CFTs and topological phases of matter, they are distinct physical systems, and so they each require their own physical understanding. Many of our general results and examples go beyond the analogous problem that has been studied in the CFT literature, for which the general results are limited. For example, much of the CFT work on orbifolding is typically focused on holomorphic CFTs, which correspond to only a small class of possible topological phases. The important classifying objects (the symmetry action, fractionalization class, and defectification class) summarized above also have not, to our knowledge, been generally discussed in the CFT literature on orbifolding. On the other hand, a CFT possesses a great deal of structure that does not exist in the corresponding MTC. As such, the orbifold construction can be applied to CFTs in ways that do not correspond to gauging the corresponding MTC. This distinction is highlighted by the fact that applying the orbifold construction multiple times to a CFT can result in the original CFT, whereas applying the gauging construction multiple times to a MTC cannot result in the original MTC.

Our work on gauging topological phases of matter is closely related to work of \Refs{kirillov2004,muger2004}, which sets out to find a mathematical formulation in terms of MTCs of the concept of orbifolding in CFTs. For example, \Ref{kirillov2004} also contains results on the extended Verlinde algebra. Again, our results extend some of these mathematical results and put them into more concrete terms with direct physical context.

In recent years, the notion of gauging symmetries of a topological phase has been increasingly studied in the condensed matter literature. The resulting non-Abelian topological phases that are obtained by gauging either the layer exchange symmetry of bilayer Abelian FQH states, or the electric-magnetic duality of $\mathbb{Z}_N$ toric code models were studied in \Refs{barkeshli2010,barkeshli2010twist,barkeshli2011orb}. In studies of SPT phases, the notion of gauging the symmetry of the system has been powerful in developing an understanding of the distinction between SPT states~\cite{levin2012,ChengPRL2014}. While those were isolated classes of examples, our work provides a concrete prescription to derive the properties obtained when any topological phase of matter $\mathcal{C}$ is gauged by any finite group $G$.

While gauging a discrete global symmetry $G$ of a topological phase $\mathcal{C}$ gives rise to a new topological phase $\gauged{C}{G}$, there is an inverse process, known as topological Bose condensation~\cite{bais2009}, which takes $\gauged{C}{G}$ to $\mathcal{C}$. The quantum phase transition between $\gauged{C}{G}$ and $\mathcal{C}$ corresponds to a confinement/deconfinement transition or, in other words, a ``gauge symmetry breaking'' transition. The notion of condensation was discussed mathematically in \Refs{bruguieres2000,muger2000}. This has been studied in the context of topological phases in \Refs{bais2009,Bais2009b,kong2014,eliens2013}. In the topological Bose condensation picture, there is an intermediate stage between $\gauged{C}{G}$ and $\mathcal{C}$, referred to as the $\mathcal{T}$-theory in \Ref{bais2009}, which includes the objects that are confined by the condensate. These confined objects are ${\bf g}$-defects and the $G$-crossed theory $\mathcal{C}_{G}^{\times}$ provides a complete description of the topological properties of the $\mathcal{T}$-theory, including their braiding transformations, which have not been previously identified. Most of the prior work along these lines has focused on the nature of the topological phase that is obtained when topologically non-trivial bosons of a topological phase are condensed. However, \Refs{barkeshli2010prl,barkeshli2011orb,burnell2012} focused on the nature of the universality class of quantum phase transitions associated with topological Bose condensation by studying some simple classes of examples when $G=\mathbb{Z}_2$. We generalize these results to an understanding of the universality class of topological Bose condensation transitions between $\gauged{C}{G}$ to $\mathcal{C}$ for general finite $G$.

\section{Review of Algebraic Theory of Anyons}
\label{sec:Anyons}

This section provides a summary review of anyon models, known in mathematical terminology as unitary braided tensor categories (UBTC)~\cite{Turaev94,Bakalov01}. We use a diagrammatic representation of anyonic states and operators acting on them, following \Refs{Preskill-lectures,Kitaev06a,Bonderson07b,Bonderson07c}. (Many relations in this review section are stated without proof. For additional details and proofs, we refer the reader to the references listed here or, in some cases, to Secs.~\ref{sec:Algebraic_Theory} and \ref{sec:G-Crossed_Modularity} where one may find the generalized versions.) This formalism encodes the purely topological properties of anyons, i.e. quasiparticle excitations of topological phases of matter, independent of any particular physical realization.

\subsection{Fusion}
\label{sec:Fusion}

In this section, we describe the properties of fusion tensor categories, and will introduce braiding in the next. We begin with a set $\mathcal{C}$ of superselection sector labels called topological or anyonic charges $a,b,c\ldots \in \mathcal{C}$.~\footnote{It is often assumed that the set of topological charges $\mathcal{C}$ is finite, but we may allow it to be infinite in the definition of a fusion tensor category or a braided tensor category, as long as fusion is finite. However, for a modular tensor category, we will require $\mathcal{C}$ to be a finite set.} (We will often also use the symbol $\mathcal{C}$ to refer the category itself.) These conserved charges obey an associative fusion algebra
\begin{equation}
\label{eq:fusion_rules}
a\times b=\sum\limits_{c\in \mathcal{C}}N_{ab}^{c}c
\end{equation}
where the fusion multiplicities $N_{ab}^{c}$ are non-negative integers which indicate the number of different ways the charges $a$ and $b$ can be
combined to produce the charge $c$. We require that fusion is finite, meaning $\sum_{c} N_{ab}^{c}$ is a finite integer for any fixed $a$ and $b$.
Associativity requires these to satisfy
\begin{equation}
\sum_{e} N_{ab}^{e} N_{ec}^{d} = \sum_{f} N_{af}^{d} N_{bc}^{f}
.
\end{equation}

In the diagrammatic formalism, each line segment is oriented (indicated with an arrow) and ascribed a value of topological charge. Each fusion product has an associated vector space $V_{ab}^{c}$ with $\dim{V_{ab}^{c}} = N_{ab}^{c}$, and its dual (splitting) space $V^{ab}_{c}$. The states in these fusion and splitting spaces are assigned to trivalent vertices with the appropriately corresponding anyonic charges, with basis states written as
\begin{equation}
\left( d_{c} / d_{a}d_{b} \right) ^{1/4}
\pspicture[shift=-0.6](-0.1,-0.2)(1.5,-1.2)
  \small
  \psset{linewidth=0.9pt,linecolor=black,arrowscale=1.5,arrowinset=0.15}
  \psline{-<}(0.7,0)(0.7,-0.35)
  \psline(0.7,0)(0.7,-0.55)
  \psline(0.7,-0.55) (0.25,-1)
  \psline{-<}(0.7,-0.55)(0.35,-0.9)
  \psline(0.7,-0.55) (1.15,-1)	
  \psline{-<}(0.7,-0.55)(1.05,-0.9)
  \rput[tl]{0}(0.4,0){$c$}
  \rput[br]{0}(1.4,-0.95){$b$}
  \rput[bl]{0}(0,-0.95){$a$}
 \scriptsize
  \rput[bl]{0}(0.85,-0.5){$\mu$}
  \endpspicture
=\left\langle a,b;c,\mu \right| \in
V_{ab}^{c} ,
\label{eq:bra}
\end{equation}
\begin{equation}
\left( d_{c} / d_{a}d_{b}\right) ^{1/4}
\pspicture[shift=-0.65](-0.1,-0.2)(1.5,1.2)
  \small
  \psset{linewidth=0.9pt,linecolor=black,arrowscale=1.5,arrowinset=0.15}
  \psline{->}(0.7,0)(0.7,0.45)
  \psline(0.7,0)(0.7,0.55)
  \psline(0.7,0.55) (0.25,1)
  \psline{->}(0.7,0.55)(0.3,0.95)
  \psline(0.7,0.55) (1.15,1)	
  \psline{->}(0.7,0.55)(1.1,0.95)
  \rput[bl]{0}(0.4,0){$c$}
  \rput[br]{0}(1.4,0.8){$b$}
  \rput[bl]{0}(0,0.8){$a$}
 \scriptsize
  \rput[bl]{0}(0.85,0.35){$\mu$}
  \endpspicture
=\left| a,b;c,\mu \right\rangle \in
V_{c}^{ab},
\label{eq:ket}
\end{equation}
where $\mu=1,\ldots ,N_{ab}^{c}$. (Many anyon models of interest have no fusion multiplicities, i.e. $N_{ab}^{c}=0$ or $1$ only, in which case the trivial vertex labels $\mu$ will usually be left implicit.) The bra/ket basis vectors are orthonormal. The normalization factors $\left( d_{c}/d_{a}d_{b}\right) ^{1/4}$ are included so that diagrams will be in the isotopy invariant convention, as will be explained in the following. Isotopy invariance means that the value of a (labeled) diagram is not changed by continuous deformations, so long as open endpoints are held fixed and lines are not passed through each other or around open endpoints. Open endpoints should be thought of as ending on some boundary (e.g. a timeslice or an edge of the system) through which isotopy is not permitted. We note that the diagrammatic expressions of states and operators are, by design, reminiscent of particle worldlines, but there is not a strict identification between the two. The anyonic charge lines are only a diagrammatic expression of the algebraic encoding of the topological properties of anyons, and interpreting them as worldlines is not always correct.

Diagrammatically, inner products are formed by stacking vertices so the fusing/splitting lines connect
\begin{equation}
\label{eq:inner_product}
  \pspicture[shift=-0.95](-0.2,-0.35)(1.2,1.75)
  \small
  \psarc[linewidth=0.9pt,linecolor=black,border=0pt] (0.8,0.7){0.4}{120}{240}
  \psarc[linewidth=0.9pt,linecolor=black,arrows=<-,arrowscale=1.4,
    arrowinset=0.15] (0.8,0.7){0.4}{165}{240}
  \psarc[linewidth=0.9pt,linecolor=black,border=0pt] (0.4,0.7){0.4}{-60}{60}
  \psarc[linewidth=0.9pt,linecolor=black,arrows=->,arrowscale=1.4,
    arrowinset=0.15] (0.4,0.7){0.4}{-60}{15}
  \psset{linewidth=0.9pt,linecolor=black,arrowscale=1.5,arrowinset=0.15}
  \psline(0.6,1.05)(0.6,1.55)
  \psline{->}(0.6,1.05)(0.6,1.45)
  \psline(0.6,-0.15)(0.6,0.35)
  \psline{->}(0.6,-0.15)(0.6,0.25)
  \rput[bl]{0}(0.07,0.55){$a$}
  \rput[bl]{0}(0.94,0.55){$b$}
  \rput[bl]{0}(0.26,1.25){$c$}
  \rput[bl]{0}(0.24,-0.05){$c'$}
 \scriptsize
  \rput[bl]{0}(0.7,1.05){$\mu$}
  \rput[bl]{0}(0.7,0.15){$\mu'$}
  \endpspicture
=\delta _{c c ^{\prime }}\delta _{\mu \mu ^{\prime }} \sqrt{\frac{d_{a}d_{b}}{d_{c}}}
  \pspicture[shift=-0.95](0.15,-0.35)(0.8,1.75)
  \small
  \psset{linewidth=0.9pt,linecolor=black,arrowscale=1.5,arrowinset=0.15}
  \psline(0.6,-0.15)(0.6,1.55)
  \psline{->}(0.6,-0.15)(0.6,0.85)
  \rput[bl]{0}(0.75,1.25){$c$}
  \endpspicture
,
\end{equation}%
which can be applied inside more complicated diagrams. Note that this diagrammatically encodes charge conservation. Since we want to use this to describe the states associated with anyonic quasiparticles (in a topological phase of matter), we require the inner product to be positive definite, i.e. $d_a$ are required to be real and positive.

With this inner product, the identity operator on a pair of anyons with charges $a$ and $b$ is written (diagrammatically) as the partition of unity
\begin{equation}
\label{eq:Id}
\openone_{ab} =
\pspicture[shift=-0.65](0,-0.5)(1.1,1.1)
  \small
  \psset{linewidth=0.9pt,linecolor=black,arrowscale=1.5,arrowinset=0.15}
  \psline(0.3,-0.45)(0.3,1)
  \psline{->}(0.3,-0.45)(0.3,0.50)
  \psline(0.8,-0.45)(0.8,1)
  \psline{->}(0.8,-0.45)(0.8,0.50)
  \rput[br]{0}(1.05,0.8){$b$}
  \rput[bl]{0}(0,0.8){$a$}
  \endpspicture
 = \sum\limits_{c,\mu }
\sqrt{\frac{d_{c}}{d_{a}d_{b}}} \;
 \pspicture[shift=-0.6](-0.1,-0.45)(1.4,1)
  \small
  \psset{linewidth=0.9pt,linecolor=black,arrowscale=1.5,arrowinset=0.15}
  \psline{->}(0.7,0)(0.7,0.45)
  \psline(0.7,0)(0.7,0.55)
  \psline(0.7,0.55) (0.25,1)
  \psline{->}(0.7,0.55)(0.3,0.95)
  \psline(0.7,0.55) (1.15,1)
  \psline{->}(0.7,0.55)(1.1,0.95)
  \rput[bl]{0}(0.38,0.2){$c$}
  \rput[br]{0}(1.4,0.8){$b$}
  \rput[bl]{0}(0,0.8){$a$}
  \psline(0.7,0) (0.25,-0.45)
  \psline{-<}(0.7,0)(0.35,-0.35)
  \psline(0.7,0) (1.15,-0.45)
  \psline{-<}(0.7,0)(1.05,-0.35)
  \rput[br]{0}(1.4,-0.4){$b$}
  \rput[bl]{0}(0,-0.4){$a$}
\scriptsize
  \rput[bl]{0}(0.85,0.4){$\mu$}
  \rput[bl]{0}(0.85,-0.03){$\mu$}
  \endpspicture
\; .
\end{equation}
A similar decomposition applies for an arbitrary number of anyons.

More complicated diagrams can be constructed by connecting lines of matching charge. The resulting vector spaces obey a notion of associativity given by isomorphisms, which can be reduced using the expression of three anyon splitting/fusion spaces in terms of two anyon splitting/fusion
\begin{equation}
V_{d}^{abc}\cong \bigoplus\limits_{e}V_{e}^{ab}\otimes V_{d}^{ec}\cong
\bigoplus\limits_{f}V_{f}^{bc}\otimes V_{d}^{af},
\end{equation}
to isomorphisms called $F$-moves, which are written diagrammatically as
\begin{equation}
\scalebox{.8}{\pspicture[shift=-1.0](0,-0.45)(1.8,1.8)
  \small
  \psset{linewidth=0.9pt,linecolor=black,arrowscale=1.5,arrowinset=0.15}
  \psline(0.2,1.5)(1,0.5)
  \psline(1,0.5)(1,0)
  \psline(1.8,1.5) (1,0.5)
  \psline(0.6,1) (1,1.5)
   \psline{->}(0.6,1)(0.3,1.375)
   \psline{->}(0.6,1)(0.9,1.375)
   \psline{->}(1,0.5)(1.7,1.375)
   \psline{->}(1,0.5)(0.7,0.875)
   \psline{->}(1,0)(1,0.375)
   \rput[bl]{0}(0.05,1.6){$a$}
   \rput[bl]{0}(0.95,1.6){$b$}
   \rput[bl]{0}(1.75,1.6){${c}$}
   \rput[bl]{0}(0.5,0.5){$e$}
   \rput[bl]{0}(0.9,-0.3){$d$}
 \scriptsize
   \rput[bl]{0}(0.3,0.8){$\alpha$}
   \rput[bl]{0}(0.7,0.25){$\beta$}
  \endpspicture
}
= \sum_{f,\mu,\nu} \left[F_d^{abc}\right]_{(e,\alpha,\beta)(f,\mu,\nu)}
\scalebox{.8}{\pspicture[shift=-1.0](0,-0.45)(1.8,1.8)
  \small
  \psset{linewidth=0.9pt,linecolor=black,arrowscale=1.5,arrowinset=0.15}
  \psline(0.2,1.5)(1,0.5)
  \psline(1,0.5)(1,0)
  \psline(1.8,1.5) (1,0.5)
  \psline(1.4,1) (1,1.5)
   \psline{->}(0.6,1)(0.3,1.375)
   \psline{->}(1.4,1)(1.1,1.375)
   \psline{->}(1,0.5)(1.7,1.375)
   \psline{->}(1,0.5)(1.3,0.875)
   \psline{->}(1,0)(1,0.375)
   \rput[bl]{0}(0.05,1.6){$a$}
   \rput[bl]{0}(0.95,1.6){$b$}
   \rput[bl]{0}(1.75,1.6){${c}$}
   \rput[bl]{0}(1.25,0.45){$f$}
   \rput[bl]{0}(0.9,-0.3){$d$}
 \scriptsize
   \rput[bl]{0}(1.5,0.8){$\mu$}
   \rput[bl]{0}(0.7,0.25){$\nu$}
  \endpspicture
}
.
\end{equation}
The $F$-moves can be viewed as changes of bases for the states associated with quasiparticles. To describe topological phases, these are required to be unitary transformations, i.e.
\begin{eqnarray}
\left[ \left( F_{d}^{abc}\right) ^{-1}\right] _{\left( f,\mu
,\nu \right) \left( e,\alpha ,\beta \right) }
&=& \left[ \left( F_{d}^{abc}\right) ^{\dagger }\right]
_{\left( f,\mu ,\nu \right) \left( e,\alpha ,\beta \right) }
\notag \\
&=& \left[ F_{d}^{abc}\right] _{\left( e,\alpha ,\beta \right) \left( f,\mu
,\nu \right) }^{\ast }
.
\end{eqnarray}%

\begin{center}
\begin{figure}[t!]
\includegraphics[scale=0.5]{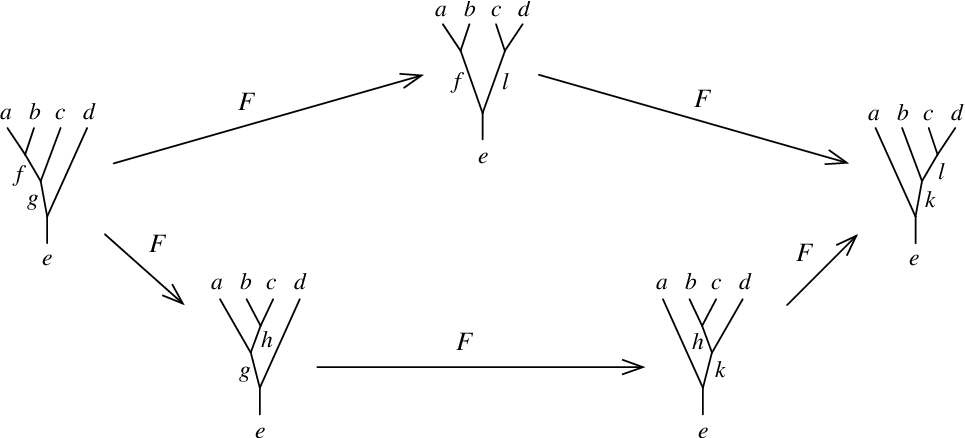}
\caption{The Pentagon equation enforces the condition that different sequences
of $F$-moves from the same starting fusion basis decomposition to the same
ending decomposition gives the same result. Eq.~(\ref{eq:pentagon}) is obtained
by imposing the condition that the above diagram commutes.}
\label{fig:pentagon}
\end{figure}
\end{center}

In order for this notion of associativity to be self-consistent, any two sequences of $F$-moves applied within an arbitrary diagram which start from the same state space and end in the same state space must be equivalent.
MacLane's coherence theorem~\cite{MacLane98} establishes that this consistency can be achieved by imposing the constraint called the Pentagon equation%
\begin{widetext}
\begin{equation}
\sum_{\delta }\left[ F_{e}^{fcd}\right] _{\left( g,\beta ,\gamma \right)
\left( l,\nu, \delta \right) }\left[ F_{e}^{abl}\right]
_{\left( f,\alpha ,\delta \right) \left( k,\mu, \lambda \right) }
=\sum_{h,\sigma ,\psi ,\rho }\left[ F_{g}^{abc}\right] _{\left( f,\alpha
,\beta \right) \left( h,\psi,\sigma \right) }\left[ F_{e}^{ahd}\right]
_{\left( g,\sigma ,\gamma \right) \left( k,\rho, \lambda \right) }%
\left[ F_{k}^{bcd}\right] _{\left( h,\psi ,\rho \right) \left(
l,\nu ,\mu \right) }
\label{eq:pentagon}
\end{equation}
\end{widetext}
which equates the two sequences of $F$-moves shown in Fig.~\ref{fig:pentagon}. In other words, given a set of fusion rules, one can find all consistent fusion categories by solving the Pentagon equations for all consistent sets of $F$-symbols.

We require the existence of a unique ``vacuum'' charge $0 \in \mathcal{C}$ for which fusion (and braiding) is trivial. In particular, the fusion coefficients must satisfy $N_{a0}^{c}=N_{0a}^{c}=\delta _{ac}$, charge lines can be added and removed from diagram at will (in other words, there are canonical isomorphisms between $V_{a}^{a0}$, $V_{a}^{0a}$, and $\mathbb{C}$), and the associativity relations must obey $\left[F_d^{abc}\right]=\openone$ if any one of $a$, $b$, or $c$ equals $0$ when the involved fusions are allowed (this enforces the compatibility of $F$-moves with the previously mentioned canonical isomorphism and corresponds to choosing the basis vectors of $V_{a}^{a0}$ and $V_{a}^{0a}$ such that they map to $1$ in the canonical isomorphisms mentioned above). Note that it is not required that $\left[F_d^{abc}\right]=\openone$ when $d=0$, nor is this even generally possible. We often specially denote vacuum lines as dotted lines.

For each $a\in \mathcal{C}$, we require the existence of a conjugate charge, or ``antiparticle,'' $\bar{a}\in \mathcal{C}$, for which $[F^{a \bar{a} a}_a]_{(0,\alpha)(0,\mu)} \neq 0$. It follows that $N_{ab}^{0}=\delta _{b\bar{a}}$, i.e. $\bar{a}$ is unique and $\dim V^{0}_{a \bar{a}} =1$. Also, $0=\bar{0}$ and $\bar{\bar{a}}=a$. Thus, we can write
\begin{equation}
\label{eq:FSInd}
[F^{a \bar{a} a}_a]_{00} = \frac{\varkappa_{a}}{d_a}
,
\end{equation}
where we have defined the quantum dimension $d_a$ of charge $a$ to be
\begin{equation}
\label{eq:quantum_dimension}
d_a = \left| [F^{a \bar{a} a}_a]_{00} \right|^{-1}
\end{equation}
and $\varkappa_{a}$ is a phase. It follows that $d_0 =1$, and
\begin{equation}
\label{eq:loop=d}
d_{a}=d_{\bar{a}} =
\pspicture[shift=-0.35](-0.08,0.25)(1.55,1.25)
  \small
  \psarc[linewidth=0.9pt,linecolor=black,arrows=<-,arrowscale=1.5,
    arrowinset=0.15] (0.8,0.7){0.5}{165}{363}
  \psarc[linewidth=0.9pt,linecolor=black,border=0pt]
(0.8,0.7){0.5}{0}{170}
  \rput[bl]{0}(-0.03,0.55){$a$}
 \endpspicture
.
\end{equation}
Here we have introduced the convention of smoothing out the charge $a$ line at $|a,\bar{a};0\rangle$ vertices to form a ``cup'' when we remove the vacuum charge $0$ line, and similarly forming a ``cap'' from $\langle a,\bar{a};0 |$.

We also define the total quantum dimension of $\mathcal{C}$ to be
\begin{equation}
\mathcal{D} = \sqrt{ \sum_{a \in \mathcal{C}} d_{a}^{2} } .
\end{equation}

In the diagrammatic formalism, reversing the orientation of a line is equivalent to conjugating the charge labeling it, i.e.
\begin{equation}
\pspicture[shift=-0.7](0.4,-0.1)(1.05,1.5)
  \small
  \psset{linewidth=0.9pt,linecolor=black,arrowscale=1.5,arrowinset=0.15}
  \psline(0.6,0.1)(0.6,1.3)
  \psline{->}(0.6,0.1)(0.6,0.9)
  \rput[bl]{0}(0.75,0.41){$a$}
  \endpspicture
=
\pspicture[shift=-0.7](0.4,-0.1)(1.05,1.5)
  \small
  \psset{linewidth=0.9pt,linecolor=black,arrowscale=1.5,arrowinset=0.15}
  \psline(0.6,0.1)(0.6,1.3)
  \psline{-<}(0.6,0.1)(0.6,0.8)
  \rput[bl]{0}(0.75,0.41){$\bar{a}$}
  \endpspicture
.
\end{equation}

Isotopy invariance is essentially the ability to introduce and remove bends in a line. Bending a line horizontally (so that the line always
flows upward) is trivial (in that it utilizes the canonical isomorphisms of adding/removing vacuum lines), but a complication arises when a line is bent
vertically. To understand this, consider the $F$-move associated with this type of bending%
\begin{equation}
\pspicture[shift=-1.15](-1.2,-1.3)(1.2,1.3)
  \small
\pspolygon[linecolor=black,linewidth=0.8pt,linestyle=dotted](0,0.8)(0.8,0)
    (0,-0.8)(-0.8,0)
  \psset{linewidth=0.9pt,linecolor=black,arrowscale=1.5,arrowinset=0.15}
  \psline(0,-0.8)(0.8,0)
  \psline{->}(0,-0.8)(0.5,-0.3)
  \psline(-0.8,0)(0,0.8)
  \psline{->}(-0.8,0)(-0.3,0.5)
  \psline(-0.4,-0.4)(0.4,0.4)
  \psline{->}(-0.4,-0.4)(0.1,0.1)
  \psline(0.8,0)(0.4,0.4)
  \psline{->}(0.8,0)(0.5,0.3)
  \psline(-0.4,-0.4)(-0.8,0)
  \psline{->}(-0.4,-0.4)(-0.7,-0.1)
  \psline(0,0.8)(0,1.2)
  \psline{->}(0,0.8)(0,1.13)
  \psline(0,-0.8)(0,-1.2)
  \psline{->}(0,-1.2)(0,-0.85)
  \rput[bl]{0}(-1.1,-0.2){$a$}
  \rput[bl]{0}(0.1,-0.2){$\bar{a}$}
  \rput[bl]{0}(0.9,-0.2){$a$}
  \rput[bl]{0}(0.25,0.6){$0$}
  \rput[tr]{0}(-0.2,-0.65){$0$}
  \endpspicture
= \varkappa_{a}
\pspicture[shift=-1.15](-0.2,-1.3)(0.5,1.3)
  \small
  \psset{linewidth=0.9pt,linecolor=black,arrowscale=1.5,arrowinset=0.15}
  \psline(0,-1.2)(0,1.2)
  \psline{->}(0,-1.2)(0,0.2)
  \rput[bl]{0}(0.2,-0.2){$a$}
  \endpspicture
.
\end{equation}
(Notice the vertex normalization comes into play here.) In general, the phase $\varkappa _{a} = \varkappa _{\bar{a}}^{\ast}$ is not equal to $1$, but for $a \neq \bar{a}$, it is gauge dependent and can be fixed to $1$ by a gauge choice. For $a = \bar{a}$, $\varkappa _{a} =\pm 1$ is a gauge invariant quantity, known as the Frobenius-Schur indicator. Thus, we see that one needs more than just diagrammatic vertex normalization to produce isotopy invariance
for this kind of bending. This can be dealt with using flags that keep track of nontrivial $\varkappa _{a}$ phases and unitary transformations (which can be defined in terms of the $F$-symbols) when the legs of a vertex are bent up or down, which can be used, for example, to prove the pivotal property. (We refer the reader to \Refs{Kitaev06a,Bonderson07b} for details.) It follows that the dimension of fusion/splitting spaces related by bending lines are equal, so
\begin{equation}
N_{ab}^{c} = N_{\bar{a}c}^{b}= N_{c \bar{b}}^{a} = N_{b \bar{c} }^{\bar{a}} = N_{\bar{c} a}^{\bar{b}} = N_{\bar{b} \bar{a} }^{\bar{c}}
.
\end{equation}

We can also define a diagrammatic trace of operators (known as the ``quantum trace'') by closing the diagram with loops that match the outgoing lines with the respective incoming lines at the same position%
\begin{equation}
\widetilde{\text{Tr}}X =
\widetilde{\text{Tr}}
\left[
\psscalebox{.7}{
 \pspicture[shift=-1.0](-1,-1.1)(1,1.1)
  \small
  \psframe[linewidth=0.9pt,linecolor=black,border=0](-0.8,-0.5)(0.8,0.5)
  \rput[bl]{0}(-0.15,-0.1){$X$}
  \rput[bl]{0}(-0.22,0.7){$\mathbf{\ldots}$}
  \rput[bl]{0}(-0.22,-0.75){$\mathbf{\ldots}$}
  \psset{linewidth=0.9pt,linecolor=black,arrowscale=1.5,arrowinset=0.15}
  \psline(0.6,0.5)(0.6,1)
  \psline(-0.6,0.5)(-0.6,1)
  \psline(0.6,-0.5)(0.6,-1)
  \psline(-0.6,-0.5)(-0.6,-1)
  \psline{->}(0.6,0.5)(0.6,0.9)
  \psline{->}(-0.6,0.5)(-0.6,0.9)
  \psline{-<}(0.6,-0.5)(0.6,-0.9)
  \psline{-<}(-0.6,-0.5)(-0.6,-0.9)
\endpspicture
}
\right]
= \sum_{a_1 , \ldots , a_n}
\psscalebox{.7}{
\pspicture[shift=-1.1](-1.0,-1.2)(2.3,1.1)
  \small
  \psframe[linewidth=0.9pt,linecolor=black,border=0](-0.8,-0.5)(0.8,0.5)
  \rput[bl]{0}(-0.15,-0.1){$X$}
  \rput[bl]{0}(-0.4,0.7){$\mathbf{\ldots}$}
  \rput[bl]{0}(-0.22,-0.75){$\mathbf{\ldots}$}
  \rput[bl]{0}(1.52,0){$\mathbf{\ldots}$}
  \psset{linewidth=0.9pt,linecolor=black,arrowscale=1.5,arrowinset=0.15}
  \psarc(1.0,0.5){0.4}{0}{180}
  \psarc(1.0,-0.5){0.4}{180}{360}
  \psarc(0,0.5){0.6}{90}{180}
  \psarc(0,-0.5){0.6}{180}{270}
  \psarc(1.5,0.5){0.6}{0}{90}
  \psarc(1.5,-0.5){0.6}{270}{360}
  \psline(1.4,-0.5)(1.4,0.5)
  \psline(0,1.1)(1.5,1.1)
  \psline(0,-1.1)(1.5,-1.1)
  \psline(2.1,-0.5)(2.1,0.5)
  \psline{->}(1.4,0.2)(1.4,-0.1)
  \psline{->}(2.1,0.2)(2.1,-0.1)
  \rput[bl](-1.07,0.6){$a_1$}
  \rput[bl](0.1,0.6){$a_n$}
 \endpspicture
}
.
\end{equation}%
Connecting the endpoints of two lines labeled by different topological charge values violates charge conservation, so such diagrams evaluate to zero. One can equivalently take the trace either by looping the lines around to the right (as shown above) or to the left (with their equality following from $d_{a} = d_{\bar{a}}$).

By taking the trace of $\openone_{ab}$ and using isotopy, together with Eqs.~(\ref{eq:inner_product}) and (\ref{eq:Id}), we obtain the important relation
\begin{equation}
\label{eq:d_relation}
d_a d_b = \sum_{c} N_{ab}^{c} d_c
.
\end{equation}

Let us define fusion matrices $\mathbf{N}_{a}$ using the fusion coefficients to be $\left[ \mathbf{N}_{a} \right]_{bc} =  N_{ba}^{c}$. We note that the bending relations indicate that $\mathbf{N}_{a}^{T} = \mathbf{N}_{\bar{a}}$. From Eq.~(\ref{eq:d_relation}), we see that the vector $\mathbf{v}$ with components $v_c = d_c / \mathcal{D}$ is a normalized eigenvector of each matrix $\mathbf{N}_a$ with corresponding eigenvalue $d_a$. Moreover, the Perron-Frobenius theorem assures us that $\mathbf{v}$ is the only eigenvector (up to overall multiplicative factors) of $\mathbf{N}_a$ with all positive components and that $d_a$ is the largest (in absolute value) eigenvalue of $\mathbf{N}_a$. Thus, the dimension of the state space asymptotically grows as powers of $d_a$ as one increases the number $n$ of $a$ quasiparticles, i.e. $\sum_{c} \dim V^{a\ldots a}_{c} = \sum_{c} \left[ \mathbf{N}_{a}^{n} \right]_{0c} \sim d_{a}^{n}$ as $n\rightarrow \infty$. If $d_{a}=1$, we call charge $a$ Abelian, which is equivalent to saying it has unique fusion with all other charges ($\sum_{c} N_{ab}^{c} =1$ for all $b$). Otherwise, $d_{a} > 1$ and we call it non-Abelian.

Given fusion rules specified by $N_{ab}^{c}$, we can define the corresponding Verlinde algebra spanned by elements ${\bf v}_a$ which satisfy ${\bf v}_{\bar{a}} = {\bf v}_a^{\dagger}$ and
\begin{equation}
{\bf v}_a {\bf v}_b = \sum_{c} N_{ab}^{c} {\bf v}_c
.
\end{equation}
Notice that ${\bf v}_a$ may be (faithfully) represented by $\mathbf{N}_{a}$.

\subsection{Braiding}
\label{sec:Braiding}

The theory described in the previous subsection defined a unitary fusion tensor category with positive-definite inner product. We now wish to introduce braiding. For this, we require the fusion algebra to also be commutative, i.e.
\begin{equation}
N_{ab}^{c}=N_{ba}^{c}
,
\end{equation}
so that the dimension of the state space is unaltered when the positions of anyons are interchanged.

We note that this, together with associativity, implies $\mathbf{N}_{a} \mathbf{N}_{b} = \mathbf{N}_{b} \mathbf{N}_{a}$, i.e. all of the fusion matrices commute with each other. Hence, the fusion matrices are also normal and simultaneously diagonalizable by a unitary matrix $\mathbf{P}$. Specifically, $\mathbf{N}_{a} = \mathbf{P} \Lambda^{(a)} \mathbf{P}^{-1}$, where $[\Lambda^{(a)}]_{bc} = \lambda^{(a)}_{b} \delta_{bc}$ and the eigenvalues are $\lambda^{(a)}_{b} = {P_{ab}}/{P_{0b}}$. The eigenvalues form the fusion characters of the Verlinde algebra, i.e. for each $b$ the map $\lambda_{b}: a \mapsto \lambda_{b}^{(a)}$ is a fusion character satisfying the relations
\begin{eqnarray}
\lambda^{(a)}_{e} \lambda^{(b)}_{e} &=& \sum_{c} N_{ab}^{c} \lambda^{(c)}_{e}, \\
\sum_{a} \lambda^{(a)}_{b} \lambda^{(a) \ast}_{c} &=& \delta_{bc} \left|P_{0b} \right|^{-2}
.
\end{eqnarray}
Moreover, we have the relation
\begin{equation}
N_{ab}^{c}= \sum_{x} \frac{P_{ax} P_{bx} P_{cx}^{\ast} }{P_{0x}}
.
\end{equation}

The counterclockwise braiding exchange operator of two anyons is represented diagrammatically by
\begin{equation}
R^{ab}=
\pspicture[shift=-0.55](-0.1,-0.2)(1.3,1.05)
\small
  \psset{linewidth=0.9pt,linecolor=black,arrowscale=1.5,arrowinset=0.15}
  \psline(0.96,0.05)(0.2,1)
  \psline{->}(0.96,0.05)(0.28,0.9)
  \psline(0.24,0.05)(1,1)
  \psline[border=2pt]{->}(0.24,0.05)(0.92,0.9)
  \rput[bl]{0}(-0.02,0.8){$a$}
  \rput[br]{0}(1.2,0.8){$b$}
  \endpspicture
=\sum\limits_{c,\mu ,\nu }\sqrt{\frac{d_{c}}{d_{a}d_{b}}}\left[
R_{c}^{ab}\right] _{\mu \nu }
\pspicture[shift=-0.7](-0.1,-0.6)(1.2,1)
  \small
  \psset{linewidth=0.9pt,linecolor=black,arrowscale=1.5,arrowinset=0.15}
  \psline{->}(0.7,0)(0.7,0.45)
  \psline(0.7,0)(0.7,0.55)
  \psline(0.7,0.55) (0.25,1)
  \psline{->}(0.7,0.55)(0.3,0.95)
  \psline(0.7,0.55) (1.15,1)
  \psline{->}(0.7,0.55)(1.1,0.95)
  \rput[bl]{0}(0.38,0.2){$c$}
  \rput[br]{0}(1.4,0.8){$b$}
  \rput[bl]{0}(0,0.8){$a$}
  \psline(0.7,0) (0.25,-0.45)
  \psline{-<}(0.7,0)(0.35,-0.35)
  \psline(0.7,0) (1.15,-0.45)
  \psline{-<}(0.7,0)(1.05,-0.35)
  \rput[br]{0}(1.4,-0.4){$a$}
  \rput[bl]{0}(0,-0.4){$b$}
\scriptsize
  \rput[bl]{0}(0.85,0.4){$\nu$}
  \rput[bl]{0}(0.85,-0.03){$\mu$}
\endpspicture
,
\end{equation}
where the $R$-symbols are the maps $R^{ab}_{c} : V^{ba}_{c} \rightarrow V^{ab}_{c}$ that result from exchanging two anyons of charges $b$ and $a$, respectively, which are in the charge $c$ fusion channel. This can be written as
\begin{equation}
\pspicture[shift=-0.65](-0.1,-0.2)(1.5,1.2)
  \small
  \psset{linewidth=0.9pt,linecolor=black,arrowscale=1.5,arrowinset=0.15}
  \psline{->}(0.7,0)(0.7,0.43)
  \psline(0.7,0)(0.7,0.5)
 \psarc(0.8,0.6732051){0.2}{120}{240}
 \psarc(0.6,0.6732051){0.2}{-60}{35}
  \psline (0.6134,0.896410)(0.267,1.09641)
  \psline{->}(0.6134,0.896410)(0.35359,1.04641)
  \psline(0.7,0.846410) (1.1330,1.096410)	
  \psline{->}(0.7,0.846410)(1.04641,1.04641)
  \rput[bl]{0}(0.4,0){$c$}
  \rput[br]{0}(1.35,0.85){$b$}
  \rput[bl]{0}(0.05,0.85){$a$}
 \scriptsize
  \rput[bl]{0}(0.82,0.35){$\mu$}
  \endpspicture
=\sum\limits_{\nu }\left[ R_{c}^{ab}\right] _{\mu \nu}
\pspicture[shift=-0.65](-0.1,-0.2)(1.5,1.2)
  \small
  \psset{linewidth=0.9pt,linecolor=black,arrowscale=1.5,arrowinset=0.15}
  \psline{->}(0.7,0)(0.7,0.45)
  \psline(0.7,0)(0.7,0.55)
  \psline(0.7,0.55) (0.25,1)
  \psline{->}(0.7,0.55)(0.3,0.95)
  \psline(0.7,0.55) (1.15,1)	
  \psline{->}(0.7,0.55)(1.1,0.95)
  \rput[bl]{0}(0.4,0){$c$}
  \rput[br]{0}(1.4,0.8){$b$}
  \rput[bl]{0}(0,0.8){$a$}
 \scriptsize
  \rput[bl]{0}(0.82,0.37){$\nu$}
  \endpspicture
.
\end{equation}
Similarly, the clockwise braiding exchange operator is
\begin{equation}
\left(R^{ab}\right)^{-1}=
\pspicture[shift=-0.55](-0.1,-0.2)(1.3,1.05)
\small
  \psset{linewidth=0.9pt,linecolor=black,arrowscale=1.5,arrowinset=0.15}
  \psline{->}(0.24,0.05)(0.92,0.9)
  \psline(0.24,0.05)(1,1)
  \psline(0.96,0.05)(0.2,1)
  \psline[border=2pt]{->}(0.96,0.05)(0.28,0.9)
  \rput[bl]{0}(-0.01,0.8){$b$}
  \rput[bl]{0}(1.06,0.8){$a$}
  \endpspicture
.
\end{equation}

\begin{figure*}[t!]
\begin{center}
\includegraphics[scale=0.88]{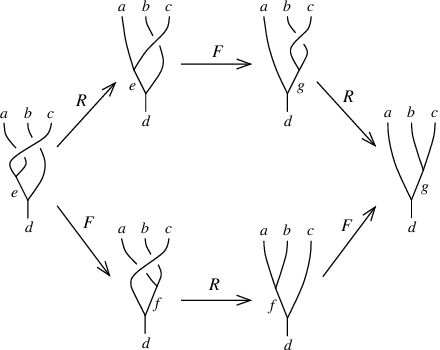}
\hspace{1cm}
\includegraphics[scale=0.88]{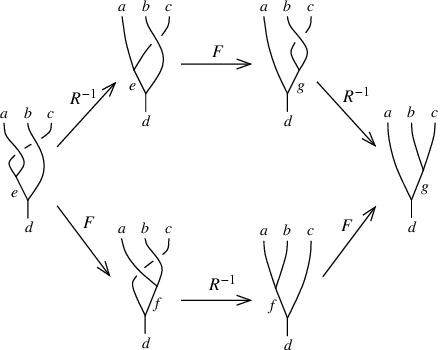}
\caption{The Hexagon equations enforce the condition that braiding is compatible
with fusion, in the sense that different sequences of $F$-moves and $R$-moves
from the same starting configuration to the same ending configuration give
the same result. Eqs. (\ref{eq:hexagon+}) and (\ref{eq:hexagon-}) are obtained
by imposing the condition that the above diagram commutes.}
\label{fig:hexagon}
\end{center}
\end{figure*}

In order for braiding to be compatible with fusion, we require that the two operations commute. Diagrammatically, this means we can freely slide lines over or under fusion/splitting vertices
\begin{equation}
\psscalebox{.6}{
\pspicture[shift=-1.15](-0.8,-0.2)(2.2,2.2)
  \small
  \psset{linewidth=0.9pt,linecolor=black,arrowscale=1.5,arrowinset=0.15}
  \psline{->}(0.7,0)(0.7,0.45)
  \psline(0.7,0)(0.7,1.55)
  \psline(0.7,1.55)(0.25,2)
  \psline{->}(0.7,1.55)(0.3,1.95)
  \psline(0.7,1.55) (1.15,2)	
  \psline{->}(0.7,1.55)(1.1,1.95)
  \psline[border=2pt](-0.65,0)(2.05,2)
  \psline{->}(-0.65,0)(0.025,0.5)
  \rput[bl]{0}(-0.4,0.6){$x$}
  \rput[bl]{0}(0.4,0){$c$}
  \rput[br]{0}(1.4,1.8){$b$}
  \rput[bl]{0}(0,1.8){$a$}
 \scriptsize
  \rput[bl]{0}(0.85,1.35){$\mu$}
  \endpspicture
}
=
\psscalebox{.6}{
\pspicture[shift=-1.15](-1,-0.2)(2.4,2.2)
  \small
  \psset{linewidth=0.9pt,linecolor=black,arrowscale=1.5,arrowinset=0.15}
  \psline{->}(0.7,0)(0.7,0.45)
  \psline(0.7,0)(0.7,0.55)
  \psline(0.7,0.55)(0.25,1)
  \psline(0.7,0.55)(1.15,1)	
  \psline(0.25,1)(0.25,2)
  \psline{->}(0.25,1)(0.25,1.9)
  \psline(1.15,1)(1.15,2)
  \psline{->}(1.15,1)(1.15,1.9)
  \psline[border=2pt](-0.65,0)(2.05,2)
  \psline{->}(-0.65,0)(0.025,0.5)
  \rput[bl]{0}(-0.4,0.6){$x$}
  \rput[bl]{0}(0.4,0){$c$}
  \rput[br]{0}(1.45,1.8){$b$}
  \rput[bl]{0}(-0.05,1.8){$a$}
 \scriptsize
  \rput[bl]{0}(0.85,0.35){$\mu$}
  \endpspicture
}
\end{equation}

\begin{equation}
\psscalebox{.6}{
\pspicture[shift=-1.15](-0.8,-0.2)(2.2,2.2)
  \small
  \psset{linewidth=0.9pt,linecolor=black,arrowscale=1.5,arrowinset=0.15}
  \psline(-0.65,0)(2.05,2)
  \psline[border=2pt](0.7,0)(0.7,1.55)
  \psline{->}(0.7,0)(0.7,0.45)
  \psline(0.7,1.55)(0.25,2)
  \psline{->}(0.7,1.55)(0.3,1.95)
  \psline(0.7,1.55) (1.15,2)	
  \psline{->}(0.7,1.55)(1.1,1.95)
  \psline{->}(-0.65,0)(0.025,0.5)
  \rput[bl]{0}(-0.4,0.6){$x$}
  \rput[bl]{0}(0.4,0){$c$}
  \rput[br]{0}(1.4,1.8){$b$}
  \rput[bl]{0}(0,1.8){$a$}
 \scriptsize
  \rput[bl]{0}(0.85,1.35){$\mu$}
  \endpspicture
}
=
\psscalebox{.6}{\pspicture[shift=-1.15](-1,-0.2)(2.4,2.2)
  \small
  \psset{linewidth=0.9pt,linecolor=black,arrowscale=1.5,arrowinset=0.15}
  \psline(-0.65,0)(2.05,2)
  \psline[border=2pt](0.7,0.55)(0.25,1)
  \psline[border=2pt](1.15,1)(1.15,2)
  \psline(0.7,0.55)(1.15,1)	
  \psline{->}(0.7,0)(0.7,0.45)
  \psline(0.7,0)(0.7,0.55)
  \psline(0.25,1)(0.25,2)
  \psline{->}(0.25,1)(0.25,1.9)
  \psline{->}(1.15,1)(1.15,1.9)
  \psline{->}(-0.65,0)(0.025,0.5)
  \rput[bl]{0}(-0.4,0.6){$x$}
  \rput[bl]{0}(0.4,0){$c$}
  \rput[br]{0}(1.45,1.8){$b$}
  \rput[bl]{0}(-0.05,1.8){$a$}
 \scriptsize
  \rput[bl]{0}(0.85,0.35){$\mu$}
  \endpspicture
}
.
\end{equation}
These relations imply the Yang-Baxter equations for braiding operators,
$R_{j,j+1} R_{j-1,j} R_{j,j+1} = R_{j-1,j} R_{j,j+1} R_{j-1,j}$, where $R_{j,j+1}$ is the operator that braids the strands in the $j$th and $(j+1)$th positions in the counterclockwise sense, which are equivalent to the property that lines can slide over braids, since the ability to freely slide lines over/under vertices allows lines to slide over/under braiding operators. Diagrammatically, this is written as
\begin{equation}
\psscalebox{.6}{
\pspicture[shift=-1](-0.7,-0.1)(1.8,2.3)
		\psbezier[linewidth=1pt](1,0)(1,0.6) (0,1.2)(0,2.2)
		\psbezier[linewidth=1pt,border=2pt](0,0) (0,0.6) (1,1.2)  (1,2.2)
		\psbezier[linewidth=1pt,border=2pt] (1.5,2.2) (1.2,1.5) (-0.4, 1.3) (-0.5,0)
		\endpspicture
}
=
\psscalebox{.6}{
\pspicture[shift=-1](-0.7,-0.1)(1.8,2.3)
		\psbezier[linewidth=1pt](1,0)(1,0.6) (0,1.2)(0,2.2)
		\psbezier[linewidth=1pt,border=2pt](0,0) (0,0.6) (1,1.2)  (1,2.2)
		\psbezier[linewidth=1pt,border=2pt] (1.5,2.2) (1.1,-0.3) (0, 1.1) (-0.5,0)
\endpspicture
}
.
\end{equation}

Requiring consistency between fusion and braiding, we find conditions that must be satisfied by the $F$-symbols and $R$-symbols, which may be expressed as the Hexagon equations
\begin{widetext}
\begin{eqnarray}
\sum_{\lambda ,\gamma }\left[ R_{e}^{ac}\right] _{\alpha \lambda }\left[
F_{d}^{acb}\right] _{\left( e,\lambda ,\beta \right) \left( g,\gamma, \nu
\right) }\left[ R_{g}^{bc}\right] _{\gamma \mu }
&=&\sum_{f,\sigma ,\delta ,\psi }\left[ F_{d}^{cab}\right] _{\left( e,\alpha
,\beta \right) \left( f ,\delta ,\sigma \right) }\left[ R_{d}^{fc}\right]
_{\sigma \psi }\left[ F_{d}^{abc}\right] _{\left( f,\delta ,\psi \right)
\left( g,\mu ,\nu \right) }
,
\label{eq:hexagon+}
\\
\sum_{\lambda ,\gamma }\left[ \left( R_{e}^{ca}\right) ^{-1}\right]
_{\alpha \lambda }\left[ F_{d}^{acb}\right] _{\left( e,\lambda ,\beta
\right) \left( g,\gamma, \nu \right) }\left[ \left( R_{g}^{cb}\right) ^{-1}\right] _{\gamma \mu }
&=&\sum_{f,\sigma ,\delta ,\psi }\left[ F_{d}^{cab}\right] _{\left( e,\alpha
,\beta \right) \left( f ,\delta,\sigma \right) }\left[ \left(
R_{d}^{cf}\right) ^{-1}\right] _{\sigma \psi }\left[ F_{d}^{abc}\right]
_{\left( f,\delta ,\psi \right) \left( g,\mu ,\nu \right) }
\label{eq:hexagon-}
.
\end{eqnarray}
\end{widetext}
These relations are represented diagrammatically in Fig.~\ref{fig:hexagon}. MacLane's coherence theorem~\cite{MacLane98} establishes that if the Pentagon equation and Hexagon equations are satisfied, then any two sequences of $F$-moves and $R$-moves (braiding) applied within an arbitrary diagram which start from the same state space and end in the same state space are equivalent, which is to say that fusion and braiding are consistent. The $F$-symbols and $R$-symbols completely specify a braided tensor category (BTC).

Given the trivial associativity of the vacuum charge $0$ ($F^{abc}_{d} = \openone$ when $a$, $b$, or $c=0$), the Hexagon equations imply that braiding with the vacuum is trivial, i.e. $R^{a0}_{a} = R^{0a}_{a} = \left( R^{a0}_{a} \right)^{-1} = \left(R^{0a}_{a}\right)^{-1} = 1$.

If we further require unitarity of the theory, then $\left(R^{ab}\right)^{-1}= \left( R^{ab} \right)^{\dag}$, which can be expressed in terms of $R$-symbols as $\left[ \left( R_{c}^{ab}\right)^{-1} \right] _{\mu \nu} =
\left[ R_{c}^{ab}\right] _{\nu \mu}^{\ast}$ (which are simply phases when $N^{c}_{ab}=1$).

An important quantity derived from braiding is the topological twist (or topological spin) of charge $a$
\begin{equation}
\theta _{a}=\theta _{\bar{a}}
=\sum\limits_{c,\mu } \frac{d_{c}}{d_{a}}\left[ R_{c}^{aa}\right] _{\mu \mu }
= \frac{1}{d_{a}}
\pspicture[shift=-0.5](-1.3,-0.6)(1.3,0.6)
\small
  \psset{linewidth=0.9pt,linecolor=black,arrowscale=1.5,arrowinset=0.15}
  \psarc[linewidth=0.9pt,linecolor=black] (0.7071,0.0){0.5}{-135}{135}
  \psarc[linewidth=0.9pt,linecolor=black] (-0.7071,0.0){0.5}{45}{315}
  \psline(-0.3536,0.3536)(0.3536,-0.3536)
  \psline[border=2.3pt](-0.3536,-0.3536)(0.3536,0.3536)
  \psline[border=2.3pt]{->}(-0.3536,-0.3536)(0.0,0.0)
  \rput[bl]{0}(-0.2,-0.5){$a$}
  \endpspicture
,
\end{equation}
which is a root of unity~\cite{Vafa88}. This can be used to show that the $R$-symbols satisfy the ``ribbon property''
\begin{equation}
\label{eq:ribbon}
\sum_{\lambda} \left[ R^{ab}_{c} \right]_{\mu \lambda} \left[R^{ba}_{c} \right]_{\lambda \nu} = \frac{\theta_{c}}{\theta_{a} \theta_{b}} \delta_{\mu \nu}
.
\end{equation}

Another important quantity is the topological $S$-matrix
\begin{equation}
S_{ab} =\mathcal{D}^{-1}\sum%
\limits_{c}N_{\bar{a} b}^{c}\frac{\theta _{c}}{\theta _{a}\theta _{b}}d_{c}
=\frac{1}{\mathcal{D}}
\pspicture[shift=-0.4](0.0,0.2)(2.6,1.3)
\small
  \psarc[linewidth=0.9pt,linecolor=black,arrows=<-,arrowscale=1.5,arrowinset=0.15] (1.6,0.7){0.5}{167}{373}
  \psarc[linewidth=0.9pt,linecolor=black,border=3pt,arrows=<-,arrowscale=1.5,arrowinset=0.15] (0.9,0.7){0.5}{167}{373}
  \psarc[linewidth=0.9pt,linecolor=black] (0.9,0.7){0.5}{0}{180}
  \psarc[linewidth=0.9pt,linecolor=black,border=3pt] (1.6,0.7){0.5}{45}{150}
  \psarc[linewidth=0.9pt,linecolor=black] (1.6,0.7){0.5}{0}{50}
  \psarc[linewidth=0.9pt,linecolor=black] (1.6,0.7){0.5}{145}{180}
  \rput[bl]{0}(0.1,0.45){$a$}
  \rput[bl]{0}(0.8,0.45){$b$}
  \endpspicture
.
\label{eqn:mtcs}
\end{equation}
It is clear that $S_{ab} = S_{ba} = S_{\bar{a} b}^{\ast}$ and $S_{0a} = d_a / \mathcal{D}$. A related invariant quantity
\begin{equation}
M_{ab}=\frac{S_{ab}^{\ast} S_{00}}{S_{0a} S_{0b}}
\end{equation}
is the monodromy scalar component, which plays an important role in anyonic interferometry~\cite{Bonderson06b,Bonderson07a,Bonderson07c} and which will show up later in the classification of symmetry fractionalizations and group extensions of categories. If $M_{ab} = e^{i \phi (a,b)}$ is a phase, then the braiding of $a$ with $b$ is Abelian in the sense that
\begin{equation}
  \pspicture[shift=-0.6](0.0,-0.05)(1.1,1.45)
  \small
  \psarc[linewidth=0.9pt,linecolor=black,border=0pt] (0.8,0.7){0.4}{120}{225}
  \psarc[linewidth=0.9pt,linecolor=black,arrows=<-,arrowscale=1.4,
    arrowinset=0.15] (0.8,0.7){0.4}{165}{225}
  \psarc[linewidth=0.9pt,linecolor=black,border=0pt] (0.4,0.7){0.4}{-60}{45}
  \psarc[linewidth=0.9pt,linecolor=black,arrows=->,arrowscale=1.4,
    arrowinset=0.15] (0.4,0.7){0.4}{-60}{15}
  \psarc[linewidth=0.9pt,linecolor=black,border=0pt]
(0.8,1.39282){0.4}{180}{225}
  \psarc[linewidth=0.9pt,linecolor=black,border=0pt]
(0.4,1.39282){0.4}{-60}{0}
  \psarc[linewidth=0.9pt,linecolor=black,border=0pt]
(0.8,0.00718){0.4}{120}{180}
  \psarc[linewidth=0.9pt,linecolor=black,border=0pt]
(0.4,0.00718){0.4}{0}{45}
  \rput[bl]{0}(0.1,1.2){$a$}
  \rput[br]{0}(1.06,1.2){$b$}
  \endpspicture
= e^{i \phi (a,b)}
\pspicture[shift=-0.6](-0.2,-0.45)(1.0,1.1)
  \small
  \psset{linewidth=0.9pt,linecolor=black,arrowscale=1.5,arrowinset=0.15}
  \psline(0.3,-0.4)(0.3,1)
  \psline{->}(0.3,-0.4)(0.3,0.50)
  \psline(0.7,-0.4)(0.7,1)
  \psline{->}(0.7,-0.4)(0.7,0.50)
  \rput[br]{0}(0.96,0.8){$b$}
  \rput[bl]{0}(0,0.8){$a$}
  \endpspicture
.
\end{equation}
Moreover, when this is true, it follows that $M_{ab} M_{ac} = M_{ae}$ whenever $N_{bc}^{e}\neq 0$.

An important property that follows from the definition of the $S$-matrix is the ability to remove closed loops that encircle other line, which is done by acquiring an amplitude determined by the $S$-matrix. In particular, we have
\begin{equation}
\pspicture[shift=-0.55](-0.25,-0.1)(0.9,1.3)
\small
  \psset{linewidth=0.9pt,linecolor=black,arrowscale=1.5,arrowinset=0.15}
  \psline(0.4,0)(0.4,0.22)
  \psline(0.4,0.45)(0.4,1.2)
  \psellipse[linewidth=0.9pt,linecolor=black,border=0](0.4,0.5)(0.4,0.18)
  \psset{linewidth=0.9pt,linecolor=black,arrowscale=1.4,arrowinset=0.15}
  \psline{-<}(0.2,0.37)(0.3,0.34)
\psline[linewidth=0.9pt,linecolor=black,border=2.5pt,arrows=->,arrowscale=1.5,
arrowinset=0.15](0.4,0.5)(0.4,1.1)
  \rput[bl]{0}(-0.15,0.15){$a$}
  \rput[tl]{0}(0.55,1.2){$b$}
  \endpspicture
=\frac{S_{ab}}{S_{0b}}
\pspicture[shift=-0.55](0.05,-0.1)(1,1.3)
\small
  \psset{linewidth=0.9pt,linecolor=black,arrowscale=1.5,arrowinset=0.15}
  \psline(0.4,0)(0.4,1.2)
\psline[linewidth=0.9pt,linecolor=black,arrows=->,arrowscale=1.5,
arrowinset=0.15](0.4,0.5)(0.4,0.9)
  \rput[tl]{0}(0.52,1.0){$b$}
  \endpspicture
\label{eq:loopaway}
\end{equation}%
which can be verified by taking the trace of both sides, closing the $b$ charge line into a loop.

Using Eq.~(\ref{eq:loopaway}) for a diagram with two loops of topological charge $a$ and $b$, respectively, linked on a line of topological charge $x$, together with Eqs.~(\ref{eq:inner_product}) and (\ref{eq:Id}) and isotopy, we obtain the important relation
\begin{equation}
\frac{S_{ax}}{S_{0x}} \frac{S_{bx}}{S_{0x}} = \sum_{c} N_{ab}^{c} \frac{S_{cx}}{S_{0x}}
.
\label{eq:preVerlinde}
\end{equation}
This relation shows that $\lambda^{(a)}_{[x]} = {S_{a x}}/{S_{0 x}}$ is a character of the Verlinde algebra. Here, we wrote $[x]$ to indicate an equivalence class of topological charges that correspond to the same character, reflecting the fact that the $S$-matrix may be degenerate.

When the $S$-matrix is non-degenerate it is unitary, and this is equivalent to the condition that braiding is non-degenerate, which means that for each topological charge $a\neq 0$ there is some charge $b$ such that $R_{ab}R_{ba} \neq \openone_{ab}$.

Indeed, when the $S$-matrix is unitary, the equivalence classes $[x]$ of topological charges corresponding to the same Verlinde algebra character are singletons and all the fusion characters of the Verlinde algebra are specified by the $S$-matrix and given by $\lambda^{(a)}_{x} = {S_{a x}}/{S_{0 x}}$. In this case, we can also write $P_{ab} = S_{ab}$, which is often phrased as ``the $S$-matrix diagonalizes the fusion rules.'' In this case, we can use the inverse of the $S$-matrix with Eq.~(\ref{eq:preVerlinde}) to determine the fusion rules from the $S$-matrix, as specified by the Verlinde formula~\cite{Verlinde88}
\begin{equation}
N_{ab}^{c} = \sum_{x\in\mathcal{C}} \frac{S_{ax} S_{bx} S_{cx}^{\ast}}{S_{0x}}
.
\end{equation}

When the $S$-matrix is unitary, the braided tensor category is called a modular tensor category (MTC). Such theories can be consistently defined for 2D manifolds of arbitrary genus and are related to $(2+1)$D TQFTs. In this case, the $S$-matrix together with the $T$-matrix, $T_{ab} = \theta_a \delta_{ab}$, and the charge conjugation matrix $C_{ab} =\delta_{a \bar{b}}$ obey the modular relations
\begin{equation}
(ST)^3 = \Theta C, \quad S^2 = C, \quad C^2 = \openone
\end{equation}
where
\begin{equation}
\Theta = \frac{1}{\mathcal{D}} \sum_{a\in \mathcal{C}} d_a^2 \theta_a = e^{i \frac{2\pi}{8} c_{-}}
\end{equation}
is a root of unity and $c_{-} \equiv c - \bar{c}$ is the chiral central charge. These correspond to the TQFT's projective representation of the respective modular transformations on a torus.

Another useful property of a UMTC is that, if a given topological charge $a$ has Abelian braiding with all other charges, i.e. if $M_{ab} = e^{i \phi (a,b)}$ is a phase for all charges $b\in \mathcal{C}$, then $a$ is Abelian in the sense that it has $d_a =1$ (and hence Abelian fusion and associativity). This follows from unitarity of the $S$-matrix, which implies that
\begin{eqnarray}
1 &=& \sum_{b} \left| S_{ab} \right|^2 = \sum_{b} \left| \frac{S_{0a} S_{0b}}{S_{00}} M_{ab} \right|^2 \notag \\
&=& \sum_{b} \left| \frac{d_{a} d_{b}}{\mathcal{D}} e^{i \phi (a,b)} \right|^2 = d_{a}^{2}
.
\label{eq:MphaseAbelian}
\end{eqnarray}
In other words, non-Abelian topological charges (those with $d_a>1$) necessarily have non-Abelian braiding in a UMTC.

Finally, we establish the following property for MTCs, which will be very useful for establishing the classification of symmetry fractionalization. If there are phase factors $e^{i \phi_{a}}$ (defined for all charge values) that satisfy the relation
\begin{equation}
e^{i \phi_{a}} e^{i \phi_{b}}= e^{i \phi_{c}}
\end{equation}
whenever $N_{ab}^{c} \neq 0$, then it must be the case that
\begin{equation}
e^{i \phi_{a}} = M_{a e}^{\ast}
\end{equation}
for some Abelian topological charge $e$. To verify this claim, we write $\lambda^{(a)} = d_a e^{i \phi_{a}}$ and notice that
\begin{equation}
\lambda^{(a)} \lambda^{(b)} =  \sum_{c} N_{ab}^{c} \lambda^{(c)}
.
\end{equation}
Hence, it is a fusion character and must be given by $\lambda^{(a)} = {S_{a e}}/{S_{0 e}}$ for some topological charge $e$. Thus, we have
\begin{equation}
e^{i \phi_{a}} = \frac{\lambda^{(a)}}{d_a} = \frac{S_{a e } S_{00}}{S_{0 e} S_{0 a }}= M_{a e}^{\ast}
,
\end{equation}
and since this makes $M_{a e}$ a phase for all values of $a$, it follows that $e $ must be an Abelian topological charge. In this case, $M_{a e}^{\ast} = S_{a e } / S_{0 a }$.

\subsection{Gauge Transformations}
\label{sec:Gauge_Trans}

Distinct sets of $F$-symbols and $R$-symbols describe equivalent theories if they can be related by a gauge transformation given by unitary transformations acting on the fusion/splitting state spaces $V_{c}^{ab}$ and $V^{c}_{ab}$, which can be though of as a redefinition of the basis states as
\begin{equation}
\label{eq:gauge}
\widetilde{ \left| a,b;c,\mu \right\rangle } = \sum_{\mu'} \left[\Gamma^{ab}_{c}\right]_{\mu \mu'} \left| a,b;c,\mu' \right\rangle
\end{equation}
where $\Gamma^{ab}_{c}$ is the unitary transformation. Such gauge transformations modify the $F$-symbols as
\begin{widetext}
\begin{equation}
\label{eq:gauge_F}
\left[\widetilde{F}_d^{abc}\right]_{(e,\alpha,\beta)(f,\mu,\nu)} = \sum_{\alpha', \beta', \mu' ,\nu'} \left[\Gamma^{ab}_{e}\right]_{\alpha \alpha'} \left[\Gamma^{ec}_{d}\right]_{\beta \beta'}  \left[F_d^{abc}\right]_{(e,\alpha',\beta')(f,\mu',\nu')} \left[\left(\Gamma^{bc}_{f} \right)^{-1} \right]_{\mu' \mu}  \left[ (\Gamma^{af}_{d} )^{-1} \right]_{\nu' \nu}
\end{equation}
\end{widetext}
and the $R$-symbols as
\begin{equation}
\left[ \widetilde{R}_{c}^{ab} \right]_{\mu \nu} = \sum_{\mu',\nu'} \left[\Gamma^{ba}_{c} \right]_{\mu \mu'}  \left[ R_{c}^{ab} \right]_{\mu' \nu'} \left[\left( \Gamma^{ab}_{c} \right)^{-1} \right]_{\nu' \nu}
.
\end{equation}
One must be careful not to use the gauge freedom associated with $\Gamma^{a 0}_{a}$ and $\Gamma^{0 b}_{b}$ to ensure that fusion and braiding with the vacuum $0$ remain trivial. More specifically, one should fix $\Gamma^{a 0}_{a}=\Gamma^{0 b}_{b}=\Gamma^{0 0}_{0}$. (One can think of this as respecting the canonical isomorphisms that allow one to freely add and remove vacuum lines. Alternatively, one could allow the use of these gauge factors and compensate by similarly modifying the canonical isomorphisms.)
It is often useful to consider quantities of the anyon model that are invariant under such gauge transformation. The most relevant gauge invariant quantities are the quantum dimensions $d_a$ and topological twist factors $\theta_a$, since these, together with the fusion coefficients $N_{ab}^{c}$, usually uniquely specify the theory (there are no known counterexamples).

\section{Symmetry of Topological Phases}
\label{sec:Symmetry}

We would like to consider a system that realizes a topological phase described by a UMTC $\mathcal{C}$ and which has a global unitary or anti-unitary symmetry of the microscopic Hamiltonian described by a group $G$. In this section, we do not require $G$ to be discrete, nor do we assume that the symmetry is on-site.
In order to characterize the interplay of symmetry and topological order, we first define the notion of the ``topological symmetry'' of $\mathcal{C}$, which is independent of the group $G$. We then consider the action of the global symmetry on the topological properties through its relation to the topological symmetry (via a homomorphism from the global symmetry group to the topological symmetry group).

\subsection{Topological Symmetry}
\label{sec:Topological_Symmetry}

The symmetries of a category $\mathcal{C}$ are described by invertible maps $\varphi:\mathcal{C} \rightarrow \mathcal{C}$ from the category to itself. Each such map $\varphi$ can be classified according to whether it is unitary or anti-unitary, and whether it preserves or reverses the spatial parity. We will first consider unitary, parity-preserving symmetries. Such maps are called auto-equivalences, or braided auto-equivalences for a BTC, and may permute the topological charge labels
\begin{equation}
\varphi(a) = a',
\end{equation}
in such a way that all of the topological properties are left invariant. In particular, the vacuum must always be left invariant under symmetry, so $0' = 0$, and gauge invariant quantities will be left invariant under these permutations of topological charge, so that
\begin{eqnarray}
\label{eq:phi_N'}
N_{a' b'}^{c'} &=& N_{a b}^{c} \\
\label{eq:phi_d}
d_{a'} &=& d_a \\
\label{eq:phi_theta}
\theta_{a'} &=& \theta_{a} \\
\label{eq:phi_S}
S_{a' b'} &=& S_{a b}
\end{eqnarray}
under auto-equivalence maps.

Quantities in the theory that are not gauge invariant must be left invariant by auto-equivalence maps, up to some gauge transformation. At a more detailed level, an auto-equivalence $\varphi$ maps basis state vectors of fusion/splitting spaces to (possibly different) basis state vectors of the corresponding fusion/splitting spaces
\begin{eqnarray}
\varphi\left( \left| a,b;c,\mu \right\rangle \right) &=& \widetilde{ \left| a',b';c',\mu \right\rangle } \notag \\
&=& \sum_{\mu'} \left[u^{a'b'}_{c'}\right]_{\mu \mu'} \left| a',b';c',\mu' \right\rangle
,
\label{eq:phi_vertex}
\end{eqnarray}
where $\left[u^{a'b'}_{c'}\right]$ is a unitary transformation that is included so that the map will leave the basic data exactly invariant, rather than just gauge equivalent to their original values. Notice that this mapping to new basis states is generally the same as applying a permutation of labels together with a gauge transformation, so we have used a similar notation to that of the previous section describing fusion/splitting vertex basis gauge transformations.

Under such mappings of the fusion/splitting basis states, the basic data map to
\begin{eqnarray}
\label{eq:phi_N}
\varphi \left( N_{ab}^{c} \right) &=& N_{a' b'}^{c'} = N_{a b}^{c} \\
\label{eq:phi_F}
\varphi \left( \left[{F}_d^{abc}\right]_{(e,\alpha,\beta)(f,\mu,\nu)} \right) &=& \left[\widetilde{F}_{d'}^{a'b'c'}\right]_{(e',\alpha,\beta)(f',\mu,\nu)} \notag \\
&=& \left[F_{d}^{abc}\right]_{(e,\alpha,\beta)(f,\mu,\nu)} \\
\label{eq:phi_R}
\varphi \left( \left[{R}_{c}^{ab} \right]_{\mu \nu} \right) &=& \left[ \widetilde{R}_{c'}^{a'b'} \right]_{\mu \nu} = \left[ R_{c}^{ab} \right]_{\mu \nu}
.
\end{eqnarray}
We see that this would generally result in gauge equivalent values of the $F$-symbols and $R$-symbols without the factors $u^{a'b'}_{c'}$, but including these factors in the definition of symmetry maps gives the stronger condition that the $F$-symbols and $R$-symbols are left exactly invariant.

The collection of all such maps $\varphi$ that leave all properties of $\mathcal{C}$ invariant form the set of braided auto-equivalences of $\mathcal{C}$. However, there is redundancy in these maps given by the ``natural isomorphisms,'' which, in this context, are the braided auto-equivalence maps of the form
\begin{eqnarray}
\Upsilon \left( a \right) &=& a \\
\Upsilon \left( \left| a,b;c,\mu \right\rangle \right) &=& \frac{ \gamma_a \gamma_b}{\gamma_c} \left| a,b;c,\mu \right\rangle
\label{eq:tau_gamma}
,
\end{eqnarray}
for some phases $\gamma_a$. It is straightforward to see that such maps always leave all the basic data exactly invariant. Hence, one can think of these natural isomorphisms as vertex basis gauge transformations of the form $[ \Gamma_{c}^{ab} ]_{\mu \nu}= \frac{ \gamma_a \gamma_b}{\gamma_c} \delta_{\mu \nu}$, which leave the basic data unchanged.~\footnote{In fact, we believe that all vertex basis gauge transformations of a BTC that leave the basic data unchanged must take this form and thus they are actually natural isomorphisms, so that the two concepts are synonymous for BTCs. This property is equivalent to the statement that specifying the permutation of topological charges $\varphi (a)$ for an auto-equivalence map uniquely determines the associated unitary transformation $u^{ab}_{c}$ up to natural isomorphism. We assume that this property is true for all BTCs in this paper. If it were not true, statements of equivalence up to natural isomorphism would need to be modified to hold up to braided auto-equivalences that leave the basic data invariant.
\newline
It is straightforward to verify that this property is true for some simple non-Abelian BTCs, such as the Fibonacci and Ising BTCs. We can also prove it is true for a general Abelian BTC in the following way. Let us view the topological charges as elements of an Abelian group $\mathcal{A}$ whose multiplication is defined by the fusion rules and the vertex basis gauge transformation phase factors $\Gamma^{ab}_{a \times b}$ as 2-cochains with $\text{U}(1)$ coefficients. The condition that a gauge transformation leaves the $F$-symbols unchanged is simply the cocycle condition $\Gamma^{ab}\Gamma^{(a\times b)c} = \Gamma^{a(b \times c)}\Gamma^{bc}$. The condition that a gauge transformation is a natural isomorphism is simply the coboundary condition that $\Gamma^{ab} = \frac{\gamma_{a} \gamma_{b}}{ \gamma_{a\times b}}$ for some 1-cochain $\gamma_{a}$. It follows that taking the quotient of gauge transformations that leave the $F$-symbols unchanged by those that are natural isomorphisms results in the 2nd cohomology group $H^{2}(\mathcal{A}, \text{U}(1))$, which is known to classify the projective representation of the group $\mathcal{A}$. The condition that a gauge transformation leaves the $R$-symbols unchanged, i.e. $\Gamma^{ab} = \Gamma^{ba}$, implies that multiplication in the corresponding projective irrep $\pi$ (for which $\Gamma^{ab}$ is the factor set) is strictly commutative, i.e. $\pi (a) \pi(b) = \Gamma^{ab} \pi (a\times b)= \Gamma^{ba} \pi (b \times a) =\pi (b) \pi(a)$. By Shur's Lemma, it follows that $\pi (a)$ are scalars for all $a\in\mathcal{A}$, and we can thus write $\Gamma^{ab} = \frac{\pi (a) \pi(b)}{\pi (a \times b)}$, which shows that the vertex basis gauge transformation is a natural isomorphism.}

Consequently, we wish to consider braided auto-equivalence maps as equivalent if they are related by a natural isomorphism, and doing so defines a group, which we denote as Aut$_{0,0}(\mathcal{C})$. (The $0,0$ here indicates unitary and parity preserving, as we will further explain.) In particular, if $\check{\varphi} = \Upsilon \circ \varphi$ for a natural isomorphism $\Upsilon$, then the braided auto-equivalence maps $\check{\varphi}$ and $\varphi$ represent the same equivalence class $[\check{\varphi}] = [\varphi]$. In this way, group multiplication in Aut$_{0,0}(\mathcal{C})$ is defined by composition up to natural isomorphism $[\varphi_1] \cdot [\varphi_2] = [\varphi_1 \circ \varphi_2]$. In other words, $[\varphi_3] = [\varphi_1] \cdot [\varphi_2]$  if for any representatives $\varphi_1$, $\varphi_2$, and $\varphi_3$ of the corresponding equivalence classes, there are natural isomorphisms $\Upsilon_1$, $\Upsilon_2$, and $\Upsilon_3$ such that $\Upsilon_3 \circ \varphi_3 = \Upsilon_1 \circ \varphi_1 \circ \Upsilon_2 \circ \varphi_2$, or, equivalently, if there is a natural isomorphism $\kappa$ such that $\varphi_{3} = \kappa \circ \varphi_1 \circ \varphi_2$. (These definitions are related by $\kappa = \Upsilon_{3}^{-1} \circ \Upsilon_1 \circ \varphi_1 \circ \Upsilon_2 \circ \varphi_{1}^{-1}$.)

There is yet another level of redundancy that arises in the decomposition of the natural isomorphisms into topological charge dependent phase factors, as in Eq.~(\ref{eq:tau_gamma}). Specifically, there is freedom to equivalently choose
\begin{eqnarray}
\Upsilon \left( \left| a,b;c,\mu \right\rangle \right) &=& \frac{ \breve{\gamma}_{a} \breve{\gamma}_{b} }{ \breve{\gamma}_{c} } \left| a,b;c,\mu \right\rangle \\
 \breve{\gamma}_{a} &=&  \zeta_{a} \gamma_{a}
\label{eq:vgamma}
,
\end{eqnarray}
for phases $\zeta_{a}$ that satisfy $\zeta_{a} \zeta_{b} = \zeta_{c}$ whenever $N_{ab}^{c} \neq 0$. In other words, the phase factors $\zeta_{a}$ that obey this condition provide a way of decomposing the completely trivial natural isomorphism $\Upsilon = \openone$ into topological charge dependent phase factors. As explained at the end of Sec.~\ref{sec:Braiding}, phase factors that obey this condition are related to some Abelian topological charge $z$ through the relation
\begin{equation}
\zeta_{a} = M^{\ast}_{a z}
.
\end{equation}
As such, this redundancy of natural isomorphisms between braided auto-equivalence maps (the natural isomorphisms themselves being a redundancy of the braided auto-equivalences) is classified by the subset $\mathcal{A} \subset \mathcal{C}$ of Abelian topological charges of the UMTC $\mathcal{C}$, which can also be considered an Abelian group where multiplication in this group is given by the fusion rules.~\footnote{In mathematical parlance, the braided auto-equivalence maps $\varphi$ are $1$-automorphism functors, the natural isomorphisms $\Upsilon$ are $2$-isomorphisms between the auto-equivalence functors, and the decomposition freedom of natural isomorphisms (given by the phase factors $\zeta_{a}$) are the automorphisms of the identity functor.}

We may also consider anti-unitary symmetries of the BTC $\mathcal{C}$, which we called braided anti-auto-equivalences. These were previously examined in the context of time-reversal symmetries in \Refs{Fidkowski13,Bonderson13d}. For anti-unitary symmetries, the map $\varphi$ is anti-unitary, which means it is a bijective, anti-linear map, i.e.
\begin{equation}
\varphi\left( C_{\alpha} | \alpha \rangle + C_{\beta} | \beta \rangle \right) =  C_{\alpha}^{\ast} \, \varphi\left( | \alpha \rangle \right)  + C_{\beta}^{\ast} \, \varphi\left( | \beta \rangle \right),
\end{equation}
for any states $| \alpha \rangle$ and $| \beta \rangle$ and complex numbers $C_{\alpha},C_{\beta} \in \mathbb{C}$, that also obeys the condition
\begin{equation}
\langle \varphi(\alpha) |  \varphi(\beta) \rangle =  \langle \alpha |  \beta \rangle ^{\ast}
.
\end{equation}
Any anti-unitary operator $A$ can be written as $A=UK$, where $U$ is a unitary operator and $K$ is the complex conjugation operator. Its inverse is $A^{-1}=A^{\dagger}=K U^{-1} = K U^{\dagger}$.

The vertex basis states transform as in Eq.~(\ref{eq:phi_vertex}) when $\varphi$ is anti-unitary, though any (complex-valued) coefficients in front of such states would be complex conjugated. Under such anti-auto-equivalence mappings of the fusion/splitting basis states, the basic data map to
\begin{eqnarray}
\label{eq:phi_N_anti}
\varphi \left( N_{ab}^{c} \right) &=& N_{a' b'}^{c'} = N_{a b}^{c} \\
\label{eq:phi_F_anti}
\varphi \left( \left[{F}_d^{abc}\right]_{(e,\alpha,\beta)(f,\mu,\nu)} \right) &=& \left[\widetilde{F}_{d'}^{a'b'c'}\right]_{(e',\alpha,\beta)(f',\mu,\nu)} \notag \\
&=& \left[F_{d}^{abc}\right]_{(e,\alpha,\beta)(f,\mu,\nu)}^{\ast} \\
\label{eq:phi_R_anti}
\varphi \left( \left[{R}_{c}^{ab} \right]_{\mu \nu} \right) &=& \left[ \widetilde{R}_{c'}^{a'b'} \right]_{\mu \nu} = \left[ R_{c}^{ab} \right]_{\mu \nu}^{\ast}
.
\end{eqnarray}
Anti-unitarity similarly introduces complex conjugation for the gauge invariant quantities, so that
\begin{eqnarray}
\theta_{a'} &=& \theta_{a}^{\ast} \\
S_{a' b'} &=& S_{a b}^{\ast}
.
\end{eqnarray}

As mentioned above, when including both unitary and anti-unitary topological symmetries (braided auto-equivalences), it is useful to define a function
\begin{equation}
q\left( \varphi \right) = \left\{
\begin{array}{lll}
0  & & \text{ if $\varphi$ is unitary} \\
1  & & \text{ if $\varphi$ is anti-unitary}
\end{array}
\right.
,
\end{equation}
which specifies when a braided auto-equivalence map is unitary or anti-unitary. When we form equivalence classes of maps related by natural isomorphism, the combined set of unitary and anti-unitary topological symmetries is again a group. The function $q$ provides a homomorphism from this group to $\mathbb{Z}_2$, i.e. $q\left( [\varphi_1 \circ \varphi_2] \right) = q\left( [\varphi_1] \right) q\left( [\varphi_2] \right)$, since the composition of a unitary transformation and an anti-unitary transformation is anti-unitary and the composition between two anti-unitary transformations is unitary. This homomorphism defines a $\mathbb{Z}_2$-grading of the group of unitary and anti-unitary auto-equivalences.

We can also include spatial parity symmetry, which is a unitary symmetry, by introducing an additional $\mathbb{Z}_2$ grading structure. The action of spatial parity on the topological state space and basic data is a somewhat complicated matter, because the quasiparticles may, in principle, exist in a 2D surface of arbitrary topology, and the action of parity depends on both how one chooses to linearly order the quasiparticles for the purposes of writing a fusion tree decomposition of the states, and what is the line across which one performs the parity reflection. The full details of such parity transformations will not be used in this paper, so we will not present them here. However, it is simple to state the transformation of the gauge invariant quantities
\begin{eqnarray}
\label{eq:phi_N_parity}
\varphi \left( N_{ab}^{c} \right) &=& N_{a' b'}^{c'} = N_{a b}^{c} \\
\label{eq:phi_theta_parity}
\varphi \left( \theta_{a} \right) &=& \theta_{a'} = \theta_{a}^{\ast} \\
\varphi \left( S_{a b} \right) &=&  S_{a' b'} = S_{a b}^{\ast}
,
\end{eqnarray}
which holds for any parity reflection transformation, regardless of the details of quasiparticle ordering or reflection line.

With this in mind, we introduce the function
\begin{equation}
p\left( \varphi \right) = \left\{
\begin{array}{lll}
0  & & \text{ if $\varphi$ is spatial parity even} \\
1  & & \text{ if $\varphi$ is spatial parity odd}
\end{array}
\right.
.
\end{equation}
Forming equivalence classes of symmetry transformations under natural isomorphisms, this provides another $\mathbb{Z}_2$-grading of the resulting group, since the composition of two parity reversing (odd) transformations is obviously parity preserving (even), and thus $p\left( [\varphi_1 \circ \varphi_2] \right) = p\left( [\varphi_1] \right) p\left( [\varphi_2] \right)$.

We write the full group of quantum symmetries of the topological theory as
\begin{equation}
\text{Aut}(\mathcal{C}) = \bigsqcup_{q,p \in \{0,1\}} \text{Aut}_{q,p}(\mathcal{C})
,
\end{equation}
where $\text{Aut}_{q,p}(\mathcal{C})$ is the set of equivalence classes (under natural isomorphisms) of braided auto-equivalence maps that are unitary for $q=0$ or anti-unitary for $q=1$, and parity preserving for $p=0$ or parity reversing for $p=1$.

We consider Aut$(\mathcal{C})$ to be the \emph{topological symmetry} group of $\mathcal{C}$, because it describes the symmetry of the emergent topological quantum numbers of the topological phase, as described by $\mathcal{C}$. This is in contrast to and independent of any global symmetry of the underlying physical system, as described by the microscopic Hamiltonian.

\subsection{Global Symmetry Action on the Topological State Space}
\label{sec:Global_Symmetry}

We now consider the case where a physical system that realizes a topological phase described by the UMTC $\mathcal{C}$, has a
global symmetry group $G$ of the microscopic Hamiltonian. We restrict our attention to the case where the elements of $G$ correspond
to symmetries that preserve the orientation of space, i.e. those with $p=0$. Since the elements of $G$ act as symmetries on $\mathcal{C}$, their action must correspond to a group homomorphism
\begin{align}
[\rho] : G \rightarrow \text{Aut}(\mathcal{C}),
\end{align}
to the topological symmetry group $\text{Aut}(\mathcal{C})$, which is to say that $[ \rho_{\bf g} ] \cdot [\rho_{\bf h}] = [\rho_{\bf gh}]$. In other words, for each element ${\bf g} \in G$, the action of ${\bf g}$ can be described by a (unitary or anti-unitary) braided auto-equivalence map $\rho_{\bf g}$, which is a topological symmetry of $\mathcal{C}$, that respects group multiplication by satisfying
\begin{equation}
\label{eq:rho_compose}
\kappa_{\bf g,h} \circ \rho_{\bf g} \circ \rho_{\bf h} = \rho_{\bf gh},
\end{equation}
where $\kappa_{\bf g,h}$ is the corresponding natural isomorphism necessary to equate $\rho_{\bf g} \circ \rho_{\bf h}$ with $\rho_{\bf gh}$. We denote the identity element of $G$ as ${\bf 0}$ and let $\rho_{\bf 0} = \openone$ be the completely trivial transformation. Clearly, this gives $\kappa_{\bf g,0} = \kappa_{\bf 0,h} = \openone$.

The group action on topological charge labels is simply permutation [with $\rho_{\bf gh}(0)=0$], and so must satisfy $\rho_{\bf g} \circ \rho_{\bf h}(a) = \rho_{\bf gh}(a)$. Consequently, $\kappa_{\bf g,h}$ is trivial with respect to the action on topological charge labels, i.e. $\kappa_{\bf g,h}(a)=a$. It will be convenient to introduce the shorthand notations
\begin{eqnarray}
\label{eq:rho}
^{\bf g}a &=& \rho_{\bf g}(a)
\end{eqnarray}
\begin{eqnarray}
{\bf \bar g} &=& {\bf g}^{-1} \\
q({\bf g}) &=& q(\rho_{\bf g}).
\end{eqnarray}

We emphasize that the transformation factors $u^{a'b'}_{c'}$ associated with $\rho_{\bf g}$ acting on vertices need not be the same for different ${\bf g}$, and, in general, may require nontrivial action of the natural isomorphism $\kappa_{\bf g,h}$ in order to respect the group multiplication. We denote the transformation factors $u^{a'b'}_{c'}$ for a given $\rho_{\bf g}$ that leaves the basic data invariant as $U_{\bf g}\left(\,^{\bf g}a,\,^{\bf g}b;\,^{\bf g}c \right)$. Thus, with this symmetry action, we have
\begin{widetext}
\begin{eqnarray}
&& \rho_{\bf g}\left( \left| a,b;c,\mu \right\rangle \right) = \sum_{\mu'} \left[U_{\bf g}(\,^{\bf g}a,\,^{\bf g}b ;\,^{\bf g}c)\right]_{\mu \mu'} K^{q({\bf g})} \left|\,^{\bf g}a,\,^{\bf g}b ;\,^{\bf g}c,\mu' \right\rangle , \label{eq:rho_g_U_g} \\
&& \rho_{\bf g} \left( N_{ab}^{c} \right) = N_{\,^{\bf g}a \,^{\bf g}b}^{\,^{\bf g}c} = N_{a b}^{c} , \label{eq:rho_g_N}\\
&& \rho_{\bf g} \left( \left[F_{d}^{abc}\right]_{(e,\alpha,\beta)(f,\mu,\nu)} \right) = \sum_{\alpha',\beta',\mu'\nu'} \left[U_{\bf g}(\,^{\bf g}a,\,^{\bf g}b ;\,^{\bf g}e )\right]_{\alpha \alpha'} \left[U_{\bf g}(\,^{\bf g}e, \,^{\bf g}c ;\,^{\bf g}d)\right]_{\beta \beta'} \left[F_{\,^{\bf g}d}^{ \,^{\bf g}a \,^{\bf g}b \,^{\bf g}c}\right]_{( ^{\bf g}e,\alpha',\beta')(^{\bf g}f,\mu',\nu')} \notag \\
&& \qquad \qquad \qquad \qquad \qquad \qquad  \times \left[U_{\bf g}(\,^{\bf g}b,\,^{\bf g}c ;\,^{\bf g}f)^{-1}\right]_{\mu' \mu} \left[U_{\bf g}(\,^{\bf g}a,\,^{\bf g}f ;\,^{\bf g}d)^{-1}\right]_{\nu' \nu} = K^{q({\bf g})} \left[F_{d}^{abc}\right]_{(e,\alpha,\beta)(f,\mu,\nu)}  K^{q({\bf g})} \qquad
\label{eq:rho_g_F} \\
&& \rho_{\bf g} \left( \left[{R}_{c}^{ab} \right]_{\mu \nu} \right)
= \sum_{\mu',\nu'} \left[U_{\bf g}(\,^{\bf g}b,\,^{\bf g}a ;\,^{\bf g}c)\right]_{\mu \mu'}  \left[R_{\,^{\bf g}c}^{\,^{\bf g}a \,^{\bf g}b} \right]_{\mu' \nu'} \left[U_{\bf g}(\,^{\bf g}a,\,^{\bf g}b ;\,^{\bf g}c)^{-1}\right]_{\nu' \nu}
= K^{q({\bf g})} \left[ R_{c}^{ab} \right]_{\mu \nu} K^{q({\bf g})} ,
\label{eq:rho_g_R}
\end{eqnarray}
which produce the corresponding
\begin{eqnarray}
&& \kappa_{\bf g,h} \left( \left| a,b;c,\mu \right\rangle \right) = \sum_{\nu} \left[ \kappa_{\bf g,h} (a,b;c) \right]_{\mu \nu} \left| a,b;c,\nu \right\rangle ,
\label{eq:beta_vertex}
\\
&& \left[ \kappa_{\bf g,h} (a,b;c) \right]_{\mu \nu} = \sum_{\alpha , \beta} \left[U_{\bf g}(a,b;c)^{-1}\right]_{\mu \alpha} K^{q({\bf g})} \left[U_{\bf h}(\,^{\bf \bar g}a,\,^{\bf \bar g}b ;\,^{\bf \bar g}c)^{-1}\right]_{\alpha \beta} K^{q({\bf g})} \left[U_{\bf gh}(a,b;c)\right]_{\beta \nu}
\label{eq:beta_gh}
.
\end{eqnarray}
\end{widetext}
We note that, to account for the possibility of anti-unitary symmetries, we have inserted the complex conjugation operators $K$ in such a way that has the effect of complex conjugating the $F$-symbol, $R$-symbol, or $U_{\bf h}$-symbol that is sandwiched between a pair of $K$ operators when ${\bf g}$ corresponds to an anti-unitary symmetry, which has ${q({\bf g})}=1$. It is often convenient to choose basis states such that $K \left|a,b ;c,\mu \right\rangle = \left|a,b ;c,\mu \right\rangle$, which can be done with a vertex basis gauge transformation.

Since $\kappa_{\bf g,h}$ is a natural isomorphism, its action on vertices takes the form
\begin{equation}
\left[ \kappa_{\bf g,h} (a,b;c) \right]_{\mu \nu} = \frac{ \beta_{a}({\bf g,h}) \beta_{b}({\bf g,h}) }{\beta_{c}({\bf g,h})} \delta_{\mu \nu}
\label{eq:beta_nat_iso}
,
\end{equation}
where $\beta_{a}({\bf g,h})$ are phases that only depend on the topological charge $a$ and the group elements ${\bf g}$ and ${\bf h}$.

As discussed in the previous subsection, there is redundancy due to the freedom of choosing how one decomposes a natural isomorphism into the topological charge dependent phase factors. Specifically, it is always possible to transform the $\beta_{a}({\bf g,h})$ phases into
\begin{equation}
\label{eq:tilde_eta}
\breve{\beta}_{a}({\bf g,h}) = \nu_{a}({\bf g,h}) \beta_{a}({\bf g,h})
,
\end{equation}
while leaving $\breve{\kappa}_{\bf g,h} (a,b;c) = \kappa_{\bf g,h} (a,b;c)$ unchanged, if the phases $\nu_{a}({\bf g,h})$ satisfy $\nu_{a}({\bf g,h}) \nu_{b}({\bf g,h}) = \nu_{c}({\bf g,h})$ whenever $N_{ab}^{c} \neq 0$. Moreover, it is clear that whenever two sets of phase factors $\beta_{a}({\bf g,h})$ and $\breve{\beta}_{a}({\bf g,h})$ give the same $\kappa_{\bf g,h} (a,b;c)$, they must be related by $\nu_{a}({\bf g,h})$ of this form. Therefore, the derived properties of $\beta_{a}({\bf g,h})$ and $\breve{\beta}_{a}({\bf g,h})$ related in this manner should be considered equivalent, and this redundancy should be viewed as a sort of gauge freedom.

Requiring the symmetry action on vacuum to be trivial imposes the conditions
\begin{equation}
\label{eq:U_vacuum}
U_{\bf g}(0,0;0)= U_{\bf g}(a,0;a) = U_{\bf g}(0,a;a) =1
,
\end{equation}
which makes the symmetry action compatible with introducing and removing vacuum lines at will. Clearly, $\rho_{\bf 0} = \openone$ requires $U_{\bf 0}(a,b;c) = \openone$.

Eq.~(\ref{eq:U_vacuum}) requires
\begin{equation}
\kappa_{\bf g,h}(0,0;0)= \beta_{0}({\bf g,h})=1.
\end{equation}
Since $\kappa_{\bf g,0} = \kappa_{\bf 0,h} = \openone$, it follows that
\begin{eqnarray}
\beta_{a}({\bf g,0}) \beta_{b}({\bf g,0}) &=& \beta_{c}({\bf g,0}) \\
\beta_{a}({\bf 0,h}) \beta_{b}({\bf 0,h}) &=& \beta_{c}({\bf 0,h})
\end{eqnarray}
whenever $N_{ab}^{c} \neq 0$. Given the gauge freedom described in Eq.~(\ref{eq:tilde_eta}), it is always possible to freely modify such terms to be trivial, so we will always impose on them the simplifying condition
\begin{equation}
\beta_{a}({\bf 0,0}) = \beta_{a}({\bf g,0}) = \beta_{a}({\bf 0,h}) = 1
,
\end{equation}
as a choice of gauge.

We can use Eq.~(\ref{eq:rho_compose}) to write the decomposition of $\rho_{\bf ghk}$ in the two equivalent ways related by associativity (leaving the $\circ$ symbols implicit from now on)
\begin{eqnarray}
\rho_{\bf ghk} &=& \kappa_{\bf g,hk} \rho_{\bf g} \rho_{\bf hk} \notag \\
&=& \kappa_{\bf g,hk} \rho_{\bf g} \kappa_{\bf h,k} \rho_{\bf h} \rho_{\bf k} \notag \\
&=& \kappa_{\bf g,hk} \rho_{\bf g} \kappa_{\bf h,k} \rho_{\bf g}^{-1} \rho_{\bf g} \rho_{\bf h} \rho_{\bf k} \notag \\
&=& \kappa_{\bf gh,k} \rho_{\bf gh} \rho_{\bf k} \notag \\
&=& \kappa_{\bf gh,k} \kappa_{\bf g,h} \rho_{\bf g} \rho_{\bf h} \rho_{\bf k}
.
\end{eqnarray}
This gives the consistency condition on $\kappa_{\bf g,h}$
\begin{equation}
\label{eq:beta_cocycle}
\kappa_{\bf g,hk} \rho_{\bf g} \kappa_{\bf h,k} \rho_{\bf g}^{-1} = \kappa_{\bf gh,k} \kappa_{\bf g,h}
.
\end{equation}
We emphasize that the $\rho_{\bf g}$ transformation here may be anti-unitary, so that it applies complex conjugation (as well as the topological charge permutation) to the $\kappa_{\bf h,k}$ which it conjugates.

Since we consider braided auto-equivalence maps to be equivalent when they are related by natural isomorphisms, we may equivalently choose to use the auto-equivalence maps $\check{\rho}_{\bf g} = \Upsilon_{\bf g} \circ \rho_{\bf g}$ for the global symmetry action. With this choice of action, we have the redefined quantities
\begin{equation}
\left[\check{U}_{\bf g}(a,b;c)\right]_{\mu \mu'} = \frac{\gamma_{a}({\bf g}) \gamma_{b}({\bf g})}{\gamma_{c}({\bf g})} \left[U_{\bf g}(a,b;c)\right]_{\mu \mu'}
\label{eq:U'_g}
.
\end{equation}
These result in a correspondingly redefined $\check{\kappa}_{\bf g,h}$, for which we may choose the redefined vertex decomposition factors
\begin{equation}
\check{\beta}_{a}({\bf g,h})= \frac{\gamma_{a}({\bf gh}) } { K^{q({\bf g})} \gamma_{\,^{\bf \bar{g}}a}({\bf h}) K^{q({\bf g})} \gamma_{a}({\bf g})  } \beta_{a}({\bf g,h})
\label{eq:eta'_gh}
.
\end{equation}
We emphasize that it is not always possible to set $\left[U_{\bf g}(a,b;c)\right]_{\mu \mu'} = \delta_{\mu \mu'}$ by using gauge transformations, see e.g. Eq.~(\ref{eq:SU(3)3_nontrivial_symm_action}).
We also emphasize that the transformation of the $F$-symbols and $R$-symbols are precisely the same for $\check{\rho}_{\bf g}$ and $\rho_{\bf g}$, since they are related by a natural isomorphism. In order to preserve the trivial action on the vacuum charge and the triviality of the factor $\beta_{a}({\bf 0,0})=1$, we must fix $\gamma_{0}({\bf g})=\gamma_{a}({\bf 0}) =1$. We may think of the relation between auto-equivalence maps by natural isomorphisms as a sort of gauge transformation for the symmetry action, which is a notion that will be made more clear in Sec.~\ref{sec:G-crossed_gauge_transformations}.

\subsection{$H^3_{[\rho]}(G, \mathcal{A})$ Invariance Class of the Symmetry Action}
\label{sec:H3Invariance}

Given the global symmetry action $[\rho]$ described in Sec.~\ref{sec:Global_Symmetry}, we wish to find an invariant that would allow us to determine whether or not it would be possible to fractionalize the symmetry action. In this subsection, we will define such an invariant $[\coho{O}] \in H^3_{[\rho]}(G, \mathcal{A})$, and in the following section, we will demonstrate that the symmetry can be fractionalized when $[\coho{O}]=[0]$, whereas $[\coho{O}] \neq [0]$ indicates that there is an obstruction to fractionalizing the symmetry. (See Appendix~\ref{app:cohomology} for a review of group cohomology.)

We begin by defining (for a particular choice of $\rho \in [\rho]$) the quantity
\begin{equation}
\Omega_{a}({\bf g,h,k}) = \frac{  K^{q({\bf g})} \beta_{\rho_{\bf g}^{-1}(a)}({\bf h,k})  K^{q({\bf g})}  \beta_{a}({\bf g,hk})}{\beta_{a}({\bf gh,k}) \beta_{a}({\bf g,h})}
,
\end{equation}
which is a phase from which we will obtain the desired invariant. From this definition, it immediately follows that
\begin{equation}
\label{eq:Omega_cocycle}
\frac{  K^{q({\bf g})} \Omega_{\rho_{\bf g}^{-1}(a)}({\bf h,k,l})  K^{q({\bf g})} \Omega_{a}({\bf g,hk,l}) \Omega_{a}({\bf g,h,k})}{ \Omega_{a}({\bf gh,k,l}) \Omega_{a}({\bf g,h,kl})} = 1
.
\end{equation}
By using Eqs.~(\ref{eq:beta_nat_iso}) and (\ref{eq:beta_cocycle}), we see that
\begin{equation}
\Omega_{a}({\bf g,h,k}) \Omega_{b}({\bf g,h,k}) = \Omega_{c}({\bf g,h,k})
\end{equation}
whenever $N_{ab}^{c} \neq 0$. As explained in the end of Sec.~\ref{sec:Braiding}, this implies
\begin{equation}
\Omega_{a}({\bf g,h,k}) = M^{\ast}_{a \cohosub{O} ({\bf g,h,k})}
\end{equation}
for some $\coho{O} ({\bf g,h,k}) \in \mathcal{A}$, where $\mathcal{A} \subset \mathcal{C}$ is the subset of topological charges in $\mathcal{C}$ that are Abelian. (One can also think of $\mathcal{A} \subset \mathcal{C}$ as a subcategeory of $\mathcal{C}$.) More precisely, $\coho{O} ({\bf g,h,k})\in C^{3}(G,\mathcal{A})$ is a $3$-cochain, since it is a function of three group elements ${\bf g,h,k} \in G$ to $\mathcal{A}$, which we can now consider to be the Abelian group whose elements are the Abelian topological charges of $\mathcal{C}$ with group multiplication given by their corresponding fusion rules. Moreover, through this relation, Eq.~(\ref{eq:Omega_cocycle}) maps to the condition
\begin{widetext}
\begin{eqnarray}
1 &=&  K^{q({\bf g})} M_{\rho_{\bf g}^{-1}(a) \cohosub{O} ({\bf h,k,l}) } K^{q({\bf g})}  M^{\ast}_{a \cohosub{O} ({\bf gh,k,l})} M_{a \cohosub{O} ({\bf g,hk,l})} M^{\ast}_{a \cohosub{O} ({\bf g,h,kl})} M_{a \cohosub{O} ({\bf g,h,k})} \notag \\
&=&  M_{a \rho_{\bf g}[\cohosub{O} ({\bf h,k,l})] } M^{\ast}_{a \cohosub{O} ({\bf gh,k,l})} M_{a \cohosub{O} ({\bf g,hk,l})} M^{\ast}_{a \cohosub{O} ({\bf g,h,kl})} M_{a \cohosub{O} ({\bf g,h,k})} \notag \\
&=& M_{a \rho_{\bf g}[\cohosub{O} ({\bf h,k,l})] } M_{a \overline{\cohosub{O} ({\bf gh,k,l})}} M_{a \cohosub{O} ({\bf g,hk,l})} M_{a \overline{\cohosub{O} ({\bf g,h,kl})}} M_{a \cohosub{O} ({\bf g,h,k})} \notag \\
&=& M_{a, \rho_{\bf g}[\cohosub{O} ({\bf h,k,l})] \times \overline{\cohosub{O} ({\bf gh,k,l})} \times \cohosub{O} ({\bf g,hk,l}) \times \overline{\cohosub{O} ({\bf g,h,kl})} \times  \cohosub{O} ({\bf g,h,k})}
,
\label{eq:O_cocycle}
\end{eqnarray}
Here, we used the symmetry property $S_{\rho_{\bf g}(a)\rho_{\bf g}(b)} = K^{q({\bf g})} S_{ab} K^{q({\bf g})}$, the relation $S_{ab}^{\ast}=S_{a\bar{b}}$, and the fact that if $M_{ab}$ is a phase, then $M_{ab} M_{ac} = M_{ae}$ whenever $N_{bc}^{e}\neq 0$. Since this condition holds for all $a$, the non-degeneracy of braiding implies that
\begin{equation}
\text{d} \coho{O} ({\bf g,h,k,l}) =\rho_{\bf g}[\coho{O} ({\bf h,k,l})] \times \overline{\coho{O} ({\bf gh,k,l})} \times \coho{O} ({\bf g,hk,l}) \times \overline{\coho{O} ({\bf g,h,kl})} \times  \coho{O} ({\bf g,h,k}) = 0
.
\end{equation}
\end{widetext}
In other words, $\coho{O} ({\bf g,h,k})$ satisfies the $3$-cocycle condition, when treated as a 3-cochain. Thus, there is an invertible map between the phase $\Omega_{a}({\bf g,h,k})$ and the $3$-cocycle $\coho{O} ({\bf g,h,k}) \in Z^{3}_{\rho}(G,\mathcal{A})$.

As explained in the discussion around Eq.~(\ref{eq:tilde_eta}), there is gauge freedom to modify the phases $\beta_{a}({\bf g,h})$ to $\breve{\beta}_{a}({\bf g,h})=\nu_{a}({\bf g,h}) \beta_{a}({\bf g,h})$, for phase factors $\nu_{a}({\bf g,h})$ that satisfy $\nu_{a}({\bf g,h}) \nu_{b}({\bf g,h}) = \nu_{c}({\bf g,h})$ whenever $N_{ab}^{c} \neq 0$. The correspondingly modified
\begin{eqnarray}
\label{eq:Omega_tilde}
&&\breve{\Omega}_{a}({\bf g,h,k}) = \frac{ K^{q({\bf g})} \breve{\beta}_{\rho_{\bf g}^{-1}(a)}({\bf h,k}) K^{q({\bf g})} \breve{\beta}_{a}({\bf g,hk})}{\breve{\beta}_{a}({\bf gh,k}) \breve{\beta}_{a}({\bf g,h})} \notag \\
&& = \frac{ K^{q({\bf g})} \nu_{\rho_{\bf g}^{-1}(a)}({\bf h,k}) K^{q({\bf g})} \nu_{a}({\bf g,hk})}{\nu_{a}({\bf gh,k}) \nu_{a}({\bf g,h})} \Omega_{a}({\bf g,h,k})
\qquad
\end{eqnarray}
is to be considered in the same equivalence class as $\Omega_{a}({\bf g,h,k})$ and obeys the same properties as $\Omega_{a}({\bf g,h,k})$, except $\breve{\Omega}_{a}({\bf g,h,k}) = M^{\ast}_{a \breve{\cohosub{O}} ({\bf g,h,k})}$ maps to a potentially different $\breve{\coho{O}} ({\bf g,h,k})$, which should therefore be considered to be in the same equivalence class as $\coho{O} ({\bf g,h,k})$. To find the relation between these, we note that we similarly have the condition that
\begin{equation}
\nu_{a}({\bf g,h}) = M^{\ast}_{a \cohosub{v}({\bf g,h})}
,
\end{equation}
where $\coho{v}({\bf g,h}) \in C^{2}(G,\mathcal{A})$ is a $2$-cochain taking values in the set of Abelian topological charges. Using this in Eq.~(\ref{eq:Omega_tilde}) and employing the same properties utilized in Eq.~(\ref{eq:O_cocycle}), we obtain the corresponding relation
\begin{eqnarray}
\breve{\coho{O}} ({\bf g,h,k}) &=& \rho_\mb{g}[\coho{v}({\bf h,k})] \times \overline{\coho{v}({\bf gh,k})} \notag \\
&& \quad \times \coho{v}({\bf g,hk}) \times \overline{\coho{v}({\bf g,h})} \times \coho{O} ({\bf g,h,k}) \notag \\
&=& \text{d}\coho{v}({\bf g,h,k}) \times  \coho{O} ({\bf g,h,k})
\label{eq:O_tilde}
,
\end{eqnarray}
which shows that $\coho{O} ({\bf g,h,k})$ and $\breve{\coho{O}} ({\bf g,h,k})$ in the same equivalence class are related by fusion with a $3$-coboundary $\text{d}\coho{v}({\bf g,h,k}) \in B^{3}_{\rho}(G,\mathcal{A})$. Thus, the equivalence classes $[\coho{O}]$ are elements of the $3$rd cohomology group given by taking the quotient of $3$-cocycles by $3$-coboundaries
\begin{equation}
[\coho{O} ] \in  H^3_{\rho}(G, \mathcal{A}) = \frac{ Z^3_{\rho}(G, \mathcal{A})} {B^3_{\rho}(G, \mathcal{A})}
.
\end{equation}
We emphasize that the equivalence class $[\coho{O}]$ is uniquely defined entirely in terms of $\rho$ (which defines $\kappa_{\bf g,h}$). We further emphasize that $[\coho{O} ]=[0]$ does not necessarily imply that $\beta_{a}({\bf g,h}) \beta_{b}({\bf g,h}) = \beta_{c}({\bf g,h})$ whenever $N_{ab}^{c} \neq 0$ nor, equivalently, that $\kappa_{\bf g,h}=\openone$.

We can also see from the definitions that the equivalence class $[\coho{O}]$ is actually an invariant of the equivalence class $[\rho]$ of symmetry actions that are related by natural isomorphisms. In particular, if we instead used the action $\check{\rho}_{\bf g} = \Upsilon_{\bf g}\rho_{\bf g}$, where $\Upsilon_{\bf g}$ is a natural isomorphism, and the corresponding modified vertex decomposition factors $\check{\beta}_{a}({\bf g})$ as given in Eq.~(\ref{eq:eta'_gh}), then we would find that the corresponding quantity $\check{\Omega}_{a}({\bf g,h,k}) = \Omega_{a}({\bf g,h,k})$ is unchanged. Thus, any such symmetry actions related by natural isomorphisms define the same equivalence class $[\check{\coho{O}}]=[\coho{O}]$, so we actually have
\begin{equation}
[\coho{O} ] \in  H^3_{[\rho]}(G, \mathcal{A})
.
\end{equation}

We note that if the symmetry action is unitary and does not permute topological charges, i.e. $\rho_{\bf g}(a)=a$ for all $a$ and ${\bf g}$, then it is always the case that $[\coho{O} ]=[0]$. To demonstrate this property, we observe that $\rho_{\bf g}$ are actually natural isomorphisms when this is the case. It follows that we can write $[U_{\bf g}(a,b;c)]_{\mu \nu} = \frac{\chi_a ({\bf g}) \chi_b ({\bf g})}{\chi_c ({\bf g})} \delta_{\mu \nu}$, where $\chi_a ({\bf g})$ are phases, and that we can make a choice within the equivalence class for which $\beta_{a}({\bf g,h}) = \frac{\chi_a ({\bf gh})}{  \chi_{a} ({\bf h}) \chi_a ({\bf g})}$. Using this with the definition, we find $\Omega_{a}({\bf g,h,k}) =1$ and hence $[\coho{O}]=[0]$. Alternatively, we could have used the gauge transformation of Eqs.~(\ref{eq:U'_g}) and (\ref{eq:eta'_gh}) with $\gamma_{a}({\bf g}) = \chi_a ({\bf g})^{-1}$ to set $\left[\check{U}_{\bf g}(a,b;c)\right]_{\mu \nu}=\delta_{\mu \nu}$, and $\check{\beta}_{a}({\bf g,h})=1$, which obviously gives $\check{\Omega}_{a}({\bf g,h,k}) =1$.

Given $\mathcal{C}$ and $G$, there are many different possible choices of $\rho$. These different choices correspond to different ways
that the global symmetry (of the microscopic Hamiltonian) and the topological order can interplay with each other. From the above discussion, we see that
clearly the first important choice is how $\rho_{\bf g}$ permutes the various anyons. The next important choice
depends on more subtle properties of the gauge transformations that
are required when implementing $\rho_{\bf g}$. In the next section, we examine how these properties lead to a concept known
as symmetry fractionalization, whereby the quasiparticles have the ability to form a sort of projective representation of the symmetry group.
We will classify the ways in which the symmetry can fractionalize and, in doing so, find that $[\coho{O}] \neq [0]$ indicates that there is an obstruction to fractionalizing the symmetry.

\section{Symmetry Fractionalization}
\label{sec:symmetryfrac}

Before carrying out the detailed derivation, we will state the result of this section and provide a summary overview of the arguments (and direct the reader to Appendix~\ref{app:cohomology}, if a review of group cohomology is needed):

For a system that realizes a topological phase described by the UMTC $\mathcal{C}$ and which has the global symmetry group $G$ with corresponding group action $[\rho] : G\rightarrow \text{Aut}(\mathcal{C})$:
\begin{enumerate}
\item There is an obstruction to symmetry fractionalization if $[\coho{O} ] \neq [0]$, where $[\coho{O} ] \in  H^3_{[\rho]}(G, \mathcal{A})$  was the invariant of $[\rho]$ defined in Sec.~\ref{sec:H3Invariance}.
\item When $[\coho{O} ] = [0]$, symmetry fractionalization may occur and is classified by the cohomology group $H^2_{[\rho]}(G,\mathcal{A})$, where $\mathcal{A}$ is defined to be the finite group whose elements are the Abelian topological charges of $\mathcal{C}$ with group multiplication given by their corresponding fusion rules. More precisely, the set of distinct symmetry fractionalization classes is an $H^2_{[\rho]}(G,\mathcal{A})$ torsor.~\footnote{Given a group $G$, a $G$ torsor is a non-empty set $X$ upon which $G$ acts freely and transitively. In other words, it is what you get if the group $G$ had lost its identity. In the context of classification, this means that distinct symmetry fractionalization classes are related to each other by the action of distinct elements of $H^2_{[\rho]}(G,\mathcal{A})$.}
\end{enumerate}

We emphasize that $H^3_{[\rho]}(G, \mathcal{A})$ does not classify obstructions to fractionalization. In particular, the object $[\coho{O} ]$ is uniquely defined by the symmetry action $[\rho] : G \rightarrow \text{Aut}(\mathcal{C})$, and it indicates whether or not the symmetry can be fractionalized when $\mathcal{C}$, $G$, and $[\rho]$ are specified.

In this section, we assume that the global symmetry acts in an on-site or ``locality preserving'' fashion on the underlying physical system, where locality preserving action is a generalization of the notion of on-site action that may include symmetries that act non-locally, such as anti-unitary, time-reversal, translation, rotation, and other spacetime symmetries. The on-site and locality preserving properties of symmetry actions are fundamental requirements for symmetry fractionalization, so we will define precisely what we mean when we use these terms. We do not restrict the symmetry group $G$ to be discrete.

In order to explain the above mathematical statement of symmetry fractionalization,
we begin by examining the action of a unitary on-site symmetry on the physical Hilbert space of the underlying physical system and its microscopic Hamiltonian. We argue that, for on-site symmetry, the action of the global symmetry operator $R_\mb{g}$ on the physical states $\ket{\Psi_{a_1 , \ldots , a_n }}$, corresponding to the system with $n$ quasiparticles carrying topological charges $a_1, \dots, a_n$, respectively, (which must collectively fuse to vacuum topological charge $0$,) can always be written as
\begin{equation}
R_\mb{g}\ket{\Psi_{a_1 , \ldots , a_n } }=\prod_{j=1}^{n} U_\mb{g}^{(j)}\rho_\mb{g}\ket{\Psi_{a_1 , \ldots , a_n } }.
\end{equation}
Here, we have separated local unitary transformations $U_\mb{g}^{(j)}$ from the non-local unitary transformation $\rho_\mb{g}$ that acts as the symmetry action on the topological quantum numbers.

Since $R_\mb{g}$ are the physical symmetry transformations, $R_\mb{g}R_\mb{h}=R_\mb{gh}$ (at least projectively). Writing out the localized forms explicitly leads to the relation
\begin{equation}
\prod_{j=1}^{n} U_\mb{g}^{(j)} \rho_\mb{g} U_\mb{h}^{(j)} \rho_\mb{g}^{-1}  = \prod_{j=1}^{n} U_{\mb{gh}}^{(j)}\kappa_{\mb{g},\mb{h}}
.
\end{equation}

We can also argue that the local operators $U_\mb{g}^{(j)}$ satisfy the projective multiplication relation
\begin{equation}
U_\mb{g}^{(j)} \rho_\mb{g} U_\mb{h}^{(j)} \rho_\mb{g}^{-1} \ket{\Psi_{a_1 , \ldots , a_n }} = \eta_{a_j}(\mb{g},\mb{h}) U_{\mb{gh}}^{(j)} \ket{\Psi_{a_1 , \ldots , a_n }}
\end{equation}
when acting on quasiparticle states, for some phase factors $\eta_{a_j}(\mb{g},\mb{h})$ that only depend on the topological charge $a_j$ and group elements $\mb{g}$ and $\mb{h}$. Then the condition $R_\mb{g}R_\mb{h}=R_{\mb{gh}}$ yields
\begin{equation}
\kappa_{\bf g,h} (a_1, \ldots , a_n ) = \prod_{j=1}^{n} \beta_{a_j}(\mb{g},\mb{h}) = \prod_{j=1}^{n} \eta_{a_j}(\mb{g},\mb{h}),
	\label{eqn:easy_omega_cond}
\end{equation}
where $\beta_{a}({\bf g,h})$ are the phase factors that decompose the natural isomorphism $\kappa_{\bf g,h}$, as in Sec.~\ref{sec:Global_Symmetry}. This provides a strong constraint relating the phases $\eta_{a}(\mb{g},\mb{h})$ and $\beta_{a}({\bf g,h})$ for different values of topological charges $a$.

The associativity of the local operators leads to the cocycle-like relation
\begin{equation}
\frac{ \eta_{\rho_{\bf g}^{-1}(a)}({\bf h,k})  \eta_{a}({\bf g,hk}) }{ \eta_{a}({\bf gh,k})  \eta_{a}({\bf g,h}) } = 1
.
\label{eqn:easy_phi_cocy}
\end{equation}
This imposes a required condition on $\beta_{a}({\bf g,h})$ factors, which defines an obstruction given by the previously described invariance class $[\coho{O}] \in H^3_{[\rho]}(G,\mathcal{A})$.

When the obstruction class is trivial, one is guaranteed to have at least one set of $\eta_{a}({\bf g,h})$ which can satisfy both Eq.~(\ref{eqn:easy_omega_cond}) and Eq.~(\ref{eqn:easy_phi_cocy}). It follows that there are actually many solutions, since, given one solution with phase factors $\eta_{a}({\bf g,h})$, another solution $\eta'_{a}({\bf g,h}) = \tau_a(\mb{g},\mb{h})^{-1} \eta_{a}({\bf g,h}) $ is obtained from it by dividing by phases $\tau_a(\mb{g},\mb{h})$ that satisfy the conditions
\begin{eqnarray}
&& \frac{ \tau_{\rho_{\bf g}^{-1}(a)}({\bf h,k})  \tau_{a}({\bf g,hk}) }{ \tau_{a}({\bf gh,k})  \tau_{a}({\bf g,h}) } = 1, \\
&& \tau_{a}(\mb{g},\mb{h}) \tau_{b}(\mb{g},\mb{h}) = \tau_{c}(\mb{g},\mb{h}) , \quad \text{if} \quad N_{ab}^c \neq 0.
\end{eqnarray}

However, there is some redundancy in these solutions that is due to the freedom to redefine the operators $U_\mb{g}^{(j)}$ by local operators $Z_\mb{g}^{(j)}$ that do not affect $R_\mb{g}$, which means $\prod_j Z_\mb{g}^{(j)}=\openone$. This property requires that the action on quasiparticle state $Z_\mb{g}^{(j)} \ket{\Psi_{\{a_j\}}} = \zeta_{a_j}(\mb{g}) \ket{\Psi_{\{a_j\}}}$, where $\zeta_{a}(\mb{g})$ are phases that satisfy $\zeta_{a}(\mb{g}) \zeta_{b}(\mb{g})=\zeta_{c}(\mb{g})$ whenever $N_{ab}^{c} \neq 0$. This redefinition of local operators changes the phases $\eta_{a}({\bf g,h})$ in the following way
\begin{equation}
\hat{\eta}_{a}({\bf g,h}) = \frac{ \zeta_{a}({\bf gh} ) }{\zeta_{\rho_{\bf g}^{-1} (a)}({\bf h} ) \zeta_{a}({\bf g} ) } \eta_{a}({\bf g,h}) .
\label{eqn:easy_coboundary}
\end{equation}
Thus, if two sets of solutions are related by such a transformation, they should be considered physically indistinguishable, so they belong to a single equivalence class of solutions.

Since $\mathcal{C}$ is modular, the factors $\tau_{a}(\mb{g},\mb{h})$ uniquely define a 2-cocycle $\coho{t} \in Z^{2}_{\rho}(G,\mathcal{A})$ and the factors $\zeta_{a}({\bf g} )$ uniquely define a 1-cochain $\coho{z} \in C^1(G,\mathcal{A})$, which makes the equivalence classes related by $2$-coboundaries d$\coho{z} \in B^{2}_{\rho}(G,\mathcal{A})$. Taking the quotient (and noting the invariance of the results under the choice of $\rho \in [\rho]$) results in the classification of solutions by $H^2_{[\rho]}(G,\mathcal{A})$.

After these arguments, we will generalize the results to the case where the global symmetry action is a projective representation. Finally, we will introduce the notion of locality preserving symmetry and explain how the on-site symmetry arguments and results are generalized to apply to such symmetries.

\subsection{Physical Manifestation of On-Site Global Symmetry}
\label{sec:Physical_Manifestation}

We wish to examine the quantum states of the underlying physical system in which there are quasiparticles present. Initially, let us consider the case when there are two quasiparticles, and we will subsequently generalize to an arbitrary number. We assume the two quasiparticles possess topological charges $a$ and $\bar{a}$, respectively, and that they are respectively localized within the well-separated, simply-connected regions $\mathcal{R}_1$ and $\mathcal{R}_2$. Well-separated means that the minimum distance $r_{12} \equiv \min_{r_{j}\in \mathcal{R}_j }|r_1 - r_2|$ between any two points of the distinct regions is much larger than the correlation length $\xi$ of the system, i.e. $r_{12} \gg \xi$. (We typically think of $\mathcal{R}_j$ as a disk centered at the quasiparticle coordinate $r_j$ with a radius that is a few correlations lengths.)

For concreteness, we consider the system to be defined on a sphere (or any genus zero surface) and assume that there are no other quasiparticles nor boundaries that carry topological charge, so this pair must fuse to vacuum. (The analysis can be generalized to surfaces of arbitrary genus with any number of boundaries, but we will not do so in this paper.) In general, since $N_{a\bar{a}}^0 = 1$, there is a single topological sector in such a setup, which is described by $|a , \bar{a}; 0 \rangle$ in the topological state space. However, this topological state represents a universality class of many microscopic states that share its topological properties and which differ by the application of local operators. Such a state in this universality class can be obtained by starting from the uniform Hamiltonian $H_0$ of the system in the topological phase, adiabatically creating a pair of quasiparticles with charges $a$ and $\bar{a}$ from vacuum by tuning the Hamiltonian to locally favor the existence of such quasiparticles that are not well-separated, and then subsequently moving the quasiparticles individually to regions $\mathcal{R}_1$ and $\mathcal{R}_2$, respectively, through a sequence of similar modifications of the Hamiltonian (which return the Hamiltonian to its original form in the regions away from the quasiparticles positions).

The corresponding Hamiltonian resulting after this process is of the form
\begin{equation}
H_{a,\bar{a};0}^{\alpha} = H_{0} + h_{a;\alpha}^{(1)}+ h_{\bar{a};\alpha}^{(2)}
,
\end{equation}
where $h_{a;\alpha}^{(j)}$ is a modification of the Hamiltonian whose nontrivial action is localized within $\mathcal{R}_j$ and which favors the localization of a quasiparticle of charge $a$ in this region. The label $\alpha$ is a parameter which simply identifies these terms as one of many that favors localization of a quasiparticle of this type. We write the ground state of this Hamiltonian $H_{a,\bar{a};0}^{\alpha}$ as $|\Psi_{a , \bar{a}; 0}^{\alpha} \rangle$ (which is in the $|a , \bar{a}; 0 \rangle$ universality class). We emphasize that $|\Psi_{a , \bar{a}; 0}^{\alpha} \rangle$ with different values of the parameter $\alpha$ are not necessarily orthogonal; in fact, we expect that they may have very high overlaps for some different values of $\alpha$. However, such states with different values of topological charge will be orthogonal, up to exponentially suppressed corrections, i.e. $\langle \Psi_{a , \bar{a}; 0}^{\alpha} |\Psi_{b , \bar{b}; 0}^{\beta} \rangle \approx 0$ whenever $a \neq b$.

Let us now assume that the symmetry acts on the system in an on-site manner, with $R_{\bf g}$ being the unitary operator representing the action of ${\bf g}$.
By on-site, we mean that if we decompose the space manifold $\mathcal{M}= \bigcup_{k \in I} \mathcal{M}_{k}$ into a collection of simply connected disjoint regions $\mathcal{M}_{k}$ (a subset of which can be taken to be the regions $\mathcal{R}_j$) with index set $I$, the symmetry operators take the form
\begin{equation}
\label{eq:R_onsite}
R_{\bf g}=\prod_{k\in I} R_{\bf g}^{(k)}
,
\end{equation}
where $R_{\bf g}^{(k)}$ is a unitary operator that has nontrivial action localized in region $\mathcal{M}_{k}$.
Since ${\bf g}$ is a symmetry of the system that acts on $\mathcal{C}$ by $\rho_{\bf g}$, the Hamiltonians should transform as
\begin{eqnarray}
R_{\bf g} H_{0} R_{\bf g}^{-1} &=& H_{0} \\
R_{\bf g} H_{a,\bar{a};0}^{\alpha} R_{\bf g}^{-1} &=& H_{\,^{\bf g}a,\,^{\bf g}\bar{a};0}^{{\bf g}(\alpha)}
\end{eqnarray}
where
\begin{equation}
h_{\,^{\bf g}a;{\bf g}(\alpha)}^{(j)} = R_{\bf g} h_{a;\alpha}^{(j)} R_{\bf g}^{-1}
\end{equation}
of the new Hamiltonian remains an operator that is localized in the region $\mathcal{R}_j$, but now favors the localization of a quasiparticle of charge $^{\bf g}a = \rho_{\bf g}(a)$. Indeed, since the symmetry is on-site, any operator $\mathcal{O}^{(j)}$ whose nontrivial action is localized in a region $\mathcal{R}_{j}$ remains localized in this region when acted upon by the symmetry transformation, i.e.
\begin{equation}
^{\bf g} \mathcal{O}^{(j)} \equiv R_{\bf g} \mathcal{O}^{(j)} R_{\bf g}^{-1}=R_{\bf g}^{(j)} \mathcal{O}^{(j)} R_{\bf g}^{(j)-1}
\label{eq:g-conjugate-operator}
\end{equation}
is localized in $\mathcal{R}_{j}$. We stress that the label ${\bf g}(\alpha)$ of the Hamiltonian defined with $h_{\,^{\bf g}a;{\bf g}(\alpha)}^{(j)}$ obtained from the symmetry transformation indicates that this Hamiltonian need not equal the Hamiltonian defined with the modification $h_{\,^{\bf g}a;\alpha}^{(j)}$ for localizing a charge $^{\bf g}a$ quasiparticle, to which we already ascribed the label $\alpha$. In other words, while the universality class of states transforms as
\begin{equation}
|a , \bar{a}; 0 \rangle \mapsto \rho_{\bf g}|a , \bar{a}; 0 \rangle = U_{\bf g}(\,^{\bf g}a,\,^{\bf g}\bar{a};0) |\,^{\bf g}a,\,^{\bf g}\bar{a};0 \rangle
\end{equation}
under the action of ${\bf g}$, the ground state of the Hamiltonian transforms as
\begin{equation}
|\Psi_{a , \bar{a}; 0}^{\alpha} \rangle \mapsto R_{\bf g} |\Psi_{a , \bar{a}; 0}^{\alpha} \rangle = |\Psi_{ \,^{\bf g}a,\,^{\bf g}\bar{a};0}^{{\bf g}(\alpha)} \rangle
,
\end{equation}
where $|\Psi_{ \,^{\bf g}a,\,^{\bf g}\bar{a};0}^{{\bf g}(\alpha)} \rangle$ is not necessarily equal (nor proportional) to $|\Psi_{ \,^{\bf g}a,\,^{\bf g}\bar{a};0}^{\alpha} \rangle$.

In fact, we have not yet made clear what it even means to have states $|\Psi_{a , \bar{a}; 0}^{\alpha} \rangle$ and $|\Psi_{ \,^{\bf g}a,\,^{\bf g}\bar{a};0}^{\alpha} \rangle$ in different topological charge sectors with the same label $\alpha$. For this, we make a choice of complete orthonormal basis states $|\varphi_{a , \bar{a}; 0}^{s} \rangle$ for each topological charge sector. Then, given a state
\begin{equation}
|\Psi_{ a,\bar{a};0}^{\alpha} \rangle = \sum_{s} A_{s} |\varphi_{a , \bar{a}; 0}^{s} \rangle
\end{equation}
we identify the corresponding state in the different topological charge sector to be
\begin{equation}
|\Psi_{ \,^{\bf g}a,\,^{\bf g}\bar{a};0}^{\alpha} \rangle = \sum_{s} A_{s} |\varphi_{\,^{\bf g}a,\,^{\bf g}\bar{a};0}^{s} \rangle
.
\end{equation}

We can now define the unitary operator $U_{\bf g}$ via the basis states of each subspace (with respect to which the operator is block diagonal)
\begin{eqnarray}
&& \langle \varphi_{\,^{\bf g}a,\,^{\bf g}\bar{a};0}^{r} | U_{\bf g} |\varphi_{\,^{\bf g}a,\,^{\bf g}\bar{a};0}^{s} \rangle \notag \\
&& \qquad \qquad = \langle \varphi_{\,^{\bf g}a,\,^{\bf g}\bar{a};0}^{r} | R_{\bf g} |\varphi_{a , \bar{a}; 0}^{s} \rangle
.
\end{eqnarray}
This gives the relation
\begin{equation}
R_{\bf g} |\Psi^{\alpha}_{a , \bar{a};0} \rangle = U_{\bf g} |\Psi^{\alpha}_{\,^{\bf g}a,\,^{\bf g}\bar{a};0} \rangle
\end{equation}
for any state $|\Psi^{\alpha}_{a , \bar{a};0} \rangle$ in the $|a , \bar{a};0 \rangle$ universality class. We emphasize that $U_{\bf g}$ is independent of $\alpha$, but it does depend on the choice of basis, and simply provides the relation between the orthonormal basis given by the states $|\varphi_{\,^{\bf g}a,\,^{\bf g}\bar{a};0}^{s} \rangle$ and the orthonormal basis given by the states $R_{\bf g} |\varphi_{a , \bar{a}; 0}^{s} \rangle$.

Since the quasiparticles are localized at well-separated positions, the system has exponentially decaying correlations, and the system is locally uniform and symmetric away from the quasiparticles, the states in the $|a , \bar{a}; 0 \rangle$ universality class will be locally indistinguishable from the ground state $|\Psi_0 \rangle$ of $H_0$ in (simply-connected) regions well-separated from the quasiparticles' $\mathcal{R}_1$ and $\mathcal{R}_2$. More specifically, we expect that any two such states $|\Psi_{a , \bar{a}; 0}^{\alpha} \rangle$ and $|\Psi_{a , \bar{a}; 0}^{\beta} \rangle$ in this universality class can be related by unitary operators acting independently in regions $\mathcal{R}_1$ and $\mathcal{R}_2$, i.e. there exist unitary operators $V^{(j)}$ whose nontrivial action is localized within $\mathcal{R}_{j}$ such that
\begin{equation}
|\Psi^{\beta}_{a , \bar{a};0} \rangle \approx V^{(1)} V^{(2)} |\Psi^{\alpha}_{a , \bar{a};0} \rangle
.
\end{equation}
The approximation in this expression is up to $O(e^{-r_{12}/\xi})$ corrections, which we will leave implicit in the following.~\footnote{We note that, given $H_{a,\bar{a};0}^{\alpha}$ and its ground state $|\Psi^{\alpha}_{a , \bar{a};0} \rangle$, it is always possible to construct a Hamiltonian for which another state $|\Psi^{\beta}_{a , \bar{a};0} \rangle$ in this universality class is the ground state. In particular, one can use
$h^{(j)}_{a;\beta}= V^{(j)} \left[h^{(j)}_{a;\alpha} + H_0^{(j)}  \right] V^{(j)-1} - H_0^{(j)}$,
where $H_0^{(j)}$ is the sum of the terms in $H_0$ that act nontrivially in $\mathcal{R}_j$.}

Thus, it follows that we can write the symmetry action as
\begin{align}
\label{eq:symmetryLocalization}
R_{\bf g} |\Psi_{a ,\bar{a};0} \rangle = U^{(1)}_{\bf g} U^{(2)}_{\bf g} U_{\bf g}(\,^{\bf g}a,\,^{\bf g}\bar{a};0) | \Psi_{\,^{\bf g}a,\,^{\bf g}\bar{a};0} \rangle,
\end{align}
for any state $|\Psi_{a ,\bar{a};0} \rangle$ in the $|a , \bar{a};0 \rangle$ universality class (we now drop the inconsequential label $\alpha$). In this expression, $U^{(1)}_{\bf g}$ and $U^{(2)}_{\bf g}$ are unitary operators whose nontrivial action is localized within $\mathcal{R}_1$ and $\mathcal{R}_2$, respectively. The quantity $U_{\bf g}(\,^{\bf g}a,\,^{\bf g}\bar{a};0)$ is precisely the transformation on the topological state space from Eqs.~(\ref{eq:rho_g_U_g})-(\ref{eq:beta_gh}) that leaves the basic data invariant. In particular, $U_{\bf g}(\,^{\bf g}a,\,^{\bf g}\bar{a};0)$ is an overall phase that depends only on the universality class of the state. Normally, one would safely ignore such an overall phase,  but we include it here to match with the symmetry action on the topological degrees of freedom, as this will play an essential role in the subsequent generalization to $n$ quasiparticles. In this way, we have decomposed $U_{\bf g}=U^{(1)}_{\bf g}U^{(2)}_{\bf g}U_{\bf g}(\,^{\bf g}a,\,^{\bf g}\bar{a};0)$ into terms that act locally around the quasiparticles and the term that acts on the topological state space. Clearly, $U^{(1)}_{\bf g}$ and $U^{(2)}_{\bf g}$ commute with each other, since their respective nontrivial actions are in two well-separated regions.

Given Eq.~(\ref{eq:symmetryLocalization}), we can define the operator
\begin{equation}
\rho_{\bf g} = U^{(1)-1}_{\bf g} U^{(2)-1}_{\bf g} R_{\bf g}
\end{equation}
acting on the physical Hilbert space that has the same action on states $|\Psi_{a ,\bar{a};0} \rangle$ in the $|a , \bar{a};0 \rangle$ universality class as does the previously defined symmetry operator $\rho_{\bf g}$ ( see Sec.~\ref{sec:Global_Symmetry} ) acting on $|a , \bar{a};0 \rangle$ in the topological state space, i.e.
\begin{equation}
\rho_{\bf g} |\Psi_{a ,\bar{a};0} \rangle = U_{\bf g}(^{\bf g}a, \,^{\bf g}\bar{a};0) | \Psi_{\,^{\bf g}a,\,^{\bf g}\bar{a};0} \rangle
.
\end{equation}
We note that, similar to $R_{\bf g}$, this operator also has the form $\rho_{\bf g} = \prod_{k \in I} \rho_{\bf g}^{(k)}$.~\footnote{From this, we can see that $ |\Psi^{\alpha}_{\rho_{\bf g}(a) , \rho_{\bf g}(\bar{a});0} \rangle$ is the ground state of the Hamiltonian
$H_{\rho_{\bf g}(a),\rho_{\bf g}(\bar{a});0}^{\alpha} = \rho_{\bf g} H_{a,\bar{a};0}^{\alpha} \rho_{\bf g}^{-1}$,
for which the corresponding
$h_{\rho_{\bf g}(a);\alpha}^{(j)} = \rho_{\bf g} \left[ h_{a;\alpha}^{(j)} +H_{0}^{(j)} \right] \rho_{\bf g}^{-1} - H_{0}^{(j)}$
is again localized within $\mathcal{R}_{j}$, where $H_0^{(j)}$ is the sum of the terms in $H_0$ that act nontrivially in $\mathcal{R}_j$.}

\begin{figure}[t!]
\begin{center}
\includegraphics[scale=0.23]{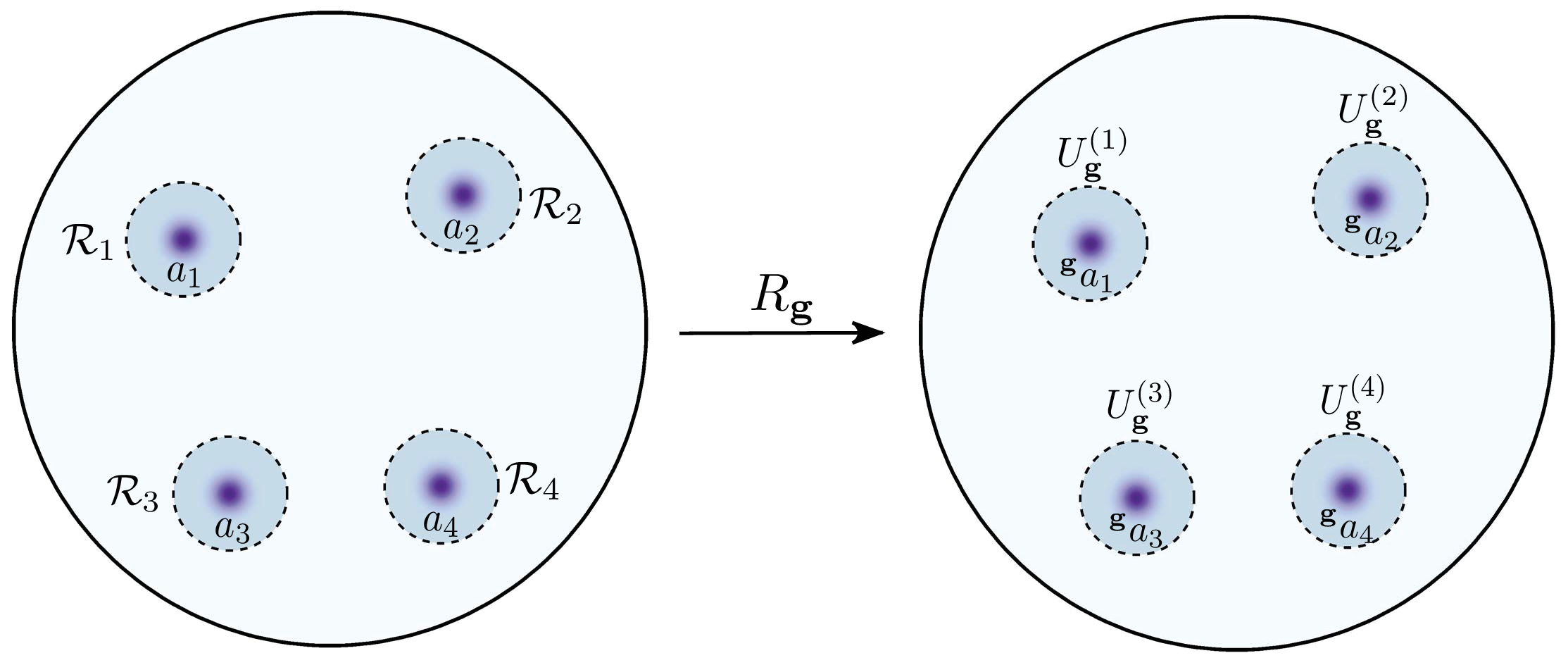}
\caption{The global on-site symmetry action on states containing quasiparticles takes the form given in Eq.~(\ref{eq:symmetryLocalization_n}), where the global action $R_{\bf g}$ factorizes into the global symmetry action operator $\rho_{\bf g}$, which acts only on the topological quantum numbers, and local transformations $U^{(j)}_{\bf g}$, each of which only acts nontrivially within a region $\mathcal{R}_{j}$ well-localized around the $j$th quasiparticle carrying topological charge $a_j$.}
\label{fig:frac}
\end{center}
\end{figure}

We now generalize to consider the system in a configuration with $n$ quasiparticles with corresponding topological charges $a_j$ localized in well-separated regions $\mathcal{R}_{j}$ (for $j=1,\ldots,n$), with corresponding Hamiltonians
\begin{equation}
H_{a_1 , \ldots , a_n ;0}^{\alpha} = H_{0} + \sum_{j=1}^{n} h_{a_j ;\alpha}^{(j)}
.
\end{equation}
The same steps can be followed as above, though one must be more careful to properly account for fusion degeneracies. In particular, there will be $N_{ab}^{c}$ distinct ways to create two quasiparticles with respective charges $a$ and $b$ from a single quasiparticle of topological charge $c$, and this will be reflected in the corresponding states and Hamiltonians. For a system with $n$ quasiparticles, the topological state space may be degenerate, with the dimensionality given by
\begin{equation}
N_{a_1 \ldots a_n}^{0}= \sum_{c_{12},c_{123},\ldots,c_{1 \ldots n-1}} N_{a_1 a_2}^{c_{12}} N_{c_{12} a_3}^{c_{123}} \ldots N_{c_{1 \ldots n-1} a_n}^{0}
,
\end{equation}
where here we use the standard basis decomposition of the topological state space where topological charges are fused together successively in increasing order of $j$, and $c_{1 \ldots k}$ is the collective topological charge of quasiparticles $1, \ldots, k$. The states will correspondingly carry the labels $c_{12},\ldots,c_{1 \ldots n-1}$, as well as the fusion space basis labels $\mu_{12},\ldots,\mu_{1\ldots n-1}$. (We can, of course, write the states in a different basis related by $F$-moves.) We write all these topological charges and fusion basis labels of the state collectively as $\{a;c,\mu\}$, with the understanding that the overall fusion channel of the $n$ quasiparticles is vacuum (i.e. $c_{1 \ldots n}=0$), so we can more compactly write a state in this universality class as $|\Psi_{\{a;c,\mu \}} \rangle$. Following the same arguments given above, we find that the symmetry action on such states will take the form
\begin{equation}
\label{eq:symmetryLocalization_n}
R_{\bf g} |\Psi_{\{a;c,\mu \}} \rangle
= U^{(1)}_{\bf g} \ldots U^{(n)}_{\bf g} \rho_{\bf g}|\Psi_{\{a;c,\mu \}} \rangle,
\end{equation}
where the unitary operator $U^{(j)}_{\bf g}$ has its nontrivial action localized within $\mathcal{R}_j$. This is shown schematically in Fig.~\ref{fig:frac}. (Again, the operators $U^{(j)}_{\bf g}$ depend on a choice of basis within the universality class, but not on the particular state it is acting upon.) Here, we use the generalized definition of the operator (in the physical Hilbert space)
\begin{equation}
\rho_{\bf g} = \prod_{j=1}^{n} U^{(j)-1}_{\bf g} R_{\bf g}
\label{eq:rho_g-R_g-relation}
\end{equation}
which acts on physical states $|\Psi_{\{a;c,\mu \}} \rangle$ in the universality class $|\{a;c,\mu \} \rangle$ precisely as the operator $\rho_{\bf g}$ acts on states $|\{a;c,\mu \} \rangle$ in the topological state space. Explicitly, this is given by
\begin{widetext}
\begin{eqnarray}
\rho_{\bf g} |\Psi_{\{a;c,\mu \}} \rangle &=& \sum_{\mu_{12}',\ldots , \mu_{1 \ldots n-1}' } \left[ U_{\bf g}( \,^{\bf g} a_1 , \,^{\bf g}a_2;\,^{\bf g}c_{12})\right]_{\mu_{12} \mu_{12}'} \left[ U_{\bf g}( \,^{\bf g}c_{12} , \,^{\bf g}a_3; \,^{\bf g}c_{123})\right]_{\mu_{123} \mu_{123}'} \times \ldots \notag \\
&& \qquad \quad \ldots \times \left[ U_{\bf g}(\,^{\bf g}c_{1\ldots n-2 } , \,^{\bf g}a_{n-1}; \,^{\bf g}c_{1\ldots n-1 })\right]_{\mu_{1\ldots n-1} \mu_{1\ldots n-1}'}  U_{\bf g}(\,^{\bf g}c_{1\ldots n-1 }, \,^{\bf g}a_n ;0) |\Psi_{\{\,^{\bf g}a; \,^{\bf g}c,\mu' \}} \rangle . \quad
\label{eq:rho_g_action}
\end{eqnarray}
\end{widetext}
Given the physical states containing quasiparticles and the symmetry transformations $R_{\bf g}$ acting upon them, one may use these expressions as a means of determining the global symmetry action $\rho_{\bf g}$ on the topological state space.

We can consider symmetry transformations taking the form in Eq.~(\ref{eq:symmetryLocalization_n}) when acting on states in the physical Hilbert space containing quasiparticles in a topological phase to be the fundamental condition from which the symmetry fractionalization arguments follow, regardless of the particular form of the Hamiltonian.

\subsection{Obstruction to Fractionalization}
\label{sec:Obstruction}

We will allow the global symmetry action to form either linear or projective representations of the symmetry group when acting on the physical Hilbert space, but first consider the case of linear representations of the global symmetry, and then return to the case of projective representations in Sec.~\ref{sec:Proj_Rep_Glob_Symm}. For linear representations, the symmetry operators will satisfy $R_{\bf gh} = R_{\bf g} R_{\bf h} $. However, the local operators $U^{(j)}_{\bf g}$ can nonetheless take a projective form, and we wish to classify the types of projective forms that they can realize. We compare the action of ${\bf gh}$, which is given by
\begin{eqnarray}
R_{\bf gh}|\Psi_{\{a;c,\mu \}} \rangle &=&  \prod_{j=1}^{n} U^{(j)}_{\bf gh} \rho_{\bf gh}|\Psi_{\{a;c,\mu \}} \rangle \notag \\
&=&  \prod_{j=1}^{n} U^{(j)}_{\bf gh} \kappa_{\bf g,h} \rho_{\bf g} \rho_{\bf h}|\Psi_{\{a;c,\mu \}} \rangle
,
\end{eqnarray}
where $\kappa_{\bf g,h} = \rho_{\bf gh} \rho_{\bf h}^{-1} \rho_{\bf g}^{-1}$ (as in Sec.~\ref{sec:Global_Symmetry}), and the successive actions of ${\bf g}$ and ${\bf h}$, which is given by
\begin{eqnarray}
&& R_{\bf g} R_{\bf h} |\Psi_{\{a;c,\mu \}} \rangle = R_{\bf g} \prod_{j=1}^{n} U^{(j)}_{\bf h} \rho_{\bf h}|\Psi_{\{a;c,\mu \}} \rangle
\nonumber \\
&&= R_{\bf g} \prod_{j=1}^{n} U^{(j)}_{\bf h} R^{-1}_{\bf g}R_{\bf g} \rho_{\bf h}|\Psi_{\{a;c,\mu \}} \rangle \notag \\
&&= R_{\bf g} \prod_{j=1}^{n} U^{(j)}_{\bf h} R^{-1}_{\bf g} \prod_{k=1}^{n} U^{(k)}_{\bf g} \rho_{\bf g} \rho_{\bf h} |\Psi_{\{a;c,\mu \}} \rangle  \notag \\
&& = \prod_{j=1}^{n} \,^{\bf g} U^{(j)}_{\bf h} U^{(j)}_{\bf g} \rho_{\bf g} \rho_{\bf h} |\Psi_{\{a;c,\mu \}}  \rangle ,
\end{eqnarray}
where $^{\bf g} U^{(j)}_{\bf h} = R_{\bf g} U^{(j)}_{\bf h} R^{-1}_{\bf g}= R^{(j)}_{\bf g} U^{(j)}_{\bf h} R^{(j)-1}_{\bf g}$ has its nontrivial action localized within the region $\mathcal{R}_{j}$, and we used the fact that operators whose nontrivial actions are localized in different regions commute with each other.
Comparing these expressions, we see that
\begin{equation}
\prod_{j=1}^{n} U^{(j)-1}_{\bf g} \,^{\bf g} U^{(j)-1}_{\bf h}  U^{(j)}_{\bf gh} \kappa_{\bf g,h} = \openone
\end{equation}
when acting in the subspace of states of the form $|\Psi_{\{a;c,\mu \}}  \rangle$ corresponding to the system with $n$ quasiparticles.
We note that
\begin{equation}
^{\bf g} \mathcal{O}^{(j)} U^{(j)}_{\bf g} = U^{(j)}_{\bf g} \rho_{\bf g} \mathcal{O}^{(j)} \rho_{\bf g}^{-1}
,
\label{eq:R_g-to-rho_g-conj}
\end{equation}
for any operator $\mathcal{O}^{(j)}$ localized in $\mathcal{R}_{j}$, so we could rewrite these expressions using $U^{(j)}_{\bf g} \rho_{\bf g} U^{(j)}_{\bf h} \rho_{\bf g}^{-1}$ instead of $\,^{\bf g} U^{(j)}_{\bf h} U^{(j)}_{\bf g}$, if desired.

Since the action of $\rho_{\bf g}$ on the physical states of the form $|\Psi_{\{a;c,\mu \}} \rangle$ is precisely the same as the action of $\rho_{\bf g}$ on the states $|\{a;c,\mu \} \rangle$ in the topological state space, we know that the action of $\kappa_{\bf g,h}$ on physical states of the form $|\Psi_{\{a;c,\mu \}} \rangle$ also matches the action of $\kappa_{\bf g,h}$ in the topological state space, and thus takes the form
\begin{equation}
\kappa_{\bf g,h}|\Psi_{\{a;c,\mu \}} \rangle = \prod_{j=1}^{n} \beta_{a_j}({\bf g,h}) |\Psi_{\{a;c,\mu \}} \rangle
,
\end{equation}
where $\beta_{a}({\bf g,h})$ are the phases defined in Sec.~\ref{sec:Global_Symmetry} that depends only on the topological charge value $a$, and group elements ${\bf g}$ and ${\bf h}$. Let us define a unitary operator $B^{(j)}_{\bf g,h}$ localized in region $\mathcal{R}_{j}$ whose action on a quasiparticle state produces the phase $\beta_{a_j}({\bf g,h})$ of the topological charge contained in the region $\mathcal{R}_{j}$, that is~\footnote{Such an operator localized in $\mathcal{R}_{j}$ can be defined, for example, by taking
$B^{(j)}_{\bf g,h} = \sum_{a,b} \beta_{a}({\bf g,h}) S_{0 a }S_{ab} W_{b}(\partial \mathcal{R}_{j})$, where $W_{b}(\partial \mathcal{R}_{j})$ is a Wilson loop of topological charge $b$ whose path follows the loop delineated by the boundary $\partial \mathcal{R}_{j}$ (or just inside the boundary) of the region in a counterclockwise fashion.}
\begin{equation}
B^{(j)}_{\bf g,h} |\Psi_{\{a;c,\mu \}} \rangle = \beta_{a_j}({\bf g,h}) |\Psi_{\{a;c,\mu \}} \rangle
.
\end{equation}

We can now define the unitary operators
\begin{eqnarray}
W^{(j)}_{\bf g,h} &=&  U^{(j)-1}_{\bf g} \,^{\bf g} U^{(j)-1}_{\bf h}  U^{(j)}_{\bf gh} B^{(j)}_{\bf g,h} \notag \\
&=& \rho_{\bf g} U^{(j)-1}_{\bf h} \rho_{\bf g}^{-1} U^{(j)-1}_{\bf g} U^{(j)}_{\bf gh} B^{(j)}_{\bf g,h}
.
\end{eqnarray}
Since the $U^{(j)}_{\bf g}$ and $B^{(j)}_{\bf g,h}$  are all unitary operators with nontrivial action localized within the region $\mathcal{R}_{j}$, this is also true for $W^{(j)}_{\bf g,h}$. From the above relations, we see that
\begin{equation}
\label{eq:prod_W_j}
\prod_{j=1}^{n} W^{(j)}_{\bf g,h} = \openone
\end{equation}
when acting in the subspace of $n$ quasiparticles states of the form $|\Psi_{\{a;c,\mu \}} \rangle$, for any values of $\{a;c,\mu \}$.

Since the respective regions $\mathcal{R}_{j}$ where $W^{(j)}_{\bf g,h}$ act nontrivially are well-separated from each other, each one of these operators can, at most, change a state of the form $|\Psi_{\{a;c,\mu \}} \rangle$ by an overall phase factor. Hence, we have
\begin{equation}
\label{eq:phase_omega_j}
\langle \Psi_{\{a;c,\mu \}} | W^{(j)}_{\bf g,h} | \Psi_{\{b;e,\nu \}} \rangle = \omega_{a_j}({\bf g,h} ) \delta_{\{a;c,\mu \}\{b;e,\nu \}} ,
\end{equation}
where the phase $\omega_{a_j}({\bf g,h} )$ only depends on the topological charge $a_j$ contained in the region $\mathcal{R}_j$.

In order to see that the phases $\omega_{a_j}({\bf g,h} )$ do not depend on anything else, we first note that the phase factor can obviously depend, at most, on the group elements ${\bf g}$ and ${\bf h}$, and the properties of the state $| \Psi_{\{a;c,\mu \}} \rangle$ that are local to the region $\mathcal{R}_{j}$. In order to see that the only property of the state that the phase depends on is the topological charge contained in the region $\mathcal{R}_{j}$, we must show that the phase is actually independent of the specific state $| \Psi_{\{a;c,\mu \}} \rangle$ taken from the $|\{a;c,\mu \} \rangle$ universality class. For this, assume that the phase may depend on the specific state, which we indicate by writing it as $\omega({\bf g,h} ;\Psi_{\{a;c,\mu \}} )$. Then consider any two orthonormal states $|\Psi^{\alpha}_{\{a;c,\mu \}} \rangle$ and $|\Psi^{\beta}_{\{a;c,\mu \}} \rangle$ from this universality class, and their normalized superposition $|\Psi^{\gamma}_{\{a;c,\mu \}} \rangle = C_{\alpha} |\Psi^{\alpha}_{\{a;c,\mu \}} \rangle +C_{\beta} |\Psi^{\beta}_{\{a;c,\mu \}} \rangle$. The above expression yields the relation
\begin{eqnarray}
\omega({\bf g,h} ;\Psi^{\gamma}_{\{a;c,\mu \}} ) &=& |C_{\alpha}|^2 \omega({\bf g,h} ;\Psi^{\alpha}_{\{a;c,\mu \}} ) \notag \\
&&  +|C_{\beta}|^2 \omega({\bf g,h} ;\Psi^{\beta}_{\{a;c,\mu \}} )
\end{eqnarray}
which can only be true for arbitrary $C_{\alpha}$ and $C_{\beta}$ if
\begin{equation}
\omega({\bf g,h} ;\Psi^{\gamma}_{\{a;c,\mu \}} ) =  \omega({\bf g,h} ;\Psi^{\alpha}_{\{a;c,\mu \}} ) =\omega({\bf g,h} ;\Psi^{\beta}_{\{a;c,\mu \}} )
\end{equation}
which shows that the phase is the same for all states in the universality class. Since the only universal property of the state that is local to the region $\mathcal{R}_{j}$ is the topological charge $a_j$ contained in that region, this establishes the claimed dependence of the phase.

It follows that, within the subspace of states of the form $|\Psi_{\{a;c,\mu \}} \rangle$, the operators $W^{(j)}_{\bf g,h}$, $W^{(j)}_{\bf k,l}$, $B^{(j)}_{\bf g,h}$, and $B^{(j)}_{\bf k,l}$ all commute with each other. It also follows that
\begin{eqnarray}
\eta_{a_j}({\bf g,h}) U^{(j)}_{\bf gh} |\Psi_{\{a;c,\mu \}} \rangle &=& U^{(j)}_{\bf g} \rho_{\bf g} U^{(j)}_{\bf h} \rho_{\bf g}^{-1} |\Psi_{\{a;c,\mu \}} \rangle  \notag \\
&=& \,^{\bf g} U^{(j)}_{\bf h} U^{(j)}_{\bf g} |\Psi_{\{a;c,\mu \}} \rangle
\label{eq:U_ghU_gU_h}
,
\end{eqnarray}
where the projective phases are given by
\begin{equation}
\eta_{a}({\bf g,h}) = \frac{ \beta_{a}({\bf g,h}) }{ \omega_{a}({\bf g,h} ) }
\label{eq:eta_defn}
.
\end{equation}
Eq.~(\ref{eq:U_ghU_gU_h}) exhibits a characteristic property of symmetry fractionalization, which is that the action of the symmetry can be broken up into topological and local actions, where the local actions are locally consistent in a projective fashion. Of course, the topological action is topologically consistent, and the local and topological actions must also be consistent with each other. For this, we have already decomposed the consistency of the topological action into terms $\beta_{a_j}({\bf g,h})$ that only depend on the localized topological charge values, and must now examine the phases $\omega_{a_j}({\bf g,h} )$ to analyze the consistency of the interplay between the local and topological actions of the symmetry.

It is clear that we should have
\begin{equation}
\eta_{0}({\bf g,h}) =1
,
\end{equation}
since the symmetry action on the ground state is trivial (and any region $\mathcal{R}_{j}$ containing total topological charge $a_j=0$ can be locally transformed into the ground state). Additionally, we will always fix
\begin{equation}
\eta_{a}({\bf 0,0})=\eta_{a}({\bf g,0})=\eta_{a}({\bf 0,h}) =1
,
\end{equation}
since we can always freely set $U^{(j)}_{\bf 0} = \openone$ as a gauge choice, which we will describe in more detail in Sec.~\ref{sec:Frac_Gauge_Trans}. It follows that we also have $\omega_{a}({\bf 0,0})=\omega_{a}({\bf g,0})=\omega_{a}({\bf 0,h}) =1$.

Given Eq.~(\ref{eq:prod_W_j}), the phases $\omega_{a_j}({\bf g,h} )$ must obey the constraint
\begin{equation}
\label{eq:prod_omega_j}
\prod_{j=1}^{n} \omega_{a_j}({\bf g,h} ) =1
.
\end{equation}
We emphasize that this does not mean that the product of the phases $\prod_{j=1}^{n} \eta_{a_j}({\bf g,h}) $ is equal to 1, nor that the product of the phases $\prod_{j=1}^{n}\beta_{a_j}({\bf g,h})$ is equal to 1. These products would only individually equal 1 when $\kappa_{\bf g,h}=\openone$, which is not generally true (though, this condition is often satisfied by examples of physical interest).

Considering the case of $n=2$ quasiparticles with respective topological charges $a$ and $\bar{a}$, we find the relation
\begin{equation}
\omega_{\bar{a}}({\bf g,h} ) = \omega_{a}({\bf g,h} )^{-1}
.
\end{equation}
Considering the case of $n=3$ quasiparticles, with respective topological charges $a$, $b$, and $\bar{c}$, for which $N_{ab}^{c}\neq 0$, and using the result from the $n=2$ case, we find the relation
\begin{equation}
\omega_{a}({\bf g,h} ) \omega_{b}({\bf g,h} ) = \omega_{c}({\bf g,h} )
\label{eq:omega_a_b_c relation}
\end{equation}
for any charges $a$, $b$, and $c$ with $N_{ab}^{c} \neq 0$.
Thus, as explained at the end of Sec.~\ref{sec:Braiding}, the phase factors are given by
\begin{equation}
\omega_{a}({\bf g,h} ) = M_{a \cohosub{w}\left( {\bf g,h} \right)}^{\ast}
,
\end{equation}
for some Abelian topological charge value $\coho{w}\left( {\bf g,h} \right)\in \mathcal{A} \subset \mathcal{C}$. Through this relation, the object $\coho{w}\left( {\bf g,h} \right)$ provides a consistent specification of the phases $\omega_{a}({\bf g,h} )$ for all values of topological charges $a$ simultaneously.

It also follows from Eqs.~(\ref{eq:eta_defn}) and (\ref{eq:omega_a_b_c relation}) that
\begin{equation}
\label{eq:eta_kappa_relation}
\frac{ \eta_{a}({\bf g,h}) \eta_{b}({\bf g,h}) }{ \eta_{c}({\bf g,h}) } = \frac{ \beta_{a}({\bf g,h}) \beta_{b}({\bf g,h}) }{ \beta_{c}({\bf g,h}) } = \kappa_{\bf g,h}(a,b;c)
\end{equation}
whenever $N_{ab}^{c}\neq 0$.

Next, we consider the product of three symmetry operations and apply the relation $U^{(j)}_{\bf gh} =  \,^{\bf g}U^{(j)}_{\bf h} U^{(j)}_{\bf g} W^{(j)}_{\bf g,h} B^{(j)-1}_{\bf g,h}$ in the two distinct, but equivalent orders to obtain
\begin{widetext}
\begin{eqnarray}
U^{(j)}_{\bf ghk} &=& \,^{\bf gh} U^{(j)}_{\bf k} U^{(j)}_{\bf gh} W^{(j)}_{\bf gh,k} B^{(j)-1}_{\bf gh,k}
= \,^{\bf gh} U^{(j)}_{\bf k} \,^{\bf g} U^{(j)}_{\bf h} U^{(j)}_{\bf g} W^{(j)}_{\bf g,h} B^{(j)-1}_{\bf g,h} W^{(j)}_{\bf gh,k} B^{(j)-1}_{\bf gh,k} \notag \\
&=& \,^{\bf g} U^{(j)}_{\bf hk} U_{\bf g} W^{(j)}_{\bf g,hk} B^{(j)-1}_{\bf g,hk}
= \,^{\bf gh} U^{(j)}_{\bf k} \,^{\bf g} U^{(j)}_{\bf h} \,^{\bf g} W^{(j)}_{\bf h,k} \,^{\bf g} B^{(j)-1}_{\bf h,k}  U^{(j)}_{\bf g} W^{(j)}_{\bf g,hk} B^{(j)-1}_{\bf g,hk} \notag \\
&=& \,^{\bf gh} U^{(j)}_{\bf k} \,^{\bf g} U^{(j)}_{\bf h} U^{(j)}_{\bf g} \rho_{\bf g} W^{(j)}_{\bf h,k} \rho_{\bf g}^{-1} \rho_{\bf g} B^{(j)-1}_{\bf h,k} \rho_{\bf g}^{-1} W^{(j)}_{\bf g,hk} B^{(j)-1}_{\bf g,hk}
.
\label{eq:U_{ghk}}
\end{eqnarray}
This gives the relation
\begin{equation}
\label{eq:WB_cocycle}
\rho_{\bf g} W^{(j)}_{\bf h,k} \rho_{\bf g}^{-1} \rho_{\bf g} B^{(j)-1}_{\bf h,k} \rho_{\bf g}^{-1} W^{(j)}_{\bf g,hk} B^{(j)-1}_{\bf g,hk}
= W^{(j)}_{\bf g,h} B^{(j)-1}_{\bf g,h} W^{(j)}_{\bf gh,k} B^{(j)-1}_{\bf gh,k},
\end{equation}
which, when applied to a state $|\Psi_{\{a;c,\mu \}} \rangle$, yields the crucial relation
\begin{eqnarray}
\Omega_{a}({\bf g,h,k}) &=& \beta_{\rho_{\bf g}^{-1}(a)}({\bf h,k}) \beta_{a}({\bf gh,k})^{-1}  \beta_{a}({\bf g,hk}) \beta_{a}({\bf g,h})^{-1} \notag \\
&=& \omega_{\rho^{-1}_{\bf g}(a)}({\bf h,k} ) \omega_{a}({\bf gh,k} )^{-1} \omega_{a}({\bf g,hk} ) \omega_{a}({\bf g,h} )^{-1}
\label{eq:omega_a_relation}
\end{eqnarray}
where we use the definition of $\Omega_{a}({\bf g,h,k})$ from Sec.~\ref{sec:H3Invariance}.
This relation is equivalent to the condition
\begin{equation}
\eta_{\rho_{\bf g}^{-1}(a)}({\bf h,k}) \eta_{a}({\bf gh,k})^{-1}  \eta_{a}({\bf g,hk}) \eta_{a}({\bf g,h})^{-1} = 1
\label{eq:eta_twisted_cocycle}
\end{equation}
on the projective phases of the local terms, which is a sort of twisted $2$-cocycle condition. From Eq.~(\ref{eq:eta_twisted_cocycle}), one might na\"{i}vely expect a classification of fractionalization by something like a separate $H^2(G,\text{U}(1))$ for each topological charge value $a$, particularly when the symmetry action does not permute topological charge types. However, the relation between $\eta_{a}({\bf g,h})$, $\beta_{a}({\bf g,h})$, and $\omega_{a}({\bf g,h} )$, as well as a potentially nontrivial group action, introduce additional structure. Specifically, the relation between the phases with different values of topological charge given by Eq.~(\ref{eq:prod_omega_j}) requires consistency of the fractionalization phases for different values of topological charge in a way that leads to classification through the objects $\coho{w}\left( {\bf g,h} \right) \in C^2(G, \mathcal{A})$, as we will now describe.

Using $\Omega_{a}({\bf g,h,k}) = M_{a \cohosub{O}\left( {\bf g,h,k} \right)}^{\ast}$ and $\omega_{a}({\bf g,h} ) = M_{a \cohosub{w}\left( {\bf g,h} \right)}^{\ast}$, where $\coho{O}\left( {\bf g,h,k} \right) \in Z^3_{\rho}(G, \mathcal{A})$ and $\coho{w}\left( {\bf g,h} \right) \in C^2(G, \mathcal{A})$ are Abelian topological charges, together with the relation $S_{ab}^{\ast}=S_{a\bar{b}}$ and the symmetry property $S_{\rho_{\bf g}(a)\rho_{\bf g}(b)} = S_{ab}$, Eq.~(\ref{eq:omega_a_relation}) becomes
\begin{eqnarray}
M_{a \cohosub{O}\left( {\bf g,h,k} \right)}
&=& M_{a \rho_{\bf g}[\cohosub{w}\left( {\bf h,k} \right)]} M_{a \cohosub{w}\left( {\bf gh,k} \right)}^{\ast} M_{a \cohosub{w}\left( {\bf g,hk} \right)} M_{a \cohosub{w}\left( {\bf g,h} \right)}^{\ast} \notag \\
&=&  M_{a \rho_{\bf g}[\cohosub{w}\left( {\bf h,k} \right)]} M_{a \overline{\cohosub{w}\left( {\bf gh,k} \right)}} M_{a \cohosub{w}\left( {\bf g,hk} \right)} M_{a \overline{\cohosub{w}\left( {\bf g,h} \right)}} \notag \\
&=&  M_{a ,\rho_{\bf g}[\cohosub{w}\left( {\bf h,k} \right)]\times \overline{\cohosub{w}\left( {\bf gh,k} \right)} \times \cohosub{w}\left( {\bf g,hk} \right) \times \overline{\cohosub{w}\left( {\bf g,h} \right)}}
.
\end{eqnarray}
In the last line, we used the fact that if $M_{ab}$ is a phase and $N_{bc}^{e}\neq 0$, then it follows that $M_{ab} M_{ac} = M_{ae}$.
Finally, the non-degeneracy of braiding in a MTC makes this equivalent to the condition
\begin{equation}
\label{eq:kappa_depsilon}
\coho{O}\left( {\bf g,h,k} \right) = \rho_{\bf g}[\coho{w}\left( {\bf h,k} \right)]\times \overline{\coho{w}\left( {\bf gh,k} \right)} \times \coho{w}\left( {\bf g,hk} \right) \times \overline{\coho{w}\left( {\bf g,h} \right)}= \text{d} \coho{w} \left( {\bf g,h,k} \right)
.
\end{equation}
\end{widetext}
Thus, we have found that consistency between the local and topological portions of the symmetry action requires that $\coho{O}\left( {\bf g,h,k} \right)$ is necessarily a 3-coboundary, which is to say that $\coho{O} \in B^3_{\rho}(G, \mathcal{A})$ and its equivalence class is $[\coho{O}] = [0]$.
This establishes the first statement regarding symmetry fractionalization, which was that $[\coho{O}] \neq [0]$ indicates that there is an obstruction to fractionalizing the symmetry, since this would contradict the result in Eq.~(\ref{eq:kappa_depsilon}). In particular, such an obstruction implies that it is not actually possible for the symmetry of the system to take the assumed on-site form of Eq.~(\ref{eq:R_onsite}) with the corresponding action on quasiparticle states given in Eq.~(\ref{eq:symmetryLocalization_n}), as the symmetry action cannot be consistently split into local and topological components.

When the symmetry action does not permute topological charge values, one can interpret Eq.~(\ref{eq:U_ghU_gU_h}) as indicating that the local operators $U^{(j)}_{\bf g}$ provide projective representations of the group $G$. In particular, the equivalence class $[\eta_{a}({\bf g,h})] \in H^2(G,\text{U}(1))$ defined by the phases $\eta_{a}({\bf g,h})$ identified under multiplication by $B^2(G,\text{U}(1))$ coboundaries specifies the projective representation. However, the allowed projective representations for a given topological charge value must be consistent with those of the other topological charge values, so, even in this case, the classification is more complicated than simply taking the product of $|\mathcal{C}|$ independent projective representations.

\subsection{Gauge Transformations}
\label{sec:Frac_Gauge_Trans}

There is gauge freedom to redefine the local operators $U^{(j)}_{\bf g}$ by the local transformations
\begin{equation}
\check{U}^{(j)}_{\bf g} = U^{(j)}_{\bf g} Y_{\bf g}^{(j)-1}
\end{equation}
where $Y_{\bf g}^{(j)}$ are unitary operators whose nontrivial action is localized in region $\mathcal{R}_{j}$. In order to leave the global operator $R_{\bf g}$ unchanged, there must be a corresponding transformation of the symmetry action operator
\begin{equation}
\check{\rho}_{\bf g} = \prod_{j=1}^{n} Y^{(j)}_{\bf g} \rho_{\bf g}
.
\end{equation}
In order for this operator to again act on the physical states with quasiparticles as does a symmetry action on the topological state space, we require it to only depend on the topological quantum numbers (and the group element ${\bf g}$). Since $Y^{(j)}_{\bf g}$ acts locally in region in the region $\mathcal{R}_{j}$, the only topological quantum number it can depend upon is the topological charge $a_j$ in that region. Thus, we must have
\begin{equation}
\langle \Psi_{\{a;c,\mu \}} | Y_{\bf g}^{(j)} | \Psi_{\{b;e,\nu \}} \rangle = \gamma_{a_j}({\bf g} ) \delta_{\{a;c,\mu \}\{b;e,\nu \}} ,
\end{equation}
where $\gamma_{a_j}({\bf g} )$ is some phase factor that depends only on the topological charge $a_j$ and the group element ${\bf g}$. Of course, the notation we used here anticipated the fact that these gauge transformations have precisely the form of natural isomorphisms, as described in Sec.~\ref{sec:Symmetry} by
\begin{equation}
\check{\rho}_{\bf g} = \Upsilon_{\bf g} \rho_{\bf g}
,
\end{equation}
with corresponding decomposition into the phase factors $\gamma_{a}({\bf g} )$ when acting on fusion vertex states.

We notice that, under these transformations, the projective phases transform as
\begin{equation}
\label{eq:check_eta}
\check{\eta}_{a}({\bf g,h})= \frac{\gamma_{a}({\bf gh}) } { \gamma_{\,^{\bf \bar{g}}a}({\bf h}) \gamma_{a}({\bf g})  } \eta_{a}({\bf g,h})
.
\end{equation}
For the choice of
\begin{equation}
\check{\beta}_{a}({\bf g,h})= \frac{\gamma_{a}({\bf gh}) } { \gamma_{\,^{\bf \bar{g}}a}({\bf h}) \gamma_{a}({\bf g})  } \beta_{a}({\bf g,h})
,
\end{equation}
as in Eq.~(\ref{eq:eta'_gh}), this exactly cancels to leave $\check{\omega}_{a}({\bf g,h} ) = \omega_{a}({\bf g,h} )$ unchanged. As previously mentioned, it also leaves $\check{\Omega}_{a}({\bf g,h,k})=\Omega_{a}({\bf g,h,k})$ and hence $\check{\coho{O}}\left( {\bf g,h,k} \right) = \coho{O}\left( {\bf g,h,k} \right) $ unchanged.

In this way, the nontrivial transformations of these quantities are relegated to the transformations
\begin{equation}
\breve{\beta}_{a}({\bf g,h})= \nu_{a}({\bf g,h}) \beta_{a}({\bf g,h})
,
\label{eq:breve_eta}
\end{equation}
where $\nu_{a}({\bf g,h}) \nu_{b}({\bf g,h}) = \nu_{c}({\bf g,h})$ whenever $N_{ab}^{c}\neq 0$, corresponding to the freedom of decomposing the action of $\kappa_{\bf g,h}$ on vertices into factors $\beta_{a}({\bf g,h})$. These transformations give $\breve{\omega}_{a}({\bf g,h} ) =  \nu_{a}({\bf g,h}) \omega_{a}({\bf g,h} )$, while $\breve{\Omega}_{a}({\bf g,h,k})$ and $\breve{\coho{O}}\left( {\bf g,h,k} \right)$ are given in Eqs.~(\ref{eq:Omega_tilde}) and (\ref{eq:O_tilde}). We emphasize that the projective phases are left unchanged by these transformations, i.e. $\breve{\eta}_{a}({\bf g,h})= \eta_{a}({\bf g,h})$, so they do not change the symmetry fractionalization.

\subsection{Classification of Symmetry Fractionalization}
\label{sec:Classification_Frac}

We now wish to classify the different ways in which the symmetry can fractionalize, when there is no obstruction. For this, we must analyze the solutions of Eq.~(\ref{eq:kappa_depsilon}) for a given $\rho$ and $\coho{O}$.

Since $[\coho{O}] = [0]$, there must exist some $\coho{v}({\bf g,h}) \in C^{2}(G,\mathcal{A})$ such that $\coho{O} = \text{d}\bar{\coho{v}}$. This is just the equivalence class statement that one can use the gauge transformation in Eq.~(\ref{eq:breve_eta}) for some $\nu_{a}({\bf g,h})=M^{\ast}_{a \cohosub{v}({\bf g,h}) }$ which results in $\breve{\Omega}_{a}({\bf g,h,k})=1$ and $\breve{\coho{O}} = 0$. Thus, we are guaranteed to have at least one solution of Eq.~(\ref{eq:kappa_depsilon}) given by $\coho{w} = \coho{v}$.

Given a solution $\coho{w}({\bf g,h})$ of Eq.~(\ref{eq:kappa_depsilon}), it is straightforward to see that another solution
\begin{equation}
\coho{w}'({\bf g,h}) = \coho{t}({\bf g,h}) \times \coho{w}({\bf g,h})
\end{equation}
can be obtained from it by multiplying by a $2$-cocycle $\coho{t}({\bf g,h}) \in Z^2_{\rho}(G, \mathcal{A})$. In fact, it should be clear that all solutions of Eq.~(\ref{eq:kappa_depsilon}) may be obtained from any given solution in this way.

Assuming $\beta_{a}({\bf g,h})$ is fixed, this way of obtaining different solutions of Eq.~(\ref{eq:kappa_depsilon}) yields different solutions of Eq.~(\ref{eq:eta_twisted_cocycle}) for the local projective phases, that is
\begin{equation}
\eta'_{a}({\bf g,h})= \tau_{a}({\bf g,h})^{-1} \eta_{a}({\bf g,h})
,
\end{equation}
where $\tau_{a}({\bf g,h}) = M^{\ast}_{a \cohosub{t}({\bf g,h})}$ are phases that satisfy the condition that $\tau_{a}({\bf g,h}) \tau_{b}({\bf g,h}) = \tau_{c}({\bf g,h})$ whenever $N_{ab}^{c}\neq 0$, but which are also required to satisfy the additional condition
\begin{equation}
\tau_{^{\mb{\bar{g}}}a}(\mb{h},\mb{k}) \tau_{a}(\mb{g},\mb{hk}) = \tau_{a}(\mb{g},\mb{h}) \tau_{a}(\mb{gh},\mb{k})
.
\end{equation}

There is, however, a sense in which na\"ively different solutions should be considered equivalent. In particular, if we locally redefine the operators $U^{(j)}_{\bf g}$ by a transformation
\begin{align}
\widehat{U}^{(j)}_{\bf g} =  U^{(j)}_{\bf g} Z^{(j)-1}_{\bf g},
\end{align}
where $Z^{(j)}_{\bf g}$ are unitary operators whose nontrivial action is localized within $\mathcal{R}_{j}$, this redefinition will not change the global action $R_{\bf g}$ on states as long as these operators satisfy
\begin{equation}
\prod_{j=1}^{n} Z^{(j)}_{\bf g} = \openone
\label{eq:Z_gauge_prod}
\end{equation}
when acting in the subspace of quasiparticle states of the form $|\Psi_{\{a;c,\mu \}} \rangle$. These are gauge transformations, and so they should be treated as trivial modifications of the operators $U^{(j)}_{\bf g}$, i.e. all operators related by such a transformation are in the same equivalence class.

By similar arguments as used for $W^{(j)}_{\bf g,h}$, it follows from Eq.~(\ref{eq:Z_gauge_prod}) that
\begin{equation}
\langle \Psi_{\{a;c,\mu \}}  | Z^{(j)}_{\bf g} | \Psi_{\{b;e,\nu \}}  \rangle = \zeta_{a_j}({\bf g} )  \delta_{\{a;c,\mu \}\{b;e,\nu \}} ,
\end{equation}
where $\zeta_{a_j}({\bf g} )$ is a phase that only depends on the topological charge $a_j$ contained in the region $\mathcal{R}_j$ and that these phases obey the constraint
\begin{equation}
\prod_{j=1}^{n} \zeta_{a_j}({\bf g} ) =1
.
\end{equation}

This similarly leads to the property that $\zeta_{a}({\bf g} ) \zeta_{b}({\bf g} ) = \zeta_{c}({\bf g} )$
whenever $N_{ab}^{c} \neq 0$, which, in turn, gives the relation
\begin{equation}
\zeta_{a}({\bf g} ) = M_{a \cohosub{z}\left( {\bf g} \right)}^{\ast}
,
\end{equation}
for some Abelian topological charge $\coho{z}({\bf g})\in C^1(G, \mathcal{A})$. These are precisely the same redundancies that arose due to the freedom to decompose the trivial natural isomorphism into topological charge dependent phase factors, as described in Sec.~\ref{sec:Symmetry}.

Under such transformations, the operators $W^{(j)}_{\bf g,h}$ transform into
\begin{eqnarray}
\widehat{W}^{(j)}_{\bf g,h} &=& Z^{(j)}_{\bf g} U^{(j)-1}_{\bf g} \,^{\bf g}Z^{(j)}_{\bf h} \,^{\bf g} U^{(j)-1}_{\bf h} U^{(j)}_{\bf gh} Z^{(j)-1}_{\bf gh} B^{(j)}_{\bf g,h} \notag \\
&=& Z^{(j)}_{\bf g} \rho_{\bf g} Z^{(j)}_{\bf h}  \rho_{\bf g}^{-1} W^{(j)}_{\bf g,h} Z^{(j)-1}_{\bf gh} .
\end{eqnarray}
Acting on states of the form $|\Psi_{\{a;c,\mu \}} \rangle$, this produces the equivalent relations
\begin{eqnarray}
\widehat{\omega}_{a}({\bf g,h} ) &=& \frac{\zeta_{\rho_{\bf g}^{-1} (a)}({\bf h} ) \zeta_{a}({\bf g} ) } { \zeta_{a}({\bf gh} ) }  \omega_{a}({\bf g,h} )  , \\
M_{a \widehat{\cohosub{w}}({\bf g,h} )} &=& M_{a \rho_{\bf g}[ \cohosub{z}({\bf h} )]} M_{a \cohosub{z}({\bf gh} )}^{\ast} M_{a \cohosub{z}({\bf g} )}  M_{a {\cohosub{w}}({\bf g,h} )} \notag \\
&=& M_{a, \rho_{\bf g}[\cohosub{z}({\bf h} )] \times \overline{\cohosub{z}({\bf gh} )} \times \cohosub{z}({\bf g} ) \times \cohosub{w}({\bf g,h} )}
\end{eqnarray}
from which we obtain
\begin{eqnarray}
\widehat{\coho{w}}({\bf g,h} ) &=& \rho_{\bf g}[\coho{z}({\bf h} )] \times \overline{\coho{z}({\bf gh} )} \times \coho{z}({\bf g} ) \times \coho{w}({\bf g,h} ) \notag \\
&=& \text{d} \coho{z}({\bf g,h} ) \times \coho{w}({\bf g,h} )
,
\end{eqnarray}
showing that $\coho{w}({\bf g,h} )$ and $\widehat{\coho{w}}({\bf g,h} )$ that are related by fusion with a 2-coboundary $ \text{d} \coho{z}({\bf g,h} ) \in B^2_{\rho}(G, \mathcal{A})$ correspond precisely to operators $W^{(i)}_{\bf g,h}$ and $\widehat{W}^{(i)}_{\bf g,h}$ that are related by gauge transformations, and so should be considered equivalent, i.e. one should take the quotient by $B^2_{\rho}(G, \mathcal{A})$.

In terms of the local projective phases (for fixed $\beta_{a}({\bf g,h})$), this translates into the equivalence of between symmetry fractionalization described by $\eta_{a}({\bf g,h})$ and
\begin{equation}
\widehat{\eta}_{a}({\bf g,h} ) = \frac{ \zeta_{a}({\bf gh} ) }{\zeta_{\rho_{\bf g}^{-1} (a)}({\bf h} ) \zeta_{a}({\bf g} ) }  \eta_{a}({\bf g,h} )
.
\end{equation}
We emphasize that, despite the similar appearance to Eq.~(\ref{eq:check_eta}), this transformation and corresponding equivalence is distinct from the symmetry action gauge transformation, because of how the two transformations act on $\coho{w}({\bf g,h} )$ and $\beta_{a}({\bf g,h} )$, as well as the additional condition that $\zeta_{a}({\bf g} )$ must respect the fusion rules.

Thus, the solutions of Eq.~(\ref{eq:kappa_depsilon}) for the $[\coho{O}]=[0]$ equivalence class are classified by
\begin{equation}
H^2_{\rho}(G, \mathcal{A}) = \frac{ Z^2_{\rho}(G, \mathcal{A})}{ B^2_{\rho}(G, \mathcal{A})}
.
\end{equation}
One should not, however, think of the set of solutions itself as being equal to $H^2_{\rho}(G, \mathcal{A})$; rather, the set of solutions is an $H^2_{\rho}(G, \mathcal{A})$ torsor. In particular, the distinct cohomology classes $[\coho{t}] \in H^2_{\rho}(G, \mathcal{A})$ relate distinct equivalence classes of solutions $[\coho{w}]$, with different solutions being related by $\coho{w}'({\bf g,h}) = \coho{t}({\bf g,h}) \times \coho{w}({\bf g,h})$. In terms of the local projective phases, the $H^2_{\rho}(G, \mathcal{A})$ action relates distinct symmetry fractionalization classes by $\eta'_{a}({\bf g,h})= \tau_{a}({\bf g,h})^{-1} \eta_{a}({\bf g,h})$. The number of inequivalent symmetry fractionalization classes is thus equal to $|H^2_{\rho}(G, \mathcal{A})|$. In this sense, there is no notion of an ``identity'' or ``zero'' element of the set of fractionalization classes. (Had one chosen to use the representative $\coho{O} =0$ of the $[\coho{O}]=[0]$ equivalence class, Eq.~(\ref{eq:kappa_depsilon}) becomes a cocycle condition on $\coho{w}({\bf g,h})$, so, in this case, $[\coho{w}] \in H^2_{\rho}(G, \mathcal{A})$, though this is not an invariant statement.)

Once again, symmetry actions in the same equivalence class related by natural isomorphisms lead to the same results here, so this classification of solutions is actually independent of the choice $\rho \in [\rho]$. Thus, the symmetry fractionalization is classified by $H^2_{[\rho]}(G, \mathcal{A})$.

We reemphasize the contrast between $H^2_{[\rho]}(G, \mathcal{A})$, which classifies the symmetry fractionalization, and $H^3_{[\rho]}(G, \mathcal{A})$, which contains $[\coho{O} ]$, the fractionalization obstruction class of $[\rho]$. Since $[\coho{O} ]$ is uniquely defined by $[\rho]$, it is only possible to realize exactly one element of $H^3_{[\rho]}(G, \mathcal{A})$ for specified $\mathcal{C}$, $G$, and $[\rho]$.

\subsubsection{Charge conjugation symmetry fractionalization}

It is worth considering fractionalization in more detail for the case of a unitary $\mathbb{Z}_2$ symmetry whose nontrivial element ${\bf C}$ acts as charge conjugation on the topological charges, i.e. ${\bf C}^{2} = {\bf 0}$ and $\rho_{\bf C}(a) = \bar{a}$. When there are topological charges that are not self-dual ($\bar{a} \neq a$), if topological charge conjugation is a braided autoequivalence of the MTC $\mathcal{C}$, then it corresponds represents a subgroup $\mathbb{Z}_{2}$ of $\text{Aut}(\mathcal{C})$. When all topological charges are self-dual ($\bar{a}=a$), then topological charge conjugation is clearly the trivial autoequivalence.

The symmetry action on the topological state space of a global $\mathbb{Z}_2$ symmetry that conjugates all topological charges is specified by the action on fusion vertex states
\begin{equation}
\rho_{\bf C} |a,b;c,\mu \rangle = \sum_{\nu} [ U_{{\bf C}}(\bar{a}, \bar{b}; \bar{c}) ]_{\mu \nu} |\bar{a}, \bar{b}; \bar{c} ,\nu \rangle
.
\end{equation}
It follows that
\begin{eqnarray}
[\kappa_{{\bf C}, {\bf C}}(a,b;c)]_{\mu \nu} &=&  \sum_{ \lambda} [ U_{{\bf C}} (a,b;c) ]^{\ast}_{\lambda \mu} [ U_{{\bf C}}(\bar{a}, \bar{b}; \bar{c})]^{\ast}_{\nu \lambda}
\notag \\
&=& \frac{\beta_{a}({\bf C},{\bf C}) \beta_{b}({\bf C},{\bf C})}{\beta_{c}({\bf C},{\bf C})} \delta_{\mu \nu}
\end{eqnarray}
and the obstruction class is defined by
\begin{equation}
\Omega_{a}({\bf C},{\bf C},{\bf C}) = \frac{ \beta_{ \bar{a} }({\bf C},{\bf C})}{ \beta_{a}({\bf C},{\bf C}) }
.
\end{equation}

We now assume the obstruction vanishes and that global symmetry $R_{\bf C}$ acts in an on-site fashion, for which we have
\begin{equation}
R_{\bf C} |\Psi_{\{a;c,\mu \}} \rangle
= U^{(1)}_{\bf C} \ldots U^{(n)}_{\bf C} \rho_{\bf C}|\Psi_{\{a;c,\mu \}} \rangle
.
\end{equation}
The localized symmetry action operators $U^{(j)}_{\bf C}$ have the projective consistency relation
\begin{eqnarray}
\eta_{a_j}({\bf C},{\bf C}) |\Psi_{\{a;c,\mu \}} \rangle &=& \,^{\bf C} U^{(j)}_{\bf C} U^{(j)}_{\bf C} |\Psi_{\{a;c,\mu \}} \rangle \notag \\
&=& R_{\bf C} U^{(j)}_{\bf C} R_{\bf C}^{-1} U^{(j)}_{\bf C} |\Psi_{\{a;c,\mu \}} \rangle \notag \\
&=& U^{(j)}_{\bf C} \rho_{\bf C} U^{(j)}_{\bf C} \rho_{\bf C}^{-1} |\Psi_{\{a;c,\mu \}} \rangle
\quad
\end{eqnarray}
where the projective phases $ \eta_{a}({\bf C,C})$ satisfy Eq.~(\ref{eq:eta_twisted_cocycle}), which, in this case, is simply the condition that
\begin{equation}
\eta_{\bar{a}}({\bf C,C}) = \eta_{a}({\bf C,C})
,
\end{equation}
and Eq.~(\ref{eq:eta_kappa_relation}), which requires
\begin{equation}
\label{eq:eta_kappa_relation_cc}
\frac{\eta_{c}({\bf C},{\bf C})}{\eta_{a}({\bf C},{\bf C}) \eta_{b}({\bf C},{\bf C})} \openone = U_{{\bf C}} (a,b;c) U_{{\bf C}}(\bar{a}, \bar{b}; \bar{c})
,
\end{equation}
when $N_{ab}^{c} \neq 0$.

We can define an invariant of topological charge conjugation symmetry fractionalization for all $a \in \mathcal{C}$ by
\begin{equation}
\label{eq:cc_eta_invariant}
\eta_{a}^{\bf C} = \varkappa_{a} U_{{\bf C}} (a,\bar{a};0) \eta_{a}({\bf C},{\bf C})
,
\end{equation}
where we recall that
\begin{equation}
\varkappa_{a} = d_{a} [F^{a \bar{a} a}_a]_{00} = \theta_{a} R^{\bar{a} a}_{0}
.
\end{equation}
It is straightforward to check that $\eta_{a}^{\bf C}$ is invariant under both vertex basis gauge transformations and symmetry action gauge transformations, i.e. $\eta_{a}^{\bf C} =\tilde{\eta}_{a}^{\bf C} =\check{\eta}_{a}^{\bf C}$. We also notice, using Eq.~(\ref{eq:eta_kappa_relation_cc}), that
\begin{equation}
\eta_{\bar{a}}^{\bf C} = (\eta_{a}^{\bf C})^{\ast}
.
\end{equation}
When $a=\bar{a}$, this condition implies
\begin{equation}
\eta_{a}^{\bf C} = \pm 1
.
\end{equation}

The classification of the symmetry fractionalization is given by $H^2_{[\rho]}(\mathbb{Z}_{2}, \mathcal{A})$. In order to compute this cohomology group, we first considering the $2$-cocycle condition, which for this symmetry is simply $\coho{w}({\bf C},{\bf C} ) = \overline{\coho{w}({\bf C},{\bf C} )}$. The 2-coboundaries are trivial, since $\text{d}\coho{z}({\bf C},{\bf C} ) = \overline{\coho{z}({\bf C})} \times \coho{z}({\bf C}) = 0$. Thus, we find that symmetry fractionalization is classified (torsorially) by
\begin{equation}
H^2_{[\rho]}(\mathbb{Z}_{2}, \mathcal{A}) = \mathcal{A}^{\bf C} = \mathbb{Z}_{2}^{k}
,
\end{equation}
where $\mathcal{A}^{\bf C} = \{ a \in \mathcal{A} | a = \bar{a} \}$ is the group defined by the set of self-dual Abelian topological charges, with group multiplication given by the fusion rules. Since every self-dual Abelian topological charge obeys $a \times a = 0$, each of them defines a $\mathbb{Z}_{2}$ group element, and, hence, $\mathcal{A}^{\bf C}= \mathbb{Z}_{2}^{k}$ for some non-negative integer $k$. Specifying $\eta_{a}^{\bf C}$ for all $a\in \mathcal{C}$ distinguishes between all the fractionalization classes.

\subsection{Projective Representations of the Global Symmetry}
\label{sec:Proj_Rep_Glob_Symm}

In the above discussion, we assumed that the local Hilbert space on each site transforms in a linear representation of the global symmetry $G$. However this is not fully general, and it is possible that instead the local Hilbert space on each site transforms according to a projective representation of $G$. The canonical example is a spin-$\frac{1}{2}$ system. While the global symmetry of spin rotation is $G = \text{SO}(3)$, each site contains a spin-$\frac{1}{2}$ which transforms in a projective representation of SO$(3)$. Describing symmetry fractionalization when the local Hilbert space already forms a projective representation of $G$ requires some minor modifications of the previous arguments. In particular, the action of a projective symmetry representation on the ground state will take the form
\begin{equation}
R_{\bf gh} |\Psi_{0} \rangle = e^{i \Phi_{\bf g,h}} R_{\bf g} R_{\bf h} |\Psi_{0} \rangle
,
\end{equation}
where $e^{i \Phi_{\bf g,h}}$ are the projective representation phase factors. The projective representations are classified by $H^{2}\left( G , \text{U}(1) \right)$. In particular, the phases $e^{i \Phi_{\bf g,h}}$ must satisfy the 2-cocycle condition
\begin{equation}
e^{i \Phi_{\bf h,k}} e^{-i \Phi_{\bf gh,k}} e^{i \Phi_{\bf g,hk}} e^{-i \Phi_{\bf g,h}} =1
\label{eq:Phi_2cocycle}
\end{equation}
in order for the two different, but equivalent ways of relating $R_{\bf ghk}$ and $R_{\bf g}R_{\bf h}R_{\bf k}$ to be consistent. Additionally, different projective phase factors $e^{i \Phi_{\bf g,h}}$ and $e^{i \tilde{\Phi}_{\bf g,h}}$ are considered equivalent if they are related by a 2-coboundary
\begin{equation}
e^{i \tilde{\Phi}_{\bf g,h}} = e^{i f_{\bf h}} e^{-i f_{\bf gh}} e^{i f_{\bf g}} e^{i \Phi_{\bf g,h}}
\end{equation}
for some phase function $e^{i f_{\bf g}}$ of the group elements of $G$, since their difference could simply be absorbed into the operator $R_{\bf g}$ by the trivial redefinition $\tilde{R}_{\bf g} = e^{i f_{\bf g}} R_{\bf g}$. The equivalence class $[e^{i \Phi_{\bf g,h}}] \in H^{2}\left( G , \text{U}(1) \right)$ of the projective representation is a global property of the system that does not change under application of local operations, such as those that create quasiparticles. Thus, we also have
\begin{equation}
R_{\bf gh} |\Psi_{\{a;c,\mu \}} \rangle = e^{i \Phi_{\bf g,h}} R_{\bf g} R_{\bf h} |\Psi_{\{a;c,\mu \}} \rangle
\end{equation}
with the same $e^{i \Phi_{\bf g,h}}$ for any state of the form $|\Psi_{\{a;c,\mu \}} \rangle$ obtainable from the ground state $|\Psi_{0} \rangle$ through adiabatic creation and manipulation of quasiparticles.
We now define $W^{(j)}_{\bf g,h}$ as before for $j=2,\ldots,n$, while for $j=1$ we slightly modify the definition to be
\begin{equation}
W^{(1)}_{\bf g,h} = e^{-i \Phi_{\bf g,h}}  U^{(1)-1}_{\bf g} \,^{\bf g} U^{(1)-1}_{\bf h}  U^{(1)}_{\bf gh} B^{(1)}_{\bf g,h}
.
\end{equation}
With this definition, we retain the properties that $W^{(j)}_{\bf g,h}$ is localized in region $\mathcal{R}_{j}$, and that the $W^{(j)}_{\bf g,h}$ satisfy Eqs.~(\ref{eq:prod_W_j}) and (\ref{eq:phase_omega_j}). This allows the argument relating the eigenvalues of $W^{(j)}_{\bf g,h}$ to Abelian topological charges to go through unaltered. To see that the cocycle relations are unchanged, we only need to check that Eq.~(\ref{eq:WB_cocycle}) remains the same for $W^{(1)}_{\bf g,h}$. This follows from the previous argument, together with the fact that $e^{i \Phi_{\bf g,h}}$ itself satisfies the 2-cocycle condition of Eq.~(\ref{eq:Phi_2cocycle}). Thus, the same cohomological relations hold and all the arguments go through as before to give the same results for obstruction and classification of symmetry fractionalization.

\subsection{Locality Preserving Symmetry}
\label{sec:QuasiOnSite}

There are a number of symmetries, such as time-reversal symmetry and translation symmetry, that do not act in a strictly on-site fashion, but which may nonetheless be fractionalized. In order to understand fractionalization of such symmetries, we must generalize the notion of symmetries acting in an on-site fashion so as to include the possibility of anti-unitary symmetries and other nonlocal symmetries.

We call a (unitary or anti-unitary) symmetry operator $R_{\bf g}$ ``locality preserving'' if it acts in the following manner. For any operators $\mathcal{O}^{(j)}$ localized in the simply connected regions $\mathcal{R}_j$, the operators
\begin{equation}
^{\bf g} \mathcal{O}^{{\bf g}(j)} \equiv R_{\bf g} \mathcal{O}^{(j)} R_{\bf g}^{-1}
\end{equation}
are localized in the (possibly distinct) simply connected regions that we denote as $^{\bf g}\mathcal{R}_{j}$, and whenever two such simply connected regions $\mathcal{R}_j$ and $\mathcal{R}_k$ are disjoint, i.e. $\mathcal{R}_j \cap \mathcal{R}_k = \emptyset$, the corresponding regions $^{\bf g}\mathcal{R}_{j}$ and $^{\bf g}\mathcal{R}_{k}$ are disjoint, i.e. $^{\bf g}\mathcal{R}_j \cap \,^{\bf g}\mathcal{R}_k = \emptyset$.

\begin{figure}[t!]
	\centering
	\includegraphics[width=\columnwidth]{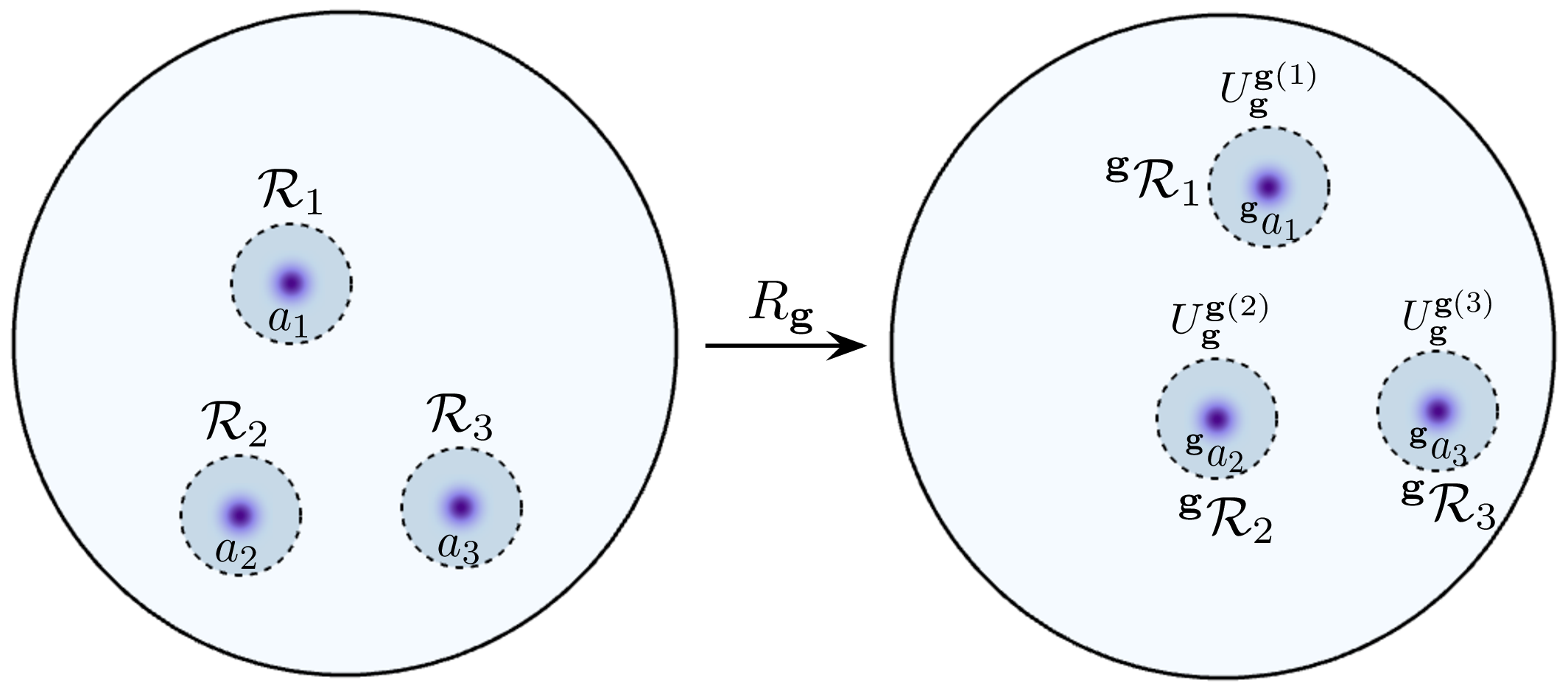}
	\caption{The action of a global ``locality preserving'' symmetry operator $R_\mb{g}$ on a state with quasiparticles may move the locations where the quasiparticles are localized, from the regions $\mathcal{R}_{j}$ to the regions $^{\bf g}\mathcal{R}_{j}$. The locality preserving property ensures that the regions $^{\bf g}\mathcal{R}_{j}$ are mutually disjoint for distinct $j$ whenever the regions $\mathcal{R}_{j}$ are mutually disjoint for distinct $j$. Additionally, the locality preserving symmetry action induces unitary transformations $U_\mb{g}^{{\bf g}(j)}$ that are, respectively, localized in the regions $^{\bf g}\mathcal{R}_{j}$, together with a global transformation $\rho_\mb{g}$ which strictly acts on the topological quantum numbers.}
	\label{fig:quasi-on-site-sym-frac}
\end{figure}

Specific examples that we have in mind for such symmetry operators include the complex conjugation operator $K$, in which case $^{\bf g}\mathcal{R}_{j} = \mathcal{R}_{j}$, or a translation operator $T_{\vec{x}}$ (in a translationally invariant system), in which case $^{\bf g}\mathcal{R}_{j}$ is the region $\mathcal{R}_{j}$ translated by the vector $\vec{x}$. Clearly, on-site symmetries satisfy the above locality preserving condition.

We can now repeat the entire analysis of this section with a few small, but important modifications to account for the generalization to locality preserving symmetries. We note that our treatment here requires that the symmetries also leave the spatial orientation of the fusion/splitting spaces invariant. Consequently, we omit spatial symmetries involving rotations or parity reversal.

The first modification is to the conjugation of local operators by $R_{\bf g}$. In particular, given the above locality preserving property of $R_{\bf g}$, we generalize the definition in Eq.~(\ref{eq:g-conjugate-operator}) to
\begin{equation}
^{\bf g} \mathcal{O}^{(j)} \equiv R_{\bf g} \mathcal{O}^{{\bf \bar{g}}(j)} R_{\bf g}^{-1}
,
\label{eq:quasi-g-conjugate-operator}
\end{equation}
which is thus an operator whose nontrivial action is localized in the region $\mathcal{R}_{j}$.

The next modification is that when the $j$th quasiparticle of the state $|\Psi_{\{a;c,\mu \}} \rangle$ is localized in region $\mathcal{R}_j$, it follows that the $j$th quasiparticle of the state $R_{\bf g} |\Psi_{\{a;c,\mu \}} \rangle$ is localized in the region $^{\bf g}\mathcal{R}_j$. Consequently, the action of $R_{\bf g}$ on states in the physical Hilbert space containing quasiparticles, as in Eq.~(\ref{eq:symmetryLocalization_n}), is modified to
\begin{equation}
\label{eq:symmetryquasiLocalization_n}
R_{\bf g} |\Psi_{\{a;c,\mu \}} \rangle
= U^{{\bf g}(1)}_{\bf g} \ldots U^{{\bf g}(n)}_{\bf g} \rho_{\bf g}|\Psi_{\{a;c,\mu \}} \rangle,
\end{equation}
where $U^{{\bf g}(j)}_{\bf g}$ is a unitary operator whose nontrivial action is localized in the region $^{\bf g}\mathcal{R}_{j}$, and we have defined $\rho_{\bf g}$ exactly as before, i.e.
\begin{equation}
\rho_{\bf g} = \prod_{j=1}^{n} U^{{\bf g}(j)-1}_{\bf g} R_{\bf g}
,
\end{equation}
which now makes it a locality preserving operator, in accord with $R_{\bf g}$. Now, $\rho_{\bf g}$ acts on the physical Hilbert space as does the symmetry action on the topological state space, so it may potentially move the regions where quasiparticles are localized or complex conjugate coefficients in front of the state, depending on ${\bf g}$. This is shown schematically in Fig.~\ref{fig:quasi-on-site-sym-frac}.
We can leave Eq.~(\ref{eq:rho_g_action}) unmodified, with the understanding that if the $j$th quasiparticle of the state $|\Psi_{\{a;c,\mu \}} \rangle$ is localized is region $\mathcal{R}_j$, then the $j$th quasiparticle of the state $|\Psi_{\{\,^{\bf g}a; \,^{\bf g}c,\mu' \}} \rangle$ (and the state $\rho_{\bf g} |\Psi_{\{a;c,\mu \}} \rangle$) is localized in the region $^{\bf g}\mathcal{R}_j$.

With these modifications, one must be careful to modify the localization regions of the operators appropriately in all steps of the arguments of the previous sections, but, in the end, this dependence drops out entirely. In particular, we note that we should modify Eq.~(\ref{eq:R_g-to-rho_g-conj}) to
\begin{equation}
^{\bf g} \mathcal{O}^{(j)} U^{(j)}_{\bf g} = U^{(j)}_{\bf g} \rho_{\bf g} \mathcal{O}^{{\bf \bar{g}}(j)} \rho_{\bf g}^{-1}
,
\label{eq:R_g-to-rho_g-conj-quasi}
\end{equation}
the definition of the operator $W^{(j)}_{\bf g,h}$, which has its nontrivial action localized in the region $\mathcal{R}_{j}$, to
\begin{widetext}
\begin{equation}
W^{(j)}_{\bf g,h} =  U^{(j)-1}_{\bf g} \,^{\bf g} U^{(j)-1}_{\bf h}  U^{(j)}_{\bf gh} B^{(j)}_{\bf g,h}
= \rho_{\bf g} U^{{\bf \bar{g}}(j)-1}_{\bf h} \rho_{\bf g}^{-1} U^{(j)-1}_{\bf g} U^{(j)}_{\bf gh} B^{(j)}_{\bf g,h}
,
\end{equation}
and the relation of Eq.~(\ref{eq:U_ghU_gU_h}) to
\begin{equation}
\eta_{a_j}({\bf g,h}) U^{(j)}_{\bf gh} |\Psi_{\{a;c,\mu \}} \rangle = \,^{\bf g} U^{(j)}_{\bf h} U^{(j)}_{\bf g} |\Psi_{\{a;c,\mu \}} \rangle = U^{(j)}_{\bf g} \rho_{\bf g} U^{{\bf \bar{g}}(j)}_{\bf h} \rho_{\bf g}^{-1} |\Psi_{\{a;c,\mu \}} \rangle
\label{eq:U_ghU_gU_h-quasi}
.
\end{equation}

The final relation in terms of operators, given in Eq.~(\ref{eq:WB_cocycle}), is modified to
\begin{equation}
\label{eq:WB_cocycle-quasi}
\rho_{\bf g} W^{{\bf \bar{g}}(j)}_{\bf h,k} \rho_{\bf g}^{-1} \rho_{\bf g} B^{{\bf \bar{g}}(j)-1}_{\bf h,k} \rho_{\bf g}^{-1} W^{(j)}_{\bf g,hk} B^{(j)-1}_{\bf g,hk}
= W^{(j)}_{\bf g,h} B^{(j)-1}_{\bf g,h} W^{(j)}_{\bf gh,k} B^{(j)-1}_{\bf gh,k}.
\end{equation}
Applying this relation to a state $|\Psi_{\{a;c,\mu \}} \rangle$, we find that the dependence on localization regions drops out of the resulting relation in terms of (eigenvalue) phases, and the only modification that we must now account for is the potential complex conjugation due to ${\bf g}$ being an anti-unitary symmetry (which was encoded in the operator $\rho_{\bf g}$). Specifically, this yields the modification of Eq.~(\ref{eq:omega_a_relation}) to the relation
\begin{eqnarray}
\Omega_{a}({\bf g,h,k}) &=& K^{q({\bf g})} \beta_{\rho_{\bf g}^{-1}(a)}({\bf h,k}) K^{q({\bf g})} \beta_{a}({\bf gh,k})^{-1}  \beta_{a}({\bf g,hk}) \beta_{a}({\bf g,h})^{-1} \notag \\
&=& K^{q({\bf g})} \omega_{\rho^{-1}_{\bf g}(a)}({\bf h,k} ) K^{q({\bf g})} \omega_{a}({\bf gh,k} )^{-1} \omega_{a}({\bf g,hk} ) \omega_{a}({\bf g,h} )^{-1}
,
\label{eq:omega_a_relation-quasi}
\end{eqnarray}
and the modification of Eq.~(\ref{eq:eta_twisted_cocycle}) to
\begin{equation}
K^{q({\bf g})} \eta_{\rho_{\bf g}^{-1}(a)}({\bf h,k}) K^{q({\bf g})} \eta_{a}({\bf gh,k})^{-1}  \eta_{a}({\bf g,hk}) \eta_{a}({\bf g,h})^{-1} = 1
.
\label{eq:eta_twisted_cocycle-quasi}
\end{equation}
\end{widetext}
Using $\Omega_{a}({\bf g,h,k}) = M_{a \cohosub{O}\left( {\bf g,h,k} \right)}^{\ast}$ and $\omega_{a}({\bf g,h} ) = M_{a \cohosub{w}\left( {\bf g,h} \right)}^{\ast}$ exactly as before, though with the relation $S_{\rho_{\bf g}(a)\rho_{\bf g}(b)} = K^{q({\bf g})} S_{ab} K^{q({\bf g})}$ that applies for unitary and anti-unitary symmetries, we obtain precisely the same consistency condition
\begin{equation}
\coho{O}\left( {\bf g,h,k} \right) = \text{d} \coho{w} \left( {\bf g,h,k} \right)
\end{equation}
of Eq.~(\ref{eq:kappa_depsilon}). We emphasize that the complex conjugations due to symmetries being anti-unitary dropped out in the process of mapping the relation of phases into the relation of $C^{n}(G,\mathcal{A})$ cochains.

The remaining arguments that lead to the classification results are similarly modified. Similar to the steps described above, the localization region dependence drops out when the operator relations are converted into phase relations by applying them to states of the form $|\Psi_{\{a;c,\mu \}} \rangle$, and the complex conjugations that occur for anti-unitary symmetries drop out when these phase relations are converted into cochain relations. Thus, the obstruction of fractionalization by nontrivial $[\coho{O}] \in H^3_{[\rho]}(G,\mathcal{A})$ and the classification of symmetry fractionalization (when the obstruction vanishes) in terms of the cohomology class $H^2_{[\rho]}(G, \mathcal{A})$ is precisely the same for unitary and anti-unitary locality preserving symmetries as it was for unitary on-site symmetries.

We note that the projective representation analysis of Sec.~\ref{sec:Proj_Rep_Glob_Symm} must include the complex conjugation of anti-unitary symmetries, so they are classified by $H^2_{q}(G, \text{U}(1))$, which includes complex conjugation from anti-unitary symmetry action. In particular, the boundary operator includes the complex conjugation through the $\rho_{g}$ action, so the $2$-cocycle condition on the projective phases becomes
\begin{equation}
e^{i (-1)^{q({\bf g})} \Phi \left( {\bf h,k} \right)} e^{-i \Phi \left( {\bf gh,k} \right)} e^{i \Phi \left( {\bf g,hk} \right)} e^{-i \Phi \left( {\bf g,h} \right)}=1
,
\label{eq:anti-projective-cocycle}
\end{equation}
and the projective phase is a $2$-coboundary when
\begin{equation}
e^{i \Phi \left( {\bf g,h} \right)}= e^{i (-1)^{q({\bf g})} f\left( {\bf h} \right)} e^{-i f\left( {\bf gh} \right)} e^{i f\left( {\bf g} \right)}
\label{eq:anti-projective-coboundary}
\end{equation}
for some phase $e^{i f\left( {\bf g} \right)} \in C(G,\text{U}(1))$. These modifications do not affect the symmetry fractionalization results.

When we specify a fusion basis decomposition of the topological state space of $n$ quasiparticles, we first specify an order in which to place the quasiparticles from left to right at the top of a fusion tree. Specifying an order in which one lists the quasiparticles is equivalent to specifying a line in the 2D manifold that passes through the quasiparticles in that order. The inclusion of rotational and spatial parity symmetry is complicated by the fact that these symmetry operations generally change the positions of the quasiparticles with respect to their ordering line. For spatial parity symmetries, we note that one can repeat the analysis above, with the modification that when phases in the analysis are mapped to $C^{n}(G,\mathcal{A})$ cochains, the action of $\rho_{\bf g}$ on the group elements $\mathcal{A}$ is modified to include topological charge conjugation whenever $p(\rho_{\bf g}) = 1$. This modification follows from the relation $S_{\rho_{\bf g}(a)\rho_{\bf g}(b)} = K^{q({\bf g})+p({\bf g})} S_{ab} K^{q({\bf g})+p({\bf g})}$, which modifies the $\rho$ action in the cohomology structure, i.e. in the coboundary operator and the groups $H^n_{[\rho]}(G,\mathcal{A})$, whenever $\rho_{\bf g}$ corresponds to a spatial orientation reversing symmetry.

Before concluding this section, we note that the above considerations provide a framework to classify the different possible types of symmetry fractionalization. However, not all elements of the $H^2_{[\rho]}(G, \mathcal{A})$ classes will be allowed in general. When $G$ corresponds to a spatial symmetry, there can be additional constraints that rule out certain types of fractionalization~\cite{oshikawa2000,paramekanti2004,hastings2004,hastings2005,zaletel2014}. Even for on-site symmetries, as we will see, some of the fractionalization classes are anomalous and cannot be realized in a purely $2+1$ dimensional system.

\subsubsection{Time reversal symmetry fractionalization and local Kramers degeneracy}

It is worth considering fractionalization in more detail for the case of time reversal symmetry, or, rather, a group element ${\bf T} \in G$ such that ${\bf T}^{2} = {\bf 0}$ and $q({\bf T})=1$, i.e. it is an anti-unitary $\mathbb{Z}_2$ symmetry.~\footnote{A more general definition of time reversal is possible, wherein a $\mathbb{Z}_2$ grading specifies whether elements of the symmetry group $G$ reverse time. When there is a single time reversing group element, it must be a $\mathbb{Z}_2$ element, as considered in this subsection. Moreover, if the spectrum of the Hamiltonian is symmetric about $0$, it is possible for time-reversing symmetry group elements to be represented by unitary operators, in which case one must specify separate $\mathbb{Z}_2$ gradings for anti-unitarity and time-reversal. We do not consider such Hamiltonians in this paper.}

We first note that the state of the system can either form a linear representation with $R_{\bf T}^2 |\Psi \rangle= |\Psi \rangle$ or a projective representation with $R_{\bf T}^2 |\Psi \rangle = -|\Psi \rangle$. This follows from the $H^2_{q}(\mathbb{Z}_{2}^{{\bf T}}, \text{U}(1))= \mathbb{Z}_{2}$ classification of projective representations. In particular, the modified $2$-cocycle condition of Eq.~(\ref{eq:anti-projective-cocycle}) is simply the condition $e^{i 2 \Phi_{{\bf T},{\bf T}}} =1$, and the modified $2$-coboundary condition of Eq.~(\ref{eq:anti-projective-coboundary}) is $e^{i \Phi_{{\bf T},{\bf T}}} =1$. The projective representation $e^{i \Phi_{{\bf T},{\bf T}}} =-1$ gives the usual degeneracy from Kramers theorem, where $|\Psi \rangle$ and $R_{\bf T} |\Psi \rangle$ are necessarily orthogonal and degenerate in energy for any state $|\Psi \rangle$ when $R_{\bf T}$ commutes with the Hamiltonian. Physically, this corresponds to the case where the system has half-integer angular momentum, i.e. an odd number of electrons in the system.

The symmetry action on the topological state space is specified by the action on fusion vertex states
\begin{equation}
\rho_{\bf T} |a,b;c,\mu \rangle = \sum_{\nu} [ U_{{\bf T}}(\,^{{\bf T}}a,\,^{{\bf T}}b; \,^{{\bf T}}c) ]_{\mu \nu} K |\,^{{\bf T}}a, \,^{{\bf T}}b; \,^{{\bf T}}c,\nu \rangle.
\end{equation}
Since this is an anti-unitary symmetry, it follows that
\begin{eqnarray}
[\kappa_{{\bf T}, {\bf T}}(a,b;c)]_{\mu \nu} &=&  \sum_{ \lambda} [ U_{{\bf T}} (a,b;c) ]^{\ast}_{\lambda \mu} [ U_{{\bf T}}(\,^{{\bf T}}a,\,^{{\bf T}}b;\,^{{\bf T}}c)]_{\nu \lambda}
\notag \\
&=& \frac{\beta_{a}({\bf T},{\bf T}) \beta_{b}({\bf T},{\bf T})}{\beta_{c}({\bf T},{\bf T})} \delta_{\mu \nu} \notag \\
&=& \frac{\eta_{a}({\bf T},{\bf T}) \eta_{b}({\bf T},{\bf T})}{\eta_{c}({\bf T},{\bf T})} \delta_{\mu \nu}
.
\end{eqnarray}

The obstruction class is defined by
\begin{equation}
\Omega_{a}({\bf T},{\bf T},{\bf T}) = \frac{1}{ \beta_{\,^{{\bf T}}a }({\bf T},{\bf T}) \beta_{a}({\bf T},{\bf T}) }
.
\end{equation}
The condition that the obstruction vanishes is equivalent to there being some $\omega_{a}({\bf T},{\bf T})$ such that
\begin{eqnarray}
&& \beta_{\,^{{\bf T}}a }({\bf T},{\bf T}) \beta_{a}({\bf T},{\bf T}) = \omega_{\,^{{\bf T}}a }({\bf T},{\bf T}) \omega_{a}({\bf T},{\bf T}), \\
&& \omega_{a}({\bf T},{\bf T})\omega_{b}({\bf T},{\bf T})=\omega_{c}({\bf T},{\bf T}) , \quad \text{if} \quad N_{ab}^c \neq 0. \qquad
\end{eqnarray}

We now assume that the obstruction vanishes and that this anti-unitary symmetry $R_{\bf T}$ acts in a locality preserving fashion, for which we have
\begin{equation}
R_{\bf T} |\Psi_{\{a;c,\mu \}} \rangle
= U^{(1)}_{\bf T} \ldots U^{(n)}_{\bf T} \rho_{\bf T}|\Psi_{\{a;c,\mu \}} \rangle
.
\end{equation}
The localized symmetry action operators $U^{(j)}_{\bf T}$ have the projective consistency relation
\begin{eqnarray}
\eta_{a_j}({\bf T,T}) |\Psi_{\{a;c,\mu \}} \rangle &=& \,^{\bf T} U^{(j)}_{\bf T} U^{(j)}_{\bf T} |\Psi_{\{a;c,\mu \}} \rangle \notag \\
&=& R_{\bf T} U^{(j)}_{\bf T} R_{\bf T}^{-1} U^{(j)}_{\bf T} |\Psi_{\{a;c,\mu \}} \rangle \notag \\
&=& U^{(j)}_{\bf T} \rho_{\bf T} U^{(j)}_{\bf T} \rho_{\bf T}^{-1} |\Psi_{\{a;c,\mu \}} \rangle
\quad
\end{eqnarray}
where the projective phases $ \eta_{a}({\bf T,T})$ satisfy Eq.~(\ref{eq:eta_twisted_cocycle-quasi}), which, in this case, is simply the condition that
\begin{equation}
\eta_{^{{\bf T}}a}({\bf T,T}) = \eta_{a}({\bf T,T})^{\ast}
.
\end{equation}

When $^{{\bf T}}a = a$, this condition implies
\begin{equation}
\eta_{a}({\bf T,T})=\pm 1
,
\end{equation}
and we interpret $\eta_{a}({\bf T,T})$ as the ``local ${\bf T}^{2}$'' value ascribed to the topological charge $a$. We notice that, for $^{{\bf T}}a = a$, this quantity is an invariant under both vertex basis and symmetry action gauge transformations, i.e. $\eta_{a}({\bf T,T}) = \tilde{\eta}_{a}({\bf T,T}) = \check{\eta}_{a}({\bf T,T})$ for such $a$.  When $\eta_{a}({\bf T,T})=- 1$, there is also a local Kramers degeneracy~\cite{levinPRB2012} associated with the topological charge $a$. In other words, quasiparticles that carry topological charge $a$ also carry a local degenerate state space in physical systems that possess this symmetry. We also emphasize that $\theta_{\,^{{\bf T}}a}=\theta_{a}^{\ast}$, so, when $^{{\bf T}}a = a$, we also have $\theta_{a} = \pm 1$. However, we stress that it is not necessarily the case that $\theta_{a}$ equals $\eta_{a}({\bf T,T})$, as one might have na\"{i}vely expected from the usual understanding of Kramers degeneracy in terms of spin and fermionic parity.

When $^{{\bf T}}a = a$, $^{{\bf T}}b = b$, $^{{\bf T}}c = c$, and $N_{ab}^{c}\neq 0$, we have
\begin{eqnarray}
\pm \delta_{\mu \nu} &=& \frac{\eta_{a}({\bf T},{\bf T}) \eta_{b}({\bf T},{\bf T})}{\eta_{c}({\bf T},{\bf T})} \delta_{\mu \nu}  \notag \\
&=&  \sum_{ \lambda} [ U_{{\bf T}}(a,b;c) ]_{\mu \lambda} [ U_{{\bf T}}(a,b;c) ]^{\ast}_{\lambda \nu}
.
\label{eq:special_etaTT}
\end{eqnarray}
We note that, if $U U^{\ast} = \pm \openone$ for a unitary $m \times m$ matrix $U$, then $\det[U]=\det[\pm U^{T}]=(\pm 1)^{m} \det[U]$, which must also be nonzero.
Thus, when $N_{ab}^{c}$ is odd, the second line of Eq.~(\ref{eq:special_etaTT}) is simply equal to $1$, which implies the relation
\begin{equation}
\eta_{a}({\bf T},{\bf T}) \eta_{b}({\bf T},{\bf T}) = \eta_{c}({\bf T},{\bf T})
,
\label{eq:T2_relation1}
\end{equation}
when $^{{\bf T}}a = a$, $^{{\bf T}}b = b$, $^{{\bf T}}c = c$, and $N_{ab}^{c}$ is odd. When $N_{ab}^{c}$ is even, this relation may include a relative sign, which would be a gauge invariant quantity.

When $^{{\bf T}}c = c$ and $N_{a \,^{{\bf T}}a}^{c} \neq 0$, the ribbon property gives
\begin{equation}
\sum_{\lambda} \left[R^{a \,^{{\bf T}}a}_{c} \right]_{\mu \lambda} \left[ R^{\,^{{\bf T}}a a}_{c} \right]_{\lambda \nu} = \theta_{c} \delta_{\mu \nu}
,
\end{equation}
and the transformation of the $R$-symbols under ${\bf T}$ gives
\begin{widetext}
\begin{eqnarray}
\rho_{{\bf T}}\left( \left[R^{a \,^{{\bf T}}a}_{c} \right]_{\mu \nu} \right) &=& \left[R^{a \,^{{\bf T}}a }_{c} \right]_{\mu \nu}^{\ast}
= \sum_{\mu', \nu'} \left[ U_{{\bf T}}(a, \,^{{\bf T}}a ; c) \right]_{\mu \mu'} \left[ R^{\,^{{\bf T}}a a }_{c} \right]_{\mu' \nu'} \left[U_{{\bf T}}(\,^{{\bf T}}a , a ; c)^{-1}\right]_{\nu' \nu}
.
\end{eqnarray}
\end{widetext}
Combining these with
\begin{eqnarray}
&& [\kappa_{{\bf T},{\bf T}}( \,^{{\bf T}}a , a ; c)]_{\mu \nu} = \frac{1}{\eta_{c}({\bf T},{\bf T})} \delta_{\mu \nu} \notag \\
&& \quad = \sum_{\lambda}  \left[U_{{\bf T}}(\,^{{\bf T}}a , a ; c)\right]^{\ast}_{\lambda \mu } \left[ U_{{\bf T}}(a, \,^{{\bf T}}a ; c) \right]_{\nu \lambda}
,
\end{eqnarray}
it follows that
\begin{eqnarray}
&& \sum_{\alpha, \beta, \lambda} \left[ U_{{\bf T}}(a, \,^{{\bf T}}a ; c) \right]_{\mu \alpha} \left[ R^{\,^{{\bf T}}a a }_{c} \right]_{\alpha \beta} \notag \\
&&  \quad \times \left[ U_{{\bf T}}(a, \,^{{\bf T}}a ; c) \right]_{\beta \lambda}^{\ast} \left[ R^{\,^{{\bf T}}a a }_{c} \right]_{\lambda \nu}^{\ast} = \frac{\eta_{c}({\bf T},{\bf T})}{\theta_{c}} \delta_{\mu \nu}
. \qquad
\end{eqnarray}
The right hand side of this expression is equal to $\pm \delta_{\mu \nu}$. Thus, using the same argument leading to Eq.~(\ref{eq:T2_relation1}), when $N_{a \,^{{\bf T}}a}^{c}$ is odd, the left hand side (which is a unitary operator times it complex conjugate) must equal $\delta_{\mu \nu}$, which implies the relation
\begin{equation}
\eta_{c}({\bf T},{\bf T}) = \theta_{c} = \pm 1
,
\label{eq:T2_relation2}
\end{equation}
when $^{{\bf T}}c = c$ and $N_{a \,^{{\bf T}}a}^{c}$ is odd. When $N_{a \,^{{\bf T}}a}^{c}$ is even, there may be a relative sign relating $\eta_{c}({\bf T},{\bf T})$ and $\theta_{c}$, which would be a gauge invariant quantity.

The properties given in Eqs.~(\ref{eq:T2_relation1}) and (\ref{eq:T2_relation2}) are useful for determining the local ${\bf T}^2$ values of quasiparticle excitations in typical time-reversal invariant topological phases, see e.g. \Refs{Fidkowski13,Bonderson13d,chen2014b}.

The analysis of fractionalization of time reversal symmetry presented in this section precisely matches that of \Ref{Bonderson13d}. In contrast with \Ref{levinPRB2012}, our definition of local ${\bf T}^{2}$ for the $j$th quasiparticle of a state $|\Psi_{\{a;c,\mu \}} \rangle$ (which carries topological charge $a_j$) is the corresponding eigenvalue $\eta_{a_j}({\bf T,T})$ of the operator $R_{\bf T} U^{(j)}_{\bf T} R_{\bf T}^{-1} U^{(j)}_{\bf T}$ whose nontrivial action is localized in the region $\mathcal{R}_{j}$ containing the $j$th quasiparticle. In particular, this definition applies to the general case where there are an arbitrary number of regions/quasiparticles that transform nontrivially under ${\bf T}$ and where the entire system may transform projectively with $R_{\bf T}^{2}=-1$. In considering the case where there are only two regions that transform nontrivially under ${\bf T}$ and where and the entire system transforms as $R_{\bf T}^{2}=1$, \Ref{levinPRB2012} interprets the operator $R_{\bf T} U^{(1)}_{\bf T}$ as the ``local ${\bf T}$'' operator for region $\mathcal{R}_{2}$ and $R_{\bf T} U^{(2)}_{\bf T}$ as the ``local ${\bf T}$'' operator for region $\mathcal{R}_{1}$. We avoid interpreting the operator $R_{\bf T} U^{(j)}_{\bf T}$ as a ``local ${\bf T}$'' operator (of some complementary region), as it is not a local operator and even its action on a quasiparticle state, which is given by
\begin{equation}
R_{\bf T} U^{(j)}_{\bf T} |\Psi_{\{a;c,\mu \}} \rangle = \eta_{^{\bf T}a_j}({\bf T,T}) \prod_{k\neq j} U^{(k)}_{\bf T} \rho_{\bf T} |\Psi_{\{a;c,\mu \}} \rangle
,
\end{equation}
is generally not localized in one region (even when all the topological charges involved are ${\bf T}$-invariant).

\section{Extrinsic Defects}

Given the existence of a global symmetry $G$, we can introduce point-like defects that carry flux associated with
the group elements ${\bf g} \in G$. In this section, we will describe a way to create such defects and some of their basic properties. We first give a prescription for creating ${\bf g}$-defects in some simple lattice model systems, and subsequently generalize this construction to an arbitrary system in a topological phase. At the end of this discussion, we will briefly discuss the case where there is no global symmetry, which
still allows nontrivial point-like defects as long as Aut$(\mathcal{C})$ is nontrivial. In the following section (Sec.~\ref{sec:Algebraic_Theory}), we will build upon the physical motivation of this section and provide a detailed presentation of the algebraic theory of extrinsic defects, which is known in the mathematical literature as
$G$-crossed braided tensor category theory~\cite{turaev2000,turaev2010}.

\subsection{Physical Realization of ${\bf g}$-Defects}

\subsubsection{Simple lattice model}

We begin by considering a concrete model system, in which we can precisely describe the general idea we wish to abstract.
In particular, we consider a system with a local Hilbert space defined on the sites of a square lattice, whose Hamiltonian $H_0$
has a local on-site unitary symmetry $G$. For simplicity, we restrict to the case where the interactions
in $H_0$ are just nearest neighbor or plaquette interactions, so that the Hamiltonian takes the form
\begin{align}
H_0 = \sum_i h_i + \sum_{\langle i j \rangle} h_{ij} + \sum_{[ ijkl ]} h_{ijkl},
\end{align}
where $h_i$ consists of local operators that act on site $i$, $h_{ij}$ consists of local operators that act
on a pair of neighboring sites $i$ and $j$ connected by the link $\langle i j \rangle$, and $h_{ijkl}$ consists of local operators that act
on a plaquette $[ ijkl ]$ defined by the sites $i$, $j$, $k$, and $l$.

\begin{figure}[t!]
\includegraphics[scale=0.53]{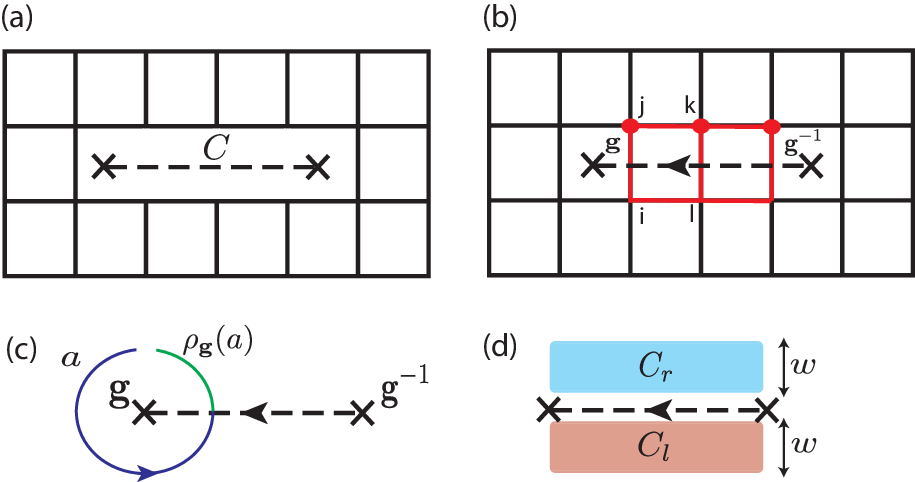}
\caption{\label{fig:defect} (a) When the system is cut along a line $C$, quasiparticles cannot propagate across the cut.
(b) The system can be reglued together along $C$ in a manner that conjugates bond/plaquette operators straddling the cut by a local ${\bf g}$-symmetry action on one side of the cut, as indicated by red dots. The result is a ${\bf g}$ and ${\bf g}^{-1}$ pair of defects at the end-points of a defect branch line (replacing the cut).
(c) Such a construction effectively implements a ${\bf g}$-symmetry transformation on quasiparticles that propagate across the defect branch line, around the defects.
For example, a quasiparticle $a$ will be transformed into $\rho_{\bf g} (a)$ when it encircles the ${\bf g}$-defect in a counterclockwise fashion.
For symmetries that are not on-site, such as translational or rotational symmetries, ${\bf g}$-defects correspond to lattice dislocations or disclinations, respectively.
(d) For more general systems, the defect construction can be generalized by defining regions $C_{l}$ and $C_{r}$ on either side of the cut line, such that terms in the Hamiltonian that straddle the cut line are localized within $C_{l} \cup C_{r}$. These regions will typically have width $w$ that is a few correlation lengths $\xi$.
}
\end{figure}

A pair of defects carrying fluxes ${\bf g}$ and ${\bf g}^{-1}$, respectively, can be created and localized at a well-separated pair of plaquettes by modifying the Hamiltonian as follows. Imagine a line $C$ emanating from the center of one of the defect's corresponding plaquette, cutting across a set of links of the lattice, and terminating at the center of the other defect's plaquette, as shown in Fig.~\ref{fig:defect}. We modify the original Hamiltonian by replacing each term in $H_0$ that straddles the line $C$ with the corresponding operator obtained from that term by acting with the symmetry locally on the sites only on one side of the line $C$.

In order to make this procedure well-defined, we first ascribe an orientation of the line $C$, indicated by an arrow pointing from the ${\bf g}^{-1}$-defect endpoint towards the ${\bf g}$-defect endpoint. (If ${\bf g}={\bf g}^{-1}$, it will not matter which orientation we choose.) This provides a well-defined notion of sites being immediately to the left or to the right of the line $C$. Specifically, the site $i$ is immediately to the left of $C$ and the site $j$ is immediately to the right of the line $C$, if $C$ crosses the link $\langle ij \rangle$ of the lattice connecting sites $i$ and $j$ with $i$ to the left and $j$ to the right, with respect to the orientation of the line $C$. We denote the set of all sites immediately to the left of $C$ as $C_l$ and the set of all sites immediately to the right as $C_r$. We can now define a term in the Hamiltonian to be straddling the line $C$ if it only acts nontrivially on sites in the union $C_l \cup C_r$ and it has nontrivial action on sites in both $C_l$ and $C_r$.~\footnote{We could modify this definition slightly to include also the plaquettes that contain the end-points of the line $C$. Such a modification corresponds to a local change in the Hamiltonian and would also describe a ${\bf g}$ and ${\bf g}^{-1}$ pair of defects.} Finally, we conjugate such terms by the operator $R_{\bf g}^{(C_r)} = \prod_{j \in C_r} R_{\bf g}^{(j)}$, where $R_{\bf g}^{(j)}$ represents the local action of ${\bf g}\in G$ acting on site $j$. (Recall that the global on-site symmetry action can be written as the product of local operators $R_{\bf g} = \prod_{k\in I} R_{\bf g}^{(k)}$, where $I$ in this example is simply the set of all sites.)

Thus, the modified Hamiltonian is given by
\begin{eqnarray}
&& H_{{\bf g},{\bf g}^{-1}} = H_0 + \sum_{ \substack{ \langle ij \rangle : \\ i \in C_l ; j \in C_r }} [R_{\bf g}^{(j)} h_{ij} R_{\bf g}^{(j)-1} - h_{ij} ]
\notag \\
&&+\!\!\!\!\!\! \sum_{ \substack { [ijkl] : \\ i,l \in C_l ; j,k \in C_r }} \!\!\!\!\!\!\!\![R_{\bf g}^{(j)}R_{\bf g}^{(k)} h_{ijkl} R_{\bf g}^{(j)-1} R_{\bf g}^{(k)-1} - h_{ijkl} ]
.
\label{eq:H_{g,g^-1}}
\end{eqnarray}
Here, we have assumed that the line $C$ is straight for simplicity. If $C$ was not a straight line, the last line in this Hamiltonian would include plaquette terms with one site on one side of $C$ and three sites on the other side of $C$, corresponding to the plaquettes where $C$ makes turns. This Hamiltonian $H_{{\bf g},{\bf g}^{-1}}$ defines a line defect associated with the line $C$. The two end points of $C$ are codimension-2 point defects which carry flux ${\bf g}$ and ${\bf g}^{-1}$, respectively. We refer to the line $C$ as a ${\bf g}$-defect branch line.

\subsubsection{${\bf g}$-conjugation of quasiparticles across defect line}

When a quasiparticle is adiabatically transported around a ${\bf g}$-defect, it will be transformed by the symmetry action of the group element ${\bf g}$, as a consequence of crossing the ${\bf g}$-defect branch line. When the action $\rho_{\bf g}$ on topological charges is non-trivial, as a quasiparticle with topological charge $a$ encircles the point-like ${\bf g}$-defect at the end of the defect line $C$, the quasiparticle is transformed into one that carries topological charge $\rho_{\bf g} (a)$. Defects that permute the topological charge values of quasiparticles are sometimes referred to as ``twist defects.''

In order to understand this property, it is useful to first consider starting from the uniform system with Hamiltonian $H_0$, and introducing some quasiparticles using local potentials of the form $h_{a_j}^{(j)}$, as described in Sec.~\ref{sec:Physical_Manifestation}, with the corresponding Hamiltonian $H_{a_1 , \ldots , a_n ;0}$. We now consider an operator $T_{a_k}(k,k')$ that moves the quasiparticle of charge $a_k$ from site $k$ on one side of the line $C$ (which at this point is simply an imaginary line drawn on the system) to the site $k'$ on the other side of $C$ in a manner that crosses the line $C$. Such an operator annihilates a quasiparticle of topological charge $a_k$ at site $k$, creates a quasiparticle of charge $a_k$ at site $k'$, and commutes with the Hamiltonian away from the sites $k$ and $k'$. (One may think of this as a ``string operator.'') Thus, if $|\Psi_{\{a;c,\mu \}} \rangle$ were the ground states of the Hamiltonian $H_{a_1 , \ldots , a_n ;0}$ with $h_{a_j}^{(j)}$ localizing the quasiparticle at site $j$, then $|\Psi'_{\{a;c,\mu \}} \rangle = T_{a_k}(k,k') |\Psi_{\{a;c,\mu \}} \rangle$ are the ground states of the Hamiltonian $H'_{a_1 , \ldots , a_n ;0}$ with $k$th term changed to $h_{a_k}^{(k')}$ localizing the quasiparticle at site $k'$ (perhaps up to some additional unitary transformations localized around the sites $k$ and $k'$). Consequently, it is possible to adiabatically change the Hamiltonian between these configurations and, in doing so, adiabatically move the quasiparticle of charge $a_k$ from site $k$ to site $k'$.

We next imagine cutting all bonds of the system along the line $C$, as indicated in Fig.~\ref{fig:defect}(a). The corresponding Hamiltonian is
\begin{align}
H_{\text{cut}(C)} &= H_0 - \sum_{ \substack{ \langle ij \rangle : \\ i \in C_l ; j \in C_r }}  h_{ij} -\sum_{ \substack { [ijkl] : \\ i,l \in C_l ; j,k \in C_r }} h_{ijkl} ,
\end{align}
where we have again assumed $C$ is a straight line for simplicity. In this system, it is no longer possible to adiabatically move a quasiparticle across the line $C$ (without reintroducing the excised terms in the Hamiltonian), because there are no terms in the Hamiltonian that connect the system across $C$. If we introduce quasiparticles away from the line $C$ using local potentials to similarly produce a Hamiltonian $H_{a_1 , \ldots , a_n ;\text{cut}(C)}$, we would find that the operator $T_{a_k}(k,k')$ does not commute with the Hamiltonian $H_{\text{cut}(C)}$ in the vicinity of $C$ (nor in the vicinity of the sites $k$ and $k'$), hence it will create quasiparticles there. Consequently, this operator would now correspond to moving the quasiparticle from site $k$ to its nearer side the cut line $C$, pair creating quasiparticles of charge $a_k$ and $\bar{a}_{k}$ on the other side of the cut line $C$, and moving the charge $a_k$ of that pair to site $k'$, while leaving the charge $\bar{a}_{k}$ quasiparticle next to the cut line $C$ on the opposite side from the original quasiparticle. Such a process involves more than just adiabatically transporting the quasiparticle, since one must either introduce additional local potentials for the extra quasiparticles, or cost energy above the gap for creating the additional quasiparticles.

We now imagine reintroducing the bond/plaquette operators that connect the system across the cut line $C$ with a conjugation of these operators by the symmetry action of ${\bf g}$ acting locally only on the sites on one side of the cut, to obtain the Hamiltonian $H_{{\bf g },{\bf g}^{-1} }$. Then we introduce quasiparticles away from $C$ using local potentials to similarly produce a Hamiltonian $H_{a_1 , \ldots , a_n ;{\bf g },{\bf g}^{-1}}$. We similarly find that the operator $T_{a_k}(k,k')$ will, in general, not commute with the Hamiltonian $H_{\text{cut}(C)}$ in the vicinity of $C$ (nor in the vicinity of the sites $k$ and $k'$), and, therefore, must create extra quasiparticles there.

However, in this case, the line $C$ is not an untraversable cut line, and one can actually construct an operator that corresponds to adiabatically transporting a quasiparticle across $C$ (without creating extra quasiparticles). For this, we start from the operator $T_{a_k}(k,k')$, which can be written as a product of local operators, and modify it in the following way. The local terms in the product whose nontrivial action is entirely on the left side of $C$ are left unaltered, the local terms in the product whose nontrivial action is entirely on the right side of $C$ are conjugated by $R_{\bf g}$, and the local terms in the product that straddle $C$ are conjugated by $R_{\bf g}^{(C_r)}$. The resulting operator, which we denote $T_{a_k;{\bf g}}(k,k')$, annihilates a quasiparticle of topological charge $a_k$ at site $k$, creates a quasiparticle of charge $^{\bf g} a_k$ at site $k'$, and commutes with the Hamiltonian $H_{{\bf g },{\bf g}^{-1} }$ away from the sites $k$ and $k'$. (Note that if the unmodified operator $T_{a_k}(k,k')$ commutes with $H_0$ away from the sites $k$ and $k'$, then so does $R_{\bf g} T_{a_k}(k,k') R_{\bf g}^{-1}$.) Thus, if $|\Psi_{\{a;c,\mu \}; {\bf g },{\bf g}^{-1}} \rangle$ were the ground states of the Hamiltonian $H_{a_1 , \ldots , a_n ;{\bf g },{\bf g}^{-1}}$ with $h_{a_j}^{(j)}$ localizing the quasiparticle at site $j$, then $|\Psi'_{\{a';c',\mu' \};{\bf g },{\bf g}^{-1}} \rangle = T_{a_k;{\bf g}}(k,k') |\Psi_{\{a;c,\mu \};{\bf g },{\bf g}^{-1}} \rangle$ are the ground states of the Hamiltonian $H'_{a_1 , \ldots, \,^{\bf g}a_k ,\ldots , a_n ;{\bf g },{\bf g}^{-1}}$ with the $k$th term changed to $h_{\,^{\bf g}a_k}^{(k')}$ localizing a quasiparticle of charge $^{\bf g}a_k$ at site $k'$ (perhaps up to some additional unitary transformations localized around the sites $k$ and $k'$). Consequently, it is possible to adiabatically change the Hamiltonian between these configurations (without creating extra quasiparticles), and, in doing so, adiabatically move the quasiparticle from site $k$ to site $k'$, while also transforming its topological charge from $a_k$ to $\rho_{\bf g}(a_k)$ as it crosses the ${\bf g}$-defect branch line.

\subsubsection{General construction of ${\bf g}$-defects}

We can generalize the above discussion and prescription for creating defects to a general topologically ordered system with a local Hamiltonian $H_0$. Again, we first draw an oriented line $C$ in the system. We then define regions $C_l$ and $C_r$, which are ``immediately'' to the left and right of the line $C$, respectively. These regions should have width $w$ such that any term in the Hamiltonian that straddles the line $C$ has nontrivial action that is localized (perhaps up to exponentially damped tails) in the union $C_l \cup C_r$. Typically, this will require the width $w$ to be a few correlation lengths $\xi$. The precise details of how these regions, $C_l$ and $C_r$, terminate near the endpoints of the line $C$ is unimportant for establishing that there is a ${\bf g}$-defect (though it may play a role in determining which type of ${\bf g}$-defect is preferred, as we will explain below). We next identify the terms in $H_0$ whose nontrivial action is localized entirely within $C_l \cup C_r$, and denote the sum of these terms as $H_0(C)$. We define the operator $R^{(C_r)}_{\bf g} = \prod_{j : \mathcal{M}_{j} \subset C_r} R_{\bf g}^{(j)}$, where we decompose the space manifold $\mathcal{M} = \cup_{k\in I} \mathcal{M}_{j}$ into a collection of simply connected disjoint regions $\mathcal{M}_{j}$, none of which straddle the line $C$, i.e. $C \cap \text{int}(\mathcal{M}_{j}) = \emptyset$ for all $j$. Finally, we define the defect Hamiltonian
\begin{equation}
H_{{\bf g},{\bf g}^{-1}} = H_0 + [R^{(C_r)}_{\bf g} H_0(C) R^{(C_r)-1}_{\bf g}  - H_0(C)] .
\end{equation}

It should be clear that these constructions can also be generalized to describe the system with an arbitrary number $n$ of defects which carry group elements ${\bf g}_{1}, \ldots , {\bf g}_{n}$ whose product is identity $\prod_{j=1}^{n} {\bf g}_{j} = {\bf 0}$.

\subsubsection{Point-like nature and confinement of ${\bf g}$-defects}

When $G$ is continuous or is physically obtained by spontaneously breaking a larger continuous symmetry, the ${\bf g}$-defects can be created gradually. This property is familiar in the case of superfluid vortices, where the phase of the order parameter rotates continuously by $2\pi$. For symmetries that are not on-site, such as translational or rotational symmetries, the defects correspond to lattice dislocations or disclinations. In all of these cases, the ${\bf g}$-defects are well-defined even though there is no specific ${\bf g}$-branch line across which the ${\bf g}$-action takes place. In other words, the ${\bf g}$-defects are truly point-like objects.

In fact, from the perspective of the topological order and quantum numbers, the defect branch lines are completely invisible in general. There are no local measurements one can perform using topological properties and operations, such as quasiparticle braiding, that can identify the location of a defect branch line. Only the end-points of the branch lines, where the ${\bf g}$-defects are localized, are locally detectable by topological objects or operations. We stress that this does not necessarily mean that the branch lines are invisible to all forms of local measurements. Depending on the physical realization, the branch lines may or may not be a physically well-localized and measurable object. For example, in superconductor-semiconductor heterostructure-based realizations of Majorana and
parafendleyon wires~\cite{Kitaev01,Lutchyn10,Oreg10,clarke2013,lindner2012,cheng2012,Fendley12},
the defect branch lines are the segments of nanowires in the topological phase, and are clearly locally measurable and identifiable. On the other hand, for multi-layer systems with genons~\cite{barkeshli2012a,barkeshli2013genon}, which are defects whose group action transfers quasiparticles from one layer to another, abstractly there may be no precise, well-defined location of the branch lines, whereas there may be in some experimental realizations~\cite{barkeshli2013}.

The ${\bf g}$-defects defined above are \it extrinsic defects \rm in the system, in the sense that they are imposed by deforming the uniform Hamiltonian to the defect Hamiltonian $H_{ {\bf g}, {\bf g}^{-1}}$. The locations of the ${\bf g}$-defects are classical parameters in $H_{ {\bf g}, {\bf g}^{-1}}$ and thus do not fluctuate quantum mechanically. However, if we allow the defects to become dynamical objects, whose positions do fluctuate quantum mechanically, then there is a question of whether they are confined or deconfined. If they are confined, then the energy cost to separating the dynamical ${\bf g}$-defects will grow with their separation. If they are deconfined, then the energy cost for separating the ${\bf g}$-defects will be finite and independent of their separation, up to exponentially small corrections. Given the Hamiltonian of the system, diagnosing whether the ${\bf g}$-defects correspond to confined or deconfined excitations may be a non-trivial task. We expect that one possible way to do this would be to obtain the ground state $|\Psi_{{\bf g}, {\bf g}^{-1}} \rangle$ of $H_{ {\bf g}, {\bf g}^{-1}}$, and then to compute the average energy of this ground state with respect to the original Hamiltonian: $E^0_{ {\bf g}, {\bf g}^{-1}} = \langle \Psi_{{\bf g}, {\bf g}^{-1}} | H_0 | \Psi_{{\bf g}, {\bf g}^{-1}}  \rangle$. The confinement/deconfinement of the defects would then correspond to whether $E^0_{ {\bf g}, {\bf g}^{-1}}$ diverges with the separation between the defects or is bounded by a finite value, respectively, in the limit of large separations.

If the ${\bf g}$-defects are deconfined, as described above, then they correspond to quasiparticle excitations of the phase $\mathcal{C}$. In such a case, the global $G$ symmetry effectively becomes an emergent local gauge invariance with gauge group $G$ at long wavelengths. In what follows, we
focus on the case where the ${\bf g}$-defects correspond to \it confined \rm objects, and in fact we will reserve the term ${\bf g}$-defect for this case. The case where $G$ is promoted to a local gauge invariance is described in Sec.~\ref{sec:gauging}.

\subsubsection{Aut$(\mathcal{C})$ defects without global symmetry}

It is important to note that even when the underlying physical system has no exact global symmetry of its microscopic Hamiltonian (i.e. $G$ is trivial), the existence of nontrivial topological symmetry Aut$(\mathcal{C})$ of the emergent topological phase $\mathcal{C}$ implies the possibility of nonetheless being able to support defects that effect Aut$(\mathcal{C})$ action on quasiparticles. In particular, one can potentially have point-like defects associated with nontrivial group elements in Aut$(\mathcal{C})$. However, without any global symmetries, the microscopic Hamiltonian constructions of defects previously described in this section cannot be applied. As such, creating Aut$(\mathcal{C})$ defects with a generic microscopic Hamiltonian without global symmetry is a more complicated issue, which we do not address here.~\footnote{This requires a detailed understanding of gapped line defects; see, e.g.,~\Refs{kitaev2012,barkeshli2013defect2}.}

As a simple example of the realization of Aut$(\mathcal{C})$ defects, without a global symmetry, consider the defects associated with layer exchange in a double-layer topological phase~\cite{barkeshli2012a}. These defects are well-defined even in the absence of an exact layer-exchange symmetry. Therefore, the concept of an Aut$(\mathcal{C})$ defect is not logically dependent on the global symmetry of the microscopic Hamiltonian. In what follows, we focus on extrinsic point-like defects that are associated with elements of a global symmetry $G$. This is because we wish to develop a complete characterization of symmetry-enriched topological phases associated with a global symmetry $G$, and we also wish to study the mechanism of gauging the global symmetry $G$, which requires us to start with a system where $G$ is an exact microscopic global symmetry. We will still be able to consider Aut$(\mathcal{C})$ defects in the absence of global symmetries using the same formalism that we will subsequently develop by taking a \emph{fictitious} symmetry group $G = \text{Aut}(\mathcal{C})$ with corresponding symmetry action that is the trivial isomorphism $[\rho ] : \text{Aut}(\mathcal{C}) \rightarrow \text{Aut}(\mathcal{C})$, specified by $\rho_{[\varphi]} = \varphi$. However, doing this may also require a modified understanding of which properties of the resulting defect theory are well-defined and which are not physical, when there is no global symmetry.

\subsection{Topologically Distinct Types of ${\bf g}$-Defects}
\label{sec:Topo_Distinct_Defects}

In the previous subsection, we provided an example of how to modify the Hamiltonian to realize
${\bf g}$-defects. However, it is not necessarily the case that there is a unique type of ${\bf g}$-defect that may be physically realized in a given topological phase. In principle, a topological
phase may support multiple types of ${\bf g}$-defects that cannot be transformed into one another by the application of a local operator. In these cases, there would be topologically distinct types of ${\bf g}$-defects.

As a simple example, we may consider a Hamiltonian which makes it locally
preferable for a quasiparticle with topological charge $b$ to be bound to the ${\bf g}$-defect. Under certain circumstances,
this composite object might correspond to a topologically distinct type of ${\bf g}$-defect as
compared to the original one. Indeed, as we will explain in the next subsection, two topologically distinct types of ${\bf g}$-defects can always be obtained from each other by fusion with a quasiparticle carrying an appropriate value of topological charge. This can be understood intuitively, since topologically distinct types of ${\bf g}$-defects can only differ by topological properties of the topological phase that can be point-like localized at the defect (endpoint of a ${\bf g}$-branch line). While there is no preference between topologically distinct ${\bf g}$-defects when considered in the topological context, it will generically be the case that there will be an energetic preference between distinct ${\bf g}$-defects, as they will have different energy costs for a given physical realization.

\begin{figure}[t!]
\includegraphics[scale=0.35]{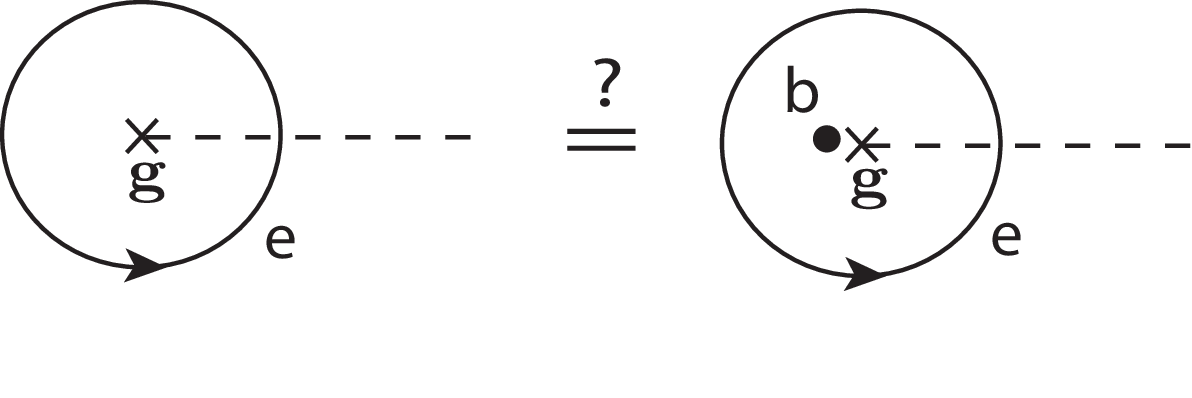}
\caption{\label{fig:defectType} A ${\bf g}$-defect can possibly be altered into a topologically distinct type of
${\bf g}$-defect by fusing it with a quasiparticle carrying nontrivial topological charge $b \in \mathcal{C}$.
Whether the original $g$-defect and the $b$-${\bf g}$ composite object are topologically distinct depends
on whether there is some topological charge $e \in \mathcal{C}$, whose Wilson loop around the defects
can distinguish them. Such a topological charge $e$ must be ${\bf g}$-invariant, $\rho_{\bf g}(e) = e$,
otherwise its Wilson loop could not close upon itself after crossing the ${\bf g}$-defect branch line. Trying to close the loop would necessarily result in a quasiparticle when $\rho_{\bf g}(e) \neq e$.
}
\end{figure}

If two ${\bf g}$-defects are topologically distinct, then there must be a topological process that can distinguish
them. This process corresponds to the Wilson loop operator $W_e$ associated with a $\rho_{\bf g}$-invariant
topological charge $e$ encircling the ${\bf g}$-defect, as shown schematically in Fig.~\ref{fig:defectType}. Different possible eigenvalues of $W_e$ can be used
to distinguish topologically distinct types of defects. In fact, we will later show that this statement can be made more precise. In particular, for a modular theory $\mathcal{C}$, we will show that one can write a linear combinations of such Wilson loop operators which acts as orthogonal projectors on the enclosed area onto each topologically distinct type of ${\bf g}$-defect. (We will also show that the number of topologically distinct types of ${\bf g}$-defects is equal to the number of $\rho_{\bf g}$ invariant topological charges in the original topological phase $\mathcal{C}$.)

In order to refer to topologically distinct types of ${\bf g}$-defects, we must use a more refined labeling system than simply assigning them the group element ${\bf g}$. We give each topologically distinct type of defect its own label $a$, which, in accord with prior terminology, we call topological charge. We write the set of topological charges corresponding to distinct types of ${\bf g}$-defects as $\mathcal{C}_{\bf g}$. We will often use the notation $a_{\bf g}$ as a shorthand to indicate that $a \in \mathcal{C}_{\bf g}$. We emphasize that this does not mean $a_{\bf g}$ is a composite object formed by a ${\bf g}$-defect and a topological charge $a \in \mathcal{C}$ from the original topological phase. In this notation, the topological charge set labeled by the identity group element ${\bf 0}$ is equal to the original set of topological charges of the topological phase, i.e. $\mathcal{C}_{\bf 0} = \mathcal{C}$. We write the set of all topological charges as $\mathcal{C}_G$.

\section{Algebraic Theory of Defects}
\label{sec:Algebraic_Theory}

We now wish to develop a mathematical description of the topological properties, such as fusion and braiding, of ${\bf g}$-defects in a topological phase $\mathcal{C}$ with global on-site symmetry $G$, that generalizes (and includes) the UBTC theory used to describe (deconfined) quasiparticle excitations of the topological phase. The proper mathematical description of such defects is known as a $G$-crossed braided tensor category~\cite{turaev2000,turaev2010}. In this section, we present the $G$-crossed theory, starting with $G$-graded fusion and then introducing $G$-crossed braiding. We derive the consistency conditions and a number of important properties for such theories. In Appendix~\ref{sec:Cat_Formulation}, we provide a concise presentation of $G$-crossed categories more properly using the abstract formalism of category theory.

\subsection{$G$-Graded Fusion}
\label{sec:G-graded_Fusion}

It is clear that combining a ${\bf g}$-defect with an ${\bf h}$-defect should yield a ${\bf gh}$-defect. Hence, the
fusion of defects must respect the group multiplication structure of $G$, leading to the notion of $G$-graded fusion.

A fusion category $\mathcal{C}_{G}$ is $G$-graded if it can be written as
\begin{equation}
\mathcal{C}_{G} = \bigoplus_{{\bf g} \in G} \mathcal{C}_{\bf g}
.
\end{equation}
In particular, this means each topological charge $a\in \mathcal{C}_{G}$ is assigned a unique group element ${\bf g} \in G$ and corresponding charge subset $\mathcal{C}_{\bf g}$ to which it belongs, such that fusion respects the group multiplication of $G$, i.e. if $a\in\mathcal{C}_{\bf g}$ and $b\in\mathcal{C}_{\bf h}$, then $N_{ab}^{c}$ can only be nonzero if $c\in\mathcal{C}_{\bf gh}$.

We recall the shorthand notation $a_{\bf g}$ used to indicate that $a \in \mathcal{C}_{\bf g}$. With this, we can write the fusion rules [of Eq.~(\ref{eq:fusion_rules})] as
\begin{equation}
\label{eq:G-graded_fusion}
a_{\bf g} \times b_{\bf h} = \sum_{c\in \mathcal{C}_{G}} N_{ab}^{c} c = \sum_{c\in \mathcal{C}_{\bf gh}} N_{ab}^{c} c = \sum_{c} N_{ab}^{c} c_{\bf gh}
.
\end{equation}

All the properties and constraints of fusion categories from Sec.~\ref{sec:Fusion} carry over directly to $G$-graded fusion categories.
Clearly, the vacuum charge $0 \in \mathcal{C}_{\bf 0}$, where we write the identity element of the group $G$ as ${\bf 0}$. It should be clear that $\mathcal{C}_{\bf 0}$ is itself a fusion category, since it is closed under fusion. As such, we consider a $G$-graded category $\mathcal{C}_{G}$ to be a ``$G$-extension'' of its subcategory $\mathcal{C}_{\bf 0}$.

The unique charge conjugate of a topological charge $a_{\bf g} $ is denoted $\overline{a_{\bf g}} \in \mathcal{C}_{{\bf g}^{-1}}$.
Since $\overline{a_{\bf g}}$ is the unique topological charge with which $a_{\bf g}$ can fuse into vacuum, i.e. $N_{a_{\bf g} b_{\bf h}}^{0} = \delta_{\overline{a_{\bf g}} b_{\bf h}}$, it follows
that for any two distinct topological charges $a_{\bf g}, c_{\bf g} \in \mathcal{C}_{\bf g}$, there must exist some
nontrivial topological charges $b_{\bf 0},b'_{\bf 0} \in \mathcal{C}_{\bf 0}$ such that $c_{\bf g}$ is one of the
fusion outcomes obtained from fusing $a_{\bf g}$ with $b_{\bf 0}$ or fusing $b'_{\bf 0}$ with $a_{\bf g}$, i.e.  $N_{a_{\bf g} b_{\bf 0}}^{c_{\bf g}} = N_{\overline{a_{\bf g}} c_{\bf g}}^{b_{\bf 0}} \neq 0$ and $N_{ b'_{\bf 0} a_{\bf g}}^{c_{\bf g}} = N_{ c_{\bf g} \overline{a_{\bf g}}}^{b'_{\bf 0}} \neq 0$. Physically, this means that different types of ${\bf g}$-defects in (a $G$-extension of) a topological phase described by $\mathcal{C}_{\bf 0}$ can indeed be obtained from each other by fusing quasiparticles, which carry topological charges in $\mathcal{C}_{\bf 0}$, with the ${\bf g}$-defects.~\footnote{After introducing $G$-crossed braiding in the next section, we will see that the same charge $b_{\bf 0}$ can always be used for either left or right fusion with $a_{\bf g}$ to obtain $c_{\bf g}$, i.e. there exists some $b_{\bf 0}$ such that $N_{a_{\bf g} b_{\bf 0}}^{c_{\bf g}} = N_{\overline{a_{\bf g}} c_{\bf g}}^{b_{\bf 0}} = N_{c_{\bf g} \overline{a_{\bf g}}}^{b_{\bf 0}}= N_{b_{\bf 0} a_{\bf g}}^{c_{\bf g}} \neq 0$.}

As before, the quantum dimensions (which are defined in the same way) obey the relation
\begin{equation}
\label{eq:d_g_relation}
d_{a_{\bf g}} d_{b_{\bf h}} = \sum_{c} N_{ab}^{c} d_{c_{\bf gh}}
.
\end{equation}

We define the (total) quantum dimension of $\mathcal{C}_{\bf g}$ to be
\begin{equation}
\mathcal{D}_{\bf g} = \sqrt{ \sum_{a \in \mathcal{C}_{\bf g}} d_{a_{\bf g}}^2 }
.
\end{equation}
Using Eq.~(\ref{eq:d_g_relation}) and the fact that $N_{ab}^{c} = N_{c \bar{b}}^{a}$, we see, by picking some arbitrary $b \in \mathcal{C}_{\bf g}$, that
\begin{eqnarray}
\mathcal{D}_{\bf 0}^2 &=& \sum_{a \in \mathcal{C}_{\bf 0}} d_{a_{\bf 0}}^2 = \sum_{\substack{ a \in \mathcal{C}_{\bf 0} \\ c\in\mathcal{C}_{\bf g} }} d_{a_{\bf 0}} d_{b_{\bf g}}^{-1} N_{a b}^{c} d_{c_{\bf g}}
\notag \\
&=& \sum_{\substack{ a \in \mathcal{C}_{\bf 0} \\ c\in\mathcal{C}_{\bf g} }}  d_{b_{\bf g}}^{-1} d_{c_{\bf g}} N_{c \bar{b}}^{a} d_{a_{\bf 0}}
= \sum_{c\in\mathcal{C}_{\bf g}} d_{c_{\bf g}}^{2} = \mathcal{D}_{\bf g}^{2}
\end{eqnarray}
for any ${\bf g} \in G$ with nonempty $\mathcal{C}_{\bf g}\neq \emptyset$. In particular, the quantum dimension of every nonempty $\mathcal{C}_{\bf g}$ is
\begin{equation}
\mathcal{D}_{\bf g} = \mathcal{D}_{\bf 0} = |H|^{-\frac{1}{2}} \mathcal{D}_{\mathcal{C}_{G}}
,
\end{equation}
where $\mathcal{D}_{\mathcal{C}_{G}}$ is the total quantum dimension of $\mathcal{C}_{G}$ and we define the subgroup
\begin{equation}
H = \left\{\, {\bf h}\in G \,\, | \,\, \mathcal{C}_{\bf h} \neq \emptyset \, \right\} \leq G
.
\end{equation}
That $H$ forms a subgroup of $G$ follows from the fact that $\mathcal{C}_{\bf g},\mathcal{C}_{\bf h} \neq \emptyset$ implies that $\mathcal{C}_{\bf gh} \neq \emptyset$, together with the existence of a vacuum charge and charge conjugates.

In this paper, we will focus our attention to faithfully $G$-graded categories, i.e. those with $H=G$, so that there is no ${\bf g}\in G$ with $\mathcal{C}_{\bf g} = \emptyset$.
In other words, we study the full defect theory associated with all group elements ${\bf g} \in G$.
We note that one could instead choose to study the defect theory associated with a subgroup $H \leq G$. In this case, one can leave
$\mathcal{C}_{\bf g}$ for ${\bf g} \notin H$ empty and then study the resulting non-faithfully $G$-graded category. Such a
non-faithfully $G$-graded category would just be a faithfully $H$-graded category, with the empty
sets $\mathcal{C}_{\bf g}$ for ${\bf g} \notin H$ included formally. This is only nontrivial once we also include the symmetry action of such ${\bf g} \notin H$.

\subsection{$G$-Crossed Braiding}
\label{sec:G-crossed_Braiding}

We can consider a continuous family of Hamiltonians $H(\lambda)$ of the physical system containing defects (possibly including quasiparticles, which we consider to be ${\bf 0}$-defects), where the locations of the defects and their corresponding branch lines are changed adiabatically as a function of the parameter $\lambda$. This allows us to implement physical operations that exchange the positions of defects.

With this in mind, we wish to define a notion of braiding of defects, called ``$G$-crossed braiding,'' that includes group action and which is compatible with a $G$-graded fusion category $\mathcal{C}_{G}$. We denote such a $G$-crossed braided tensor category as $\mathcal{C}_{G}^{\times}$. This requires some modification of the usual definition of braiding. In fact, when $G$ is a non-Abelian group, fusion in a $G$-graded fusion category is not commutative, so the usual notion of braiding cannot even be applied. In particular, there must be a group action when the positions of objects (carrying nontrivial group elements) are exchanged. (Of course, the usual definition of braiding still applies within the subcategory $\mathcal{C}_{\bf 0}$, which is a BTC.)

As the mathematical formalism is developed, it will become clear that one can also physically implement braiding transformations for non-Abelian defects by using topological charge measurements and/or tunable interactions, following the ``measurement-only'' methods of \Refs{Bonderson08a,Bonderson08b,Bonderson12a}. As these methods remove the need to physically move the defects, they may provide a more preferable physical implementation of braiding transformations, depending on the details of the physical system.

\begin{figure}[t!]
	\centering
	\resizebox{0.9\columnwidth}{!}{
	\pspicture[shift=-0.4](-1.2,-0.5)(0.5,2.1)
	\psset{linewidth=1.0pt,linecolor=black}
	\psline(-0.06,-0.26)(0.06,-0.14)
	\psline(-0.06,-0.14)(0.06,-0.26)
	\psline(-1.01,-0.26)(-0.89,-0.14)
	\psline(-1.01,-0.14)(-0.89,-0.26)
	\psset{linewidth=0.9pt,linecolor=red, linestyle=dashed, dash=2pt 1pt}
	\psline(0,-0.2)(0,1.8)
	\psset{linewidth=0.9pt,linecolor=blue, linestyle=dashed, dash=2pt 1pt}
	\psline(-0.95,-0.2)(-0.95,1.8)
	\psset{linewidth=0.9pt,linecolor=red,arrowscale=1.0,arrowinset=0.15}
	\psline{<-}(0,0.8)(0,0.81)
	\psline[linewidth=0.9pt,linecolor=blue,arrowscale=1.0,arrowinset=0.15]{<-}(-0.95,0.8)(-0.95,0.81)
	\psset{linewidth=0.7pt, linecolor=black, linestyle=solid}
\scriptsize
	\rput[bl]{0}(-1.2,1.2){$\mb{g}$}
	\rput[bl]{0}(0.05,1.2){$\mb{h}$}
	\rput[bl]{0}(-1.2,-0.1){$a$}
	\rput[bl]{0}(0.1,-0.1){$b$}
	\rput[bl]{0}(-1.5,1.8){(a)}
\endpspicture
\pspicture[shift=-0.4](-1.8,-0.5)(0.5,2.1)
	\psset{linewidth=1.0pt,linecolor=black}
	\psline(-0.06,-0.26)(0.06,-0.14)
	\psline(-0.06,-0.14)(0.06,-0.26)
	\psline(-1.01,-0.26)(-0.89,-0.14)
	\psline(-1.01,-0.14)(-0.89,-0.26)
	\psset{linewidth=0.9pt,linecolor=blue, linestyle=dashed, dash=2pt 1pt}
	\psline(0,-0.2)(0,1)
	\psarc(-0.2,1){0.2}{5}{85}
	\psline(-0.2,1.2)(-0.6,1.2)
	\psarc(-0.6,1){0.2}{95}{175}
	\psline(-0.8,1)(-0.8,-0.2)
	\psarc(-0.95,-0.2){0.15}{185}{355}
	\psline(-1.1,-0.2)(-1.1,1.8)
	\psset{linewidth=0.9pt,linecolor=red, linestyle=dashed, dash=2pt 1pt}
	\psline(-0.95,-0.2)(-0.95,1)
	\psarc(-0.65, 1){0.3}{95}{175}
	\psline(-0.65,1.3)(-0.2,1.3)
	\psarc(-0.2,1.5){0.2}{275}{355}
	\psline(0,1.5)(0,1.8)
	\psset{linewidth=0.9pt,linecolor=blue,arrowscale=1.0,arrowinset=0.15}
	\psline{<-}(0,0.7)(0,0.71)
	\psline{<-}(-0.8,0.85)(-0.8,0.84)
	\psline{<-}(-1.1, 0.7)(-1.1,0.71)
	\psline[linewidth=0.9pt,linecolor=red,arrowscale=1.0,arrowinset=0.15]{<-}(-0.95,0.4)(-0.95,0.41)
\scriptsize
	\rput[bl]{0}(0.05,0.6){$\mb{g}$}
	\rput[bl]{0}(0.05,1.5){$\mb{h}$}
	\rput[bl]{0}(0.1,-0.2){$a$}
    \rput[bl]{0}(-1.3,-0.25){$b$}
	\rput[bl]{0}(-1.6,1.8){(b)}
\endpspicture
\pspicture[shift=-0.4](-1.8,-0.5)(0.3,2.0)
	\psset{linewidth=1.0pt,linecolor=black}
	\psline(-0.06,-0.26)(0.06,-0.14)
	\psline(-0.06,-0.14)(0.06,-0.26)
	\psline(-1.01,-0.26)(-0.89,-0.14)
	\psline(-1.01,-0.14)(-0.89,-0.26)
	\psset{linewidth=0.9pt,linecolor=blue, linestyle=dashed, dash=2pt 1pt}
	\psline(0,-0.2)(0,1)
	\psarc(-0.2,1){0.2}{5}{85}
	\psline(-0.2,1.2)(-0.8,1.2)
	\psarc(-0.8,1.4){0.2}{185}{265}
	\psline(-1,1.4)(-1,1.8)
	\psset{linewidth=0.9pt,linecolor=purple, linestyle=dashed, dash=2pt 1pt}
	\psline(-0.95,-0.2)(-0.95,1)
	\psarc(-0.65, 1){0.3}{140}{175}
	\psline[linewidth=0.9pt,linecolor=purple,arrowscale=1.0,arrowinset=0.15]{<-}(-0.95,0.5)(-0.95,0.51)
	\psset{linewidth=0.9pt,linecolor=red, linestyle=dashed, dash=2pt 1pt}
	\psarc(-0.65, 1){0.3}{95}{138}
	\psline(-0.65,1.3)(-0.2,1.3)
	\psarc(-0.2,1.5){0.2}{275}{355}
	\psline(0,1.5)(0,1.8)
	\psset{linewidth=0.9pt,linecolor=blue,arrowscale=1.0,arrowinset=0.15}
	\psline{<-}(0,0.5)(0,0.51)
\scriptsize
	\rput[bl]{0}(-1.35,0.7){$\mb{^gh}$}
	\rput[bl]{0}(-1.35,-0.2){$^\mb{g}b$}
	\rput[bl]{0}(0.1,-0.2){$a$}
	\rput[bl]{0}(0.05,0.7){$\mb{g}$}
	\rput[bl]{0}(-1.5,1.8){(c)}
\endpspicture
}
\caption{\label{fig:branch_change} (a) Each symmetry defect is labeled by a topological charge and has a corresponding defect branch line emanating from it characterized by a symmetry group element. Here we show a ${\bf g}$-defect with charge $a$ and an ${\bf h}$-defect with charge $b$, and their corresponding branch lines in a 2D system. (b) As a ${\bf g}$-defect is braided with an ${\bf h}$-defect in the counterclockwise sense, one can imagine deforming the corresponding branch lines, so that no objects cross them. (c) In order to return to the original configuration of branch lines, one must pass the ${\bf g}$ branch line across the ${\bf h}$-defect and its branch line. As the ${\bf h}$-defect of topological charge $b$ passes through the ${\bf g}$ branch line, the topological charge $b$ is transformed to $\rho_{\bf g}(b)$ and the ${\bf h}$ branch line is transformed into a $\mb{^gh}={\bf ghg}^{-1}$ branch line. This corresponds to the $G$-crossed braiding operator $R^{^{\bf g} b_{\bf h} a_{\bf g}}$, as defined in Eq.~(\ref{eq:G-crossed_R}).}
\end{figure}

\begin{figure}[t!]
\begin{center}
\includegraphics[scale=0.4]{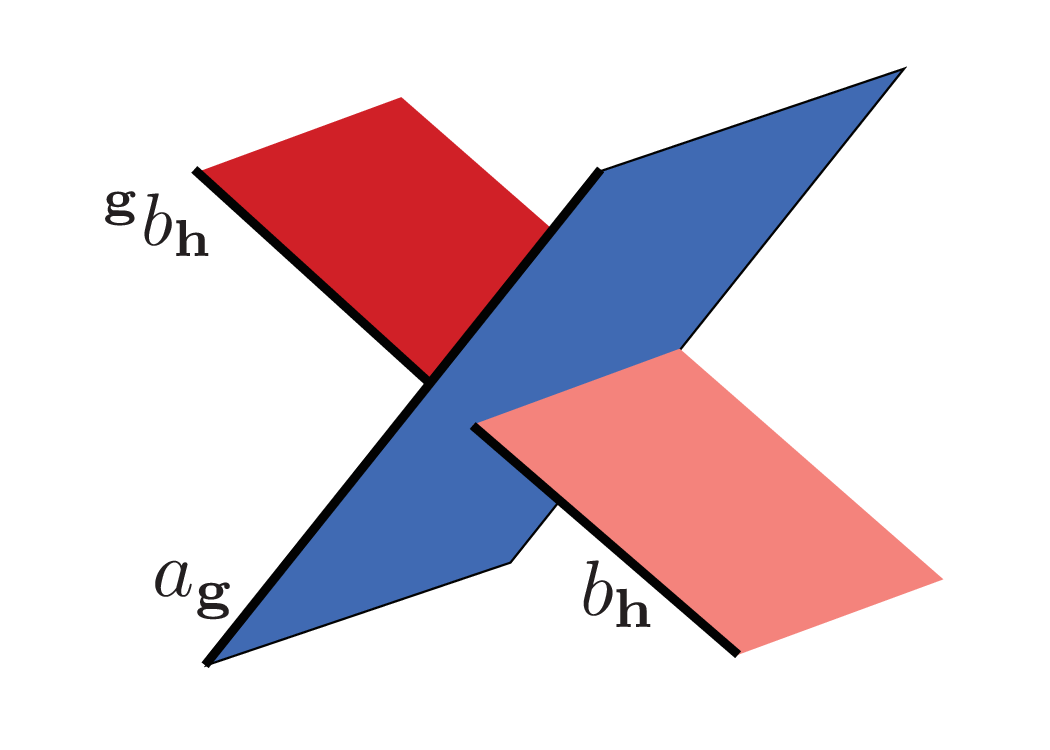}
\caption{Defect branch lines form worldsheets in spacetime. A local portion of the worldsheet diagram is shown for the G-crossed braiding of defects described in Fig.~\ref{fig:branch_change}.}
\label{defectBraid3d}
\end{center}
\end{figure}

When the objects carry non-trivial group elements, they are considered symmetry defects, which one can think of as having a branch cut line emanating from the otherwise point-like object. These branch cuts are oriented and are labeled by the group element of the object at which they terminate, so that taking an object through a ${\bf g}$-branch in the counterclockwise sense around the branch point at the corresponding ${\bf g}$-defect gives ${\bf g}$-action on that object, as shown in Fig.~\ref{fig:branch_change}. This process can be depicted in the three-dimensional space-time as shown in Fig. \ref{defectBraid3d}. In order to describe this using diagrammatics, we choose the convention where the branch lines, which form worldsheets that end on the worldlines of the defects, go into the page, and then we leave the branch line worldsheets implicit in the diagrammatics. This does not impose any restriction on how the defect branch lines must be physically configured in the actual system. Rather, it is merely a bookkeeping tool that allows us to consistently keep track of the effects of the branch lines in the diagrammatics, while only drawing the worldlines of the defects and not the branch line worldsheets. With this convention, a ${\bf g}$-defect worldline applies group action on objects when it crosses over their worldlines. In particular, we define $G$-crossed braiding by
\begin{eqnarray}
R^{a_{\bf g} b_{\bf h}}&=&
\pspicture[shift=-0.75](-0.1,-0.4)(1.3,1.4)
\small
  \psset{linewidth=0.9pt,linecolor=black,arrowscale=1.5,arrowinset=0.15}
  \psline(0.96,0.05)(0.2,1)
  \psline{->}(0.96,0.05)(0.28,0.9)
  \psline(0.24,0.05)(1,1)
  \psline[border=2pt]{->}(0.24,0.05)(0.92,0.9)
  \rput[bl]{0}(-0.1,1.1){$a_{\bf g}$}
  \rput[br]{0}(1.2,1.1){$b_{\bf h}$}
  \rput[bl]{0}(-0.15,-0.05){$b_{\bf h}$}
  \rput[br]{0}(1.55,-0.1){$^{\bf \bar h} a_{\bf g}$}
  \endpspicture
\notag \\
&=&\sum\limits_{c,\mu ,\nu }\sqrt{\frac{d_{c}}{d_{a}d_{b}}}\left[
R_{c_{\bf gh}}^{a_{\bf g} b_{\bf h}}\right] _{\mu \nu }
 \pspicture[shift=-1](-0.3,-0.85)(1.8,1.3)
  \small
  \psset{linewidth=0.9pt,linecolor=black,arrowscale=1.5,arrowinset=0.15}
  \psline{->}(0.7,0)(0.7,0.45)
  \psline(0.7,0)(0.7,0.55)
  \psline(0.7,0.55) (0.25,1)
  \psline{->}(0.7,0.55)(0.3,0.95)
  \psline(0.7,0.55) (1.15,1)
  \psline{->}(0.7,0.55)(1.1,0.95)
  \rput[bl]{0}(0.1,0.2){$c_{\bf gh}$}
  \rput[br]{0}(1.4,1.05){$b_{\bf h}$}
  \rput[bl]{0}(-0.1,1.05){$a_{\bf g}$}
  \psline(0.7,0) (0.25,-0.45)
  \psline{-<}(0.7,0)(0.35,-0.35)
  \psline(0.7,0) (1.15,-0.45)
  \psline{-<}(0.7,0)(1.05,-0.35)
  \rput[br]{0}(1.7,-0.5){$^{\bf \bar h}a_{\bf g}$}
  \rput[bl]{0}(-0.1,-0.45){$b_{\bf h}$}
\scriptsize
  \rput[bl]{0}(0.85,0.4){$\nu$}
  \rput[bl]{0}(0.85,-0.03){$\mu$}
  \endpspicture
,
\label{eq:G-crossed_R}
\end{eqnarray}%
where the $R$-symbols for a $G$-crossed theory are the maps $R^{ab}_{c} : V^{b_{\bf h} \,^{\bf \bar h}a_{\bf g}}_{c_{\bf gh}} \rightarrow V^{a_{\bf g} b_{\bf h}}_{c_{\bf gh}}$ that result from exchanging (in a counterclockwise manner) two objects of charges $b_{\bf h}$ and $^{\bf \bar h}a_{\bf g}$, respectively, which are in the charge $c_{\bf gh}$ fusion channel. We recall that
\begin{eqnarray}
\label{eq:G-crossed_rho}
^{\bf g} b_{\bf h} &=& \rho_{\bf g}(b_{\bf h}) \\
\bar{{\bf g}} &=& {\bf g}^{-1}\\
^\mb{g}\mb{h}&=& \mb{g}\mb{h}\mb{g}^{-1}
\end{eqnarray}
in the shorthand notation introduced in Sec.~\ref{sec:Global_Symmetry} for the symmetry group action on topological charges.~\footnote{Had we allowed the $G$-crossed braiding action to depend on the topological charge value, rather than only depending on the corresponding group element, i.e. if we replaced $\rho_{\bf g}$ with a more general map $\rho_{a_{\bf g}}$, compatibility with fusion would require that $\rho_{a_{\bf g}} \circ \rho_{b_{\bf h}} = \rho_{c_{\bf gh}}$ whenever $N_{ab}^{c}\neq 0$. Combining this property with an axiom that $\rho_{b_{\bf 0}}$ act trivially on all topological charges for all $b_{\bf 0} \in \mathcal{C}_{\bf 0}$, i.e. $\rho_{b_{\bf 0}}(e)=e$ for any $e\in \mathcal{C}_{G}^{\times}$, would lead back to $\rho_{a_{\bf g}} = \rho_{\bf g}$ being independent of the particular topological charge $a$ within $\mathcal{C}_{\bf g}$. In particular, for any two distinct charges $a_{\bf g} \neq c_{\bf g}$ in $\mathcal{C}_{\bf g}$ , there is always some $b_{\bf 0} \in \mathcal{C}_{\bf 0}$ with $N_{ab}^{c}\neq 0$, and hence $\rho_{a_{\bf g}} = \rho_{a_{\bf g}} \circ \rho_{b_{\bf 0}} = \rho_{c_{\bf g}}$. This axiom is physically natural, because the topological charges in $\mathcal{C}_{\bf 0}$ correspond to quasiparticles, which are truly point-like localized (they do not have defect branch cut lines) and hence should be unable to alter operators or topological charges localized in a distant region, unless it enters that region.}

The symmetry action $[\rho] : G \rightarrow$Aut$(\mathcal{C}_{\bf 0})$ on the original theory must now be self-consistently extended to an action of the symmetry group
\begin{equation}
[\rho] : G \rightarrow \text{Aut}(\mathcal{C}_{G}^{\times})
\end{equation}
that is incorporated within the structure of the extended theory. Notice, for example, that compatibility with the $G$-graded fusion rules required that $^{\bf g}b_{\bf h} \in \mathcal{C}_{{\bf ghg}^{-1}}$, i.e.
\begin{equation}
\rho_{\bf g} : \mathcal{C}_{\bf h} \rightarrow \mathcal{C}_{\bf ghg^{-1}}
.
\end{equation}
More generally, compatibility with the fusion algebra requires
\begin{equation}
N_{a_{\bf g} b_{\bf h}}^{c_{\bf gh}} = N_{^{\bf g} b_{\bf h} a_{\bf g} }^{c_{\bf gh}} = N_{b_{\bf h}  \,^{\bar{\bf h}}a_{\bf g} }^{c_{\bf gh}}.
\end{equation}
From this, together with the properties of charge conjugates, it follows that $N_{a_{\bf g} b_{\bf h}}^{0} = N_{^{\bf g} b_{\bf h} a_{\bf g} }^{0} = \delta_{b_{\bf h} \overline{a_{\bf g}} }$, and hence any topological charge in $\mathcal{C}_{\bf g}$ will be invariant under the action of the corresponding ${\bf g}$, i.e.
\begin{equation}
^{\bf g} a_{\bf g} = \:^{{\bf g}^{n}}a_{\bf g} = a_{\bf g}
,
\end{equation}
for all $n \in \mathbb{Z}$.

For some theories (this may occur also in FTCs or BTCs), it may be possible for a topological charge $a_{\bf g}$ to remain unchanged
after fusion/splitting with another nontrivial topological charge $b_{\bf 0}$. In particular, this occurs when $N_{a_{\bf g} b_{\bf 0}}^{a_{\bf g}} = N_{b_{\bf 0} a_{\bf g}}^{a_{\bf g}}\neq 0$. In this case, $b_{\bf 0}$ quasiparticles can be absorbed or emitted at
the $a_{\bf g}$-defect without changing the localized topological charge or localization energy of the defect. As such, we say that
defects (or quasiparticles) that carry charge $a_{\bf g}$ localize a ``$b_{\bf 0}$ zero mode.'' It is clear from
\begin{equation}
N_{a_{\bf g} b_{\bf 0}}^{a_{\bf g}} = N_{a_{\bf g} \overline{b_{\bf 0}}}^{a_{\bf g}} = N_{b_{\bf 0} a_{\bf g}}^{a_{\bf g}} = N_{a_{\bf g} \overline{a_{\bf g}}}^{b_{\bf 0}} = N_{\overline{a_{\bf g}} b_{\bf 0}}^{\overline{a_{\bf g}}} = N_{\:^{\bf g} b_{\bf 0} a_{\bf g}}^{a_{\bf g}}
\end{equation}
that if $a_{\bf g}$ localizes a $b_{\bf 0}$ zero mode, then: (1) $\overline{a_{\bf g}}$ also localizes a $b_{\bf 0}$ zero mode, (2) $a_{\bf g}$ and $\overline{a_{\bf g}}$ localize  $\overline{b_{\bf 0}}$  zero modes and also zero modes associated with the entire ${\bf g}$-orbit of charges $\:^{{\bf g}^{n}}b_{\bf 0}$, and (3) $b_{\bf 0}$ is one of the fusion channels of $a_{\bf g}$ with its conjugate $\overline{a_{\bf g}}$, as is $\overline{b_{\bf 0}}$ and $\:^{{\bf g}^{n}}b_{\bf 0}$.

The $G$-crossed $R$-symbols can equivalently be written in terms of the relation
\begin{equation}
\pspicture[shift=-0.8](-0.1,-0.5)(1.5,1.4)
  \small
  \psset{linewidth=0.9pt,linecolor=black,arrowscale=1.5,arrowinset=0.15}
  \psline{->}(0.7,0)(0.7,0.43)
  \psline(0.7,0)(0.7,0.5)
 \psarc(0.8,0.6732051){0.2}{120}{240}
 \psarc(0.6,0.6732051){0.2}{-60}{35}
  \psline (0.6134,0.896410)(0.267,1.09641)
  \psline{->}(0.6134,0.896410)(0.35359,1.04641)
  \psline(0.7,0.846410) (1.1330,1.096410)	
  \psline{->}(0.7,0.846410)(1.04641,1.04641)
  \rput[bl]{0}(0.5,-0.3){$c_{\bf gh}$}
  \rput[br]{0}(1.4,1.15){$b_{\bf h}$}
  \rput[bl]{0}(-0.1,1.15){$a_{\bf g}$}
 \scriptsize
  \rput[bl]{0}(0.82,0.35){$\mu$}
  \endpspicture
=\sum\limits_{\nu }\left[R_{c_{\bf gh}}^{a_{\bf g} b_{\bf h}}\right] _{\mu \nu }
\pspicture[shift=-0.8](-0.1,-0.5)(1.5,1.4)
  \small
  \psset{linewidth=0.9pt,linecolor=black,arrowscale=1.5,arrowinset=0.15}
  \psline{->}(0.7,0)(0.7,0.45)
  \psline(0.7,0)(0.7,0.55)
  \psline(0.7,0.55) (0.25,1)
  \psline{->}(0.7,0.55)(0.3,0.95)
  \psline(0.7,0.55) (1.15,1)	
  \psline{->}(0.7,0.55)(1.1,0.95)
  \rput[bl]{0}(0.5,-0.3){$c_{\bf gh}$}
  \rput[br]{0}(1.4,1.05){$b_{\bf h}$}
  \rput[bl]{0}(-0.1,1.05){$a_{\bf g}$}
 \scriptsize
  \rput[bl]{0}(0.82,0.37){$\nu$}
  \endpspicture
.
\end{equation}%

Similarly, the clockwise $G$-crossed braiding exchange operator is
\begin{eqnarray}
&& \left(R^{ a_{\bf g} b_{\bf h}}\right)^{-1} =
\pspicture[shift=-0.75](-0.1,-0.4)(1.5,1.25)
\small
  \psset{linewidth=0.9pt,linecolor=black,arrowscale=1.5,arrowinset=0.15}
  \psline{->}(0.24,0.05)(0.92,0.9)
  \psline(0.24,0.05)(1,1)
  \psline(0.96,0.05)(0.2,1)
  \psline[border=2pt]{->}(0.96,0.05)(0.28,0.9)
  \rput[bl]{0}(0.0,1.1){$b_{\bf h}$}
  \rput[br]{0}(1.3,1.05){$^{\bf \bar h}a_{\bf g}$}
  \rput[bl]{0}(0,-0.3){$a_{\bf g}$}
  \rput[br]{0}(1.3,-0.25){$b_{\bf h}$}
  \endpspicture
\notag \\
&& \quad =\sum\limits_{c,\mu ,\nu }\sqrt{\frac{d_{c}}{d_{a}d_{b}}}\left[\left(
R_{c_{\bf gh}}^{a_{\bf g} b_{\bf h}}\right)^{-1}\right] _{\mu \nu }
 \pspicture[shift=-1](-0.1,-0.85)(1.5,1.4)
  \small
  \psset{linewidth=0.9pt,linecolor=black,arrowscale=1.5,arrowinset=0.15}
  \psline{->}(0.7,0)(0.7,0.45)
  \psline(0.7,0)(0.7,0.55)
  \psline(0.7,0.55) (0.25,1)
  \psline{->}(0.7,0.55)(0.3,0.95)
  \psline(0.7,0.55) (1.15,1)
  \psline{->}(0.7,0.55)(1.1,0.95)
  \rput[bl]{0}(0.1,0.2){$c_{\bf gh}$}
  \rput[br]{0}(1.4,1.05){$^{\bf \bar h}a_{\bf g}$}
  \rput[bl]{0}(0,1.1){$b_{\bf h}$}
  \psline(0.7,0) (0.25,-0.45)
  \psline{-<}(0.7,0)(0.35,-0.35)
  \psline(0.7,0) (1.15,-0.45)
  \psline{-<}(0.7,0)(1.05,-0.35)
  \rput[br]{0}(1.4,-0.75){$b_{\bf h}$}
  \rput[bl]{0}(0,-0.8){$a_{\bf g}$}
\scriptsize
  \rput[bl]{0}(0.85,0.4){$\nu$}
  \rput[bl]{0}(0.85,-0.03){$\mu$}
  \endpspicture
.
\end{eqnarray}

In order for $G$-crossed braiding to be compatible with fusion, we again wish to have the ability to slide lines over or under fusion vertices. However, we may no longer assume that such operations are completely trivial, since one must at least account for the group action on a vertex. The appropriate relations are given by the unitary transformations
\begin{eqnarray}
\label{eq:GcrossedU}
\psscalebox{.6}{
\pspicture[shift=-1.7](-0.8,-0.8)(1.8,2.4)
  \small
  \psset{linewidth=0.9pt,linecolor=black,arrowscale=1.5,arrowinset=0.15}
  \psline{->}(0.7,0)(0.7,0.45)
  \psline(0.7,0)(0.7,0.55)
  \psline(0.7,0.55)(0.25,1)
  \psline(0.7,0.55)(1.15,1)	
  \psline(0.25,1)(0.25,2)
  \psline{->}(0.25,1)(0.25,1.9)
  \psline(1.15,1)(1.15,2)
  \psline{->}(1.15,1)(1.15,1.9)
  \psline[border=2pt](-0.65,0)(2.05,2)
  \psline{->}(-0.65,0)(0.025,0.5)
  \rput[bl]{0}(-0.4,0.6){$x_{\bf k}$}
  \rput[br]{0}(1.5,0.65){$^{\bf \bar k}b$}
  \rput[bl]{0}(0.4,-0.4){$^{\bf \bar k}c_{\bf gh}$}
  \rput[br]{0}(1.3,2.1){$b_{\bf h}$}
  \rput[bl]{0}(0.0,2.05){$a_{\bf g}$}
 \scriptsize
  \rput[bl]{0}(0.85,0.35){$\mu$}
  \endpspicture
}
&=&
\sum_{\nu} \left[ U_{\bf k}\left(a, b; c\right)\right]_{\mu \nu}
\psscalebox{.6}{
\pspicture[shift=-1.7](-0.8,-0.8)(1.8,2.4)
  \small
  \psset{linewidth=0.9pt,linecolor=black,arrowscale=1.5,arrowinset=0.15}
  \psline{->}(0.7,0)(0.7,0.45)
  \psline(0.7,0)(0.7,1.55)
  \psline(0.7,1.55)(0.25,2)
  \psline{->}(0.7,1.55)(0.3,1.95)
  \psline(0.7,1.55) (1.15,2)	
  \psline{->}(0.7,1.55)(1.1,1.95)
  \psline[border=2pt](-0.65,0)(2.05,2)
  \psline{->}(-0.65,0)(0.025,0.5)
  \rput[bl]{0}(-0.4,0.6){$x_{\bf k}$}
  \rput[bl]{0}(0.4,-0.4){$^{\bf \bar k}c_{\bf gh}$}
  \rput[bl]{0}(0.15,1.2){$c_{\bf gh}$}
  \rput[br]{0}(1.3,2.1){$b_{\bf h}$}
  \rput[bl]{0}(0.0,2.05){$a_{\bf g}$}
 \scriptsize
  \rput[bl]{0}(0.85,1.35){$\nu$}
  \endpspicture
}
\qquad
\\
\label{eq:Gcrossed_eta}
\psscalebox{.6}{
\pspicture[shift=-1.7](-0.8,-0.8)(1.8,2.4)
  \small
  \psset{linewidth=0.9pt,linecolor=black,arrowscale=1.5,arrowinset=0.15}
  \psline(-0.65,0)(2.05,2)
  \psline[border=2pt](0.7,0.55)(0.25,1)
  \psline[border=2pt](1.15,1)(1.15,2)
  \psline(0.7,0.55)(1.15,1)	
  \psline{->}(0.7,0)(0.7,0.45)
  \psline(0.7,0)(0.7,0.55)
  \psline(0.25,1)(0.25,2)
  \psline{->}(0.25,1)(0.25,1.9)
  \psline{->}(1.15,1)(1.15,1.9)
  \psline{->}(-0.65,0)(0.025,0.5)
  \rput[bl]{0}(-0.5,0.6){$x_{\bf k}$}
  \rput[bl]{0}(0.5,1.3){$^{\bf \bar g}x$}
  \rput[bl]{0}(1.6,2.1){$^{\bf \bar h \bar g}x_{\bf k}$}
  \rput[bl]{0}(0.4,-0.35){$c_{\bf gh}$}
  \rput[br]{0}(1.3,2.1){$b_{\bf h}$}
  \rput[bl]{0}(0.0,2.05){$a_{\bf g}$}
 \scriptsize
  \rput[bl]{0}(0.85,0.35){$\mu$}
  \endpspicture
}
&=&
\eta_{x}\left({\bf g},{\bf h}\right)
\psscalebox{.6}{
\pspicture[shift=-1.7](-0.8,-0.8)(1.8,2.4)
  \small
  \psset{linewidth=0.9pt,linecolor=black,arrowscale=1.5,arrowinset=0.15}
  \psline(-0.65,0)(2.05,2)
  \psline[border=2pt](0.7,0)(0.7,1.55)
  \psline{->}(0.7,0)(0.7,0.45)
  \psline(0.7,1.55)(0.25,2)
  \psline{->}(0.7,1.55)(0.3,1.95)
  \psline(0.7,1.55) (1.15,2)	
  \psline{->}(0.7,1.55)(1.1,1.95)
  \psline{->}(-0.65,0)(0.025,0.5)
  \rput[bl]{0}(-0.5,0.6){$x_{\bf k}$}
  \rput[bl]{0}(1.6,2.1){$^{\bf \bar h \bar g}x_{\bf k}$}
  \rput[bl]{0}(0.4,-0.35){$c_{\bf gh}$}
  \rput[br]{0}(1.3,2.1){$b_{\bf h}$}
  \rput[bl]{0}(0.0,2.05){$a_{\bf g}$}
 \scriptsize
  \rput[bl]{0}(0.85,1.35){$\mu$}
  \endpspicture
}
\qquad
.
\end{eqnarray}
We have used the same notation $\left[ U_{\bf k}\left(a,b;c\right)\right]_{\mu \mu'}$ and $\eta_{x}\left({\bf g},{\bf h}\right)$ that we previously introduced for the global symmetry action on the topological degrees of freedom in Sec.~\ref{sec:Global_Symmetry} and the fractionalized (projective) local symmetry action in Sec.~\ref{sec:Obstruction}, because, as we will see, these are precisely the same quantities extended to the entire $G$-crossed theory. Intuitively, it should be clear why this is the case, since an $a_{\bf g}$ line in the $G$-crossed braided diagrammatics has an implicit ${\bf g}$ branch sheet extending from behind it, which applies a ${\bf g}$ action to any object passing through it, i.e. everything that the $a_{\bf g}$ line passes over. Hence, sliding an $x_{\bf k}$ line over a vertex, as in Eq.~(\ref{eq:GcrossedU}), passes the vertex through the ${\bf k}$ branch sheet, and should result in the ${\bf k}$ action on that vertex. Similarly, passing a $|a_{\bf g},b_{\bf h};c_{\bf gh},\mu \rangle$ vertex over an $x_{\bf k}$ line, as in Eq.~(\ref{eq:Gcrossed_eta}), should capture the local projective relation of equating ${\bf gh}$ action on charge $x$ with successively applied ${\bf g}$ and ${\bf h}$ actions on charge $x$, as the vertex indicates where the ${\bf gh}$ branch sheet splits into a ${\bf g}$ branch sheet and an ${\bf h}$ branch sheet. The validity of this claim will be established through the following consistency arguments and conditions. The quantity $\left[ U_{\bf k}\left(a,b;c\right)\right]_{\mu \mu'}$ here corresponds to a specific choice of $\rho \in [\rho]$, and we will see that the relation between choices within a symmetry action equivalence class (related by natural isomorphisms) will take the form of a gauge transformation in this theory.

We begin by arguing that the factors in these expressions must have the given dependence on the various topological and group quantities. In particular, in Eq.~(\ref{eq:GcrossedU}), we see that the nontrivial interaction is between the ${\bf k}$-branch line and the vertex, hence there may be dependence on ${\bf k}$, but not the specific topological charge $x \in \mathcal{C}_{\bf k}$, and the transformation on the fusion state space may be nontrivial, so it may depend on all the vertex labels. For Eq.~(\ref{eq:Gcrossed_eta}), we see that the nontrivial interaction is between the ${\bf g}$, ${\bf h}$, and ${\bf gh}$ branch lines and the topological charge $x$, so this expression may depend on ${\bf g}$ and ${\bf h}$, but should not depend on the specific topological charge values $a$, $b$, or $c$, nor should it have any effect within the fusion state space of the fusion vertex.

Sliding a line over a vertex, as in Eq.~(\ref{eq:GcrossedU}) is a unitary transformation between $V^{\,^{\bf \bar k}a \,^{\bf \bar k}b}_{\,^{\bf \bar k}c}$ and $V^{ab}_c$, as specified by the unitary operators $U_{\bf k}\left(a,b;c\right)$. This requires the dimensionality of the fusion spaces to be preserved under the corresponding symmetry action, giving
\begin{equation}
N_{ \,^{\bf k}a_{\bf g} \,^{\bf k}b_{\bf h}}^{\,^{\bf k}c_{\bf gh}} = N_{a_{\bf g} b_{\bf h}}^{c_{\bf gh}}
\end{equation}
for any ${\bf k}$ acting on a vertex. It follows that the quantum dimensions are also invariant
\begin{equation}
d_{a_{\bf g}} = d_{\,^{\bf k}a_{\bf g} }
.
\end{equation}

Clearly, if the sliding line has vacuum charge $x_{\bf k} =0$, the sliding transformations should be trivial, so
 \begin{eqnarray}
\left[ U_{\bf 0}\left(a,b;c\right)\right]_{\mu \nu} &=& \delta_{\mu \nu} \\
\eta_{0}\left({\bf g},{\bf h}\right) &=& 1
.
\end{eqnarray}

We require that the sliding rules are compatible with the property that vacuum lines can be freely added or removed from a diagram, i.e. sliding over/under a vertex $|a,b;c\rangle$ with $a=0$ or $b=0$ should be trivial, since it is equivalent to simply sliding over a line. This imposes the conditions
\begin{eqnarray}
U_{\bf k}\left(0,0;0\right) &=& U_{\bf k}\left(a,0;a\right) = U_{\bf k}\left(0,b;b\right) = 1 \\
\eta_{x}\left({\bf 0},{\bf 0}\right) &=& \eta_{x}\left({\bf g},{\bf 0}\right) = \eta_{x}\left({\bf 0},{\bf h}\right) = 1
.
\end{eqnarray}

Combining Eqs.~(\ref{eq:GcrossedU}) and (\ref{eq:Gcrossed_eta}) with trivial braidings, such as
\begin{equation}
\pspicture[shift=-0.6](-0.2,-0.45)(1.0,1.1)
  \small
  \psset{linewidth=0.9pt,linecolor=black,arrowscale=1.5,arrowinset=0.15}
  \psline(0.3,-0.4)(0.3,1)
  \psline{->}(0.3,-0.4)(0.3,0.50)
  \psline(0.7,-0.4)(0.7,1)
  \psline{->}(0.7,-0.4)(0.7,0.50)
  \rput[br]{0}(0.96,0.8){$b$}
  \rput[bl]{0}(0,0.8){$a$}
  \endpspicture
=
\pspicture[shift=-0.6](0.0,-0.05)(1.1,1.45)
  \small
  \psarc[linewidth=0.9pt,linecolor=black,border=0pt] (0.8,0.7){0.4}{120}{225}
  \psarc[linewidth=0.9pt,linecolor=black,arrows=<-,arrowscale=1.4,arrowinset=0.15] (0.8,0.7){0.4}{165}{240}
  \psarc[linewidth=0.9pt,linecolor=black,border=0pt] (0.4,0.7){0.4}{-45}{45}
  \psarc[linewidth=0.9pt,linecolor=black,arrows=->,arrowscale=1.4,arrowinset=0.15] (0.4,0.7){0.4}{-45}{15}
  \psarc[linewidth=0.9pt,linecolor=black,border=0pt](0.8,1.39282){0.4}{180}{225}
  \psarc[linewidth=0.9pt,linecolor=black,border=0pt](0.4,1.39282){0.4}{-60}{0}
  \psarc[linewidth=0.9pt,linecolor=black,border=0pt](0.8,0.00718){0.4}{135}{180}
  \psarc[linewidth=0.9pt,linecolor=black,border=0pt](0.4,0.00718){0.4}{0}{60}
  \rput[bl]{0}(0.1,1.2){$a$}
  \rput[br]{0}(1.06,1.2){$b$}
  \endpspicture
,
\end{equation}
we see that sliding lines over or under vertices with the opposite braiding are given by
\begin{eqnarray}
\psscalebox{.6}{
\pspicture[shift=-1.7](-0.8,-0.8)(1.8,2.4)
  \small
  \psset{linewidth=0.9pt,linecolor=black,arrowscale=1.5,arrowinset=0.15}
  \psline{->}(0.7,0)(0.7,0.45)
  \psline(0.7,0)(0.7,1.55)
  \psline(0.7,1.55)(0.25,2)
  \psline{->}(0.7,1.55)(0.3,1.95)
  \psline(0.7,1.55) (1.15,2)	
  \psline{->}(0.7,1.55)(1.1,1.95)
  \psline[border=2pt](-0.65,2)(2.05,0)
  \psline{->}(2.05,0)(1.375,0.5)
  \rput[bl]{0}(1.4,0.6){$x_{\bf k}$}
  \rput[bl]{0}(0.4,-0.4){$^{\bf k}c_{\bf gh}$}
  \rput[bl]{0}(0.85,1.2){$c$}
  \rput[br]{0}(1.3,2.1){$b_{\bf h}$}
  \rput[bl]{0}(0.0,2.05){$a_{\bf g}$}
 \scriptsize
  \rput[bl]{0}(0.45,1.35){$\mu$}
  \endpspicture
}
&=&
\sum_{\nu} \left[U_{\bf k}\left(\,^{\bf k}a,\,^{\bf k}b;\,^{\bf k}c\right)\right]_{\mu \nu}
\psscalebox{.6}{
\pspicture[shift=-1.7](-0.8,-0.8)(1.8,2.4)
  \small
  \psset{linewidth=0.9pt,linecolor=black,arrowscale=1.5,arrowinset=0.15}
  \psline{->}(0.7,0)(0.7,0.45)
  \psline(0.7,0)(0.7,0.55)
  \psline(0.7,0.55)(0.25,1)
  \psline(0.7,0.55)(1.15,1)	
  \psline(0.25,1)(0.25,2)
  \psline{->}(0.25,1)(0.25,1.9)
  \psline(1.15,1)(1.15,2)
  \psline{->}(1.15,1)(1.15,1.9)
  \psline[border=2pt](-0.65,2)(2.05,0)
  \psline{->}(2.05,0)(1.375,0.5)
  \rput[bl]{0}(1.4,0.6){$x_{\bf k}$}
  \rput[bl]{0}(0.4,-0.4){$^{\bf k}c_{\bf gh}$}
  \rput[br]{0}(1.3,2.1){$b_{\bf h}$}
  \rput[bl]{0}(0.0,2.05){$a_{\bf g}$}
  \rput[bl]{0}(-0.2,0.8){$^{\bf k}a$}
  \rput[bl]{0}(0.8,0.3){$^{\bf k}b$}
 \scriptsize
  \rput[bl]{0}(0.3,0.3){$\nu$}
  \endpspicture
}
\qquad
\label{eqn:sliding_fusion3}
\\
\psscalebox{.6}{
\pspicture[shift=-1.7](-0.8,-0.8)(1.8,2.4)
  \small
  \psset{linewidth=0.9pt,linecolor=black,arrowscale=1.5,arrowinset=0.15}
  \psline(2.05,0)(-0.65,2)
  \psline[border=2pt](0.7,0)(0.7,1.55)
  \psline{->}(0.7,0)(0.7,0.45)
  \psline(0.7,1.55)(0.25,2)
  \psline{->}(0.7,1.55)(0.3,1.95)
  \psline(0.7,1.55) (1.15,2)	
  \psline{->}(0.7,1.55)(1.1,1.95)
  \psline{->}(2.05,0)(1.375,0.5)
  \rput[bl]{0}(-0.8,2.1){$x_{\bf k}$}
  \rput[bl]{0}(1.4,0.6){$^{\bf \bar h \bar g} x_{\bf k}$}
  \rput[bl]{0}(0.4,-0.35){$c_{\bf gh}$}
  \rput[br]{0}(1.3,2.1){$b_{\bf h}$}
  \rput[bl]{0}(0.0,2.05){$a_{\bf g}$}
 \scriptsize
  \rput[bl]{0}(0.85,1.35){$\mu$}
  \endpspicture
}
&=&
\eta_{x}\left({\bf g},{\bf h}\right)
\psscalebox{.6}{
\pspicture[shift=-1.7](-0.8,-0.8)(1.8,2.4)
  \small
  \psset{linewidth=0.9pt,linecolor=black,arrowscale=1.5,arrowinset=0.15}
  \psline(2.05,0)(-0.65,2)
  \psline[border=2pt](0.7,0.55)(1.15,1)
  \psline[border=2pt](0.25,1)(0.25,2)
  \psline[border=2pt](1.15,1)(1.15,2)
  \psline(0.7,0.55)(0.25,1)
  \psline{->}(0.7,0)(0.7,0.45)
  \psline(0.7,0)(0.7,0.55)
  \psline(0.25,1)(0.25,2)
  \psline{->}(0.25,1)(0.25,1.9)
  \psline{->}(1.15,1)(1.15,1.9)
  \psline{->}(2.05,0)(1.375,0.5)
  \rput[bl]{0}(-0.8,2.1){$x_{\bf k}$}
  \rput[bl]{0}(1.4,0.6){$^{\bf \bar h \bar g} x_{\bf k}$}
  \rput[bl]{0}(0.5,1.3){$^{\bf \bar g}x$}
  \rput[bl]{0}(0.4,-0.35){$c_{\bf gh}$}
  \rput[br]{0}(1.3,2.1){$b_{\bf h}$}
  \rput[bl]{0}(0.0,2.05){$a_{\bf g}$}
 \scriptsize
  \rput[bl]{0}(0.85,0.35){$\mu$}
  \endpspicture
}
\qquad
.
\label{eqn:sliding_fusion4}
\end{eqnarray}

Compatibility of the sliding moves with the inner product Eq.~(\ref{eq:inner_product}) is obtained by sliding a line over a bubble diagram, as in Eq.~(\ref{eq:inner_product}). In this way, we obtain the corresponding relations for sliding over fusion (rather than splitting) vertices
\begin{eqnarray}
\psscalebox{.6}{
\pspicture[shift=-1.7](-0.8,-0.8)(1.8,2.4)
  \small
  \psset{linewidth=0.9pt,linecolor=black,arrowscale=1.5,arrowinset=0.15}
  \psline{<-}(0.7,1.9)(0.7,1.5)
  \psline(0.7,2)(0.7,1.45)
  \psline(0.7,1.45)(0.25,1)
  \psline(0.7,1.45)(1.15,1)	
  \psline(0.25,1)(0.25,0)
  \psline{<-}(0.25,0.4)(0.25,0.0)
  \psline(1.15,1)(1.15,0)
  \psline{<-}(1.15,0.4)(1.15,0.0)
  \psline[border=2pt](-0.65,0)(2.05,2)
  \psline{->}(-0.65,0)(0.025,0.5)
  \rput[bl]{0}(-0.4,0.6){$x_{\bf k}$}
  \rput[bl]{0}(0.4,2.1){$c_{\bf gh}$}
  \rput[bl]{0}(-0.1,1.2){$a$}
 \rput[bl]{0}(0.85,1.4){$b$}
  \rput[br]{0}(1.4,-0.4){$^{\bf \bar k} b_{\bf h}$}
  \rput[bl]{0}(0.0,-0.45){$^{\bf \bar k} a_{\bf g}$}
 \scriptsize
  \rput[bl]{0}(0.4,1.4){$\mu$}
  \endpspicture
}
&=&
\sum_{\nu} \left[ U_{\bf k}\left(a,b;c\right)\right]_{\nu \mu}
\psscalebox{.6}{
\pspicture[shift=-1.7](-0.8,-0.8)(1.8,2.4)
  \small
  \psset{linewidth=0.9pt,linecolor=black,arrowscale=1.5,arrowinset=0.15}
  \psline{<-}(0.7,1.9)(0.7,1.5)
  \psline(0.7,2)(0.7,0.45)
  \psline(0.7,0.45)(0.25,0)
  \psline{<-}(0.55,0.3)(0.25,0)
  \psline(0.7,0.45) (1.15,0)	
  \psline{<-}(0.85,0.3)(1.15,0)
  \psline[border=2pt](-0.65,0)(2.05,2)
  \psline{->}(-0.65,0)(0.025,0.5)
  \rput[bl]{0}(-0.4,0.6){$x_{\bf k}$}
  \rput[bl]{0}(0.4,2.1){$c_{\bf gh}$}
  \rput[bl]{0}(0.8,0.5){$^{\bf \bar k}c$}
  \rput[br]{0}(1.4,-0.4){$^{\bf \bar k}b_{\bf h}$}
  \rput[bl]{0}(0.0,-0.45){$^{\bf \bar k}a_{\bf g}$}
 \scriptsize
  \rput[bl]{0}(0.35,0.4){$\nu$}
  \endpspicture
}
\qquad
\label{eqn:sliding_fusion5}
\\
\psscalebox{.6}{
\pspicture[shift=-1.7](-0.8,-0.8)(1.8,2.4)
  \small
  \psset{linewidth=0.9pt,linecolor=black,arrowscale=1.5,arrowinset=0.15}
  \psline{<-}(0.7,1.9)(0.7,1.5)
  \psline(0.7,2)(0.7,0.45)
  \psline(0.7,0.45)(0.25,0)
  \psline{<-}(0.55,0.3)(0.25,0)
  \psline(0.7,0.45) (1.15,0)	
  \psline{<-}(0.85,0.3)(1.15,0)
  \psline[border=2pt](-0.65,2)(2.05,0)
  \psline{->}(2.05,0)(1.375,0.5)
  \rput[bl]{0}(1.4,0.7){$x_{\bf k}$}
  \rput[bl]{0}(0.4,2.1){$ c_{\bf gh}$}
  \rput[bl]{0}(0.25,0.6){$^{\bf k}c$}
  \rput[br]{0}(1.4,-0.4){$^{\bf k}b_{\bf h}$}
  \rput[bl]{0}(0.0,-0.4){$^{\bf k}a_{\bf g}$}
 \scriptsize
  \rput[bl]{0}(0.85,0.4){$\mu$}
  \endpspicture
}
&=&
\sum_{\nu} \left[ U_{\bf k}\left(\,^{\bf k}a,\,^{\bf k}b;\,^{\bf k}c\right)\right]_{\nu \mu}
\psscalebox{.6}{
\pspicture[shift=-1.7](-0.8,-0.8)(1.8,2.4)
  \small
  \psset{linewidth=0.9pt,linecolor=black,arrowscale=1.5,arrowinset=0.15}
  \psline{<-}(0.7,1.9)(0.7,1.5)
  \psline(0.7,2)(0.7,1.45)
  \psline(0.7,1.45)(0.25,1)
  \psline(0.7,1.45)(1.15,1)	
  \psline(0.25,1)(0.25,0)
  \psline{<-}(0.25,0.4)(0.25,0.0)
  \psline(1.15,1)(1.15,0)
  \psline{<-}(1.15,0.4)(1.15,0.0)
  \psline[border=2pt](-0.65,2)(2.05,0)
  \psline{->}(2.05,0)(1.375,0.5)
  \rput[bl]{0}(1.6,0.6){$x_{\bf k}$}
  \rput[bl]{0}(0.4,2.1){$ c_{\bf gh}$}
  \rput[bl]{0}(0.3,1.4){$a$}
 \rput[bl]{0}(1.2,1.1){$b$}
  \rput[br]{0}(1.4,-0.4){$^{\bf k} b_{\bf h}$}
  \rput[bl]{0}(0.0,-0.4){$^{\bf k} a_{\bf g}$}
 \scriptsize
  \rput[bl]{0}(0.85,1.4){$\nu$}
  \endpspicture
}
.
\qquad
\label{eqn:sliding_fusion6}
\end{eqnarray}
A similar calculation gives the relations for sliding lines under fusion vertices
\begin{eqnarray}
\psscalebox{.6}{
\pspicture[shift=-1.7](-0.8,-0.8)(1.8,2.4)
  \small
  \psset{linewidth=0.9pt,linecolor=black,arrowscale=1.5,arrowinset=0.15}
  \psline(-0.65,0)(2.05,2)
  \psline[border=2pt](0.25,1)(0.25,0)
  \psline[border=2pt](0.7,1.45)(1.15,1)
  \psline{<-}(0.7,1.9)(0.7,1.5)
  \psline(0.7,1.45)(0.25,1)
  \psline(0.7,2)(0.7,1.45)	
  \psline{<-}(0.25,0.4)(0.25,0.0)
  \psline(1.15,1)(1.15,0)
  \psline{<-}(1.15,0.4)(1.15,0.0)
  \psline{->}(-0.65,0)(0.025,0.5)
  \rput[bl]{0}(1.6,2.1){$^{\bf \bar h \bar g} x_{\bf k}$}
  \rput[bl]{0}(-0.5,0.6){$x_{\bf k}$}
  \rput[bl]{0}(0.4,0.35){$^{\bf \bar g}x_{\bf k}$}
  \rput[bl]{0}(0.4,2.1){$ c_{\bf gh}$}
  \rput[br]{0}(1.4,-0.35){$ b_{\bf h}$}
  \rput[bl]{0}(0.0,-0.35){$ a_{\bf g}$}
 \scriptsize
  \rput[bl]{0}(0.85,1.4){$\mu$}
  \endpspicture
}
&=&
\eta_{x}\left({\bf g},{\bf h}\right)
\psscalebox{.6}{
\pspicture[shift=-1.7](-0.8,-0.8)(1.8,2.4)
  \small
  \psset{linewidth=0.9pt,linecolor=black,arrowscale=1.5,arrowinset=0.15}
  \psline(-0.65,0)(2.05,2)
  \psline{<-}(0.7,1.9)(0.7,1.5)
  \psline[border=2pt](0.7,2)(0.7,0.45)
  \psline(0.7,0.45)(0.25,0)
  \psline{<-}(0.55,0.3)(0.25,0)
  \psline(0.7,0.45) (1.15,0)	
  \psline{<-}(0.85,0.3)(1.15,0)
  \psline{->}(-0.65,0)(0.025,0.5)
  \rput[bl]{0}(1.6,2.1){$^{\bf \bar h \bar g} x_{\bf k}$}
  \rput[bl]{0}(-0.5,0.6){$x_{\bf k}$}
  \rput[bl]{0}(0.4,2.1){$ c_{\bf gh}$}
  \rput[br]{0}(1.4,-0.35){$ b_{\bf h}$}
  \rput[bl]{0}(0.0,-0.35){$ a_{\bf g}$}
 \scriptsize
  \rput[bl]{0}(0.85,0.4){$\mu$}
  \endpspicture
}
\qquad
\label{eqn:sliding_fusion7}
\\
\psscalebox{.6}{
\pspicture[shift=-1.7](-0.8,-0.8)(1.8,2.4)
  \small
  \psset{linewidth=0.9pt,linecolor=black,arrowscale=1.5,arrowinset=0.15}
  \psline(-0.65,2)(2.05,0)
  \psline[border=2pt](0.7,2)(0.7,0.45)
  \psline{<-}(0.7,1.9)(0.7,1.5)
  \psline(0.7,0.45)(0.25,0)
  \psline{<-}(0.55,0.3)(0.25,0)
  \psline(0.7,0.45) (1.15,0)	
  \psline{<-}(0.85,0.3)(1.15,0)
  \psline{->}(2.05,0)(1.375,0.5)
  \rput[bl]{0}(-0.8,2.1){$x_{\bf k}$}
  \rput[bl]{0}(1.4,0.6){$^{\bf \bar h \bar g} x_{\bf k}$}
  \rput[bl]{0}(0.4,2.1){$ c_{\bf gh}$}
  \rput[br]{0}(1.4,-0.35){$b_{\bf h}$}
  \rput[bl]{0}(0.0,-0.35){$a_{\bf g}$}
 \scriptsize
  \rput[bl]{0}(0.85,0.4){$\mu$}
  \endpspicture
}
&=& \eta_{x}\left({\bf g},{\bf h}\right)
\psscalebox{.6}{
\pspicture[shift=-1.7](-0.8,-0.8)(1.8,2.4)
  \small
  \psset{linewidth=0.9pt,linecolor=black,arrowscale=1.5,arrowinset=0.15}
  \psline(-0.65,2)(2.05,0)
  \psline{<-}(0.7,1.9)(0.7,1.5)
  \psline(0.7,2)(0.7,1.45)
  \psline[border=2pt](0.7,1.45)(0.25,1)
  \psline[border=2pt](1.15,1)(1.15,0)
  \psline(0.7,1.45)(1.15,1)	
  \psline(0.25,1)(0.25,0)
  \psline{<-}(0.25,0.4)(0.25,0.0)
  \psline{<-}(1.15,0.4)(1.15,0.0)
  \psline{->}(2.05,0)(1.375,0.5)
  \rput[bl]{0}(0.4,0.45){$^{\bf \bar g}x_{\bf k}$}
  \rput[bl]{0}(-0.8,2.1){$x_{\bf k}$}
  \rput[bl]{0}(1.4,0.6){$^{\bf \bar h \bar g} x_{\bf k}$}
  \rput[bl]{0}(0.4,2.1){$ c_{\bf gh}$}
  \rput[br]{0}(1.4,-0.35){$ b_{\bf h}$}
  \rput[bl]{0}(0.0,-0.35){$ a_{\bf g}$}
 \scriptsize
  \rput[bl]{0}(0.85,1.4){$\mu$}
  \endpspicture
}
\qquad
\label{eqn:sliding_fusion8}
\end{eqnarray}

Consistency of the sliding moves with each other can be achieved by equating the two different sequences of sliding moving shown in Fig.~\ref{fig:UVdiagrams}, which yields the relation
\begin{widetext}
\begin{equation}
\eta_{b}\left({\bf k}, {\bf l} \right) \eta_{a}\left({\bf k}, {\bf l} \right)  \sum_{\lambda} \left[ U_{\bf l}\left( \,^{\bf \bar k}a ,\,^{\bf \bar k}b ;\,^{\bf \bar k}c \right) \right]_{\mu \lambda} \left[ U_{\bf k}\left( a ,b ;c \right) \right]_{\lambda \nu}
= \left[ U_{\bf kl}\left( a ,b ;c \right) \right]_{\mu \nu} \eta_{c}\left({\bf k}, {\bf l} \right)
.
\label{eqn:symmetry_action_consistency}
\end{equation}

\begin{figure}[t!]
\begin{center}
\includegraphics[scale=0.6]{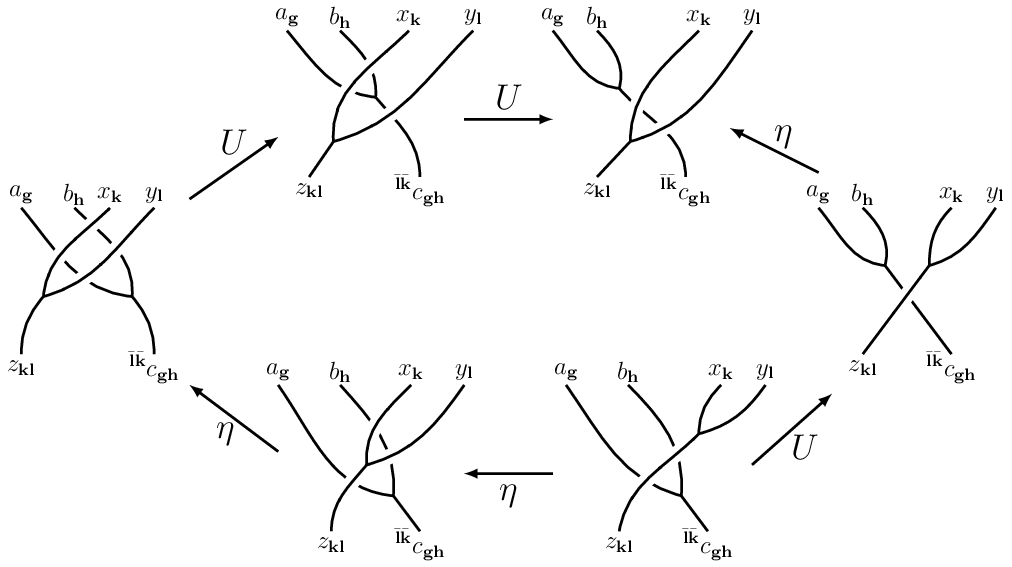}
\caption{The $G$-crossed symmetry action consistency equation provides consistency between the sliding moves, which implement the $U$ and $\eta$ transformations associated with the global and fractionalized (local projective) symmetry action. Eq.~(\ref{eq:U_eta_consistency}) is obtained by imposing the condition that the above diagram commutes.}
\label{fig:UVdiagrams}
\end{center}
\end{figure}

If we define $\kappa_{\bf k,l} = \rho_{\bf kl} \rho_{\bf l}^{-1} \rho_{\bf k}^{-1}$ and $\kappa_{\bf k,l} |a_{\bf g},b_{\bf h};c_{\bf gh},\mu \rangle =  \sum_{\nu} \left[\kappa_{\bf k,l} (a ,b ;c) \right]_{\mu \nu} |a_{\bf g},b_{\bf h};c_{\bf gh},\nu \rangle$,
we see that this condition can be rewritten as the symmetry action consistency equation
\begin{equation}
\label{eq:U_eta_consistency}
\left[\kappa_{\bf k,l} (a ,b ;c) \right]_{\mu \nu} = \sum_{\alpha, \beta} \left[ U_{\bf k}\left( a ,b ;c \right)^{-1} \right]_{\mu \alpha} \left[ U_{\bf l}\left( \,^{\bf \bar k}a ,\,^{\bf \bar k}b ;\,^{\bf \bar k}c \right)^{-1} \right]_{\alpha \beta} \left[ U_{\bf kl}\left( a ,b ;c \right) \right]_{\beta \nu}
= \frac{ \eta_{a}\left({\bf k}, {\bf l} \right) \eta_{b}\left({\bf k}, {\bf l} \right)}{\eta_{c}\left({\bf k}, {\bf l} \right)} \delta_{\mu \nu}
.
\end{equation}

Using this condition to decompose $U_{\bf klm}\left( a ,b ;c \right)$ in the two equivalent ways related by associativity, one obtains the following consistency condition on the $\kappa_{\bf k,l}$
\begin{equation}
\kappa_{\bf l,m} ( \,^{\bf \bar k}a ,\,^{\bf \bar k}b ;\,^{\bf \bar k}c ) \kappa_{\bf k,lm} (a ,b ;c) = \kappa_{\bf k,l} (a ,b ;c) \kappa_{\bf kl,m} (a ,b ;c)
.
\end{equation}
Thus, we see that sliding an $x_{\bf k}$ line over a vertex or operator can indeed be thought of as implementing the $G$-crossed extension of the symmetry action $\rho_{\bf k}$, with $U_{\bf k}\left( a ,b ;c \right)$ playing the same role as in Sec.~\ref{sec:Global_Symmetry}. Similarly, sliding an $x_{\bf k}$ line under a $|a_{\bf g},b_{\bf h};c_{\bf gh},\mu \rangle$ vertex can be thought of as implementing the $G$-crossed extension of the projective phases $\eta_{x}\left({\bf g}, {\bf h} \right)$ relating the local symmetry action of ${\bf g}$ and ${\bf h}$ to ${\bf gh}$.

We continue expounding the relation of the sliding moves to the symmetry action by next requiring consistency between the sliding moves and the $F$-moves. Sliding a line over a fusion tree before or after application of an $F$-move gives
\begin{eqnarray}
\psscalebox{.6}{
  \pspicture[shift=-1.5](-0.5,-0.45)(2.5,2.8)
  \small
  \psset{linewidth=0.9pt,linecolor=black,arrowscale=1.5,arrowinset=0.15}
  \psline(0.2,1.5)(0.2,2.5)
  \psline(1,1.5)(1,2.5)
  \psline(1.8,1.5)(1.8,2.5)
  \psline[border=2pt](-0.5,1.5)(2.5,2.5)
  \psline{->}(-0.5,1.5)(0.1,1.7)
  \psline(0.2,1.5)(1,0.5)
  \psline(1,0.5)(1,0)
  \psline(1.8,1.5) (1,0.5)
  \psline(0.6,1) (1,1.5)
   \psline{->}(0.6,1)(0.3,1.375)
   \psline{->}(0.6,1)(0.9,1.375)
   \psline{->}(1,0.5)(1.7,1.375)
   \psline{->}(1,0.5)(0.7,0.875)
   \psline{->}(1,0)(1,0.375)
   \rput[bl]{0}(-0.5,1.8){$x_{\bf k}$}
   \rput[bl]{0}(-0.1,2.6){$^{\bf k}a$}
   \rput[bl]{0}(0.8,2.6){$^{\bf k}b$}
   \rput[bl]{0}(1.6,2.6){$^{\bf k}c$}
   \rput[bl]{0}(0.1,1.0){$a$}
   \rput[bl]{0}(0.95,1.0){$b$}
   \rput[bl]{0}(1.75,1.0){$c$}
   \rput[bl]{0}(0.6,0.5){$e$}
   \rput[bl]{0}(0.9,-0.3){$d$}
 \scriptsize
   \rput[bl]{0}(0.4,0.8){$\alpha$}
   \rput[bl]{0}(1.2,0.25){$\beta$}
  \endpspicture
}
&=&  \sum_{\alpha',\beta', f,\mu',\nu'} \left[U_{\bf k}\left( \,^{\bf k}a ,\,^{\bf k}b ;\,^{\bf k}e \right)\right]_{\alpha \alpha'} \left[U_{\bf k}\left( \,^{\bf k}e ,\,^{\bf k}c ;\,^{\bf k}d \right)\right]_{\beta \beta'}  \left[F_{\,^{\bf k}d}^{\,^{\bf k}a \,^{\bf k}b \,^{\bf k}c}\right]_{(\,^{\bf k}e,\alpha',\beta')(\,^{\bf k}f,\mu',\nu')}
\psscalebox{.6}{
\pspicture[shift=-1.5](-0.5,-1.45)(2.5,1.8)
  \small
  \psset{linewidth=0.9pt,linecolor=black,arrowscale=1.5,arrowinset=0.15}
  \psline(0.2,1.5)(1,0.5)
  \psline(1,0.5)(1,0)
  \psline(1.8,1.5) (1,0.5)
  \psline(1.4,1) (1,1.5)
   \psline{->}(0.6,1)(0.3,1.375)
   \psline{->}(1.4,1)(1.1,1.375)
   \psline{->}(1,0.5)(1.7,1.375)
   \psline{->}(1,0.5)(1.3,0.875)
   \psline{->}(1,0)(1,0.375)
   \psline(1,0)(1,-1)
  \psline[border=2pt](-0.5,-1)(2.5,0)
  \psline{->}(-0.5,-1)(0.1,-0.8)
   \rput[bl]{0}(-0.2,-0.7){$x_{\bf k}$}
   \rput[bl]{0}(-0.1,1.6){$^{\bf k}a$}
   \rput[bl]{0}(0.8,1.6){$^{\bf k}b$}
   \rput[bl]{0}(1.6,1.6){$^{\bf k}c$}
   \rput[bl]{0}(1.3,0.4){$^{\bf k}f$}
   \rput[bl]{0}(1.15,-0.1){$^{\bf k}d$}
  \rput[bl]{0}(0.7,-1.3){$d$}
 \scriptsize
   \rput[bl]{0}(1.5,0.8){$\mu'$}
   \rput[bl]{0}(0.65,0.35){$\nu'$}
  \endpspicture
}
\notag \\
&=& \sum_{f,\mu,\nu,\mu',\nu'} \left[F_d^{abc}\right]_{(e,\alpha,\beta)(f,\mu,\nu)} \left[U_{\bf k}\left( \,^{\bf k}b , \,^{\bf k}c ;\,^{\bf k}f \right)\right]_{\mu \mu'} \left[U_{\bf k}\left( \,^{\bf k}a ,\,^{\bf k}f ;\,^{\bf k}d \right)\right]_{\nu \nu'}
\psscalebox{.6}{
\pspicture[shift=-1.5](-0.5,-1.45)(2.5,1.8)
  \small
  \psset{linewidth=0.9pt,linecolor=black,arrowscale=1.5,arrowinset=0.15}
  \psline(0.2,1.5)(1,0.5)
  \psline(1,0.5)(1,0)
  \psline(1.8,1.5) (1,0.5)
  \psline(1.4,1) (1,1.5)
   \psline{->}(0.6,1)(0.3,1.375)
   \psline{->}(1.4,1)(1.1,1.375)
   \psline{->}(1,0.5)(1.7,1.375)
   \psline{->}(1,0.5)(1.3,0.875)
   \psline{->}(1,0)(1,0.375)
   \psline(1,0)(1,-1)
  \psline[border=2pt](-0.5,-1)(2.5,0)
  \psline{->}(-0.5,-1)(0.1,-0.8)
   \rput[bl]{0}(-0.2,-0.7){$x_{\bf k}$}
   \rput[bl]{0}(-0.1,1.6){$^{\bf k}a$}
   \rput[bl]{0}(0.8,1.6){$^{\bf k}b$}
   \rput[bl]{0}(1.6,1.6){$^{\bf k}c$}
   \rput[bl]{0}(1.3,0.4){$^{\bf k}f$}
   \rput[bl]{0}(1.15,-0.1){$^{\bf k}d$}
  \rput[bl]{0}(0.7,-1.3){$d$}
 \scriptsize
   \rput[bl]{0}(1.5,0.8){$\mu'$}
   \rput[bl]{0}(0.65,0.35){$\nu'$}
  \endpspicture
}
\qquad
,
\end{eqnarray}
which yields the consistency condition
\begin{eqnarray}
&&\sum_{\alpha',\beta',\mu'\nu'} \left[U_{\bf k}(\,^{\bf k}a,\,^{\bf k}b ;\,^{\bf k}e )\right]_{\alpha \alpha'} \left[U_{\bf k}(\,^{\bf k}e, \,^{\bf k}c ;\,^{\bf k}d)\right]_{\beta \beta'} \left[F_{\,^{\bf k}d}^{ \,^{\bf k}a \,^{\bf k}b \,^{\bf k}c}\right]_{(^{\bf k}e,\alpha',\beta')(^{\bf k}f,\mu',\nu')} \notag \\
&& \qquad \qquad \qquad \qquad \qquad \qquad \qquad \qquad  \times \left[U_{\bf k}(\,^{\bf k}b,\,^{\bf k}c ;\,^{\bf k}f)^{-1}\right]_{\mu' \mu} \left[U_{\bf k}(\,^{\bf k}a,\,^{\bf k}f ;\,^{\bf k}d)^{-1}\right]_{\nu' \nu} = \left[F_{d}^{abc}\right]_{(e,\alpha,\beta)(f,\mu,\nu)}
. \quad
\label{eq:G-crossed_F_consistency}
\end{eqnarray}
This condition is the statement of invariance of the $F$-symbols (of the $G$-crossed theory) under the symmetry action.

Similarly, sliding a line under a fusion tree before or after application of an $F$-move gives
\begin{eqnarray}
\psscalebox{.6}{
  \pspicture[shift=-1.5](-0.5,-0.45)(2.5,2.8)
  \small
  \psset{linewidth=0.9pt,linecolor=black,arrowscale=1.5,arrowinset=0.15}
  \psline(-0.5,1.5)(2.5,2.5)
  \psline[border=2pt](0.2,1.5)(0.2,2.5)
  \psline[border=2pt](1,1.5)(1,2.5)
  \psline[border=2pt](1.8,1.5)(1.8,2.5)
  \psline{->}(-0.5,1.5)(0.1,1.7)
  \psline(0.2,1.5)(1,0.5)
  \psline(1,0.5)(1,0)
  \psline(1.8,1.5) (1,0.5)
  \psline(0.6,1) (1,1.5)
   \psline{->}(0.6,1)(0.3,1.375)
   \psline{->}(0.6,1)(0.9,1.375)
   \psline{->}(1,0.5)(1.7,1.375)
   \psline{->}(1,0.5)(0.7,0.875)
   \psline{->}(1,0)(1,0.375)
   \rput[bl]{0}(-0.4,1.8){$x$}
   \rput[bl]{0}(0,2.6){$a_{\bf g}$}
   \rput[bl]{0}(0.95,2.6){$b_{\bf h}$}
   \rput[bl]{0}(1.75,2.6){$c_{\bf k}$}
   \rput[bl]{0}(0.3,0.4){$e_{\bf gh}$}
   \rput[bl]{0}(0.8,-0.4){$d_{\bf ghk}$}
 \scriptsize
   \rput[bl]{0}(0.4,0.8){$\alpha$}
   \rput[bl]{0}(1.2,0.25){$\beta$}
  \endpspicture
}
&=&  \eta_{x}\left({\bf g}, {\bf h} \right) \eta_{x}\left({\bf gh}, {\bf k} \right) \sum_{f,\mu,\nu} \left[F_d^{abc}\right]_{(e,\alpha,\beta)(f,\mu,\nu)}
\psscalebox{.6}{
\pspicture[shift=-1.5](-0.5,-1.45)(2.5,1.8)
  \small
  \psset{linewidth=0.9pt,linecolor=black,arrowscale=1.5,arrowinset=0.15}
  \psline(0.2,1.5)(1,0.5)
  \psline(1,0.5)(1,0)
  \psline(1.8,1.5) (1,0.5)
  \psline(1.4,1) (1,1.5)
   \psline{->}(0.6,1)(0.3,1.375)
   \psline{->}(1.4,1)(1.1,1.375)
   \psline{->}(1,0.5)(1.7,1.375)
   \psline{->}(1,0.5)(1.3,0.875)
   \psline{->}(1,0)(1,0.375)
  \psline(-0.5,-1)(2.5,0)
  \psline{->}(-0.5,-1)(0.1,-0.8)
   \psline[border=2pt](1,0)(1,-1)
   \rput[bl]{0}(-0.2,-0.7){$x$}
   \rput[bl]{0}(0,1.6){$a_{\bf g}$}
   \rput[bl]{0}(0.95,1.6){$b_{\bf h}$}
   \rput[bl]{0}(1.75,1.6){$c_{\bf k}$}
   \rput[bl]{0}(1.3,0.4){$f_{\bf hk}$}
  \rput[bl]{0}(0.7,-1.3){$d_{\bf ghk}$}
 \scriptsize
   \rput[bl]{0}(1.5,0.8){$\mu$}
   \rput[bl]{0}(0.65,0.35){$\nu$}
  \endpspicture
}
\notag \\
&=&  \sum_{f,\mu,\nu} \left[F_d^{abc}\right]_{(e,\alpha,\beta)(f,\mu,\nu)} \eta_{^{\bf \bar{g}}x}\left({\bf h}, {\bf k} \right) \eta_{x}\left({\bf g}, {\bf hk} \right)
\psscalebox{.6}{
\pspicture[shift=-1.5](-0.5,-1.45)(2.5,1.8)
  \small
  \psset{linewidth=0.9pt,linecolor=black,arrowscale=1.5,arrowinset=0.15}
  \psline(0.2,1.5)(1,0.5)
  \psline(1,0.5)(1,0)
  \psline(1.8,1.5) (1,0.5)
  \psline(1.4,1) (1,1.5)
   \psline{->}(0.6,1)(0.3,1.375)
   \psline{->}(1.4,1)(1.1,1.375)
   \psline{->}(1,0.5)(1.7,1.375)
   \psline{->}(1,0.5)(1.3,0.875)
   \psline{->}(1,0)(1,0.375)
  \psline(-0.5,-1)(2.5,0)
  \psline{->}(-0.5,-1)(0.1,-0.8)
   \psline[border=2pt](1,0)(1,-1)
   \rput[bl]{0}(-0.2,-0.7){$x$}
   \rput[bl]{0}(0,1.6){$a_{\bf g}$}
   \rput[bl]{0}(0.95,1.6){$b_{\bf h}$}
   \rput[bl]{0}(1.75,1.6){$c_{\bf k}$}
   \rput[bl]{0}(1.3,0.4){$f_{\bf hk}$}
  \rput[bl]{0}(0.7,-1.3){$d_{\bf ghk}$}
 \scriptsize
   \rput[bl]{0}(1.5,0.8){$\mu$}
   \rput[bl]{0}(0.65,0.35){$\nu$}
  \endpspicture
}
\qquad
,
\end{eqnarray}
which yields the consistency condition
\begin{equation}
\eta_{^{\bf \bar{g}}x}\left({\bf h}, {\bf k} \right) \eta_{x}\left({\bf g}, {\bf hk} \right) = \eta_{x}\left({\bf g}, {\bf h} \right) \eta_{x}\left({\bf gh}, {\bf k} \right)
.
\label{eq:eta_consistency}
\end{equation}
This is the statement of fractionalization being consistent in the $G$-crossed theory. Recall from Sec.~\ref{sec:symmetryfrac} that this relation translates into the condition that the obstruction to fractionalization vanishes, so here we see a direct way in which a nontrivial obstruction would make it impossible to consistently extend the original theory $\mathcal{C}_{\bf 0}$ to a $G$-crossed theory $\mathcal{C}_{G}^{\times}$.

Sliding a line under a $G$-crossed braiding operation gives the $G$-crossed Yang-Baxter equation
\begin{equation}
\pspicture[shift=-1.7](-0.7,-0.6)(1.8,2.7)
\rput[tl]{0} (-0.2, -0.05) {$b_\mathbf{h}$}
\rput[tl]{0} (0.6, -0.0) {$\,^{\mb{\bar{h}}}a_\mathbf{g}$}
\rput[tl]{0} (-0.8, -0.15) {$x_\mathbf{k}$}
\rput[bl]{0} (-0.4, 2.25) {$\,^{\bf k}a_\mathbf{g}$}
\rput[bl]{0} (0.6, 2.3) {$\,^{\bf k}b_\mathbf{h}$}
\rput[bl]{0} (1.45, 2.3) {$x_\mathbf{k}$}
	\psbezier[linewidth=0.9pt](1,0) (1,1) (0,1.2) (0,2.2)
	\psbezier[linewidth=0.9pt, border=2pt](0,0) (0,1) (1,1.2) (1,2.2)
	\psbezier[linewidth=0.9pt, border=2pt](1.5,2.2) (1.0,1.2) (-0.1,2.3) (-0.5,0)
	\psline[linewidth=0.9pt, arrows=->, arrowscale=1.5, arrowinset=0.15](-0.355, 0.6)(-0.33, 0.685)
	\psline[linewidth=0.9pt, arrows=->, arrowscale=1.5, arrowinset=0.15](0.85, 0.6)(0.81, 0.685)
	\psline[linewidth=0.9pt, arrows=->, arrowscale=1.5, arrowinset=0.15](0.15, 0.6)(0.19, 0.685)
	\endpspicture
=\frac{\eta_{\,^{\bf k}a}({\bf kh\bar{k}},{\bf k}) }{\eta_{\,^{\bf k}a}({\bf k},{\bf h})}
\pspicture[shift=-1.7](-0.7,-0.6)(1.8,2.6)
\rput[tl]{0} (-0.2, -0.05) {$b_\mathbf{h}$}
\rput[tl]{0} (0.6, -0.0) {$\,^{\bf \bar{h}}a_\mathbf{g}$}
\rput[tl]{0} (-0.8, -0.15) {$x_\mathbf{k}$}
\rput[bl]{0} (-0.4, 2.25) {$\,^{\bf k}a_\mathbf{g}$}
\rput[bl]{0} (0.6, 2.3) {$\,^{\bf k}b_\mathbf{h}$}
\rput[bl]{0} (1.45, 2.3) {$x_\mathbf{k}$}		
        \psbezier[linewidth=1pt](1,0)(1,0.6) (0,1.2)(0,2.2)
		\psbezier[linewidth=1pt,border=2pt](0,0) (0,0.6) (1,1.2)  (1,2.2)
		\psbezier[linewidth=1pt,border=2pt] (1.5,2.2) (1.1,-0.3) (0, 1.1) (-0.5,0)
\psline[linewidth=0.9pt, arrows=->, arrowscale=1.5, arrowinset=0.15](0.13, 1.6)(0.09, 1.685)
\psline[linewidth=0.9pt, arrows=->, arrowscale=1.5, arrowinset=0.15](0.88, 1.6)(0.91, 1.685)
\psline[linewidth=0.9pt, arrows=->, arrowscale=1.5, arrowinset=0.15](1.375, 1.6)(1.40, 1.685)
\endpspicture
\label{eq:G-crossed_Yang_Baxter}
.
\end{equation}
Here, we slid the $a$ line under the $R^{bx}$ braiding operator and obtained the $\eta_a$ factors by expanding the $R^{bx}$ braiding operator in terms of fusion and splitting vertices.

Alternatively, we can obtain a similar relation by sliding the $x$ line over the $R^{ab}$ braiding operator. In this case, there will be symmetry action applied to the braiding operation, so we must explicitly expand it, giving
\begin{eqnarray}
\pspicture[shift=-1.7](-0.7,-0.6)(1.8,2.7)
\rput[tl]{0} (-0.2, -0.05) {$b_\mathbf{h}$}
\rput[tl]{0} (0.6, -0.0) {$\,^{\bar{\mb{h}}}a_\mathbf{g}$}
\rput[tl]{0} (-0.8, -0.15) {$x_\mathbf{k}$}
\rput[bl]{0} (-0.4, 2.25) {$\,^{\bf k}a_\mathbf{g}$}
\rput[bl]{0} (0.6, 2.3) {$\,^{\bf k}b_\mathbf{h}$}
\rput[bl]{0} (1.45, 2.3) {$x_\mathbf{k}$}
	\psbezier[linewidth=0.9pt](1,0) (1,1) (0,1.2) (0,2.2)
	\psbezier[linewidth=0.9pt, border=2pt](0,0) (0,1) (1,1.2) (1,2.2)
	\psbezier[linewidth=0.9pt, border=2pt](1.5,2.2) (1.0,1.2) (-0.1,2.3) (-0.5,0)
	\psline[linewidth=0.9pt, arrows=->, arrowscale=1.5, arrowinset=0.15](-0.355, 0.6)(-0.33, 0.685)
	\psline[linewidth=0.9pt, arrows=->, arrowscale=1.5, arrowinset=0.15](0.85, 0.6)(0.81, 0.685)
	\psline[linewidth=0.9pt, arrows=->, arrowscale=1.5, arrowinset=0.15](0.15, 0.6)(0.19, 0.685)
	\endpspicture
&=& \sum_{c,\mu,\nu} \sqrt{\frac{d_{c}}{d_{a}d_{b}}} \left[ R^{ab}_{c} \right]_{\mu\nu}
\pspicture[shift=-1.7](-0.7,-0.6)(1.8,2.7)
\rput[tl]{0} (-0.2, -0.05) {$b_\mathbf{h}$}
\rput[tl]{0} (0.6, -0.0) {$\,^\mb{\bar{h}}a_\mathbf{g}$}
\rput[tl]{0} (-0.8, -0.15) {$x_\mathbf{k}$}
\rput[bl]{0} (-0.4, 2.25) {$\,^{\bf k}a_\mathbf{g}$}
\rput[bl]{0} (0.6, 2.3) {$\,^{\bf k}b_\mathbf{h}$}
\rput[bl]{0} (1.45, 2.3) {$x_\mathbf{k}$}
\rput[tl]{0}(0.55, 0.95) {$c_{\bf gh}$}
 \scriptsize
\rput[tl]{0}(0.25, 1.2) {$\nu$}
\rput[tl]{0}(0.25, 0.7) {$\mu$}
\psline[linewidth=0.9pt](0,0)(0.5,0.6)
		\psline[linewidth=0.9pt](1,0) (0.5,0.6)
		 \psline[linewidth=0.9pt](0.5, 0.6)(0.5, 1.2)
		\psline[linewidth=0.9pt](0.5, 1.2)(0, 2.2)
		\psline[linewidth=0.9pt](0.5, 1.2)(1, 2.2)
	\psbezier[linewidth=0.9pt,border=2pt] (1.5,2.2)(1.0,1.4)(-0.1,2.3)(-0.5,0)
\psline[linewidth=0.9pt, arrows=->, arrowscale=1.5, arrowinset=0.15](-0.355, 0.6)(-0.33, 0.685)
\endpspicture
\notag \\
&=& \sum_{c, \mu,\nu,\mu',\nu'} \sqrt{\frac{d_{c}}{d_{a}d_{b}}} \left[U_{\bf k}(\,^{\bf k}b,\,^{\bf k\bar{h}}a;\,^{\bf k}c)^{-1} \right]_{\mu' \mu} \left[ R^{ab}_{c} \right]_{\mu\nu} \left[U_{\bf k}(\,^{\bf k}a,\,^{\bf k}b;\,^{\bf k}c) \right]_{\nu \nu'}
\pspicture[shift=-1.7](-0.7,-0.6)(1.8,2.7)
\rput[tl]{0} (-0.2, -0.05) {$b_\mathbf{h}$}
\rput[tl]{0} (0.6, -0.0) {$\,^{\mb{\bar{h}}}a_\mathbf{g}$}
\rput[tl]{0} (-0.8, -0.15) {$x_\mathbf{k}$}
\rput[bl]{0} (-0.4, 2.25) {$\,^{\bf k}a_\mathbf{g}$}
\rput[bl]{0} (0.6, 2.3) {$\,^{\bf k}b_\mathbf{h}$}
\rput[bl]{0} (1.45, 2.3) {$x_\mathbf{k}$}
\rput[tl]{0}(0.55, 1.4) {\mbox{$^\mb{k}c$}}
 \scriptsize
\rput[tl]{0}(0.2,1.6) {$\nu'$}
\rput[tl]{0}(0.2,1.2) {$\mu'$}
		\psline[linewidth=0.9pt] (0,0)(0.5,1)
		\psline[linewidth=0.9pt] (1,0)(0.5,1)
		\psline[linewidth=0.9pt] (0.5, 1)(0.5, 1.6)
		\psline[linewidth=0.9pt] (0.5, 1.6)(0, 2.2)
	\psline[linewidth=0.9pt](0.5, 1.6)(1, 2.2)
\psbezier[linewidth=0.9pt,border=2pt] (1.5,2.2)(1.2,0)(-0.3,0.6)(-0.5,0)
\psline[linewidth=0.9pt, arrows=->, arrowscale=1.5, arrowinset=0.15](1.375, 1.6)(1.40, 1.685)
\endpspicture
.
\end{eqnarray}
Comparing this relation with the G-crossed Yang-Baxter equation by expanding the $R^{ab}$ braiding operator in Eq.~(\ref{eq:G-crossed_Yang_Baxter}), we obtain the consistency condition between braiding and sliding moves
\begin{equation}
\frac{\eta_{\,^{\bf k}a}({\bf kh\bar{k}},{\bf k}) }{ \eta_{\,^{\bf k}a}({\bf k},{\bf h} )} \sum_{\mu',\nu'} \left[U_{\bf k}(\,^{\bf k}b , \,^{\bf k \bar{h}}a ; \,^{\bf k} c ) \right]_{\mu \mu'} \left[ R^{ \,^{\bf k}a \,^{\bf k}b}_{\,^{\bf k}c} \right]_{\mu' \nu'} \left[ U_{\bf k} (\,^{\bf k}a , \,^{\bf k}b ; \,^{\bf k}c )^{-1} \right]_{\nu' \nu} =   \left[ R^{a b}_{c} \right]_{\mu \nu}
\end{equation}
This is the $G$-crossed generalization of the statement that the $R$-symbols are invariant under the symmetry action. Notice the presence of the $\eta$ factors, as compared to Eq.~(\ref{eq:rho_g_R}), to which this expression reduces when $a,b,c \in \mathcal{C}_{\bf 0}$.

We reemphasize the fact that imposing consistency on the sliding moves has resulted in consistency conditions that precisely replicate the symmetry action constraints and properties described in Secs.~\ref{sec:Symmetry} and \ref{sec:symmetryfrac}, and extend them from acting on the $\mathcal{C}_{\bf 0}$ theory to its $G$-crossed extensions. This justifies our use of the same symbols $\left[ U_{\bf k}\left(a,b;c\right)\right]_{\mu \mu'}$ and $\eta_{x}\left({\bf g},{\bf h}\right)$ for the transformations associated with the sliding moves.

We note, for future use, that sliding a line under and another line over a vertex gives the relation
\begin{equation}
\eta_{^{\bf k}x}\left(\,^{\bf k}{\bf g}, \,^{\bf k}{\bf h} \right) = \frac{ \eta_{^{\bf \bar{g}}x}\left( {\bf \bar{k}} , {\bf kh\bar{k}} \right) }{ \eta_{^{\bf \bar{g}}x}\left({\bf h},{\bf \bar{k}} \right) } \frac{ \eta_{x}\left({\bf gh},{\bf \bar{k}} \right) }{ \eta_{x}\left( {\bf \bar{k}} , {\bf kgh\bar{k}} \right) }  \frac{ \eta_{x}\left( {\bf \bar{k}} , {\bf kg\bar{k}} \right) }{ \eta_{x}\left({\bf g},{\bf \bar{k}} \right) }  \eta_{x}\left({\bf g},{\bf h} \right)
\label{eq:eta_k-action}
\end{equation}
for how $\eta_{x}\left({\bf g}, {\bf h} \right)$ transforms under ${\bf k}$-action. This can be obtained from
\begin{eqnarray}
\psscalebox{.6}{
\pspicture[shift=-2.2](-0.9,-0.8)(2.0,3.4)
  \small
  \psset{linewidth=0.9pt,linecolor=black,arrowscale=1.5,arrowinset=0.15}
  \psline(-0.65,0)(2.05,2)
  \psline[border=2pt](0.7,0.55)(0.25,1)
  \psline[border=2pt](1.15,1)(1.15,2)
  \psline(0.7,0.55)(1.15,1)	
  \psline{->}(0.7,0)(0.7,0.45)
  \psline(0.7,0)(0.7,0.55)
  \psline(0.25,1)(0.25,2)
  \psline{->}(0.25,1)(0.25,1.8)
  \psline{->}(1.15,1)(1.15,1.8)
  \psline{->}(-0.65,0)(0.025,0.5)
  \psline(0.25,2)(0.25,3)
  \psline(1.15,2)(1.15,3)
  \psline[border=2pt](-0.65,3)(2.05,1.5)
  \psline{-<}(-0.65,3)(-0.11,2.7)
  \rput[bl]{0}(-0.4,0.5){$x$}
  \rput[bl]{0}(-0.5,2.4){$y_{\bf \bar{k}}$}
  \rput[bl]{0}(0.4,-0.35){$c_{\bf gh}$}
  \rput[br]{0}(1.3,3.1){$^{\bf k}b_{\bf h}$}
  \rput[bl]{0}(0.0,3.05){$^{\bf k}a_{\bf g}$}
  \rput[br]{0}(1.0,1.6){$b_{\bf h}$}
  \rput[bl]{0}(-0.2,1.6){$a_{\bf g}$}
 \scriptsize
  \rput[bl]{0}(0.85,0.35){$\mu$}
  \endpspicture
}
&=&
\eta_{x}\left({\bf g},{\bf h}\right) \left[U_{\bf \bar{k}}(\,^{\bf \bar{k}}a,\,^{\bf \bar{k}}b;\,^{\bf \bar{k}}c)^{-1} \right]_{\mu \nu} \frac{\eta_{x}\left({\bf gh},{\bf \bar{k}}\right)}{\eta_{x}\left({\bf \bar{k}},{\bf kgh\bar{k}}\right)}
\psscalebox{.6}{
\pspicture[shift=-2.2](-0.9,-1.8)(1.8,2.4)
  \small
  \psset{linewidth=0.9pt,linecolor=black,arrowscale=1.5,arrowinset=0.15}
  \psline(-0.65,0)(2.05,2)
  \psline[border=2pt](0.7,0)(0.7,1.55)
  \psline(0.7,0)(0.7,-1)
  \psline{->}(0.7,0)(0.7,0.45)
  \psline(0.7,1.55)(0.25,2)
  \psline{->}(0.7,1.55)(0.3,1.95)
  \psline(0.7,1.55) (1.15,2)	
  \psline{->}(0.7,1.55)(1.1,1.95)
  \psline{->}(-0.65,0)(0.16,0.6)
  \psline[border=2pt](-0.65,0.5)(2.05,-1.0)
  \psline{-<}(-0.65,0.5)(-0.11,0.2)
  \rput[bl]{0}(-0.3,0.7){$^{\bf k}x$}
  \rput[bl]{0}(-0.3,-0.3){$y_{\bf \bar{k}}$}
  \rput[bl]{0}(0.4,-1.35){$c_{\bf gh}$}
  \rput[bl]{0}(0.8,0.35){$^{\bf k}c_{\bf gh}$}
  \rput[br]{0}(1.3,2.1){$^{\bf k}b_{\bf h}$}
  \rput[bl]{0}(0.0,2.05){$^{\bf k}a_{\bf g}$}
 \scriptsize
  \rput[bl]{0}(0.85,1.35){$\nu$}
\endpspicture
}
\\
&=&  \frac{ \eta_{^{\bf \bar{g}}x}\left({\bf h},{\bf \bar{k}} \right) }{ \eta_{^{\bf \bar{g}}x}\left( {\bf \bar{k}} , {\bf kh\bar{k}} \right) }  \frac{ \eta_{x}\left({\bf g},{\bf \bar{k}} \right) }{ \eta_{x}\left( {\bf \bar{k}} , {\bf kg\bar{k}} \right) } \left[U_{\bf \bar{k}}(\,^{\bf \bar{k}}a,\,^{\bf \bar{k}}b;\,^{\bf \bar{k}}c)^{-1} \right]_{\mu \nu} \eta_{^{\bf k}x}\left(\,^{\bf k}{\bf g}, \,^{\bf k}{\bf h} \right)
\psscalebox{.6}{
\pspicture[shift=-2.2](-0.9,-1.8)(1.8,2.4)
  \small
  \psset{linewidth=0.9pt,linecolor=black,arrowscale=1.5,arrowinset=0.15}
  \psline(-0.65,0)(2.05,2)
  \psline[border=2pt](0.7,0)(0.7,1.55)
  \psline(0.7,0)(0.7,-1)
  \psline{->}(0.7,0)(0.7,0.45)
  \psline(0.7,1.55)(0.25,2)
  \psline{->}(0.7,1.55)(0.3,1.95)
  \psline(0.7,1.55) (1.15,2)	
  \psline{->}(0.7,1.55)(1.1,1.95)
  \psline{->}(-0.65,0)(0.16,0.6)
  \psline[border=2pt](-0.65,0.5)(2.05,-1.0)
  \psline{-<}(-0.65,0.5)(-0.11,0.2)
  \rput[bl]{0}(-0.3,0.7){$^{\bf k}x$}
  \rput[bl]{0}(-0.3,-0.3){$y_{\bf \bar{k}}$}
  \rput[bl]{0}(0.4,-1.35){$c_{\bf gh}$}
  \rput[bl]{0}(0.8,0.35){$^{\bf k}c_{\bf gh}$}
  \rput[br]{0}(1.3,2.1){$^{\bf k}b_{\bf h}$}
  \rput[bl]{0}(0.0,2.05){$^{\bf k}a_{\bf g}$}
 \scriptsize
  \rput[bl]{0}(0.85,1.35){$\nu$}
\endpspicture
}
\end{eqnarray}
where the two lines in this expression correspond to the two orders in which one can slide the $x$ and $y$ lines.

\begin{figure}[t!]
\begin{center}
\includegraphics[scale=0.55]{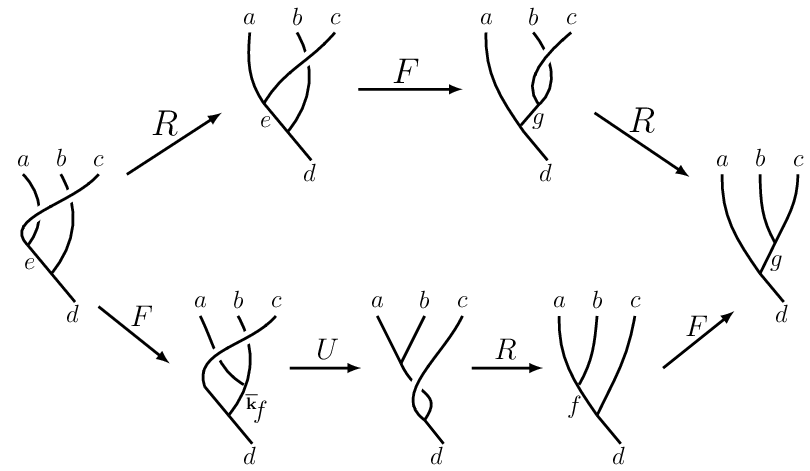}
\qquad \qquad
\includegraphics[scale=0.55]{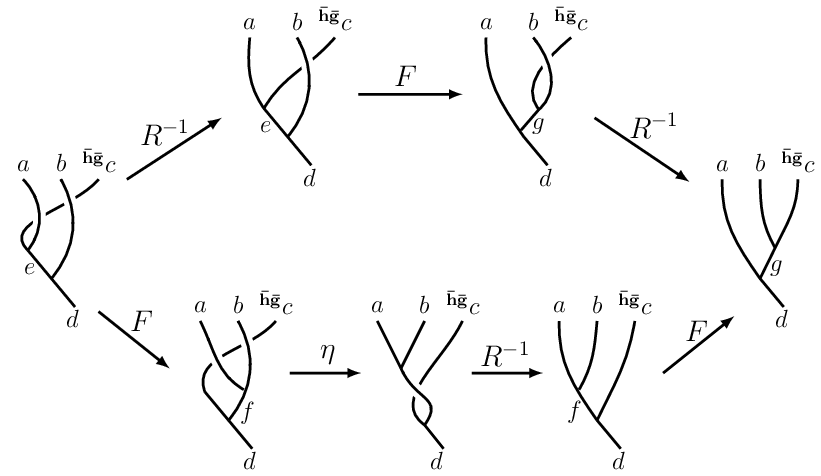}
\caption{The $G$-crossed Heptagon equations provide consistency conditions between $G$-crossed braiding, fusion, and sliding moves. Eqs.~(\ref{eq:heptagon+}) and (\ref{eq:heptagon-}) are obtained by imposing the conditions that the above diagrams commute.}
\label{fig:heptagon}
\end{center}
\end{figure}

Finally, we require consistency between $G$-crossed braiding and fusion, as well as the sliding moves, so that any two sequences of moves that start from the same configuration and end in the same configuration must be equivalent. This is achieved by imposing the following $G$-crossed Heptagon equations, which are analogous to the Hexagon equations of BTCs, a diagrammatic representation of which is shown in Fig.~\ref{fig:heptagon}. The Heptagon equation for counterclockwise braiding exchanges is
\begin{eqnarray}
&& \sum_{\lambda ,\gamma }
\left[ R_{e}^{ac}\right] _{\alpha \lambda }
\left[F_{d}^{ac \,^{\bf \bar{k}}b}\right] _{\left( e,\lambda ,\beta \right) \left( g,\gamma, \nu \right) }
\left[ R_{g}^{bc}\right] _{\gamma \mu } \notag \\
&&\quad =\sum_{f,\sigma ,\delta, \eta ,\psi }
\left[ F_{d}^{c \,^{\bf \bar{k}}a \,^{\bf \bar{k}}b}\right] _{\left( e,\alpha,\beta \right) \left( \,^{\bf \bar{k}}f ,\delta ,\sigma \right) }
\left[ U_{\bf k}\left( a ,b ;f \right) \right]_{\delta \eta }
\left[ R_{d}^{fc}\right]_{\sigma \psi }
\left[ F_{d}^{abc}\right] _{\left(f, \eta ,\psi \right) \left( g,\mu ,\nu \right) }
,
\label{eq:heptagon+}
\end{eqnarray}
in which we left the group labels for $a_{\bf g}$, $b_{\bf h}$, $c_{\bf k}$, $d_{\bf ghk}$, $e_{\bf gk}$, $f_{\bf gh}$, and $g_{\bf hk}$ implicit. Similarly, the Heptagon equation for clockwise braiding exchanges is
\begin{eqnarray}
&&\sum_{\lambda ,\gamma }
\left[ \left( R_{e}^{ca}\right) ^{-1}\right]_{\alpha \lambda }
\left[ F_{d}^{a \,^{\bf \bar{g}}cb}\right] _{\left( e,\lambda ,\beta \right) \left( g,\gamma, \nu \right) }
\left[ \left( R_{g}^{\,^{\bf \bar{g}}c b}\right) ^{-1}\right] _{\gamma \mu } \notag \\
&& \quad =\sum_{f,\sigma ,\delta ,\psi }
\left[ F_{d}^{cab}\right] _{\left( e,\alpha,\beta \right) \left( f ,\delta,\sigma \right) }
\eta_{c}\left({\bf g},{\bf h}\right)
\left[ \left(R_{d}^{cf}\right) ^{-1}\right] _{\sigma \psi }
\left[ F_{d}^{ab\,^{\bf \bar{h} \bar{g}}c}\right]_{\left( f,\delta ,\psi \right) \left( g,\mu ,\nu \right) }
\label{eq:heptagon-}
,
\end{eqnarray}
\end{widetext}
in which we left the group labels for $a_{\bf g}$, $b_{\bf h}$, $c_{\bf k}$, $d_{\bf kgh}$, $e_{\bf kg}$, $f_{\bf gh}$, and $g_{\bf \bar{g}kgh}$ implicit (the differences being due to how the group action enters braiding in the counterclockwise vs. clockwise braiding operators).

Given the trivial associativity of the vacuum charge $0$ ($F^{abc}_{d} = \openone$ when $a$, $b$, or $c=0$), the Heptagon equations imply that braiding with the vacuum is trivial, i.e. $R^{a0}_{a} = R^{0a}_{a} = \left( R^{a0}_{a} \right)^{-1} = \left(R^{0a}_{a}\right)^{-1} = 1$ for any value of $a \in \mathcal{C}_{G}$.

If we further require unitarity of the theory, then $\left(R^{ab}\right)^{-1}= \left( R^{ab} \right)^{\dag}$, which can be expressed in terms of $R$-symbols as $\left[ \left( R_{c}^{ab}\right)^{-1} \right] _{\mu \nu} = \left[ R_{c}^{ab}\right] _{\nu \mu}^{\ast}$.

\subsection{Gauge Transformations}
\label{sec:G-crossed_gauge_transformations}

The basic data given by $N_{ab}^{c}$, $F^{abc}_{d}$, $R^{ab}_{c}$, $\rho_{\bf k}$ [which includes $U_{\bf k}(a,b;c)$], and $\eta_{a}({\bf g,h})$ that satisfy the consistency conditions described in the previous subsections defines a $G$-crossed braided tensor category, which we can consider to be a generalized anyon and defect model. There is, however, some redundancy between different collections of basic data due to gauge freedom, similar to the case of BTCs. Thus, we again wish to characterize theories as equivalent when they are related by gauge transformations. For $G$-crossed BTCs, it is useful to separate gauge transformations into two classes.

The first type of gauge transformation is familiar from BTCs. In particular, these gauge transformations derive from the redundancy of redefining the fusion/splitting vertex basis states
\begin{equation}
\widetilde{ \left| a,b;c,\mu \right\rangle } = \sum_{\mu'} \left[\Gamma^{ab}_{c}\right]_{\mu \mu'} \left| a,b;c,\mu' \right\rangle
\end{equation}
where $\Gamma^{ab}_{c}$ is a unitary transformation. Such gauge transformations modify the $F$-symbols in precisely the same way we have previously seen in Eq.~(\ref{eq:gauge_F}).
The transformation of $G$-crossed $R$-symbols is slightly modified from that of BTCs to accommodate the symmetry actions that are incorporated in braiding, and is given by
\begin{equation}
\left[ \widetilde{R}_{c_{\bf gh}}^{a_{\bf g} b_{\bf h}} \right]_{\mu \nu} = \sum_{\mu',\nu'} \left[\Gamma^{b\,^{\bf \bar{h}}a}_{c} \right]_{\mu \mu'}  \left[ R_{c_{\bf gh}}^{a_{\bf g} b_{\bf h}} \right]_{\mu' \nu'} \left[\left( \Gamma^{ab}_{c} \right)^{-1} \right]_{\nu' \nu}
.
\end{equation}
The symmetry action transformation become
\begin{eqnarray}
&& \left[ \widetilde{U}_{\bf k}\left( a ,b ;c \right) \right]_{\mu \nu} = \\
&& \qquad \sum_{\mu',\nu'} \left[\Gamma^{\,^{\bf \bar{k}}a \,^{\bf \bar{k}}b}_{\,^{\bf \bar{k}}c} \right]_{\mu \mu'}  \left[ U_{\bf k}\left( a ,b ;c \right) \right]_{\mu' \nu'} \left[\left( \Gamma^{ab}_{c} \right)^{-1} \right]_{\nu' \nu}
.
\notag
\end{eqnarray}
These gauge transformations leave $\widetilde{\eta}_{x}({\bf g,h})=\eta_{x}({\bf g,h})$ unchanged, and consequently $\widetilde{\kappa}_{\bf g,h}=\kappa_{\bf g,h}$ is also unchanged.

The second type of gauge transformation is derived from the equivalence of symmetry actions by natural isomorphisms, i.e. $\check{\rho}_{\bf g} = \Upsilon_{\bf g} \rho_{\bf g}$, which we discussed in Secs.~\ref{sec:Symmetry} and \ref{sec:symmetryfrac}. In particular, these gauge transformations enact the following modifications of the basic data
\begin{eqnarray}
\left[\check{F}_{d}^{abc}\right]_{(e,\alpha,\beta)(f,\mu,\nu)} &=& \left[F_{d}^{abc}\right]_{(e,\alpha,\beta)(f,\mu,\nu)} , \\
\left[\check{R}_{c_{\bf gh}}^{a_{\bf g} b_{\bf h}} \right]_{\mu \nu} &=& \gamma_{a}({\bf h}) \left[ R_{c_{\bf gh}}^{a_{\bf g} b_{\bf h}} \right]_{\mu \nu} , \\
\left[\check{U}_{\bf k}\left( a ,b ;c \right) \right]_{\mu \nu} &=& \frac{\gamma_{a}({\bf k}) \gamma_{b}({\bf k}) }{ \gamma_{c}({\bf k}) }  \left[ U_{\bf k}\left( a ,b ;c \right) \right]_{\mu \nu}, \quad \\
\check{\eta}_{x}({\bf g,h}) &=& \frac{ \gamma_{x}({\bf gh}) }{ \gamma_{\,^{\bf \bar{g}}x}({\bf h}) \gamma_{x}({\bf g})} \eta_{x}({\bf g,h})
,
\label{eqn:gauge}
\end{eqnarray}
which leave the $F$-symbols unchanged, since the symmetry action is incorporated through braiding. [The symmetry action on topological charge labels is unchanged $\check{\rho}_{\bf g}(a) = \rho_{\bf g} (a)$.] Thus, theories with different choices of $\rho \in [\rho]$ are equivalent under this type of gauge transformation.

We refer to these two types of gauge transformations as vertex basis gauge transformations and symmetry action gauge transformations, respectively. It is straightforward to check that all the consistency conditions are left invariant under both types of gauge transformations.

As before, one must be careful not to use the gauge freedom associated with the canonical gauge choices associated with making fusion, braiding, and sliding with the vacuum trivial, and respecting the canonical isomorphisms that allow one to freely add and remove vacuum lines. In particular, one must fix $\Gamma^{a 0}_{a}=\Gamma^{0 b}_{b}=\Gamma^{0 0}_{0}$, as in the case of BTCs, and also fix $\gamma_{0}({\bf h}) = \gamma_{a}({\bf 0}) =1$.

\subsection{$G$-Crossed Invariants, Twists, and $S$-Matrix }
\label{sec:G-Crossed_Invariants}

It is useful to consider quantities of a $G$-crossed theory that are invariant under gauge transformations, as we did for BTCs. (In this section, we will discuss invariants that are straightforward to obtain in the $G$-crossed theory, e.g. using diagrammatics, but we will later see that another class of invariants can be constructed by gauging the symmetry of the theory.) Clearly, invariants derived from fusion and $F$-symbols alone are the same in both BTCs and $G$-crossed BTCs, since the new symmetry action gauge transformations do not affect the $F$-symbols. In particular, the quantum dimensions $d_{a}=d_{\bar{a}} = d_{\,^{\bf k}a}$ are invariants.

Eq.~(\ref{eq:G-crossed_F_consistency}) with $e=f=0$ yields the relation
\begin{equation}
\frac{\varkappa_{\,^{\bf k}a} }{ \varkappa_{a} } =\frac{ \left[F_{\,^{\bf k}a}^{ \,^{\bf k}a \,^{\bf k}\bar{a} \,^{\bf k}a}\right]_{00} } {\left[F_a^{a\bar{a}a}\right]_{00}}
=\frac{U_{\bf k}(\,^{\bf k}\bar{a},\,^{\bf k}a ;0)}{ U_{\bf k}(\,^{\bf k}a, \,^{\bf k}\bar{a};0) }
.
\label{eq:G-crossed_FS}
\end{equation}
When $a = \bar{a}$, the Frobenius-Shur indicator $\varkappa_{a} = \pm
1$ is a gauge invariant quantity and it follows from
Eq.~(\ref{eq:G-crossed_FS}) that  $\varkappa_{a}=\varkappa_{\,^{\bf
    k}a}$. (We recall that, more generally, $\varkappa_{a}
=\varkappa_{\bar{a}}^{-1}$.) When $^{\bf k}a = a$ is ${\bf
  k}$-invariant, it follows from Eq.~(\ref{eq:G-crossed_FS}) that
\begin{align}
U_{\bf k}(a,\bar{a};0) = U_{\bf k}(\bar{a},a;0).
\label{eq:U_k_relation}
\end{align}

On the other hand, we must be more careful when trying to carry over gauge invariant quantities that are derived from braiding operations, such as the twist factors and $S$-matrix, as these may no longer be gauge invariant in a $G$-crossed theory. Consequently, we will examine these in more detail.

The topological twists are defined the same way as before by taking the quantum trace of a counterclockwise braid of a topological charge with itself
\begin{equation}
\theta _{a}
= \frac{1}{d_{a}}
\pspicture[shift=-0.5](-1.3,-0.6)(1.3,0.6)
\small
  \psset{linewidth=0.9pt,linecolor=black,arrowscale=1.5,arrowinset=0.15}
  \psarc[linewidth=0.9pt,linecolor=black] (0.7071,0.0){0.5}{-135}{135}
  \psarc[linewidth=0.9pt,linecolor=black] (-0.7071,0.0){0.5}{45}{315}
  \psline(-0.3536,0.3536)(0.3536,-0.3536)
  \psline[border=2.3pt](-0.3536,-0.3536)(0.3536,0.3536)
  \psline[border=2.3pt]{->}(-0.3536,-0.3536)(0.0,0.0)
  \rput[bl]{0}(-0.2,-0.5){$a$}
  \endpspicture
=\sum\limits_{c,\mu } \frac{d_{c}}{d_{a}}\left[ R_{c}^{aa}\right] _{\mu \mu }
.
\end{equation}
We immediately see that $\theta_{a_{\bf g}}$ is always invariant under the vertex basis gauge transformations, but is only invariant under the symmetry action gauge transformations if ${\bf g} = {\bf 0}$, since
\begin{equation}
\check{\theta}_{a_{\bf g}} = \gamma_{a_{\bf g}} ({\bf g}) \theta_{a_{\bf g}}
.
\end{equation}
This corroborates the interpretation of topological charges $a_{\bf  g}$ with ${\bf g}\neq {\bf 0}$ as describing extrinsic defects, for
which one should not expect invariant braiding or exchange statistics in the usual sense, since they are not true quasiparticles (deconfined
topological excitations) of the system. We will examine this matter in more detail.

We can immediately notice that
\begin{equation}
\frac{ \sum\limits_{\mu} \left[ R_{c}^{aa}\right] _{\mu \mu } } { \sum\limits_{\mu'} \left[ R_{c'}^{aa}\right] _{\mu' \mu' } }
\label{eqn:projexchange}
\end{equation}
is gauge invariant under both types of gauge transformations.

Using Eq.~(\ref{eq:G-crossed_Yang_Baxter}) with the definition of the twist, we find the general relation between $\theta _{a}$ and $\theta _{\,^{\bf k}a}$ is
\begin{equation}
\theta_{a_{\bf g}} = \frac{\eta_{\,^{\bf k}a_{\bf g}}({\bf kg \bar{k} , k}) }{\eta_{\,^{\bf k}a_{\bf g}}({\bf k, g})} \theta_{\,^{\bf k}a_{\bf g}} = \frac{\eta_{a_{\bf g}}({\bf \bar{k} , kg\bar{k} }) }{\eta_{a_{\bf g}}({\bf g, \bar{k}})} \theta_{\,^{\bf k}a_{\bf g}}
.
\label{eqn:action_on_twist}
\end{equation}
When $^{\bf k}a = a$, it follows that
\begin{equation}
\eta_{a_{\bf g}}({\bf g, k}) = \eta_{a_{\bf g}}({\bf k, g}).
\label{eqn:etasym}
\end{equation}
We also note that Eq.~(\ref{eq:eta_consistency}) gives $\eta_{^{\bf k}x} (\mb{k},\mb{ \bar{k} }) = \eta_{x} (\mb{ \bar{k} } , \mb{k})$ for any $x$ and ${\bf k}$, so we also have
\begin{equation}
\eta_{a_{\bf g}}({\bf k,\bar{k}}) = \eta_{a_{\bf g}}({\bf \bar{k},k})
\label{eqn:etasym_2}
\end{equation}
when $^{\bf k}a = a$.

The definition of topological twists can also be written in the form
\begin{equation}
\pspicture[shift=-1.3](1.0,-1.4)(2.2,1.1)
  \small
  \psset{linewidth=0.9pt,linecolor=black,arrowscale=1.5,arrowinset=0.15}
\rput[bl]{0}(1.4,0.8){$a_\mb{g}$}
\rput[bl]{0}(1.4,-1.2){$a_\mb{g}$}
  \psbezier[linewidth=0.9pt,linecolor=black,border=0.1](1.5,0.7)(1.5,-0.5)(2.0,-0.4)(2.0,0.0)
  \psbezier[linewidth=0.9pt,linecolor=black,border=2.1pt](1.5,-0.6)(1.5,0.5)(2.0,0.4)(2.0,0.0)
   \psline{->}(1.5,-0.7)(1.5,-0.5)
   \psline(1.5,-0.9)(1.5,-0.7)
\endpspicture
= \theta_{a}
\pspicture[shift=-1.3](1.3,-1.4)(1.8,1.1)
  \small
  \psset{linewidth=0.9pt,linecolor=black,arrowscale=1.5,arrowinset=0.15}
\rput[bl]{0}(1.4,0.8){$a_\mb{g}$}
\rput[bl]{0}(1.4,-1.2){$a_\mb{g}$}
   \psline{->}(1.5,-0.7)(1.5,-0.5)
   \psline(1.5,-0.9)(1.5,0.7)
\endpspicture
=
\pspicture[shift=-1.3](0.8,-1.4)(1.8,1.1)
  \small
  \psset{linewidth=0.9pt,linecolor=black,arrowscale=1.5,arrowinset=0.15}
\rput[bl]{0}(1.4,0.8){$a_\mb{g}$}
\rput[bl]{0}(1.4,-1.2){$a_\mb{g}$}
  \psbezier[linewidth=0.9pt,linecolor=black,border=2.1pt](1.5,-0.6)(1.5,0.5)(1.0,0.4)(1.0,0.0)
  \psbezier[linewidth=0.9pt,linecolor=black,border=2.0pt](1.5,0.7)(1.5,-0.5)(1.0,-0.4)(1.0,0.0)
   \psline{->}(1.5,-0.7)(1.5,-0.5)
   \psline(1.5,-0.9)(1.5,-0.7)
\endpspicture
,
\end{equation}
as is the case with BTCs. It is clear that the inverse topological twists are similarly obtained from clockwise braidings
\begin{equation}
\pspicture[shift=-1.3](1.0,-1.4)(2.2,1.1)
  \small
  \psset{linewidth=0.9pt,linecolor=black,arrowscale=1.5,arrowinset=0.15}
\rput[bl]{0}(1.4,0.8){$a_\mb{g}$}
\rput[bl]{0}(1.4,-1.2){$a_\mb{g}$}
  \psbezier[linewidth=0.9pt,linecolor=black,border=2.0pt](1.5,-0.6)(1.5,0.5)(2.0,0.4)(2.0,0.0)
  \psbezier[linewidth=0.9pt,linecolor=black,border=2.0pt](1.5,0.7)(1.5,-0.5)(2.0,-0.4)(2.0,0.0)
     \psline{->}(1.5,-0.7)(1.5,-0.5)
   \psline(1.5,-0.9)(1.5,-0.7)
\endpspicture
= \theta_{a}^{-1}
\pspicture[shift=-1.3](1.3,-1.4)(1.8,1.1)
  \small
  \psset{linewidth=0.9pt,linecolor=black,arrowscale=1.5,arrowinset=0.15}
\rput[bl]{0}(1.4,0.8){$a_\mb{g}$}
\rput[bl]{0}(1.4,-1.2){$a_\mb{g}$}
   \psline{->}(1.5,-0.7)(1.5,-0.5)
   \psline(1.5,-0.9)(1.5,0.7)
\endpspicture
=
\pspicture[shift=-1.3](0.8,-1.4)(1.8,1.1)
  \small
  \psset{linewidth=0.9pt,linecolor=black,arrowscale=1.5,arrowinset=0.15}
\rput[bl]{0}(1.4,0.8){$a_\mb{g}$}
\rput[bl]{0}(1.4,-1.2){$a_\mb{g}$}
    \psbezier[linewidth=0.9pt,linecolor=black,border=0.1](1.5,0.7)(1.5,-0.5)(1.0,-0.4)(1.0,0.0)
  \psbezier[linewidth=0.9pt,linecolor=black,border=2.0pt](1.5,-0.6)(1.5,0.5)(1.0,0.4)(1.0,0.0)
     \psline{->}(1.5,-0.7)(1.5,-0.5)
   \psline(1.5,-0.9)(1.5,-0.7)
\endpspicture
.
\end{equation}
For unitary theories, it is straightforward to see that $\theta_a^{-1} = \theta_a^{\ast}$, and hence the topological twist factors must be phases.

Unlike a BTC, it is not necessarily the case that $\theta_{a_{\bf g}}$ and $\theta_{\overline{a_{\bf g}}}$ are equal in a $G$-crossed BTC. In particular, we find the relations
\begin{equation}
\theta_{a_{\bf g}}= U_\mathbf{g}(\overline{a_{\bf g}},a_{\bf g};0) \eta_{\ol{a_{\bf g}}}(\mathbf{\bar{g}},{\mb{g}}) \theta_{\ol{a_{\bf g}}} .
\label{eq:lemmatwist}
\end{equation}
and
\begin{eqnarray}
\theta_{a_{\bf g}} &=& U_\mathbf{g}(a_{\bf g},\ol{a_{\bf g}};0) \varkappa_{a_{\bf g}} \left( R^{ \overline{a_{\bf g}} a_{\bf g} }_{0}\right)^{-1} \\
&=& \eta_{a_{\bf g}}({\bf g} , {\bf \bar{g}})^{-1} \varkappa_{a_{\bf g}}^{-1} \left( R^{a_{\bf g} \overline{a_{\bf g}}  }_{0}\right)^{-1}
\end{eqnarray}
from the following diagrammatic manipulations
\begin{eqnarray}
&&
\psscalebox{1}{
\pspicture[shift=-1.5](-2.3,-1.6)(-0.1,1.5)
  \small
  \psset{linewidth=0.9pt}
    \psbezier[linewidth=0.9pt,linecolor=black,border=2.0pt](-0.5,-0.5)(-0.6,0.9)(-1.3,0.6)(-1.3,0.0)		
	\psbezier[linewidth=0.9pt,linecolor=black,border=2.0pt](-0.5,0.5)(-0.6,-0.9)(-1.3,-0.6)(-1.3,0.0)
    \psline(-0.5,0.5)(-0.5,0.7)
    \psarc(-1.2,0.7){0.7}{0}{180}
    \psline(-1.9,0.7)(-1.9,-0.7)
    \psline[arrows=->, arrowscale=1.5,arrowinset=0.15](-1.9,-0.68)(-1.9,-0.7)
    \psline(-0.5,-0.5)(-0.5,-0.7)
    \psarc(-1.2,-0.7){0.7}{180}{360}
    \rput[bl]{0}(-1.8,-0.8){$a_\mb{g}$}
\endpspicture
}
= U_\mathbf{g}(\overline{a_{\bf g}},a_{\bf g};0)
\psscalebox{1}{
\pspicture[shift=-1.6](0,-2.0)(2.6,1)
	\small
	  \psset{linewidth=0.9pt}
	  \psbezier(1.7,0.7)(1.2,0.7)(0.8,-0.3)(1.3,-0.3)
	  \psbezier[border=2.0pt](0.9,0.7)(1.4,0.7)(1.8,-0.3)(1.3,-0.3)
	  \psbezier(1.7,0.7)(2.1,0.7)(2.2,0.1)(2.2,-0.4)
	  \psbezier(0.9,0.7)(0.5,0.7)(0.4,0.1)(0.4,-0.4)
	  \psbezier(0.4,-0.4)(0.4,-1.5)(2.2,-1.5)(2.2,-0.4)
   \psline[arrows=->, arrowscale=1.5,arrowinset=0.15](0.4,-0.38)(0.4,-0.4)
    \rput[bl]{0}(0.5,-0.5){$a_\mb{g}$}
\endpspicture
}
\\
&&\qquad =  U_\mathbf{g}(\overline{a_{\bf g}},a_{\bf g};0) \eta_{\ol{a_{\bf g}}}(\mathbf{\bar{g}},{\mb{g}})
\psscalebox{1}{
\pspicture[shift=-1.7](0,-1.8)(2.3,1.5)
  \small
  \psset{linewidth=0.9pt}
	\psbezier[linewidth=0.9pt,linecolor=black,border=2.0pt](0.5,0.5)(0.6,-0.9)(1.3,-0.6)(1.3,0.0)
    \psbezier[linewidth=0.9pt,linecolor=black,border=2.0pt](0.5,-0.5)(0.6,0.9)(1.3,0.6)(1.3,0.0)		
    \psline(0.5,0.5)(0.5,0.7)
    \psarc(1.2,0.7){0.7}{0}{180}
    \psline(1.9,0.7)(1.9,-0.7)
    \psline(0.5,-0.5)(0.5,-0.7)
    \psarc(1.2,-0.7){0.7}{180}{360}
    \psline[arrows=->, arrowscale=1.5,arrowinset=0.15](0.5,-0.78)(0.5,-0.8)
\rput[bl]{0}(0.6,-0.9){$a_\mb{g}$}
\endpspicture
} .
\end{eqnarray}

We can now derive the $G$-crossed generalization of the ribbon property by using the following diagrammatic relations
\begin{widetext}
\begin{eqnarray}
\pspicture[shift=-1.3](0.5,-0.9)(2.4,1.6)
  \small
  \psset{linewidth=0.9pt,linecolor=black,arrowscale=1.5,arrowinset=0.15}
  \rput[bl]{0}(1.0,1.6){$a_\mb{g}$}
\rput[bl]{0}(1.5,1.6){$b_\mb{h}$}
  \rput[bl]{0}(0.7,-0.9){$c_\mathbf{gh}$}
  \psbezier[linewidth=0.9pt,linecolor=black,border=0.1](1.45,0.1)(1.45,-0.2)(2.4,-0.5)(2.4,0.5)
  \psline(1.45,0.1)(1.3,0.3)
  \psline(1.45,0.1)(1.6,0.3)
  \psline(1.3,0.3)(1.3, 1.5)
  \psline(1.6,0.3)(1.6,1.5)
  \psbezier[linewidth=0.9pt,linecolor=black,border=2.0pt](1.1,0.1)(1.2,1.5)(2.4,1.2)(2.4,0.5)
   \psline{->}(1.1,-0.1)(1.1,-0.0)
   \psline(1.1,-0.6)(1.1,0.1)
 \scriptsize
  \rput[bl]{0}(1.55,0){$\mu$}
\endpspicture
&=& \sum_{\nu} \left[ U_{\bf gh}(a,b;c) \right]_{\mu \nu}
\pspicture[shift=-1.7](0.5,-1.7)(2.4,1.6)
  \small
  \psset{linewidth=0.9pt,linecolor=black,arrowscale=1.5,arrowinset=0.15}
  \rput[bl]{0}(1.1,1.3){$a_\mb{g}$}
\rput[bl]{0}(1.7,1.3){$b_\mb{h}$}
  \rput[bl]{0}(0.9,-1.0){$c_\mathbf{gh}$}
  \psbezier[linewidth=0.9pt,linecolor=black,border=0.1](1.5,0.7)(1.5,-0.5)(2.0,-0.4)(2.0,0.0)
  \psbezier[linewidth=0.9pt,linecolor=black,border=2.1pt](1.5,-0.6)(1.5,0.5)(2.0,0.4)(2.0,0.0)
   \psline{->}(1.5,-0.7)(1.5,-0.5)
   \psline(1.5,-0.9)(1.5,-0.7)
\psline(1.5,0.7)(1.2,1.2)
\psline(1.5,0.7)(1.8,1.2)
 \scriptsize
  \rput[bl]{0}(1.58,0.6){$\nu$}
\endpspicture
= \theta_{c} \sum_{\nu} \left[ U_{\bf gh}(a,b;c) \right]_{\mu \nu}
\pspicture[shift=-0.8](-0.1,-0.5)(1.5,1.4)
  \small
  \psset{linewidth=0.9pt,linecolor=black,arrowscale=1.5,arrowinset=0.15}
  \psline{->}(0.7,0)(0.7,0.45)
  \psline(0.7,0)(0.7,0.55)
  \psline(0.7,0.55) (0.25,1)
  \psline{->}(0.7,0.55)(0.3,0.95)
  \psline(0.7,0.55) (1.15,1)	
  \psline{->}(0.7,0.55)(1.1,0.95)
  \rput[bl]{0}(0.5,-0.3){$c_{\bf gh}$}
  \rput[br]{0}(1.4,1.05){$b_{\bf h}$}
  \rput[bl]{0}(-0.1,1.05){$a_{\bf g}$}
 \scriptsize
  \rput[bl]{0}(0.82,0.37){$\nu$}
  \endpspicture
\\
&=&
\eta_{b}({\bf \,^{\bar{h}}g, \,^{\bar{h}\bar{g}}h}) \eta_{a}({\bf \,^{\bar{h}}g, \,^{\bar{h}\bar{g}}h})
\pspicture[shift=-1.1](0.5,-1.5)(2.6,1.5)
  \small
  \psset{linewidth=0.9pt,linecolor=black,arrowscale=1.5,arrowinset=0.15}
  \psbezier[linewidth=0.9pt,linecolor=black,border=0.1](1.5,1.2)(1.5,-0.7)(2.3,-0.5)(2.3,0.0)
  \psbezier[linewidth=0.9pt,linecolor=black,border=0.1](1.14,1.2)(1.14,-1.1)(2.6,-1.0)(2.6,0.0)
  \psbezier[linewidth=0.9pt,linecolor=black,border=2pt](1.5,-0.8)(1.5,0.9)(2.3,0.5)(2.3,0.0)
  \psbezier[linewidth=0.9pt,linecolor=black,border=2pt](1.14,-0.8)(1.14,1.3)(2.6,1.0)(2.6,0.0)
  \psline{->}(1.15,1.0)(1.14,1.1)
\psline{->}(1.508,1.0)(1.5,1.1)
  \psline(1.14,-0.8)(1.32,-1.1)
	\psline(1.5,-0.8)(1.32,-1.1)
	\psline(1.32,-1.6)(1.32,-1.1)
\rput[bl]{0}(1.0,1.3){$a_\mb{g}$}
\rput[bl]{0}(1.45,1.3){$b_\mb{h}$}
\rput[bl]{0}(0.8,-1.5){$c_\mathbf{gh}$}
   \scriptsize
  \rput[bl]{0}(1.4,-1.3){$\mu$}
\endpspicture
= \eta_{a}({\bf g,h}) \eta_{b}({\bf h, \bar{h}gh})
\pspicture[shift=-1.8](0.5,-1.7)(2.4,2.6)
  \small
  \psset{linewidth=0.9pt,linecolor=black,arrowscale=1.5,arrowinset=0.15}
  \rput[bl]{0}(1.2,-1.3){$c_\mathbf{gh}$}
   \psline(1.5,-0.6)(1.5,-0.5)
   \psline(2.0,-0.6)(2.0,-0.5)
   \psline(1.75, -0.85)(1.5,-0.6)
\psline(1.75, -0.85)(2.0,-0.6)
\psline(1.75,-1.3)(1.75,-0.85)
 \psbezier[linewidth=0.9pt] (2.0, -0.5) (2.0, -0.2) (1.5, -0.1) (1.5, 0.3)
\psbezier[linewidth=0.9pt, border=2.2pt] (1.5, -0.5) (1.5, -0.2) (2.0, -0.1) (2.0, 0.3)
 \psbezier[linewidth=0.9pt] (2.0, 0.3) (2.0, 0.6) (1.5, 0.7) (1.5, 1.0)
\psbezier[linewidth=0.9pt, border=2.2pt] (1.5, 0.3) (1.5, 0.6) (2.0, 0.7) (2.0, 1.0)
  \psbezier[linewidth=0.9pt,linecolor=black,border=0.1](1.5,2.0)(1.5,1.2)(1.8,1.1)(1.8,1.5)
  \psbezier[linewidth=0.9pt,linecolor=black,border=1.8pt](1.5,1.0)(1.5,1.8)(1.8,1.9)(1.8,1.5)
  \psbezier[linewidth=0.9pt,linecolor=black,border=0.1](2.0,2.0)(2.0,1.2)(2.3,1.1)(2.3,1.5)
  \psbezier[linewidth=0.9pt,linecolor=black,border=1.8pt](2.0,1.0)(2.0,1.8)(2.3,1.9)(2.3,1.5)
\rput[bl]{0}(1.3,2.1){$a_\mb{g}$}
\rput[bl]{0}(1.9,2.1){$b_\mb{h}$}
   \scriptsize
  \rput[bl]{0}(1.8,-1.0){$\mu$}
\endpspicture
\notag \\
&=& \theta_{a} \theta_{b} \eta_{a}({\bf g,h}) \eta_{b}({\bf h, \bar{h}gh})  \sum_{\nu, \lambda}  \left[R^{b_{\bf h} \,^{\bf \bar{h}}a_{\bf g}}_{c_{\bf gh}}\right]_{\mu \lambda} \left[R^{a_{\bf g} b_{\bf h}}_{c_{\bf gh}} \right]_{\lambda \nu}
\pspicture[shift=-0.8](-0.1,-0.5)(1.5,1.4)
  \small
  \psset{linewidth=0.9pt,linecolor=black,arrowscale=1.5,arrowinset=0.15}
  \psline{->}(0.7,0)(0.7,0.45)
  \psline(0.7,0)(0.7,0.55)
  \psline(0.7,0.55) (0.25,1)
  \psline{->}(0.7,0.55)(0.3,0.95)
  \psline(0.7,0.55) (1.15,1)	
  \psline{->}(0.7,0.55)(1.1,0.95)
  \rput[bl]{0}(0.5,-0.3){$c_{\bf gh}$}
  \rput[br]{0}(1.4,1.05){$b_{\bf h}$}
  \rput[bl]{0}(-0.1,1.05){$a_{\bf g}$}
 \scriptsize
  \rput[bl]{0}(0.82,0.37){$\nu$}
  \endpspicture
.
\end{eqnarray}
Notice that the first and second lines are related using the pivotal property and we used the Yang-Baxter relation and the fact that lines can slide freely under a twist. This yields the $G$-crossed ribbon property
\begin{equation}
\sum_\lambda \left[R^{b_{\bf h} \,^{\bf \bar{h}}a_{\bf g}}_{c_{\bf gh}}\right]_{\mu \lambda} \left[R^{a_{\bf g} b_{\bf h}}_{c_{\bf gh}} \right]_{\lambda \nu} =
\frac{\theta_{c}}{\theta_{a}\theta_{b}} \frac{\left[ U_{\bf gh}(a,b;c)\right]_{\mu\nu} }{\eta_{a}({\bf g,h}) \eta_{b}({\bf h},\,^{\bf \bar{h}}\mb{g})}
.
\label{eq:G-crossed_ribbon}
\end{equation}
\end{widetext}

Clearly, the operator $R^{a_{\bf g} b_{\bf h}} R^{b_{\bf h} \,^{\bf \bar{h}}a_{\bf g}}$ is not gauge invariant, unless ${\bf g} = {\bf h} = {\bf 0}$. However, when $^{\bf h}a_{\bf g} = a_{\bf g}$ and $^{\bf g}b_{\bf h} = b_{\bf h}$, the quantities
\begin{equation}
\frac{ \sum\limits_{\mu ,\nu }  \left[R^{b_{\bf h} a_{\bf g}}_{c_{\bf gh}}\right]_{\mu \nu} \left[R^{a_{\bf g} b_{\bf h}}_{c_{\bf gh}} \right]_{\nu \mu}  } {\sum\limits_{\mu' , \nu' }  \left[R^{b_{\bf h} a_{\bf g}}_{c'_{\bf gh}}\right]_{\mu' \nu'} \left[R^{a_{\bf g} b_{\bf h}}_{c'_{\bf gh}} \right]_{\nu' \mu'}}
=
\frac{ \theta_{c} \sum\limits_{\mu } \left[ U_{\bf gh}(a,b;c)\right]_{\mu\mu} } { \theta_{c'} \sum\limits_{\mu' } \left[ U_{\bf gh}(a,b;c')\right]_{\mu' \mu'} }
\label{eqn:projbraiding}
\end{equation}
are invariant under both types of gauge transformations.

More generally, when $^{\bf k} a_{\bf g} = a_{\bf g}$ and $^{\bf k} b_{\bf h} = b_{\bf h}$ for ${\bf k} = ({\bf gh})^{n}$, the quantities
\begin{equation}
\frac{ \sum\limits_{\mu }  \left[\left( R^{2n} \right)^{a_{\bf g} b_{\bf h}}_{c_{\bf gh}} \right]_{\mu \mu}  } {\sum\limits_{\mu' }  \left[\left( R^{2n} \right)^{a_{\bf g} b_{\bf h}}_{c'_{\bf gh}} \right]_{\mu' \mu'} }
\label{eqn:projbraiding_2n}
\end{equation}
are invariant under both types of gauge transformations (where the notation $R^{2n}$ indicates the operator for $2n$ successive counter-clockwise exchanges).

Once again, we define the topological $S$-matrix by
\begin{eqnarray}
\label{eqn:topoSmatrix}
S_{a_{\bf g} b_{\bf h} }&=&\frac{1}{\mathcal{D}_\mb{0}}
\pspicture[shift=-0.4](0.0,0.2)(2.3,1.3)
\small
  \psarc[linewidth=0.9pt,linecolor=black,arrows=<-,arrowscale=1.5,arrowinset=0.15] (1.6,0.7){0.5}{167}{373}
  \psarc[linewidth=0.9pt,linecolor=black,border=3pt,arrows=<-,arrowscale=1.5,arrowinset=0.15] (0.9,0.7){0.5}{167}{373}
  \psarc[linewidth=0.9pt,linecolor=black] (0.9,0.7){0.5}{0}{180}
  \psarc[linewidth=0.9pt,linecolor=black,border=3pt] (1.6,0.7){0.5}{45}{150}
  \psarc[linewidth=0.9pt,linecolor=black] (1.6,0.7){0.5}{0}{50}
  \psarc[linewidth=0.9pt,linecolor=black] (1.6,0.7){0.5}{145}{180}
  \rput[bl]{0}(0.1,0.45){$a$}
  \rput[bl]{0}(0.8,0.45){$b$}
  \endpspicture
\notag \\
&=&\frac{1}{\mathcal{D}_\mb{0}}\sum_{c,\mu,\nu}d_c \left[ R^{b \bar{a} }_c \right]_{\mu \nu} \left[R_c^{ \bar{a} b} \right]_{\nu\mu}
\notag\\
&=&\frac{1}{\mathcal{D}_\mb{0}}\sum_{c,\mu} d_c \frac{\theta_{c}}{\theta_{\bar{a} }\theta_{b}} \frac{\left[ U_{\bf \bar{g}h}(\bar{a},b;c)\right]_{\mu \mu} }{\eta_{\bar{a}}({\bf \bar{g},h}) \eta_{b}({\bf h , \bar{g} })}
.
\end{eqnarray}
We emphasize that, when $a\in\mathcal{C}_\mb{g}$ and $b\in\mathcal{C}_\mb{h}$, the $S$-matrix is only well-defined if $^\mb{h}a=a$ and $^\mb{g}b=b$, and consequently $\mb{gh}=\mb{hg}$. Otherwise, the topological charge values would change in the braiding and one would not be able to close the lines back upon themselves. We note that we have used $\mathcal{D}_\mb{0} = \mathcal{D}_\mb{g}$, the total quantum dimension of each subsector $\mathcal{C}_{\bf g}$, rather than the total quantum dimension $\mathcal{D}_{\mathcal{C}_{G}} = |G|^{\frac{1}{2}} \mathcal{D}_\mb{0}$ of the entire $G$-crossed theory $\mathcal{C}_{G}^{\times}$ for reasons that will be made clear later.

The elements of the $S$-matrix do not obey all the same relations as that of a BTC, nor are they gauge invariant, unless ${\bf g} = {\bf h} = {\bf 0}$, or unless either $a = 0$ or $b = 0$ (in which case $S_{ab} = d_{a} d_{b} / \mathcal{D}_{\bf 0}$), since
\begin{equation}
\check{S}_{a_{\bf g} b_{\bf h} } = \gamma_{\bar{a}}({\bf h}) \gamma_{b}({\bf \bar{g}})S_{a_{\bf g} b_{\bf h} }
.
\end{equation}
Nonetheless, the $S$-matrix will be an important quantity that again plays an important role in defining the system and modular transformations on higher genus surfaces, so we will examine its properties in detail.

We first note that
\begin{equation}
S_{\,^{\bf k}a_{\bf g} \,^{\bf k} b_{\bf h} }= \frac{\eta_{\,^{\bf k}\bar{a}}({\bf k,h}) \eta_{\,^{\bf k}b}({\bf k , \bar{g} })} {\eta_{\,^{\bf k}\bar{a}}({\bf kh\bar{k},k}) \eta_{\,^{\bf k}b}({\bf k\bar{g}\bar{k} , k }) }S_{a_{\bf g} b_{\bf h} }
,
\end{equation}
which follows from the definition and Eq.~(\ref{eq:G-crossed_Yang_Baxter}). It follows that, when $^{\bf k}a_{\bf g} = \,^{\bf h}a_{\bf g}=a_{\bf g}$ and $^{\bf k} b_{\bf h}=\,^{\bf g} b_{\bf h}=b_{\bf h}$, we have
\begin{equation}
\frac{\eta_{\bar{a}}({\bf k,h}) \eta_{b}({\bf k , \bar{g} })} {\eta_{\bar{a}}({\bf h,k}) \eta_{b}({\bf \bar{g} , k }) } =1
.
\end{equation}

It is straightforward to see that
\begin{equation}
S_{a_{\bf g} b_{\bf h} }^{\ast} =\frac{1}{\mathcal{D}_\mb{0}}
\pspicture[shift=-0.4](0.0,0.2)(2.3,1.3)
\small
  \psarc[linewidth=0.9pt,linecolor=black] (1.6,0.7){0.5}{45}{160}
  \psarc[linewidth=0.9pt,linecolor=black,border=3pt] (0.9,0.7){0.5}{0}{180}
  \psarc[linewidth=0.9pt,linecolor=black,arrows=<-,arrowscale=1.5,arrowinset=0.15] (0.9,0.7){0.5}{167}{373}
  \psarc[linewidth=0.9pt,linecolor=black,border=3pt,arrows=<-,arrowscale=1.5,arrowinset=0.15] (1.6,0.7){0.5}{167}{373}
  \psarc[linewidth=0.9pt,linecolor=black] (1.6,0.7){0.5}{0}{50}
  \psarc[linewidth=0.9pt,linecolor=black] (1.6,0.7){0.5}{155}{180}
  \rput[bl]{0}(0.1,0.45){$a$}
  \rput[bl]{0}(0.8,0.45){$b$}
  \endpspicture
\end{equation}
for a unitary theory. It also follows immediately from the definition (and the cyclic property of the trace) that
\begin{equation}
S_{a_{\bf g} b_{\bf h} } = S_{ \overline{b_{\bf h}} \overline{a_{\bf g} }}.
\label{eq:S_2}
\end{equation}
While these $S$-matrix relations are the same as for UBTCs, we must be more careful with properties obtained by deforming lines, because of the nontrivial sliding rules of a $G$-crossed theory.

When $^{\bf h}a_{\bf g} = a_{\bf g}$ and $^{\bf g}b_{\bf h} = b_{\bf h}$ (and hence ${\bf gh=hg}$), so that the corresponding $S$-matrix element is well-defined, we have the loop-removal relation
\begin{equation}
\pspicture[shift=-1.0](-0.5,-0.9)(1.5,1.5)
\small
  \psset{linewidth=0.9pt,linecolor=black,arrowscale=1.5,arrowinset=0.15}
  \psline(0.75,-0.7)(0.75,-0.15)
  \psline(0.75,0.15)(0.75,1.3)
  \psellipse[linewidth=0.9pt,linecolor=black,border=0](0.4,0.3)(0.8,0.35)
  \psline{-<}(0.2,-0.027)(0.3,-0.04)
\psline[linewidth=0.9pt,linecolor=black,border=2.2pt,arrows=->,arrowscale=1.5,
arrowinset=0.15](0.75,0.1)(0.75,1.05)
\rput[bl]{0}(0.15,-0.4){$a_\mb{g}$}
  \rput[tl]{0}(0.9,1.1){$b_\mathbf{h}$}
\endpspicture
=\frac{S_{ab}}{S_{0b}}
\pspicture[shift=-1.0](0.4,-0.9)(1.5,1.5)
\small
  \psset{linewidth=0.9pt,linecolor=black,arrowscale=1.5,arrowinset=0.15}
  \psline(0.75,-0.7)(0.75,1.3)
\psline[linewidth=0.9pt,linecolor=black,border=2.2pt,arrows=->,arrowscale=1.5,
arrowinset=0.15](0.75,0.1)(0.75,1.05)
  \rput[tl]{0}(0.9,1.1){$b_\mathbf{h}$}
 \endpspicture
,
\label{eq:G-crossed_loop_removal}
\end{equation}
which can be verified by closing the $b$ line upon itself in this expression. In fact, if either $^{\bf h}a_{\bf g} \neq a_{\bf g}$ or $^{\bf g}b_{\bf h} \neq b_{\bf h}$, then left hand side of the equation evaluates to zero, so, for these purposes, we can consider $S_{ab}=0$ when it is not well-defined.

In writing this relation, we must be more careful than in a BTC to indicate clearly where the lines are drawn with respect to vertices, including local minima and maxima (cups and caps). Recall that the minima/maxima of the cups/caps correspond to splitting/fusion vertices, respectively, between a topological charge, its conjugate, and the vacuum. Therefore, we see that
\begin{equation}
\pspicture[shift=-1.0](-0.5,-0.9)(1.5,1.5)
\small
  \psset{linewidth=0.9pt,linecolor=black,arrowscale=1.5,arrowinset=0.15}
  \psline(0.75,-0.7)(0.75,-0.15)
  \psline(0.75,0.15)(0.75,1.3)
  \psellipse[linewidth=0.9pt,linecolor=black,border=0](0.4,0.3)(0.8,0.35)
  \psline{-<}(0.2,-0.027)(0.3,-0.04)
\psline[linewidth=0.9pt,linecolor=black,border=2.2pt,arrows=->,arrowscale=1.5,
arrowinset=0.15](0.75,0.1)(0.75,1.05)
\rput[bl]{0}(0.15,-0.4){$a_\mb{g}$}
  \rput[tl]{0}(0.9,1.1){$b_\mathbf{h}$}
\endpspicture
=\frac{U_{\bf h} (a,\bar{a};0) }{ \eta_{b} ({\bf g , \bar{g}}) }
\pspicture[shift=-1.0](-0.5,-0.9)(1.5,1.5)
\small
  \psset{linewidth=0.9pt,linecolor=black,arrowscale=1.5,arrowinset=0.15}
  \psline(0.05,-0.7)(0.05,-0.15)
  \psline(0.05,0.15)(0.05,1.3)
  \psellipse[linewidth=0.9pt,linecolor=black,border=0](0.4,0.3)(0.8,0.35)
  \psline{-<}(0.2,-0.027)(0.3,-0.04)
\psline[linewidth=0.9pt,linecolor=black,border=2.2pt,arrows=->,arrowscale=1.5,arrowinset=0.15](0.05,0.1)(0.05,1.05)
\rput[bl]{0}(0.15,-0.4){$a_\mb{g}$}
  \rput[tl]{0}(0.2,1.1){$b_\mathbf{h}$}
\endpspicture
.
\label{eq:G-crossed_loop_slide}
\end{equation}

Since one can equivalently take the trace of Eq.~(\ref{eq:G-crossed_loop_slide}) by closing the $b$-line on itself into a loop to the left or right, it leads to the relation
\begin{equation}
S_{a_{\bf g} b_{\bf h} } = \frac{U_\mb{h}(a,\bar{a};0)}{ \eta_b(\mb{g},\mb{\bar{g}})} S_{\overline{b_{\bf h}} a_{\bf g} }^{\ast}
.
\label{eq:S_3}
\end{equation}
Combining Eqs.~(\ref{eq:S_2}) and (\ref{eq:S_3}) yields a relation between the $S$-matrix and its transpose
\begin{equation}
S_{a_{\bf g} b_{\bf h} }= \frac{ U_{\mb{h}}(a,\bar{a};0) \eta_{a}( \mb{\bar{h}} , \mb{h}) }{ U_{\mb{g}}(b,\bar{b};0) \eta_{b}(\mb{g},\mb{ \bar{g} })} S_{b_{\bf h} a_{\bf g} }
.
\label{eq:S_4}
\end{equation}

Another useful relation allows us to flip the tilt of a loop encircling another line, as follows
\begin{widetext}
\begin{eqnarray}
\pspicture[shift=-1.0](-0.5,-0.9)(1.5,1.5)
\small
  \psset{linewidth=0.9pt,linecolor=black,arrowscale=1.5,arrowinset=0.15}
  \psline(0.05,-0.7)(0.05,-0.15)
  \psline(0.05,0.15)(0.05,1.3)
  \psellipse[linewidth=0.9pt,linecolor=black,border=0](0.4,0.3)(0.8,0.35)
  \psline{-<}(0.2,-0.027)(0.3,-0.04)
\psline[linewidth=0.9pt,linecolor=black,border=2.2pt,arrows=->,arrowscale=1.5,arrowinset=0.15](0.05,0.1)(0.05,1.05)
\rput[bl]{0}(0.3,-0.4){$a_\mb{g}$}
  \rput[tl]{0}(0.2,1.1){$b_\mathbf{h}$}
\endpspicture
&=& \theta_{a}
\pspicture[shift=-1.0](-0.5,-0.9)(1.7,1.5)
\small
  \psset{linewidth=0.9pt,linecolor=black,arrowscale=1.5,arrowinset=0.15}
  \psarc(0.1, 0.2){0.4}{90}{270}
  \psarc(1.4,0.2){0.3}{-90}{90}
  \psbezier(0.1, -0.2)(0.4,-0.2)(1.0,0.5)(1.4,0.5)
\psline[border=2.1pt](1.1, -0.7)(1.1,0.2)
  \psbezier[linewidth=0.9pt, border=2.0pt](0.1, 0.6)(0.4,0.6)(1.0,-0.1)(1.4,-0.1)
  \psline[linewidth=0.9pt,arrows=->, arrowscale=1.5, arrowinset=0.1](0.5,0.41)(0.4,0.48)
  \psline[linewidth=0.9pt,arrows=->, arrowscale=1.5, arrowinset=0.1, border=2.1pt](1.1, 0.2)(1.1,1.0)
  \psline(1.1,0.9)(1.1, 1.3)
\rput[bl]{0}(0.3,0.6){$a_\mathbf{g}$}
  \rput[tl]{0}(1.2,1.1){$b_\mathbf{h}$}
\endpspicture
= \theta_a
\pspicture[shift=-1.0](-0.5,-0.9)(1.8,1.8)
\small
  \psset{linewidth=0.9pt,linecolor=black,arrowscale=1.5,arrowinset=0.15}
  \psarc(0.1, 0.2){0.4}{90}{270}
  \psarc(1.4,0.2){0.3}{-90}{90}
  \psbezier(0.1, -0.2)(0.4,-0.2)(1.0,0.5)(1.4,0.5)
  \psbezier[linewidth=0.9pt, border=2.0pt](0.1, 0.6)(0.4,0.6)(1.0,-0.1)(1.4,-0.1)
  \psline[linewidth=0.9pt,arrows=->, arrowscale=1.5, arrowinset=0.1](1.0,0.06)(0.9,0.12)
  \psline[border=2.1pt](0.4, -0.7)(0.4,0.35)
  \psline[linewidth=0.9pt,arrows=->, arrowscale=1.5, arrowinset=0.1](0.4, 0.55)(0.4,1.0)
  \psline(0.4,0.9)(0.4, 1.3)
\rput[bl]{0}(0.8,-0.3){$a_\mathbf{g}$}
  \rput[tl]{0}(0.5,1.1){$b_\mathbf{h}$}
\endpspicture
=
\pspicture[shift=-1.0](-0.5,-0.9)(1.3,1.5)
\small
  \psset{linewidth=0.9pt,linecolor=black,arrowscale=1.5,arrowinset=0.15}
  \psellipse[linewidth=0.9pt,linecolor=black,border=0](0.4,0.3)(0.8,0.35)
  \psline[linewidth=0.9pt,linecolor=black,border=2.2pt](0.75,-0.7)(0.75,0.5)
  \psline[linewidth=0.9pt, arrows=->,arrowscale=1.5, arrowinset=0.15] (0.75,0.65) (0.75, 1.05)
  \psline{-<}(0.2,0.627)(0.3,0.64)
\psline(0.75,0.7)(0.75,1.1)
\psline (0.75, 1.0) (0.75, 1.3)
\rput[bl]{0}(-0.2,0.7){$a_\mathbf{g}$}
  \rput[tl]{0}(0.9,1.1){$b_\mathbf{h}$}
  \endpspicture
,
\end{eqnarray}
in which we used Eqs.~(\ref{eq:G-crossed_Yang_Baxter}) and (\ref{eqn:etasym}).

An important diagrammatic relation, which is the precursor of the $G$-crossed Verlinde formula, is obtained by putting two loops on a line and using a partition of identity to relate it to a single loop on the line
\begin{eqnarray}
\psscalebox{.6}{
\pspicture[shift=-1.6](-1.4,-1.0)(0.9,2.6)
	\small
	\psset{linewidth=0.9pt,linecolor=black,arrowscale=1.5,arrowinset=0.15}
	\psline[linewidth=0.9pt](0,-0.7)(0,1.0)
	\psellipse[linewidth=0.9pt,border=2.1pt](-0.15,0.8)(0.4,0.5)
	\psellipse[linewidth=0.9pt,border=2.1pt](-0.2,0.8)(0.8,1.0)
	\psline[border=2.1pt](0,1.0)(0,2.2)
	\psline{->}(0,1.8)(0,2.1)
	\psline{->}(-0.53,0.79)(-0.53,0.89)
	\psline{->}(-0.98,0.75)(-0.98,0.85)
	\rput[bl]{0}(-1.4,0.7){$a_{\bf g}$}
	\rput[bl]{0}(-0.45,0.7){$b_{\bf h}$}
	\rput[tl]{0}(-0.1,2.45){$x_\mb{k}$}
\endpspicture
}
&=&\sum_{c,\mu}\sqrt{\frac{d_c}{d_ad_b}}
\psscalebox{.6}{
\pspicture[shift=-2.0](-1.5,-1.5)(1.1,2.6)
	\small
	\psset{linewidth=0.9pt,linecolor=black,arrowscale=1.5,arrowinset=0.15}
	\psline[linewidth=0.9pt](0.2,-1.0)(0.2,1.0)
	\psline(0.6,0.4)(0.6,0.8)
	\psline(0.6,0.8)(0.8,1.1)
	\psline(0.6,0.8)(0.4,1.1)
	\psline(0.6,0.4)(0.8,0.1)
	\psline(0.6,0.4)(0.4,0.1)
	\psline(0.4,0.1)(0.4,0.0)
	\psline(0.8,0.1)(0.8,0.0)
	\psline(0.4,1.1)(0.4,1.2)
	\psline(0.8,1.1)(0.8,1.2)
	\psbezier(0.4,1.2)(0.4,1.7)(-0.5,1.7)(-0.5,1.2)
	\psbezier(0.8,1.2)(0.8,2.2)(-1.0,2.2)(-1.0,1.2)
	\psline[border=2.1pt](0.2,1.0)(0.2,2.2)
	\psbezier[border=2.0pt](0.4,0.0)(0.4,-0.5)(-0.5,-0.5)(-0.5,0.0)
	\psbezier[border=2.0pt](0.8,0.0)(0.8,-1.0)(-1.0,-1.0)(-1.0,0.0)
	\psline(-0.5,0)(-0.5,1.2)
	\psline(-1.0, 0)(-1.0,1.2)
	\psline{->}(-0.5,0.6)(-0.5,0.8)
	\psline{->}(-1.0,0.6)(-1.0,0.8)
	\psline{->}(0.2,0.7)(0.2,1.3)
	\psline{->}(0.6,0.6)(0.6,0.45)
	\rput[bl]{0}(0.35,0.7){\mbox{\tiny $\mu$}}
	\rput[bl]{0}(0.35,0.35){\mbox{\tiny $\mu$}}
	\rput[br]{0}(-0.6,0.7){$b$}
	\rput[br]{0}(-1.1,0.7){$a$}
	\rput[bl]{0}(0.75,0.5){$c$}
	\rput[tl]{0}(-0.0,2.45){$x_\mb{k}$}
\endpspicture
}
=\sum_{c,\mu,\nu}\sqrt{\frac{d_c}{d_ad_b}}\frac{[U_\mb{k}(\bar{b}, \bar{a} ;\bar{c})]_{\mu \nu}} {\eta_{x}({\bf \bar{h}},{\bf \bar{g}})}
\psscalebox{.6}{
\pspicture[shift=-2.0](-1.3,-1.5)(1.5,2.6)
	\small
	\psset{linewidth=0.9pt,linecolor=black,arrowscale=1.5,arrowinset=0.15}
	\psline[linewidth=0.9pt](0.2,-1.0)(0.2,1.0)
	\psline(0.8,0.1)(0.8,0.0)
	\psline(0.8,1.1)(0.8,1.2)
	\psbezier(0.8,1.2)(0.8,2.2)(-1.0,2.2)(-1.0,1.2)
	\psline[border=2.1pt](0.2,1.0)(0.2,2.2)
	\psbezier[border=2.0pt](0.8,0.0)(0.8,-1.0)(-1.0,-1.0)(-1.0,0.0)
	\psline(-1.0, 0)(-1.0,1.2)
	\psbezier(0.8,0.1)(0.5,0.3)(0.5,0.9)(0.8,1.1)
	\psbezier(0.8,0.1)(1.1,0.3)(1.1,0.9)(0.8,1.1)
	\psline{->}(-1.0,0.4)(-1.0,0.8)
	\psline{->}(0.2,1.5)(0.2,1.8)
	\psline{-<}(1.01,0.61)(1.01,0.65)
	\psline{-<}(0.59,0.61)(0.59,0.65)
	\rput[bl]{0}(0.9,1.1){\mbox{\tiny $\mu$}}
	\rput[bl]{0}(0.9,-0.0){\mbox{\tiny $\nu$}}
	\rput[br]{0}(-0.7,0.7){$c$}
	\rput[br]{0}(0.45,0.5){$b$}
	\rput[br]{0}(1.3,0.5){$a$}
	\rput[tl]{0}(-0.0,2.45){$x_\mb{k}$}
\endpspicture
}
=\sum_{c\in \mathcal{C}_{\bf gh}^{\bf k},\mu}\frac{[U_\mb{k}(\bar{b},\bar{a};\bar{c})]_{\mu \mu}}{\eta_{x}({\bf \bar{h}},{\bf \bar{g}})}
\psscalebox{.6}{
\pspicture[shift=-1.6](-1.1,-1.0)(1,2.6)
	\small
	\psset{linewidth=0.9pt,linecolor=black,arrowscale=1.5,arrowinset=0.15}
	\psline[linewidth=0.9pt](0.2,-0.7)(0.2,1.0)
	\psellipse[linewidth=0.9pt,border=2.1pt](-0.18,0.8)(0.6,0.9)
	\psline[border=2.1pt](0.2,1.0)(0.2,2.2)
	\psline{->}(0.2,2.0)(0.2,2.1)
	\psline{->}(-0.75,0.79)(-0.75,0.89)
	\rput[br]{0}(-0.2,0.7){$c_{\bf gh}$}
	\rput[tl]{0}(0.1,2.45){$x_\mb{k}$}
\endpspicture
}
\label{eq:preVerlinde_diagram}
\end{eqnarray}
\end{widetext}

Combining Eqs.~(\ref{eq:preVerlinde_diagram}) and (\ref{eq:G-crossed_loop_removal}), we find that when $^{\bf k}a_{\bf g} = a_{\bf g}$, $^{\bf k}b_{\bf h} = b_{\bf h}$, and $^{\bf g}x_{\bf k} = \,^{\bf h}x_{\bf k} = x_{\bf k}$, we have the important relation
\begin{equation}
\frac{S_{a_{\bf g} x_{\bf k}}}{S_{0 x_{\bf k}}} \frac{S_{b_{\bf h} x_{\bf k}}}{S_{0x_{\bf k}}} = \sum_{c_{\bf gh}\in \mathcal{C}_{\bf gh}^{\bf k} , \mu} \frac{[U_\mb{k}(\bar{b},\bar{a};\bar{c})]_{\mu\mu}}{\eta_{x}({\mb{\bar{h}}},{\mb{\bar{g}}})} \frac{S_{c_{\bf gh} x_{\bf k}}}{S_{0x_{\bf k}}}
.
\label{eq:preVerlinde1}
\end{equation}
We can similarly obtain
\begin{equation}
\frac{S_{ x_{\bf k} a_{\bf g}}}{S_{x_{\bf k} 0}} \frac{S_{ x_{\bf k} b_{\bf h} }}{S_{ x_{\bf k} 0 }} = \sum_{c_{\bf gh}\in \mathcal{C}_{\bf gh}^{\bf k},\mu} \frac{[U_\mb{\bar{k} }(a,b;c)]_{\mu\mu}}{\eta_{\bar{x}}({\mb{g}},{\mb{h}})} \frac{S_{ x_{\bf k} c_{\bf gh} }}{S_{ x_{\bf k} 0}}
.
\label{eq:preVerlinde_2}
\end{equation}

If we take $x\in \mathcal{C}_{\bf 0}$, these expressions become
\begin{eqnarray}
\label{eq:preVerlinde3}
\frac{S_{a_{\bf g} x_{\bf 0}} S_{b_{\bf h} x_{\bf 0}} }{S_{0 x_{\bf 0}}} \eta_{x}({\mb{\bar{h}}},{\mb{\bar{g}}}) &=& \sum_{c_{\bf gh}} N_{ab}^{c} S_{c_{\bf gh} x_{\bf 0}} \\
\label{eq:preVerlinde4}
\frac{S_{ x_{\bf 0} a_{\bf g}} S_{ x_{\bf 0} b_{\bf h}} }{S_{x_{\bf 0} 0}} \eta_{\bar{x}}({\mb{g}},{\mb{h}}) &=& \sum_{c_{\bf gh}} N_{ab}^{c} S_{ x_{\bf 0} c_{\bf gh}}
,
\end{eqnarray}
which show that one may think of $S_{a_{\bf g} x_{\bf 0}} / S_{0 x_{\bf 0}}$ (or, equivalently, $S_{x_{\bf 0} a_{\bf g}} / S_{ x_{\bf 0} 0}$) as projective characters of the extended (non-commutative) Verlinde algebra.

We will now establish several interesting relations that we will find particularly useful for the discussion of modularity. We first define
\begin{equation}
\Theta_{\bf 0} = \frac{1}{\mathcal{D}_\mb{0}}  \sum_{c \in \mathcal{C}_{\bf 0}} d_c^2 \theta_c
\end{equation}
to be the normalized Gauss sum of the $\mathcal{C}_0$ BTC.
Then, we have the relation
\begin{equation}
\sum_{a \in \mathcal{C}_{\bf g}} \frac{d_a \theta_a}{ U_\mathbf{g}(\bar{a},a;0)}
\pspicture[shift=-1.2](-0.35,-0.9)(0.9,1.1)
\small
  \psset{linewidth=0.9pt,linecolor=black,arrowscale=1.5,arrowinset=0.15}
  \psline(0.58,-0.5)(0.58,0.05)
  \psline(0.58,0.25)(0.58,1.3)
  \psellipse[linewidth=0.9pt,linecolor=black,border=0](0.4,0.4)(0.5,0.28)
  \psline{-<}(0.2,0.175)(0.3,0.15)
\psline[linewidth=0.9pt,linecolor=black,border=2.5pt,arrows=->,arrowscale=1.5,arrowinset=0.15](0.58,0.3)(0.58,1.1)
  \rput[bl]{0}(-0.15,-0.1){$a_\mathbf{g}$}
  \rput[tl]{0}(0.7,-0.2){$b_\mathbf{g}$}
  \endpspicture
= \frac{ \mathcal{D}_\mb{0} \Theta_{\bf 0} }{ \eta_b({\bf g , \bar{g} }) \theta_{b} }
\pspicture[shift=-1.2](0.05,-0.9)(1,1.1)
\small
  \psset{linewidth=0.9pt,linecolor=black,arrowscale=1.5,arrowinset=0.15}
  \psline(0.4, -0.5)(0.4,1.3)
\psline[linewidth=0.9pt,linecolor=black,arrows=->,arrowscale=1.5,
arrowinset=0.15](0.4,0.5)(0.4,0.7)
\rput[tl]{0}(0.52,-0.2){$b_\mathbf{g}$}
  \endpspicture
\label{eq:loop-twist}
.
\end{equation}
In order to obtain this relation, we use the fact that when $a,b \in \mathcal{C}_{\bf g}$, the $S$-matrix takes the form
\begin{equation}
S_{a_{\bf g} b_{\bf g} } = \frac{1} { \eta_{\bar{a}}({\bf \bar{g}, g}) \eta_{b}({\bf g , \bar{g} }) }  \frac{1}{\mathcal{D}_\mb{0}}   \sum_{c \in \mathcal{C}_{\bf 0}} N_{\bar{a} b}^{c} d_c \frac{\theta_{c}}{\theta_{\bar{a} }\theta_{b}}
\end{equation}
and therefore obeys the property
\begin{eqnarray}
\sum_{a_{\bf g}} \frac{d_{a} \theta_{a}}{ U_\mathbf{g}(\bar{a},a;0)  }  S_{a_{\bf g} b_{\bf g} } &=& \frac{1 }{ \eta_b({\bf g , \bar{g} }) \theta_{b} }\frac{1}{\mathcal{D}_\mb{0}} \sum_{a_{\bf g},c_{\bf 0}} N_{\bar{a} b}^{c} d_a d_c \theta_{c} \notag\\
&=& \frac{d_b \Theta_{\bf 0} }{ \eta_b({\bf g , \bar{g} }) \theta_{b} }
,
\end{eqnarray}
which is established using Eqs.~(\ref{eq:d_g_relation}) and (\ref{eq:lemmatwist}).

The next relation (which holds even when $^{\bf h}a_{\bf g} \neq a_{\bf g}$ or $^{\bf g}b_{\bf h} \neq b_{\bf h}$) is
\begin{widetext}
\begin{eqnarray}
\sum_{x \in \mathcal{C}_\mathbf{gh}} \frac{ d_x \theta_x  \eta_{\bar{x}}({ \bf ^{\bar{h}} g, \,^{\bar{h} \bar{g} } h} ) }{ U_\mathbf{gh}(\bar{x},x;0) }
\pspicture[shift=-1.4](-0.5,-1.3)(1.5,1.5)
\small
  \psset{linewidth=0.9pt,linecolor=black,arrowscale=1.5,arrowinset=0.15}
  \psline(0.65,-0.7)(0.65,-0.15)
    \psline{->}(0.65,-0.7)(0.65,-0.3)
  \psline(0.65,0.15)(0.65,1.3)
  \psline(0.95,-0.7)(0.95,-0.05)
    \psline{->}(0.95,-0.7)(0.95,-0.3)
  \psline(0.95,0.18)(0.95,1.3)
  \psellipse[linewidth=0.9pt,linecolor=black,border=0](0.4,0.3)(0.8,0.35)
  \psline{-<}(0.2,-0.027)(0.3,-0.04)
  \psline[linewidth=0.9pt,linecolor=black,border=2.2pt,arrows=->,arrowscale=1.5,arrowinset=0.15](0.65,0.1)(0.65,1.0)
  \psline[linewidth=0.9pt,linecolor=black,border=2.2pt,arrows=->,arrowscale=1.5,arrowinset=0.15](0.95,0.2)(0.95,1.0)
  \psline[linewidth=0.9pt,linecolor=black,border=2.2pt,arrows=->,arrowscale=1.5,arrowinset=0.15](0.65,0.1)(0.65,1.0)
  \psline[linewidth=0.9pt,linecolor=black,border=2.2pt,arrows=->,arrowscale=1.5,arrowinset=0.15](0.95,0.2)(0.95,1.0)
  \rput[bl]{0}(-0.15,-0.4){$x_{\bf gh}$}
  \rput[bl]{0}(0.45,-1.05){$a_{\bf g}$}
  \rput[bl]{0}(0.95,-1){$b_{\bf h}$}
  \rput[bl]{0}(1.05,0.95){$^{\bf \bar{h} \bar{g}}b_{\bf h}$}
  \rput[bl]{0}(-0.0,0.9){$^{\bf \bar{h}}a_{\bf g}$}
  \endpspicture
&=&  \sum_{\substack{ x,c \in \mathcal{C}_\mathbf{gh} \\ \mu, \nu} } \frac{ d_x \theta_x  }{ U_\mathbf{gh}(\bar{x},x;0) } \sqrt{ \frac{d_{c}}{ d_{a} d_{b} } } \left[ U_{\bf \bar{h}\bar{g}}(\,^{\bf \bar{h}}a,\,^{\bf \bar{h}\bar{g}}b;c) \right]_{\mu \nu}
\pspicture[shift=-1.4](-0.5,-1.3)(1.5,1.5)
\small
  \psset{linewidth=0.9pt,linecolor=black,arrowscale=1.5,arrowinset=0.15}
  \psline(0.65,-0.7)(0.65,-0.35)
  \psline(0.95,-0.7)(0.95,-0.35)
  \psline(0.8,-0.2)(0.8,0.4)
  \psline(0.8,-0.2)(0.65,-0.35)
  \psline(0.8, 0.75)(0.65, 0.95)
  \psline(0.8, 0.75)(0.95, 0.95)
  \psline(0.8,-0.2)(0.95,-0.35)
  \psellipse[linewidth=0.9pt,linecolor=black,border=2.1pt](0.4,0.3)(0.8,0.35)
  \psline{-<}(0.2,-0.027)(0.3,-0.04)
  \psline[border=2.1pt](0.8, 0.4)(0.8,0.75)
   \psline{->}(0.8,0.4)(0.8,0.5)
\psline[linewidth=0.9pt,linecolor=black,border=2.1pt](0.65,0.95)(0.65,1.3)
\psline[linewidth=0.9pt,arrows=->,arrowscale=1.5, arrowinset=0.15](0.65, 1.15)(0.65,1.23)
\psline[linewidth=0.9pt,linecolor=black,border=2.1pt](0.95, 0.95)(0.95,1.3)
\psline[linewidth=0.9pt,arrows=->,arrowscale=1.5, arrowinset=0.15](0.95,1.15)(0.95,1.23)
    \rput[bl]{0}(0.9,0.7){\mbox{\tiny $\nu$}}
    \rput[bl]{0}(0.9,-0.2){\mbox{\tiny $\mu$}}
    \rput[bl]{0}(0.5,0.25){$c$}
  \rput[bl]{0}(-0.15,-0.4){$x_{\bf gh}$}
  \rput[bl]{0}(0.45,-1.05){$a_{\bf g}$}
  \rput[bl]{0}(0.95,-1){$b_{\bf h}$}
  \rput[bl]{0}(1.05,0.95){$^{\bf \bar{h} \bar{g}}b_{\bf h}$}
  \rput[bl]{0}(-0.0,0.9){$^{\bf \bar{h}}a_{\bf g}$}
\endpspicture	
\notag \\
&=&  \sum_{c_{\bf gh}, \mu, \nu}  \sqrt{ \frac{d_{c}}{ d_{a} d_{b} } } \left[ U_{\bf \bar{h}\bar{g}}(\,^{\bf \bar{h}}a,\,^{\bf \bar{h}\bar{g}}b;c) \right]_{\mu \nu}  \frac{ \mathcal{D}_\mb{0} \Theta_{\bf 0} }{ \eta_c({\bf gh , \bar{h} \bar{g} }) \theta_{c} }
 \pspicture[shift=-1.3](-0.3,-1.05)(1.8,1.4)
  \small
  \psset{linewidth=0.9pt,linecolor=black,arrowscale=1.5,arrowinset=0.15}
  \psline{->}(0.7,0)(0.7,0.45)
  \psline(0.7,0)(0.7,0.55)
  \psline(0.7,0.55) (0.25,1)
  \psline{->}(0.7,0.55)(0.3,0.95)
  \psline(0.7,0.55) (1.15,1)
  \psline{->}(0.7,0.55)(1.1,0.95)
  \rput[bl]{0}(0.1,0.2){$c_{\bf gh}$}
  \rput[br]{0}(1.4,1.1){$^{\bf \bar{h} \bar{g}}b_{\bf h}$}
  \rput[bl]{0}(-0.1,1.05){$^{\bf \bar{h}}a_{\bf g}$}
  \psline(0.7,0) (0.25,-0.45)
  \psline{-<}(0.7,0)(0.35,-0.35)
  \psline(0.7,0)(1.15,-0.45)
  \psline{-<}(0.7,0)(1.05,-0.35)
  \rput[br]{0}(1.4,-0.75){$b_{\bf h}$}
  \rput[bl]{0}(0.1,-0.8){$a_{\bf g}$}
\scriptsize
  \rput[bl]{0}(0.85,0.4){$\nu$}
  \rput[bl]{0}(0.85,-0.03){$\mu$}
  \endpspicture
\notag \\
&=& \frac{ \mathcal{D}_\mb{0} \Theta_{\bf 0} }{ \theta_{a} \theta_{b} \eta_{a}({\bf gh , \bar{h} \bar{g} }) \eta_{b}({\bf gh , \bar{h} \bar{g} }) \eta_{a}({\bf g , h }) \eta_{b}({\bf h , \,^{\bar{h}} g }) }
 \pspicture[shift=-1.4](-0.3,-0.95)(1.6,1.8)
  \small
   \psset{linewidth=0.9pt,linecolor=black,arrowscale=1.5,arrowinset=0.15}
  \psarc[linewidth=0.9pt,linecolor=black,border=0pt](0.8,0.7){0.4}{135}{240}
  \psarc[linewidth=0.9pt,linecolor=black,border=0pt](0.4,0.7){0.4}{-45}{60}
  \psarc[linewidth=0.9pt,linecolor=black,border=0pt](0.8,1.39282){0.4}{180}{240}
  \psarc[linewidth=0.9pt,linecolor=black,border=0pt](0.4,1.39282){0.4}{-40}{0}
  \psarc[linewidth=0.9pt,linecolor=black,border=0pt](0.8,0.00718){0.4}{135}{180}
  \psarc[linewidth=0.9pt,linecolor=black,border=0pt](0.4,0.00718){0.4}{0}{60}
    \psline{->}(0.8,-0.3)(0.8,0.05)
    \psline{->}(0.4,-0.3)(0.4,0.05)
    \psline{>-}(0.8,1.35)(0.8,1.7)
    \psline{>-}(0.4,1.35)(0.4,1.7)
  \rput[bl]{0}(0.15,-0.65){$a_{\bf g}$}
  \rput[bl]{0}(0.8,-0.6){$b_{\bf h}$}
  \rput[bl]{0}(0.95,1.35){$^{\bf \bar{h} \bar{g}}b_{\bf h}$}
  \rput[bl]{0}(-0.25,1.3){$^{\bf \bar{h}}a_{\bf g}$}
  \endpspicture
,
\label{eq:G-crossed-loop-braid}
\end{eqnarray}
which is obtained by using Eq.~(\ref{eq:loop-twist}), the relation
\begin{equation}
\left[ U_{\bf \bar{k}}(\,^{\bf \bar{k}}a,\,^{\bf \bar{k}}b;\,^{\bf \bar{k}}c) \right]_{\mu \nu} = \frac{ \eta_{c}({\bf k , \bar{k} }) }{ \eta_{a}({\bf k , \bar{k} }) \eta_{b}({\bf k , \bar{k} })  } \left[ U_{\bf k}(a,b;c)^{-1} \right]_{\mu \nu}
,
\end{equation}
[which is the sliding move consistency Eq.~(\ref{eq:U_eta_consistency}) with ${\bf l = \bar{k}}$,] and the (inverse of the) the ribbon property given in Eq.~(\ref{eq:G-crossed_ribbon}).

Finally, when $^{\bf h}a_{\bf g} = a_{\bf g}$ (which requires ${\bf gh}={\bf hg}$), we have the relation
\begin{eqnarray}
\sum_{x \in \mathcal{C}_{\bf gh}} \frac{ d_{x} \theta_{x}  \eta_{\bar{x}}(\mathbf{g},\mathbf{h}) }{ U_{\bf gh}( \bar{x},x; 0)}
\pspicture[shift=-1.7](-1.0,-1.6)(1.5,2.2)
\small
  \psset{linewidth=0.9pt,linecolor=black,arrowscale=1.5,arrowinset=0.15}
  \psline(0.65,-0.7)(0.65,-0.15)
  \psline(0.65,0.15)(0.65,1.3)
  \psline(0.95,-1.3)(0.95,-0.05)
  \psline(0.95,0.18)(0.95,1.9)
  \psellipse[linewidth=0.9pt,linecolor=black,border=0](0.4,0.3)(0.8,0.35)
  \psline{-<}(0.2,-0.027)(0.3,-0.04)
\psline[linewidth=0.9pt,linecolor=black,border=2.2pt,arrows=->,arrowscale=1.5,arrowinset=0.15](0.65,0.1)(0.65,1.2)
\psline[linewidth=0.9pt,linecolor=black,border=2.2pt,arrows=->,arrowscale=1.5,arrowinset=0.15](0.95,0.2)(0.95,1.6)
\psline[linewidth=0.9pt,linecolor=black,border=2.2pt,arrows=->,arrowscale=1.5,arrowinset=0.15](0.95,-1.1)(0.95,-0.8)
\psarc(0,1.3){0.65}{0}{180}
\psline(-0.65,1.3)(-0.65,-0.7)
\psarc(0,-0.7){0.65}{-180}{-0.001}
  \rput[bl]{0}(-0.15,-0.4){$x_{\bf gh}$}
  \rput[tl]{0}(0.2,1.3){$a_{\bf g}$}
  \rput[tl]{0}(0.95,-1.35){$b_{\bf h}$}
  \rput[tl]{0}(1.1,1.7){$^{\bf \bar{g}}b_{\bf h}$}
  \endpspicture
&=& \sum_{x \in \mathcal{C}_{\bf gh}} \frac{ d_{x} \theta_{x}  \eta_{\bar{x}}(\mathbf{g},\mathbf{h}) }{ U_{\bf gh}(\bar{x},x; 0)} \frac{ U_{\bf g}(x, \bar{x}; 0) }{\eta_{a}({\bf gh,\bar{h}\bar{g}})} \frac{\eta_{x}({\bf g,\bar{g}})}{ U_{\bf gh}(a, \bar{a}; 0) }
\pspicture[shift=-1.7](-1.7,-1.6)(1.5,2.2)
\small
  \psellipse[linewidth=0.9pt,linecolor=black,border=0](0.4,0.3)(0.8,0.35)
  \psset{linewidth=0.9pt,linecolor=black,arrowscale=1.5,arrowinset=0.15}
  \psline[border=2.2pt](-0.15,-0.7)(-0.15,0.4)
  \psline(-0.15, 0.7)(-0.15,1.3)
  \psline(0.95,-1.3)(0.95,-0.05)
  \psline(0.95,0.18)(0.95,1.9)
  \psline{-<}(0.2,-0.027)(0.3,-0.04)
\psline[linewidth=0.9pt,linecolor=black,border=2.2pt,arrows=->,arrowscale=1.5, arrowinset=0.15](-1.45,0.65)(-1.45,1.2)
\psline[linewidth=0.9pt,linecolor=black,border=2.2pt,arrows=->,arrowscale=1.5,arrowinset=0.15](0.95,0.2)(0.95,1.6)
\psline[linewidth=0.9pt,linecolor=black,border=2.2pt,arrows=->,arrowscale=1.5,arrowinset=0.15](0.95,-1.1)(0.95,-0.8)
\psarc(-0.8,1.3){0.65}{0}{180}
\psline(-1.45,1.3)(-1.45,-0.7)
\psarc(-0.8,-0.7){0.65}{-180}{-0.001}
  \rput[bl]{0}(0.0,-0.4){$x_{\bf gh}$}
  \rput[tl]{0}(-1.3,1.3){$a_{\bf g}$}
  \rput[tl]{0}(0.95,-1.35){$b_{\bf h}$}
  \rput[tl]{0}(1.1,1.7){$^{\bf \bar{g}}b_{\bf h}$}
  \endpspicture
\notag \\
&=& \frac{ \mathcal{D}_\mb{0} \Theta_{\bf 0} \eta_{b}({\bf g ,\bar{g} }) }{ \theta_{a} \theta_{b} U_{\bf h}(a, \bar{a}; 0) \eta_{a}({\bf gh , \bar{h} \bar{g} }) \eta_{b}({\bf gh , \bar{h} \bar{g} }) \eta_{a}({\bf g , h }) \eta_{b}({\bf h , g }) }
\pspicture[shift=-1.2](-0.5,-1.1)(1.5,1.5)
\small
  \psset{linewidth=0.9pt,linecolor=black,arrowscale=1.5,arrowinset=0.15}
  \psline(0.75,-0.7)(0.75,-0.15)
  \psline(0.75,0.15)(0.75,1.3)
  \psellipse[linewidth=0.9pt,linecolor=black,border=0](0.4,0.3)(0.8,0.35)
  \psline{-<}(0.2,-0.027)(0.3,-0.04)
\psline[linewidth=0.9pt,linecolor=black,border=2.2pt,arrows=->,arrowscale=1.5,arrowinset=0.15](0.75,0.1)(0.75,1.05)
\psline[linewidth=0.9pt,linecolor=black,border=2.2pt,arrows=->,arrowscale=1.5,arrowinset=0.15](0.75,-0.4)(0.75,-0.25)
\rput[bl]{0}(-0.15,-0.35){$a_\mb{g}$}
  \rput[tl]{0}(0.9,1.1){$^{\bf \bar{g}}b_\mathbf{h}$}
  \rput[tl]{0}(0.7,-0.8){$b_\mathbf{h}$}
\endpspicture
.
\label{eq:diag-STS-TST}
\end{eqnarray}
To obtain these relations, we used Eqs.~(\ref{eq:G-crossed_loop_slide}) and (\ref{eq:loop-twist}) in both lines, though, in the second line, we first applied Eq.~(\ref{eq:G-crossed-loop-braid}). We emphasize that the individual diagrams in this equation evaluate to zero, unless $^{\bf h}a_{\bf g}=a_{\bf g}$, $^{\bf g}b_{\bf h} = b_{\bf h}$, and $^{\bf g}x_{\bf gh} = \,^{\bf h}x_{\bf gh} = x_{\bf gh}$. In particular, the sum here can be taken to be over $x_{\bf gh} \in \mathcal{C}_{\bf gh}^{\bf g,h} = \mathcal{C}_{\bf gh}^{\bf g} \cap \mathcal{C}_{\bf gh}^{\bf h}$, the topological charges in $\mathcal{C}_{\bf gh}$ that are both ${\bf g}$-invariant and ${\bf h}$-invariant, where we define the invariant topological charge subsets
\begin{equation}
\mathcal{C}_{\bf g}^{\bf h} = \{ \,\, a \in \mathcal{C}_{\bf g} \,\, | \,\, ^{\bf h}a = a \,\, \}
.
\end{equation}

Taking the trace of Eq.~(\ref{eq:diag-STS-TST}), i.e. closing the $b$-line back on itself (which requires $^{\bf g}b_{\bf h} = b_{\bf h}$), we finally obtain the important relation
\begin{equation}
\sum_{x \in \mathcal{C}_{\bf gh}} \eta_{a}({\bf g,h}) \theta_{a_{\bf g}} \frac{S_{a_{\bf g} x_{\bf gh}}}{U_{\bf gh}(a, \bar{a}; 0)}  \eta_{x}({\bf gh, \bar{g}}) \theta_{x_{\bf gh}} \frac{S_{x_{\bf gh} b_{\bf h}}}{U_{\bf h}(x, \bar{x}; 0)}  \eta_{b}({\bf h, \bar{h}\bar{g} }) \theta_{b_{\bf h}} = \Theta_{\bf 0} \frac{S_{a_{\bf g} b_{\bf h}}}{U_{\bf h}(a, \bar{a}; 0)}
.
\label{eq:TSTST-S}
\end{equation}
\end{widetext}
In order to manipulate the trace of Eq.~(\ref{eq:diag-STS-TST}) into this form, we have used Eqs.~(\ref{eq:U_k_relation}), (\ref{eqn:etasym}), and (\ref{eq:S_4}), together with the relations
\begin{eqnarray}
\eta_{a}({\bf k , l }) \eta_{\bar{a}}({\bf k , l }) &=& \frac{ U_{\bf kl}(a,\bar{a};0)} {U_{\bf k}(a,\bar{a};0) U_{\bf l}(\,^{\bf \bar{k}}a,\,^{\bf \bar{k}}\bar{a};0)}
,
\label{eq:sliding_consistency_c0}
\\
\eta_{^{\bf g}x}({\bf g ,h }) \eta_{x}({\bf \bar{g} , gh })  &=&  \eta_{x}({\bf \bar{g} , g })  ,
\label{eq:eta_relation_1}
\\
\eta_{^{\bf \bar{g}}b}({\bf h , \bar{h}\bar{g} }) \eta_{b}({\bf g , \bar{g} })&=& \eta_{b}({\bf gh , \bar{h}\bar{g} }) \eta_{b}({\bf g , h})  ,
\label{eq:eta_relation_2}
\end{eqnarray}
the first of which is the sliding move consistency Eq.~(\ref{eq:U_eta_consistency}) with $c=0$, while the second and third are special cases of Eq.~(\ref{eq:eta_consistency}).

We conclude this section by noting that a number of additional $G$-crossed gauge invariant quantities will naturally arise in the context of modular transformations of the $G$-crossed theory and gauging the symmetry of theory. As these quantities would be somewhat out of context and mysterious here, we leave their discussion for the subsequent Secs.~\ref{sec:G-Crossed_Modularity} and \ref{sec:gauging}.

\section{$G$-Crossed Modularity}
\label{sec:G-Crossed_Modularity}

An important property of a topological phase of matter is the ground state degeneracy when the system inhabits manifolds with different topologies. For a $2+1$ dimensional topological phase, the ground state degeneracy will depend on the genus $g$ of the surface inhabited by the system and the topological charge values of the quasiparticles (and boundaries) of the system. More generally, it is important that the theory describing a topological phase is well-defined and consistent for the system on arbitrary topologies. In other words, the topological properties of the system are described by a TQFT. In terms of the BTC theory, this is achieved by requiring the theory to be a modular tensor category (MTC), i.e. to have unitary $S$-matrix. In this case, the $S$-matrix and $T$-matrix provide a projective representation of the modular transformations for the system on the torus. (More general modular transformations for the system on a manifold of arbitrary topology and quasiparticle content can similarly be defined in terms of the MTC properties.)

We wish to establish a similar notion of modularity for $G$-crossed BTCs, which allows one to relate the theory to a $G$-crossed TQFT that describes the topological phase with defects on arbitrary 2D surfaces. The $G$-crossed extended defect theory $\ext{C}{G}$ admits a richer set of possibilities, as defect branch lines can wrap the nontrivial cycles of surfaces with genus $g>0$, thus giving rise to ``defect sectors.'' For $G$-crossed modularity, we will require that the set of ${\bf g}$-defect topological charges $\mathcal{C}_{\bf g}$ is finite for each ${\bf g} \in G$ (though not necessarily that $G$ is finite or even discrete). Some special cases of $G$-crossed modular transformations have been studied recently in \Refs{Zaletel_arxiv,HungPRB2014}.

In this section, we will develop an understanding of the defect sectors and their associated topological ground state degeneracies. We also establish the notion of $G$-crossed modularity and the corresponding modular transformations for the system when it includes defect sectors. The topological ground state degeneracies of the defect sectors, together with the $G$-crossed modular transformations, can provide valuable information about the symmetry-enriched topological order.

\subsection{$G$-Crossed Verlinde Formula and $\omega_{a}$-Loops}

Before considering the $G$-crossed theory and modular transformations for a system on surfaces with genus $g>0$, we first investigate some properties that are closely related to modularity, namely the Verlinde formula and $\omega_{a}$-loops. For this, we begin with the minimal assumption that the original theory $\mathcal{C}_{\bf 0}$ is a MTC, which is to say that its $S$-matrix is unitary. From this assumption and Eqs.~(\ref{eq:preVerlinde3}) and (\ref{eq:preVerlinde4}), we obtain the formula
\begin{eqnarray}
N_{a_{\bf g} b_{\bf \bar{g}} }^{c_{\bf 0}} &=& \sum_{x_{\bf 0} \in \mathcal{C}_{\bf 0}^{\bf g}} \frac{S_{a_{\bf g} x_{\bf 0}} S_{b_{\bf \bar{g}} x_{\bf 0}} S^{\ast}_{c_{\bf 0} x_{\bf 0}} }{S_{0 x_{\bf 0}}} \eta_{x}({\bf g, \bar{g} }) \\
&=& \sum_{x_{\bf 0} \in \mathcal{C}_{\bf 0}^{\bf g}} \frac{ S_{x_{\bf 0} a_{\bf g}} S_{ x_{\bf 0} b_{\bf \bar{g}}} S^{\ast}_{ x_{\bf 0} c_{\bf 0}} }{S_{ x_{\bf 0} 0}} \eta_{\bar{x}}({\bf g, \bar{g} })
,
\end{eqnarray}
where the sums in these expressions are over the subset $\mathcal{C}_{\bf 0}^{\bf g}$ of ${\bf g}$-invariant topological charges in $\mathcal{C}_{\bf 0}$. (Actually, we could let the sums go over the entire $\mathcal{C}_{\bf 0}$ if we consider the $S$-matrices to be equal to zero when $^{\bf g}x \neq x $.)

Setting $c=0$ in these expressions and using Eqs.~(\ref{eq:S_2}) and (\ref{eq:S_3}), we obtain
\begin{equation}
\label{eq:S_0_ortho}
\delta_{ a_{\bf g} a'_{\bf g} } =  \sum_{x_{\bf 0} \in \mathcal{C}_{\bf 0}^{\bf g} } S_{a_{\bf g} x_{\bf 0}} S^{\ast}_{ a'_{\bf g} x_{\bf 0}}
= \sum_{x_{\bf 0} \in \mathcal{C}_{\bf 0}^{\bf g} }  S_{ x_{\bf 0} a_{\bf g}} S^{\ast}_{ x_{\bf 0} a'_{\bf g}}
.
\end{equation}

Now, we can use Eq.~(\ref{eq:S_0_ortho}) with Eqs.~(\ref{eq:preVerlinde3}) and (\ref{eq:preVerlinde4}) to obtain the $G$-crossed Verlinde formula
\begin{eqnarray}
N_{a_{\bf g} b_{\bf h} }^{c_{\bf gh}} &=& \sum_{x_{\bf 0} \in \mathcal{C}_{\bf 0}^{\bf g,h}} \frac{S_{a_{\bf g} x_{\bf 0}} S_{b_{\bf h} x_{\bf 0}} S^{\ast}_{c_{\bf gh} x_{\bf 0}} }{S_{0 x_{\bf 0}}} \eta_{x}({\bf \bar{h}, \bar{g} })
\label{eq:G-crossed_Verlinde_1}
\\
&=& \sum_{x_{\bf 0} \in \mathcal{C}_{\bf 0}^{\bf g,h}} \frac{ S_{x_{\bf 0} a_{\bf g}} S_{ x_{\bf 0} b_{\bf h}} S^{\ast}_{ x_{\bf 0} c_{\bf gh}} }{S_{ x_{\bf 0} 0}} \eta_{\bar{x}}({\bf g, h})
,
\label{eq:G-crossed_Verlinde_2}
\end{eqnarray}
where $\mathcal{C}_{\bf 0}^{\bf g,h} = \mathcal{C}_{\bf 0}^{\bf g} \cap \mathcal{C}_{\bf 0}^{\bf h}$ is the subset of topological charges in $\mathcal{C}_{\bf 0}$ that are both ${\bf g}$-invariant and ${\bf h}$-invariant.

Moreover, we may use these properties to define $\omega_{a_{\bf g}}$-loops, which are linear combinations of loops of topological charge lines that act as topological charge projectors on the collection of topological charge lines passing through them. [These should not to be confused with the $\omega_{a}({\bf g,h})$ phase factors associated with symmetry fractionalization in Sec.~\ref{sec:symmetryfrac},
nor should $\omega_{a_{\bf g}}$ be confused with an element of $\mathcal{C}_{\bf g}$.] Similar to the definition in a MTC, we can define the $\omega_{a_{\bf g}}$-loop enclosing a single defect line for a $G$-crossed theory by
\begin{equation}
\psscalebox{.7}{
\pspicture[shift=-1.0](-0.5,-0.9)(1.5,1.5)
\small
  \psset{linewidth=0.9pt,linecolor=black,arrowscale=1.5,arrowinset=0.15}
  \psline(0.75,-0.7)(0.75,-0.15)
  \psline(0.75,0.15)(0.75,1.3)
  \psellipse[linewidth=0.9pt,linecolor=black,border=0](0.4,0.3)(0.8,0.35)
  \psline{-<}(0.2,-0.027)(0.3,-0.04)
\psline[linewidth=0.9pt,linecolor=black,border=2.2pt,arrows=->,arrowscale=1.5,
arrowinset=0.15](0.75,0.1)(0.75,1.05)
\rput[bl]{0}(-0.15,-0.4){$\omega_{a_\mb{g}}$}
  \rput[tl]{0}(0.9,1.1){$b_\mathbf{g}$}
\endpspicture
}
= \sum_{x_{\bf 0} \in \mathcal{C}_{\bf 0}^{\bf g}} S_{0 a_{\bf g}} S^{\ast}_{x_{\bf 0} a_{\bf g}}
\psscalebox{.7}{
\pspicture[shift=-1.0](-0.5,-0.9)(1.5,1.5)
\small
  \psset{linewidth=0.9pt,linecolor=black,arrowscale=1.5,arrowinset=0.15}
  \psline(0.75,-0.7)(0.75,-0.15)
  \psline(0.75,0.15)(0.75,1.3)
  \psellipse[linewidth=0.9pt,linecolor=black,border=0](0.4,0.3)(0.8,0.35)
  \psline{-<}(0.2,-0.027)(0.3,-0.04)
\psline[linewidth=0.9pt,linecolor=black,border=2.2pt,arrows=->,arrowscale=1.5,
arrowinset=0.15](0.75,0.1)(0.75,1.05)
\rput[bl]{0}(-0.1,-0.4){$x_\mb{0}$}
  \rput[tl]{0}(0.9,1.1){$b_\mathbf{g}$}
\endpspicture
}
=\delta_{a_{\bf g} b_{\bf g}}
\psscalebox{.7}{
\pspicture[shift=-1.0](0.4,-0.9)(1.5,1.5)
\small
  \psset{linewidth=0.9pt,linecolor=black,arrowscale=1.5,arrowinset=0.15}
  \psline(0.75,-0.7)(0.75,1.3)
\psline[linewidth=0.9pt,linecolor=black,border=2.2pt,arrows=->,arrowscale=1.5,
arrowinset=0.15](0.75,0.1)(0.75,1.05)
  \rput[tl]{0}(0.9,1.1){$b_\mathbf{g}$}
 \endpspicture
}
,
\label{eq:omega_a-loop}
\end{equation}
where the first equality is a definition, and the last step used Eqs.~(\ref{eq:G-crossed_loop_removal}) and (\ref{eq:S_0_ortho}) to show that act on ${\bf g}$-defects as projectors that distinguish between the different topological charge values of ${\bf g}$-defects. Eq.~(\ref{eq:omega_a-loop}) establishes our previous claim in Sec.~\ref{sec:Topo_Distinct_Defects} that, when the original theory $\mathcal{C}_{\bf 0}$ is modular, there are physical processes involving the ${\bf g}$-invariant topological charges in $\mathcal{C}_{\bf 0}$ which are able to distinguish between the distinct types of ${\bf g}$-defects.

When an $\omega_{a_{\bf g}}$-loop (as previously defined) encloses multiple defect lines, it is not quite equal to the desired projection operator of the collective charge of the enclosed defect charge lines. In particular, for $n$ defects with topological charges $b_{j} \in \mathcal{C}_{{\bf g}_{j}}$ for $j=1,\ldots,n$, respectively, with $\prod_{j=1}^{n}{\bf g}_{j} ={\bf g}$, the collective charge projection is
\begin{equation}
\Pi_{a_{\bf g}}^{(1 \ldots n)} = \sum_{\substack{ e_{2},\ldots,e_{n-1} \\ \mu_{2},\ldots,\mu_{n} }} \sqrt{ \frac{d_{a}}{ d_{b_{1}} \cdots d_{b_{n}} } }
\psscalebox{.8}{
 \pspicture[shift=-2.05](-0.35,-1.9)(2.5,2.0)
  \small
  \psset{linewidth=0.9pt,linecolor=black,arrowscale=1.5,arrowinset=0.15}
  \psline(0.0,1.75)(1,0.5)
  \psline(2.0,1.75)(1,0.5)
  \psline(0.4,1.25)(0.8,1.75)
   \psline{->}(0.4,1.25)(0.1,1.625)
   \psline{->}(0.4,1.25)(0.7,1.625)
   \psline{->}(1,0.5)(1.9,1.625)
   \psline{->}(1,0.5)(0.5,1.125)
   \rput[bl]{0}(-0.15,1.85){$b_1$}
   \rput[bl]{0}(0.75,1.85){$b_2$}
   \rput[bl]{0}(1.95,1.85){$b_n$}
\rput[bl](1.25,1.85){$\cdots$}
\rput{-45}(0.9,1.05){$\cdots$}
   \rput[bl]{0}(0.25,0.65){$e_2$}
  \psset{linewidth=0.9pt,linecolor=black,arrowscale=1.5,arrowinset=0.15}
  \psline(0.0,-1.45)(1,-0.2)
  \psline(2.0,-1.45)(1,-0.2)
  \psline(0.4,-0.95)(0.8,-1.45)
   \psline{-<}(0.4,-0.95)(0.1,-1.325)
   \psline{-<}(0.4,-0.95)(0.7,-1.325)
   \psline{-<}(1,-0.2)(1.9,-1.325)
   \psline{-<}(1,-0.2)(0.5,-0.825)
   \rput[bl]{0}(-0.15,-1.75){$b_1$}
   \rput[bl]{0}(0.75,-1.75){$b_2$}
   \rput[bl]{0}(1.95,-1.75){$b_n$}
   \rput[bl]{0}(0.25,-0.6){$e_2$}
   \rput[bl](1.25,-1.75){$\cdots$}
   \rput{45}(0.9,-0.75){$\cdots$}
  \psline(1,-0.2)(1,0.5)
  \psline{->}(1,-0.2)(1,0.3)
   \rput[bl]{0}(1.15,0.0){$a$}
 \scriptsize
   \rput[bl]{0}(0.1,1){$\mu_2$}
   \rput[bl]{0}(0.05,-0.95){$\mu_2$}
   \rput[bl]{0}(0.55,0.35){$\mu_n$}
   \rput[bl]{0}(0.6,-0.2){$\mu_n$}
  \endpspicture
}
.
\label{eq:n_line_c_projector}
\end{equation}
In order to be equal this $n$ defect line projection, we define the $n$ defect line $\omega_{a_{\bf g}}$-loop to be
\begin{widetext}
\begin{equation}
\Pi_{a_{\bf g}}^{(1 \ldots n)} =
\psscalebox{.8}{
\pspicture[shift=-2.5](-1.5,-2)(2.8,2.4)
  \small
  \psset{linewidth=0.9pt,linecolor=black,arrowscale=1.5,arrowinset=0.15}
  \psline(0.0,1.75)(0.0,-1.0)
  \psline(2.0,1.75)(2.0,-1.0)
  \psline(0.8,1.75)(0.8,-1.0)
   \psline{->}(0.0,1.25)(0.0,1.625)
   \psline{->}(0.8,1.25)(0.8,1.625)
   \psline{->}(2.0,1.25)(2.0,1.625)
   \rput[bl]{0}(-0.2,1.85){$b_1$}
   \rput[bl]{0}(0.6,1.85){$b_2$}
   \rput[bl]{0}(1.25,1.85){$\ldots$}
   \rput[bl]{0}(1.8,1.85){$b_n$}
  \psline(-0.5,1.25)(2.5,0.75)
  \psline[linewidth=0.9pt,linecolor=black,border=0.1](-0.5,-0.25)(2.5,0.25)
  \pscurve(2.5,0.75)(2.6,0.5)(2.5,0.25)
  \pscurve(-0.5,-0.25)(-0.8,0.5)(-0.5,1.25)
  \psline[linewidth=0.9pt,linecolor=black,border=0.1](0.0,0.5)(0.0,1.25)
  \psline[linewidth=0.9pt,linecolor=black,border=0.1](0.8,0.5)(0.8,1.25)
  \psline[linewidth=0.9pt,linecolor=black,border=0.1](2.0,0.5)(2.0,1.25)
  \psset{linewidth=0.9pt,linecolor=black,arrowscale=1.4,arrowinset=0.15}
  \psline{->}(-0.8,0.6)(-0.8,0.65)
  \rput[bl]{0}(-1.4,0.4){$\omega_{a_{\bf g}}$}
\endpspicture
}
= \sum_{x_{\bf 0} \in \mathcal{C}_{\bf 0}^{\bf g}} \eta_{\overline{x_{\bf 0}}}({\bf g}_{1},\ldots,{\bf g}_{n}) S_{0 a_{\bf g}} S^{\ast}_{x_{\bf 0} a_{\bf g}}
\psscalebox{.8}{
\pspicture[shift=-2.5](-1.5,-2)(2.8,2.4)
  \small
  \psset{linewidth=0.9pt,linecolor=black,arrowscale=1.5,arrowinset=0.15}
  \psline(0.0,1.75)(0.0,-1.0)
  \psline(2.0,1.75)(2.0,-1.0)
  \psline(0.8,1.75)(0.8,-1.0)
   \psline{->}(0.0,1.25)(0.0,1.625)
   \psline{->}(0.8,1.25)(0.8,1.625)
   \psline{->}(2.0,1.25)(2.0,1.625)
   \rput[bl]{0}(-0.2,1.85){$b_1$}
   \rput[bl]{0}(0.6,1.85){$b_2$}
   \rput[bl]{0}(1.25,1.85){$\ldots$}
   \rput[bl]{0}(1.8,1.85){$b_n$}
  \psline(-0.5,1.25)(2.5,0.75)
  \psline[linewidth=0.9pt,linecolor=black,border=0.1](-0.5,-0.25)(2.5,0.25)
  \pscurve(2.5,0.75)(2.6,0.5)(2.5,0.25)
  \pscurve(-0.5,-0.25)(-0.8,0.5)(-0.5,1.25)
  \psline[linewidth=0.9pt,linecolor=black,border=0.1](0.0,0.5)(0.0,1.25)
  \psline[linewidth=0.9pt,linecolor=black,border=0.1](0.8,0.5)(0.8,1.25)
  \psline[linewidth=0.9pt,linecolor=black,border=0.1](2.0,0.5)(2.0,1.25)
  \psset{linewidth=0.9pt,linecolor=black,arrowscale=1.4,arrowinset=0.15}
  \psline{->}(-0.8,0.6)(-0.8,0.65)
  \rput[bl]{0}(-1.25,0.4){$x_{\bf 0}$}
\endpspicture
}
,
\end{equation}
\end{widetext}
where we have defined
\begin{equation}
\label{eq:eta_n_generalization}
\eta_{x}({\bf g}_{1},\ldots,{\bf g}_{n}) = \prod_{j=2}^{n} \eta_{x}({\bf g}_{1}\cdots {\bf g}_{j-1},{\bf g}_{j})
.
\end{equation}
We note that the quantity $\eta_{x}({\bf g}_{1},\ldots,{\bf g}_{n})$ does not depend on the particular values of topological charge $b_{j}$ carried by the defects, only their group element labels ${\bf g}_{j}$. The fact that the group element labels ${\bf g}_{j}$ of the defects enter the definition of the $n$ defect line $\omega_{a_{\bf g}}$-loop is not problematic, since the defects are extrinsic objects with definite values of ${\bf g}_{j}$ (superpositions of different values of ${\bf g}_{j}$ are not possible). Additionally, Eq.~(\ref{eq:eta_consistency}) guarantees that this quantity is independent of the order of fusion used in the fusion tree of the defect charge lines, i.e. it commutes with the $F$-moves. Thus, the $n$ defect line $\omega_{a_{\bf g}}$-loop defined here applies for all configurations of ${\bf g}_{1},\ldots,{\bf g}_{n}$ defects, including when there are superpositions of topological charge values $b_{j} \in \mathcal{C}_{{\bf g}_{j}}$.

It is worth re-emphasizing that, so far, we have only assumed that $\mathcal{C}_{\bf 0}$ is modular (i.e. has unitary $S$-matrix), and made no further assumption about the $S$-matrix of the extended $G$-crossed theory. The results here seem to suggest that it may be the case that requiring $\mathcal{C}_{\bf 0}$ to be modular would be sufficient to obtain a notion of modularity of the $G$-crossed theory. Indeed, by combining theorems from \Refs{Etingof05,kirillov2004} that relate $\mathcal{C}_{\bf 0}$ and $\mathcal{C}_{G}^{\times}$ to the theory $\gauged{C}{G}$ obtained by gauging the symmetry, one has the property that $\mathcal{C}_{\bf 0}$ is a MTC if and only if $\mathcal{C}_{G}^{\times}$ is $G$-crossed modular (both of which are true if and only if $\gauged{C}{G}$ is a MTC). We now define the notion of a $G$-crossed BTC being $G$-crossed modular in the following subsection.

\subsection{Torus Degeneracy and $G$-Crossed Modular Transformations}
\label{sec:higher_genus}

\begin{figure}[t!]
	\centering
	\includegraphics[width=0.59\columnwidth]{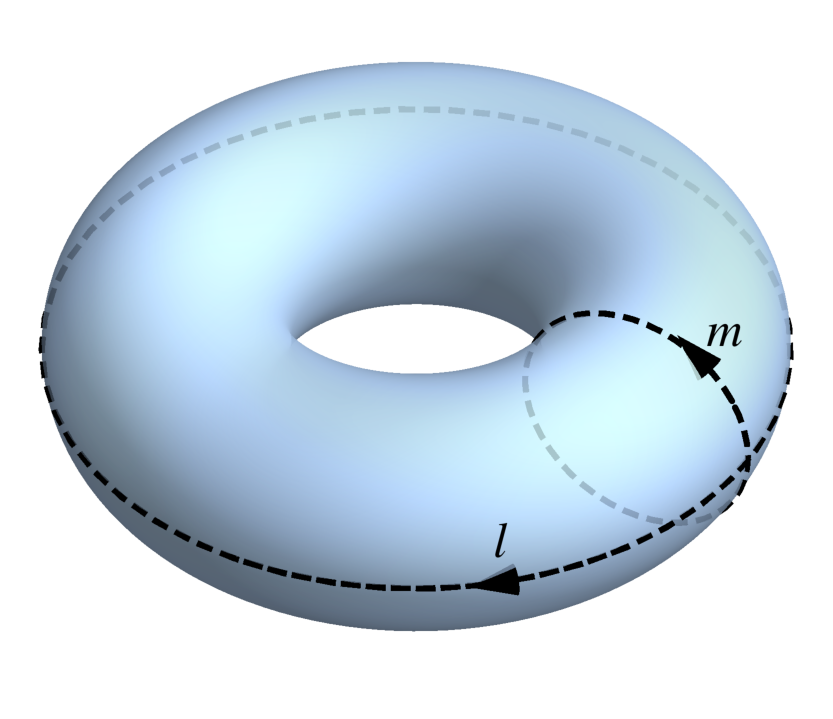}
	\caption{The generating cycles $l$ and $m$ of a torus representing the longitudinal and meridional cycles for a particular embedding in 3D space.}
	\label{fig:toruslm}
\end{figure}

\begin{figure}[t!]
	\centering
	\includegraphics[width=0.59\columnwidth]{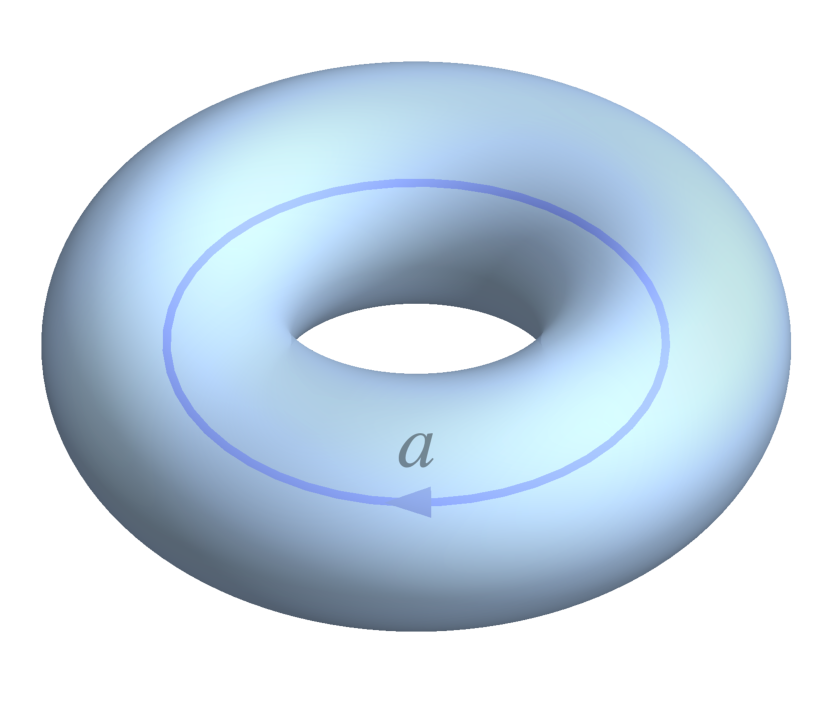}
	\caption{A topological phase described by the MTC $\mathcal{C}$ on a torus has ground state degeneracy equal to the number of distinct topological charge types $|\mathcal{C}|$. A basis for the degenerate ground state subspace is provided by the states $| a \rangle_{(l,m)}$ for $a\in \mathcal{C}$, which have topological flux $a$ threading the interior of the torus along the $l$ direction, with no twisting around the $m$ direction. For the state $| a \rangle_{(l,m)}$, a topological charge measurement around a meridional loop (cycle $m$) yields the measurement result $a$.}
	\label{fig:torusL}
\end{figure}

\begin{figure}[t!]
	\centering
	\includegraphics[width=0.59\columnwidth]{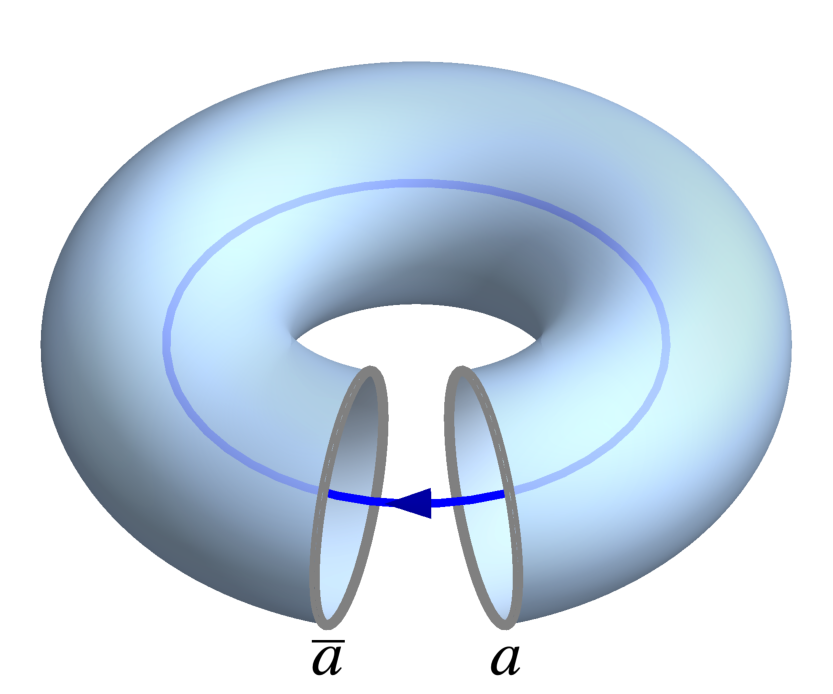}
	\caption{When the system on the torus is in the state $| a \rangle_{(l,m)}$, cutting open the torus along the meridian cycle yields two boundaries with charges $a$ and $\bar{a}$, respectively.}
	\label{fig:torusLcut}
\end{figure}

When a topological phase of matter characterized by a UMTC $\mathcal{C}$ inhabits a torus, it possesses a topologically protected ground state degeneracy equal to the number of distinct topological charges in $\mathcal{C}$. More specifically, an orthonormal basis for this degenerate ground state subspace on the torus is given by the states $|a\rangle_{(l,m)}$, for $a \in \mathcal{C}$, where $(l,m)$ specifies an ordered pair of generating cycles of the torus with intersection number $+1$. We can think of the cycles $l$ and $m$ as representing the longitudinal and meridional cycles of the torus for a particular embedding in 3D space, as shown in Fig.~\ref{fig:toruslm}. These states are defined such that a topological charge measurement performed around the cycle $m$ yields the measurement outcome $a$, and the state $\ket{a}_{(l,m)}$ is obtained from $\ket{0}_{(l,m)}$ by pair-creating quasiparticles carrying topological charge $a$ and $\bar{a}$, transporting the quasiparticle of charge $a$ around the cycle $l$, and then pair-annihilating the quasiparticles.  The state $\ket{a}_{(l,m)}$ for the torus can be thought of as having a topological flux $a$ threading the interior of the torus along the $l$ direction, with no twisting around the $m$ direction, as shown in Fig.~\ref{fig:torusL}. Here, the flux line should be thought of as a ribbon with no twisting, i.e. with both edges running parallel to the $l$ cycle. The statement regarding topological charge measurement along the cycle $m$ can be interpreted as saying that if the torus were cut open along the cycle $m$, the resulting boundaries would be found to carry topological charges $a$ and $\bar{a}$ for the basis state $|a\rangle_{(l,m)}$, as shown in Fig.~\ref{fig:torusLcut}.

\begin{figure}[t!]
	\centering
	\includegraphics[width=0.65\columnwidth]{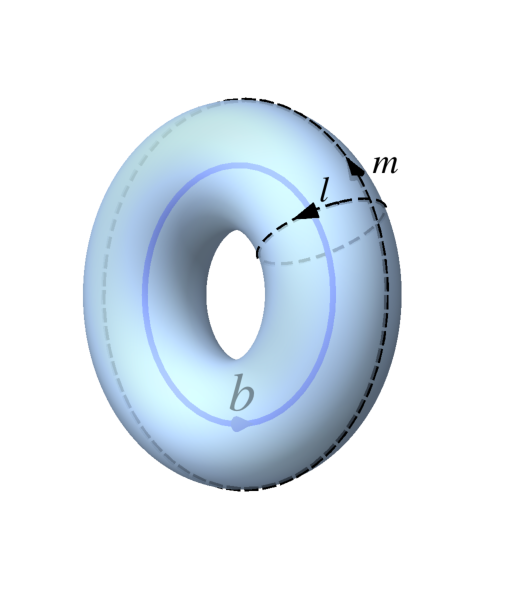}
	\caption{The basis states $\ket{b}_{(m,-l)}$ for the torus have topological flux $b$ threading the interior of a different embedding of the torus, along the $m$ direction, with no twisting around the $l$ direction. A topological charge measurement around the cycle $-l$ yields the measurement result $b$.}
	\label{fig:torusMrotated}
\end{figure}

Alternatively, one may interchange the roles of the longitudinal and meridional cycles of the torus, while maintaining the relative orientation (intersection number $+1$), which introduces a relative minus sign between the cycles. In this way, we can equivalently define a basis for the ground state subspace by $|b\rangle_{(m,-l)}$, as indicated in Fig.~\ref{fig:torusMrotated}. These basis states are defined as having the definite topological charge value $b \in \mathcal{C}$ when measured around the cycle $-l$ of the torus, and can be obtained from $\ket{0}_{(m,-l)}$ by pair-creating quasiparticles carrying topological charge $b$ and $\bar{b}$, transporting $b$ around the meridional cycle $m$, and then pair-annihilating.

\begin{figure}[t!]
	\centering
	\includegraphics[width=0.69\columnwidth]{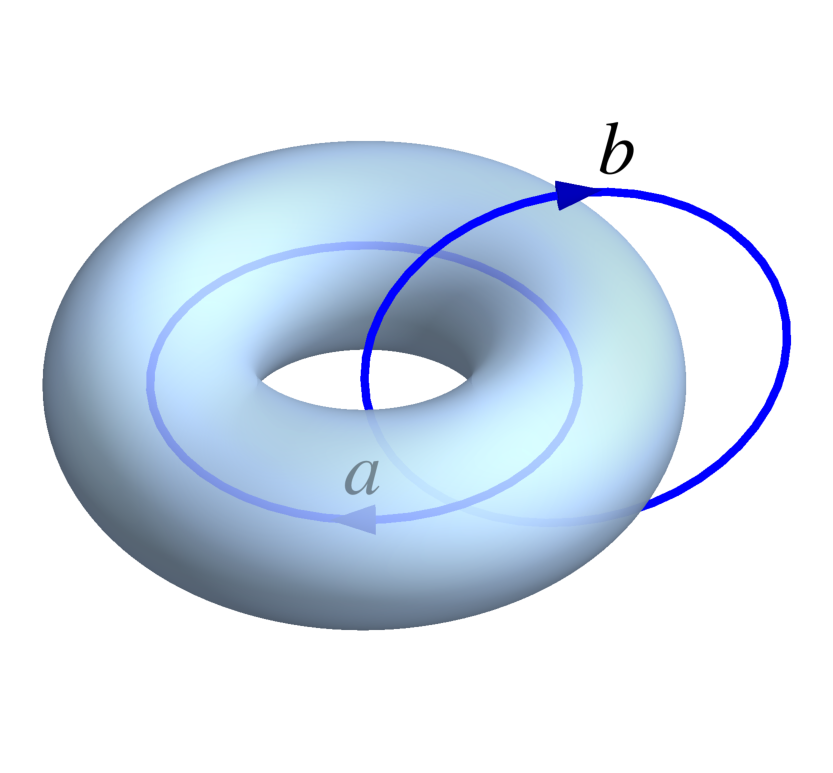}
	\caption{The two bases are related by the modular $S$ transformation, which is represented in a MTC by the topological $S$-matrix, giving $ | a \rangle_{(l,m)} = \sum\limits_{b \in \mathcal{C}} S_{ab} | b \rangle_{(m,-l)}$. This can be seen by computing the inner product $_{(m,-l)}\langle b | a \rangle_{(l,m)}$ using TQFT cutting and gluing operations. For this, the $\ket{b}_{(m,-l)}$ torus is embedded such that its interior is the exterior of the $| a \rangle_{(l,m)}$ torus. Viewing these as solid tori, gluing them together along their boundary surfaces would yield a 3-sphere $S^3$ containing the linked topological flux loops. The conjugate state $_{(m,-l)}\bra{b}$ is obtained from $\ket{b}_{(m,-l)}$ by reflecting the cycle $-l$ and orientation of the flux line $b$. Evaluating the inner product this way amounts to evaluating the resulting diagram of flux lines, which is the topological $S$-matrix.}
	\label{fig:MTC_torus}
\end{figure}

These two bases are related by the modular $S$ transformation, which interchanges the cycles of the torus (and flips the direction of one of them). As mentioned in Sec.~\ref{sec:Braiding}, the topological $S$-matrix of a MTC provides a (projective) representation of the modular $S$ transformation, where the bases are related by
\begin{equation}
| a \rangle_{(l,m)} = \sum\limits_{b \in \mathcal{C}} S_{ab} | b \rangle_{(m,-l)}
.
\end{equation}
This relation is motivated by the observation in Fig.~\ref{fig:MTC_torus} that, for the inner product $_{(m,-l)}\langle b | a \rangle_{(l,m)}$, the $a$ and $b$ topological flux lines passing around the complementary cycles of the torus forming linked loops, as in the topological $S$-matrix (the arrow of the $b$ flux line is reversed when a state is conjugated).

\begin{figure}[t!]
	\centering
	\includegraphics[width=0.59\columnwidth]{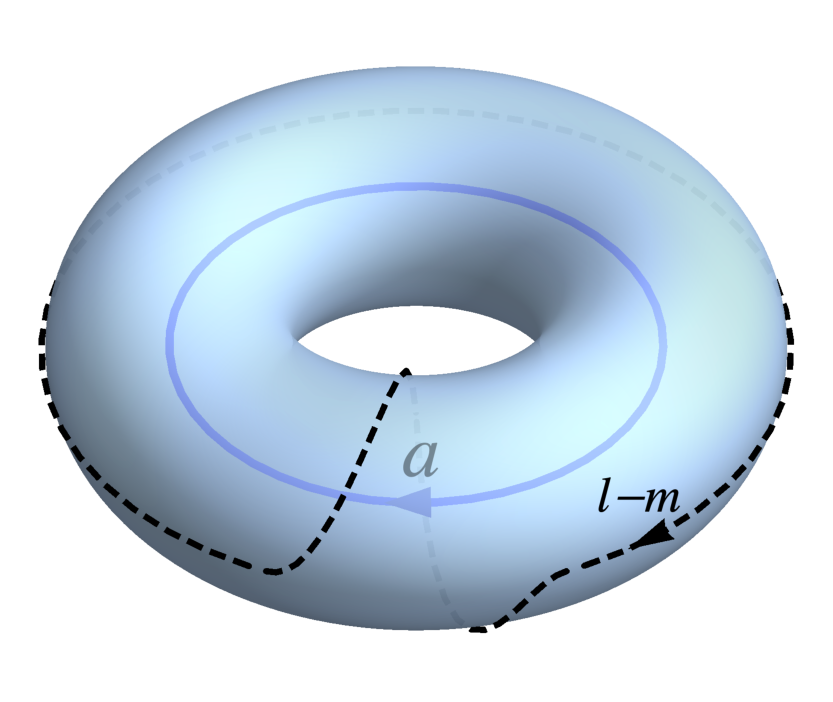}
	\caption{The modular $T$ transformation, known as a Dehn twist, replaces the longitudinal cycle $l$ with the cycle $l-m$ that wraps once around the longitude and once (negatively) around the meridian. Such transformations are represented in a MTC by the topological twists, i.e. $T_{ab} = \theta_{b} \delta_{ab}$, and relate the basis states $| a \rangle_{(l,m)}$ to the basis states $| a \rangle_{(l-m,m)}$ through the relation $| a \rangle_{(l,m)} = \sum\limits_{b \in \mathcal{C}} T_{ab} | b \rangle_{(l-m,m)}$.}
	\label{fig:Dehn_twist}
\end{figure}

In order to generate all modular transformations on the torus, we additionally consider the modular $T$ transformations, known as Dehn twists. This transformation replaces the longitudinal cycle around the torus with one that wraps once around the longitude and once negatively around the meridian, as shown in Fig.~\ref{fig:Dehn_twist}. Providing the topological flux line $a$ with a framing, which is equivalent to drawing a line on the surface of the torus running parallel to the $a$ line around the longitudinal cycle, we see that this transformation puts a twist in the framing ribbon of the flux line. This ribbon twist, which one can equate to the topological twist, motivates the definition $T_{ab} = \theta_{b} \delta_{ab}$ in the transformation
\begin{equation}
| a \rangle_{(l,m)} = \sum\limits_{b \in \mathcal{C}} T_{ab} | b \rangle_{(l-m,m)}
,
\end{equation}
since a $2 \pi$ twist around $m$ must be introduced to go from the basis states $|a\rangle_{(l-m,m)}$ to the basis states $|a\rangle_{(l,m)}$, and such a twist does not change the topological charge as measured around the meridional cycle $m$.

As mentioned in Sec.~\ref{sec:Braiding}, when the $S$-matrix of a UBTC is unitary, the theory is considered modular, as the $S$ and $T$ matrices provide a projective representation of SL$(2,\mathbb{Z})$, the modular transformations on a torus, i.e.
\begin{equation}
(ST)^3 = \Theta C, \quad S^2 = C, \quad C^2 = \openone
,
\end{equation}
where $C_{ab}= \delta_{a \bar{b}}$ is the topological charge conjugation operator. In this case, one may also define the corresponding modular transformations for punctured tori, and consequently, the theory can be consistently defined on arbitrary surfaces.

In the defect theory described by a $G$-crossed BTC $\ext{C}{G}$, the situation becomes more complicated. Clearly, the $\mathcal{C}_{\bf 0}$ subcategory, which describes the original topological phase without defects, must behave exactly the same as described above. In other words, when $S_{a_{\bf 0} b_{\bf 0}}$ is a unitary matrix (when restricted to topological charge labels $a,b \in \mathcal{C}_{\bf 0}$), so that $\mathcal{C}_{\bf 0}$ is a UMTC, the ground states on a torus without defect branches are described exactly as above and the operators $S_{a_{\bf 0} b_{\bf 0}}$ and $T_{a_{\bf 0} b_{\bf 0}}= \theta_{b_{\bf 0}} \delta_{a_{\bf 0} b_{\bf 0}}$ provide a projective representation of the modular transformations in the subtheory without defects. We call this restriction to the defect-free theory on the torus the $({\bf 0,0})$-sector and denote the corresponding modular transformations defined in this way as $\mathcal{S}^{({\bf 0,0})}$ and $\mathcal{T}^{({\bf 0,0})}$.

\begin{figure}[t!]
	\centering
	\includegraphics[width=0.59\columnwidth]{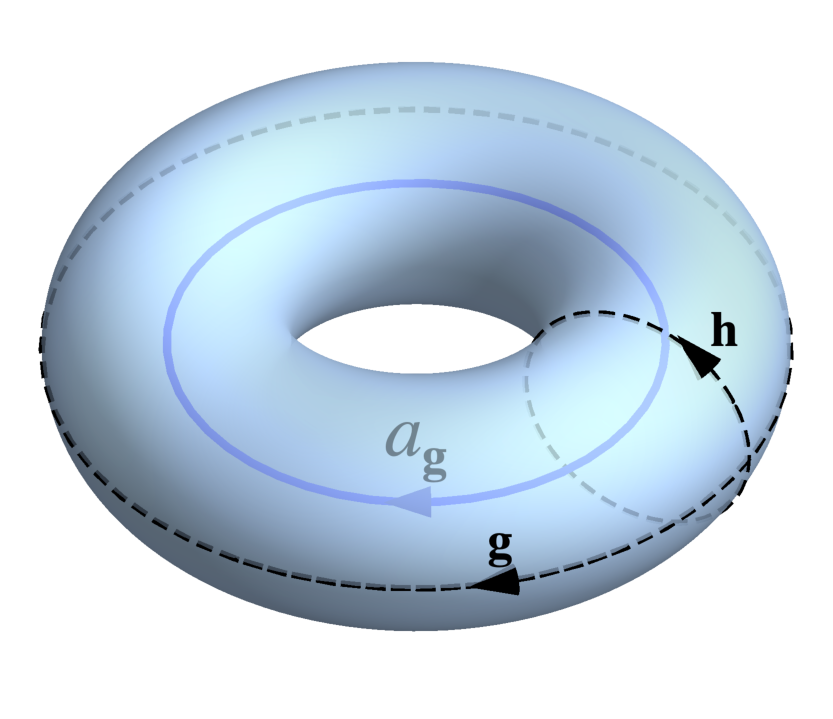}
	\caption{The $({\bf g, h})$-sector on a torus, where a closed ${\bf g}$-defect branch line wraps around the longitudinal cycle of the torus and a ${\bf h}$-defect branch line wraps around the meridional cycle of the torus. A basis for the degenerate ground state subspace of the $({\bf g, h})$-sector is given by the states $| a_{\bf g}^{({\bf g,h})} \rangle_{(l,m)}$ corresponding to definite topological charge value $a_{\bf g} \in \mathcal{C}_{\bf g}^{\bf h}$ ascribed to a charge line passing through the interior of the torus around the longitudinal cycle. }
	\label{fig:gTorusL}
\end{figure}

When we allow for the inclusion of defects in the theory, we can produce defect sectors on the torus, each of which is labeled by two group elements ${\bf g}, {\bf h} \in G$, which correspond to the accumulated defect branch lines that, respectively, wind around the $l$ and $m$ cycles of the torus, as shown in Fig.~\ref{fig:gTorusL}. The original UMTC $\mathcal{C}_{\bf 0}$ is described by the trivial defect sector $({\bf g,h})=({\bf 0,0})$. One can obtain a state in the $({\bf g,h})$-sector from the $({\bf 0,0})$-sector by adiabatically creating a ${\bf h , h}^{-1}$ defect pair from vacuum, transporting the ${\bf h}$-defect around the cycle $m$ (in the positive sense), pair annihilating the defect pair, and then doing the same process with an ${\bf g, g}^{-1}$ defect pair winding around the cycle $l$. This is only possible when
\begin{equation}
\bf g \bf h = \bf h \bf g
,
\end{equation}
since otherwise the group element ascribed to the defects would necessarily change type as they crossed the other defect branch line wrapping around the complementary cycle, making it impossible to pair-annihilate the defects or close the branch line on itself.

In this way, we see that the topological (defect) flux line that runs through the interior of the torus around the cycle $l$ can only take values in $\mathcal{C}_{\bf g}$, since it must be created by a ${\bf g}$-defect encircling the cycle. Moreover, this topological charge must be ${\bf h}$-invariant, since the flux lines cross the ${\bf h}$-branch. Similarly, the topological flux line that runs through the exterior of the torus around the cycle $m$ can only take ${\bf g}$-invariant topological charge values in $\mathcal{C}_{\bf h}$. It is clear that states from different $(\bf g,\bf h)$-sectors cannot be superposed, since the defects are extrinsic, confined objects, which can be thought of as defining distinct superselection sectors.

We label the ground state subspace associated with the $(\bf g,\bf h)$-sector of the system on a torus as $\mathcal{V}_{({\bf g, h})}$. Similar to UMTCs, a basis for $\mathcal{V}_{({\bf g, h})}$ is given by orthonormal states $| a_{\bf g}^{({\bf g,h})} \rangle_{(l,m)}$ for $a_{\bf g} \in \mathcal{C}_{\bf g}^{\bf h}$, as shown in Fig.~\ref{fig:gTorusL}. In general, the notation used here means that: (1) as the cycle $l$ is traversed in the positive sense, a ${\bf \bar{h}}$-branch is crossed in the positive sense and as the cycle $m$ is traversed in the positive direction, a ${\bf g}$-branch is crossed in the positive sense~\footnote{Our notion of crossing a ${\bf g}$-branch line ``in the positive sense'' means crossing it in the direction that enacts ${\bf g}$-action on the object crossing it. Crossing a ${\bf g}$-branch line in the negative sense enacts ${\bf \bar{g}}$-action.}; (2) a topological charge measurement performed around the cycle $m$ yields the value $a_{\bf g}$; (3) one can obtain different basis states from one another by pair-creating quasiparticles, transporting one around the cycle $l$, and pair-annihilating; and (4) one can switch between states in different defect sectors by pair-creating defects, transporting one around a nontrivial cycle, and pair-annihilating.

\begin{figure}[t!]
	\centering
	\includegraphics[width=0.69\columnwidth]{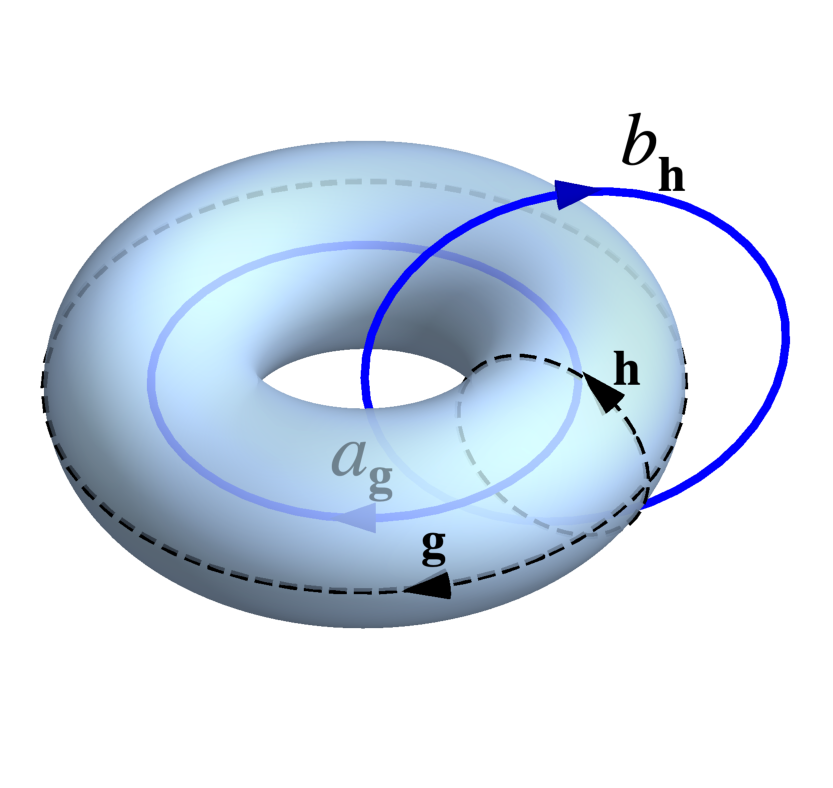}
	\caption{The $({\bf g, h})$-sector on a torus may be considered to be a $({\bf h,\bar{g}})$-sector on a torus by interchanging the roles of the longitudinal and meridional cycles. In this case, a basis for the ground state subspace is given by the states $| b_{\bf h}^{({\bf h,\bar{g}})} \rangle_{(m,l)}$ corresponding to definite topological charge $b_{\bf h} \in \mathcal{C}_{\bf h}^{\bf g}$ ascribed to the charge line passing through the exterior of the torus around the meridional cycle. These two bases are related by the modular $\mathcal{S}$ transformation  $ | a_{\bf g}^{({\bf g,h})} \rangle_{(l,m)} = \sum\limits_{b \in \mathcal{C}_{\bf h}^{\bf g}} \mathcal{S}_{a_{\bf g} b_{\bf h}}^{({\bf g,h})} | b_{\bf h}^{({\bf h,\bar{g}})} \rangle_{(m,-l)}$.}
	\label{fig:gTorusS}
\end{figure}

For a $G$-crossed theory, if we interchange the roles of the longitudinal and meridional cycles (and flip one of their directions), corresponding to a modular $\mathcal{S}$ transformation, then the system belongs to the $({\bf h,\bar{g}})$-sector on a torus with associated state space $\mathcal{V}_{({\bf h,\bar{g}})}$. In this case, a basis for the ground state subspace is given by the states $| b_{\bf h}^{({\bf h,\bar{g}})} \rangle_{(m,-l)}$ corresponding to definite topological charge $b_{\bf h} \in \mathcal{C}_{\bf h}^{\bf g}$ ascribed to the topological flux line passing through the exterior of the torus around the cycle $m$. We emphasize that we have not changed the system, so the configuration of defect branch lines is the same as before. In particular, the notation means that as the cycle $m$ is traversed, a ${\bf g}$-branch is crossed and as the cycle $-l$ is traversed, a ${\bf h}$-branch is crossed, which is just a different way of describing the torus with a ${\bf g}$-branch wrapping around the cycle $l$ and a ${\bf h}$-branch wrapping around the cycle $m$. Thus, there must be a unitary operator relating these two bases which represents the modular $\mathcal{S}$ transformation between the $({\bf g,h})$ and $({\bf h,\bar{g}})$ sectors, as shown in Fig.~\ref{fig:gTorusS}. In particular, this takes the form
\begin{equation}
| a_{\bf g}^{({\bf g,h})} \rangle_{(l,m)} = \sum\limits_{b \in \mathcal{C}_{\bf h}^{\bf g}} \mathcal{S}_{a_{\bf g} b_{\bf h}}^{({\bf g,h})} | b_{\bf h}^{({\bf h,\bar{g}})} \rangle_{(m,-l)}
.
\end{equation}

\begin{figure}[t!]
	\centering
	\includegraphics[width=0.59\columnwidth]{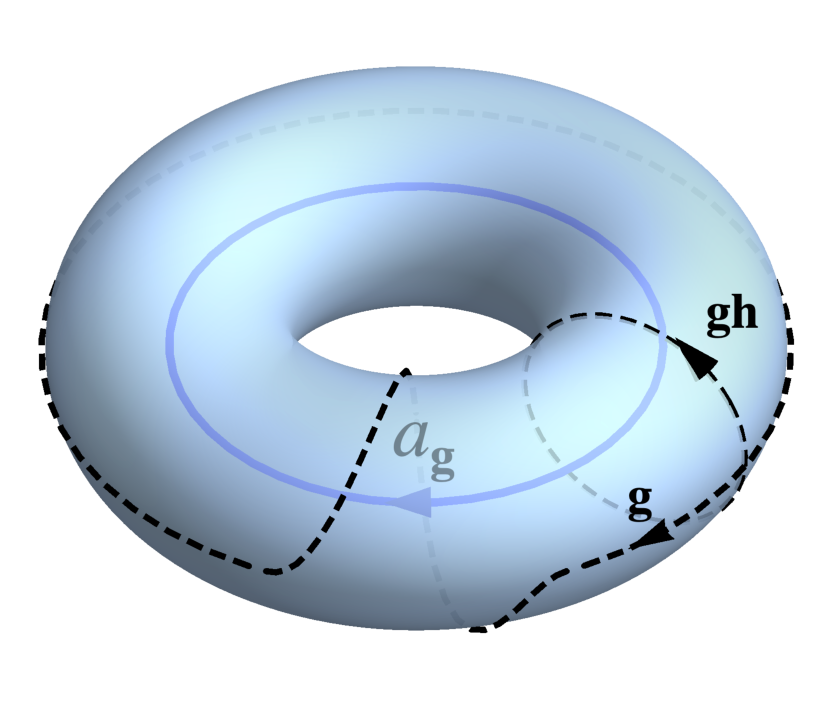}
	\caption{The modular $\mathcal{T}$ transformation (Dehn twist) maps between the $({\bf g, h})$-sector on a torus to the $({\bf g, gh})$-sector shown here. This transformation acts diagonally (i.e. with relative phases) between bases for the $({\bf g, h})$ and $({\bf g, gh})$ sectors, both of which are labeled by topological charge values $a_{\bf g}$ ascribed to the topological flux line passing through the interior of the torus, though this topological flux is interpreted as winding around the cycle $l$ in the former basis and around the cycle $l-m$ in the latter basis. These two bases are related by $| a_{\bf g}^{({\bf g,h})} \rangle_{(l,m)} = \sum\limits_{b \in \mathcal{C}_{\bf g}^{\bf h}} \mathcal{T}_{a_{\bf g} b_{\bf g}}^{({\bf g,h})} | b_{\bf g}^{({\bf g,gh})} \rangle_{(l-m,m)}$.}
	\label{fig:G_Dehn_twist}
\end{figure}

Similarly, the modular $\mathcal{T}$ transformation (Dehn twist) takes the system between the $({\bf g, h})$ and $({\bf g,gh})$ sectors, as indicated in Fig.~\ref{fig:G_Dehn_twist}, with basis states related by
\begin{equation}
| a_{\bf g}^{({\bf g,h})} \rangle_{(l,m)} = \sum\limits_{b \in \mathcal{C}_{\bf g}^{\bf h}} \mathcal{T}_{a_{\bf g} b_{\bf g}}^{({\bf g,h})} | b_{\bf g}^{({\bf g,gh})} \rangle_{(l-m,m)}
.
\end{equation}
In this case, the notation for the basis states means that as the cycle $l-m$ is traversed, a ${\bf \bar{h}\bar{g}}$-branch is crossed and as the cycle $m$ is traversed, a ${\bf g}$-branch is crossed. We emphasize that these states still describes the system with a ${\bf g}$-branch line winding around the cycle $l$ and a ${\bf h}$-branch line winding around $m$. We can, however, continuously deform the branch line configuration (without introducing new defects or branch lines) so that the system has a ${\bf g}$-branch line winding around the cycle $l-m$ and a ${\bf gh}$-branch line winding around the cycle $m$, as shown in Fig.~\ref{fig:G_Dehn_twist}.

Thus, we can write the modular $\mathcal{S}$ and $\mathcal{T}$ transformations for a $G$-crossed theory in the form
\begin{eqnarray}
\mathcal{S} &=& \bigoplus_{\{({\bf g,h}) \,| \, {\bf gh=hg}\}} \mathcal{S}^{({\bf g,h})} \\
\mathcal{T} &=& \bigoplus_{\{({\bf g,h}) \,| \,  {\bf gh=hg}\}} \mathcal{T}^{({\bf g,h})}
,
\end{eqnarray}
where these transformations map from one defect sector to another (without mixing sectors)
\begin{eqnarray}
\mathcal{S}^{({\bf g,h})} &:& \; \mathcal{V}_{({\bf h},{\bf \bar{g}})}  \rightarrow  \mathcal{V}_{({\bf g},{\bf h})} \\
\mathcal{T}^{({\bf g,h})} &:& \; \mathcal{V}_{({\bf g}, {\bf gh}) } \rightarrow \mathcal{V}_{({\bf g},{\bf h})}
.
\end{eqnarray}

For example, the $G$-crossed modular transformations for $G=\mathbb{Z}_2=\{0,1\}$ take the block form
 \begin{equation}
	 \mathcal{S}=\left[
	\begin{array}{c|c|c|c}
		\mathcal{S}^{({\bf 0,0})} & 0 & 0 & 0 \\ \hline
		0 & 0 & \mathcal{S}^{({\bf 0,1})} & 0 \\ \hline
		0 & \mathcal{S}^{({\bf 1,0})} & 0 & 0\\ \hline
		0 & 0 & 0 & \mathcal{S}^{({\bf 1,1})}
\end{array}
\right],
 \end{equation}
\begin{equation}
 \mathcal{T}=\left[
	\begin{array}{c|c|c|c}
		\mathcal{T}^{({\bf 0,0})} & 0 & 0 & 0 \\ \hline
		0 &  \mathcal{T}^{({\bf 0,1})} & 0 & 0 \\ \hline
		0 & 0 & 0 & \mathcal{T}^{({\bf 1,0})}\\ \hline
		0 & 0 & \mathcal{T}^{({\bf 1,1})} & 0
\end{array}
\right],
\end{equation}
where the rows and columns are separated into $({\bf 0,0})$, $({\bf 0,1})$, $({\bf 1,0})$, and $({\bf 1,1})$ sectors, in that order.

Thus, imposing unitarity on the representations of the modular $\mathcal{S}$ and $\mathcal{T}$ transformations amounts to imposing unitarity on their restricted actions $\mathcal{S}^{({\bf g,h})}$ and $\mathcal{T}^{({\bf g,h})}$ for each $({\bf g,h})$-sector individually. Since the system in the $({\bf g,h})$-sector has a ground state degeneracy
\begin{equation}
\mathcal{N}_{({\bf g,h})} = \dim \mathcal{V}_{({\bf g,h})} = |\mathcal{C}_{\bf g}^{\bf h}|
\end{equation}
equal to the number of ${\bf h}$-invariant topological charges in $\mathcal{C}_{\bf g}$, it follows that requiring the modular transformations to be unitary gives the condition that
\begin{equation}
|\mathcal{C}_{\bf h}^{\bf g}| = |\mathcal{C}_{\bf g}^{\bf h}|
,
\end{equation}
whenever ${\bf gh=hg}$. In particular, for ${\bf h=0}$, this gives us the important property
\begin{equation}
|\mathcal{C}_{\bf g}| = |\mathcal{C}_{\bf 0}^{\bf g}|,
\end{equation}
which says the number of topologically distinct types of ${\bf g}$-defects (i.e. the topological charge types in $\mathcal{C}_{\bf g}$) is equal to the number of ${\bf g}$-invariant topological charges in $\mathcal{C}_{\bf 0}$.

We now wish to provide a projective representation of the modular transformations that are defined by the $G$-crossed UBTC data. Let us take the representation of the modular transformations defined by
\begin{eqnarray}
\mathcal{S}^{({\bf g,h})}_{a_{\bf g} b_{\bf h}} &=& \frac{ S_{a_{\bf g} b_{\bf h}}}{ U_{\bf h}(a,\bar{a};0) }
\label{eq:modularS}
\\
\mathcal{T}^{({\bf g,h})}_{a_{\bf g} b_{\bf g}} &=& \eta_{a}({\bf g}, {\bf h} ) \theta_{a_{\bf g}} \delta_{a_{\bf g} b_{\bf g} }
\label{eq:modularT}
.
\end{eqnarray}
Recall that $S_{a_{\bf g} b_{\bf h}}$ is the topological $S$-matrix defined in Eq.~(\ref{eqn:topoSmatrix}).
It is convenient for us to also define the $G$-crossed ``charge conjugation'' transformation
\begin{eqnarray}
C &=& \bigoplus_{\{({\bf g,h}) : \, {\bf gh=hg}\}} C^{({\bf g,h})} \\
C^{({\bf g,h})} &:& \;  \mathcal{V}_{({\bf \bar{g}},{\bf \bar{h}})} \rightarrow  \mathcal{V}_{({\bf g}, {\bf h}) }  \\
| a_{\bf g}^{({\bf g,h})} \rangle_{(l,m)} &=& \sum\limits_{b \in \mathcal{C}_{\bf \bar{g}}^{\bf h}} C_{a_{\bf g} b_{\bf \bar{g}}}^{({\bf g,h})} | b_{\bf \bar{g}}^{({\bf \bar{g},\bar{h}})} \rangle_{(-l,-m)} \\
C^{({\bf g,h})}_{a_{\bf g} b_{\bf \bar{g}}} &=& \frac{1}{ U_{\bf h}(\bar{b},b;0)  \eta_{b}({\bf h}, {\bf \bar{h}} ) } \delta_{a_{\bf g} \overline{b_{\bf \bar{g}}} }
\label{eq:modularC}
.
\end{eqnarray}

Given the properties derived for a general $G$-crossed UBTC in Sec.~\ref{sec:Algebraic_Theory}, we can obtain the relation
\begin{equation}
\sum_{w,x,y,z}\mathcal{T}^{({\bf g,h})}_{a_{\bf g} w_{\bf g}} \mathcal{S}^{({\bf g,gh})}_{w_{\bf g} x_{\bf gh}} \mathcal{T}^{({\bf gh,\bar{g} })}_{x_{\bf gh} y_{\bf gh}} \mathcal{S}^{({\bf gh,h})}_{y_{\bf gh} z_{\bf h}} \mathcal{T}^{({\bf h,\bar{h}\bar{g}})}_{z_{\bf h} b_{\bf h}} = \Theta_{\bf 0} \mathcal{S}^{({\bf g,h})}_{a_{\bf g} b_{\bf h}}
\end{equation}
from Eq.~(\ref{eq:TSTST-S}), where $\Theta_{\bf 0}=\frac{1}{\mathcal{D}_{\bf 0}}\sum\limits_{c \in \mathcal{C}_{\bf 0}} d_{c}^{2} \theta_{c}$, the relation
\begin{equation}
\mathcal{S}^{({\bf g,h})}_{a_{\bf g} b_{\bf h}} = \sum_{x} \left[ \mathcal{S}^{({\bf \bar{h},g})}_{ x_{\bf \bar{h} }  a_{\bf g} } \right]^{\ast} \, C^{({\bf \bar{h},g})}_{x_{\bf \bar{h}} b_{\bf h}}
\end{equation}
from Eq.~(\ref{eq:S_3}), and the relation
\begin{equation}
\sum_{x} C^{({\bf g,h})}_{a_{\bf g} x_{\bf \bar{g}}} C^{({\bf \bar{g},\bar{h}})}_{ x_{\bf \bar{g}} b_{\bf g }} = \delta_{a_{\bf g } b_{\bf g }}
,
\end{equation}
from Eq.~(\ref{eq:U_eta_consistency}).
Thus, without imposing unitarity of the topological $S$-matrix nor any other extra conditions on a $G$-crossed UBTC, the transformations defined by Eqs.~(\ref{eq:modularS}), (\ref{eq:modularT}), and (\ref{eq:modularC}) obey the relations
\begin{eqnarray}
\left(\mathcal{S}\mathcal{T} \right)^3 &=& \Theta_{\bf 0} \mathcal{S}^2
\label{eq:G-crossed-ST3}
\\
\mathcal{S} &=& \mathcal{S}^{\dagger} C
\label{eq:S-S*C}
\\
C^2 &=& \openone
.
\end{eqnarray}
We can also show that
\begin{equation}
C \mathcal{S} = \mathcal{S} C
\end{equation}
using Eqs.~(\ref{eq:S_2}) and (\ref{eq:S_4}), and that
\begin{equation}
C \mathcal{T} = \mathcal{T} C
\end{equation}
using Eqs.~(\ref{eq:lemmatwist}), (\ref{eq:U_k_relation}), (\ref{eqn:etasym}), and (\ref{eq:sliding_consistency_c0})-(\ref{eq:eta_relation_2}).

It is clear from these relations that all that is needed for these operators to provide a projective representation of the modular transformations is to impose a condition on the topological $S$-matrix that makes the modular operator $\mathcal{S}$ defined by Eq.~(\ref{eq:modularS}) unitary, in which case Eq.~(\ref{eq:S-S*C}) would become
\begin{equation}
\mathcal{S}^2 = C
.
\end{equation}
We can see that requiring $\mathcal{S}$ to be unitary is equivalent to the condition that the topological $S$-matrix of the $G$-crossed UBTC gives unitary matrices when it is $G$-graded, by which we mean that for any fixed pair of group elements ${\bf g}$ and ${\bf h}$, the matrix defined by $S_{a_{\bf g} b_{\bf h}}$ with indices $a \in \mathcal{C}_{\bf g}^{\bf h}$ and $b \in \mathcal{C}_{\bf h}^{\bf g}$ is a unitary matrix. Thus, when the topological $S$-matrix of a $G$-crossed UBTC $\mathcal{C}_{G}^{\times}$ is $G$-graded unitary (in the fashion described here), we say that $\mathcal{C}_{G}^{\times}$ is $G$-crossed modular or that it is a $G$-crossed modular tensor category.

We note that, for a modular theory, the quantity
\begin{equation}
\Theta_{\bf 0}=\frac{1}{\mathcal{D}_{\bf 0}} \sum\limits_{c \in \mathcal{C}_{\bf 0}} d_{c}^{2} \theta_{c} = e^{i \frac{2\pi}{8} c_{-}}
\end{equation}
is a phase related to $c_{-}$ which is the chiral central charge of the topological phase described by the UMTC $\mathcal{C}_{\bf 0}$. Thus, we can ascribe the same chiral central charge to the $G$-crossed extensions of a topological phase.

It follows from the definition of $G$-crossed modularity that the $\mathcal{C}_{\bf 0}$ subcategory of a $G$-crossed MTC is a MTC. As previously mentioned, the converse is also true, as can be shown by combining highly nontrivial theorems from \Refs{Etingof05,kirillov2004}. Thus, the conditions of modularity of a UBTC and its $G$-crossed extensions are equivalent, i.e. $\mathcal{C}_{G}^{\times}$ is a $G$-crossed UMTC if and only if $\mathcal{C}_{\bf 0}$ is a UMTC.

We note that, as was the case for a UMTC, the same arguments used in Eq.~(\ref{eq:MphaseAbelian}) apply to a $G$-crossed UMTC, implying that a defect topological charge $a_{\bf g}$ with $d_{a_{\bf g}} > 1$ necessarily has non-Abelian braiding.

We also note that, just as in the case of a MTC, we could actually obtain a linear (rather than projective) representation of the modular transformations on the torus if we instead defined the Dehn twist transformation to be given by
\begin{equation}
\mathcal{T}^{({\bf g,h})}_{a_{\bf g} b_{\bf g}} = e^{-i \frac{2 \pi}{24} c_{-} } \eta_{a}({\bf g}, {\bf h} ) \theta_{a_{\bf g}} \delta_{a_{\bf g} b_{\bf g} }
,
\end{equation}
as this would give the relation $\left(\mathcal{S}\mathcal{T} \right)^3 = \mathcal{S}^2$. This convention may be more useful when performing concrete calculations or physical simulations on the torus. However, it is not generally possible to trivialize the projective phases for the representations of modular transformations for higher genus surfaces, so we will not generally include the central charge dependent phase.

An important distinction from MTCs is that the quantities representing the $G$-crossed modular $\mathcal{S}$ and $\mathcal{T}$ transformations defined here are not gauge invariant, except in the $({\bf 0,0})$-sector (which was also the case with the topological twists and $S$-matrix in the $G$-crossed theory). In particular, while they are invariant under vertex basis gauge transformations, they transform under symmetry action gauge transformations as
\begin{eqnarray}
\check{\mathcal{S}}^{({\bf g,h})}_{a_{\bf g} b_{\bf h}} &=& \frac{\gamma_{b}({\bf \bar{g}})}{\gamma_{a}({\bf h})} \mathcal{S}^{({\bf g,h})}_{a_{\bf g} b_{\bf h}}
\label{eq:modularS_symm_gauge_trans}
\\
\check{\mathcal{T}}^{({\bf g,h})}_{a_{\bf g} b_{\bf g}} &=& \frac{\gamma_{b}({\bf gh})}{\gamma_{a}({\bf h})}  \mathcal{T}^{({\bf g,h})}_{a_{\bf g} b_{\bf g}}
\label{eq:modularT_symm_gauge_trans}
.
\end{eqnarray}
This is not unexpected, since these two modular transformations map the $({\bf h,\bar{g}})$-sector and the $({\bf g,gh})$-sector to the $({\bf g,h})$-sector, respectively, and there is no well-defined gauge invariant notion of comparing distinct superselection sectors, i.e. there is no canonical map between different sectors. (This is related to the fact that the defects are extrinsic objects which define different superselection sectors for different group elements and for which one should not expect overall phases to be well-defined.) As such, it is important to be careful with the details of how one sets up configurations and analyzes their modular transformations when working on a torus or higher genus system.

On the other hand, we may expect some modular transformations to be gauge invariant [in addition to those of the $({\bf 0,0})$-sectors]. From Eqs.~(\ref{eq:modularS_symm_gauge_trans}) and (\ref{eq:modularT_symm_gauge_trans}), and the fact that $\mathcal{S}$ and $\mathcal{T}$ generate the modular transformations on the torus, it follows that a general modular transformation $\mathcal{Q}$ that maps the $({\bf g,h})$-sector to the $({\bf g',h'})$-sector, i.e.
\begin{equation}
\mathcal{Q}^{({\bf g,h})} \, : \; \mathcal{V}_{({\bf g}', {\bf h}') }  \rightarrow \mathcal{V}_{({\bf g},{\bf h})}
,
\label{eqn:modularinv}
\end{equation}
transforms under symmetry action gauge transformations as
\begin{equation}
\check{\mathcal{Q}}^{({\bf g,h})}_{a_{\bf g} b_{\bf g'}} = \frac{\gamma_{b}({\bf h'})}{\gamma_{a}({\bf h})} \mathcal{Q}^{({\bf g,h})}_{a_{\bf g} b_{\bf g'}}
.
\end{equation}
From this expression, it is easy to see that (a) if a modular transformation $\mathcal{Q}$ maps a $({\bf g,h})$-sector to itself, then $\mathcal{Q}^{({\bf g,h})}_{a_{\bf g} a_{\bf g}}$ is a gauge invariant quantity and (b) if $\mathcal{Q}$ maps a $({\bf g,0})$-sector to itself, then $\mathcal{Q}^{({\bf g,0})}_{a_{\bf g} b_{\bf g}}$ is a gauge invariant quantity.

For example, if ${\bf g}^{n} = {\bf 0}$, then $\mathcal{T}^{n}$ will map a $({\bf g,h})$-sector to itself, and the coefficients
\begin{equation}
\left[\mathcal{T}^{n} \right]^{({\bf g,h})}_{a_{\bf g} a_{\bf g}} =  \theta_{a_{\bf g}}^{n} \prod_{j=0}^{n-1} \eta_{a_{\bf g}}({\bf g , g}^{j}{\bf h})
= \theta_{a_{\bf g}}^{n} \prod_{j=1}^{n-1} \eta_{a_{\bf g}}({\bf g}^{j},{\bf g})
\label{eq:T^n_invariant}
\end{equation}
provide gauge invariant quantities of the $G$-crossed theory. We note that these quantities are independent of ${\bf h}$.

If ${\bf g}^{2} = {\bf 0}$, we see that
\begin{eqnarray}
\mathcal{S}^{({\bf g,g})}_{a_{\bf g} a_{\bf g}} &=& \frac{S_{a_{\bf g} a_{\bf g}}}{ U_{\bf g}(a, \bar{a};0) } \\
\left[\mathcal{S} \mathcal{T} \mathcal{S} \right]^{({\bf g,0})}_{a_{\bf g} b_{\bf g}} &=& \sum_{x_{\bf 0}} \frac{ S_{a_{\bf g} x_{\bf 0}} \theta_{x_{\bf 0}} S_{x_{\bf 0} b_{\bf g}}}  { U_{\bf g}(x, \bar{x};0) } \\
\left[\mathcal{T} \mathcal{S} \mathcal{T} \right]^{({\bf g,0})}_{a_{\bf g} b_{\bf g}} &=& \frac{ \theta_{a_{\bf g}} S_{a_{\bf g} b_{\bf g}} \theta_{b_{\bf g}}  \eta_{b}({\bf g , g})} { U_{\bf g}(a, \bar{a};0) }
\end{eqnarray}
are also gauge invariant quantities [the last two are, of course, not independent of each other, given Eq.~(\ref{eq:G-crossed-ST3})].

\subsection{Higher Genus Surfaces}
\label{sec:generalGenus}

\begin{figure}[t!]
	\centering
	\includegraphics[width=1\columnwidth]{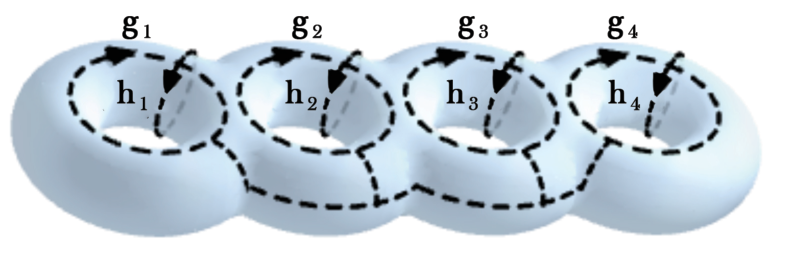}
	\resizebox{0.9\columnwidth}{!}{
  \pspicture[shift=0](-1.7,-1.5)(8.0,1.5)
  \psset{linewidth=1.3pt,linecolor=red,arrowscale=1.2,arrowinset=0.15, linestyle=dashed, dash=3pt 1.8pt}
	\psellipse(0.15,1)(0.5,0.15)
	\psellipse(2.05,1)(0.5,0.15)
	\psellipse(5.25,1)(0.5,0.15)
	\psellipse(7.15,1)(0.5,0.15)
	\psline{->}(0.07,1.14)(0.12,1.148)
	\psline{->}(1.97,1.14)(2.02,1.148)
	\psline{->}(5.17,1.14)(5.22,1.148)
	\psline{->}(7.07,1.14)(7.12,1.148)
	\psset{linewidth=1.0pt,linecolor=black,arrowscale=1.3,arrowinset=0.15,linestyle=solid, border=1.5pt}
	\psarc{<-}(-0.3,1){0.45}{-178}{4}
	\psarc{<-}(1.6,1){0.45}{-178}{4}
	\psarc{<-}(4.8,1){0.45}{-178}{4}
	\psarc{<-}(6.7,1){0.45}{-178}{4}
	\psset{linewidth=1.0pt,linecolor=black,arrowscale=1.3,arrowinset=0.15,linestyle=solid, border=0}
	\psarc(-0.3,1){0.45}{28}{215}
	\psarc(1.6,1){0.45}{28}{215}
	\psarc(4.8,1){0.45}{28}{215}
	\psarc(6.7,1){0.45}{28}{215}
	\psline(-0.3,-0.1)(-0.3,0.55)
	\psline{->}(-0.3,0.25)(-0.3,0.3)
	\psline(1.6,-0.3)(1.6,0.55)
	\psline{->}(1.6,0.20)(1.6,0.25)
	\psline(4.8,-0.7)(4.8,0.55)
	\psline{->}(4.8,0.05)(4.8,0.1)
	\psline(6.7,-0.9)(6.7,0.55)
	\psline{->}(6.7,-0.05)(6.7,-0.0)
	\psline(-0.3,-0.1)(6.7,-0.9)
	\rput[tr]{0}(-0.65,0.7){\scalebox{0.8}{$a_{1}$}}
	\rput[tr]{0}(1.22,0.7){\scalebox{0.8}{$a_{2}$}}
	\rput[tr]{0}(4.37,0.7){\scalebox{0.8}{$a_{g-1}$}}
	\rput[tr]{0}(6.35,0.7){\scalebox{0.8}{$a_{g}$}}
	\rput[tl]{0}(-0.2,0.3){$z_1$}
	\rput[tl]{0}(1.7,0.2){$z_2$}
	\rput[tl]{0}(4.9,0.1){$z_{g-1}$}
	\rput[tl]{0}(6.8,0.0){$z_{g}$}
	\rput[br]{0}(1.0,1.1){$\mathbf{h}_1$}
	\rput[br]{0}(2.9,1.1){$\mathbf{h}_2$}
	\rput[br]{0}(6.1,1.15){$\mathbf{h}_{g\sminus 1}$}
	\rput[br]{0}(8.0,1.1){$\mathbf{h}_{g}$}
	\rput[tl]{0}(1.8,-0.5){$c_{12}$}
	\rput[tl]{0}(5.1,-0.9){$c_{1 \ldots g-1}$}
	\rput{352}(2.8,-0.7){\scalebox{1.1}{$\cdots$}}
	\rput{0}(3.3, 1){\scalebox{1.2}{$\cdots$}}
	\endpspicture
}
\caption{\label{fig:genusgLoops}
The defect sectors on a genus $g$ surface can be labeled by $2g$ group elements $\{{\bf{g}}_j, {\bf{h}}_j\}$ for $j=1,\ldots,g$, which are ascribed to the defect branch lines around two independent non-contractible cycles associated with the $j$th handle.  In this case, one does not require that ${\bf g}_{j}$ and ${\bf h}_{j}$ commute, so one of the branch lines at a given handle may change as it crosses the complementary branch line at that handle. We pick the ${\bf h}_{j}$-branch lines to close around their cycles unchanged, while the ${\bf g}_{j}$ branch lines transform into ${\bf h}_{j}^{-1}{\bf g}_{j}{\bf h}_{j}$ branch lines when they cross the ${\bf h}_{j}$-branch. This requires a ${\bf k}_{j}$-branch line, where ${\bf k}_{j} = {\bf g}_{j} {\bf h}_{j}^{-1}  {\bf g}_{j}^{-1} {\bf h}_{j}$, to enter the handle to cancel the leftover branch. Similarly, the $a_j \in \mathcal{C}_{{\bf g}_{j}}$ charge lines used to define basis state may also transform nontrivially as $^{{\bf \bar{h}}_{j}}a_{j}$ when it crosses the ${\bf h}_j$-branch loop. This requires a line of charge $z_{j} \in \mathcal{C}_{{\bf k}_{j}}$ with $N_{a_j \,^{{\bf \bar{h}}_{j}}\overline{a_j} }^{z_j } \neq 0$ to enter the handle to cancel the leftover topological charge. The $z_j$ charge lines from different handles form a fusion tree. These charge line configurations, together with the fusion vertex state labels, provide a basis of states for the genus $g$ surface in the $\{{\bf{g}}_j, {\bf{h}}_j\}$-sector.
}
\end{figure}

When the system is on a genus $g$ surface, the topological ground state degeneracy is more complicated. In general, it can be obtained by summing over the possible states associated with a fusion tree of topological charge lines that pass through either the interior or exterior of the surface, and which encircle independent non-contractible cycles, as shown in Fig.~\ref{fig:genusgLoops}. For a UMTC (without defects), this leads to the ground state degeneracy
\begin{align}
\mathcal{N}_{g} &= \sum_{\{b,z,c\} \in\mathcal{C}}  N_{z_1 z_2}^{c_{12}} N_{c_{12} z_3}^{c_{123}} \cdots N_{c_{1 \ldots g-1} z_g}^0 \prod_{j=1}^{g} N_{a_j \overline{a_{j}} }^{z_j}
\nonumber \\
&= \mathcal{D}_{\bf 0}^{2g-2} \sum_{x \in\mathcal{C}} d_{x}^{-(2g-2)},
\label{eq:genus_g_Verlinde}
\end{align}
where the evaluation may be carried out using the Verlinde formula.

For a topological phase with defects, described by a $G$-crossed UMTC $\mathcal{C}_{G}^{\times}$, the system on a genus $g$ surface may have defect branch lines around any non-contractible loop, similar to the case of the torus. In this case, we can label the distinct defect sectors of a genus $g$ surface by $2g$ group elements, $\{{\bf g}_j,{\bf h}_j\}$, $j= 1, \ldots , g$, each of which corresponds to a defect branch line wrapping around an independent generating cycle. We write the corresponding ground state subspace as $\mathcal{V}_{\{ {\bf g}_{j}, {\bf h}_{j} \}}$. The group elements $\{ {\bf g}_j,{\bf h}_j\}$ must satisfy relations to ensure that
the corresponding defect branch lines can close consistently upon themselves. In this case, we do not require that ${\bf g}_{j}$ and ${\bf h}_{j}$ necessarily commute. When they do not, one of the branch lines at a given handle may have its group element label change as it crosses the other branch line. If we pick the ${\bf h}_{j}$-branch lines to close around their cycles unchanged, then the ${\bf g}_{j}$ branch lines transform into ${\bf h}_{j}^{-1}{\bf g}_{j}{\bf h}_{j}$ branch lines when they cross the ${\bf h}_{j}$-branch. When this branch line loops back on itself, we are left with a nontrivial branch line, which requires a ${\bf k}_{j}$-branch line, where
\begin{equation}
{\bf k}_{j} = {\bf g}_{j} {\bf h}_{j}^{-1}  {\bf g}_{j}^{-1} {\bf h}_{j}
,
\end{equation}
to enter the handle and cancel this off, as shown in Fig.~\ref{fig:genusgLoops}.

Thus, while we do not require ${\bf g}_{j}$ and ${\bf h}_{j}$ to commute, we do, however, require that the product of their commutators ${\bf k}_j$ equals the identity group element, that is
\begin{equation}
\prod_{j=1}^g {\bf k}_{j} = \prod_{j=1}^g  {\bf g}_{j} {\bf h}_{j}^{-1}  {\bf g}_{j}^{-1} {\bf h}_{j} = {\bf 0}
,
\end{equation}
as this condition is necessary for a consistent configuration of branch lines that do not contain any free endpoints, as can be seen from Fig.~\ref{fig:genusgLoops}.

A basis for the ground state subspace $\mathcal{V}_{\{ {\bf g}_{j}, {\bf h}_{j} \}}$ of the $\{{\bf g}_j, {\bf h}_j\}$-sector can be given in terms of fusion trees of topological charge lines passing through the interior of the surface, as shown in Fig.~\ref{fig:genusgLoops}. Using the choice where the ${\bf h}_j$-branch lines loop around their cycles unchanged, we may have a charge line $a_j \in \mathcal{C}_{{\bf g}_{j}}$ that winds around the complementary cycle of the $j$th handle and transforms into $^{{\bf \bar{h}}_{j}}a_{j}$ when it crosses the ${\bf h}_j$-branch loop. In closing back on itself, this topological charge loop must fuse with a possibly nontrivial line of charge $z_j \in \mathcal{C}_{{\bf k}_j}$ such that $N_{a_j \,^{{\bf \bar{h}}_{j}}\overline{a_j} }^{z_j } \neq 0$. The charge $z_{j}$ lines from the different handles then form a fusion tree that must terminate in the trivial topological charge.

In particular, the basis states described in this way can be written as
\begin{equation}
\label{eq:genusgstate}
\bigotimes\limits_{j=1}^g | a_j , \,^{{\bf \bar{h}}_{j}}\overline{a_j} ; z_j , \mu_{j} \rangle | c_{1 \ldots j-1} , z_{j} ; c_{1 \ldots j} , \nu_{1 \ldots j}  \rangle,
\end{equation}
for all possible values (allowed by fusion) of topological charges $a_j \in \mathcal{C}_{{\bf g}_j}$, $z_j \in \mathcal{C}_{{\bf k}_j}$, and $c_{1\ldots j} \in \mathcal{C}_{{\bf l}_j}$ for ${\bf l}_j = \prod\limits_{i=1}^{j} {\bf k}_i$, and fusion vertex basis labels $\mu_{j} = 1,\ldots, N_{a_j \,^{{\bf \bar{h}}_{j}}\overline{a_j} }^{z_j }$, and $\nu_{1 \ldots j} = 1,\ldots, N_{c_{1 \ldots j-1}  z_{j} }^{ c_{1 \ldots j}}$. We set $c_{\emptyset}=c_{1 \ldots g}=0$ (which gives $c_{1}=z_{1}$) and ${\bf l}_g = {\bf 0}$, in order to let $j=1,\ldots,g$ for all these quantities.

We note that the states in Eq.~(\ref{eq:genusgstate}) may transform non-trivially under the symmetry action of ${\bf q} \in G$. In particular,
\begin{eqnarray}
\rho_{\bf q} \,: \, \mathcal{V}_{\{ {\bf g}_j, {\bf h}_j \}} &\rightarrow& V_{\{ {\bf q} {\bf g}_j {\bf q}^{-1}, {\bf q}{\bf h}_j {\bf q}^{-1} \}} \\
|\psi\rangle &\mapsto& \rho_{\bf q}( |\psi \rangle)
\end{eqnarray}
This symmetry action will play a crucial role when $G$ is promoted to a local gauge invariance.

In order to obtain the number of ground states in the $\{ {\bf g}_j,{\bf h}_j\}$-sector
\begin{align}
\mathcal{N}_{\{ {\bf g}_j, {\bf h}_j \}} = \text{dim } \mathcal{V}_{\{ {\bf g}_j, {\bf h}_j \}},
\end{align}
we can sum over the fusion channels
\begin{equation}
\mathcal{N}_{\{ {\bf g}_j,{\bf h}_j\} } = \sum_{ \substack{ a_j \in \mathcal{C}_{{\bf g}_{j} } \\ z_j \in \mathcal{C} _{{\bf k}_j}  }} N_{z_1 z_2 \cdots z_g}^{0} \prod_{j=1}^{g} N_{a_j \,^{{\bf \bar{h}}_{j}}\overline{a_j} }^{z_j },
\end{equation}
where
\begin{align}
N_{z_1 z_2 \cdots z_g}^{0} = \sum_{ c_{1\ldots j} \in \mathcal{C}_{{\bf l}_j}} N_{z_1 z_2}^{c_{12}} N_{c_{12} z_3}^{c_{123}}  \cdots N_{c_{1 \ldots g-1} z_g}^{0}
\end{align}
is the number of ways the topological charges $z_1,\ldots,z_{g}$ can fuse to $0$.
We can evaluate these expressions using the $G$-crossed Verlinde formula Eq.~(\ref{eq:G-crossed_Verlinde_2}), together with $G$-graded modularity and other properties that we derived for the $S$-matrix in Sec.~\ref{sec:G-Crossed_Invariants}, which yields
\begin{eqnarray}
\label{eqn:gDeg}
&& \mathcal{N}_{\{ {\bf g}_j, {\bf h}_j \}} = \mathcal{D}_{\bf 0}^{2g-2} \sum_{x \in \mathcal{C}_{\bf 0}^{ \{ {\bf g}_j, {\bf h}_j \} }} d_{x}^{-(2g-2)}
\notag \\
&& \qquad \times \prod_{j=1}^{g} \frac{ \eta_{x}( {\bf \bar{h} }_j , {\bf \bar{g} }_j  ) \eta_{x}( {\bf g}_j , {\bf {\bf \bar{h}}_j \bar{g} }_j {\bf h}_j )  }  { \eta_{x}( {\bf g}_j , {\bf \bar{g} }_j )  \eta_{x}( {\bf \bar{h}}_j {\bf \bar{g} }_j {\bf h }_j , {\bf \bar{h}}_j ) } \eta_{x}( {\bf l}_{j-1} , {\bf k}_j )
,\qquad
\end{eqnarray}
where $\mathcal{C}_{\bf 0}^{ \{ {\bf g}_j, {\bf h}_j \} }$ is the set of all topological charges in $\mathcal{C}_{\bf 0}$ that are ${\bf g}_j$-invariant and ${\bf h}_j$-invariant for all $j=1,\ldots ,g$.
When ${\bf h}_j = {\bf 0}$ for all $j$, this expression simplifies to
\begin{eqnarray}
\mathcal{N}_{\{ {\bf g}_j, {\bf h}_j =0 \}} = \mathcal{D}_{\bf 0}^{2g-2} \sum_{x \in \mathcal{C}_{\bf 0}^{ \{ {\bf g}_j \} }} d_{x}^{-(2g-2)}
,
\end{eqnarray}
which clearly satisfies $\mathcal{D}_{\bf 0}^{2g-2} \leq \mathcal{N}_{\{ {\bf g}_j, {\bf h}_j={\bf 0} \}} \leq \mathcal{D}_{\bf 0}^{2g}$. From Eq.~(\ref{eqn:gDeg}), we see that, in general, the genus $g$ degeneracy
$\mathcal{N}_{\{ {\bf g}_j, {\bf h}_j \}} \leq  \mathcal{N}_{\{ {\bf 0}_j, {\bf 0}_j \}} $, and generally scales as $\mathcal{N}_{\{ {\bf g}_j, {\bf h}_j \}} \sim \mathcal{D}_{\bf 0}^{2g}$ in the large $g$ limit,
regardless of the defect sector. This provides a physical interpretation of the total quantum dimension $\mathcal{D}_{\bf 0} = \mathcal{D}_{\bf g}$ of each $\mathcal{C}_{\bf g}$ subsector.

We note that another physical interpretation of the total quantum dimension $\mathcal{D}_{\bf 0}$ is given by the topological entanglement entropy~\cite{Kitaev06b,levin06a}. One can use the properties of $G$-crossed modularity to compute the topological entanglement entropy of a region, following the arguments of \Ref{Kitaev06b}. Unsurprisingly, this yields the same result as for MTCs
that $S_{\text{topo}} = - n \log \mathcal{D}_{\bf 0}$, where $n$ is the number of connected components of the boundary of the region in question,
regardless of the number of branch lines passing through the region. There are also anyonic contributions $S_{a} = \log d_{a}$ to the entanglement entropy
when there are quasiparticles or ${\bf g}$-defects within the region whose collective topological charge is $a$ (see also \Ref{brown2013}).

\subsubsection{Dehn twists on high genus surfaces}
\label{eq:Dehn_twists_appendix}

Another powerful method of computing $\mathcal{N}_{\{ {\bf g}_j, {\bf h}_j \}}$ on a genus $g$ surface is to make use of modular transformations. Similar to the case of the torus, we can define operators using the data of a $G$-crossed UMTC $\mathcal{C}_{G}^{\times}$, that provide a projective representation of the modular transformations of the genus $g$ surface. We will not go into these details here, but, instead, will simply utilize the property that the modular transformations can be used to interchange, combine, and twist the various non-contractible cycles of the surface, as we saw for the torus. Unitarity of the modular transformations implies that when two different defect sectors $\{{\bf{g}}_j, {\bf{h}}_j\}$ and $\{{\bf{g}}_j', {\bf{h}}_j'\}$ can be related by such modular transformations, they must have the same ground state degeneracy.

\begin{figure}[t!]
	\centering
	\includegraphics[width=1\columnwidth]{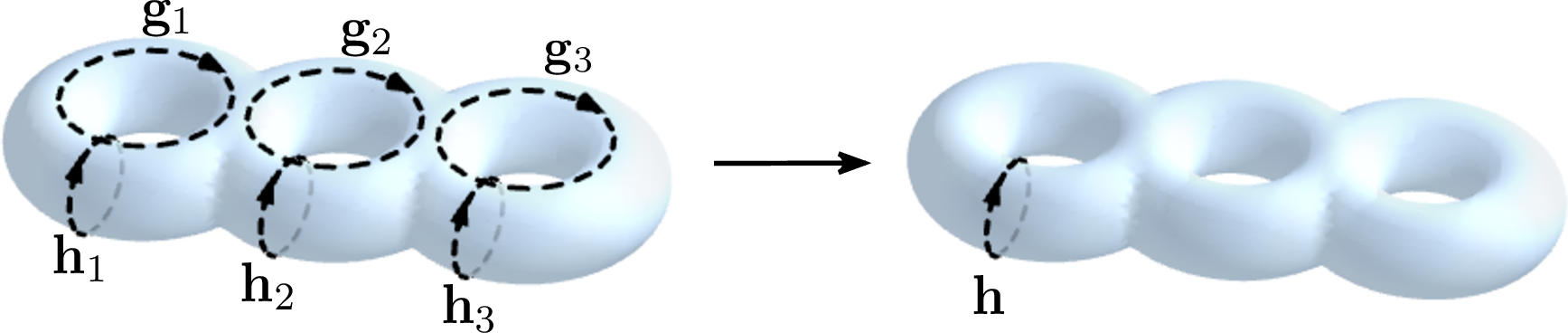}
\caption{When $G = \mathbb{Z}_p$ for $p$ prime, any defect sector can be mapped via Dehn twists to the sector with a single defect branch line corresponding to a element ${\bf h} \in \mathbb{Z}_p$, which generates the group. Thus, all ${\{ {\bf g}_j, {\bf h}_j \}}$-sectors that are not completely trivial must have the same ground state degeneracy.
}
\label{fig:dehnZp}
\end{figure}

As a simple example, let us consider $G = \mathbb{Z}_p$ and take ${\bf g}$ to be a generator of this group. When $p$ is prime, any element ${\bf h} \in \mathbb{Z}_p$ generates the group. In this case, every nontrivial ${\{ {\bf g}_j, {\bf h}_j \}}$-sector can be related by Dehn twists to the sector with only a single ${\bf h}$-defect branch line wrapped around a single cycle (see Fig.~\ref{fig:dehnZp}). The proof of this statement, and some generalizations, is given below. 

Specifically, in the following we show that for a genus $g$ surface, when $G = \mathbb{Z}_N$, the $(N^{2g} - 1)$ non-trivial defect sectors can all be obtained by Dehn twists from a small ``generating'' set of generating sectors (the case $N = 2$ was proven in \Ref{barkeshli2010}). We start by examining a torus ($g=1$). Since we are considering a cyclic group, group multiplication will be denoted additively. We label the defect sector by $({\bf g,h})=(m,n)$ where $m,n=0,1, \dots, N-1$. There are $N^2-1$ nontrivial sectors in total.

An arbitrary modular transformation acts on a defect sector $(m,n)$ by a SL$(2, \mathbb{Z})$ matrix
\begin{equation}
\left[
\begin{matrix}
		a & b\\
		c & d
\end{matrix}
\right]
	\begin{pmatrix}
		m \\ n
	\end{pmatrix}=
	\begin{pmatrix}
		a m + bn\\
		c m+dn
	\end{pmatrix}.
	\label{eqn:modular}
\end{equation}
Here $ad-bc=1, a,b,c,d\in\mathbb{Z}$. Letting $r=\text{gcd}(m,n)$, we now show that $(m,n)$ can be obtained from $(r,0)$ by a modular transformation. To see this, we set $a=\frac{m}{r},c=\frac{n}{r}$ in the SL$(2, \mathbb{Z})$ matrix. We then need to find $b,d$ such that $\frac{m}{r}d-\frac{n}{r}b=1$. Since gcd$\left( \frac{m}{r},\frac{n}{r} \right)=1$, this equation has integral solutions.

Next, we show that for arbitrary $m$, $(m,0)$ can be obtained from $(s,0)$ where $s=\text{gcd}(m,N)$. From \eqref{eqn:modular}, we see that we need to find an SL$(2,\mathbb{Z})$ matrix with $a=\frac{m}{s}$ and $c=\frac{N}{s}$. We need to find integers $b,d$ such that $ad-bc=\frac{m}{s}d-\frac{N}{s}b=1$, which is solvable since gcd$(m,N)=s$. Therefore we have established that \textit{the defect sectors $(r,0)$, where $r$ is a divisor of $N$, is a generating set.} That is, the number of generating defect sectors is equal to the number of divisors of $N$.

\begin{figure}[t!]
  \begin{center}
	\includegraphics[width=0.6\columnwidth]{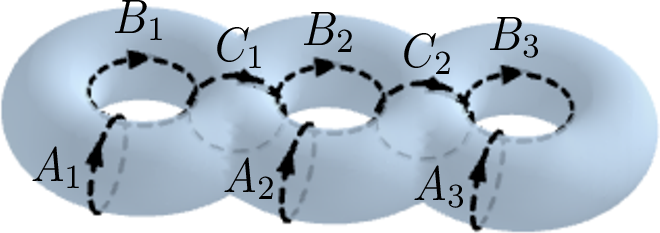}
  \end{center}
  \caption{Non-contractible cocycles on a $g=3$ surface.}
  \label{fig:genus3}
\end{figure}

We now consider a genus $g$ surface. A similar reduction of a general defect sector to a small number of generating defect sectors is also possible. The inequivalent cycles associated with each handle are labeled by $A_i, B_i$ where $i=1,\dots,g$ (see Fig. \ref{fig:genus3}). The defect sectors are now labeled by $2g$ integers (mod $N$) $\{(m_1, n_1), \dots, (m_g, n_g)\}$.
We note that applying a Dehn twist along $C_1$ has the following effect:
\begin{equation}
	\begin{split}
	A_1 &\rightarrow A_1\\
	B_1 &\rightarrow B_1+C_1=B_1-A_1+A_2\\
	A_2 &\rightarrow A_2\\
	B_2 &\rightarrow B_2-A_1+A_2.
	\end{split}
\end{equation}
The configuration then becomes $\{(m_1, n_1-m_1+m_2), (m_2, n_2-m_1+m_2), \dots\}$.

The arguments from the genus $g = 1$ case above imply that by applying Dehn twists along $A_i$ or $B_i$,
we can always map any general defect sector to the form $\{(m_1, 0), (m_2, 0), \dots, (m_g, 0)\}$.
If at least one  of the $m_i$'s is coprime with $N$, we can further perform Dehn twists to reduce the configuration to a defect branch line along a single cycle.
To see this, let us assume gcd$(m_1,N)=1$. We can do an S transformation to map to the configuration $\{(m_1, 0), (0, m_2),\dots\}$. After applying $k$ Dehn twists along $-C_1$, we get $\{(m_1, k m_1), (0, m_2+k m_1), \dots\}$. Since gcd$(m_1, N)=1$, there exists a $k$ such that $m_2+k m_1 \equiv 0\,(\text{mod } N)$, resulting in the sector $\{(m_1, k m_1), (0, 0), \dots\}$. This can be further reduced to $\{(m_1, 0), (0, 0), \dots\}$ by Dehn twists. A similar argument can be applied in the case when $m_1=m_2=\cdots=m_g$, without the need to assume gcd$(m_i,N)=1$.

In particular, the above arguments imply that when $N$ is prime, then the general defect sector can always be mapped to a sector with a single elementary defect branch line along only one cycle of the genus $g$ surface.

\section{Gauging the Symmetry}
\label{sec:gauging}

We have, so far, studied the properties of the defects, which correspond to extrinsically imposed (confined) fluxes of the symmetry group $G$, as described by a $G$-crossed theory $\ext{C}{G}$. In this section, we consider the nature of the phase that results when the global symmetry $G$ is promoted to a local gauge invariance -- ``gauging the symmetry.'' This is also referred to as ``equivariantization'' in the mathematical literature. A physical consequence of gauging the symmetry is that the confined ${\bf g}$-defects become deconfined quasiparticle excitations of the gauged phase. As such, the resulting phase is described by a topological phase described by a UMTC, which we denote as $\gauged{C}{G}$, conveying $G$-equivariantization of the $G$-crossed theory. We would like to understand how to obtain the properties and basic data of the gauged theory $\gauged{C}{G}$ from the $G$-crossed extension $\ext{C}{G}$ of the UMTC $\mathcal{C}$ describing the original topological phase.

Given the complete data of the $G$-crossed UMTC $\ext{C}{G}$, we will demonstrate how to obtain the quasiparticle content, fusion rules, quantum dimensions, and topological twists of the corresponding UMTC $\gauged{C}{G}$. We also use these results to provide an expression for the topological $S$-matrix of $\gauged{C}{G}$ in terms of that of $\ext{C}{G}$. The gauging procedure that we describe in this section does not require modularity and can, thus, be applied to a non-modular $G$-crossed UBTC $\ext{C}{G}$, in which case the original UBTC $\mathcal{C}$ and the resulting UBTC $\gauged{C}{G}$ will also be non-modular. For modular theories, we further use our results to show that the chiral central charges of $\mathcal{C}$, $\ext{C}{G}$, and $\gauged{C}{G}$ are all equal and we explain how to obtain the ground state degeneracies of $\gauged{C}{G}$ on higher genus surfaces from the $\ext{C}{G}$ theory.

We can also consider the inverse of the gauging construction. Starting from the gauged theory $\gauged{C}{G}$, we can tune the interactions so that the ``charged'' matter, which transforms under irreducible representations of $G$, condenses, and the system undergoes a continuous confinement-deconfinement transition into the Higgs phase. The resulting topological order can be analyzed using the theory of topological Bose condensation~\cite{bais2009}, where the subcategory, known as Rep$(G)$, consisting of gauge charges of $G$ condenses. In short, condensing Rep$(G)$ results in $\ext{C}{G}$; all defects with ${\bf g} \neq {\bf 0}$ in $\gauged{C}{G}$ become confined, while the deconfined remnants give rise to $\mathcal{C} = \mathcal{C}_{\bf 0}$. The algebraic theory of topological defects that we have developed in this paper provides a complete topological description of the system after topological Bose condensation of $\mathrm{Rep}(G)$, in particular providing the previously unknown braiding and modular transformations of the confined sectors, which is called the $\mathcal{T}$-theory in \Ref{bais2009}.

We summarize the relation between $\mathcal{C}$, $\ext{C}{G}$, and $\gauged{C}{G}$ by the following diagram:
\begin{equation*}
	\pspicture[shift=0](-3,-1.0)(3.3,1.0)
	\small
	\rput[br]{0}(-2.6,0){\scalebox{1.6}{$\mathcal{C}$}}
	\rput[bl]{0}(0.1,-0.1){\scalebox{1.5}{$\ext{C}{G}$}}
	\rput[bl]{0}(3.1,-0.1){\scalebox{1.5}{$\gauged{C}{G}$}}
	\psset{linewidth=0.9pt,arrowscale=1.5, arrowinset=0.15}
	\psline{->}(-2.3,0.25)(-0.2,0.25)
	\psline{<-}(-2.3,0)(-0.2,0)
	\psline{->}(0.9,0.25)(2.8,0.25)
	\psline{<-}(0.9,0)(2.8,0)
	\rput[bl]{0}(-2.16,0.45){\scalebox{0.9}{Defectification}}
	\rput[tl]{0}(-2.0,-0.15){\scalebox{0.9}{Confinement}}
	\rput[bl]{0}(1.25,0.37){\scalebox{0.9}{Gauging}}
	\rput[tl]{0}(1.15,-0.15){\scalebox{0.9}{Condensation}}
	\endpspicture
	\qquad
\end{equation*}

In general, distinct (gauge-inequivalent) $G$-crossed extensions $\ext{C}{G}$ always lead to distinct $\gauged{C}{G}$ as topological gauge theories. However, when viewed as UMTCs in which we neglect the origin of the charge and flux labels of the quasiparticles in $\gauged{C}{G}$, different $G$-crossed extensions can potentially result in the same $\gauged{C}{G}$. Examples of such phenomena have been noticed for gauging bosonic SPT phases in \Refs{ChengPRL2014, lu2013}.

Another notation for the gauged theory that is sometimes used in the literature is $\mathcal{C}/G$, which comes from applying category theory to the study of CFT orbifold models~\cite{kirillov2004}. It is worth stressing that there are important distinctions between gauging a symmetry in a topological phase or MTC and the closely related concept of orbifolding a rational CFT~\cite{difrancesco,dijkgraaf1989} (which may be viewed as gauging a symmetry in the CFT). While there is MTC structure in a rational CFT, there is additional structure in a CFT that does not exist in its corresponding MTC. Because of this property, certain applications of orbifolding in a CFT have an analogous realization as gauging a symmetry in the corresponding MTC, but others do not. For example, repeatedly applying the orbifold construction may return the original CFT, in which case orbifolding is analogous, in some sense, to both symmetry gauging and topological Bose condensing in topological phases.

In this section, we consider only finite symmetry groups $G$. We will first examine the problem of how to modify a microscopic Hamiltonian that realizes a topological phase $\mathcal{C}$ and has an on-site symmetry $G$ in a manner that gauges the symmetry and realizes the topological phase $\gauged{C}{G}$. Then we will study how to derive the mathematical properties of the gauged phase's UBTC $\gauged{C}{G}$ from the corresponding $G$-crossed UBTC $\ext{C}{G}$.

\subsection{Microscopic Models}

Gauging a symmetry of a microscopic Hamiltonian is a well-known
notion in physics.  However, a gauge theory does not, in general, have a local Hilbert space.
Suppose we are given a $G$-symmetric microscopic Hamiltonian $H$ that (1) is defined on a Hilbert space that decomposes
into a tensor product of local Hilbert spaces on each site, (2) has local interactions,
and (3) realizes a topological phase $\mathcal{C}$ at long-wavelengths. Here, we address the question of whether or not we can produce
a new Hamiltonian $H_G$ that also satisfies (1) and (2) above, but realizes $\gauged{C}{G}$
at long-wavelengths.

We will briefly describe the case where $G = \mathbb{Z}_2$. Suppose that the Hamiltonian consists of nearest neighbor interactions on a two-dimensional lattice. We assume that there is a finite-dimensional bosonic Hilbert space at each site of the lattice, and there is a global on-site $\mathbb{Z}_2$ symmetry with $R_{\bf g}=\prod_j R^{(j)}_{\bf g}$. Such a $\mathbb{Z}_2$ symmetric
Hamiltonian can generically be written as
\begin{align}
H = &\sum_{\langle i j \rangle} J_{+,ij}^{\alpha \beta} \mathcal{O}^{+,\alpha}_i \mathcal{O}^{+,\beta}_j
+ J_{-,ij}^{\alpha \beta} \mathcal{O}^{-,\alpha}_i \mathcal{O}^{-,\beta}_j
\nonumber \\
&+ \sum_i m_i^\alpha \mathcal{O}_i^{+,\alpha} + \text{H.c.},
\end{align}
where $\{\mathcal{O}^{\pm,\alpha}_{j} \}$ are a complete set of $\mathbb{Z}_2$ even/odd local operators at site $j$. In particular, these operators satisfy $R^{(j)}_{\bf g} \mathcal{O}_j^{\pm} R^{(j)-1}_{\bf g}= \pm \mathcal{O}_j^\pm$.

Now, let us introduce a two-dimensional Hilbert space on each bond $\langle i j \rangle$
of the lattice. The gauged Hamiltonian is defined as
\begin{align}
H_{\mathbb{Z}_2} &= \sum_{\langle i j \rangle} J_{+,ij}^{\alpha \beta} \mathcal{O}^{+,\alpha}_i \mathcal{O}^{+,\beta}_j
+ \sum_i m_i^\alpha \mathcal{O}_i^{+,\alpha} \nonumber \\
&+\sum_{\langle ij\rangle}J_{-,ij}^{\alpha \beta} \mathcal{O}^{-,\alpha}_i \mathcal{O}^{-,\beta}_j \sigma_{ij}^z +\text{H.c.}
\nonumber\\
&- K \sum_{\square} \prod_{\langle i j \rangle \in \square} \sigma_{ij}^z
- \Gamma\sum_{\langle ij\rangle} \sigma^x_{ij}- U \sum_{+} R^{(i)}_{\bf g} \prod_{\langle i j \rangle \in +} \sigma_{ij}^x .
\end{align}
We always assume that $U$ is the largest energy scale, which effectively imposes a $\mathbb{Z}_2$ analog of Gauss's law in the low-energy Hilbert space: $\prod_{\langle i j \rangle \in +} \sigma_{ij}^x=R^{(i)}_{\bf g}$. It is straightforward to extend the construction to Hamiltonians involving longer-range interactions.

We notice that the full gauged Hamiltonian (not just the low-energy subspace) still preserves the $\mathbb{Z}_2$ symmetry $R_{\bf g}$. In the low-energy subspace $U\rightarrow \infty$ where the dynamics can be described by a $\mathbb{Z}_2$ gauge theory with matter, the global symmetry is enhanced to a local gauge symmetry generated by precisely the local conserved quantity $R^{(i)}_{\bf g} \prod_{\langle i j \rangle \in +} \sigma_{ij}^x$. The gauged Hamiltonian has the feature that when $\Gamma=0$ and $K, U$ are both much larger than any energy scale in $H$, the low-energy spectrum without any $\mathbb{Z}_2$ fluxes is identical to that of $H$. However, the states must be projected to the gauge-invariant Hilbert space.

We now review the phase diagram of the gauge theory~\cite{FradkinPRD1979}, focusing on the three parameters $J_{-}, K$ and $\Gamma$. Three limiting cases can be easily identified. When $J_-, \Gamma\ll K$, the gauge field is in the deconfined phase. When $J_-\gg K, \Gamma$, the gauge theory is in the Higgs phase and the $\mathbb{Z}_2$ fluxes (i.e. visons) are linearly confined. If $\Gamma$ is dominant, $\mathbb{Z}_2$ charges are linearly confined. It is however well-known that the Higgs and the confinement phases are smoothly connected. Hence there are only two phases which are separated by a second-order phase transition belonging to the 3D Ising universality class~\cite{FradkinPRD1979}.

The above construction can straightforwardly be generalized to the case $G = \mathbb{Z}_N$. The generalization to a general
finite group $G$ is technically more involved and will be left for future work.

\subsection{Topological Properties of the Gauged Theory}
\label{sec:gaugeQP}

We now derive the topological properties of the gauged theory $\gauged{C}{G}$, described by a new UBTC, which are specified in terms of properties of the corresponding $G$-crossed extension $\ext{C}{G}$. We will not specify the $F$-symbols and $R$-symbols, but simply focus on the gauge invariant data given by the topological charges and their fusion rules, quantum dimensions, and topological twists. It is a conjecture that this gauge invariant topological data, which is equivalent to specifying the modular $S$ and $T$ transformations, uniquely characterizes a MTC, i.e. that it uniquely specifies the $F$-symbols and $R$-symbols, up to gauge equivalence. We also examine the relation between the topological $S$-matrix of the gauged theory and that of the $G$-crossed theory, as well as the ground state degeneracy on higher genus surfaces (when the theory is modular).

\subsubsection{Topological charges}

The simplest information about the gauged theory $\gauged{C}{G}$ that we can read off from $\ext{C}{G}$ is the topological charge content. The mathematical description of this was provided in \Ref{Burciu2013}.

For each topological charge (simple object) $a \in \ext{C}{G}$ of the $G$-crossed theory, including those of defects, we define its orbit under $G$ to be the set of charges
\begin{align}
[a] = \{ ^\mb{g}a, \forall \mathbf{g} \in G\}.
\end{align}
(We will often leave the corresponding group element labels of topological charges in $\ext{C}{G}$ implicit in this section, except when it is necessary or useful.)
Heuristically, the reason for considering $G$ orbits is that, under the $G$ action, all topological charges within an orbit must combine into a single object by ``quantum superposition'' once the global symmetry is promoted to a local gauge invariance. In this way, the original topological charges in $\ext{C}{G}$ become internal degrees of freedom. In particular, if we ignore the topological charge labels within each $\mathcal{C}_\mb{g}$ and only focus on the group elements, the orbit would simply be a conjugacy class of $G$, which is what labels gauge fluxes in a discrete gauge theory. Keeping track of the topological charge labels, it is clear that there can be multiple orbits associated with a given conjugacy class of $G$.

Additionally, we need to take into account the different representations of the symmetry, which thus allows us to include the gauge charges and flux-charge composites. For this, we do not consider the full symmetry group $G$, but rather the subgroups that keep the relevant topological charge labels invariant. More precisely, for a given $[a]$, we choose a representative element $a \in [a]$, and define its stabilizer subgroup
\begin{equation}
G_a=\{\, \mb{g} \in G  \,|\,^\mb{g}a=a\,\}.
\end{equation}
The topological charges of $\gauged{C}{G}$ are then defined to be the pairs
\begin{equation}
([a], \pi_a),
\label{eq:charges_gauged}
\end{equation}
where $\pi_a$ is an irreducible projective representation of $G_a$ with the factor set given by $\eta_a$, i.e.
\begin{align}
\pi_a({\bf g}) \pi_a({\bf h}) = \eta_a({\bf g},{\bf h}) \pi_a({\bf gh}), \;\; {\bf g, h} \in G_a.
\end{align}
We will refer to such an irreducible projective $\eta_a$-representation as an $\eta_a$-irrep.
The phases $\eta_a({\bf g},{\bf h})$ here are precisely the projective symmetry fractionalization phases of the $G$-crossed theory, defined in Sec.~\ref{sec:Algebraic_Theory}. Thus, we see that the data $\eta_a$ are essential in defining the quasiparticles of the gauged theory.

In this way, the topological charges of $\gauged{C}{G}$ are essentially dyonic excitations, very much like ``flux-charge'' composites in discrete gauge theories, but generalized to account for distinct types of ${\bf g}$-flux defects $a \in \mathcal{C}_{\bf g}$.~\footnote{Discrete gauge theories $\text{D}^{[\alpha]}(G)$ are obtained by gauging the $G$ symmetry in a trivial topological phase $\mathcal{C}_{\bf 0}=\{ 0 \}$, as described in Sec.~\ref{sec:trivial}.} The $G$-orbits $[a]$ here play the same role as the conjugacy classes $[{\bf g}]$, describing fluxes in discrete gauge theories; the projective $\eta_a$-irreps $\pi_a$ of the stabilizer subgroup $G_{a}$ for some $a \in [a]$ here play the same role as the irreps $\pi_{\bf g}$ of the centralizer $C_{G}({\bf g})$ for some ${\bf g} \in [{\bf g}]$, describing charges in discrete gauge theories.

In order for this definition of topological charge to be well-defined, the specific choice of $a$ within the conjugacy class $[a]$ should not lead to essential differences in the corresponding projective representations. To make this notion more precise, we first notice that conjugation by $\mb{k}\in G$ provides a canonical isomorphism between $G_{a}$ and $G_{^{\bf k}a}$
\begin{eqnarray}
{\bf k} : G_{a} &\rightarrow& G_{^{\bf k}a}
\notag \\
{\bf g} &\mapsto& \,^{\bf k} {\bf g}
.
\end{eqnarray}
Next, from Eq.~(\ref{eq:eta_consistency}), we see that, for group elements $\mb{g},\mb{h} \in G_a$, we have the cocycle condition
\begin{equation}
\frac{ \eta_{a}\left({\bf h}, {\bf k} \right) \eta_{a}\left({\bf g}, {\bf hk} \right) }{ \eta_{a}\left({\bf g}, {\bf h} \right) \eta_{a}\left({\bf gh}, {\bf k} \right) } =1
,
\end{equation}
so $\eta_{a} \in Z^{2}(G_{a},\text{U}(1))$. From Eq.~(\ref{eq:eta_k-action}), we see that, for $\mb{g},\mb{h} \in G_a$, we have the relation
\begin{eqnarray}
\eta_{^\mb{k}a}(^\mb{k}\mb{g},\!\!\,^\mb{k}\mb{h} ) &=& \frac{ \eta_a(\mb{\bar{k}},\mb{^kh}) }{ \eta_a(\mb{h}, \mb{\bar{k}})}
\frac{ \eta_a(\mb{gh}, \mb{\bar{k}})}{ \eta_a(\mb{\bar{k}},\mb{^kg^kh}) }  \frac{ \eta_a(\mb{\bar{k}},\mb{^kg}) }{ \eta_a(\mb{g}, \mb{\bar{k}})}
\eta_a(\mb{g},\mb{h}) \notag \\
&=& \text{d} \varepsilon_{a,{\bf k}}({\bf g,h } )  \eta_a(\mb{g},\mb{h}),
\label{eqn:action_on_rep}
\end{eqnarray}
where we have defined the $1$-cochain $\varepsilon_{a,{\bf k}} \in C^{1}(G_{a},\text{U}(1) )$ to be $\varepsilon_{a,{\bf k}}({\bf g } ) = \frac{ \eta_a(\mb{\bar{k}},\mb{^kg}) }{ \eta_a(\mb{g}, \mb{\bar{k}})}$. Thus, when viewed in terms of cohomology, we see that the ${\bf k}$-action does not change the cohomology class of $\eta_{a}$, i.e.
\begin{equation}
[\eta_{^\mb{k}a}(^\mb{k}\mb{g},\!\!\,^\mb{k}\mb{h} )] = [ \eta_a(\mb{g},\mb{h})] \in H^{2}(G_{a},\text{U}(1) )
.
\end{equation}
Moreover, it is clear that we then also have
\begin{equation}
[\eta_{^\mb{k}a}(^\mb{k}\mb{g},\!\!\,^\mb{k}\mb{h} )] = [ \eta_a(\mb{g},\mb{h})] \in H^{2}(G_{^{\bf k}a},\text{U}(1) )
.
\end{equation}
As discussed in Appendix~\ref{sec:Projective_reps}, this implies that there is a canonical one-to-one correspondence between the set of $\eta_a$-irreps of $G_a$ and the set of $\eta_{^{\bf k}a}$-irreps of $G_{^\mb{k}a}$. We will write
\begin{equation}
^\mb{k}\pi_a (^\mb{k}{\bf g}) = \varepsilon_{a,{\bf k}} ({\bf g}) \, \pi_a ({\bf g})
\label{eq:k_pi_a}
\end{equation}
to denote the $\eta_{^{\bf k}a}$-irrep of $G_{^\mb{k}a}$ which is canonically isomorphic to the $\eta_{a}$-irrep $\pi_a$ of $G_{a}$ under this mapping.

\subsubsection{Quantum dimensions}

With the definition of the topological charges of $\gauged{C}{G}$ specified in Eq.~(\ref{eq:charges_gauged}), it is straightforward to determine the corresponding quantum dimensions. In particular, we just sum over the quantum dimensions of all the charges in the orbit and multiply by the dimension of the attached $\eta_a$-irrep, so that $([a], \pi_a)$ has quantum dimension given by
\begin{equation}
d_{([a], \pi_a)} = d_a \cdot \big|[a]\big| \cdot  \text{dim}(\pi_a),
\label{eq:q_dim_gauged}
\end{equation}
where $d_a$ is the quantum dimension of $a$ (which is the same for all $a \in [a]$), $\big|[a]\big|$ the number of elements in the orbit $[a]$, and $\text{dim}(\pi_a)$ the
dimension of the $\eta_{a}$-irrep $\pi_a$.

Having specified the topological charges of $\gauged{C}{G}$ and their quantum dimensions, it is straightforward to prove that the total quantum dimension is
\begin{align}
\mathcal{D}_{\gauged{C}{G}}= |G|^{\frac{1}{2}} \mathcal{D}_{\mathcal{C}_{G}} = |G|\mathcal{D}_{\bf 0}= |G|\mathcal{D}_{\mathcal{C}}.
\end{align}
For this, we first consider the different $\eta_{a}$-irreps of the stabilizer subgroup $G_a$ of $a$ in a given orbit $[a]$. It is known that $\sum_{\pi_a}|\text{dim}(\pi_a)|^2=|G_{a}|$ for such $\eta_{a}$-irreps, as shown in Appendix~\ref{sec:Projective_reps}. With this, and the fact that $\big|[a]\big||G_{a}| = |G|$, we obtain the result
\begin{eqnarray}
\mathcal{D}_{\gauged{C}{G}}^{2} &=& \sum_{([a], \pi_a) \in \gauged{C}{G}} d_{([a], \pi_a)}^{2} \notag \\
&=& \sum_{([a], \pi_a)} d_a^{2} \big|[a]\big|^{2}  |\text{dim}(\pi_a)|^2 = \sum_{ [a] } d_a^{2} \big|[a]\big|^{2}  |G_{a}|
\notag \\
&=& |G| \sum_{[a]} d_a^{2} \big|[a]\big| = |G|  \sum_{a \in \mathcal{C}_{G}^{\times}} d_{a}^{2}
\notag \\
&=&  |G| \mathcal{D}_{\mathcal{C}_{G}}^{2} = |G|^{2} \mathcal{D}_{\mathcal{C}_{\bf 0}}^{2}
.
\end{eqnarray}

\subsubsection{Fusion rules}

The fusion rules for the topological charges of $\gauged{C}{G}$ have also been recently described in the mathematical literature~\cite{Burciu2013}. To obtain these, we need to understand both how to fuse two $G$-orbits and how to fuse two $\eta_{a}$-irreps. For pedagogical reasons, we will give a heuristic discussion to justify the fusion rules of $\gauged{C}{G}$ before presenting the actual expression.

We first consider a very coarse version of the problem. In particular, we suppress the topological charge label associated with an orbit and multiply two conjugacy classes $C_1$ and $C_2$ of $G$. For this, we first form the product set
\begin{equation*}
\{\mb{g}_1\mb{g}_2 \,|\, \mb{g}_1\in C_1, \mb{g}_2\in C_2\},
\end{equation*}
which can be equivalently expressed using representative elements $\mathbf{g}_1\in C_1$ and $\mathbf{g}_2\in C_2$ as
\begin{equation*}
\{\mb{h}\mb{g}_1\mb{h}^{-1}\mb{k}\mb{g}_2\mb{k}^{-1} \,|\, \mb{h}\in G/N_{\mb{g}_1}, \mb{k}\in G/N_{\mb{g}_2}\},
\end{equation*}
where
\begin{equation}
N_\mathbf{g} = \{ \mb{h} \in G \, | \, \mb{g}\mb{h}=\mb{h}\mb{g} \},
\end{equation}
denotes the centralizer of $\mathbf{g}$ in $G$. Now the problem is to decompose the product set into conjugacy classes. To this end, we observe that if $\mb{h}'=\mb{l}\mb{h}$ and $\mb{k}'=\mb{l}\mb{k}$, then $\mb{h}'\mb{g}_1\mb{h'}^{-1}\mb{k}'\mb{g}'_2\mb{k'}^{-1} =\mb{l}(\mb{h}\mb{g}_1\mb{h}^{-1}\mb{k}\mb{g}_2\mb{k}^{-1})\mb{l}^{-1}$, i.e. the two elements are in the same coset. Hence, we are naturally led to conclude that the conjugacy classes contained in the product set are given by the coset of diagonal left multiplication on $G/N_{\mathbf{g}_1}\times G/N_{\mathbf{g}_2}$, which is the double coset $N_{\mathbf{g}_1}\backslash G/N_{\mathbf{g}_2}$.

We now return to the problem of the fusion of two orbits $[a]$ and $[b]$, neglecting for the moment the $\eta_{a}$-irreps attached to them. Selecting representative elements $a\in [a]$ and $b\in [b]$, the fusion of the two orbits give a direct sum of all the elements in the set
\begin{equation*}
\{{\rho}_\mathbf{g}(a)\times\rho_\mathbf{h}(b)\,|\,\mathbf{g}\in G/G_a, \mathbf{h}\in G/G_b\}
,
\end{equation*}
where we take the coset over $G_a$ and $G_b$ here, since these subgroups do not modify the corresponding labels. We now need to decompose this set further into $G$-orbits. For this, we have the similar property that if $\mathbf{g}'=\mathbf{k}\mathbf{g}$ and $ \mathbf{h}'=\mathbf{k}\mathbf{g}$, then
\begin{eqnarray}
{\rho}_{\mathbf{g}'}(a)\times\rho_{\mathbf{h}'}(b)&=& \rho_{\mathbf{k}}\big(\rho_\mathbf{g}(a)\big)\times \rho_{\mathbf{k}}\big(\rho_\mathbf{h}(b)\big)
\notag \\
&=& \rho_{\mathbf{k}}\big(\rho_\mathbf{g}(a)\times \rho_\mathbf{h}(b)\big).
\end{eqnarray}
This essentially says that the fusion channels of ${\rho}_{\mathbf{g}'}(a)\times\rho_{\mathbf{h}'}(b)$ are exactly the image of those of $\rho_\mathbf{g}(a)\times \rho_\mathbf{h}(b)$ under the action of $\mathbf{k}$. Therefore, fusion of orbits correspond to the equivalence classes of $G/G_a\times G/G_b$ under diagonal left (or right) multiplication, which is known to be isomorphic to the double coset $G_a\backslash G/G_b$.

Next, we consider how the $\eta_{a}$-irreps attached to the defects should be combined. Na\"ively, one would expect that we just take the tensor product of the representations and decompose it as a direct sum of irreps. However, an important subtlety in this case is that the fusion/splitting spaces of the defects can transform nontrivially under the symmetry group action, and this should also be taken into account in the fusion. More explicitly, we consider the fusion/splitting vertex state spaces  $V_{a_\mb{g} b_\mb{h}}^{c_{\mb{gh}}}$ and $V^{a_\mb{g} b_\mb{h}}_{c_{\mb{gh}}}$, and we define the stablizer subgroup for this space as $H_{(a,b;c)}= G_a \cap G_b \cap G_c$. The symmetry action (sliding moves) consistency Eq.~(\ref{eqn:symmetry_action_consistency}) tells us that
\begin{multline}
\sum_{\lambda,\delta}[U_{\mb{l}}(a,b;c)]_{\mu\lambda}[U_{\mb{k}}(a,b;c)]_{\lambda\nu}\\
 =\frac{\eta_c(\mb{k},\mb{l})}{\eta_a(\mb{k},\mb{l})\eta_b(\mb{k},\mb{l})}[U_{\mb{kl}}(a,b;c)]_{\mu\nu}
\end{multline}
for $\mb{k},\mb{l}\in H_{(a,b;c)}$. We notice that the $U$ transformations can be thought of as being associated with the action on the splitting spaces $V^{ab}_c$, while the transpose $U^{\text{T}}$ corresponds to the action on the fusion spaces $V_{ab}^c$, as seen in Eqs.~(\ref{eqn:sliding_fusion5}) and (\ref{eqn:sliding_fusion6}). The symmetry action consistency implies that $U^{\text{T}}$ form a projective representation of $H_{(a,b;c)}$, with a factor set given by
\begin{equation}
\kappa_{\bf k,l} (a,b;c)^{-1} = \frac{\eta_c ({\bf k,l})}{\eta_a ({\bf k,l}) \eta_b ({\bf k,l})}
\end{equation}
restricted to ${\bf k,l} \in H_{(a,b;c)}$. We will denote this projective representation of $H_{(a,b;c)}$ by $U^{\text{T}}$ as $\pi_{(a,b;c)}$ and its character is given by
\begin{equation}
\chi_{\pi_{(a,b;c)}}(\mb{k}) =\sum_{\mu}\big[U_\mb{k}(a,b;c)\big]_{\mu \mu}
.
\label{eq:chi_pi_abc}
\end{equation}

With the above discussion as justification, we present the formula for the fusion coefficients of the $\gauged{C}{G}$ MTC~\cite{Burciu2013}
\begin{widetext}
\begin{equation}
  \begin{split}
	  N_{([a], \pi_a) ([b], \pi_b)}^{([c],\pi_c)} &=
\sum_{(\mathbf{t},\mathbf{s})\in G_a\backslash G/G_b} m \left( \pi_c\big|_{H_{(^\mb{t}a, \,^\mb{s}b ; c)}} \, , \,^\mb{t} \pi_a \big|_{H_{(^\mb{t}a, \,^\mb{s}b ; c)}} \otimes \,^\mb{s} \pi_b \big|_{H_{(^\mb{t}a, \,^\mb{s}b ; c)}} \otimes \pi_{(^\mb{t}a, \,^\mb{s}b ; c)} \right)
,
  \end{split}
\label{eqn:fusion}
\end{equation}
\end{widetext}
where $H_{(^\mb{t}a, \,^\mb{s}b ; c)}=G_{^\mb{t} a} \cap G_{^\mb{s} b} \cap G_c$ and the notation $\pi\big|_{H_{(^\mb{t}a, \,^\mb{s}b ; c)}}$ means the restriction of the irrep $\pi$ to the subgroup $H_{(^\mb{t}a, \,^\mb{s}b ; c)}$. As we discussed above, the tensor product $\,^\mb{t} \pi_a \big|_{H_{(^\mb{t}a, \,^\mb{s}b ; c)}} \otimes \,^\mb{s} \pi_b \big|_{H_{(^\mb{t}a, \,^\mb{s}b ; c)}} \otimes \pi_{(^\mb{t}a, \,^\mb{s}b ; c)}$ has the factor set given by
\begin{equation}
\eta_a ({\bf k,l}) \eta_b ({\bf k,l}) \kappa_{\bf k,l} (a,b;c)^{-1} = \eta_c ({\bf k,l})
\end{equation}
for ${\bf k,l} \in H_{(^\mb{t}a, \,^\mb{s}b ; c)}$, which is precisely the same factor set as $\pi_c\big|_{H_{(^\mb{t}a, \,^\mb{s}b ; c)}}$. We note that the restriction of an irrep to a subgroup is not necessarily an irrep of the subgroup. Finally, $m(\cdot, \cdot)$ is a sort of integer-valued inner product that, in some sense, measures the multiplicity of the entries with respect to each other. If one of the entries is an irrep, then this multiplicity function simply counts the number of times this irrep occurs in the other entry's irrep decomposition. However, the general description of the multiplicity function is more complicated than the statement that it counts the number of times one entry occurs in the other. The precise definition of this multiplicity function $m$ is given in Appendix~\ref{sec:Projective_reps}. For practical purposes, it may be computed in terms of the projective characters of the projective representations, as in Eq.~(\ref{eqn:multiplicity}).

The formula in Eq.~(\ref{eqn:fusion}) may appear obtuse without some experience in using it for concrete computations. For this, we refer the reader to Sec.~\ref{exampleSec}, where this formula is utilized to derive the fusion rules of the gauged theory for several examples.

As the first application of this formula, we determine the topological charge conjugate (antiparticle) of $([a], \pi_a)$. It should be clear that if $([b], \pi_b)$ is the charge conjugate $([a], \pi_a)$, then $[b] = [\bar{a}]$, since, for each $a \in [a]$, there must be an element $b \in [b]$ such that $N_{ab}^{0} \neq 0$. Regarding the $\eta_{\bar{a}}$-irrep of the conjugate charge, a natural guess would be the conjugate irrep $\pi_a^*$, since $\pi_a \otimes \pi_a^* =\openone \oplus \cdots$. However, the factor set of $\pi_a^*$ is $\eta_a^*$, which is in general only gauge-equivalent to $\eta_{\bar{a}}$. In fact, from the symmetry action consistency Eq.~(\ref{eqn:symmetry_action_consistency}), we have the relation
\begin{equation}
\eta_a(\mb{k},\mb{l})\eta_{\ol{a}}(\mb{k},\mb{l})=\frac{U_{\mb{kl}}(a,\ol{a};0)}{U_{\mb{k}}(a,\ol{a};0)U_{\mb{l}}(a,\ol{a};0)}
,
\end{equation}
for $\mb{k},\mb{l}\in G_a$. It follows that we should define the charge conjugate's irrep to be
\begin{equation}
\ol{\pi}_a (\mb{k}) = U_\mb{k}(a,\bar{a};0)^{-1} \pi_a^*(\mb{k}) .
\label{eq:conjugate_pi}
\end{equation}
This is, indeed, an $\eta_{\bar{a}}$-irrep of $G_{\bar{a}}$, i.e. it has the factor set $\eta_{\bar{a}}$. Thus, the topological charge conjugate of $([a], \pi_a) \in \gauged{C}{G}$ is
\begin{equation}
\overline{ ([a], \pi_a) } = ([\bar{a}],\ol{\pi}_a)
\end{equation}
with $\bar{a}$ the charge conjugate of $a \in \ext{C}{G}$ and $\ol{\pi}_a$ the  $\eta_{\bar{a}}$-irrep of $G_{\bar{a}}$ defined in Eq.~(\ref{eq:conjugate_pi}).
We can verify this by plugging $([a], \pi_a)$ and $\overline{ ([a], \pi_a) } $ into Eq.~(\ref{eqn:fusion}), where we would find that the tensor product in the second entry of $m$ simply becomes $\pi_a\otimes \pi_a^*$ which contains the trivial representation $\openone$ precisely once.

\subsubsection{Topological twists}

As we have discussed above, a topological charge in $\gauged{C}{G}$ has the form of a generalized dyon, the ``flux'' being a $G$-orbit of defects and the ``charge'' being a projective $\eta$-irrep. Thus, we expect that the topological twist of such objects will receive a contribution from the defect's twist (carrying over from the $\ext{C}{G}$ theory), as well as an Aharonov-Bohm type phase from the (internal) braiding of the object's flux and charge around each other. The latter contribution is roughly given by the character of the projective irreps, as it is in discrete gauge theories. Therefore, we have the following formula for the topological twists of topological charges in $\gauged{C}{G}$
\begin{equation}
\theta_{([a],\pi_a)}=\theta_{a_\mb{g}} \frac{\chi_{\pi_a}(\mb{g})}{\chi_{\pi_a}(\mb{0})}.
\label{eqn:twist}
\end{equation}
In this expression, $\theta_{a_\mb{g}}$ is the topological twist of $a_{\bf g} \in \ext{C}{G}$ and
\begin{equation}
\chi_{\pi_a}(\mb{g}) = \tr\big[\pi_a (\mb{g})\big]
\end{equation}
is the projective character of the $\eta_{a}$-irrep $\pi_a$ (see Appendix~\ref{sec:Projective_reps}). $\chi_{\pi_a}(\mb{0}) =\dim(\pi_a)$ is equal to the dimension of $\pi_a$.

It is straightforward to see that this expression for $\theta_{([a],\pi_a)}$ is indeed equal to a phase. Specifically, since $\eta_{a_{\bf g}}(\mb{g},\mb{h})=\eta_{a_{\bf g}}(\mb{h},\mb{g})$ for all $\mb{h}\in G_a$, it follows that $\pi_a(\mb{g})\pi_a(\mb{h})=\pi_a(\mb{h})\pi_a(\mb{g})$. Using Schur's lemma, we deduce that $\pi_a(\mb{g})\propto \openone$. Since the representations are unitary, it follows that $\frac{\chi_{\pi_a}(\mb{g})}{\chi_{\pi_a}(\mb{0})}$ is a $\mathrm{U}(1)$ phase.

We stress that the projective character depends on the particular factor set $\eta_{a}$, not just the equivalence class to which it belongs. While neither $\theta_{a_{\bf g}}$ nor $\chi_{\pi_a}(\mb{g})$ is individually invariant under the symmetry action gauge transformations, their product actually is invariant under such gauge transformations. More explicitly, under a symmetry action gauge transformation, as in \eqref{eqn:gauge}, the projective character transforms as
\begin{equation}
\check{\chi}_{\pi_a}(\mb{g})=\gamma_a^{-1}(\mb{g})\chi_{\pi_a}(\mb{g})
\end{equation}
and $\check{\theta}_{a_\mb{g}}=\gamma_a(\mb{g})\theta_{a_\mb{g}}$. Thus,
\begin{equation}
\check{\chi}_{\pi_a}(\mb{g})\check{\theta}_{a_\mb{g}}=\chi_{\pi_a}(\mb{g})\theta_{a_\mb{g}}.
\end{equation}
We also notice that vertex basis transformations leave both $\theta_{a_\mb{g}}$ and $\chi_{\pi_a}$, and hence $\theta_{([a],\pi_a)}$ invariant.

We must also check that $\theta_{([a_\mb{g}],\pi_a)}$ does not depend upon the choice of $a_{\bf g} \in [a]$. Consider a different representative element $^\mb{k}a_\mb{g}$ with $\mb{k}\in G/G_a$. In Eq.~(\ref{eqn:action_on_twist}), we saw that
\begin{equation}
\theta_{^\mb{k}a_{\bf g}}=\frac{\eta_a(\mb{g},\mb{\bar{k}})}{\eta_a(\mb{\bar{k}},\mb{^kg})}\theta_{a_{\bf g}}.
\end{equation}
As shown in the previous subsection, there is a canonical correspondence between the projective representations of $G_{^\mb{k}a}$ and $G_a$. Thus, we choose the projective representation for $^\mb{k}a$ to be $^\mb{k}\pi_a$. According to Eq.~(\ref{eq:k_pi_a}), we have
\begin{equation}
\chi_{^\mb{k}\pi_a}(\mb{^kg})=\frac{\eta_a(\mb{\bar{k}}, \mb{^kg})}{\eta_a(\mb{g},\mb{\bar{k}})}\chi_{\pi_a}(\mb{g})
.
\end{equation}
This results in the relation
\begin{equation}
\theta_{^\mb{k}a_{\bf g}}\chi_{^\mb{k}\pi_a}(\mb{^kg})=\theta_{a_{\bf g}} \chi_{\pi_a}(\mb{g})
,
\end{equation}
which demonstrates that the expression for the topological twist is indeed independent of the choice of $a \in [a]$.

As a special case, we notice that if $[a]\in \mathcal{C}_\mb{0}$, then $\theta_{[a]}=\theta_a$, which is expected from the theory of topological Bose condensation.

\subsubsection{Topological $S$-matrix}

Given the topological twists and fusion rules, we can compute the modular data of a UMTC, since the quantum dimensions can be obtained from the fusion rules and the $S$-matrix is defined in terms of these quantities by Eq.~(\ref{eqn:mtcs}). However, it is illuminating to obtain an expression for the $S$-matrix of $\gauged{C}{G}$ in terms of the $G$-crossed $S$-matrix of $\ext{C}{G}$. We begin the derivation with the expression for $S_{\overline{([a],\pi_a)} ([b],\pi_b)}$ in terms of the gauged theory's data (where we use the topological charge conjugate of $([a],\pi_a)$ to simplify the subsequent expressions)
\begin{widetext}
\begin{eqnarray}
S_{\overline{([a],\pi_a)} ([b],\pi_b)} &=& \frac{1}{\mathcal{D}_{\gauged{C}{G}}} \sum_{([c],\pi_c)} N_{([a],\pi_a) ([b],\pi_b)}^{([c],\pi_c)} d_{([c],\pi_c)}\frac{\theta_{([c],\pi_c)}}{\theta_{([a],\pi_a)}\theta_{([b],\pi_b)}} \notag \\
&=& \frac{1 } {|G|\mathcal{D}_\mb{0}}  \sum_{(\mb{t},\mb{s})\in G_a\backslash G/G_b} \sum_{[c]} \sum_{\pi_c} d_c \big|[c]\big|  \frac{ \theta_{c_{\mb{\,^{t} g \,^{s}h}}} }{\theta_{^\mb{t}a_{\bf g}} \theta_{^\mb{s}b_{\bf h}}} \frac{ \chi_{\pi_c}(\mb{\,^{t} g \,^{s}h})} { \chi_{^\mb{t}\pi_{a}}(\,^\mb{t}\mb{g})\chi_{^\mb{s}\pi_{b}}(\,^\mb{s}\mb{h})} \notag \\
&& \qquad\qquad\qquad\times \frac{n_{\,^{\bf t}\pi_a} n_{\,^{\bf s}\pi_b}}{|H_{(\,^\mb{t}a,\,^\mb{s}b;c)}|} \sum_{\mb{k}\in H_{(\,^\mb{t}a,\,^\mb{s}b;c)}}\chi_{\pi_c}^*(\mb{k})\chi_{^\mb{t}\pi_a}(\mb{k})\chi_{^\mb{s}\pi_b}(\mb{k}) \sum_{\mu} [U_{\mb{k}}(\,^\mb{t}a_{\bf g},\,^\mb{s}b_{\bf h} ; c_{\bf \,^{t}g \,^{s}h})]_{\mu\mu}.
\label{eq:gauged_S1}
\end{eqnarray}
Here $n_\pi \equiv \chi_\pi(\mb{0})=\mdim \pi$, and we used Eqs.~(\ref{eq:q_dim_gauged}), (\ref{eq:chi_pi_abc}), (\ref{eqn:fusion}), (\ref{eqn:twist}), and (\ref{eqn:multiplicity}). We may chose to use any representatives of the topological charge orbits in this expression, but we have specifically chosen to use $c \in [c]$ such that $c \in \mathcal{C}_{\bf \,^{t}g \,^{s}h}$ (corresponding to the choices $a_{\bf g} \in [a]$ and $b_{\bf h} \in [b]$) in order to make the evaluation more direct. The sum breaks into three parts: (1) a sum over $(\mb{t},\mb{s})\in G_a\backslash G/G_b$, (2) a sum over $G$-orbits $[c]$, and (3) a sum over irreducible $\eta_c$-representations $\pi_c$. We first carry out the sum over $\pi_c$. In order to apply the orthogonality relation \eqref{eqn:ortho2}, we notice that, in $G_c$, $\mb{^{t}g\,^{s}h}$ by itself forms an $\eta_{c}$-regular conjugacy class and its centralizer is $G_c$. Thus, we can apply \eqref{eqn:ortho2} to evaluate the sum
\begin{equation}
\sum_{\pi_c}\chi_{\pi_c}(\mb{^{t}g\,^{s}h}) \chi_{\pi_c}^*(\mb{k}) = |G_c|\delta_{\mb{^{t}g\,^{s}h},\mb{k}}.
\end{equation}
Since $\mb{k}\in H_{(\,^\mb{t}a,\,^\mb{s}b;c)}$, we conclude that in order to have $\mb{k}=\mb{^{t}g \,^{s}h}$ (so that the sum is non-vanishing), we must have $^\mb{t}\mb{g} \in G_{^\mb{s}{b}}$ and $^\mb{s}\mb{h} \in G_{^\mb{t}{a}}$. Using these properties to evaluate the sums over $\pi_c$ and $\mb{k}$, we obtain
\begin{equation}
S_{\overline{([a],\pi_a)} ([b],\pi_b)} =
\frac{1 } {\mathcal{D}_\mb{0}}  \sum_{(\mb{t},\mb{s})\in G_a\backslash G/G_b} \sum_{\substack{[c]\\ c\in \mathcal{C}_{\mb{^tg^sh}}}}d_c \frac{ n_{\,^{\bf t}\pi_a} n_{\,^{\bf s}\pi_b} }{|H_{(\,^\mb{t}a,\,^\mb{s}b;c)}|} \frac{ \theta_c}{\theta_{^\mb{t}a} \theta_{^\mb{s}b}} \frac{\chi_{^\mb{t}\pi_a}(\mb{^{t}g\,^{s}h})\chi_{^\mb{s}\pi_b}(\mb{^{t}g\,^{s}h})}{\chi_{^\mb{t}\pi_{a}}(^\mb{t}\mb{g})\chi_{^\mb{s}\pi_{b}}(^\mb{s}\mb{h})}
\sum_\mu [U_{\mb{\mb{^{t}g \,^{s}h}}}(^\mb{t}a,\,^\mb{s}b;c)]_{\mu\mu}
,
\end{equation}
where we indicate the choice $c\in \mathcal{C}_{\mb{^tg^sh}}$ on the $[c]$ sum in order to reduce clutter.
We further notice that
\begin{equation}
\chi_{^\mb{t}\pi_a}(\mb{^{t}g \,^{s}h}) = \tr\left[ \pi_{^\mb{t}a}(\mb{^{t}g \,^{s}h})\right] = \tr\left[ \eta_{^\mb{t}a}(\mb{^{t}g},\mb{\,^sh})^{-1} \pi_{^\mb{t}a}(\mb{^{t}g}) \pi_{^\mb{t}a}(^\mb{s}\mb{h}) \right]
=\eta_{^\mb{t}a}(\mb{^tg},\mb{\,^sh})^{-1} \frac{\chi_{^\mb{t}\pi_a}(\mb{^{t}g}) }{n_{^\mb{t}\pi_a}} \chi_{\mb{^t}\pi_a}(\mb{^{s}h})
,
\end{equation}
where we have used the fact that $\pi_{^\mb{t}a}(\mb{^{t}g})\propto \openone$. There is a similar relation for $\chi_{^\mb{s}\pi_b}(\mb{^tg\,^sh})$.
From these relations, we obtain
\begin{eqnarray}
S_{\overline{([a],\pi_a)} ([b],\pi_b)}&=&
\frac{1} {\mathcal{D}_\mb{0}}  \sum_{(\mb{t},\mb{s})\in G_a\backslash G/G_b} \sum_{\substack{[c]\\ c\in \mathcal{C}_{\mb{^tg^sh}}}}\frac{d_c }{|H_{(\,^\mb{t}a,\,^\mb{s}b;c)}|}
\frac{\theta_c}{\theta_{^\mb{t}a} \theta_{^\mb{s}b}} \frac{ \sum_{\mu} [U_{\mb{^tg^sh}}(^\mb{t}a,\,^\mb{s}b;c)]_{\mu\mu}} {\eta_{^\mb{t}a}(^\mb{t}\mb{g},\,^\mb{s}\mb{h}) \eta_{^\mb{s}b}(^\mb{t}\mb{g}, \,^\mb{s}\mb{h})} \chi_{\mb{^t}\pi_a}(\mb{^{s}h})\chi_{\mb{^s}\pi_b}(\mb{^{t}g}) \notag \\
&=& \frac{1} {\mathcal{D}_\mb{0}}  \sum_{(\mb{t},\mb{s})\in G_a\backslash G/G_b} \sum_{[c]}
\frac{d_c}{|H_{(\,^\mb{t}a,\,^\mb{s}b;c)}|} \sum_{\mu, \nu} [R^{^\mb{t}a \,^\mb{s}b}_c]_{\mu\nu}[R^{^\mb{s}b \,^\mb{t}a}_c]_{\nu\mu} \chi_{\mb{^t}\pi_a}(\mb{^sh})\chi_{\mb{^s}\pi_b}(\mb{^tg}),	
\end{eqnarray}
where we used the $G$-crossed ribbon identity \eqref{eq:G-crossed_ribbon} in the last step. (We can now drop the $c\in \mathcal{C}_{\mb{^tg^sh}}$, since this condition is implicitly enforced by the $R$-symbols.)

Thus, we have found
\begin{equation}
  S_{\ol{([a],\pi_a)},([b],\pi_b)} = \frac{1}{\mathcal{D}_\mb{0}} \sum_{\substack{ (\mb{t,s})\in G_a\backslash G/G_b} } \sum_{[c]} \frac{d_c}{|H_{(\,^\mb{t}a,\,^\mb{s}b;c)}|}  \tr\left[R^{^\mb{t}a \,^\mb{s}b}_c R^{^\mb{s}b \,^\mb{t}a}_c\right] \chi_{\mb{^t}\pi_a}(\mb{^sh})\chi_{\mb{^s}\pi_b}(\mb{^tg})
.
\end{equation}
By sliding a line over a double braid and applying \eqref{eq:G-crossed_Yang_Baxter}, we can show (when $^{\bf h}a=a$ and $^{\bf g}b=b$) that
\begin{equation}
\tr\left[R^{\,^{\bf k}a_{\bf g} \,^{\bf k}b_{\bf h}}_{\,^{\bf k}c} R^{\,^{\bf k}b_{\bf h} \,^{\bf k}a_{\bf g}}_{\,^{\bf k}c} \right]
= \frac{\eta_b(\mb{g},\mb{\bar{k}})\eta_a(\mb{h},\mb{\bar{k}})}{\eta_b(\mb{\bar{k}},\mb{^kg})\eta_a(\mb{\bar{k}},\mb{^kh})}\tr\left[R^{a_{\bf g} b_{\bf h}}_c R^{b_{\bf h} a_{\bf g}}_c \right].
\end{equation}
Using \eqref{eq:k_pi_a}, we also have
\begin{equation}
	 \chi_{^{\bf k}\pi_a}(^{\bf k}\mb{h}) = \frac{\eta_a(\mb{\bar{k}},\mb{^kh})}{\eta_a(\mb{h},\mb{\bar{k}})}\chi_{\pi_a}(\mb{h}), \qquad
\chi_{^{\bf k}\pi_b}(^{\bf k}\mb{g}) = \frac{\eta_b(\mb{\bar{k}},\mb{^kg})} {\eta_b(\mb{g},\mb{\bar{k}})} \chi_{\pi_b}(\mb{g}).
\end{equation}
Putting these together, we find the relation
\begin{equation}
\tr\left[R^{\,^{\bf k}a_{\bf g} \,^{\bf k}b_{\bf h}}_{\,^{\bf k}c} R^{\,^{\bf k}b_{\bf h} \,^{\bf k}a_{\bf g}}_{\,^{\bf k}c} \right] \chi_{^{\bf k}\pi_a}(^{\bf k}\mb{h}) \chi_{^{\bf k}\pi_b}(^{\bf k}\mb{g}) = \tr\left[R^{a_{\bf g} b_{\bf h}}_c R^{b_{\bf h} a_{\bf g}}_c \right] \chi_{\pi_a}(\mb{h}) \chi_{\pi_b}(\mb{g})
,
\end{equation}
which shows that this quantity is invariant under $G$ action.

Finally, we carry out the sum over the orbits $[c]$, replacing it with a sum over the actual topological charges $c \in \ext{C}{G}$ to obtain
\begin{eqnarray}
S_{\ol{([a_{\bf g}],\pi_a)} ([b_{\bf h}],\pi_b)} &=& \frac{1}{\mathcal{D}_\mb{0}} \sum_{\substack{ (\mb{t,s}) \in G_a\backslash G/G_b} } \sum_{[c]} \frac{1}{|G|} \sum_{\mb{k}\in G} \frac{d_{^\mb{k}c}}{|H_{(\,^\mb{t}a,\,^\mb{s}b;\,c)}|}
\tr\left[R^{^\mb{kt}a \,^\mb{ks}b}_{^{\bf k}c} R^{^\mb{ks}b \,^\mb{kt}a}_{^{\bf k}c} \right] \chi_{\mb{^{kt}}\pi_a}(\mb{^{ks}h}) \chi_{\mb{^{ks}}\pi_b}(\mb{^{kt}g})
\notag \\
&=& \frac{1}{|G| \mathcal{D}_\mb{0}} \sum_{\substack{ (\mb{t,s})\in G_a\backslash G/G_b} } \sum_{[c]} \sum_{\mb{k}\in G/H_{(\,^\mb{t}a,\,^\mb{s}b;\,c)} } d_{^\mb{k}c} \tr\left[R^{^\mb{kt}a \,^\mb{ks}b}_{^{\bf k}c} R^{^\mb{ks}b \,^\mb{kt}a}_{^{\bf k}c} \right] \chi_{\mb{^{kt}}\pi_a}(\mb{^{ks}h}) \chi_{\mb{^{ks}}\pi_b}(\mb{^{kt}g}).
\end{eqnarray}
We write $\mb{k} \in G/H_{(\,^\mb{t}a,\,^\mb{s}b;\,c)}$ as $\mb{k}=\mb{l}\mb{k}_1$ where $\mb{k}_1\in G_{^{\mb{t}}a}\cap\, G_{^{\mb{s}}b} \cap\,  G/H_{(\,^\mb{t}a,\,^\mb{s}b;\,c)} \equiv M_{(^\mb{t}a,^\mb{s}b;c)}$ and $\mb{l}\in [G/H_{(\,^\mb{t}a,\,^\mb{s}b;\,c)}]/M_{(^{\mb{t}}a, ^\mb{s}b;c)}\equiv L_{(^\mb{t}a,^\mb{s}b)}$. We purposefully drop the index $c$ in the definition of $L$, since $L$ contains cosets of elements that at least change one of $^\mb{t}a$ and $^\mb{s}b$, without referencing to $c$. In other words, $\mb{k}_1$ are all elements in $G/H_{(\,^\mb{t}a,\,^\mb{s}b;\,c)}$ that keep both $^\mb{t}a$ and $^\mb{s}b$ invariant and by definition necessarily transforms $c$ nontrivially within the same orbit. Once we sum over those $\mb{k}_1$ and $[c]$, we actually have a sum over all $c$ in $^\mb{t}a\times\, ^\mb{s}b$:
\begin{eqnarray}
S_{\ol{([a_{\bf g}],\pi_a)} ([b_{\bf h}],\pi_b)} &=& \frac{1}{|G| \mathcal{D}_\mb{0}}
\sum_{\substack{ (\mb{t,s})\in G_a\backslash G/G_b} } \sum_{\mb{l}\in L_{(\,^\mb{t}a,\,^\mb{s}b)} }\sum_{[c]} \sum_{\mb{k}_1\in M_{(^\mb{t}a,^\mb{s}b;c)}} d_{^\mb{lk}_{1}c} \tr\left[R^{^\mb{lt}a \,^\mb{ls}b}_{^{\bf lk}_{1}c} R^{^\mb{ls}b \,^\mb{lt}a}_{^{\bf lk}_{1}c} \right] \chi_{\mb{^{lt}}\pi_a}(\mb{^{ls}h}) \chi_{\mb{^{ls}}\pi_b}(\mb{^{lt}g})
\notag \\
&=&\frac{1}{|G| }
\sum_{\substack{ (\mb{t,s})\in G_a\backslash G/G_b} } \sum_{\mb{l}\in L_{(\,^\mb{t}a,\,^\mb{s}b)} }S_{^\mb{lt}a\,^\mb{ls}b}\, \chi_{\mb{^{lt}}\pi_a}(\mb{^{ls}h}) \chi_{\mb{^{ls}}\pi_b}(\mb{^{lt}g})
.
\end{eqnarray}
Now recall that the double coset $G_a\backslash G/G_b$ is defined as the equivalence classes of elements in $G/G_a\times G/G_b$, under the diagonal left multiplication. Therefore carrying out the sum over $\mb{l}$ is equivalent to lift the double coset back to $G/G_a\times G/G_b$. Finally we arrive at the following expression:
\begin{equation}
	\begin{split}
	S_{\ol{([a_{\bf g}],\pi_a)} ([b_{\bf h}],\pi_b)} &=
\frac{1}{|G| }
\sum_{\substack{ \substack{\mb{t}\in G/G_a \\ \mb{s}\in G/G_b} }} S_{^\mb{t}a\,^\mb{s}b}\, \chi_{\mb{^{t}}\pi_a}(\mb{^{s}h}) \chi_{\mb{^{s}}\pi_b}(\mb{^{t}g}).
	\end{split}
\end{equation}
We can now use Eqs.~(\ref{eq:S_3}), (\ref{eq:conjugate_pi}), and (\ref{eq:chi_pi_g_inverse}) to rewrite this final expression as
\begin{equation}
S_{([a_{\bf g}],\pi_a) ([b_{\bf h}],\pi_b)} = \frac{1}{|G|} \sum_{\substack{\mb{t}\in G/G_a \\ \mb{s}\in G/G_b}} S_{^\mb{t}a_{\bf g} \,^\mb{s}b_{\bf h}} \, \chi_{\overline{\mb{^t}\pi_a}}(\mb{^sh}) \, \chi_{\mb{^s}\pi_b}\left(\mb{\overline{^t g}}\right)
.
\label{eq:gauged_S}
\end{equation}
\end{widetext}
Thus, we have found that the $S$-matrix of the gauged UBTC $\gauged{C}{G}$ can be obtained from the $S$-matrix of the corresponding $G$-crossed UBTC $\ext{C}{G}$ by taking a linear combination of $S$-matrix elements that is weighted by the projective characters of the corresponding projective irreps.

\subsubsection{Chiral central charge}

Given our formula in Eq.~(\ref{eqn:twist}) for the topological twists of $\gauged{C}{G}$, we can prove that the chiral central charge $c_{-}$ (mod $8$) of the gauged theory is the same as that of $\mathcal{C}_{\bf 0}$, when these theories are modular. To see this, we first evaluate the Gauss sum for a specific $G$-orbit $[a]$, summing over $\eta_a$-irreps
\begin{eqnarray}
\sum_{\pi_a} d_{([a],\pi_a)}^2\theta_{([a],\pi_a)} &=& \sum_{\pi_a} d_a^2 \big|[a]\big|^2  |\chi_{\pi_a}(\mb{0})|^2 \theta_{a_\mb{g}} \frac{\chi_{\pi_a}(\mb{g})}{\chi_{\pi_a}(\mb{0})}
\notag \\
&=& d_a^2 \big|[a]\big|^2 \theta_{a_\mb{g}} \sum_{\pi_a}{\chi_{\pi_a}(\mb{g})}\chi_{\pi_a}(\mb{0})
\notag \\
&=& d_a^2 \big|[a]\big|^2 \theta_{a_\mb{g}} |G_a| \delta_{\bf g , 0}
.
\end{eqnarray}

Using this result in the full Gauss sum and noting that $|G| = |[a]| \cdot |G_a|$, we obtain
\begin{eqnarray}
\Theta_{\gauged{C}{G}} &=& \frac{1}{\mathcal{D}_{\gauged{C}{G}}} \sum_{([a_{\bf g}],\pi_a) \in \gauged{C}{G} } d_{([a],\pi_a)}^2 \theta_{([a],\pi_a)}
\notag \\
&=& \frac{1}{\mathcal{D}_{\gauged{C}{G}}} \sum_{[a_{\bf g}]} d_a^2 |G| \cdot \big|[a]\big| \theta_{a_\mb{g}} \delta_{\bf g,0}
\notag \\
&=& \frac{1}{\mathcal{D}_\mb{0}} \sum_{[a_{\bf 0}]} d_a^2 \big|[a]\big| \theta_a = \frac{1}{\mathcal{D}_\mb{0}} \sum_{a \in \mathcal{C}_\mb{0}} d_a^2 \theta_a
\notag \\
&=& \Theta_{\bf 0} = \Theta_{\mathcal{C}}
.
\end{eqnarray}
Thus $\gauged{C}{G}$ has the same chiral central charge mod $8$ as $\mathcal{C}$ and $\ext{C}{G}$.

\subsubsection{Genus $g$ Ground State Degeneracy}
\label{sec:gaugeGenus}

An alternative way of computing a number of properties of $\gauged{C}{G}$, when it is a MTC, is by computing the ground state degeneracy $\mathcal{N}_g$ of the theory on a genus $g$ surface. It is well-known that this is related to the quantum dimensions of $\gauged{C}{G}$ via the formula (which can be derived using the Verlinde formula)
\begin{align}
\mathcal{N}_g = \mathcal{D}^{2g-2} \sum_{A \in \gauged{C}{G}} d_A^{-(2g-2)}.
\end{align}
Therefore, knowledge of $\mathcal{N}_g$ for enough values of $g$ can be used to extract the quantum dimensions $d_j$ for every topological charge $A \in \gauged{C}{G}$.

The ground state degeneracy $\mathcal{N}_g$ of $\gauged{C}{G}$ can also be obtained from the genus $g$ ground state degeneracy of $\ext{C}{G}$, which was discussed in Sec.~\ref{sec:higher_genus}, by projecting onto the $G$-invariant subspace of states. In other words, we consider every state $| \psi \rangle \in \mathcal{V}_{\{ {\bf g}_j, {\bf h}_j \}}$ for every $\{{\bf g}_j, {\bf h}_j \}$-sector. As discussed in Sec.~\ref{sec:generalGenus}, these states transform under the $G$ action. The projection keeps only the subspace of states that are invariant under this $G$-action. That is, one takes
\begin{align}
  |\psi^G\rangle = \sum_{\mb{g} \in G} \rho_\mb{g}( |\psi \rangle) ,
\end{align}
for each state $| \psi \rangle$, belonging to any $\{{\bf g}_j, {\bf h}_j \}$-sector of the $G$-crossed theory. The ground state degeneracy $\mathcal{N}_g$ is then the dimension of the space spanned by such $G$-invariant states $|\psi^G \rangle$.

\subsection{Universality Classes of Topological Phase Transitions}
\label{sec:Universality_Classes}

A wide class of quantum phase transitions between topologically distinct phases of matter can be understood in terms of the
condensation of some set of ``bosonic'' quasiparticles~\cite{bais2009,kong2014}, i.e. those whose topological charge $a$
has trivial topological twist $\theta_{a}=1$. In these cases, the topological properties of the resulting phase can be derived
from those of the parent phase. Some of the topological charge values (quasiparticle types) become confined due to the new
condensate, some are equated with other topological charges, related to each other by fusion with the condensed bosons,
but otherwise go through the transition essentially unmodified, and others may split into multiple distinct types of topological
charge when going through the transition. We note that the mathematics underlying these transitions was initially developed
in \Refs{muger2000,bruguieres2000}.

Most of the current understanding of such topological phase transitions focuses on the formal mathematical structure, such as
the nature of the topological order on the two sides of the transition. However, another very important property of a phase
transition is its universality class. For the simplest cases, where only one boson $a$ with fusion $a \times a  =0$ condenses,
it has been shown that the resulting phase transitions can be understood as $\mathbb{Z}_2$ gauge-symmetry breaking
transitions~\cite{barkeshli2010prl,burnell2012}. Here, we will extend these results to a more general understanding of the
universality classes of topological bose condensation transitions.

Let us consider a topological phase of matter described by a UMTC $\mathcal{M}$ that contains a subtheory $\mathcal{B}$,
which is itself a UBTC (i.e. it contains topological charges that are closed under fusion) in which all the topological twists
are trivial, i.e. $\theta_a = 1, \; \forall a \in \mathcal{B}$. It follows that the subcategory $\mathcal{B}$ is symmetric,
i.e. $R^{ab}R^{ba}=\openone$ and $\mathcal{D} S_{ab}= d_a d_b$ for all $a,b \in \mathcal{B}$.

When these conditions are satisfied, a theorem due to Deligne~\cite{Deligne2002} guarantees that $\mathcal{B}$ is
gauge-equivalent to the category $\text{Rep}(G)$ for some finite group $G$. This category $\text{Rep}(G)$ has its
topological charges given by all irreducible linear representations of $G$, with the fusion rules being precisely given
by the tensor product of the irreducible representations and the $F$-symbols being given by the corresponding
Wigner $6j$-symbols. The topological charges of $\text{Rep}(G)$ are all bosons and their braiding is symmetric,
i.e. $\theta_a =1$ and $R^{ab}R^{ba}=\openone$ for all topological charges $a,b \in \text{Rep}(G)$. We notice, however,
that one generally does need the full knowledge of $F$-symbols and $R$-symbols of $\mathcal{B}$ to unambiguously
recover the group $G$ from the representation category~\cite{Etingof2000}.

In such a case, one can always condense the quasiparticles belonging to $\mathcal{B}$ following the formal rules
given in \Refs{bais2009,eliens2013,kong2014}. Let $\mathcal{C}$ denote the phase obtained by condensing the
$\mathcal{B}$ quasiparticles in $\mathcal{M}$. It was proven in \Ref{DGNO2009} that $\mathcal{M}$ can always
be obtained by starting from $\mathcal{C}$ and then gauging a symmetry group $G$. This implies the following
property of the topological phase transition:

\noindent \textit{When the topological quantum phase transition corresponding to the condensation of a Rep(G) subset of a
UMTC is continuous, its universality class can be understood in terms of the discrete gauge symmetry breaking transition
associated with the finite group $G$.}

\noindent This property follows from the fact that the universality class of the phase transition depends only on the
objects that are Bose condensing, since these are the only degrees of freedom that are becoming gapless and contributing
to the low-energy physics at the phase transition. Whether or not the phase transition is continuous generally depends
on the microscopic details of its physical realization.

Since discrete gauge symmetry breaking transitions are well-understood and can be simulated easily using numerical
methods or, in simple cases, through analytical methods~\cite{FradkinPRD1979}. This means that we can immediately understand the
critical exponents for local correlations of this much wider class of topological quantum phase
transitions.~\footnote{We note that while critical exponents for local correlation functions can be captured in this way,
non-local correlations and the topological structure of the critical points may have subtle differences depending on the
rest of the structure of the topological phases in question.}

\section{Classification of Symmetry Enriched Topological Phases}
\label{sec:Classification}

We have developed a general framework to understand the interplay of symmetry and topological order in $2+1$ dimensions. Our work leads to a systematic classification and characterization of SET phases in $2+1$ dimensions, for unitary symmetry groups $G$, which describe on-site and/or translation symmetries, based on inequivalent solutions of the defect theory $\ext{\mathcal{C}}{G}$. Our formalism for $\ext{\mathcal{C}}{G}$ encapsulates in detail the properties of the extrinsic $\textbf{g}$-defects and the way in which symmetries relate to the topological order. Below we will describe the classification of $\ext{\mathcal{C}}{G}$ and discuss the relation to the PSG framework for classifying SET phases. The extension to continuous, other spatial (non-on-site), or anti-unitary symmetries will also be be briefly discussed below.

\subsection{Classification of $G$-Crossed Extensions}
\label{sec:Classification_GCrossed}

One can, in principle, obtain all $G$-crossed BTCs by solving the consistency equations. In practice, this can quickly become computationally intractable. Fortunately, addressing this problem is aided by the theorems of \Ref{ENO2009}, which classify the $G$-crossed extensions of a BTC $\mathcal{C}_{\bf 0}$ for finite groups $G$ (and also extensions of fusion categories). In our paper, we restrict our attention to the case where $\mathcal{C}_{\bf 0}$ is a UMTC.

We have already examined part of this classification in detail in our paper. The most basic part of the classification, discussed in Sec.~\ref{sec:Symmetry}, is the choice of the symmetry action $[\rho] : G \rightarrow \text{Aut}(\mathcal{C}_{\bf 0})$, which is incorporated as a fundamental property of the defects of the extended theory.

The next part of the classification was discussed in Sec.~\ref{sec:symmetryfrac}, where we showed that, given a specific symmetry action $[\rho]$, the symmetry fractionalization is classified by $H^2_{[\rho]}(G, \mathcal{A})$. This required that the obstruction class $[\coho{O}] \in H^3_{[\rho]}(G, \mathcal{A})$ be trivial $[\coho{O}]=[0]$, since, otherwise, there would be no solutions. More precisely, the symmetry fractionalization classes were specified by the equivalence classes of the local projective phases $\eta_{x_{\bf 0}}({\bf g,h})$, and these classes are elements of an $H^2_{[\rho]}(G, \mathcal{A})$ torsor. This means distinct classes of solutions are obtained from each other by action of distinct elements of $H^2_{[\rho]}(G, \mathcal{A})$. In particular, the $U_{\bf g}(a,b;c)$ and $\eta_{x}({\bf g,h})$ transformations of a $G$-crossed MTC $\mathcal{C}_{G}^{\times}$ (or, rather, their restriction to the $\mathcal{C}_{\bf 0}$ sector) are precisely the symmetry action transformations of fusion vertex states and symmetry fractionalization projective phases, respectively, that encoded symmetry fractionalization. Similarly, the $G$-crossed consistency relations of the $U_{\bf g}(a,b;c)$ and $\eta_{x}({\bf g,h})$ transformations are precisely the corresponding consistency relations that arose in the fractionalization analysis. Thus, the $H^2_{[\rho]}(G, \mathcal{A})$ classification of symmetry fractionalization carries over to the $G$-crossed extensions of $\mathcal{C}_{\bf 0}$, where the defects in the extended theory incorporate the symmetry action through the braiding operations.

In this sense, the set of gauge inequivalent $G$-crossed MTCs that are extensions of a MTC $\mathcal{C}_{\bf 0}$ with specified $[\rho]$ is an $H^2_{[\rho]}(G, \mathcal{A})$ torsor. By this, we mean that, given a $G$-crossed MTC $\mathcal{C}_{G}^{\times}$, each element $[\coho{t}] \in H^2_{[\rho]}(G, \mathcal{A})$ specifies a potential way of modifying $\mathcal{C}_{G}^{\times}$ to obtain a distinct, gauge inequivalent $G$-crossed MTC $\hat{\mathcal{C}}_{G}^{\times}$, with a different fractionalization class. From the above discussion, it is clear that an important property of a $G$-crossed extension that is modified by $[\coho{t}]$ in this way is the symmetry action and fractionalization that is encoded in the defects, particularly their action with respect to the $\mathcal{C}_{\bf 0}$ sector.

We can also see that, for a choice of $\coho{t} \in [\coho{t}]$, the $G$-graded fusion rules of the defects in $\mathcal{C}_{G}^{\times}$ are modified to become
\begin{equation}
\label{eq:fusion_symm_frac_change}
a_\mb{g}\times b_\mb{h} = \coho{t}(\mb{g},\mb{h}) \times \sum_{c_\mb{gh}} N_{a_\mb{g}b_\mb{h}}^{c_\mb{gh}}c_{\mb{gh}},
\end{equation}
so that the fusion coefficients of the modified theory $\hat{\mathcal{C}}_{G}^{\times}$ are given by
\begin{equation}
\hat{N}_{a_\mb{g}b_\mb{h}}^{c_\mb{gh}} = N_{a_\mb{g}b_\mb{h}}^{ \overline{\cohosub{t}(\mb{g},\mb{h})} \times c_\mb{gh}}
.
\end{equation}
It follows from the $2$-cocycle condition on $\coho{t}$ that these modified fusion coefficients automatically provide an associative fusion algebra. We note that such a modification may or may not actually give a distinct fusion algebra. Clearly, the rest of the basic data of $\mathcal{C}_{G}^{\times}$ will also be modified by $[\coho{t}]$, but we will not go into these details here. (The dependence of the basic data of $\mathcal{C}_{G}^{\times}$ on the fractionalization class can be seen very explicitly for the relatively simple class of examples given in Sec.~\ref{sec:no_perm}.)

Importantly, while there is a different symmetry fractionalization class for each element $[\coho{t}] \in H^2_{[\rho]}(G, \mathcal{A})$, it is not guaranteed that each class can be consistently extended to define a full $G$-crossed defect theory $\hat{\mathcal{C}}_{G}^{\times}$, i.e. that each pair of $\rho$ and $\eta$ acting on $\mathcal{C}_{\bf 0}$ can consistently be incorporated in a $G$-crossed theory. In fact, the symmetry fractionalization class defines a new obstruction class $[\mathcal{O}] \in H^4(G, \mathrm{U}(1))$~\cite{ENO2009}, which we refer to as the ``defectification obstruction.'' Only when this obstruction class is trivial can a $G$-crossed BTC $\hat{\mathcal{C}}_{G}^{\times}$ be constructed, as there would otherwise be no solutions to the $G$-crossed consistency conditions. For the case where $\mathcal{C}_{\bf 0}$ is a MTC and the symmetry action does not permute quasiparticle types, we have derived an expression for this obstruction in Eq.~(\ref{eq:obstruction}) by directly solving the $G$-crossed consistency equations.

When the defectification obstruction $[\mathcal{O}]$ vanishes, the classification theorem established in \Ref{ENO2009} says that the remaining multiplicity of $G$-crossed extensions (after specifiying $[\rho]$ and $[\eta]$) is classified by $H^3(G, \mathrm{U}(1))$. The set of $G$-crossed extensions (with specified symmetry action and symmetry fractionalization class) is an $H^3(G, \mathrm{U}(1))$ torsor in a similar sense as above. In particular, given a $G$-crossed MTC $\mathcal{C}_{G}^{\times}$, each element $[\alpha] \in H^3(G, \mathrm{U}(1))$ specifies a way of modifying $\mathcal{C}_{G}^{\times}$ to obtain a distinct, gauge inequivalent $G$-crossed MTC $\hat{\mathcal{C}}_{G}^{\times}$ with the same symmetry action and fractionalization class.

We now describe how one may modify a particular $G$-crossed theory $\mathcal{C}_{G}^{\times}$ to obtain another $G$-crossed theory $\hat{\mathcal{C}}_{G}^{\times}$, for a given $[\alpha] \in H^3(G, \mathrm{U}(1))$. We first note that the bosonic SPT states for symmetry group $G$ are completely classified by the elements $[\alpha] \in H^3(G, \mathrm{U}(1))$, as discussed in Sec.~\ref{sec:trivial}. We will denote these states as SPT$_{G}^{[\alpha]}$. Then it is easy to see that, for each $[\alpha]$, we can produce another $G$-crossed theory by factoring in SPT states in such a way that the group element labels match up with those of $\mathcal{C}_{G}^{\times}$, i.e. we take the restricted product
\begin{equation}
\hat{\mathcal{C}}_{G}^{\times} = \left. \text{SPT}_{G}^{[\alpha]} \boxtimes \mathcal{C}_{G}^{\times} \right|_{( {\bf g} , a_{\bf g} )}
,
\end{equation}
where topological charges in $\mathcal{C}_{\bf g}$ from the $G$-crossed theory are paired up with ${\bf g}$-defects from the SPT. To be more explicit, for a choice of $\alpha \in [\alpha]$, that is, a $3$-cocycle $\alpha \in Z^3(G, \mathrm{U}(1))$, and the choice of gauge, given in Sec.~\ref{sec:trivial}, that makes all the braiding phases trivial for SPT$_{G}^{[\alpha]}$, the basic data of $\mathcal{C}_{G}^{\times}$ can be modified as
\begin{widetext}
\begin{eqnarray}
\hat{N}_{a_\mb{g}b_\mb{h}}^{c_\mb{gh}} &=& N_{a_\mb{g}b_\mb{h}}^{ c_\mb{gh}} \\
\left[ \hat{F}^{a_{\mb{g}}b_{\mb{h}}c_{\mb{k}}}_{d_{\mb{ghk}}}\right]_{(e,\alpha,\beta)(f,\mu,\nu)} &=& \alpha(\mb{g},\mb{h},\mb{k}) \left[ F^{a_{\mb{g}}b_{\mb{h}}c_{\mb{k}}}_{d_{\mb{ghk}}} \right]_{(e,\alpha,\beta)(f,\mu,\nu)} \\
\left[ \hat{R}^{a_{\mb{g}}b_{\mb{h}}}_{c_{\mb{gh}}}\right]_{\mu \nu} &=&  \left[ R^{a_{\mb{g}}b_{\mb{h}}}_{c_{\mb{gh}}}\right]_{\mu \nu} \\
\left[ \hat{U}_\mb{k}(a_\mb{g},b_{\mb{h}};c_{\mb{gh}}) \right]_{\mu \nu} &=& \frac{\alpha(\mb{g},\mb{k}, \,^{\ol{\mb{k}}}\mb{h})}{ \alpha(\mb{g},\mb{h},\mb{k}) \alpha(\mb{k}, \,^\mb{\bar{k}}\mb{g}, \,^\mb{\bar{k}}\mb{h})} \left[U_\mb{k}(a_\mb{g},b_{\mb{h}};c_{\mb{gh}})\right]_{\mu \nu} \\
\hat{\eta}_{x_\mb{k}}(\mb{g},\mb{h}) &=& \frac{\alpha( \mb{g}, \,^\mb{\bar{g}}\mb{k} , \mb{h}) } { \alpha(\mb{g},\mb{h}, \,^\mb{\bar{h} \bar{g}}\mb{k}) \alpha(\mb{k},\mb{g},\mb{h})} \eta_{x_\mb{k}}(\mb{g},\mb{h})
\end{eqnarray}
\end{widetext}
to give the basic data of $\hat{\mathcal{C}}_{G}^{\times}$, which automatically satisfies $G$-crossed consistency conditions. We note that, since $\alpha(\mb{g},\mb{h},\mb{k})=1$ if ${\bf g}$, ${\bf h}$, or ${\bf k} = {\bf 0}$, the line sliding transformations with respect to the $\mathcal{C}_{\bf 0}$ sector are unchanged by the above modification, that is $[\hat{U}_\mb{k}(a_\mb{0},b_{\mb{0}};c_{\mb{0}})]_{\mu \nu} = [U_\mb{k}(a_\mb{0},b_{\mb{0}};c_{\mb{0}})]_{\mu \nu}$ and $\hat{\eta}_{x_\mb{0}}(\mb{g},\mb{h}) = \eta_{x_\mb{0}}(\mb{g},\mb{h})$. Thus, such modifications of a $G$-crossed theory leaves the symmetry action $[\rho] : G \rightarrow \text{Aut}(\mathcal{C}_{\bf 0})$ and symmetry fractionalization class fixed.

We believe modifications of this type precisely give the $H^3(G, \mathrm{U}(1))$ classification, or, in other words, they generate all gauge inequivalent $G$-crossed MTCs for a specified symmetry action and symmetry fractionalization class. We refer to such distinct $G$-crossed theories with the same symmetry action and fractionalization class as having different defectification classes.

\begin{table*}[t!]
	\begin{tabular}{|c|c|}
		\hline
		{\bf Invariant} & {\bf Expression}\\
		\hline \hline
		Fusion coefficients, Eqs.~(\ref{eq:fusion_rules}) and (\ref{eq:G-graded_fusion})  & $N_{a_\mb{g}b_\mb{h}}^{c_{\mb{gh}}}$\\[4pt]
    	\hline
		Quantum dimensions, Eq.~(\ref{eq:quantum_dimension}) & $d_{a_\mb{g}} = \left| [F^{a_\mb{g} \overline{a_\mb{g}} a_\mb{g}}_{a_\mb{g}}]_{00} \right|^{-1}$\\[4pt]
		\hline
		Frobenius-Schur indicator, Eq.~(\ref{eq:FSInd}) & $\varkappa_{a_\mb{g}} = d_{a_\mb{g}} [F^{a_\mb{g} a_\mb{g} a_\mb{g}}_{a_\mb{g}}]_{00} $, when $a_\mb{g} = \overline{a_\mb{g}}$ \\[4pt]
		\hline
		Action on topological charge, Eqs.~(\ref{eq:rho}) and (\ref{eq:G-crossed_rho}) & $\rho_{\bf g}(a)$\\[4pt]
		\hline
		Projective exchange, Eq.~(\ref{eqn:projexchange}) & $\frac{ \sum\limits_{\mu} \left[ R_{c}^{aa}\right] _{\mu \mu } } { \sum\limits_{\mu'} \left[ R_{c'}^{aa}\right] _{\mu' \mu' } }$\\[4pt]
		\hline
		Projective braiding, Eq.~(\ref{eqn:projbraiding_2n}) & $\frac{ \sum\limits_{\mu }  \left[\left( R^{2n} \right)^{a_{\bf g} b_{\bf h}}_{c_{\bf gh}} \right]_{\mu \mu}  } {\sum\limits_{\mu' }  \left[\left( R^{2n} \right)^{a_{\bf g} b_{\bf h}}_{c'_{\bf gh}} \right]_{\mu' \mu'} }$, when $^\mb{k}a_{\bf g}=a_{\bf g}$ and $^\mb{k}b_{\bf h}=b_{\bf h}$ for ${\bf k} = ({\bf gh})^{n}$.\\[4pt]
		\hline
		$G$-crossed modular transformations, Eq.~(\ref{eqn:modularinv}) &
\begin{tabular}{l} $\mathcal{Q}^{({\bf g,h})}_{a_{\bf g} a_{\bf g}}$, when $\mathcal{Q}^{({\bf g,h})} \, : \; \mathcal{V}_{({\bf g}, {\bf h}) }  \rightarrow \mathcal{V}_{({\bf g},{\bf h})}$\\
$\mathcal{Q}^{({\bf g,0})}_{a_{\bf g} b_{\bf g}}$, when $\mathcal{Q}^{({\bf g,0})} \, : \; \mathcal{V}_{({\bf g}, {\bf 0}) }  \rightarrow \mathcal{V}_{({\bf g},{\bf 0})}$
\end{tabular}\\
		\hline
        Modular twisting, Eq.~(\ref{eq:T^n_invariant}) & $\theta_{a_{\bf g}}^{n} \prod\limits_{j=1}^{n-1} \eta_{a_{\bf g}}({\bf g}^{j},{\bf g})$, when ${\bf g}^{n} = {\bf 0}$ \\
        \hline
		Fusion rules of the gauged theory, \eqref{eqn:fusion} & $N_{([a], \pi_a) ([b], \pi_b)}^{([c],\pi_c)} $\\
		\hline
		Topological twists of the gauged theory, Eq.~(\ref{eqn:twist}) & $\theta_{a_\mb{g}}\frac{\chi_{\pi_a}(\mb{g})}{\chi_{\pi_a}(\mb{0})}$ \\
		\hline
	\end{tabular}
	\caption{Topological invariants of $G$-crossed modular tensor categories. (This is not an exhaustive list of invariants.)}
	\label{tab:inv}
\end{table*}

It is straightforward to check that when $\alpha \in B^{3}(G,\text{U}(1))$ is a $3$-coboundary, i.e. when
\begin{equation}
\alpha({ \bf g,h,k }) = \text{d} \varepsilon({ \bf g,h,k }) = \frac{\varepsilon({\bf h,k }) \varepsilon({\bf g,hk })}{\varepsilon({\bf gh,k })\varepsilon({\bf g, h })}
\end{equation}
for some $\varepsilon \in C^{2}(G,\text{U}(1))$, that the above modification of the $G$-crossed theory by $\alpha$ produces a $\hat{\mathcal{C}}_{G}^{\times}$ that is gauge equivalent to $\mathcal{C}_{G}^{\times}$ through the vertex basis and symmetry action gauge transformations
\begin{eqnarray}
\left[ \Gamma^{a_{\bf g} b_{\bf h}}_{c_{\bf gh}} \right]_{\mu \nu} &=& \varepsilon({\bf g, h }) \delta_{\mu \nu} \\
\gamma_{a_{\bf g}}({\bf k}) &=& \frac{ \varepsilon({\bf g, k }) }{ \varepsilon({\bf k, \bar{k}gk }) }
.
\end{eqnarray}
This establishes the fact that one should take a quotient by $B^{3}(G,\text{U}(1))$, since such modifications are just gauge transformations. What remains to be shown is that every pair of $G$-crossed extensions of $\mathcal{C}_{\bf 0}$ with the same symmetry action $[\rho]$ and fractionalization class $[\eta]$ is related by such a modification for some $3$-cocycle $\alpha \in Z^{3}(G,\text{U}(1))$ and that distinct cohomology classes $[\alpha]$ give gauge inequivalent solutions (up to identification under relabeling of topological charges). We do not establish this here, but note that it may be partially verified (or wholly verified for simple enough examples) using invariants of the $G$-crossed theory and/or the corresponding gauged theory, and it is true for all the examples we study in Sec.~\ref{exampleSec}. We present the more prominent invariants of $G$-crossed theories in Table~\ref{tab:inv}. The classification is established in \Ref{ENO2009} by working at a higher category level, with the subsectors $\mathcal{C}_{\bf g}$ (which are $\mathcal{C}_{\bf 0}$ bimodules)  playing the role of objects.

In summary, the $G$-crossed extensions of a MTC $\mathcal{C}_{\bf 0}$ for finite group $G$ are classified by the symmetry action, the symmetry fractionalization class, which is an element of an $H^2_{[\rho]}(G, \mathcal{A})$ torsor, and the defectification class, which is an element of an $H^3(G, \mathrm{U}(1))$ torsor. This yields the classification of $2+1$ dimensional SET phases for a system in a topological phase described by a UMTC $\mathcal{C}_{\bf 0}$ and an on-site global unitary symmetry described by a finite group $G$. Based on the classification theorem of \Ref{ENO2009}, we believe that all of the inequivalent $G$-crossed extensions can be parameterized in this way.

\subsubsection{Equivalent SET phases by relabeling objects in $\ext{C}{G}$}

While the $G$-crossed extensions $\ext{C}{G}$ of $\mathcal{C}_{\bf 0}$ with a given action $[\rho]$ are classified by $\mathcal{H}^2_{[\rho]}(G,\mathcal{A})$ and
$\mathcal{H}^3(G,U(1))$ as described above, the corresponding classification of SET phases is generally not in one-to-one correspondence
with elements of $\mathcal{H}^2_{[\rho]}(G,\mathcal{A})$ and $\mathcal{H}^3(G,U(1))$. Rather, distinct elements of the $\mathcal{H}^2_{[\rho]}(G,\mathcal{A})$ and $\mathcal{H}^3(G,U(1))$ torsors may represent the same SET. In particular, it is possible that two $G$-crossed extensions that are related by a nontrivial element of $\mathcal{H}^2_{[\rho]}(G,\mathcal{A})$ or $\mathcal{H}^3(G,U(1))$ may be equivalent to each other by relabeling the topological charges of anyons and/or defects. Moreover, the choice of labels of anyons and defects in $\ext{C}{G}$ is only physical up to relabelings that preserve the fusion and braiding data (up to gauge transformations). It follows that SET phases are classified by different $G$-crossed extensions only up to such relabelings. Accordingly, the number of elements in $\mathcal{H}^2_{[\rho]}(G,\mathcal{A})$ and $\mathcal{H}^3(G,U(1))$ is an upper bound on the number of distinct SET phases whose quasiparticles are described by $\mathcal{C}_{\bf 0}$, with the symmetry action given by $[\rho]$.

It is useful to consider a concrete example that exhibits this sort of equivalence of theories corresponding to different classes in $\mathcal{H}^2_{[\rho]}(G,\mathcal{A})$ and $\mathcal{H}^3(G,U(1))$. We consider the $\mathbb{Z}_2$ toric code model, for which the topological charges
are $\{I, e, m, \psi\}$, and let $G = \mathbb{Z}_2$ with the symmetry action $\rho$ acting trivially on the topological charges (i.e. no permutations). This example
is examined in Sec.~\ref{sec:ExampleD(Z2)Z2Trivial}. In this case, the symmetry fractionalization is classified by
$H^2(\mathbb{Z}_2, \mathbb{Z}_2 \times \mathbb{Z}_2) = \mathbb{Z}_2 \times \mathbb{Z}_2$, where physically the first $\mathbb{Z}_2$ corresponds to whether $e$ quasiparticles carry fractional $\mathbb{Z}_2$ charge and the second $\mathbb{Z}_2$ corresponds to whether $m$ quasiparticles carry fractional $\mathbb{Z}_2$ charge. The fractionalization class where $e$ carries half-integer $\mathbb{Z}_{2}$ charge and $m$ carries integer $\mathbb{Z}_{2}$ charge is seen to be equivalent to the one where $e$ carries integer
$\mathbb{Z}_{2}$ charge and $m$ carries half-integer $\mathbb{Z}_{2}$ charge under the relabeling $e \leftrightarrow m$. Furthermore, for the fractionalization class where both $e$ and $m$ carry half-integer $\mathbb{Z}_{2}$ charge, we obtain two $\ext{C}{G}$ theories, which are related to each other by the action of the nontrivial element $[\alpha] \in \mathcal{H}^3(\mathbb{Z}_{2},U(1)) = \mathbb{Z}_{2}$, i.e. by gluing on an SPT$_{\mathbb{Z}_{2}}^{[\alpha]}$ state. However, these two $\ext{C}{G}$ theories are seen to be equivalent under a relabeling of the ${\bf g}$-defects. In other words, allowing the relabeling of
defect types within the same $\mathcal{C}_\mb{g}$ sector (with $\mb{g}\neq \mb{0}$) can relate solutions associated with
distinct classes in $H^3(G, \text{U}(1))$. This particular example was discussed previously from a different perspective,
using Chern-Simons field theory, in \Refs{Metlitski_unpub, lu2013}. Another example of this kind is discussed in Sec.~\ref{sec:zneven}.

\subsection{Relation to PSG Framework}
\label{sec:PSG}

At this stage, it is worth understanding how the framework that we have developed for classifying and characterizing SETs
relates to the projective symmetry group (PSG) classification proposed in \Ref{wen2002psg}. A complementary discussion can also be
found in \Ref{essin2013}. In the PSG formulation, a topological phase is considered with a low-energy description in terms
of a gauge theory with gauge group $H$ and a global symmetry $G$. Different PSGs are classified by different mean-field solutions
within a slave-particle framework~\cite{wen04}. A crucial role is played by group extensions,
labeled $PSG$, of $H$ by $G$, which mathematically means $PSG/H = G$. It is not clear whether the classification of different slave-particle mean-field solutions, as originally formulated in \Ref{wen2002psg}, is equivalent to classifying the group extensions $PSG$ such that $PSG/H = G$. Nevertheless, each such mean-field solution must be described by such a group extension, even if the correspondence is not one-to-one. Here, we will briefly discuss the problem of classification of such group extensions, and compare the results to our approach.

When $G$ and $H$ are both finite, the mathematical problem of classifying group extensions has the following solution~\cite{brown}.
One first picks a homomorphism $\sigma: G \rightarrow \text{Out}(H)$, where $\text{Out}(H)$ is the group of outer automorphisms of $H$: $\text{Out}(H)=\text{Aut}(H)/\text{Inn}(H)$. Here $\text{Aut}(H)$ is the automorphism group of $H$ and $\text{Inn}(H)$ is the subgroup generated by conjugation.
Different group extensions are then classified by $H^2_{\sigma}(G, \mathrm{Z}(H))$, where $\mathrm{Z}(H)$ denotes the center of
$H$. There can also be an obstruction to the group extension, which is characterized by an element of $H^3_{\sigma}(G, \mathrm{Z}(H))$.

In order to demonstrate some shortcomings of the PSG classification, we consider the case when $H$ is finite and the topological phase is fully described by a discrete $H$ gauge theory, i.e. $\mathcal{C}_{\bf 0} = \text{D}(H)$ is the (untwisted) quantum double of $H$. For this discussion, we further develop the details of the PSG formalism in order to compare to the $G$-crossed theory. We first consider how an outer automorphism $\varphi \in \text{Out}(H)$ extends to an action on the topological charges of $\text{D}(H)$. Recall that the topological charges are dyons $([{\bf h}], \pi_{\bf h})$, where the ``magnetic flux'' is a conjugacy class $[{\bf h}]$ of $H$, and the ``electric charge'' is an irrep $\pi_{\bf h}$ of the centralizer $C_{\bf h}$ of some element ${\bf h} \in [{\bf h}]$. The effect of $\varphi$ on a flux is simply $\varphi : [{\bf h}] \mapsto [ \varphi( {\bf h} ) ]$. For the effect of $\varphi$ on a charge, we make use of the fact that an irrep $\pi_{\bf h}$ is uniquely determined by its corresponding character $\chi_{\pi_{\bf h}} ([{\bf k}])$, which allows us to define $\varphi(\pi_{\bf h})$ by $\chi_{\varphi(\pi_{\bf h})} ([{\bf k}]) = \chi_{\pi_{\bf h}} ([\varphi ({\bf k})])$. In this way, we can define the extension of the outer automorphism group action $\sigma: G \rightarrow \text{Out}(H)$ to a topological symmetry group action $\rho: G \rightarrow \text{Aut}(\text{D}(H))$ for which
\begin{equation}
\rho_{\bf g} ([{\bf h}], \pi_{\bf h}) = ([\sigma_{\bf g}({\bf h})], \sigma_{\bf g}(\pi_{\bf h}))
.
\end{equation}
We emphasize that symmetry actions obtained from outer automorphisms of $H$ in this way never interchange magnetic fluxes with electric charges.

Next, we notice that the subset of Abelian anyons of $\text{D}(H)$ are given by the dyons $([{\bf h}], \pi_{\bf h})$ for which the conjugacy class $[{\bf h}] = \{ {\bf h} \}$ is a singleton (whose corresponding centralizer is $H$) and the irrep $\pi_{\bf h}$ of $H$ is one-dimensional. [This follows from Eq.~(\ref{eq:q_dim_gauged}) and the fact that a topological charge $a$ is Abelian iff $d_{a}=1$.] We define $\mathcal{A}$ (in our usual fashion) to be the Abelian group whose elements are the Abelian topological charges of $\text{D}(H)$, with group multiplication given by the corresponding fusion rules. It is clear that this group takes the form of a direct product $\mathcal{A} = \mathcal{A}_{\text{flux}} \times \mathcal{A}_{\text{charge}}$, where $\mathcal{A}_{\text{flux}}$ is the subgroup defined by the singleton conjugacy classes and $\mathcal{A}_{\text{charge}}$ is the subgroup defined by the one-dimensional irreps of $H$. Moreover, the center of $H$ is isomorphic to the group defined by the Abelian flux sector, i.e. $Z(H) \cong \mathcal{A}_{\text{flux}}$, since a conjugacy class is a singleton iff it is an element of the center.

Thus, for a symmetry action $\rho$ on $\text{D}(H)$ obtained from the outer automorphism action $\sigma$ on $H$, we find that
\begin{eqnarray}
H^{2}_{[\rho]}(G,\mathcal{A}) &=& H^{2}_{[\rho]}(G,\mathcal{A}_{\text{flux}}) \times H^{2}_{[\rho]}(G,\mathcal{A}_{\text{charge}}) \notag \\
&=& H^{2}_{\sigma}(G,Z(H)) \times H^{2}_{[\rho]}(G,\mathcal{A}_{\text{charge}})
.
\end{eqnarray}
It follows for $\text{D}(H)$ that
\begin{enumerate}
\item PSG can describe at most a proper subset of the fractionalization classes described by $H^{2}_{[\rho]}(G,\mathcal{A})$ when $H^{2}_{[\rho]}(G,\mathcal{A}_{\text{charge}})$ is nontrivial.

\item PSG can describe at most all of the fractionalization classes described by $H^{2}_{[\rho]}(G,\mathcal{A})$ when $H^{2}_{[\rho]}(G,\mathcal{A}_{\text{charge}})$ is trivial.

\item PSG is not applicable for symmetries that interchange magnetic fluxes with electric charges.

\end{enumerate}

Furthermore, even when one specifies the symmetry fractionalization class of an SET according to $H^{2}_{[\rho]}(G,\mathcal{A})$, there are still additional possibilities for distinct
SETs, as indicated by the $H^3(G, \mathrm{U}(1))$ part of the classification of $G$-crossed extensions $\ext{C}{G}$. Through the simple example of
a topological phase described by pure discrete $H$ gauge theory, we see that these are also not captured by classifying the different group extensions $PSG$.

Another important distinction between the PSG approach and our approach is that the former requires knowledge of $H$, which is, in general, not unique for a given topological phase. This makes it unclear how to reconcile different manifestations of the same SET order using the PSG formalism.

\Ref{mesaros2013} has proposed an alternative framework, besides PSG, to classify the SET phases of quantum doubles of a discrete group (i.e. discrete gauge field theories). This classification is also incomplete for those classes of states, as it misses the full set of symmetry fractionalization classes $H^2_{[\rho]}(G, \mathcal{A})$ described in this paper.

\subsection{Continuous, Spatial, and Anti-Unitary Symmetries}

The theory that we have developed in this paper is most complete when $\mathcal{C}$ is a UMTC, and the symmetry $G$ is
a finite, on-site unitary symmetry. However, much of the framework we have developed applies more generally as well.

Our general discussion of the symmetry of topological phases in Sec.~\ref{sec:Symmetry} and symmetry fractionalization
in Sec.~\ref{sec:symmetryfrac} is valid for any general symmetry $G$. However, when space-time symmetries are considered,
the theory becomes more complex as a result of the symmetry transformations no longer being on-site, but rather being locality-preserving.
For example, for spatial reflection and/or time-reversal symmetries, defects have a non-local structure
in space-time and the formalism for the defect theory described in this paper is not directly applicable. For other crystalline symmetries, there may be additional constraints
on what types of symmetry fractionalization are allowed~\cite{oshikawa2000,paramekanti2004,hastings2004,hastings2005,zaletel2014}.
We leave a systematic study of this for future work~\cite{cheng2016,barkeshli2016tr}.

When $G$ is continuous, one also requires additional conditions that the maps $\rho: G \rightarrow \text{Aut}(\mathcal{C})$ respect the continuity of $G$
by mapping all group elements in a single connected component of $G$ to the same element of $\text{Aut}(\mathcal{C})$. The cochains valued in $\mathcal{A}$, such as the $\coho{O}$ and $\coho{w}$ described in Secs.~\ref{sec:Symmetry} and \ref{sec:symmetryfrac}, should similarly respect the continuity of $G$.

Similarly, the definition of ${\bf g}$-defects and the notion of topologically distinct types of ${\bf g}$-defects is valid (or can be straightforwardly generalized)
for any unobstructed unitary symmetry $G$, even if it is not discrete and on-site. It is unclear how to generalize the constructions and formalism of defects to include anti-unitary symmetries, as the complex conjugation operation is inherently nonlocal (except when acting on product states and operators).

When $G$ is not a finite group, our formalism for $G$-crossed UBTCs described in Sec.~\ref{sec:Algebraic_Theory} may still be applied as long as fusion is finite, meaning there are only a finite number of fusion outcomes when fusing two topological charges. The discussion of $G$-crossed modularity for general $G$ requires the further restriction that $|\mathcal{C}_{\bf g}|$ be finite for all ${\bf g}$, but again does not require $G$ to be finite.

When $G$ is a continuous group, the consistency conditions that we have described in Sec.~\ref{sec:Algebraic_Theory} are not complete. In particular, the basic data of $\mathcal{C}_G^\times$, such as the $F$, $R$, $U$, and $\eta$ symbols, must somehow reflect the topology and continuity of the group $G$. For
SPT states, which consist of the case where the original category $\mathcal{C}$ is trivial, it was argued in
\Ref{chen2013} that when $G$ is continuous the classification is given in terms of Borel cohomology $H_B^3(G, \mathrm{U}(1))$.
In our language, this amounts to the condition that the $F$-symbols of $\mathcal{C}_G^\times$ be Borel measurable functions
on the group manifold. Therefore a natural assumption is that SETs with continuous symmetry $G$ are classified by
distinct $\mathcal{C}_G^\times$, with the additional condition that $F$, $R$, $U$, and $\eta$ be Borel measurable functions on
the group manifold. However, a detailed study of $G$-crossed extensions for continuous $G$, in addition to the framework for
gauging continuous $G$, will be left for future work.

In the case where $G$ contains spatial symmetries, such as translations, rotations, and reflections, it is an open question how the basic data and consistency conditions for $\ext{C}{G}$ should be modified. A systematic study of these will also be left for future work.

Finally, we note that the classification theorems of \Ref{ENO2009} for $\ext{C}{G}$ and, in particular, the statement that distinct $\ext{C}{G}$ are fully classified by the symmetry action, fractionalization class, and defectification class require that $G$ is a finite, on-site unitary symmetry.

\section{Complete Solution of $G$-Crossed Extensions for Topological Phases (MTCs) with Trivial Symmetry Action}
\label{sec:no_perm}

In this section, we consider a general topological phase described by a MTC $\mathcal{C}_{\bf 0}$, with $G$ symmetry, for which the symmetry action does not permute the topological charges in $\mathcal{C}_{\bf 0}$. In this case, we can solve the $G$-crossed consistency equations explicitly to obtain the basic data for all $G$-crossed extensions.

We begin by noting that the symmetry fractionalization obstruction $[\coho{O}]$ automatically vanishes. We will choose a gauge in which
\begin{eqnarray}
U_{\bf k}(a_{\bf 0},b_{\bf 0};c_{\bf 0}) &=& \openone, \\
\beta_{c_{\bf 0}}({\bf g}, {\bf h}) &=& 1,\\
\coho{O}&=&0, \\
\eta_{c_{\bf 0}}({\bf g}, {\bf h}) &=& M_{c_{\bf 0} \cohosub{w}({\bf g}, {\bf h})},
\end{eqnarray}
where the symmetry fractionalization class is specified by $[\coho{w}] \in H^{2}(G,\mathcal{A})$.

We know that $|\mathcal{C}_{\bf g}| = |\mathcal{C}_{\bf 0}|$, since the theory is modular and all topological charges in $\mathcal{C}_{\bf 0}$ are fixed under symmetry action. However, to determine the properties of the topological charges, we must establish their fusion rules. For this, we first prove that $\mathcal{C}_{\bf g}$ contains at least one Abelian topological charge (which has quantum dimension equal to 1) for each ${\bf g} \in G$ in a $G$-crossed MTC when the action of the symmetry $G$ does not permute the topological charge values of quasiparticles.

Using the gauge choice $U_{\bf g}(a_{\bf 0},b_{\bf 0};c_{\bf 0} ) = \openone$, Eq.~(\ref{eq:preVerlinde1}) yields the relation
\begin{equation}
\frac{S_{a_{\bf 0} x_{\bf k}}}{S_{0 x_{\bf k}}} \frac{S_{b_{\bf 0} x_{\bf k}}}{S_{0 x_{\bf k}}} = \sum_{c_{\bf 0}} N_{a_{\bf 0} b_{\bf 0}}^{c_{\bf 0}} \frac{S_{c_{\bf 0} x_{\bf k}}}{S_{0x_{\bf k}}}
,
\end{equation}
which tells us that $\frac{S_{a_{\bf 0} x_{\bf k}}}{S_{0 x_{\bf k}}}$ is a character of the Verlinde algebra of $\mathcal{C}_{\bf 0}$. Since $\mathcal{C}_{\bf 0}$ is a MTC, the characters of its Verlinde algebra are given by $\lambda_{y_{\bf 0}}^{(a_{\bf 0})} = \frac{S_{a_{\bf 0} y_{\bf 0}}}{S_{0 y_{\bf 0}}}$ for $y_{\bf 0} \in \mathcal{C}_{\bf 0}$. It follows that, for each $x_{\bf k}$, there must be some $y_{\bf 0} \in \mathcal{C}_{\bf 0}$ such that $\frac{S_{a_{\bf 0} x_{\bf k}}}{S_{0 x_{\bf k}}} =\frac{S_{a_{\bf 0} y_{\bf 0}}}{S_{0 y_{\bf 0}}}$. In other words, there is a map $f: \mathcal{C}_{\bf k} \rightarrow \mathcal{C}_{\bf 0}$ such that $\frac{S_{a_{\bf 0} x_{\bf k}}}{S_{0 x_{\bf k}}} =\frac{S_{a_{\bf 0} f(x_{\bf k})}}{S_{0 f(x_{\bf k})}}$.

$G$-crossed modularity implies that $|\mathcal{C}_{\bf k}| = | \mathcal{C}_{\bf 0}|$ and
\begin{eqnarray}
\delta_{x_{\bf k} y_{\bf k}} &=& \sum_{a_{\bf 0}} S_{a_{\bf 0} x_{\bf k}} S_{a_{\bf 0} y_{\bf k}}^{\ast} \notag \\
&=& \frac{d_{x_{\bf k}} d_{y_{\bf k}}}{ d_{f(x_{\bf k})} d_{f(y_{\bf k})} } \sum_{a_{\bf 0}} S_{a_{\bf 0} f(x_{\bf k})} S_{a_{\bf 0} f(y_{\bf k})}^{\ast}
\notag\\
&=& \frac{d_{x_{\bf k}} d_{y_{\bf k}}}{ d_{f(x_{\bf k})} d_{f(y_{\bf k})} } \delta_{f(x_{\bf k}) f(y_{\bf k})}
.
\end{eqnarray}
Hence, the function $f$ is a bijection. Moreover, this expression tells us that $d_{x_{\bf k}} = d_{f(x_{\bf k})}$. Inverting $f$, we can now define the ${\bf k}$-defect $0_{\bf k} \equiv f^{-1}(0_{\bf 0})$, which thus has $d_{0_{\bf k}} =d_{0_{\bf 0}} = 1$.

Having established that there is at least one Abelian ${\bf g}$-defect for each ${\bf g} \in G$, we now choose one such Abelian defect from each sector to label $0_{\bf g}$. With this convention, we can label all the other defects as $a_{\bf g} = a_{\bf 0} \times 0_{\bf g}$. Since $0_{\bf g}$ is Abelian, this labeling is well-defined (i.e. each $a_{\bf g}$ is distinct and uniquely defined). This represents all defect types in $\mathcal{C}_{\bf g}$, since $|\mathcal{C}_{\bf g}| = | \mathcal{C}_{\bf 0}|$, and, moreover, produces the correct total quantum dimension $\mathcal{D}_{\bf g} = \mathcal{D}_{\bf 0}$.

Since $0_{\bf g}$ are Abelian, their fusion must take the form
\begin{equation}
0_{\bf g} \times 0_{\bf h} = \coho{w}({\bf g},{\bf h})_{\bf 0} \times 0_{\bf gh}
,
\end{equation}
for some Abelian topological charge $\coho{w}({\bf g},{\bf h}) \in \mathcal{A}$. (We use the notation $\coho{w}({\bf g},{\bf h})$ for this Abelian topological charge in anticipation of this quantity being identified as the 2-cocycle characterizing symmetry fractionalization.) Associativity of fusion requires $\coho{w}({\bf g},{\bf h}) \in Z^{2}(G,\mathcal{A})$, i.e. it must satisfy the cocycle condition
\begin{equation}
\coho{w}({\bf g},{\bf h}) \coho{w}({\bf gh},{\bf k}) = \coho{w}({\bf g},{\bf hk}) \coho{w}({\bf h},{\bf k})
.
\end{equation}
Consistently extending this to all defect topological charges, we find that the fusion rules must take the form
\begin{eqnarray}
a_{\bf g} \times b_{\bf h} &=&  \sum_{c \in \mathcal{C}_{\bf 0}} N_{a_{\bf 0} b_{\bf 0}}^{c_{\bf 0}} c_{\bf 0} \times \coho{w}({\bf g},{\bf h})_{\bf 0} \times 0_{\bf gh} \notag \\
&=& \sum_{c \in \mathcal{C}_{\bf 0}} N_{a_{\bf 0} b_{\bf 0}}^{c_{\bf 0}}  [c\coho{w}({\bf g},{\bf h})]_{\bf gh}
,
\end{eqnarray}
where we introduce the shorthand $[ab]_{\bf g} = [a\times b]_{\bf g}$ for $a,b \in \mathcal{C}_{\bf 0}$ when at least one of $a$ and $b$ is an Abelian topological charge, so there is no ambiguity in their fusion product. In other words, the fusion coefficients are given by
\begin{equation}
N_{a_{\bf g} b_{\bf h}}^{c_{\bf gh}} = N_{a_{\bf 0} b_{\bf 0}}^{ \overline{\cohosub{w}({\bf g},{\bf h})_{\bf 0}} \times c_{\bf 0}}.
\end{equation}

Of course, when $\mathcal{A}$ is nontrivial, the choice of $0_{\bf g}$ is not unique and could instead have been any other Abelian ${\bf g}$-defect. In other words, we are free to choose a different definition $\widehat{0}_{\bf g}$, which is necessarily related to the other choice as $\widehat{0}_{\bf g} = \coho{z}({\bf g}) \times 0_{\bf g}$, where $\coho{z}({\bf g}) \in \mathcal{A}$. This choice results in a redefinition $\widehat{\coho{w}}({\bf g},{\bf h}) = \coho{w}({\bf g},{\bf h}) \times \text{d} \coho{z}({\bf g},{\bf h})$. Thus, the consistent fusion rules for $G$-crossed extensions are classified by $[\coho{w}] \in H^{2}(G,\mathcal{A})$. Furthermore, we will see that solving the consistency conditions for $U$ and $\eta$ reveals that this quantity is precisely the symmetry fractionalization class, justifying our use of the same symbol. We also note that when we restrict to the Abelian topological charges of the defect theory, i.e. the quasiparticles and defects with $d_{a}=1$, the possible fusion rules correspond precisely to the possible central extensions of the group $G$ by the group $\mathcal{A}$, which are also known to be classified by $H^{2}(G,\mathcal{A})$.

Given the fusion rules, we see that consistency of the fusion rules with $G$-crossed braiding determines the symmetry action on the defect charges to be
\begin{eqnarray}
\rho_{\bf k}(a_{\bf g}) &=& [a \coho{w}({\bf k},{\bf g}) \overline{\coho{w}({\bf kg \bar{k}},{\bf k})}]_{{\bf kg\bar{k}}}
\notag \\
&=& [a \coho{w}({\bf g},{\bf \bar{k}}) \overline{\coho{w}({\bf \bar{k}},{\bf kg\bar{k}})}]_{{\bf kg\bar{k}}}
.
\end{eqnarray}

Solving the $G$-crossed consistency equations by iteratively increasing the the number of defects involved (i.e. the number of topological charge values at the top of the corresponding diagrams that are labeled by nontrivial elements of $G$), we find expressions for the basic data of the $\ext{C}{G}$ theories in terms of the $\mathcal{C}_{\bf 0}$ basic data and the fractionalization class specified by $\coho{w}$. Moreover, we find an expression for the defectification obstruction, which indicates whether a consistent $G$-crossed theory exists for a given fractionalization class.

We note that there is always at least one fractionalization class that is not obstructed, since one can always take the product of a MTC with an SPT state to produce a $G$-crossed theory, i.e. $\ext{C}{G} = \mathcal{C}_{\bf 0} \boxtimes \text{SPT}_{G}^{[\alpha]}$. These $G$-crossed theories correspond to the trivial fractionalization class $[\coho{w}]=0$.

We now restrict our attention to theories with no fusion multiplicities, i.e. $N_{ab}^{c}\leq 1$, though the general case may be similarly addressed. In order to solve the consistency equations explicitly, we use the vertex basis gauge freedom to fix
\begin{eqnarray}
&& \left[ F^{a_{\bf 0} b_{\bf 0} 0_{\bf k}}_{c_{\bf k}} \right]_{c_{\bf 0} b_{\bf k}} = \left[ F^{a_{\bf 0} 0_{\bf h} b_{\bf 0}}_{c_{\bf h}} \right]_{a_{\bf h} b_{\bf h}}=\left[F^{0_{\bf g} b_{\bf 0} 0_{\bf k}}_{[b\cohosub{w}({\bf g},{\bf k})]_{\bf gk}}\right]_{b_{\bf g} b_{\bf k}} \notag \\
&& \qquad=\left[F^{a_{\bf 0} 0_{\bf h} b_{\bf k}}_{[c \cohosub{w}({\bf h},{\bf k})]_{\bf hk}} \right]_{a_{\bf h} [b \cohosub{w}({\bf h},{\bf k})]_{\bf hk}} = R^{0_{\bf g} b_{\bf 0} }_{b_{\bf g}}=1
\qquad
\end{eqnarray}
when $N_{a_{\bf 0}b_{\bf 0}}^{c_{\bf 0}}\neq 0$,
and the symmetry action gauge freedom to fix
\begin{equation}
R^{a_{\bf g} 0_{\bf h} }_{[a \cohosub{w}({\bf g},{\bf h})]_{\bf gh}}=1.
\end{equation}
With these gauge choices, the resulting basic data for the terms allowed by fusion are found to be (for presentability, we leave charge labels implicit for fusion channels that are uniquely determined by the remaining labels)
\begin{widetext}
\begin{eqnarray}
\left[ F^{a_{\bf g} b_{\bf h} c_{\bf k}}_{[ d \cohosub{w}({\bf g},{\bf h})\cohosub{w}({\bf gh},{\bf k})]_{\bf ghk}} \right]_{[ e \cohosub{w}({\bf g},{\bf h})]_{\bf gh} [ f \cohosub{w}({\bf h},{\bf k})]_{\bf hk}}
&=& \left[ F^{a_{\bf 0} [b \cohosub{w}({\bf g},{\bf h})]_{\bf 0} [c\cohosub{w}({\bf gh},{\bf k})]_{\bf 0}}_{[ d \cohosub{w}({\bf g},{\bf h})\cohosub{w}({\bf gh},{\bf k})]_{\bf 0}} \right]_{[ e \cohosub{w}({\bf g},{\bf h})]_{\bf 0} [ f \cohosub{w}({\bf g},{\bf h}) \cohosub{w}({\bf gh},{\bf k})]_{\bf 0}}
\notag \\
\times \frac{F^{b_{\bf 0} [ \cohosub{w}({\bf g},{\bf h})]_{\bf 0} [c \cohosub{w}({\bf gh},{\bf k})]_{\bf 0}}_{[f \cohosub{w}({\bf g},{\bf h})\cohosub{w}({\bf gh},{\bf k})]_{\bf 0}}}
{F^{ b_{\bf 0} [ c\cohosub{w}({\bf h},{\bf k})]_{\bf 0} [ \cohosub{w}({\bf g},{\bf hk})]_{\bf 0}}_{[ f \cohosub{w}({\bf g},{\bf h})\cohosub{w}({\bf gh},{\bf k})]_{\bf 0}} }
&& \!\!\!\!\!\!\!\!\!\!\!\! \frac{F^{c_{\bf 0} [ \cohosub{w}({\bf g},{\bf h})]_{\bf 0} [ \cohosub{w}({\bf gh},{\bf k})]_{\bf 0} } }
{ F^{[\cohosub{w}({\bf g},{\bf h})]_{\bf 0} c_{\bf 0} [\cohosub{w}({\bf gh},{\bf k})]_{\bf 0}}
F^{ c_{\bf 0} [ \cohosub{w}({\bf h},{\bf k})]_{\bf 0} [ \cohosub{w}({\bf g},{\bf hk})]_{\bf 0} }}
\frac{1}{R^{[\cohosub{w}({\bf g},{\bf h})]_{\bf 0} c_{\bf 0}}}F^{0_{\bf g} 0_{\bf h} 0_{\bf k}}
\label{eq:F_defect_no_perm}
\\
R^{a_{\bf g} b_{\bf h} }_{[c\cohosub{w}({\bf g},{\bf h})]_{\bf gh}}
&=& \frac{F^{ a_{\bf 0} b_{\bf 0} [\cohosub{w}({\bf g},{\bf h})]_{\bf 0} }_{[c\cohosub{w}({\bf g},{\bf h})]_{\bf 0}}}
{F^{b_{\bf 0} a_{\bf 0} [\cohosub{w}({\bf g},{\bf h})]_{\bf 0} }_{[c\cohosub{w}({\bf g},{\bf h})]_{\bf 0}} }
R^{ a_{\bf 0} b_{\bf 0}}_{c_{\bf 0}}
\label{eq:R_defect_no_perm}
\\
U_{\bf k}( a_{\bf g}, b_{\bf h}; [c\coho{w}({\bf g},{\bf h})]_{\bf gh} ) &=&
\frac{F^{a_{\bf g}  0_{\bf k} [b \cohosub{w}({\bf h},{\bf k})\overline{\cohosub{w}({\bf k},{\bf \bar{k}hk})} ]_{\bf \bar{k}hk} }_{[c \cohosub{w}({\bf g},{\bf h})\cohosub{w}({\bf gh},{\bf k})]_{\bf ghk}}}
{F^{a_{\bf g} b_{\bf h} 0_{\bf k}}_{[c \cohosub{w}({\bf g},{\bf h})\cohosub{w}({\bf gh},{\bf k})]_{\bf ghk}}
F^{ 0_{\bf k} [a \cohosub{w}({\bf g},{\bf k})\overline{\cohosub{w}({\bf k},{\bf \bar{k}gk})} ]_{\bf \bar{k}gk} [b \cohosub{w}({\bf h},{\bf k})\overline{\cohosub{w}({\bf k},{\bf \bar{k}hk})} ]_{\bf \bar{k}hk} }_{[c \cohosub{w}({\bf g},{\bf h})\cohosub{w}({\bf gh},{\bf k})]_{\bf ghk}} }
\label{eq:U_defect_no_perm}
\\
\eta_{c_{\bf k}}({\bf g},{\bf h})
&=&
\frac{ F^{0_{\bf g} [c \cohosub{w}({\bf k},{\bf g})\overline{\cohosub{w}({\bf g},{\bf \bar{g}kg})}]_{\bf \bar{g}kg}  0_{\bf h} } }{ F^{ c_{\bf k} 0_{\bf g} 0_{\bf h} } F^{0_{\bf g} 0_{\bf h} [c \cohosub{w}({\bf k},{\bf gh}) \overline{\cohosub{w}({\bf gh},{\bf \bar{h}\bar{g}kgh})}]_{\bf \bar{h}\bar{g}kgh} } }
R^{c_{\bf k} [\cohosub{w}({\bf g},{\bf h})]_{\bf gh} }
\label{eq:eta_defect_no_perm}
\end{eqnarray}
where the $F^{0_{\bf g} 0_{\bf h} 0_{\bf k}}$ are solutions to the consistency condition
\begin{eqnarray}
\label{eq:obstruction_consistency}
\frac{ F^{0_{\bf gh} 0_{\bf k} 0_{\bf l}} F^{ 0_{\bf g} 0_{\bf h} 0_{\bf kl}} }
{F^{0_{\bf g} 0_{\bf h} 0_{\bf k}} F^{0_{\bf g} 0_{\bf hk} 0_{\bf l}} F^{0_{\bf h} 0_{\bf k} 0_{\bf l}} }
= \mathcal{O}({\bf g},{\bf h},{\bf k},{\bf l})
,
\end{eqnarray}
such that $F^{0_{\bf g} 0_{\bf h} 0_{\bf k}}=1$ when any of ${\bf g}$, ${\bf h}$, or ${\bf k}$ are equal to ${\bf 0}$,
and where we have defined the quantity
\begin{equation}
\label{eq:obstruction}
\mathcal{O} ({\bf g},{\bf h},{\bf k},{\bf l})
=
\frac{ F^{[\cohosub{w}({\bf g},{\bf h})]_{\bf 0} [\cohosub{w}({\bf k},{\bf l})]_{\bf 0} [\cohosub{w}({\bf gh},{\bf kl})]_{\bf 0}}
F^{ [\cohosub{w}({\bf k},{\bf l})]_{\bf 0} [ \cohosub{w}({\bf h},{\bf kl})]_{\bf 0} [ \cohosub{w}({\bf g},{\bf hkl})]_{\bf 0} }
F^{[\cohosub{w}({\bf h},{\bf k})]_{\bf 0} [ \cohosub{w}({\bf g},{\bf hk})]_{\bf 0} [ \cohosub{w}({\bf ghk},{\bf l})]_{\bf 0} }}
{F^{[\cohosub{w}({\bf g},{\bf h})]_{\bf 0} [ \cohosub{w}({\bf gh},{\bf k})]_{\bf 0} [ \cohosub{w}({\bf ghk},{\bf l})]_{\bf 0} }
F^{[\cohosub{w}({\bf k},{\bf l})]_{\bf 0} [ \cohosub{w}({\bf g},{\bf h})]_{\bf 0} [ \cohosub{w}({\bf gh},{\bf kl})]_{\bf 0} }
F^{ [\cohosub{w}({\bf h},{\bf k})]_{\bf 0} [ \cohosub{w}({\bf hk},{\bf l})]_{\bf 0} [ \cohosub{w}({\bf g},{\bf hkl})]_{\bf 0} }}
R^{[\cohosub{w}({\bf g},{\bf h})]_{\bf 0} [\cohosub{w}({\bf k},{\bf l})]_{\bf 0}}
,
\end{equation}
\end{widetext}
The left hand side of Eq.~(\ref{eq:obstruction_consistency}) is clearly a 4-coboundary in $B^{4}(G, \text{U}(1))$, so in order for it to be possible to satisfy this equation, $\mathcal{O} ({\bf g},{\bf h},{\bf k},{\bf l})$ must also be a 4-coboundary. Thus, we see that $\mathcal{O}$ will be an obstruction to having a $G$-crossed extension, i.e. to satisfying the $G$-crossed consistency conditions, when $\mathcal{O} \notin B^{4}(G, \text{U}(1))$. Moreover, it can be shown that $\mathcal{O} ({\bf g},{\bf h},{\bf k},{\bf l}) \in Z^{4}(G, \text{U}(1))$, so it defines a cohomology class $[\mathcal{O}] \in H^{4}(G, \text{U}(1))$.

We emphasize that the defectification obstruction class $[\mathcal{O}]$ is defined entirely in terms of the MTC $\mathcal{C}_{\bf 0}$ and the fractionalization class $[\coho{w}]$, and that it is independent of gauge choices. Indeed, if we modify the basic data of $\mathcal{C}_{\bf 0}$ by a vertex basis gauge transformation, so that it represents the same MTC, we see that the corresponding obstruction becomes
\begin{equation}
\widetilde{\mathcal{O}} ({\bf g},{\bf h},{\bf k},{\bf l}) = \mathcal{O}({\bf g},{\bf h},{\bf k},{\bf l}) \text{d} \mu ({\bf g},{\bf h},{\bf k},{\bf l})
\end{equation}
where
\begin{equation}
\mu({\bf g},{\bf h},{\bf k}) = \frac{\Gamma^{ [ \cohosub{w}({\bf h},{\bf k})]_{\bf 0} [ \cohosub{w}({\bf g},{\bf hk})]_{\bf 0}  }}
{\Gamma^{ [ \cohosub{w}({\bf g},{\bf h})]_{\bf 0} [ \cohosub{w}({\bf gh},{\bf k})]_{\bf 0} }}
,
\end{equation}
so this only modifies $\mathcal{O}$ by a 4-coboundary. On the other hand, if we modify $\coho{w}$ by a coboundary, so that it represents the same fractionalization class, i.e. $\widehat{\coho{w}}({\bf g},{\bf h}) = \coho{w}({\bf g},{\bf h}) \text{d}\coho{z}({\bf g},{\bf h})$, we find that the obstruction becomes
\begin{equation}
\widehat{\mathcal{O}} ({\bf g},{\bf h},{\bf k},{\bf l}) = \mathcal{O}({\bf g},{\bf h},{\bf k},{\bf l}) \text{d} \sigma ({\bf g},{\bf h},{\bf k},{\bf l})
\end{equation}
where
\begin{eqnarray}
&& \quad \sigma({\bf g},{\bf h},{\bf k}) =
\frac{R^{[\cohosub{w}({\bf g},{\bf h})]_{\bf 0} [\cohosub{z}(\bf k)]_{\bf 0}}}
{ F^{[\cohosub{z}(\bf g)]_{\bf 0} [\cohosub{z}(\bf h) \cohosub{w}({\bf g},{\bf h})]_{\bf 0} [\cohosub{z}(\bf k) \cohosub{w}({\bf gh},{\bf k})]_{\bf 0}}
}
\notag \\
&&\qquad \qquad \times
\frac{F^{ [\cohosub{z}(\bf h)]_{\bf 0} [\cohosub{z}(\bf k)\cohosub{w}({\bf h},{\bf k})]_{\bf 0} [ \cohosub{w}({\bf g},{\bf hk})]_{\bf 0}}}
{ F^{[\cohosub{z}(\bf h)]_{\bf 0} [ \cohosub{w}({\bf g},{\bf h})]_{\bf 0} [\cohosub{z}(\bf k) \cohosub{w}({\bf gh},{\bf k})]_{\bf 0}}}
\notag \\
&& \times \frac{F^{[\cohosub{w}({\bf g},{\bf h})]_{\bf 0} [\cohosub{z}(\bf k)]_{\bf 0} [\cohosub{w}({\bf gh},{\bf k})]_{\bf 0}}
F^{ [\cohosub{z}(\bf k)]_{\bf 0} [ \cohosub{w}({\bf h},{\bf k})]_{\bf 0} [ \cohosub{w}({\bf g},{\bf hk})]_{\bf 0} } }
{F^{[\cohosub{z}(\bf k)]_{\bf 0} [ \cohosub{w}({\bf g},{\bf h})]_{\bf 0} [ \cohosub{w}({\bf gh},{\bf k})]_{\bf 0} }}
,
\qquad
\label{eq:O'}
\end{eqnarray}
so this also only modifies $\mathcal{O}$ by a 4-coboundary. We obtain Eq.~(\ref{eq:O'}) by noticing that changing $\coho{w}$ by a coboundary can be viewed as choosing a different choice $\widehat{a}_{\bf g} = [a\coho{z}({\bf g})]_{\bf g}$ of the defect charge labels.

The expressions for the basic data in Eqs.~(\ref{eq:F_defect_no_perm})-(\ref{eq:eta_defect_no_perm}) are uniquely obtained (up to gauge freedom) in terms of $F^{ 0_{\bf g} 0_{\bf h} 0_{\bf k}}$, given $\mathcal{C}_{\bf 0}$ and $[\coho{w}]$, by solving the $G$-crossed consistency conditions (pentagon and heptagon equations) involving less than four defects. The remaining condition in Eq.~(\ref{eq:obstruction_consistency}) is then obtained by inserting the $F$-symbols from Eq.~(\ref{eq:F_defect_no_perm}) into the pentagon equation involving four defects. This demonstrates that a nontrivial obstruction does not simply indicate an inability for the defect theory to satisfy the pentagon equation, but rather an inability to satisfy the entire $G$-crossed consistency conditions, including the heptagon equations. Indeed, it is sometimes possible to satisfy the pentagon equations, but not the full $G$-crossed consistency conditions when a theory is obstructed, as we will see in subsequent examples.

When the defectification obstruction vanishes, i.e. when $\mathcal{O} \in B^{4}(G, \text{U}(1))$, the complete $G$-crossed consistency equations can be satisfied and solved. In this case, we can write
\begin{equation}
\mathcal{O} ({\bf g},{\bf h},{\bf k},{\bf l}) = \frac{\lambda({\bf g},{\bf h},{\bf k}) \lambda({\bf g},{\bf hk},{\bf l}) \lambda({\bf h},{\bf k},{\bf l}) }{\lambda({\bf gh},{\bf k},{\bf l}) \lambda({\bf g},{\bf h},{\bf kl}) }
\end{equation}
for some $\lambda({\bf g},{\bf h},{\bf k}) \in C^{3}(G, \text{U}(1))$, and then the solutions of Eq.~(\ref{eq:obstruction_consistency}) take the form
\begin{equation}
F^{0_{\bf g} 0_{\bf h} 0_{\bf k}} = \frac{ \alpha({\bf g},{\bf h},{\bf k}) }{\lambda({\bf g},{\bf h},{\bf k})}
\end{equation}
where $\alpha({\bf g},{\bf h},{\bf k})$ are 3-cocycles in $Z^{3}(G, \text{U}(1))$. As explained in Sec.~\ref{sec:Classification_GCrossed}, the different solutions that are related to each other by 3-coboundaries correspond to gauge equivalent $G$-crossed theories. Consequently, the gauge inequivalent solutions of Eq.~(\ref{eq:obstruction_consistency}) are classified by $[\alpha] \in H^{3}(G, \text{U}(1))$. Thus, the $G$-crossed MTCs extending a MTC $\mathcal{C}_{\bf 0}$ with symmetry group $G$ that does not permute topological charges in $\mathcal{C}_{\bf 0}$ are fully specified by the symmetry fractionalization class $[\coho{w}] \in  H^{2}(G,\mathcal{A})$ and the defectification class $[\alpha] \in H^{3}(G, \text{U}(1))$. This is consistent with the classification discussion of Sec.~\ref{sec:Classification_GCrossed}.

The expression in Eq.~(\ref{eq:obstruction}) for the defectification obstruction when the symmetry action does not permute anyons was also obtained in \Ref{chen2014} using a heuristic physical interpretation of \Ref{ENO2009}. A very different looking expression for the defectification obstruction was obtained in \Ref{ENO2009} by working at a higher category level, where the $\mathcal{C}_{\bf g}$ sectors (bimodule categories) are treated as objects of a fusion category.

\section{Examples}
\label{exampleSec}

In this section, we consider a number of examples, which we label by the initial anyon model (UBTC) $\mathcal{C}_\mb{0}$ and the symmetry group $G$. In our examples, we only consider unitary, on-site symmetries with finite symmetry group $G$. (As such, we restrict our attention to $\text{Aut}_{0,0}(\mathcal{C})$.) We obtain the data of the corresponding $G$-crossed UBTCs $\ext{C}{G}$ by solving the consistency equations of Sec.~\ref{sec:Algebraic_Theory}, using various derived properties and classification theorems when useful, and present as much of the basic data as is reasonable. We also present explicit derivations of the fusion rules and the modular data of the corresponding gauged theories $\gauged{C}{G}$.

The purpose of these examples is twofold: (1) to provide the basic data of $\ext{C}{G}$ and $\gauged{C}{G}$ for some of the more interesting and perhaps more physically relevant models, and (2) to illustrate the different types of nontrivial issues and structures that may arise when concrete calculations are performed. Most of the examples examined here have symmetry group $G= \mathbb{Z}_{2}$. In Sec.~\ref{sec:3-Fermion}, we thoroughly consider an example with a non-Abelian symmetry $G=S_3$. We note that, $H^{4}(\mathbb{Z}_{N} , \text{U}(1))=\mathbb{Z}_{1}$, so there is never an obstruction to defectification when $G = \mathbb{Z}_{N}$ (though there may be a fractionalization obstruction). In Sec.~\ref{sec:semion_Z2Z2}, we examine an example with $G=\mathbb{Z}_{2} \times \mathbb{Z}_{2}$ which exhibits nontrivial defectification obstruction for certain fractionalization classes. In Sec.~\ref{ex_sec:H3_obstruction}, we present an example with $G=\mathbb{Z}_{2}$ that exhibits nontrivial fractionalization obstruction. Partial results from some of the
examples that we examine have also been obtained in previous works~\cite{dijkgraaf1989,moore1989,dijkgraaf1990,Propitius1995,barkeshli2010,barkeshli2010twist,
barkeshli2011orb,chen2013,barkeshli2012a,cheng2012,clarke2013,lindner2012,barkeshli2013genon,lu2013, teo2013b,chen2014}, though mostly using different methods.

In the following, we adopt the convention that the vacuum topological charge is always referred to as either $0$ or $I$ and the identity element of $G$ is referred to as either $\mb{0}$ or $\openone$. We also will frequently use the notation $[a]_N \equiv a \text{ mod } N$ for the least residue modulo $N$ of $a$.

\subsubsection{Gauge choices}

In the following, we will need to make some gauge choices in order to specify the basic data. There are some relatively natural gauge choices that we describe here is some detail.

When the obstruction to fractionalization vanishes (which is the case for all but one of our examples), we will set $\coho{O} =0$, which can be done for a particular choice of $\beta_{a}({\bf g},{\bf h})$. With this choice, $\coho{w}({\bf g},{\bf h})$ is a 2-cocycle, so we have $[\coho{w}] \in H^{2}_{[\rho]}(G, \mathcal{A})$.

As noted at the end of Sec.~\ref{sec:H3Invariance}, when the unitary symmetry action $\rho_{\bf g}$ does not permute any topological charge values, its action on $\mathcal{C}_{\bf 0}$ is a natural isomorphism and the symmetry fractionalization obstruction automatically vanishes. In this case, we can set $[U_{\bf g}(a,b;c)]_{\mu \nu} = \delta_{\mu \nu}$ for $a,b,c \in \mathcal{C}_{\bf 0}$ as a gauge choice, and consequently may also choose $\beta_{a}({\bf g},{\bf h})=1$, $\coho{O} =0$, and $\eta_{a}(\mb{g},\mb{h}) = M_{a \cohosub{w}({\bf g},{\bf h})}$ for $a \in \mathcal{C}_{\bf 0}$.

In the simple case when $G=\mathbb{Z}_2$ (which we will encounter in many of our examples), we can use the symmetry action gauge freedom to pick a gauge in which $\eta_{a}({\bf g},{\bf h})=1$ for all $a \in \ext{C}{G}$. In particular, if these phases were nontrivial in such cases, we could apply a symmetry action gauge transformation that satisfies the condition $\gamma_{a}({\bf 1}) \gamma_{^{\bf 1}a}({\bf 1}) = \eta_{a}({\bf 1},{\bf 1})$ to obtain $\check{\eta}_{a}({\bf g},{\bf h})=1$. This gauge fixing leaves us with the freedom to apply an additional symmetry action gauge transformation with $\gamma'_{a}({\bf 1}) = \pm 1$ when $^{\bf 1}a=a$ and $\gamma'_{a}({\bf 1}) = \left[\gamma'_{^{\bf 1}a}({\bf 1})\right]^{-1}$ when $^{\bf 1}a \neq a$, without further changing the values of $\eta_{a}({\bf g},{\bf h})$. The gauge choice with $\eta_{a}({\bf g},{\bf h})=1$ is particularly convenient for determining the $\gauged{C}{G}$ theory, as only linear (not projective) irreps need be considered in constructing the dyons.

More generally, when the stabilizer subgroup $G_a$ of $a_\mb{g}$ has trivial $H^2(G_a, \mathrm{U}(1))= \mathbb{Z}_{1}$, we can choose a gauge in which $\eta_{a_\mb{g}}(\mb{h},\mb{k})=1$ for all $\mb{h},\mb{k}\in G_a$. This is because $\eta_{a_\mb{g}}(\mb{h},\mb{k})$ represents an element of $H^2(G_a, \mathrm{U}(1))$, i.e. it satisfies the $2$-cocycle condition for $Z^2(G_a, \mathrm{U}(1))$ and a symmetry action gauge transformation modifies it by a $2$-coboundary in $B^2(G_a, \mathrm{U}(1))$, since $\check{\eta}_{a_\mb{g}}(\mb{h},\mb{k}) = \frac{\gamma_{a}({\bf hk})}{\gamma_{a}({\bf h}) \gamma_{a}({\bf k})} \eta_{a_\mb{g}}(\mb{h},\mb{k})$ for $\mb{h},\mb{k}\in G_a$. We note that $H^{2}(\mathbb{Z}_{N} , \text{U}(1))=\mathbb{Z}_{1}$.

\subsection{Trivial Bosonic State with $G$ Symmetry}
\label{sec:trivial}

In this section, we consider the case where the starting topological phase $\mathcal{C}_{\bf 0}$ is trivial in the sense that it only contains topologically trivial bosonic excitations, i.e. $\mathcal{C}_{\bf 0} = \{ 0 \}$,
but possesses a symmetry group $G$. This describes a bosonic symmetry-protected topological (SPT) phase with symmetry group $G$.
In this case, the construction of the extended category $\ext{C}{G}$ is straightforward. Each $\mathcal{C}_\mathbf{g}$ contains a single defect type, which will be denoted by $\mathbf{g}$. Fusion of defects is given by group multiplication, that is
\begin{equation}
\mathbf{g}\times\mathbf{h}=\mathbf{gh}
.
\end{equation}
Since the fusion category $\graded{C}{G}$ will appear elsewhere, we will refer to it as $\mathrm{Vec}_G$. Mathematically, this is the category of $G$-graded vector spaces. It is a well-known result that the equivalence classes of $F$-symbols under vertex basis gauge transformations are determined by the $3$rd group cohomology $H^{3}(G,\text{U}(1))$~\cite{dijkgraaf1990,Kitaev06a}. Given a $3$-cocycle $\alpha\in Z^3(G, \text{U}(1))$, we define the $F$-symbols as
\begin{equation}
[F^{\mathbf{g},\mathbf{h},\mathbf{k}}_{\mathbf{ghk}}]_{\mathbf{gh},\mathbf{hk}}=\alpha(\mathbf{g},\mathbf{h},\mathbf{k})
.
\end{equation}
As usual, we require $F^{\mathbf{g},\mathbf{h},\mathbf{k}}=1$ whenever any of $\mb{g},\mb{h}, \mb{k}$ is $\mb{0}$, so we always impose this condition on the $3$-cocycle $\alpha$.

We can also always apply the symmetry action gauge transformation to set $R^{\mb{g},\mb{h}}_{\mb{gh}}=1$ for all values of $\mb{g}$ and $\mb{h}$. (If we started with nontrivial values of $R^{\mb{g},\mb{h}}_{\mb{gh}}$ in this example, then we would apply the symmetry action gauge transformation $\gamma_{\bf g}({\bf h}) = [R^{\mb{g},\mb{h}}_{\mb{gh}}]^{-1}$ to remove any nontrivial braiding phases.) The corresponding braiding operators simply involve the $G$-action of group elements acting by conjugation. For this gauge choice, the corresponding $U_\mb{k}$ and $\eta_\mb{k}$ are uniquely determined by the $G$-crossed consistency equations to be
\begin{eqnarray}
U_\mb{k}(\mb{g},\mb{h};\mb{gh}) &=& \frac{\alpha(\mb{g},\mb{k}, \,^{\ol{\mb{k}}}\mb{h})}{ \alpha(\mb{g},\mb{h},\mb{k}) \alpha(\mb{k}, \,^\mb{\bar{k}}\mb{g}, \,^\mb{\bar{k}}\mb{h})}, \\
\eta_\mb{k}(\mb{g},\mb{h}) &=& \frac{\alpha( \mb{g}, \,^\mb{\bar{g}}\mb{k} , \mb{h}) } { \alpha(\mb{g},\mb{h}, \,^\mb{\bar{h} \bar{g}}\mb{k}) \alpha(\mb{k},\mb{g},\mb{h})}
.
\end{eqnarray}
We denote the corresponding $G$-crossed theory as SPT$_{G}^{[\alpha]}$.

As discussed in Sec.~\ref{sec:Classification_GCrossed}, the $H^3(G, \text{U}(1))$ classification of the $F$-symbols of the $G$-crossed extensions $\mathcal{C}_G^\times$ for general $\mathcal{C}$ is in one-to-one correspondence with the classification of 2D bosonic SPT states with symmetry group $G$, described here and developed in \Ref{chen2013}. Therefore, we see that \it classifying $\mathcal{C}_G^\times$ reproduces the classification of bosonic SPT states. \rm

The $G$ action on $\ext{C}{G}$ (for $\mathcal{C}_{\bf 0}$ trivial) is obviously given simply by conjugation. Therefore, we immediately obtain the quasiparticle labels in the gauged theory as a pair $([{\bf g}], \pi_{\bf g})$ where $[{\bf g}] = \{ {\bf hgh}^{-1}, \forall {\bf h}\in G  \}$ is a conjugacy class of $G$ (i.e. an orbit under $G$ action) and $\pi_{\bf g}$ is an irreducible projective representation of the stabilizer group, i.e. the centralizer of a representative element ${\bf g} \in [{\bf g}]$. If we consider trivial $F$-symbols on $\text{Vec}_G$, we see that all $U$ and $\eta$ can be set to $1$ and the anyon content of the gauged theory agrees exactly with the well-known quantum double construction $\text{D}(G)$, describing discrete $G$ gauge theory. In general, gauging the symmetry of $\ext{C}{G}= \text{SPT}_{G}^{[\alpha]}$ results in a twisted quantum double $\gauged{C}{G} = \text{D}^{[\alpha]}(G)$~\cite{Dijkgraaf1991, Propitius1995,HuPRB2013}.

\subsubsection{$\mathbb{Z}_{N}$ symmetry}
\label{sec:Z_N_SPT}

For additional illustration, let us consider the $G$-crossed braiding for $G=\mathbb{Z}_N$. Since $G$ is Abelian, the fusion rules (i.e. group multiplication) is written as addition, that is $a \times b = [a+b]_{N}$. The $G$-extension is simply $\text{Vec}_{G}^{\alpha}$ equipped with a $3$-cocycle
\begin{equation}
\alpha(a,b,c)=e^{i \frac{2\pi p}{N^2}a(b+c-[b+c]_N)},
\label{eqn:zncocycle}
\end{equation}
where $p \in \{ 0,1,\ldots,N-1 \}$. In this case, we find it more illustrative to choose a gauge in which $\eta_{a}(b,c)= 1$, for all $a,b,c \in \mathbb{Z}_{N}$. Solving the $G$-crossed heptagon equations yields
\begin{equation}
R^{ab}_{[a+b]_N}=e^{-i \frac{2\pi p}{N^2}ab} e^{i \frac{2\pi}{N} m_a b}.
\label{eqn:znrsym}
\end{equation}
and
\begin{eqnarray}
U_c(a,b;[a+b]_{N})&=& e^{-i \frac{4\pi p}{N^2}(a+b-[a+b]_N)c} \notag \\
&& \quad \times e^{i \frac{2\pi}{N}(m_a+m_b-m_{[a+b]_{N}})c}
, \quad
\end{eqnarray}
where $m_a \in \mathbb{Z}$. Clearly, all the terms depending on $m_a$ represent a symmetry action gauge redundancy, so we could set $m_{a}=0$ as a gauge choice (specifically, by using $\gamma_{a}(b) = e^{-i \frac{2\pi}{N} m_{a} b}$), while leaving $\eta_{a}(b,c)= 1$ fixed. The topological twists and pure braid (double exchange) operations are given by
\begin{eqnarray}
\theta_a &=& e^{-i \frac{2\pi p }{N^2} a^2 }e^{i \frac{2\pi}{N} m_a a}, \\
R^{ab}_{[a+b]_{N}} R^{ba}_{[a+b]_{N}} &=& e^{-i \frac{4\pi p}{N^2} ab } e^{i \frac{2\pi}{N} ( m_b a + m_a b)}.
\end{eqnarray}
It is evident from these expressions that $m_a$ can be understood as the number of $\mathbb{Z}_N$ charges attached to the defect $a$, due to non-universal local energetics. This explains why these solutions should be considered as being gauge equivalent, since in the extended theory $\mathbb{Z}_N$ charges are still part of the vacuum sector.

We know consider the gauged theory. Since $G= \mathbb{Z}_{N}$ is Abelian, each $a\in\ext{C}{G}$ is also a $G$-orbit. They can also carry gauge charges labeled again by $m\in\mathbb{Z}_N$. We therefore obtain $|G|^2$ quasiparticles labeled by $(a,m)$. Their fusion rules are
\begin{eqnarray}
&& (a,m)\times(b,n)  \\
&& =\left( [a+b]_{N}, \left[m+n-\frac{2p}{N}(a+b-[a+b]_{N})\right]_{N}\right),
\notag
\end{eqnarray}
where the additional gauge charges come from the nontrivial symmetry action on the fusion state of the defects. The topological twist of $(a,m)$ is then
\begin{equation}
 \theta_{(a,m)}=e^{i \frac{2\pi }{N} a m} e^{-i \frac{2\pi p}{N^2} a^{2}} .
\end{equation}
These results agree exactly with the twisted quantum double D$^{[\alpha]}(\mathbb{Z}_N)$~\cite{Dijkgraaf1991, Propitius1995,levin2012, HuPRB2013, LinPRB2014}.

\subsection{Trivial Fermionic State $\mathbb{Z}_{2}^{(1)}$ with $G$ Symmetry}
\label{sec:fermions}

In this section, we consider a trivial fermionic topological phase, which is an example for which $\mathcal{C}_{\bf 0}$ is not a modular theory.
Even though the fermion is a local excitation in such a case, it is useful to
view it as a non-trivial element of the category and, therefore, to treat it as a topological charge.
To describe such a situation, we use the UBTC with $\mathcal{C}_{\bf 0} = \{I,\psi\}$ where $I$ is the vacuum charge, $\psi$ is the fermion, and $\psi\times\psi=I$.
$\mathcal{C}_{\bf 0}$ should be viewed therefore as a topological abstraction of gapped fermionic systems with
only short-range entanglement.

The $F$-symbols and $R$-symbols are
\begin{equation}
[F^{\psi\psi\psi}_\psi]_{I I}=1, \quad R^{\psi\psi}_I=-1.
\end{equation}
Notice that the theory is not modular, since
\begin{equation}
S=\frac{1}{\sqrt{2}} \left[
\begin{matrix}
		1 & 1\\
		1 & 1
\end{matrix}
\right]
\end{equation}
is singular. Using the notation of Sec.~\ref{sec:Z_N}, we will denote this BTC as $\mathcal{C}_\mb{0}=\mathbb{Z}_2^{(1)}$. We note that our results on classification of symmetry fractionalization do not directly apply, since modularity was an essential part of the argument.

However, we may still apply the general theory developed for $G$-crossed BTCs and gauging the symmetry. For fermionic SPT phases with an arbitrary finite symmetry group $G$, the resulting classification is given by the three cohomology groups $H^1(G, \mathbb{Z}_2), H^2(G,\mathbb{Z}_2)$ and $H^3(G,\mathrm{U}(1))$, as demonstrated in \Ref{fspt}. Here, we will only examine the case of $G=\mathbb{Z}_2$ in explicit detail.

The $\mathbb{Z}_2$-crossed extensions of $\mathcal{C}_\mb{0}$ will reproduce the known $\mathbb{Z}_8$ classification of interacting fermionic SPT states with a unitary on-site $\mathbb{Z}_2$ symmetry~\cite{GuPRB2014, QiNJP2012, RyuPRB2012, HongPRB2013}. Physically, these fermionic SPT phases can be realized in non-interacting spin-$\frac{1}{2}$ superconductors, where the spin up and spin down fermions form class D topological superconductors with Chern number $\nu$ and $-\nu$, respectively, where the collapse to the $\mathbb{Z}_8$ classification is given by $\nu \mod 8$. In this case, the $\mathcal{C}_\mb{0}=\mathbb{Z}_2^{(1)}$ BTC describes both spin up and down fermions and the $G=\mathbb{Z}_2$ symmetry is viewed as the fermion parity symmetry of the spin up fermions. We will refer to such a fermionic SPT phase as $(p_x +ip_y)^{\nu} \times (p_x - ip_y)^{\nu}$, as it represents $|\nu|$ copies of $p_{x}\pm ip_{y}$ superconductors.~\footnote{The $\mathbb{Z}_2$-crossed extensions and gauging discussed here should not be confused with a different concept, which is referred to as a modular extension of fermions. Physically, a modular extension of fermions corresponds to gauging the $\mathbb{Z}_2$ symmetry of fermion parity conservation, for which the $\psi$ fermions play the role of the $\mathbb{Z}_2$ charges and there are no other independent (bosonic) $\mathbb{Z}_2$ charges. The extended category in such a modular extension will, by definition, be a modular one and will be braided in the usual sense, not $G$-crossed braided. In contrast, the $\mathbb{Z}_2$-crossed extensions considered in this section allow bosonic $\mathbb{Z}_2$ charges and the braiding is $G$-crossed.}

We now examine the algebraic structures of these $G$-crossed extension and gauging for $G=\mathbb{Z}_2$. The $G$-crossed theories are $\mathbb{Z}_2$-graded: $\graded{C}{G}=\mathcal{C}_\mb{0}\bigoplus \mathcal{C}_{\mb{1}}$, where $\mb{1}$ is the non-trivial element of $\mathbb{Z}_2$. In this simple case ($G=\mathbb{Z}_2$ with an action that does not permute any topological charges), we can use a symmetry action gauge transformation to pick a gauge in which $\eta_{a}({\bf g},{\bf h})=1$ for all $a \in \mathcal{C}_{G}$, so we will use such a gauge choice in the following.

As explained in Sec.~\ref{sec:G-graded_Fusion}, we must have $\mathcal{D}_\mb{1}=\mathcal{D}_\mb{0}=2$. This allows for two distinct ways of constructing $\ext{C}{G}$: (1) there is a single defect type in the $\mathcal{C}_\mb{1}$ sector, which is non-Abelian with quantum dimension $d=\sqrt{2}$; or
(2) there are two defect types in $\mathcal{C}_\mb{1}$, which are Abelian with quantum dimension $d=1$. We study these two cases in turn. We will find that each case admits four distinct $\mathbb{Z}_2$ extensions $\ext{\mathcal{C}}{\mathbb{Z}_2}$, for a total of $8$ possible $\mathbb{Z}_2$ extensions of $\mathcal{C}_{\bf 0}$.

\subsubsection{Non-Abelian extensions}

We first consider case (1), where there is a single non-Abelian defect, which we write as $\mathcal{C}_{\mb{1}} = \{ \sigma\}$. Since we must have $\bar{\sigma}=\sigma$ and the quantum dimensions must satisfy Eq.~(\ref{eq:d_g_relation}), the fusion rules for $\sigma$ must be: $\sigma\times\sigma=I+\psi$.
We conclude that the $G$-graded fusion category $\mathcal{C}_{\mathbb{Z}_{2}}$ (underlying the $G$-crossed BTC extension) must be identical to one of the unitary fusion categories that have the Ising fusion rules. The $F$-symbols of such fusion categories are completely classified~\cite{Tambara1998}; there are two possible fusion categories (up to gauge transformations) with these fusion rules, and they are distinguished by the Frobenius-Schur indicator $\varkappa_{\sigma}=\pm 1$. The nontrivial $F$-symbols of these fusion categories are given in \eqref{eqn:IsingF}.

Solving the consistency equations for $G$-crossed braiding, in a choice of gauge for which $\eta_{\psi}({\bf g},{\bf h})=\eta_{\sigma}({\bf g},{\bf h})=1$, we obtain the braiding $R$-symbols
\begin{eqnarray}
R^{\sigma \psi}_\sigma &=& i\alpha, \\
R^{\psi \sigma}_\sigma &=& i\beta, \\
R^{\sigma\sigma}_I  &=& \lambda \sqrt{\varkappa_\sigma}e^{i\frac{\pi}{8}\alpha}, \\
R^{\sigma\sigma}_\psi &=& -i\alpha \lambda \sqrt{\varkappa_\sigma}e^{i\frac{\pi}{8}\alpha},
\end{eqnarray}
and symmetry action $U_\mb{g}$ symbols
\begin{eqnarray}
U_\mb{1}(\psi,\psi;I)= U_\mb{1}(\sigma,\sigma; I) &=& 1, \\
U_\mb{1}(\sigma,\sigma;\psi)= U_\mb{1}(\psi,\sigma; \sigma)=U_\mb{1}(\sigma,\psi;\sigma) &=& \alpha\beta,
\end{eqnarray}
where $\alpha^2=\beta^2=\lambda^{2} = 1$. Notice that $\beta=+1$ and $-1$ give equivalent solutions under symmetry action gauge transformations (related by $\gamma_{\psi}({\bf 1})=-1$), as do $\lambda = +1$ and $-1$ (related by $\gamma_{\sigma}({\bf 1})=-1$). It is convenient to choose a gauge for which $\beta = \alpha$, so that $U_\mb{k}(a,b;c)=1$ for all $a,b,c$ and $R^{\sigma \psi}_\sigma = R^{\psi \sigma}_\sigma = i \alpha$, and we may as well also gauge fix $\lambda = +1$. Thus, we find four distinct $\mathbb{Z}_2$ extensions $\ext{\mathcal{C}}{\mathbb{Z}_2}$, distinguished by $\alpha$ and $\varkappa_\sigma$, which may independently take the values $\pm 1$.

Next, we gauge the $\mathbb{Z}_2$ symmetry. The topological charges in $\gauged{C}{G}$ may be written as $(a, s)$ where $a=I$, $\psi$, or $\sigma$, while $s=+$ or $-$ denotes the trivial or alternating irrep of $\mathbb{Z}_2$, respectively. Since the mutual braiding between $\psi$ and $\sigma$ is $R^{\sigma \psi}_\sigma R^{\psi \sigma}_\sigma = -1$ (assuming the gauge in which $\alpha =\beta$), and the mutual braiding of an alternating irrep and $\sigma$ generates a $-1$ phase, it follows that the topological charge $(\psi, -)$ has trivial braiding with all topological charges in the gauged theory, including $(\sigma, \pm)$. As such, the gauged theory $\gauged{C}{G}$ comprises the direct product of a $\mathbb{Z}_2^{(1)}$ subcategory, formed by $\{(I,+),(\psi, -)\}$, with an $\text{Ising}^{(\nu)}$ MTC (using the notation in Appendix~\ref{sec:Ising}).

Thus, gauging the $\mathbb{Z}_2$ symmetry of the four distinct $\ext{C}{G}$ yields four distinct $\gauged{C}{G}$, which we can identify with $\mathbb{Z}_2^{(1)} \boxtimes \text{Ising}^{(\nu)}$ where $\nu=1,3,5,7$. More explicitly, we have
\begin{equation}
\begin{tabular}{llll}
$\alpha=1$, $\varkappa_\sigma=1$  & $\longrightarrow$ & &$\mathbb{Z}_2^{(1)} \boxtimes \text{Ising}^{(1)}$ \\
$\alpha=1$, $\varkappa_\sigma=-1$ & $\longrightarrow$ & &$\mathbb{Z}_2^{(1)} \boxtimes \text{Ising}^{(5)}$ \\
$\alpha=-1$, $\varkappa_\sigma=1$ & $\longrightarrow$ & &$\mathbb{Z}_2^{(1)} \boxtimes \text{Ising}^{(7)}$ \\	
$\alpha=-1$, $\varkappa_\sigma=-1$ & $\longrightarrow$ & & $\mathbb{Z}_2^{(1)} \boxtimes \text{Ising}^{(3)}$
\end{tabular}
\end{equation}
This exhibits the correspondence between the Chern number $\nu$ fermionic SPT phases $(p_x +ip_y)^{\nu} \times (p_x - ip_y)^{\nu}$ phases with $\nu=1,3,5,7$ and the non-Abelian $G$-crossed UBTCs described above, which characterizes these phases.

\subsubsection{Abelian extensions}

We now consider case (2), where there are two Abelian defect types, i.e. $|\mathcal{C}_\mb{1}| =2$. In this case, a defect's topological charge must change to the other type when it is fused with $\psi$. However, this allows for two possible sets of fusion rules: (a) $\mathbb{Z}_2 \times \mathbb{Z}_2$ fusion, in which case we write the defects as $\mathcal{C}_\mb{1} = \{ e , m \}$, so we have $e \times e = m\times m=I$ and $e\times m =\psi$, or (b) $\mathbb{Z}_4$ fusion, in which case we write the defects as $\mathcal{C}_\mb{1} = \{ v , \bar{v} \}$, so we have $v \times v = \bar{v} \times \bar{v} =\psi$ and $v \times \bar{v} =I$.

In case (2a), the $G$-crossed extensions can be written as the direct products $\mathbb{Z}_2^{(1)} \boxtimes \text{SPT}_{\mathbb{Z}_2}^{[\alpha]}$, where the bosonic SPT theories were described in Sec.~\ref{sec:trivial}. This gives two distinct extensions for this case, corresponding to $\alpha({\bf 1,1,1}) =+1$ and $-1$, respectively. These are, respectively, identified as the Chern number $\nu=0$ and $4$ fermionic SPT theories.

Gauging the $\mathbb{Z}_{2}$ symmetry of these theories yields $\mathbb{Z}_2^{(1)} \boxtimes \text{D}^{[\alpha]}(\mathbb{Z}_2)$. We note that $\text{D}^{+}(\mathbb{Z}_2) = \text{D}(\mathbb{Z}_2)$ is the $\mathbb{Z}_2$ discrete gauge theory (toric code), confirming the identification with $\nu=0$. On the other hand, $\text{D}^{-}(\mathbb{Z}_2) = \mathbb{Z}_{2}^{(\frac{1}{2})} \boxtimes \mathbb{Z}_{2}^{(-\frac{1}{2})}$ is the double semion theory, which confirms the identification with $\nu=4$, noting that $\mathbb{Z}_2^{(1)} \boxtimes \mathbb{Z}_{2}^{(\frac{1}{2})} \boxtimes \mathbb{Z}_{2}^{(-\frac{1}{2})} = \mathbb{Z}_2^{(1)} \boxtimes \mathbb{Z}_{2}^{(\frac{1}{2})} \boxtimes \mathbb{Z}_{2}^{(\frac{1}{2})}$.

In case (2b), we have $\graded{C}{G}=\mathrm{Vec}_{\mathbb{Z}_4}$, and we will again find two distinct $\mathbb{Z}_2$ extensions.
More explicitly, the topological charges of these $G$-crossed extensions are written in $\mathbb{Z}_4$ notation as: $I\equiv 0$, $v \equiv 1$, $\psi\equiv 2$, and $\bar{v}\equiv 3$. The $F$-symbols and the $G$-crossed braiding of $\mathrm{Vec}_{\mathbb{Z}_4}$ have been completely solved in Sec.~\ref{sec:trivial}, so we will use the same notation for that section, where the gauge is chosen to set $\eta_{a}({\bf g},{\bf h})=1$, the different $3$-cocycles $\alpha$ are labeled by $p$, as in Eq.~(\ref{eqn:zncocycle}), and $m_a$ enters the expression for the $G$-crossed $R$-symbols in Eq.~(\ref{eqn:znrsym}). However, there are now additional constraints on $p$ and $m_a$ imposed by requiring the even subsector of $\mathbb{Z}_{4}$ to be the BTC $\mathcal{C}_\mb{0} = \mathbb{Z}_2^{(1)}$. In particular, $F^{\psi\psi\psi}_\psi=1$ requires $p$ to be an even integer, $R^{\psi\psi}_I=-1$ requires that $(-i)^p (-1)^{m_2}=-1$, and $U_{2}(a,b;[a+b]_4) =1$ (since $2 \in \mathcal{C}_{\bf 0}$) requires $m_a+m_b-m_{a+b}$ to be an even integer. An immediate consequence is that $m_2$ is even, $m_1 + m_3$ is even, and $p=2$. Hence, we can parameterize $m_{a}$ for $a \in \mathbb{Z}_{4}$ as
\begin{equation}
\label{eqn:ma}
m_a=2 n_a + q [a]_2 ,
\end{equation}
where $q=0$ or $1$, while $n_a\in \mathbb{Z}$. The $R$-symbols are, thus
\begin{equation}
R^{ab}_{[a+b]_{4}} = e^{-i \frac{\pi}{4} ab}(-1)^{n_a b} i^{q [a]_2 b}
\end{equation}
and the $U$-symbols are given by
\begin{eqnarray}
U_{{\bf k}_c}(a,b;[a+b]_{4}) &=& (-1)^{(n_a+n_b-n_{[a+b]_{4}})c} \notag \\
&& \quad \times i^{q([a]_2 + [b]_2-[a+b]_2)c},
\end{eqnarray}
where we use the notation ${\bf k}_c$ to indicate the group element ${\bf k}\in G$ associated with the label $c$. (In this particular case, we can write ${\bf k}_{c} = [c]_2$ for our choice of labels.) One can easily see that a symmetry action gauge transformation [specifically, $\gamma_{a}(b) = (-1)^{n_a b}$] can be used to set $n_a=0$, so that
\begin{eqnarray}
R^{ab}_{[a+b]_4} &=& e^{-i \frac{\pi}{4} ab} i^{q [a]_2 b} \\
U_{{\bf k}_c} (a,b;[a+b]_{4}) &=& i^{q([a]_2 + [b]_2-[a+b]_2)c}.
\end{eqnarray}
This shows that the dependence on $n_a$ is merely a gauge freedom that can be removed and, hence, there are only two distinct $G$-crossed extensions of this form, specified by $q=0$ and $1$.
In this choice of gauge (with $n_a=0$), we find the topological twists and braiding statistics to be
\begin{equation}
	\begin{split}
		\theta_{v}=R^{v v}_{\psi}&=e^{-i \frac{\pi }{4}} i^{q}\\
		\theta_{\bar{v}}=R^{\bar{v} \bar{v}}_{\psi}&=e^{-i\frac{\pi }{4}}(-i)^{q}\\
		R^{v \psi}_{\bar{v}}R^{\psi v}_{\bar{v}}=R^{\bar{v} \psi}_{v} R^{\psi \bar{v}}_{v} &= -1 \\
		R^{\bar{v} v}_{I}R^{v \bar{v}}_{I}&=-i(-1)^{q}
	\end{split}
\end{equation}

Next, we gauge the $\mathbb{Z}_2$ symmetry of these theories. We label the topological charges in $\gauged{C}{G}$ by $(a, s)$ where $a\in\mathbb{Z}_4$, while $s= +$ or $-$ denotes the trivial or alternating irrep of $\mathbb{Z}_2$, respectively. We work in the gauge where $n_a=0$. First we consider the case $q=0$, for which all $U$ symbols equal $1$. The fusion rules of the gauged theory are then simply given by: $(a,s)\times (b,r)=([a+b]_4, sr)$. We observe that $(2,-)$ has trivial full braiding with all topological charge types and that we can write the topological charges with $s=-$ irreps as $(a,-) = ([a+2]_{4},+)\times(2,-)$. In this manner, we can write the gauged theory as the product $\gauged{C}{G} =\mathbb{Z}_2^{(1)} \boxtimes \mathbb{Z}_4^{(-\frac{1}{2})}$, where the $\mathbb{Z}_2^{(1)}$ corresponds to the topological charges $\{ (0,+), (2,-)\}$ and the $\mathbb{Z}_4^{(-\frac{1}{2})}$ corresponds to the topological charges $\{(a, +)|\: a = 0,1,2,3\}$. For $q=1$, we similarly obtain the gauged theory $\gauged{C}{G} = \mathbb{Z}_2^{(1)} \boxtimes \mathbb{Z}_4^{(\frac{1}{2})}$. Thus, we have two distinct gauged theories for case 2(b), one for each distinct $G$-crossed extension. Moreover, we identify the $G$-crossed theories here with $q=0$ and $1$ as the ones characterizing the $\nu = 6$ and $2$ fermionic SPT theories, respectively.

In summary, we have found four Abelian $\mathbb{Z}_2$-crossed extensions of $\mathbb{Z}_2^{(1)}$, which correspond to the Chern number $\nu$ fermionic SPT phases $(p_x +ip_y)^{\nu} \times (p_x - ip_y)^{\nu}$ phases with $\nu=0,2,4,6$. This completes the $\mathbb{Z}_8$ classification of the interacting fermionic SPTs. We notice that $\nu=0$ is the trivial extension, $\nu=4$ corresponds to taking the product of a trivial fermion theory with a nontrivial bosonic $\mathbb{Z}_2$ SPT~\cite{fbspt}, and $\nu=2$ and $6$ are nontrivial fermionic SPT phases~\cite{ChengPRL2014, GuPRB2014}.

\subsection{Semions $\mathbb{Z}_{2}^{(\pm \frac{1}{2} )}$ with $\mathbb{Z}_2$ Symmetry}
\label{sec:semz2}

In this section, we consider the semion MTC $\mathcal{C}_{\bf 0} = \mathbb{Z}_{2}^{(\pm \frac{1}{2} )}$ with symmetry $G = \mathbb{Z}_2$. The semion theory $\mathbb{Z}_{2}^{(\pm \frac{1}{2} )}$ consists of two topological charge types $\mathcal{C}_{\bf 0}= \{I, s\}$, where $s$ denotes a semion, which has $\mathbb{Z}_{2}$ fusion $s\times s=I$ and topological twist $\theta_s = \pm i$. Such a theory describes the topological properties of the bosonic $\nu = \frac{1}{2}$ Laughlin FQH state. The nontrivial $F$-symbols and $R$-symbols are $F^{sss}_s=-1$ and $R^{ss}_{I}=\pm i$. We will focus on the $\mathbb{Z}_2^{(\frac{1}{2})}$ theory ($R^{ss}_{I}=i$) in this section.

Since there is only one nontrivial topological charge type, the topological symmetry group is trivial, i.e. $\mathrm{Aut}( \mathbb{Z}_{2}^{(p)} )=\mathbb{Z}_{1}$. Clearly, the symmetry action does not permute topological charge values. It follows that the fractionalization obstruction automatically vanishes, so we set $\coho{O} = I$. The symmetry fractionalization is classified by $H^2(\mathbb{Z}_2, \mathbb{Z}_2)=\mathbb{Z}_2$, which gives two equivalence classes corresponding to $\coho{w}({\bf 1,1})=I$ and $s$, respectively. Physically, these two cohomology classes correspond to the semion carrying a $\mathbb{Z}_2$ charge of $0$ or $\frac{1}{2}$.

Since the symmetry does not permute the anyon types, we could simply apply the results of Sec.~\ref{sec:no_perm}. However, for additional illustration, we will solve for the $G$-crossed extension in a gauge where $\eta_{a}({\bf g,h})=1$. There will be two types of $\mathbb{Z}_2$ defects, so we can write $\mathcal{C}_\mb{g}=\{I_{\bf g}, s_{\bf g} \}$. The two fractionalization classes correspond to distinct fusion rules for the $\mathbb{Z}_2$ defects, which are, respectively, given by: $a_{\mb{1}}\times a_{\mb{1}}=I_{\mb{0}}$ or $a_{\mb{1}}\times a_{\mb{1}}=s_{\mb{0}}$. In the following, we focus more on the latter case and systematically work out the gauging procedure (although there are other simpler ways to get the gauged theory).

For the trivial fractionalization class $\coho{w}({\bf 1,1})=I$, it is straightforward to see that the resulting $G$-crossed theories take the form
\begin{equation}
\left[\mathbb{Z}_{2}^{(\pm \frac{1}{2})}\right]_{\mathbb{Z}_{2}}^{\times} = \mathbb{Z}_{2}^{(\pm \frac{1}{2})} \boxtimes \text{SPT}_{\mathbb{Z}_{2}}^{[\alpha]}
\end{equation}
where $[\alpha] \in H^{3}(\mathbb{Z}_{2}, \text{U}(1)) = \mathbb{Z}_{2}$ distinguishes the two defectification classes. The corresponding gauged theories will be $\mathbb{Z}_{2}^{(\pm \frac{1}{2})} \boxtimes \text{D}^{[\alpha]}(\mathbb{Z}_{2})$.

For the nontrivial symmetry fractionalization class $\coho{w}({\bf 1,1})=s$, we will construct the $\mathcal{C}_{\mathbb{Z}_2}$ theory in more detail. Since $a_{\bf 1}\times a_{\bf 1}=s$, the extended category has the same fusion rules as $\mathrm{Vec}_{\mathbb{Z}_4}$, similar to one of the $\mathbb{Z}_2$ extension of fermions discussed in Sec.~\ref{sec:fermions}. In fact, the derivation of the $\mathbb{Z}_2$-crossed extensions is very similar to that of the $\mathbb{Z}_2$-crossed extensions of $\mathbb{Z}_2^{(1)}$, with minor differences accounting for the differences in the $\mathcal{C}_\mb{0}$ sector.
As such, we again identify $I_{\bf 0}\equiv 0, s_{\bf 0}\equiv 2, I_{\bf 1}\equiv 1, s_{\bf 1}\equiv 3$.  The $F$-symbols are given in terms of the $\mathbb{Z}_{4}$ labels by \eqref{eqn:zncocycle}, that is
\begin{equation}
F^{abc} = e^{i \frac{\pi p}{8} a (b+c - [b+c]_{4})}
\end{equation}
for $p=0,1,2,3$. In order to match the $\mathcal{C}_\mb{0}$ sector, we must have $F^{2,2,2}=F_s^{sss}=-1$, and, hence, $p=1$ or $3$.

The $G$-crossed braiding of $\mathrm{Vec}_{\mathbb{Z}_4}$ was found to be
\begin{eqnarray}
{R}_{[a+b]_4}^{ab} &=& e^{-i \frac{\pi p}{8}a b }e^{i \frac{\pi}{2} m_a b}, \\
U_c (a,b;[a+b]_{4}) &=& e^{-i \frac{\pi p}{4}(a+b-[a+b]_{4})c} \notag \\
&& \quad \times  e^{i \frac{\pi}{2}(m_a+m_b-m_{[a+b]_{4}})c}.
\end{eqnarray}
Here, $m_a$ are integers (mod 4).  In order to match the requirements that $R^{a0}=R^{0b}=U_{2}(a,b)=1$ and $R^{ss}=i$, we find that we must have $m_0=0$, $p=3$, and
\begin{equation}
m_a=2n_a+q [a]_2 , \quad a=1,2,3
,
\end{equation}
where $q=0$ or $1$, and $n_a \in \mathbb{Z}_2$.
By applying the symmetry action gauge transformation $\gamma_{a}(b) = (-1)^{n_{a} b}$ (which leaves $\eta_{a}({\bf g,h})$ fixed), we see that the dependence on $n_a$ is a gauge freedom, so we set it to $n_a=0$ to obtain
\begin{eqnarray}
{R}_{[a+b]_4}^{ab} &=& e^{-i \frac{3\pi}{8}a b } i^{q [a]_{2} b}, \\
U_{{\bf k}_{c}} (a,b;[a+b]_{4}) &=& e^{-i \frac{3\pi}{4}(a+b-[a+b]_{4})c} \notag \\
&& \quad \times i^{q ([a]_2 + [b]_2 - [a+b]_2 )c},
\end{eqnarray}
where ${\bf k}_{c}$ is the group element in $G=\mathbb{Z}_{2}$ corresponding to the label $c$, i.e. ${\bf k}_{c} = [c]_{2}$.
This shows that there are only two gauge independent $\mathbb{Z}_2$-crossed theories for the choice of symmetry fractionalization class $\coho{w}({\bf 1,1})=s$, corresponding to $q=0$ and $1$, which represent the two distinct classes in $H^3(\mathbb{Z}_2, \text{U}(1)) = \mathbb{Z}_2$. In other words, $q$ labels the distinct defectification classes. For the choice of gauge with $n_a=0$, we find the topological twists and braiding statistics
\begin{eqnarray}
\theta_{I_{\bf 1}}=R^{I_{\bf 1} I_{\bf 1}}_{s_{\bf 0}} &=& e^{-i \frac{3\pi}{8}}i^{q}\\
\theta_{s_{\bf 1}}=R^{s_{\bf 1} s_{\bf 1}}_{s_{\bf 0}}&=& -e^{-i \frac{3\pi}{8}}(-i)^{q}\\
R^{I_{\bf 1} s_{\bf 0}}_{s_{\bf 1}}R^{s_{\bf 0} I_{\bf 1}}_{s_{\bf 1}} &=& i(-1)^{q}\\
R^{s_{\bf 1} s_{\bf 0}}_{I_{\bf 1}}R^{s_{\bf 0} s_{\bf 1}}_{I_{\bf 1}} &=& -i(-1)^{q}\\
R^{s_{\bf 1} I_{\bf 1}}_{I_{\bf 0}} R^{I_{\bf 1} s_{\bf 1}}_{I_{\bf 0}} &=& e^{-i \frac{\pi}{4}}
\end{eqnarray}

Next, we gauge the $\mathbb{Z}_{2}$ symmetry of these theories with $\coho{w}({\bf 1,1})=s$. The topological charges in the gauged theory are parameterized by $(a,x)$, where $a\in\mathbb{Z}_4$ and the trivial and alternating $\mathbb{Z}_2$ irreps are respectively labeled by $x = \pm 1 $. The fusion rules are given by
\begin{equation}
(a,x)\times(b,y)=([a+b]_{4},xy U_\mb{1}(a,b)).
\label{eqn:semion_Z2_gauged_fusion}
\end{equation}
We can verify that the fusion algebra \eqref{eqn:semion_Z2_gauged_fusion} is isomorphic to that of a $\mathbb{Z}_8$ theory. For the $\mathbb{Z}_2$-crossed theory with $q = 0$, the gauged theory is $\mathbb{Z}_8^{(\frac{1}{2})}$, which is equivalent to a $\text{U}(1)_8$ Chern-Simons theory. For $q = 1$, the gauged theory is $\mathbb{Z}_8^{(\frac{5}{2})}$. The original semion theory can then be obtained from these gauged theories by condensing the bosonic quasiparticle labeled $4$ in the $\mathbb{Z}_8$ theories.

Physically, the nontrivial $\mathbb{Z}_2$-crossed extension of the semion model with fractionalization class $\coho{w}({\bf 1,1})=s$ and defectification class corresponding to $q = 0$ can be constructed by starting from a bosonic $\nu=\frac{1}{2}$ Laughlin FQH state with $\text{U}(1)$ boson number conservation~\cite{Kalmeyer87}, where the semions carry half $\text{U}(1)$ charges, and then breaking the $\text{U}(1)$ down to a $\mathbb{Z}_2$ subgroup (e.g. adding perturbations that pair condense the bosons) to obtain a $\mathbb{Z}_2$ symmetry-enriched semion theory.

\subsection{Semions $\mathbb{Z}_{2}^{(\pm \frac{1}{2} )}$ with $\mathbb{Z}_2\times\mathbb{Z}_2$ Symmetry}
\label{sec:semion_Z2Z2}

In this section, we consider the semion theory $\mathcal{C}_\mb{0}=\{I,s\}$ with the symmetry group $G=\mathbb{Z}_2\times\mathbb{Z}_2\equiv\{\openone, X,Y,Z\}$. Since the symmetry does not permute the anyon types, we can use the results of Sec.~\ref{sec:no_perm} to write the full details of the $G$-crossed extensions. In this case, we will only explicitly consider the details of interest. The symmetry fractionalization obstruction class is trivial, so we set $\coho{O}=I$.

The symmetry fractionalization classes are given by $H^2(G, \mathbb{Z}_2)=\mathbb{Z}_2^3$, where the different fractionalization classes are distinguished by their values of the invariants $\coho{w}(\mathbf{g},\mathbf{g})= I$ or $s$ for $\mb{g}=X$, $Y$, and $Z$. Among the $7$ nontrivial $H^2(G, \mathbb{Z}_2)$ classes, three are described by the nontrivial class in $H^2(\mathbb{Z}_2, \mathbb{Z}_2)$ for the three $\mathbb{Z}_2$ subgroups of $G$, generated by $g=X$, $Y$, and $Z$, respectively (and are otherwise trivial). In particular, these have $\coho{w}(\mathbf{g},\mathbf{g})= I$ for exactly one of $\mb{g}=X$, $Y$, and $Z$. As these correspond to the example examined in Sec.~\ref{sec:semz2}, with additional trivial structure from the extra $\mathbb{Z}_2$, we will not consider them in detail here. We focus only on the other four nontrivial symmetry fractionalization classes.

We find that the defectification obstruction class is nontrivial for the symmetry fractionalization classes which have $\coho{w}(\mathbf{g},\mathbf{g})= s$ for exactly one of $\mb{g}=X$, $Y$, and $Z$. The other fractionalization classes permit consistent $G$-crossed extensions, which are, thus, classified by $H^{3}(G,\text{U}(1)) = \mathbb{Z}_{2}^{3}$. For all the defect theories, each ${\bf g}$-defect sector has two types of defects, which can be written as $\mathcal{C}_\mb{g}=\{I_{\bf g}, s_{\bf g} \}$.

\subsubsection{Symmetry fractionalization class with $\coho{w} (X,X)=\coho{w} (Y,Y)=\coho{w} (Z,Z)= s$}

We first consider the case where $\coho{w}(\mathbf{g},\mathbf{g})=s$ for all $\mb{g}\neq \openone$. The remaining nontrivial terms in the cohomology class can be taken to be $\coho{w}(Y,X)=\coho{w}(Z,Y)=\coho{w}(X,Z)=s$ (and the rest equal to $I$). The fusion rules of the extended theory are given by
\begin{equation}
a_{\bf g} \times b_{\bf h} = [ab\coho{w}(\mathbf{g},\mathbf{h})]_{\bf gh}
.
\end{equation}
In particular, this gives $I_{X} \times I_{X} = s_{\openonesub}$, $I_{X} \times I_{Y} = I_{Z}$, $I_{Y} \times I_{X} = s_{Z}$, and similar relations obtained by cyclic permutation of the group labels. We note that the fusion rules match exactly with the multiplication table of the quaternion group $\mathbb{Q}_8$.

In order to for fusion and $G$-crossed braiding to be consistent, the symmetry action must act nontrivially on defects. For example, we must have
\begin{equation}
\rho_{X}: \quad a_{X}\leftrightarrow a_X, \quad I_Y \leftrightarrow s_Y, \quad I_Z \leftrightarrow s_Z.
\label{eqn:semion-action}
\end{equation}
Similarly, the action of $Y$ and $Z$ are obtained by cyclic permutations.

We note that the $\eta_{s_{\openonesub}}(\mb{g},\mb{h}) = M_{s_{\openonesub} [\cohosub{w}(\mathbf{g},\mathbf{h})]_{\openonesub}} = \pm 1$ (depending on whether $\coho{w}(\mathbf{g},\mathbf{h})=I$ or $s$), so it represents a nontrivial cohomology class in $H^2(G, \text{U}(1))$. Additionally, the quantities $\eta_{c_{\bf g}}(\mb{g},\mb{g}) = M_{s_{\openonesub} c_{\openonesub}}/F^{I_{\bf g} I_{\bf g} I_{\bf g}}$ for ${\bf g} \neq \openone$ can be set to $1$ using the symmetry action gauge transformation $\gamma_{c_{\bf g}}({\bf g})$.~\footnote{In these theories, $F^{I_{\bf g} I_{\bf g} I_{\bf g}}= e^{i\frac{\pi}{4}} (-1)^{{\bf p}\cdot {\bf g}}$, where ${\bf p}$ is an element of the $\mathbb{Z}_{2}^{2}$ subgroup of $H^{3}(G,\text{U}(1))$ representing ``type I'' cocycles, i.e. it partially distinguishes the defectification classes.} As such, $\eta_{c_{\bf g}}(\mb{g},\mb{g})$ represents the trivial cohomology class in $H^2(G_{c_{\bf g}}, \text{U}(1))$, where $G_{c_{\bf g}}$ is the $\mathbb{Z}_{2}$ subgroup of $G$ generated by ${\bf g}$ (which leaves $c_{\bf g}$ invariant).

We now consider the theory obtained by gauging the $G$ symmetry. With the symmetry action given in \eqref{eqn:semion-action}, for $\mb{g}\neq \openone$, each $\mathcal{C}_\mb{g}$ forms an orbit under $G$, i.e. $[c_{\bf g}]=\mathcal{C}_\mb{g}$. The stabilizer group of the orbit $\mathcal{C}_\mb{g}$ is $G_{c_{\bf g}}$, the $\mathbb{Z}_2$ subgroup generated by $\mathbf{g}$. As mentioned, $\eta_{c_{\bf g}}(\mb{g},\mb{g})$ represents the trivial class in $H^2(G_{c_{\bf g}}, \text{U}(1))$, so these orbits will pair with trivially projective irreps of $G_{c_{\bf g}}$ corresponding to the linear irreps of $\mathbb{Z}_{2}$, i.e. trivial and alternating. Thus, for the gauged theory, each $\mathcal{C}_\mb{g}$ sector yields two topological charge types: $(\mathcal{C}_\mb{g}, +)$ and $(\mathcal{C}_\mb{g}, -)$, each of which has quantum dimension $d_{(\mathcal{C}_\mb{g}, \pm)}=2$. The topological twists of these anyons can be computed from the $G$-crossed data and are found to be $e^{i\frac{\pi}{8}} i^{n{(\mathcal{C}_\mb{g}, \pm)}}$, where $n{(\mathcal{C}_\mb{g}, \pm)}$ is an integer that depends on the charge and the particular defectification class of the theory.

For the $\mathcal{C}_{\openonesub}$ sector, the stabilizer group is the entire $G=\mathbb{Z}_2\times\mathbb{Z}_2$. The vacuum $I$ simply splits according to the linear irreducible representations of $G$, which results in four Abelian anyon types $(I,\pm,\pm)$ (which have quantum dimension $d=1$). However, the semion $s_{\openonesub}$ carries a nontrivial projective representation of $\mathbb{Z}_2\times\mathbb{Z}_2$, since the factor set $\eta_{s_{\openonesub}}$ belongs to the nontrivial cohomology class in $H^2(\mathbb{Z}_2\times\mathbb{Z}_2,\mathrm{U}(1))=\mathbb{Z}_2$, as previously mentioned. It is well-known that there is a unique two-dimensional irreducible representation with this factor set $\eta_{s_{\openonesub}}$, up to similarity, essentially given by Pauli matrices. We ascribe the label $\sigma$ to this projective irrep here. According to Sec.~\ref{sec:gaugeQP}, this implies that, in the gauged theory, the semion becomes a non-Abelian anyon $(s_{\openonesub}, \sigma)$ with quantum dimension $d_{(s_{\openonesub}, \sigma)}=2$. The topological twist for this anyon is the same as the semion, i.e. $\theta_{(s_{\openonesub}, \sigma)} = i$. Thus, we find there are $4$ Abelian anyons and $7$ non-Abelian anyons, for a total of $11$ topological charge types.

Another way to obtain the total number of anyons is to compute the ground state degeneracy of the gauged theory on the torus, as described in Sec.~\ref{sec:gaugeGenus}. For this, we enumerate the $G$-invariant states in each of the defect sectors on the torus. There are potentially 16 such sectors, labeled by the pair of group $G$ elements $({\bf g,h})$ winding around the longitudinal and meridional cycles, respectively. The trivial sector has two invariant states $| I_{\openonesub}^{(\openonesub,\openonesub)} \rangle_{(l,m)}$ and $| s_{\openonesub}^{(\openonesub,\openonesub)} \rangle_{(l,m)}$. There are $3$ defect sectors, labeled by $({\bf g},\openone)$, where there is a defect branch around the longitudinal cycle and no branch around the meridional cycle. In each such defect sector of the torus, there are 2 states. However, from the $\mathbb{Z}_2\times\mathbb{Z}_2$ action, it is easy to see that there is only one state, that is $| I_{\bf g}^{({\bf g},\openonesub)} \rangle_{(l,m)}+|s_{\bf g}^{({\bf g},\openonesub)} \rangle_{(l,m)}$, which is $G$-invariant and, thus, survives the gauging. Similarly, we expect sectors labeled $(\openone,{\bf h})$, which have a defect branch around the meridional cycle and no branch around the longitudinal cycle to have only one $G$-invariant state per sector, since they are related to the $({\bf h},\openone)$ sector by the modular $\mathcal{S}$ transformation. Indeed, the $G$-invariant state can be seen to be $|I_{\openonesub}^{(\openonesub,{\bf h})} \rangle_{(m,-1)}$, since the modular transformation is found to be
\begin{equation}
\mathcal{S}^{({\bf g},\openonesub)} = \mathcal{S}^{(\openonesub,{\bf h})} = \frac{1}{\sqrt{2}} \left[
\begin{array}{rr}
1 & 1 \\
1 & -1
\end{array}
\right]
.
\end{equation}
For sectors labeled by $({\bf g},{\bf g})$, which have the same nontrivial branch lines around both cycles, there are two states in each sector, and we similarly find that there is only one $G$-invariant state per sector. In this case, the $({\bf g},{\bf g})$ sector is related to the $({\bf g},\openone)$ sector by a modular $\mathcal{T}$ transformation. We find that $\mathcal{T}^{({\bf g},\openonesub)}_{a_{\bf g}, b_{\bf g}} = i^{a} \delta_{a,b}$ and $\mathcal{T}^{({\bf g},{\bf g})}_{a_{\bf g}, b_{\bf g}} = (-i)^{a} \delta_{a,b} / F^{0_{\openonesub} 0_{\openonesub} 0_{\openonesub}}$, so the states in the $({\bf g},{\bf g})$ sector acquire a phase under $G$-transformations, i.e. $^{\bf h}|I_{\bf g}^{({\bf g},{\bf g})} \rangle_{(l,l-m)} = i|s_{\bf g}^{({\bf g},{\bf g})} \rangle_{(l,l-m)}$ and $^{\bf h}|s_{\bf g}^{({\bf g},{\bf g})} \rangle_{(l,l-m)} = -i|I_{\bf g}^{({\bf g},{\bf g})} \rangle_{(l,l-m)}$when ${\bf h} \neq {\bf g}$, so the $G$-invariant state is $| I_{\bf g}^{({\bf g},{\bf g})} \rangle_{(l,l-m)}+ i |s_{\bf g}^{({\bf g},{\bf g})} \rangle_{(l,l-m)}$. The defect sectors of the torus labeled by $({\bf g},{\bf h})$ with ${\bf g} \neq {\bf h}$ are actually empty in the $G$-crossed theory, because there are no ${\bf h}$-invariant ${\bf g}$-defects, i.e. $\mathcal{C}^{\bf h}_{\bf g} = \emptyset$, so these have no contribution to the gauged theory either. Thus, we find $11$ $G$-invariant topological ground states on the torus for the gauged theory, which implies that there are $11$ distinct topological charges in the gauged theory.

One can obtain the fusion rules of the gauged theory using the solutions of the $G$-crossed consistency equations, but we choose to use a different method, namely gauging the symmetry sequentially. Without loss of generality, we first gauge one of the $\mathbb{Z}_2$ subgroups, say the one generated by $X$. Since $\coho{w}(X,X)=s$, from the previous subsection, we know that gauging this subgroup results in a $\text{U}(1)_8$ theory. The remaining $\mathbb{Z}_2$ symmetry that was not yet gauged then has a nontrivial action in the $\text{U}(1)_8$ theory. In particular, it acts as a charge conjugation symmetry. We will discuss the charge conjugation symmetry for $\mathbb{Z}_{N}^{(p)}$ theories, which includes $\text{U}(1)_8$, in more detail in Sec.~\ref{sec:zneven}. Here, we note that the gauged theory can be understood in terms of the $\mathbb{Z}_2$ orbifold of a $\text{U}(1)_8$ CFT, which was analyzed in \Ref{dijkgraaf1989}. Interestingly, it can also be understood as the $\mathbb{Z}_2\times\mathbb{Z}_2$ orbifold of the $\text{SU}(2)_1$ CFT, which fits naturally within our approach. The gauged theory can be identified with $\ol{\text{SO}(8)}_2$.

A physical realization of the semion theory with this $\mathbb{Z}_2\times\mathbb{Z}_2$ symmetry can be obtained by starting from a chiral spin liquid in spin-$\frac{1}{2}$ systems, which has $\text{SO}(3)$ spin rotational symmetry. The semion carries spin-$1/2$, i.e. a projective representation of the $\text{SO}(3)$ symmetry. One can then break the $\text{SO}(3)$ symmetry down to the $\mathbb{Z}_2\times\mathbb{Z}_2$ subgroup, i.e. $\pi$ rotations around three orthogonal axes. For this model, we can explicitly write the localized symmetry operators on a semion as
\begin{equation}
U_X^{(s)}=i\sigma_x, \quad U_Y^{(s)}=i\sigma_y, \quad U_Z^{(s)}=i\sigma_z
.
\end{equation}
Clearly, these satisfy $(U_\mb{g}^{(s)})^2=-1$, which is consistent with $\eta_{s}(\mb{g},\mb{g})=-1$. In the context of the $\text{SO}(3)$ symmetry, this property is the familiar fact that spin-$1/2$ objects acquire a phase of $-1$ phase when they are rotated by $2\pi$ around any axis.

\subsubsection{Symmetry fractionalization class with $\coho{w} ({\bf g},{\bf g})= s$ for one nontrivial ${\bf g}$ and defectification obstruction}
\label{sec:semion_H4_obstruction}

We now turn to the remaining three symmetry fractionalization classes, which have $\coho{w} ({\bf g},{\bf g})= I$ for exactly one of ${\bf g}\neq \openone$. Without loss of generality, we consider the case with $\coho{w}(X, X)=\coho{w}(Y,Y)=I$ and $\coho{w}(Z,Z)=s$. The remaining nontrivial terms of this cohomology class are $\coho{w}(Y,X)=\coho{w}(Z,X)=\coho{w}(Y,Z)=s$ (and the rest are equal to $I$).

If we proceeded na\"ively in an attempt to construct the $G$-crossed extension, we could define the fusion rules by $a_{\bf g} \times b_{\bf h} = [ab\coho{w}(\mathbf{g},\mathbf{h})]_{\bf gh}$. The corresponding $G$-graded fusion category is $\graded{C}{G}\simeq \text{Vec}_{\mathbb{D}_8}$, where $\mathbb{D}_8=\mathbb{Z}_4\rtimes\mathbb{Z}_2$ is the dihedral group of order $8$. Solving the pentagon equation, we could obtain $F$-symbols, which are classified by $H^3(\mathbb{D}_8, \text{U}(1))=\mathbb{Z}_2^2\times\mathbb{Z}_4$. Among the $16$ associativity classes, $8$ of them can match the $F$-symbols of the $\mathcal{C}_\mb{0}$ sector. Therefore, the extended category as a usual fusion category does exist.

However, if we try to proceed further, we find that the full set of $G$-crossed consistency equations admit no solutions, so there is no consistent $G$-crossed extension. Consequently, these theories also cannot be gauged. In other words, there is an obstruction to defectification and gauging the symmetry. A more efficient way to see this is to compute the defectification obstruction class $[\mathcal{O}]$ using Eq.~(\ref{eq:obstruction}), which only requires knowledge of $\coho{w} ({\bf g},{\bf h})$ and the $F$-symbols and $R$-symbols of the semion theory. Doing so, one obtains a nontrivial class in $H^{4}(G,\text{U}(1))= \mathbb{Z}_{2}^{2}$, which signals that a consistent $G$-crossed extension cannot exist. This is in full agreement with the result obtained in \Ref{chen2014}. The physical interpretation of this nontrivial obstruction is that the semion theory with such a symmetry and fractionalization class cannot exist in 2+1 dimensions, but could possibly exist at the surface of a 3+1 dimensional system, as discussed in \Ref{chen2014}.

\subsection{$\text{Ising}^{(\nu)}$ Anyons with $G$ Symmetry}
\label{sec:Ising}

The anyon model Ising$^{(\nu)}$ where $\nu$ is an odd integer has three topological charges $\{I, \sigma, \psi\}$, where the vacuum charge here is denoted $I$, and the nontrivial fusion rules are given by
\begin{equation}
	\psi\times\psi=I, \quad \sigma\times\psi=\sigma, \quad \sigma\times\sigma=I+\psi.
\end{equation}
The nontrivial $F$-symbols are
\begin{equation}
	\begin{gathered}
	F^{\psi\sigma\psi}_\sigma=F^{\sigma\psi\sigma}_\psi=-1\\
	\left[F^{\sigma\sigma\sigma}_\sigma\right]_{ab}=
	\frac{\varkappa_{\sigma}}{\sqrt{2}} \left[
	\begin{matrix}
		1 & 1\\
		1 & -1
	\end{matrix}
\right]_{ab}.
	\end{gathered}
	\label{eqn:IsingF}
\end{equation}
Here, the column and row labels of the matrix take values $I$ and $\psi$ (in this order). $\varkappa_{\sigma}=(-1)^{\frac{\nu^2-1}{8}}$ is the Frobenius-Schur indicator of $\sigma$.

The $R$-symbols are
\begin{equation}
	\begin{gathered}
	R^{\psi\sigma}_\sigma=R^{\sigma\psi}_\sigma=(-i)^{\nu}\\
	R^{\sigma\sigma}_{I}=\varkappa_{\sigma} e^{-i\frac{\pi}{8}\nu}, \quad R^{\sigma\sigma}_\psi=\varkappa_{\sigma} e^{i\frac{3\pi}{8}\nu}.
	\end{gathered}
\end{equation}
The twist factor $\theta_\sigma=e^{i\frac{\pi }{8}\nu}$ uniquely distinguishes the eight distinct Ising$^{(\nu)}$ anyon models, as does the chiral central charge $c_{-} \text{mod }8 = \frac{\nu}{2}$, with $\nu \sim \nu +16$, so we can restrict to $0< \nu <16$. The Ising TQFT corresponds to $\nu=1$, SU$(2)_2$ corresponds to $\nu=3$, and $\nu \geq 5$ can be realized by SO$(\nu)_1$ Chern-Simons field theory.

Clearly, the topological symmetry group of these anyon models are trivial, i.e. $\text{Aut}(\text{Ising}^{(\nu)}) = \mathbb{Z}_{1}$, since none of the topological charges may be permuted. Consequently, the symmetry action must be trivial, the symmetry fractionalization obstruction vanishes ($[\coho{O}]=0$), and symmetry fractionalization is classified by $H^{2}(G,\mathbb{Z}_{2})$. For a given symmetry fractionalization class, when the defectification obstruction class [which may be computed using Eq.~(\ref{eq:obstruction})] is trivial, the corresponding $G$-crossed extensions are classified by $H^{3}(G,\text{U}(1))$ and the full $G$-crossed data is given in Sec.~\ref{sec:no_perm}.

In the case of $G=\mathbb{Z}_{2}$, symmetry fractionalization is classified by $H^{2}(\mathbb{Z}_{2},\mathbb{Z}_{2})=\mathbb{Z}_{2}$. The two fractionalization classes are represented by the cocycles $\coho{w}({\bf 1},{\bf 1})= I$ and $\psi$, respectively. The defectification obstruction vanishes for both fractionalization classes: for $\coho{w}({\bf 1},{\bf 1})= I$, Eq.~(\ref{eq:obstruction}) gives $\mathcal{O}({\bf 1},{\bf 1},{\bf 1},{\bf 1}) =1$; for $\coho{w}({\bf 1},{\bf 1})= \psi$, Eq.~(\ref{eq:obstruction}) gives $\mathcal{O}({\bf 1},{\bf 1},{\bf 1},{\bf 1}) =-1$, which is a coboundary $\text{d}\lambda ({\bf 1},{\bf 1},{\bf 1},{\bf 1})$ for e.g. $\lambda ({\bf 1},{\bf 1},{\bf 1}) = i$. The two fractionalization classes extend to defect theories with differing fusion rules. In particular, the Abelian subcategory of the corresponding defect theories have $\mathbb{Z}_{2} \times \mathbb{Z}_{2}$ fusion rules and $\mathbb{Z}_{4}$ fusion rules, for $\coho{w}({\bf 1},{\bf 1})= I$ and $\psi$, respectively. These correspond to the two possible $\mathbb{Z}_{2}$ central extensions of $\mathbb{Z}_{2}$. The defect theories (for either fractionalization class) are then classified by $H^{3}(\mathbb{Z}_{2},\text{U}(1)) = \mathbb{Z}_{2}$, which may be distinguished by the $\sigma_{\bf 1}$ defect's Frobenius-Schur indicator, which we write as $\varkappa_{\sigma_{\bf 1}}= \alpha \varkappa_{\sigma_{\bf 0}}$, where $\alpha = \pm 1$. Thus, there are four possible $\mathbb{Z}_{2}$-crossed extensions of an $\text{Ising}^{(\nu)}$ theory. Gauging the $\mathbb{Z}_{2}$ symmetry, the $\coho{w}({\bf 1},{\bf 1})= I$ fractionalization class results in the gauged theories $\text{Ising}^{(\nu)} \boxtimes \text{D}^{[\alpha]}(\mathbb{Z}_{2})$, and the $\coho{w}({\bf 1},{\bf 1})= \psi$ fractionalization class results in the gauged theories $\text{Ising}^{(\nu - 2q)} \boxtimes \mathbb{Z}_{4}^{(\frac{q}{2})}$, where $q = \alpha (-1)^{\frac{\nu -1}{2}}$.

\subsection{$\mathbb{Z}_{N}^{(p)}$ Anyons with $G$ Symmetry}
\label{sec:Z_N}

In this section, we consider the $\mathbb{Z}_N^{(p)}$ anyon models. As UBTCs, we can have $p\in\mathbb{Z}$ for all $N$ and $p\in\mathbb{Z}+\frac{1}{2}$ for $N$ even. The $\mathbb{Z}_N^{(p)}$ anyon models have $N$ topological charges labeled by $a=0,1,\dots, N-1$ which obey the fusion rules $a\times b=[a+b]_N$. The $F$-symbols (in a particular choice of gauge) are
\begin{equation}
[F^{abc}_{[a+b+c]_N}]_{[a+b]_N,[b+c]_N}=e^{i \frac{2\pi p}{N}a(b+c-[b+c]_N)}.
\end{equation}
Notice that the $F$-symbols are all equal to $1$ when $p\in\mathbb{Z}$. For $p\in\mathbb{Z}+\frac{1}{2}$, some of the $F$-symbols are equal to $-1$, and they cannot all be set to $1$ using gauge freedom.

The $R$-symbols (in our choice of gauge) are
\begin{equation}
R^{ab}_{[a+b]_N}=e^{i \frac{2\pi p}{N}ab}.
\end{equation}
The twist factors are $\theta_a=e^{i \frac{2\pi p}{N}a^2}$.

Notice that $p$ is periodic in $N$, so we restrict to the range $0 \leq p <N$. For odd $N$, $\mathbb{Z}_N^{(p)}$ is a UMTC iff $p \neq 0$ and $\gcd(p,N)=1$. For even $N$, $\mathbb{Z}_N^{(p)}$ is a UMTC iff $p\in\mathbb{Z}+\frac{1}{2}$ and $\gcd(2p,N)=1$. The modularity condition for all cases can be written simply as $\gcd(2p,N)=1$. The case $p=1/2$ represents the topological order of the well-known $\text{U}(1)_N$ Chern-Simons theory, which, for example, describes the bosonic $\nu=\frac{1}{N}$ Laughlin FQH states. The quasiparticles of the fermionic $\nu = \frac{1}{m}$ Laughlin FQH states with $m$ odd are described by the UBTC $\mathbb{Z}_{2m}^{(1)} = \mathbb{Z}_2^{(1)} \times \mathbb{Z}_m^{(\frac{m+1}{2})}$.

It is useful to write $\mathbb{Z}_N^{(p)}$ in terms of the prime decomposition
\begin{equation}
\label{eq:Z_N_decomp}
\mathbb{Z}_N^{(p)} = \mathbb{Z}_{N_0}^{(p_0)} \boxtimes  \mathbb{Z}_{N_1}^{(p_1)} \boxtimes \cdots \boxtimes  \mathbb{Z}_{N_k}^{(p_k)}
\end{equation}
where $N_{j} = P_{j}^{r_j}$ for $P_{j}$ prime integers satisfying $P_0 = 2$ and $P_{j+1}> P_{j}$, and integers $r_0 \geq 0$ and $r_j > 0$ for $j>0$. In order to determine the coefficients $p_j$, we must specify a choice of isomorphism between $\mathbb{Z}_N^{(p)}$ and $\prod_{j=0}^{k} \mathbb{Z}_{N_{j}}^{(p_{j})}$. We will use $a \mapsto (a_0,a_1, \ldots , a_k )$ where
\begin{equation}
a_{j} = [a]_{N_{j}}
.
\end{equation}
To invert this isomorphism, we define $\hat{N}_{j} = \frac{N}{N_{j}}$ and solve $[x_{j} \hat{N}_{j}]_{N_{j}}=1$ for $x_j \in \mathbb{Z}_{N_{j}}$ (notice this implies that $x_{0}$ is odd). Then we have
\begin{equation}
a =  \left[ \sum_{j=0}^{k} a_{j} x_j \hat{N}_{j} \right]_{N}
.
\end{equation}
Expressed in terms of the unit generators, this isomorphism can be expressed as $1 \mapsto (1,1,\ldots,1)$ and $[x_{j} \hat{N}_{j}]_{N} \mapsto (0,\ldots,0,1,0,\ldots,0)$, where the $1$ in the last expression is the $j$th entry. With this isomorphism, the coefficients are given by
\begin{eqnarray}
p_{0} &=& \frac{1}{2}[ 2 p x_{0}^{2} \hat{N}_{0} ]_{2N_{0}} , \\
p_{j} &=& [ p x_{j}^{2} \hat{N}_{j} ]_{N_{j}}, \quad \text{ for } j>0
.
\end{eqnarray}
The inverse of this relation is
\begin{eqnarray}
p &=& \left[  \sum_{j=0}^{k} p_{j} \hat{N}_{j} \right]_{N} , \quad \text{ for }  p_{0} \in \mathbb{Z} \text{ or } r_{0}=0, \\
p &=& \frac{1}{2} \left[ 2 \sum_{j=0}^{k} p_{j} \hat{N}_{j} \right]_{2N}, \quad \text{ for } p_{0} \in \mathbb{Z}+\frac{1}{2}
.
\end{eqnarray}

The autoequivalence maps of $\mathbb{Z}_N^{(p)}$ are given by permutations of the topological charge labels that preserve fusion and the basic data up to gauge transformations. These are given by the maps $\varphi (a) = [q a]_{N}$, where $q$ is an integer that satisfies the conditions $0<q<N$, $\gcd(q , N)=1$, and $\frac{p (q^2 - 1)}{N} \in \mathbb{Z}$. For our choice of gauge, the unitary gauge transformations that leave the basic data exactly invariant can be taken to be $u^{ab}_{[a+b]_{N}}=1$ for $p \in \mathbb{Z}$ and $u^{ab}_{[a+b]_{N}}=(-1)^{a n_{b}}$ for $p \in \mathbb{Z}+\frac{1}{2}$, where we define $n_{b} = \frac{qb - [qb]_{N}}{N}$.

We now restrict our attention to modular theories, for which we can solve the conditions for autoequivalence maps more explicitly. Clearly, $q=1$ is the identity autoequivalence. The conditions imply that $[q^2]_{N}=1$, so every nontrivial autoequivalence has order 2. Moreover, all autoequivalence maps commute with each other, since $q q' = q' q$. Finally, we see by solving the congruence equation $[q^2]_{N} =1$ that there are $2^{k}$ distinct autoequivalence maps when $r_0=0$ or $1$, and there are $2^{k+1}$ distinct autoequivalence maps when $r_0 \geq 2$. It follows that the topological symmetry group of a modular $\mathbb{Z}_N^{(p)}$ theory is
\begin{equation}
\mathrm{Aut}( \mathbb{Z}_{N}^{(p)} )= \left\{
\begin{array}{ll}
\mathbb{Z}_2^{k} & \text{ for }\, r_0=0,1 \\
\mathbb{Z}_2^{k+1} & \text{ for }\, r_0>1
\end{array}
\right.
.
\end{equation}

In terms of the decomposition of Eq.~(\ref{eq:Z_N_decomp}), the autoequivalence maps and topological symmetry group takes a simple form. In particular, $\varphi$ maps to $\varphi_{j}(a_j) = [q_j a_j]_{N_j}$, where the autoequivalence conditions imply that $q_j = 1$ or $N_j -1$, so that $\varphi_{j}(a_j) = a_j$ or $[(N_j - 1)a]_{N_j} = \bar{a}_{j}$. Thus, an autoequivalence map either acts trivially or as topological charge conjugation within each $\mathbb{Z}_{N_{j}}^{(p_{j})}$ sector, and this contributes a $\mathbb{Z}_2$ factor for each sector, except if $r_0 = 1$, in which case the $0$th sector can only be acted on trivially. In other words, the topological symmetry group can also be factored in terms of the prime decomposition as
\begin{equation}
\mathrm{Aut}( \mathbb{Z}_{N}^{(p)} )=  \prod_{j=0}^{k} \mathrm{Aut}( \mathbb{Z}_{N_{j}}^{(p_{j})})
.
\end{equation}
We emphasize that the autoequivalence maps do not permute topological charges between different sectors, since this cannot preserve the fusion rules.

The action of the global symmetry on the topological degrees of freedom is specified by a homomorphism
\begin{equation}
[\rho] : G \rightarrow \mathrm{Aut}( \mathbb{Z}_{N}^{(p)} )
.
\end{equation}
Since the topological symmetry group factorizes, the action can similarly always be factorized into sectors as $\rho = ( \rho^{(0)}, \rho^{(1)}, \ldots , \rho^{(k)})$, where
\begin{equation}
[\rho^{(j)}] : G \rightarrow \mathrm{Aut}( \mathbb{Z}_{N_j}^{(p_j)} )
.
\end{equation}

Factoring the theory in terms of its prime decomposition, we can choose $U_{\bf g}$ to be trivial unless ${\bf g}$ acts nontrivially (as charge conjugation) on the $j=0$ sector, i.e. the $N_{0} = 2^{r_{0}}$ sector when $r_{0} > 1$, in which case we can choose $U_{\bf g}(a,b; [a+b]_{N}) = (-1)^{a_{0} (b_{0} - 1)}$ for $b_{0} \neq 0$. Since $[\rho^{(0)}]$ is a homomorphism, it gives a $\mathbb{Z}_{2}$ grading on $G$, which, for our choice of $U_{\bf g}(a,b; [a+b]_{N})$ implies that $\kappa_{{\bf g},{\bf h}} = \openone$. It follows that we can choose $\beta_{a}({\bf g}, {\bf h})=1$ and $\coho{O}=0$, so the symmetry fractionalization obstruction always vanishes for modular $\mathbb{Z}_{N}^{(p)}$ theories. It also follows that the cohomological classification structure of symmetry fractionalization factorizes, since $[\rho]$ does not mix sectors, that is
\begin{equation}
H^{2}_{[\rho]}(G,\mathcal{A}) = \prod_{j=0}^{k} H^{2}_{[\rho^{(j)}]}(G,\mathbb{Z}_{N_j})
.
\end{equation}

We now see that we can obtain all $G$-crossed extensions of modular $\mathbb{Z}_{N}^{(p)}$ by solving for the $G$-crossed extensions of each $\mathbb{Z}_{N_j}^{(p_j)}$ factor and then using the gluing construction of Appendix~\ref{sec:gluing} to produce all the theories of the form
\begin{equation}
\left[ \mathbb{Z}_{N}^{(p)} \right]_{G}^{\times} = \left[ \mathbb{Z}_{N_0}^{(p_0)} \right]_{G}^{\times} \underset{G}{\boxtimes} \left[ \mathbb{Z}_{N_1}^{(p_1)} \right]_{G}^{\times} \underset{G}{\boxtimes}\cdots \underset{G}{\boxtimes} \left[ \mathbb{Z}_{N_k}^{(p_k)} \right]_{G}^{\times}
.
\end{equation}
Given the symmetry action and fractionalization class, the existence of $G$-crossed extensions requires the vanishing of the defectification obstruction $[\mathcal{O}]$, which can also be written as the product of obstruction classes of the different sectors. Furthermore, this gluing construction reproduces the $H^{3}(G,\text{U}(1))$ classification of defectification.

We do not attempt to produce the full data of all $G$-crossed extensions of $\mathbb{Z}_{N_j}^{(p_j)}$ in this paper. When the symmetry acts trivially on (i.e. does not permute) the topological charges, the full data is given in Sec.~\ref{sec:no_perm}. In the following sections, we obtain the full data for $N$ odd when $G=\mathbb{Z}_{2}$ and [TBD] for $N$.

\subsection{$\mathbb{Z}_{N}^{(p)}$ Anyons with $N$ Odd and $\mathbb{Z}_2$ Symmetry}
\label{sec:Z_N_odd}

In this section, we consider $\mathbb{Z}_2$ symmetry in the $\mathbb{Z}_N^{(p)}$ anyon models
where $N$ is an odd integer and $p$ is an integer such that $0 < p < N$ and $\gcd (p, N) =1$, so that the theory is modular.
The $\mathbb{Z}_N^{(p)}$ theory has $N$ distinct topological charges,
labeled by $a=0,1,\dots,N-1$. The fusion rules are given by addition
modulo $N$: $a\times b=[a+b]_N$. The $F$-symbols are all trivial and the $R$-symbols are given by
\begin{equation}
R^{ab}_{[a+b]_N}=e^{i \frac{2\pi p }{N}ab}.
\end{equation}

As shown in Sec.~\ref{sec:Z_N}, the topological symmetry group is $\mathrm{Aut}( \mathbb{Z}_{N}^{(p)} )=\mathbb{Z}_2^{k}$, where $k$ is the number of distinct primes in the prime factorization of $N$, where the $2^{k}$ different classes of autoequivalence maps are specified by whether or not the map acts as topological charge conjugation on each factor. There is always a subgroup $\mathbb{Z}_{2} \lhd \mathrm{Aut}( \mathbb{Z}_{N}^{(p)} )$ that is associated with topological charge conjugation $a \mapsto \bar{a} = [-a]_{N}$. Furthermore, it was shown that the symmetry fractionalization obstruction always vanishes $[\coho{O}]=0$, the symmetry fractionalization factorizes, and the $G$-crossed extensions can be expressed in terms of gluing the $G$-crossed extensions of the factors. In light of this, for $G=\mathbb{Z}_{2}$, we only need to obtain the data for the cases where the symmetry acts trivially and as topological charge conjugation, and then we can generate the general cases from these. We will find that $H^2_{[\rho]}(\mathbb{Z}_2,\mathbb{Z}_N)=\mathbb{Z}_{1}$ and $H^3(\mathbb{Z}_2,\text{U}(1))=\mathbb{Z}_{2}$, so there is one symmetry fractionalization class and two defectification classes for $\mathbb{Z}_2$ symmetry when $N$ is odd. We now examine these in detail.

\subsubsection{Trivial symmetry action}

In the case where the symmetry acts trivially on the anyons, we see that every $a \in \mathbb{Z}_{N}$ defines both a 2-cocycle $\coho{w}(\mb{1},\mb{1})=a$ and a 2-coboundary ${\rm d}\coho{z}(\mb{1},\mb{1})=a$, by taking $\coho{z}(\mb{1}) = \frac{a}{2}$ for $a$ even and $\coho{z}(\mb{1}) = \frac{a +N}{2}$ for $a$ odd. Thus, $H^2(\mathbb{Z}_2,\mathbb{Z}_N)=\mathbb{Z}_{1}$, and there is one fractionalization class. The data of the two $G$-crossed theories, corresponding to the two defectification classes, are given by the results of Sec.~\ref{sec:no_perm}, where it is most convenient to use the gauge choice with $\coho{w}(\mb{1},\mb{1})=0$. It is straightforward to see that the resulting $G$-crossed theories have the form
\begin{equation}
\left(\mathbb{Z}_{N}^{(p)} \right)_{\mathbb{Z}_{2}}^{\times} = \mathbb{Z}_{N}^{(p)} \boxtimes \text{SPT}_{\mathbb{Z}_{2}}^{[\alpha]}
,
\end{equation}
where $[\alpha] \in H^3(\mathbb{Z}_2,\text{U}(1))=\mathbb{Z}_{2}$, i.e. it is represented by the 3-cocycle with element $\alpha(\mb{1},\mb{1},\mb{1})=\pm 1$. The corresponding gauged theories are $\mathbb{Z}_{N}^{(p)} \boxtimes \text{D}^{[\alpha]}(\mathbb{Z}_{2})$.

\subsubsection{Charge conjugation symmetry}
\label{sec:Z_N_odd_cc}

A symmetry defect associated with the topological charge conjugation symmetry subgroup can be engineered in the Laughlin state, wherein quasielectrons and quasiholes are permuted, by creating a superconducting trench in the bulk, or as a superconducting/magnetic domain wall on the edge of a fractional topological insulator~\cite{clarke2013,lindner2012,cheng2012,barkeshli2013genon}. Such defects are known as $\mathbb{Z}_{N}$-Parafendleyons or $\mathbb{Z}_{N}$ parafermion zero modes~\cite{Fendley12}. In the rest of this section, we will focus on the global $G=\mathbb{Z}_{2}$ symmetry action associated with topological charge conjugation. (For the case of trivial global symmetry action, see Sec.~\ref{sec:no_perm}.)

We now show that there is only one symmetry fractionalization class for $\rho$ corresponding to charge conjugation. We use the choice of gauge with $U_{\bf g} (a,b;[a+b]_{N})=1$ and $\beta_{a}({\bf g},{\bf h})=1$, so we have $\eta_{a}({\bf g},{\bf h})= M_{a \cohosub{w}({\bf g},{\bf h})}$ and $[\coho{w}]\in H^2_{[\rho]}(\mathbb{Z}_2,\mathbb{Z}_N)$. The cocycle condition simplifies to $\overline{\coho{w}(\mathbf{1},\mathbf{1})}=\coho{w}(\mathbf{1},\mathbf{1})$. Since $N$ is odd, only the vacuum $0$ is a fixed point under charge conjugation and, hence, $\coho{w}(\mb{g},\mb{h})=0$, $\eta_{a}({\bf g},{\bf h})=1$, and $H^2_{[\rho]}(\mathbb{Z}_2,\mathbb{Z}_N)=\mathbb{Z}_{1}$. In other words, the $\mathbb{Z}_2$ charge conjugation symmetry has only one fractionalization class.

As discussed in Sec.~\ref{sec:higher_genus}, since there
are no nontrivial $\mb{g}$-invariant topological charges in $\mathcal{C}_{\bf 0}$, we have $| \mathcal{C}_{\mathbf{1}} |=1$. In other words, there is exactly one type of
defect, which we denote as $\sigma$. This statement can also be proven directly from properties of the fusion rules, without appealing to
modularity, as follows. If there is another defect $\sigma'$, it must be related to $\sigma$ by fusing with some $a\in\mathcal{C}_\mb{0} \equiv \mathbb{Z}_N^{(n)}$. Assume
$\sigma'=\sigma\times a$. When $a$ is taken around the defect it becomes $\bar{a}$, which implies that
$\sigma\times a=\sigma\times \bar{a}=\sigma'$. Fusing with $a$ again, we determine that $\sigma'\times a=\sigma'\times \bar{a}=\sigma$ and $\sigma\times a^2 = \sigma\times [2a]_{N} =\sigma$. Using this relation, we find
\begin{equation}
\sigma' = \sigma\times [(N-1)a]_N = \sigma\times \left([2a]_{N} \right)^{\frac{N-1}{2}} = \sigma
,
\end{equation}
which proves $\sigma'=\sigma$.

The fusion rules of $\sigma$ can be easily obtained to be
\begin{eqnarray}
\sigma \times a &=& a \times \sigma =\sigma , \\
\sigma \times \sigma &=& \sum_{a \in \mathcal{C}_{\bf 0}} a
.
\end{eqnarray}
The fusion category $\graded{C}{G}=\mathcal{C}_\mb{0}\oplus \mathcal{C}_\mathbf{1}$ is known as the Tambara-Yamagami category~\cite{Tambara1998}. The fusion rules indicate that the quantum dimension of the defect is $d_{\sigma}=\sqrt{N}$. The $F$-symbols of $\graded{C}{G}$ are completely classified in \Ref{Tambara1998} and are given by the $F$-symbols of the original category $\mathcal{C}_{\bf 0}$ (which are all trivial in this example), together with
\begin{eqnarray}
[ F^{a \sigma b}_{\sigma} ]_{\sigma \sigma} &=& [ F^{\sigma a \sigma}_{b} ]_{\sigma \sigma} = \chi(a,b) , \\
\left[ F^{ \sigma \sigma \sigma}_{\sigma} \right]_{ab} &=& \frac{\varkappa_{\sigma}}{\sqrt{N}} \chi(a,b)^{-1} ,
\end{eqnarray}
and all other allowed $F$-symbols equal 1. Here $\chi$ is a $\text{U}(1)$-valued function on $\mathbb{Z}_N\times\mathbb{Z}_N$, satisfying
\begin{eqnarray}
\chi(a,b) &=& \chi(b,a) ,\\
\chi(ab,c) &=& \chi(a,c)\chi(b,c), \\
\chi(a,bc) &=& \chi(a,b)\chi(a,c),
\end{eqnarray}
together with normalization condition $\chi(0,a)=\chi(a,0)=1$. Such a $\chi$ is called a symmetric bi-character.
$\varkappa_\sigma=\pm 1$ is the Frobenius-Schur indicator of the $\mathbb{Z}_{2}$ symmetry defect. The two solutions of $F$-symbols, distinguished by $\varkappa_\sigma$, directly reflect the two defectification classes associated with $H^3(\mathbb{Z}_2, \text{U}(1))=\mathbb{Z}_2$. Interestingly, this fusion category $\graded{\mathcal{C}}{G}$ does not admit braiding in the usual sense.

We now consider $G$-crossed braiding. First we use the symmetry action gauge transformations to set $\eta_{\sigma}(\mb{1},\mb{1})=1$, so that all $\eta_{a}({\bf g},{\bf h})=1$.
With this gauge fixing, we find the following solutions to the $G$-crossed braiding consistency equations
\begin{eqnarray}
\chi(a,b) &=& e^{i \frac{2\pi p}{N}ab}, \\
R^{\sigma a}_{\sigma} &=& (-1)^{pa} e^{-i \frac{\pi p}{N} a^2}, \\
R^{a\sigma}_{\sigma} &=& r^{a} (-1)^{pa} e^{-i \frac{\pi p}{N} a^2}, \\
R^{\sigma\sigma}_a &=& \Upsilon (-1)^{pa} e^{i \frac{\pi p}{N} a^2}, \\
U_{\bf 1} (a,\sigma ; \sigma) &=& U_{\bf 1} (\sigma ,a ; \sigma) = r^{a} , \\
U_{\bf 1} (\sigma,\sigma ; a) &=& r^{-a} ,\\
\Upsilon^2 &=& \frac{\varkappa_\sigma}{\sqrt{N}} \sum_{a=0}^{N-1} (-1)^{p a} e^{-i \frac{\pi p}{N} a^2}.
\label{eq:Ups_squared}
\end{eqnarray}
Here, $r=e^{i \frac{2 \pi}{N}n}$ for some integer $n$. These $N$th roots of unity can be removed using the remaining symmetry action gauge freedom of the anyons.
Specifically, $\gamma_{a}({\bf 1}) = r^{-a} = e^{-i \frac{2 \pi}{N}n a}$ allows us to set $s=1$ in the above expressions and it does not spoil our previous gauge choices, since one can always choose $z \in \mathcal{C}_\mb{0}$ such that $M_{a z}= e^{i \frac{4 \pi p}{N}az} = e^{-i \frac{2 \pi}{N}na}$. The remaining symmetry action gauge freedom that does not spoil our gauge choices is $\gamma_{\sigma}(\mb{1})=\pm 1$, which can be used to change the (arbitrary) sign of the phase $\Upsilon$.
We notice that none of the $U_\mathbf{1}$ are intrinsic in the sense that they are all essentially maps between different splitting spaces, except $U_\mathbf{1}(\sigma,\sigma;0)=1$.

Having obtained the $G$-crossed data, we can calculate the topological twist
\begin{equation}
\theta_\sigma=\sum_a\frac{d_a}{d_\sigma}R^{\sigma\sigma}_a= \varkappa_{\sigma} \Upsilon^{-1}
\end{equation}
and the $S$-matrix elements
\begin{eqnarray}
S_{0 \sigma} &=& S_{\sigma 0} =1 , \\
S_{\sigma \sigma} &=& \frac{1}{\sqrt{N}} \sum_{a \in \mathcal{C}_\mb{0}} \left( R^{\sigma\sigma}_a \right)^{2} = \Upsilon^2 \Theta_{\bf 0} = \Theta_{\bf 0} \theta_{\sigma}^{-2} ,
\end{eqnarray}
where $\Theta_{\bf 0} = \frac{1}{\sqrt{N}} \sum_{a_{\bf 0}} \theta_{a} = e^{i \frac{\pi}{4} c_{-}}$, as usual.
These give the $G$-crossed modular transformations
\begin{eqnarray}
\mathcal{S}&=&
	\left[
	\begin{array}{c|c|c|c}
		\mathcal{S}^{(\mb{0},\mb{0})} & 0 & 0 & 0\\
		\hline
		 0 & 0 & 1 & 0\\
        \hline
		 0 & 1 & 0 & 0\\
        \hline
		 0 & 0 & 0 &  \Theta_{\bf 0} \theta_{\sigma}^{-2}
	\end{array}
	\right],\\
\mathcal{T}&=&
	\left[
	\begin{array}{c|c|c|c}
		\mathcal{T}^{(\mb{0},\mb{0})} & 0 & 0 & 0\\
		\hline
		 0 & 1 & 0 & 0\\
        \hline
		 0 & 0 & 0 & \theta_\sigma\\
        \hline
		 0 & 0 & \theta_\sigma & 0
	\end{array}
	\right]
,
\end{eqnarray}
where $\mathcal{S}^{(\mb{0},\mb{0})}$ and $\mathcal{T}^{(\mb{0},\mb{0})}$ are the topological $S$ and $T$ matrices for the $\mathbb{Z}_N^{(p)}$ theory. The basis states of the defect sectors are chosen to be $\ket{0^{(\mb{0},\mb{1})}}, \ket{\sigma^{(\mb{1},\mb{0})}}, \ket{\sigma^{(\mb{1},\mb{1})}}$.

We now proceed to derive the properties of the gauged theories. Under the group action the extended category is divided into $\frac{N+1}{2}$ orbits: $\{0 \}, \{a, \bar{a} \}, \{ \sigma \}$. The stabilizer subgroup for both $0$ and $\sigma$ is $\mathbb{Z}_2$. For the $\frac{N-1}{2}$ orbits $\{a, \bar{a} \}$, the stabilizer subgroups are trivial.

The vacuum $0$ gives rise to a $\mathbb{Z}_2$ even charge $(0,+)\equiv I$ (the vacuum in $\gauged{C}{G}$) and a $\mathbb{Z}_2$ odd charge $(0,-)\equiv z$ (the $\mathbb{Z}_2$ charge) which satisfies $z \times z =I$. These correspond to the trivial and alternating irreps of $\mathbb{Z}_2$. The quantum dimensions are $d_I = d_z =1$.

The orbits $\{a,\bar{a} \}$ become non-Abelian anyons in the gauged theory, which we label by $\phi_a$. Their quantum dimensions are $d_{\phi_{a}}=2$ and their fusion rules with each other are given by
\begin{equation}
\phi_a\times\phi_b=\begin{cases}
	\phi_{\min(a+b, N-a-b)}+\phi_{|a-b|} & \text{for } a\neq b\\
	I+z+\phi_{\min(2a,N-2a)} & \text{for } a=b
	\end{cases}
.
\end{equation}

The defect $\sigma$ gives rise to two quasiparticles $(\sigma, \pm)$ in the gauged theory, which are related to each other through fusion with $z$, i.e. $(\sigma, \pm) = (\sigma, \mp) \times z$. (The $\pm$ label here is an arbitrary choice, but their difference corresponds to the nontrivial irrep of $\mathbb{Z}_2$.) Their quantum dimensions are $d_{(\sigma, \pm)} = \sqrt{N}$ and their fusion rules with each other are given by
\begin{align}
(\sigma, +)\times (\sigma,+)&= (\sigma, -)\times (\sigma,-) = I+ \sum_{a=1}^{\frac{N-1}{2}} \phi_a, \\
(\sigma,+)\times (\sigma,-)&= z + \sum_{a=1}^{\frac{N-1}{2}} \phi_a.
\end{align}
Thus, $\gauged{C}{G}$ has $2+\frac{N-1}{2}+2=\frac{N+7}{2}$ topological charges.

To further identify the gauged theory, we calculate the topological twists of the anyons in $\gauged{C}{G}$.
The twist factors of $I$ and $z$ are clearly
\begin{equation}
\theta_{I}= \theta_{z} = 1 .
\end{equation}
The twist factors of $\phi_a$ are identical to those of $a$ and $\bar{a}$, so we have
\begin{equation}
\theta_{\phi_a}=e^{i \frac{2\pi p}{N}a^2}.
\end{equation}
The twists factors of $(\sigma, \pm)$ are
\begin{equation}
\theta_{(\sigma, \pm)}= \pm \theta_{\sigma} = \pm \varkappa_{\sigma} \Upsilon^{-1} .
\end{equation}

When $p=\frac{N-1}{2}$, i.e. $\mathcal{C}_0 = \text{SU}(N)_1$, and $\varkappa_{\sigma} = (-1)^{\frac{N^2 - 1}{8}}$, we find that $\gauged{C}{G}$ is equivalent to the MTC of $\text{SO}(N)_2$, by matching fusion and twist factors. In particular, the Gauss sum in Eq.~(\ref{eq:Ups_squared}) evaluates to $\sqrt{N}$ for $[N]_4 = 1$ and $i\sqrt{N}$ for $[N]_{4} =3$. This gives $\theta_\sigma^2 = \Upsilon^{-2}=\varkappa_{\sigma} (-1)^{\frac{N^2 - 1}{8}} i^{\frac{N-1}{2}}$, which matches $\text{SO}(N)_2$ when $\varkappa_{\sigma}=(-1)^{\frac{N^2 - 1}{8}}$. The gauged theories for other values of $p$ and $\varkappa_{\sigma}$ are metaplectic modular categories, which are close relatives of $\text{SO}(N)_2$, i.e. they are in the same Grothendieck class.

We note that the relation between the $\mathbb{Z}_{3}$ theory and the gauged theory $\text{SO}(3)_2 = \text{SU}(2)_{4}$ was previously observed in \Refs{bais2009,barkeshli2010,Clarke-private}.

\subsubsection{General symmetry action}

In the case of a general $\mathbb{Z}_{2}$ symmetry action on the anyons, we can factor the MTC into its subcategories upon which the symmetry acts trivially and as topological charge conjugation. In particular, using the prime decomposition from Eq.~(\ref{eq:Z_N_decomp}), we write $f_j =0$ or $1$ to represent whether $\rho_{\bf 1}$ acts trivially or as charge conjugation on the $\mathbb{Z}_{N_{j}}^{(p_{j})}$ sector. Then we can write $\mathbb{Z}_{N}^{(p)} = \mathbb{Z}_{N_{\text{t}}}^{(p_{\text{t}})} \times \mathbb{Z}_{N_{\text{c}}}^{(p_{\text{c}})}$, where $N_{\text{t}} = \prod_{j=0}^{k} N_{j}^{1-f_j}$, $N_{\text{c}} = \prod_{j=0}^{k} N_{j}^{f_j}$, and $a = (a^{(\text{t})},a^{(\text{c})})$, such that $\rho_{\bf 1} (a^{(\text{t})},a^{(\text{c})}) = (a^{(\text{t})},[-a^{(\text{c})}]_{N_{\text{c}}} )$. In this way, the corresponding $G$-crossed theories are obtained by applying the gluing construction of Appendix~\ref{sec:gluing} to the corresponding $G$-crossed theories for trivial and charge conjugation sectors obtained earlier in this section. From this, it is clear that there is one symmetry fractionalization class ($H^2_{[\rho]}(\mathbb{Z}_2,\mathbb{Z}_N)=\mathbb{Z}_{1}$) and two defectification classes ($H^3(\mathbb{Z}_2, \text{U}(1))=\mathbb{Z}_2$), and the resulting $G$-crossed theories can be written as the product
\begin{equation}
\left[\mathbb{Z}_{N}^{(p)}\right]_{\mathbb{Z}_{2}}^{\times} = \mathbb{Z}_{N_{\text{t}}}^{(p_{\text{t}})} \boxtimes \left[\mathbb{Z}_{N_{\text{c}}}^{(p_{\text{c}})}\right]_{\mathbb{Z}_{2}^{\text{cc}}}^{\times}
,
\end{equation}
where $\left[ \mathbb{Z}_{N_{\text{c}}}^{(p_{\text{c}})}\right]_{\mathbb{Z}_{2}^{\text{cc}}}^{\times}$ are given by the two possible $G$-crossed extensions obtained in Sec.~\ref{sec:Z_N_odd_cc}.

\subsection{$\mathbb{Z}_{N}^{(p)}$ Anyons with $N$ Even and $\mathbb{Z}_2$ Symmetry}
\label{sec:zneven}

In this section, we consider $\mathbb{Z}_2$ symmetry in the $\mathbb{Z}_N^{(p)}$ anyon model
where $N$ is even and $p$ is a half-integer such that $0 < p < N$ and $\gcd (2p, N) =1$, so that the theory is modular.
The $\mathbb{Z}_N^{(p)}$ theory has $N$ distinct topological charges,
labeled by $a=0,1,\dots,N-1$. The fusion rules are given by addition
modulo $N$: $a\times b=[a+b]_N$. The $F$-symbols and $R$-symbols are given by
\begin{align}
F^{abc}_{[a+b+c]_{N}}&=e^{i\frac{\pi}{N}a(b+c-[b+c]_{N})}, \\
R^{ab}_{[a+b]_{N}}&=e^{i \frac{2\pi p}{N}ab}.
\end{align}

As shown in Sec.~\ref{sec:Z_N}, the topological symmetry group is $\mathrm{Aut}( \mathbb{Z}_{N}^{(p)} )=\mathbb{Z}_2^{k}$ or $\mathbb{Z}_2^{k+1}$, depending whether $r_0 = 1$ or $r_0 >1$, respectively, where $k$ is the number of distinct odd primes in the prime factorization of $N$. There is always a subgroup $\mathbb{Z}_{2} \lhd \mathrm{Aut}( \mathbb{Z}_{N}^{(p)} )$ that is associated with topological charge conjugation $a \mapsto \bar{a} = [-a]_{N}$. Furthermore, it was shown that the symmetry fractionalization obstruction always vanishes $[\coho{O}]=0$, the symmetry fractionalization factorizes, and the $G$-crossed extensions can be expressed in terms of gluing the $G$-crossed extensions of the factors. In light of this, for $G=\mathbb{Z}_{2}$, we only need to obtain the data for the cases where the symmetry acts trivially and as topological charge conjugation, and then we can generate the general cases from these. (In fact, given the results obtained in Secs.~\ref{sec:no_perm} and \ref{sec:Z_N_odd}, it only remains to obtain for the $\mathbb{Z}_2$-crossed extensions of $\mathbb{Z}_N^{(p)}$ for $N = 2^{r}$, where the integer $r>1$, with charge conjugation symmetry. However, we will continue to consider more general $N$.) We will find that $H^2_{[\rho]}(\mathbb{Z}_2,\mathbb{Z}_N)=\mathbb{Z}_{2}$ and $H^3(\mathbb{Z}_2,\text{U}(1))=\mathbb{Z}_{2}$, so there are two symmetry fractionalization class and two defectification classes for $\mathbb{Z}_2$ symmetry when $N$ is even. We now examine these in detail.

\subsubsection{Trivial symmetry action}

In the case where the symmetry acts trivially on the anyons, we see that every $a \in \mathbb{Z}_{N}$ defines a 2-cocycle $\coho{w}(\mb{1},\mb{1})=a$, while only the even-valued $a \in \mathbb{Z}_{N}$ can be 2-coboundaries, since ${\rm d}\coho{z}(\mb{1},\mb{1})= \coho{z}(\mb{1}) \times \coho{z}(\mb{1}) = [2b]_{N}$, for $\coho{z}(\mb{1}) = b$. Thus, $H^2(\mathbb{Z}_2,\mathbb{Z}_N)=\mathbb{Z}_{2}$, and there are two fractionalization classes. These two fractionalization classes can be represented by the cocycles $\coho{w}(\mb{1},\mb{1})=0$ and $1$, respectively. The defectification obstruction vanishes for both fractionalization classes: for $\coho{w}({\bf 1},{\bf 1})= 0$, Eq.~(\ref{eq:obstruction}) gives $\mathcal{O}({\bf 1},{\bf 1},{\bf 1},{\bf 1}) =1$; for $\coho{w}({\bf 1},{\bf 1})= 1$, Eq.~(\ref{eq:obstruction}) gives $\mathcal{O}({\bf 1},{\bf 1},{\bf 1},{\bf 1}) =e^{i\frac{2 \pi p}{N}}$, which is a coboundary $\text{d}\lambda ({\bf 1},{\bf 1},{\bf 1},{\bf 1})$ for e.g. $\lambda ({\bf 1},{\bf 1},{\bf 1}) = e^{i\frac{\pi p}{N}}$. For a given fractionalization class, the data of the two $G$-crossed extensions, corresponding to the two defectification classes, are given by the results of Sec.~\ref{sec:no_perm}.

It is straightforward to see that the resulting $G$-crossed theories for $\coho{w}({\bf 1},{\bf 1})= 0$ have the form
\begin{equation}
(\mathbb{Z}_{N}^{(p)})_{\mathbb{Z}_{2}}^{\times} = \mathbb{Z}_{N}^{(p)} \boxtimes \text{SPT}_{\mathbb{Z}_{2}}^{[\alpha]}
,
\end{equation}
where $[\alpha] \in H^3(\mathbb{Z}_2,\text{U}(1))=\mathbb{Z}_{2}$, i.e. it is represented by the 3-cocycle with element $\alpha(\mb{1},\mb{1},\mb{1})=\pm 1$. The corresponding gauged theories are $\mathbb{Z}_{N}^{(p)} \boxtimes \text{D}^{[\alpha]}(\mathbb{Z}_{2})$.

For the fractionalization class represented by $\coho{w}({\bf 1},{\bf 1})= 1$, the $\mathbb{Z}_{2}$-crossed extensions have $\mathbb{Z}_{2N}$ fusion rules, in which the odd integers are the $\mathbb{Z}_{2}$ defects and the even integers are the quasiparticles. The corresponding gauged theories are $\mathbb{Z}_{4N}^{(p)}$ and $\mathbb{Z}_{4N}^{(p+2N)}$.

\subsubsection{Charge conjugation symmetry}
\label{sec:Z_N_even_cc}

In the case where the symmetry acts as topological charge conjugation, i.e. $\rho_{\bf 1}(a) = \bar{a}$, we can choose a gauge such that $U_{\bf g} (a,b;[a+b]_{N})=(-1)^{a(b-1)}$ and $\beta_{a}({\bf g},{\bf h})=1$.
We now show that there is only one symmetry fractionalization class for $\rho$ corresponding to charge conjugation. It follows that the symmetry fractionalization obstruction vanishes $[\coho{O}]=0$, and $\eta_{a}({\bf g},{\bf h})= M_{a \cohosub{w}({\bf g},{\bf h})}$ and $[\coho{w}]\in H^2_{[\rho]}(\mathbb{Z}_2,\mathbb{Z}_N)$. The cocycle condition simplifies to $\overline{\coho{w}(\mathbf{1},\mathbf{1})}=\coho{w}(\mathbf{1},\mathbf{1})$, and the coboundaries are trivial, since $\text{d} \coho{z}(\mathbf{1},\mathbf{1}) = \overline{\coho{z}(\mathbf{1})} \times \coho{z}(\mathbf{1}) =0$. Thus, $H^2_{[\rho]}(\mathbb{Z}_2,\mathbb{Z}_N) = \mathbb{Z}_{2}$, where the two fractionalization classes are specified by the two self-dual topological charges, i.e. $\coho{w}(\mathbf{1},\mathbf{1}) =0$ or $\frac{N}{2}$, respectively. Evaluating the invariant from Eq.~(\ref{eq:cc_eta_invariant}), the two fractionalization classes correspond to $\eta_{a}^{\text{cc}} = 1$ and $\eta_{a}^{\text{cc}} = (-1)^{a}$, respectively.

Since there are two topological charges in $\mathbb{Z}_{N}^{(p)}$ that are invariant under charge conjugation for $N$ even,
there are two types of symmetry defects, which we label as $\mathcal{C}_{\bf 1} = \{ \sigma^{+} , \sigma^{-} \}$. There are two possible sets of defect fusion rules, given either by
\begin{equation}
\label{eq:Z_N_defect_fusion_1}
\begin{gathered}
\sigma^{s} \times \sigma^{s} = \sum_{c \text{ even} \in \mathbb{Z}_{N} } c, \\
\sigma^{+} \times \sigma^{-} = \sum_{c \text{ odd} \in \mathbb{Z}_{N}} c,
\end{gathered}
\end{equation}
or by
\begin{equation}
\label{eq:Z_N_defect_fusion_2}
\begin{gathered}
\sigma^{s} \times \sigma^{s} = \sum_{c \text{ odd} \in \mathbb{Z}_{N} } c , \\
\sigma^{+} \times \sigma^{-} = \sum_{c \text{ even} \in \mathbb{Z}_{N}} c .
\end{gathered}
\end{equation}
In both cases, we have
\begin{equation}
\label{eq:Z_N_defect_fusion_both}
a \times \sigma^{s} = \sigma^{s} \times a = \left\{
\begin{array}{lll}
\sigma^{s} & \text{} & \text{ for } a \text { even} \\
\sigma^{-s} & \text{} & \text{ for } a \text { odd}
\end{array}
\right.
\end{equation}
The fusion rules indicate that the quantum dimension of the defects are $d_{\sigma^{s}}=\sqrt{N/2}$.

Considering Eq.~(\ref{eq:fusion_symm_frac_change}), we see that both sets of defect fusion rules can occur for $r_0 =1$ (when $\frac{N}{2}$ is odd), where the different fusion rules correspond to the two different fractionalization classes. On the other hand, for $r_0 >1$ (when $\frac{N}{2}$ is even), the two fractionalization classes actually have the same defect fusion rules, so only one of the possibilities will occur. Since the full data of the defect theories for $r_0 =1$ can be produced via the gluing construction from results we have already obtained, we focus on $r_0 >1$ for the remainder of this example.

\subsubsection{Charge conjugation symmetry when 4 divides $N$}
\label{sec:N_multiple4}

When $N$ is a multiple of 4 (i.e. $r_0 >1$), the defect fusion rules are given by Eqs.~(\ref{eq:Z_N_defect_fusion_1}) and (\ref{eq:Z_N_defect_fusion_both}). We can begin by considering the restriction to the even-valued quasiparticle charges and one of the two defects, i.e. $\mathbb{Z}_{N/2}^{(2p)} \oplus \{ \sigma^{s} \}$, where $\mathbb{Z}_{N/2}^{(2p)} \lhd \mathbb{Z}_{N}^{(p)}$ is the subcategory of even-valued topological charges; such a restriction forms a closed $\mathbb{Z}_{2}$-crossed subcategory of the full defect theory. Similar to Sec.~\ref{sec:Z_N_odd_cc}, solving the $G$-crossed consistency conditions (with certain gauge choices) gives the data \emph{for even-valued} $a,b \in \mathbb{Z}_{N}$
\begin{eqnarray}
[ F^{a \sigma^{s} b}_{\sigma^{s}} ]_{\sigma^{s} \sigma^{s}} &=& [ F^{\sigma^{s} a \sigma^{s}}_{b} ]_{\sigma^{s} \sigma^{s}} = e^{i \frac{2\pi p}{N}ab} , \\
\left[ F^{ \sigma^{s} \sigma^{s} \sigma^{s}}_{\sigma^{s}} \right]_{ab} &=& \frac{\varkappa_{\sigma^{s}}}{\sqrt{ N/2}} e^{-i \frac{2\pi p}{N}ab} , \\
R^{\sigma^{s} a}_{\sigma^{s}} &=& q_{s}^{\frac{a}{2}} (-1)^{p a} e^{-i \frac{\pi p}{N} a^2}, \\
R^{a \sigma^{s}}_{\sigma^{s}} &=& (-1)^{p a} e^{-i \frac{\pi p}{N}a^2}, \\
R^{\sigma^{s} \sigma^{s}}_a &=& \Upsilon q_{s}^{\frac{a}{2}} (-1)^{pa} e^{i \frac{\pi p}{N}a^2}, \\
\theta_{\sigma^{s}} &=&  \varkappa_{\sigma^{s}} \Upsilon^{-1} \\
\eta_{a}({\bf 1},{\bf 1}) &=& \eta_{\sigma^{s}}({\bf 1},{\bf 1}) = 1 , \\
U_{\bf 1} (a,\sigma^{s} ; \sigma^{s}) &=& U_{\bf 1} (\sigma^{s} ,a ; \sigma^{s}) =U_{\bf 1} (\sigma^{s},\sigma^{s} ; a)= q_{s}^{\frac{a}{2}} , \\
\Upsilon^2 &=& \frac{\varkappa_{\sigma^{s}}}{\sqrt{ N/2}} \sum_{a \text{ even } \in \mathbb{Z}_{N}} q_{s}^{\frac{a}{2}} (-1)^{pa} e^{- i \frac{\pi p}{N} a^2}, \qquad
\end{eqnarray}
and all other allowed $F$-symbols are equal to 1. Here, $q_{s} = \pm 1$ is a sign that is presumably fixed when considering the full defect theory. The quantity $\Upsilon$ is a phase whose sign can be fixed by using gauge freedom.

Since $\sigma^{+}$ and $\sigma^{-}$ are both self-dual, their Frobenius-Schur indicators $\varkappa_{\sigma^{s}} = \pm 1$ are invariants. Thus, the four different combinations of $\varkappa_{\sigma^{+}}$ and $\varkappa_{\sigma^{-}}$ correspond to the four possible $\mathbb{Z}_{2}$-crossed defect theories, as classified by $H^2_{[\rho]}(\mathbb{Z}_2,\mathbb{Z}_N)=\mathbb{Z}_{2}$ and $H^3(\mathbb{Z}_2,\text{U}(1))=\mathbb{Z}_{2}$. Moreover, we can identify the two theories with $\varkappa_{\sigma^{+}} = \varkappa_{\sigma^{-}}$ as having the same fractionalization class, while the two theories with $\varkappa_{\sigma^{+}} = -\varkappa_{\sigma^{-}}$ have the other fractionalization class. This is because the defectification classes that are related by gluing in a $\mathbb{Z}_{2}$ SPT state have the opposite signs for both defects' Frobenius-Schur indicators. The two defectification classes with $\varkappa_{\sigma^{+}} = -\varkappa_{\sigma^{-}}$ are actually the same under relabeling of the defects as $\sigma^{s \prime} = \sigma^{-s}$, (corresponding to $\sigma^{s \prime} = \coho{z}({\bf 1}) \times \sigma^{s}$, where $\coho{z}({\bf 1}) = a$ odd,) so the na\"ive classification count is reduced from four to three distinct $\mathbb{Z}_{2}$-crossed defect theories.

We now proceed to derive the properties of the gauged theories. Under the group action the extended category is divided into the orbits: $\{0 \}$, $\{ \frac{N}{2} \}$, $\{ \sigma^{+} \}$, $\{ \sigma^{-} \}$, and $\{a, \bar{a} \}$ for $a=1,\ldots,\frac{N}{2}-1$. The stabilizer subgroup for the singletons is $\mathbb{Z}_2$. For the $\frac{N}{2}-1$ orbits $\{a, \bar{a} \}$, the stabilizer subgroups are trivial.

Each of the singletons gives rise to two quasiparticle types in the gauged theory, corresponding to whether a trivial or alternating irrep of $\mathbb{Z}_2$ is attached to it. We write these as $(0, \pm)$, $(\frac{N}{2} , \pm )$, $(\sigma^{+} , \pm )$, and $(\sigma^{-} , \pm )$. Their quantum dimensions are $d_{(0, \pm)} = d_{(\frac{N}{2}, \pm)} =1$ and $d_{(\sigma^{s}, \pm)} = \sqrt{N/2}$. Their topological twists are given by $\theta_{(0, \pm)} = 1$, $\theta_{(\frac{N}{2}, \pm)} = (-1)^{\frac{N}{4}}$, and $\theta_{(\sigma^{s}, \pm)} = \pm \theta_{\sigma^{s}}$.
The orbits $\{a,\bar{a} \}$ become non-Abelian anyons in the gauged theory, which we label by $\phi_a$. Their quantum dimensions are $d_{\phi_{a}}=2$ and their topological twists are $\theta_{\phi_{a}}= e^{i \frac{2 \pi p}{N} a^2}$.
Thus, the gauged theories have $\frac{N}{2}+7$ topological charge types.

Gauging the $\mathbb{Z}_{2}$ charge conjugation symmetry of the $\mathbb{Z}_{N}^{(p)}$ theories results in the so-called metaplectic modular categories, such as $\text{SO}(N)_2$ and its close relatives in the same Grothendieck class. In the case of $N=4$, these gauged theories take the form $\text{Ising}^{(\nu^{+})} \boxtimes \text{Ising}^{(\nu^{-})}$, where $\left[\frac{\nu^{+} + \nu^{-}}{2} \right]_{8} = 2p$ and the Frobenius-Schur indicators match those of the defects, i.e. $\varkappa_{\sigma^{+}} = (-1)^{\frac{(\nu^{+})^{2} - 1}{8}}$ and $ \varkappa_{\sigma^{-}} = (-1)^{\frac{(\nu^{-})^{2} - 1}{8}}$. We notice that the equivalence between theories $(\nu^{+},\nu^{-}) \sim (\nu^{+} + 8,\nu^{-} +8)$ and $\nu^{s} \sim \nu^{s} +16$ yields four theories for each $p$. Moreover, we see that interchanging the Ising factors, i.e. relabeling $(\nu^{+ \prime},\nu^{- \prime}) = (\nu^{-} ,\nu^{+})$, equate the two theories (for each $p$) with $\varkappa_{\sigma^{+}} =- \varkappa_{\sigma^{-}}$, yielding three distinct gauged theories for each $p$. For example, when $p=\frac{1}{2}$, we have the gauged theories with $(\nu^{+},\nu^{-}) = (1,1)$, $(5,13)$, $(7,11)$, and $(3,15)$ corresponding to the defect theories with $(\varkappa_{\sigma^{+}},\varkappa_{\sigma^{-}})= (+1,+1)$, $(-1,-1)$, $(+1,-1)$, and $(-1,+1)$, respectively, with the last two being equated by interchanging the two sectors.

\subsubsection{General symmetry action}

In the case of a general $\mathbb{Z}_{2}$ symmetry action on the anyons, we can factor the MTC into its subcategories upon which the symmetry acts trivially and as topological charge conjugation. In particular, using the prime decomposition from Eq.~(\ref{eq:Z_N_decomp}), we write $f_j =0$ or $1$ to represent whether $\rho_{\bf 1}$ acts trivially or as charge conjugation on the $\mathbb{Z}_{N_{j}}^{(p_{j})}$ sector. Then we can write $\mathbb{Z}_{N}^{(p)} = \mathbb{Z}_{N_{\text{t}}}^{(p_{\text{t}})} \times \mathbb{Z}_{N_{\text{c}}}^{(p_{\text{c}})}$, where $N_{\text{t}} = \prod_{j=0}^{k} N_{j}^{1-f_j}$, $N_{\text{c}} = \prod_{j=0}^{k} N_{j}^{f_j}$, and $a = (a^{(\text{t})},a^{(\text{c})})$, such that $\rho_{\bf 1} (a^{(\text{t})},a^{(\text{c})}) = (a^{(\text{t})},[-a^{(\text{c})}]_{N_{\text{c}}} )$. In this way, the corresponding $G$-crossed theories are obtained by applying the gluing construction of Appendix~\ref{sec:gluing} to the corresponding $G$-crossed theories for trivial and charge conjugation sectors obtained earlier in this section and Sec.~\ref{sec:Z_N_odd}. From this, it is clear that there are two symmetry fractionalization classes ($H^2_{[\rho]}(\mathbb{Z}_2,\mathbb{Z}_N)=\mathbb{Z}_{2}$) and two defectification classes ($H^3(\mathbb{Z}_2, \text{U}(1))=\mathbb{Z}_2$) for each fractionalization class, though na\"ively different defectification classes may potentially be equivalent under relabeling of the defects. The resulting $G$-crossed theories are given by
\begin{equation}
\left[\mathbb{Z}_{N}^{(p)}\right]_{\mathbb{Z}_{2}}^{\times} = \left[ \mathbb{Z}_{N_{\text{t}}}^{(p_{\text{t}})} \right]_{\mathbb{Z}_{2}}^{\times} \underset{\mathbb{Z}_{2}}{\boxtimes} \left[ \mathbb{Z}_{N_{\text{c}}}^{(p_{\text{c}})} \right]_{\mathbb{Z}_{2}^{\text{cc}}}^{\times}
.
\end{equation}
When $\coho{w}^{(\text{t})}({\bf 1},{\bf 1})= 0$, these can be written as the product
\begin{equation}
(\mathbb{Z}_{N}^{(p)})_{\mathbb{Z}_{2}}^{\times} = \mathbb{Z}_{N_{\text{t}}}^{(p_{\text{t}})} \boxtimes \left[\mathbb{Z}_{N_{\text{c}}}^{(p_{\text{c}})}\right]_{\mathbb{Z}_{2}^{\text{cc}}}^{\times}
.
\end{equation}

\subsection{$\mathbb{Z}_2$-Toric Code D$(\mathbb{Z}_2)$ with $\mathbb{Z}_2$ Symmetry}
\label{sec:D(Z_2)_Z_2_symm}

In this section, we consider a $G=\mathbb{Z}_{2}$ symmetry for a system with the ``toric code'' topological order $\mathrm{D}(\mathbb{Z}_2)$, which corresponds to a discrete gauge theory or quantum double~\cite{Kitaev03}. The topological charges are gauge charges $e$, gauge fluxes $m$, and their bound state $\psi$. The theory has $\mathbb{Z}_2 \times \mathbb{Z}_2$ fusion rules, where we write $I=(0,0)$, $e=(1,0)$, $m=(0,1)$, and $\psi = (1,1)$. That is, the fusion rules are given by $e\times e = m\times m = \psi \times \psi =I$ and $e \times m = \psi$. The $F$-symbols of the theory are all trivial (i.e. they equal 1 when allowed by fusion) and the $R$-symbols are given by
\begin{equation}
R^{ab}=(-1)^{a_{\text{e}} b_{\text{m}}}
,
\end{equation}
where the notation corresponds to writing the topological charges as $a=(a_{\text{e}} , a_{\text{m}})$. From this, it is clear that $I$, $e$, and $m$ are bosons with topological twist $\theta_a =1$, and $\psi$ is a fermion with $\theta_\psi =-1$.  The Abelian anyons form a group $\mathcal{A} = \mathbb{Z}_2 \times \mathbb{Z}_2$.

The topological symmetry group is $\mathrm{Aut}(\mathrm{D}(\mathbb{Z}_2)) = \mathbb{Z}_2$, where the nontrivial element interchanges $e$ and $m$. This is known as the electric-magnetic (e-m) duality, and can be realized in a slightly different formulation of the toric code model by Wen~\cite{Wen2003} as lattice translations~\cite{bombin2010, you2012}. Alternatively, one can also realize this type of symmetry in an on-site fashion~\cite{cheng_setmodel}.

We consider the two cases where the symmetry action on the anyons is trivial and where the symmetry action interchanges $e$ and $m$.

\subsubsection{Trivial symmetry action}
\label{sec:ExampleD(Z2)Z2Trivial}

When the symmetry action is trivial, the symmetry fractionalization obstruction $[\coho{O}]$ is trivial and fractionalization is classified by $H^2(\mathbb{Z}_2, \mathbb{Z}_2 \times \mathbb{Z}_2)=\mathbb{Z}_2^{2}$. These classes can be labeled as $\coho{w}({\bf 1}, {\bf 1})=I$, $e$, $m$, and $\psi$. Since $H^3(\mathbb{Z}_2, \text{U}(1))=\mathbb{Z}_2$, each fractionalization class has two defectification classes, which we will label with $p=0$ and $1$. Since the symmetry does not permute anyons, we can obtain all the explicit data of the $G$-crossed extensions using the results of Sec.~\ref{sec:no_perm}. The fusion rules of the extended theories are given by
\begin{equation}
a_{\bf g} \times b_{\bf h} = [ab \coho{w}({\bf g} , {\bf h})]_{\bf gh}
.
\end{equation}
The $G$-crossed data greatly simplifies for this example to
\begin{eqnarray}
F^{a_{\bf g} b_{\bf h} c_{\bf k}} &=& F^{ I_{\bf g} I_{\bf h} I_{\bf k} } (R^{ \cohosub{w}({\bf g}, {\bf h})_{\bf 0} c_{\bf 0}  })^{-1} \\
R^{a_{\bf g} b_{\bf h} } &=& R^{a_{\bf 0} b_{\bf 0}} \\
U_{\bf k} (a_{\bf g}, b_{\bf h}) &=& ( F^{ I_{\bf g} I_{\bf h} I_{\bf k} } )^{-1} \\
\eta_{c_{\bf k}}({\bf g} , {\bf h}) &=&  M_{c_{\bf 0} \cohosub{w}({\bf g} , {\bf h})_{\bf 0}} ( F^{ I_{\bf g} I_{\bf h} I_{\bf k} } )^{-1}
\end{eqnarray}
where the defectification class enters these expressions through
\begin{equation}
F^{ I_{\bf 1} I_{\bf 1} I_{\bf 1} } = \left\{
\begin{array}{ll}
(-1)^{p} &  \text{ for } \coho{w}({\bf 1}, {\bf 1})=I,e,m \\
-i(-1)^{p} &  \text{ for } \coho{w}({\bf 1}, {\bf 1})=\psi
\end{array}
\right.
\end{equation}
which follows from $\mathcal{O}({\bf 1},{\bf 1},{\bf 1},{\bf 1}) = 1$ for $\coho{w}({\bf 1}, {\bf 1})=I$, $e$, and $m$, and $\mathcal{O}({\bf 1},{\bf 1},{\bf 1},{\bf 1}) = -1$ for $\coho{w}({\bf 1}, {\bf 1})=\psi$. We also find the $G$-crossed modular invariants
\begin{equation}
[\mathcal{T}^{2}]^{({\bf 1},{\bf 1})}_{a_{\bf g} a_{\bf g}} = \left\{
\begin{array}{ll}
(-1)^{p} M_{a_{\bf 0} \cohosub{w}({\bf 1}, {\bf 1})} &  \text{ for } \coho{w}({\bf 1}, {\bf 1})=I,e,m \\
i(-1)^{p} M_{a_{\bf 0} \cohosub{w}({\bf 1}, {\bf 1})} &  \text{ for } \coho{w}({\bf 1}, {\bf 1})=\psi
\end{array}
\right.
\end{equation}

When $\coho{w}({\bf 1}, {\bf 1})=I$, $e$, or $m$, the two defectification classes describe distinct $G$-crossed MTCs for each fractionalization class. However, for the symmetry fractionalization class with $\coho{w}(\mb{1},\mb{1})=\psi$, the two defectification classes actually describe the same $G$-crossed MTC. This can be seen by relabeling the defects and applying a gauge transformation in the following manner
\begin{eqnarray}
a'_{\bf 1} &=& [ea]_{\bf 1} \\
\Gamma^{a'_{\bf 1} b_{\bf 0}} = \Gamma^{a'_{\bf 1} b'_{\bf 1}} &=& (-1)^{b_{\text{m}}} \\
\gamma_{a_{\bf 0}}({\bf 1}) = \gamma_{a'_{\bf 1}}({\bf 1}) &=& (-1)^{a_{\text{m}}}
\end{eqnarray}
This transformation leaves the fractionalization class $\coho{w}(\mb{1},\mb{1})=\psi$ unchanged, and results in the same data, except with $p$ replaced by $p'= p+1$. Thus, the two $G$-crossed extensions $p=0$ and $1$ with this symmetry fractionalization class are actually equivalent under relabeling of the defect topological charges, and therefore represent the same SET order. This phenomena was observed in \Refs{Metlitski_unpub, lu2013} using a Chern-Simons field theory approach.

\subsubsection{Electric-magnetic duality symmetry}

When the symmetry action interchanges the $e$ and $m$ quasiparticle types, we can show that the symmetry fractionalization obstruction $\coho{O}$ can be set to $I$ identically. In particular, we can choose
\begin{equation}
\label{eq:TC_em_U}
U_{\bf 1}(a , b ; a\times b) = \frac{\sqrt{\theta_{a}} \sqrt{\theta_{b}}}{\sqrt{\theta_{a\times b}}} (-1)^{a_{\text{m}} b_{\text{e}}} ,
\end{equation}
where we let $\sqrt{\theta_{a}}=1$ for $a=I$, $e$, and $m$, and $\sqrt{\theta_{\psi}}=i$. (While this is, perhaps, not a natural choice for $U_{\bf 1}$, it is what we find for the gauge choices we make in solving for the data of the $G$-crossed theory.) With this $U_{\bf 1}$, we have $\kappa_{{\bf g},{\bf h}}(a,b) = 1$, and so can choose $\beta_{a}({\bf g},{\bf h}) = 1$, from which it follows that $\coho{O} =I$. The cocycle condition simplifies to $^\mathbf{1}\coho{w}(\mathbf{1},\mathbf{1})=\coho{w}(\mathbf{1},\mathbf{1})$. Thus, $\coho{w}(\mb{1},\mb{1}) = I$ or $\psi$. However, these are also coboundaries under the symmetry action, since $\psi = e \times m=\rho_\mb{1}(m)\times m$. It follows that $H^2_{[\rho]}(\mathbb{Z}_2, \mathbb{Z}_2\times\mathbb{Z}_2)=\mathbb{Z}_{1}$, so there is exactly one symmetry fractionalization class.  Since $H^3(\mathbb{Z}_2, \text{U}(1))=\mathbb{Z}_2$, we expect to find two defectification classes of $G$-crossed extensions.

The quasiparticle charges that are fixed under e-m duality are $\mathcal{C}_{\bf 0}^{\bf 1} = \{I, \psi\}$, so there are two species of twist defects, which we label as $\mathcal{C}_{\bf 1} = \{ \sigma^{+}, \sigma^{-} \}$. These defect charges differ by fusion with an $e$ or $m$ charge, and have the fusion rules:
\begin{eqnarray}
\sigma^{\mp} &=& e \times \sigma^{\pm} = m \times \sigma^{\pm}, \\
\sigma^{\pm} &=& \psi \times \sigma^{\pm} , \\
\sigma^{\pm} \times \sigma^{\pm} &=& I+\psi , \\
\sigma^{\pm} \times \sigma^{\mp} &=& e+m .
\end{eqnarray}
The $\mathbb{Z}_2$-symmetry action on the defects is necessarily trivial, i.e. $\rho_\mb{g}(a_{\bf 1})=a_{\bf 1}$. The fusion rules indicate that the quantum dimensions of the defects are $d_{\sigma^{s}} = \sqrt{2}$.

The fusion rules are that of an $\mathbb{Z}_{2} \times \text{Ising}$ FTC, where the $\mathbb{Z}_{2}$ fusion category here can be generated by either the $e$ or the $m$ topological charge. As such, we know~\cite{Bonderson07b} the $F$-symbols must be gauge equivalent to those of the product of the $\mathbb{Z}_{2}$ FTC with trivial $F$-symbols and one of the two Ising-type FTCs, which are distinguished by the Frobenius-Schur indicator of the $\sigma$ charge $\varkappa_{\sigma} = \pm 1$. Whether the $\mathbb{Z}_{2}$ FTC in this product is generated by $e$ or $m$ can be changed by a vertex basis gauge transformation. However, such a gauge transformation also changes the $R$-symbols of the quasiparticles.~\footnote{In order to see this, we notice that we need the vertex basis gauge transformation to leave $F^{\sigma^{+} e m}_{\sigma^{+}} = F^{\sigma^{-} m e}_{\sigma^{-}} =1$ fixed, while changing $F^{e \sigma^{+} \psi }_{\sigma^{-}}$ from $+1$ to $-1$, or vice-versa. Combining these conditions requires $\Gamma^{e m} / \Gamma^{m e} =-1$, which changes $R^{e m} = - R^{m e}$ from $-1$ to $+1$, or vice-versa.} This means that the defect $F$-symbols are not simply obtained by solving the pentagon equations and using the vertex gauge freedom, but are also constrained by the quasiparticle braiding through the heptagon equations. For our choice of quasiparticle data, the consistent choice of $F$-symbols corresponds to choosing the $\mathbb{Z}_{2}$ FTC to be generated by $e$. In this way, the nontrivial $F$-symbols of the extended category are given by
\begin{eqnarray}
F^{a_{\bf 0} b_{\bf 1} c_{\bf 0}}_{d_{\bf 1}} &=& (-1)^{a_{\text{m}} c_{\text{m}}}, \\
F^{a_{\bf 1} b_{\bf 0} c_{\bf 1}}_{d_{\bf 0}} &=& (-1)^{b_{\text{m}} d_{\text{m}}}, \\
\left[F^{a_{\bf 1} b_{\bf 1} c_{\bf 1}}_{d_{\bf 1}}\right]_{e_{\bf 0} f_{\bf 0}} &=& \frac{\varkappa_{\sigma}}{\sqrt{2}}(-1)^{e_{\text{m}} f_{\text{m}}},
\end{eqnarray}
where $e_{\bf 0}$ and $f_{\bf 0}$ take values in either $\{I,\psi \}$ or $\{ e,m \}$, depending on the values of $a_{\bf 1}$, $b_{\bf 1}$, and $c_{\bf 1}$. The remaining $F$-symbols allowed by fusion are equal to $1$.

We next solve the heptagon conditions, with appropriate symmetry action gauge choices, to obtain the $R$-symbols
\begin{eqnarray}
R^{e \sigma^{\pm}} &=& R^{m \sigma^{\pm}} = 1, \\
R^{\psi \sigma^{\pm}} &=& i , \\
R^{\sigma^{\pm} e} &=& s_{e}, \\
R^{\sigma^{\pm} m} &=& \pm i s_{e} s_{\psi}, \\
R^{\sigma^{\pm} \psi} &=& \pm i s_{\psi}, \\
R^{\sigma^{\pm} \sigma^{\pm}}_{I} &=& s_{e} R^{\sigma^{\pm} \sigma^{\mp}}_{e}  = \left( \sqrt{\varkappa_{\sigma}} e^{i \frac{\pi}{8} s_{\psi}} \right)^{\pm 1}, \\
R^{\sigma^{\pm} \sigma^{\pm}}_{\psi} &=& s_{e} R^{\sigma^{\pm} \sigma^{\mp}}_{m} = \left( \sqrt{\varkappa_{\sigma}} e^{i \frac{\pi}{8} s_{\psi}} \right)^{\mp 3},
\end{eqnarray}
where $s_{e}, s_{\psi} \in \{ 1, -1 \}$, and we let $\sqrt{\varkappa_{\sigma}} = 1$ and $i$, for $\varkappa_{\sigma} = 1$ and $-1$, respectively, the $\eta$-symbols
\begin{equation}
\eta_{a_{\bf k}}({\bf g},{\bf h}) =1,
\end{equation}
and the $U_{\bf k}$-symbols [in addition to Eq.~(\ref{eq:TC_em_U})]
\begin{eqnarray}
U_{\bf 1}(\sigma^{\pm} , e ; \sigma^{\mp} ) &=& U_{\bf 1}(\sigma^{\mp} , m ; \sigma^{\pm}) = \varkappa_{\sigma} s_{e} e^{\mp i \frac{\pi}{4} s_{\psi}} , \quad\\
U_{\bf 1}(e, \sigma^{\pm} ; \sigma^{\mp}) &=& U_{\bf 1}(m, \sigma^{\mp} ; \sigma^{\pm}) = \varkappa_{\sigma} s_{e} e^{\pm i \frac{\pi}{4} s_{\psi}} , \\
U_{\bf 1}(\sigma^{\pm} ,\psi ; \sigma^{\pm} ) &=& U_{\bf 1}(\psi, \sigma^{\pm} ; \sigma^{\pm}) = \pm s_{\psi} , \\
U_{\bf 1}(\sigma^{\pm} , \sigma^{\pm} ; I) &=& 1 , \\
U_{\bf 1}(\sigma^{\pm} , \sigma^{\pm} ; \psi) &=& \pm s_{\psi} , \\
U_{\bf 1}(\sigma^{\pm} , \sigma^{\mp} ; e) &=& \varkappa_{\sigma} s_{e} e^{\pm i \frac{\pi}{4} s_{\psi}} , \\
U_{\bf 1}(\sigma^{\pm} , \sigma^{\mp} ; m) &=& \varkappa_{\sigma} s_{e} e^{\mp i \frac{\pi}{4} s_{\psi}}
.
\end{eqnarray}

The $G$-crossed modular $\mathcal{S}$ and $\mathcal{T}$ matrices are given (for this gauge choice) by
\begin{equation}
\mathcal{S}=
\left[
	\begin{array}{c|cc|cc|cc}
	  \mathcal{S}^{({\bf 0},{\bf 0})} & 0 & 0 & 0 & 0 & 0 & 0\\
	\hline
	0 & 0 & 0 & \frac{1}{\sqrt{2}} &  \frac{1}{\sqrt{2}} & 0 & 0\\
	 0 & 0 & 0 & -\frac{s_\psi}{\sqrt{2}}  &  \frac{s_\psi}{\sqrt{2}}  & 0 & 0\\
	\hline
	0 & \frac{1}{\sqrt{2}} & -\frac{s_\psi}{\sqrt{2}}  & 0 & 0 & 0 & 0\\
	 0 & \frac{1}{\sqrt{2}}  & \frac{s_\psi}{\sqrt{2}}  & 0 & 0 & 0 & 0\\
	\hline
	0 & 0 & 0 & 0 & 0 & 0 & 1\\
	0 & 0 & 0 & 0 & 0 & 1 & 0
	\end{array}
	\right],
\end{equation}
\begin{equation}
\mathcal{T}=\left[
	\begin{array}{c|cc|cc|cc}
	\mathcal{T}^{({\bf 0},{\bf 0})} & 0 & 0 & 0 & 0 & 0 & 0\\
	\hline
	0 & 1 & 0 & 0 & 0 & 0 & 0\\
	0 & 0 & -1 & 0  & 0 & 0 & 0\\
	\hline
	0 & 0 & 0 & 0 & 0 & \theta_{\sigma^{+}} & 0\\
	0 & 0 & 0 & 0 & 0 & 0 & \theta_{\sigma^{-}}\\
	\hline
	0 & 0 & 0 & \theta_{\sigma^{+}} & 0 & 0 & 0 \\
	0 & 0 & 0 & 0 & \theta_{\sigma^{-}} & 0 & 0
	\end{array}
	\right].
\end{equation}
where the basis states of the defect sectors are chosen to be $\ket{I^{(\mb{0,1})}}$, $\ket{\psi^{(\mb{0,1})}}$, $\ket{\sigma^{+(\mb{1,0})}}$, $\ket{\sigma^{-(\mb{1,0})}}$, $\ket{\sigma^{+(\mb{1,1})}}$, and $\ket{\sigma^{-(\mb{1,1})}}$, in that order, and the topological twists of the defects are
\begin{equation}
\theta_{\sigma^{\pm}} = \left(\sqrt{\varkappa_{\sigma}}e^{-i\frac{\pi}{8}s_{\psi}}\right)^{\pm 1}
.
\end{equation}

We notice that the choices of $s_{e}$ and $s_{\psi}$ are actually redundant, though they could not be removed simply by applying a symmetry action gauge transformation. For $s_{e}=-1$, if we apply a vertex basis gauge transformation with nontrivial elements
\begin{eqnarray}
-1 &=& \Gamma^{\sigma^{\pm} \psi}_{\sigma^{\pm}} = \Gamma^{\psi \sigma^{\pm}}_{\sigma^{\pm}} = \Gamma^{e \sigma^{\pm}}_{\sigma^{\mp}} = \Gamma^{\sigma^{\pm} m}_{\sigma^{\mp}} \notag \\
&=& \Gamma^{\sigma^{-} \sigma^{+}}_{e} = \Gamma^{\sigma^{+} \sigma^{-}}_{m} = \Gamma^{\sigma^{-} \sigma^{-}}_{I} = \Gamma^{\sigma^{+} \sigma^{+}}_{\psi}
,
\end{eqnarray}
the basic data becomes that of the theory with $s_{e} = 1$. For $s_{\psi}=-1$, if we relabel the defect charges $\sigma^{\pm \prime} = \sigma^{\mp} = e \times \sigma^{\pm}$ and apply a symmetry action gauge transformation with nontrivial element $\gamma_{\sigma^{\pm}}({\bf 1}) = \varkappa_{\sigma}$, we obtain the theory with $s_{\psi}=1$.

Thus, there are two distinct $G$-crossed extensions for the e-m duality $\mathbb{Z}_{2}$ symmetry, which are distinguished by the defects' Frobenius-Schur indicator $\varkappa_{\sigma} = \pm 1$ (and the associated changes in the basic data). This matches the expected (torsorial) classification by $H^{3}(\mathbb{Z}_{2},\text{U}(1))=\mathbb{Z}_{2}$. For the gauge choice used here, the torsorial $\mathbb{Z}_{2}$ action relating these distinct $G$-crossed theories is given by gluing in a $\mathbb{Z}_{2}$ SPT state with $\alpha({\bf 1},{\bf 1},{\bf 1})=-1$, $R^{{\bf 1},{\bf 1}}=-i$, $\eta_{\bf 1}({\bf 1},{\bf 1})=1$, and $U_{\bf 1}({\bf 1},{\bf 1})=1$ (as in Sec.~\ref{sec:Z_N_SPT}) and applying a symmetry action gauge transformation whose nontrivial element is $\gamma_{\sigma^{\pm}}({\bf 1})=\mp \varkappa_{\sigma}$ (here, $\varkappa_{\sigma}$ is that of the theory before gluing in the $\mathbb{Z}_{2}$ SPT state).

We now consider the gauged theory, setting $s_{e} = s_{\psi}=1$ to remove the redundancy. All $\mathbb{Z}_{2}$ orbits of topological charges in the defect theory are singletons, except $[e]=\{e, m\}$. The stabilizer subgroup of the singletons is $\mathbb{Z}_{2}$ and that of $[e]$ is $\mathbb{Z}_{1}$. Since the $\eta$-symbols were chosen to be trivial, each singleton can carry an irrep of $\mathbb{Z}_{2}$, which we label $q = \pm$. Thus, the gauged theory has nine topological charges, labeled by $(I,q)$, $(\psi, q)$, $(\sigma^{+}, q)$, $(\sigma^{-}, q)$, $[e]$. The quantum dimensions are $d_{(I,q)}= d_{(\psi,q)} =1$, $d_{(\sigma^{\pm}, q)}=\sqrt{2}$, and $d_{[e]}=2$.

A straightforward application of \eqref{eqn:fusion} yields the fusion rules
\begin{eqnarray}
(I, q)\times (b, q') &=& (b, qq') , \\
(I, q)\times [e] &=& [e] , \\
(\psi, q)\times (\psi , q') &=& (I , qq') , \\
(\psi, q)\times (\sigma^{\pm}, q') &=& (\sigma^{\pm}, \pm qq') , \\
(\psi, q)\times [e] &=& [e] , \\
(\sigma^{\pm}, q) \times (\sigma^{\pm}, q')&=&(I,qq') + (\psi, \pm qq') ,\\
(\sigma^{\pm}, q)\times (\sigma^{\mp}, q')&=& [e] ,\\
(\sigma^{\pm}, q) \times [e] &=&(\sigma^{\mp}, +) + (\sigma^{\mp}, -) , \\
{}[e] \times [e] &=& (I,+) + (I,-) \notag \\
&& \quad + (\psi, +) + (\psi, -)
,
\end{eqnarray}
for any singleton $b$. We notice that these fusion rules are identical to that of a direct product of two $\text{Ising}^{(\nu)}$ MTCs, labeled L and R, once we identify the topological charges as
\begin{equation}
\begin{gathered}
(I_{\rm L} ,I_{\rm R}) = (I,+), \, (I_{\rm L} ,\sigma_{\rm R}) = (\sigma^{+}, +) , \, (I_{\rm L} ,\psi_{\rm R}) = (\psi,+), \\
(\sigma_{\rm L} ,I_{\rm R}) = (\sigma^{-}, +), \,\, (\sigma_{\rm L} ,\sigma_{\rm R}) = [e] , \,\, (\sigma_{\rm L} ,\psi_{\rm R}) = (\sigma^{-}, -), \\
(\psi_{\rm L} ,I_{\rm R}) = (\psi,-), \, (\psi_{\rm L} ,\sigma_{\rm R}) = (\sigma^{+}, -) , \, (\psi_{\rm L} ,\psi_{\rm R}) = (I,-) .
	\end{gathered}
\end{equation}

The topological twists are straightforward to compute using \eqref{eqn:twist}, which yields
\begin{eqnarray}
\theta_{(I, q)} = \theta_{[e]} &=& 1 , \\
\theta_{(\psi, q)} &=& -1 , \\
\theta_{(\sigma^{\pm}, q)} &=& q \left(\sqrt{\varkappa_{\sigma}}e^{-i\frac{\pi}{8}}\right)^{\pm 1}
.
\end{eqnarray}
The $S$-matrix is found to be
\begin{equation}
S_{(a_{\rm L},a_{\rm R} )(b_{\rm L},b_{\rm R} )} = S_{a_{\rm L} b_{\rm L}}^{\text{Ising}} S_{a_{\rm R} b_{\rm R} }^{\text{Ising}},
\end{equation}
where $S^{\text{Ising}}$ is the $S$-matrix of the Ising theory. This allows us to uniquely identify the gauged theory as $\text{Ising}\boxtimes \overline{\text{Ising}}$ for $\varkappa_{\sigma}=1$ and $\overline{\text{SU}(2)_{2}} \boxtimes \text{SU}(2)_{2}$ for $\varkappa_{\sigma}=-1$. (Recall that $\text{Ising} = \text{Ising}^{(1)}$ and $\text{SU}(2)_{2} = \text{Ising}^{(3)}$.)

As previously mentioned, the electric-magnetic duality in the $\mathbb{Z}_2$-toric code can be realized as an on-site symmetry. We now briefly describe a concrete model for doing so. We start from a spin-$1/2$ fermionic superconductor with the pairing $(p_{x}+ip_{y})^\nu_\uparrow \times (p_{x}-ip_{y})^\nu_\downarrow$ where $\nu$ is an odd integer. This is a model of the $\mathbb{Z}_2$ fermionic SPT phase discussed in Sec.~\ref{sec:fermions}. Next, we gauge the $\mathbb{Z}_2$ fermion parity of the whole system, i.e. coupling all fermions to a $\mathbb{Z}_2$ gauge field, and we obtain a $\mathbb{Z}_2$-toric code, where $m$ is the $\pi$ flux in the original superconductor and $e$ is the bound state of the $\pi$ flux and a fermion. The $\mathbb{Z}_2$ symmetry that protects the SPT phase, namely the fermion parity of the spin $\uparrow$ fermions, now becomes the e-m duality symmetry of the toric code. To see this, we first notice that  before the $\mathbb{Z}_2$ total fermion parity is gauged, a $\pi$ flux localizes two Majorana zero modes $\gamma_\uparrow$ and $\gamma_\downarrow$, since it penetrates two $p_{x} \pm ip_{y}$ superconductors. Under the $\mathbb{Z}_2$ symmetry $\gamma_\uparrow\rightarrow -\gamma_\uparrow, \gamma_\downarrow\rightarrow \gamma_\downarrow$, so the local fermion parity $i\gamma_\uparrow\gamma_\downarrow$ on the $\pi$ flux changes sign under the on-site $\mathbb{Z}_2$ symmetry, which  interchanges $e$ and $m$ after the total fermion parity is fully gauged. This provides the desired on-site realization. We can turn this model of a fermionic superconductor coupled to a $\mathbb{Z}_2$ gauge field into a Kitaev-type spin model.

In this model, gauging the $\mathbb{Z}_2$ symmetry becomes particularly easy: we simply gauge the fermion parities of the spin $\uparrow$ and $\downarrow$ fermions separately, and the result is precisely $\text{Ising}^{(\nu)}\boxtimes \ol{\text{Ising}^{(\nu)}}$. However, $\nu$ and $\nu+8$, as well as $\nu$ and $-\nu$, lead to exactly the same topological gauge theories, so we only obtain two distinct gauge theories corresponding to $\nu=1$ and $3$, in agreement with our previous analysis.

We note in passing that all of the Abelian MTCs in Kitaev's 16-fold way~\cite{Kitaev06a}, which correspond to even-valued $\nu$ in his notation, similarly have a $\mathbb{Z}_{2}$ topological symmetry that interchanges the ``vortex'' type quasiparticles, which have topological twists $\theta_{a} = e^{i \frac{\pi }{8} \nu}$. The corresponding $\mathbb{Z}_{2}$-crossed extensions and gauged theories of these MTCs with such a symmetry are very similar to those of D$(\mathbb{Z}_{2})$ (which is $\nu=0$) with electric-magnetic duality symmetry. The ones with $\mathbb{Z}_{2} \times \mathbb{Z}_{2}$ fusion rules, i.e. $\nu =0$, $4$, $8$, and $12$, will have two distinct $\mathbb{Z}_{2}$-crossed extensions, since they all have $H^2_{[\rho]}(\mathbb{Z}_2, \mathbb{Z}_{2} \times \mathbb{Z}_{2})=\mathbb{Z}_{1}$ and $H^3(\mathbb{Z}_2, \text{U}(1))=\mathbb{Z}_2$. The ones with $\mathbb{Z}_{4}$ fusion rules, i.e. $\nu =2$, $6$, $10$, and $14$, will have three distinct $\mathbb{Z}_{2}$-crossed extensions; there are two fractionalization classes, since $H^2_{[\rho]}(\mathbb{Z}_2, \mathbb{Z}_{4})=\mathbb{Z}_{2}$, and while $H^3(\mathbb{Z}_2, \text{U}(1))=\mathbb{Z}_2$ would na\"ively indicate each of these should have two corresponding defectification classes, two of them are identified for one of the fractionalization classes (see discussion in Sec.~\ref{sec:N_multiple4}). The corresponding gauged theories are all given by $\text{Ising}^{(\nu_{1})} \boxtimes \text{Ising}^{(\nu_{2})}$, where $\nu_{1}$ and $\nu_{2}$ are odd and satisfy $\nu_{1}+\nu_{2} = \nu \mod 16$. In the case of the three-fermion model SO$(8)_1$, which is $\nu=8$, the topological symmetry group is $S_{3}$, containing three such $\mathbb{Z}_{2}$ symmetries. We examine this example in detail in Sec.~\ref{sec:3-Fermion}.

\subsection{$\mathbb{Z}_N$-Toric Code D$(\mathbb{Z}_N)$ with $N>2$ and $\mathbb{Z}_2$ Symmetry}

In this section, we consider the $\mathrm{D}(\mathbb{Z}_N)$ UMTC, which corresponds to a $\mathbb{Z}_N$ discrete gauge theory or quantum double. Physically it can be realized by the $\mathbb{Z}_N$ generalization of Kitaev's toric code model~\cite{Kitaev03}, or as $\mathbb{Z}_N$ lattice gauge theory. The anyons are gauge charges (the unit of which is denoted by $e$), gauge fluxes (the unit of which is denoted by $m$) and their bound states, the dyons. We write the $N^2$ anyon labels as $a=(a_{\text{e}},a_{\text{m}}) \equiv e^{a_{\text{e}}} m^{a_{\text{m}}}$ where $a_{\text{e}},a_{\text{m}} \in \left\{0, 1, \ldots, N-1 \right\}$. The fusion rules are given by
\begin{equation}
(a_{\text{e}},a_{\text{m}}) \times (b_{\text{e}},b_{\text{m}}) = ([a_{\text{e}} + b_{\text{e}}]_{N}, [a_{\text{m}}+b_{\text{m}}]_{N})
,
\end{equation}
that is, they form a $\mathbb{Z}_N\times\mathbb{Z}_N$ fusion algebra. In a choice of gauge, the $F$-symbols of the theory are all trivial and the $R$-symbols are given by
\begin{equation}
R^{ab}=e^{i \frac{2\pi}{N} a_{\text{e}} b_{\text{m}}}
.
\end{equation}

The topological symmetry group $\mathrm{Aut}(\mathrm{D}(\mathbb{Z}_N))$ is complicated to determine for general $N>2$, but it always contains at least a $\mathbb{Z}_{2} \times \mathbb{Z}_{2}$ subgroup that is generated by electric-magnetic duality symmetry and topological charge conjugation symmetry. Electric-magnetic duality symmetry corresponds to the autoequivalence map for which $(a_{\text{e}},a_{\text{m}}) \mapsto (a_{\text{m}},a_{\text{e}})$. Topological charge conjugation symmetry corresponds to the autoequivalence map for which $a \mapsto \bar{a} = (N-a_{\text{m}},N-a_{\text{e}})$. Electric-magnetic duality symmetry can be realized as lattice translations~\cite{bombin2010, you2012} in Wen's plaquette model formulation of the toric code~\cite{Wen2003}. Alternatively, one can realize this symmetry of a $\mathbb{Z}_N$-toric code in an on-site fashion~\cite{cheng_setmodel}. Topological charge conjugation symmetry is straightforward to realize in the $\mathbb{Z}_N$-toric code model in an on-site fashion.

For the first few $N$, we have $\mathrm{Aut}(\mathrm{D}(\mathbb{Z}_3)) = \mathbb{Z}_{2}^{2}$, $\mathrm{Aut}(\mathrm{D}(\mathbb{Z}_4)) = \mathbb{Z}_{2}^{2}$, and $\mathrm{Aut}(\mathrm{D}(\mathbb{Z}_5)) = D_{8}$ (the dihedral group containing eight elements).

We consider the case of global symmetry group $G=\mathbb{Z}_{2}$. When the global symmetry acts either trivially or as topological charge conjugation, we can choose $U_{\bf 1}(a , b ; a\times b) =1$, so it is clear that $[\coho{O}]=0$ and symmetry fractionalization is not obstructed. For both of these symmetry actions, the corresponding symmetry fractionalization is classified by $H^2_{[\rho]}(\mathbb{Z}_2, \mathbb{Z}_N\times\mathbb{Z}_N)=\mathbb{Z}_{1}$ for $N$ odd and $H^2_{[\rho]}(\mathbb{Z}_2, \mathbb{Z}_N\times\mathbb{Z}_N)=\mathbb{Z}_{2} \times \mathbb{Z}_{2}$ for $N$ even, which follows from computations similar to those of Sec.~\ref{sec:Z_N}.

When the global $G=\mathbb{Z}_{2}$ symmetry acts as electric-magnetic duality, we can choose $U_{\bf 1}(a , b ; a\times b) = e^{i \frac{2 \pi}{N}  a_{\text{e}} b_{\text{m}}}$ and $\beta_{a}({\bf 1},{\bf 1})=\theta_{a}$, from which it follows that $[\coho{O}]=0$ and symmetry fractionalization is not obstructed. In this case, we have $H^2_{[\rho]}(\mathbb{Z}_2, \mathbb{Z}_N\times\mathbb{Z}_N)=\mathbb{Z}_{1}$ for any $N$, since the cocycle condition reduces to $^\mathbf{1}\coho{w}(\mathbf{1},\mathbf{1})=\coho{w}(\mathbf{1},\mathbf{1})$, which is satisfied iff $\coho{w}(\mb{1},\mb{1}) = (w,w)$ and coboundaries can take the same form ${\rm d}\coho{z}(\mb{1},\mb{1})=(w,w)$ by taking $\coho{z}(\mb{1}) = (w,0)$.

Since $H^3(\mathbb{Z}_2, \text{U}(1))=\mathbb{Z}_2$, we expect to find two defectification classes of $G$-crossed extensions for a given fractionalization class. It is significantly easier to solve for the $G$-crossed extensions when $N$ is odd, than when $N$ even, so we will only do so for $N$ odd here.

\subsubsection{$N$ odd}

In the case of $N$ odd, the problem simplifies because the $\mathcal{C}_{\bf 0}$ MTC can be written as $\mathrm{D}(\mathbb{Z}_N)=\mathbb{Z}_N^{(1)}\boxtimes\mathbb{Z}_N^{(-1)}$. More precisely, writing the topological charges as
\begin{equation}
a = (a_{\text{e}},a_{\text{m}}) = \widetilde{(a_{+},a_{-})}
,
\end{equation}
where $a_{\pm} \in \mathbb{Z}_N^{(\pm 1)}$, respectively, we have the relation between charge-flux representation and the $\pm$ chiralities representation given by
\begin{eqnarray}
a_{\pm} &=& \left[ \frac{N+1}{2}( a_{\text{e}} \pm a_{\text{m}} ) \right]_{N} ,\\
a_{\text{e}} &=& \left[ a_{+} + a_{-} \right]_{N} , \\
a_{\text{m}} &=& \left[ a_{+} - a_{-} \right]_{N}
.
\end{eqnarray}
The basic data can be transformed into the product form by the vertex basis gauge transformation $\Gamma^{ab} =e^{i \frac{2\pi}{N} a_{-} b_{+}}$, which leaves the $F$-symbols trivial and gives $\widetilde{R}^{ab}=e^{i \frac{2\pi}{N} (a_{+} b_{+} - a_{-} b_{-})}$.

In this form, it is easy to see that $\mathrm{Aut}(\mathbb{Z}_N^{(1)})\times\mathrm{Aut}(\mathbb{Z}_N^{(-1)}) = \mathbb{Z}_{2}^{2k}$ is a subgroup of $\mathrm{Aut}(\mathrm{D}(\mathbb{Z}_N))$ for $N$ odd with prime factorization $N = P_{1}^{r_{1}}\dots P_{k}^{r_{k}}$ (see Sec.~\ref{sec:Z_N}). In the $\pm$ chiralities representation, electric-magnetic duality acts as $\widetilde{(a_{+},a_{-})} \mapsto \widetilde{(a_{+},\overline{a_{-}})}$, i.e. it acts trivially on the $\mathbb{Z}_N^{(1)}$ sector and acts as topological charge conjugation on the $\mathbb{Z}_N^{(-1)}$ sector. Topological charge conjugation acts as topological charge conjugation on both sectors, i.e. $\widetilde{(a_{+},a_{-})} \mapsto \widetilde{(\overline{a_{+}},\overline{a_{-}})}$.

When the global $G$ symmetry action factorizes into $\pm$ sectors, i.e. when $\rho  : G \rightarrow \mathrm{Aut}(\mathbb{Z}_N^{(1)})\times\mathrm{Aut}(\mathbb{Z}_N^{(-1)})$, we write  $\rho = (\rho_{+} , \rho_{-})$. In this case, the symmetry fractionalization also factorizes, i.e. $H^2_{[\rho]}(G, \mathbb{Z}_N\times\mathbb{Z}_N)=H^2_{[\rho_{+}]}(G, \mathbb{Z}_N) \times H^2_{[\rho_{-}]}(G, \mathbb{Z}_N)$ and the $G$-crossed extensions can be obtained by gluing $G$-crossed extensions of the $\mathbb{Z}_N^{(\pm 1)}$ theories (see Appendix~\ref{sec:gluing}), that is
\begin{equation}
\mathrm{D}(\mathbb{Z}_N)_{G^{\rho}}^{\times} = \left[\mathbb{Z}_N^{(1)}\right]_{G^{\rho_{\scalebox{0.7}{+}}}}^{\times} \underset{G}{\boxtimes} \left[\mathbb{Z}_N^{(-1)}\right]_{G^{\rho_{\scalebox{0.7}{-}}}}^{\times}
.
\end{equation}

For the case of $G=\mathbb{Z}_{2}$ electric-magnetic duality symmetry, there is exactly one symmetry fractionalization class, i.e. $H^2_{[\rho]}(G, \mathbb{Z}_N\times\mathbb{Z}_N)=\mathbb{Z}_{1}$, and the $G$-crossed extensions can be written as
\begin{equation}
\mathrm{D}(\mathbb{Z}_N)_{\mathbb{Z}_{2}^{\text{e-m}}}^{\times} = \mathbb{Z}_{N}^{(1)} \boxtimes \left[\mathbb{Z}_N^{(-1)}\right]_{\mathbb{Z}_{2}^{\text{cc}}}^{\times}
,
\end{equation}
where $\left[\mathbb{Z}_N^{(-1)}\right]_{\mathbb{Z}_{2}^{\text{cc}}}^{\times}$ is one of the two possible $G$-crossed theories obtained in Sec.~\ref{sec:Z_N_odd_cc}. The corresponding gauged theories are, thus
\begin{equation}
\left[\mathrm{D}(\mathbb{Z}_N)_{\mathbb{Z}_{2}^{\text{e-m}}}^{\times}\right]^{\mathbb{Z}_{2}} = \mathbb{Z}_{N}^{(1)} \boxtimes \left(\left[\mathbb{Z}_N^{(-1)}\right]_{\mathbb{Z}_{2}^{\text{cc}}}^{\times}\right)^{\mathbb{Z}_{2}}
.
\end{equation}

\subsection{Double-Layer Systems $\mathcal{B} \boxtimes \mathcal{B}$ with $\mathbb{Z}_2$ Symmetry}
\label{sec:genons}

In this section, we consider a system composed of two identical, non-interacting layers of a topological phase described by the UMTC $\mathcal{B}$, with global $\mathbb{Z}_{2}$ layer-interchange symmetry. The topological order of the double-layer system is described by $\mathcal{C} = \mathcal{B}\boxtimes\mathcal{B}$. We write the quasiparticle topological charges as $\vec{a} = (a_1,a_2)$, for $a_j \in \mathcal{B}$. The topological symmetry group $\mathrm{Aut}(\mathcal{C})$ can generally be complicated, depending on $\mathcal{B}$, but it always contain a $\mathbb{Z}_2$ subgroup corresponding to layer interchange, i.e. $(a_1,a_2) \mapsto (a_2,a_1)$. If we assume that the basic data ($F$-symbols and $R$-symbols) are in the product form, with each layer's data identical to the other's, then we can choose $[U_{\bf k}(\vec{a},\vec{b};\vec{c})]_{\mu \nu}=\delta_{\mu \nu}$ and $\beta_{\vec{a}}({\bf g},{\bf h})=1$. This yields $\coho{O}=0$, so the symmetry obstruction always vanishes.

We can also generally prove that symmetry fractionalization is trivial, since $H^2_{[\rho]}(\mathbb{Z}_2, \mathcal{A}\times \mathcal{A})=\mathbb{Z}_{1}$, where $\mathcal{A}$ denotes the subcategory of Abelian anyons in $\mathcal{B}$, as follows. The $2$-cocycle condition $^{\bf 1}\coho{w}(\mathbf{1},\mathbf{1})= \coho{w}(\mathbf{1},\mathbf{1})$ constrains the cocycles to take the form $\coho{w}(\mathbf{1},\mathbf{1})=(a,a)$, where $a\in\mathcal{A}$. However, these are all 2-coboundaries, since $(a,a)=\,^\mathbf{1}(a,0)\times (a,0)$. Thus, there is only one symmetry fractionalization class. Since $H^3(\mathbb{Z}_2, \text{U}(1))=\mathbb{Z}_2$, there are two defectification classes.

Quasiparticles are invariant under layer interchange iff they have the form $(a,a)$ for $a \in \mathcal{B}$, so the number of defects is equal to $\mathcal{C}_{\bf 1}=|\mathcal{B}|$. There is a ``bare'' defect $X_0$ which has the following fusion rule
\begin{equation}
X_0\times X_0=\sum_{a\in\mathcal{B}}(a,\bar{a})
\end{equation}
and quantum dimension
\begin{equation}
d_{X_0}=\sqrt{\sum_{a \in \mathcal{B}} d_a^2}=\mathcal{D}_{\mathcal{B}}
.
\end{equation}
The other defects $X_a \in \mathcal{C}_{\bf 1}$ are labeled by $a\in \mathcal{B}$, and related to the bare defect by
\begin{equation}
X_a = X_0\times (a,0)=X_0\times (0,a),
\end{equation}
and, therefore, have $d_{X_a}= d_{a} \mathcal{D}_{\mathcal{B}}$. Given this, the fusion rules of $X_a$ can be easily deduced to be
\begin{eqnarray}
X_a\times(b,c)&=& X_0\times (a,0)\times(b,0)\times (c,0) \notag \\
&=& \sum_e N_{abc}^e X_e , \\
X_a\times X_b &=& \sum_c (c,\bar{c})\times (a,0)\times(b,0) \notag \\
&=&\sum_{c,e} N_{abc}^e (e, \bar{c}) ,
\end{eqnarray}
where we introduced $N_{abc}^e=\mathrm{dim}\, V_{abc}^e=\sum_{z}N_{ab}^z N_{zc}^e$.

We now consider the gauged theory of a given defectification class. For each $a\in \mathcal{B}$, the singleton $\{(a,a)\}$ is an orbit that yields two distinct topological charges in the gauged theory, corresponding to the two irreps of $\mathbb{Z}_{2}$ (trivial and alternating), each with quantum dimensions $d_{a}^{2}$. For each pair $a,b\in\mathcal{B}$, such that $a\neq b$, the doublet $\{(a,b), (b,a)\}$ is an orbit which yields one topological charge in the gauged theory (since its stabilizer subgroup is trivial) with quantum dimension $2d_a d_b$. Each defect $X_{a}$ is a singleton orbit that yields two topological charges, corresponding to the two irreps of $\mathbb{Z}_{2}$, with quantum dimensions $d_{a} \mathcal{D}_{\mathcal{B}}$. Altogether, the gauged theory has $\frac{|\mathcal{B}|(|\mathcal{B}|+7)}{2}$ topological charges.

We can also determine the gauged theory's ground state degeneracy on a genus $g$ surface from the defect theory. The defect theory has $2^{2g}$ different defect sectors, corresponding to whether there is a $\mathbb{Z}_2$ defect branch line along any of the $2g$ independent non-contractible cycles of the surface. The ground state degeneracy of the gauged theory is obtained by symmetrizing under the $\mathbb{Z}_2$ action for each defect sector and adding the resulting degeneracies together. Since $[U_{\bf k}(\vec{a},\vec{b};\vec{c})]_{\mu \nu}=\delta_{\mu \nu}$ for the quasiparticle types, we only need to consider symmetrization of the charge labels, and do not need to worry about states acquiring additional phases due to the symmetry action. Let us denote the ground state degeneracy for the defect sector $s$ by $\mathcal{N}_{g}(s)$. Picking a basis for the sector $s$, we denote the number of basis states that are invariant under the $\mathbb{Z}_2$ global symmetry action by $\mathcal{I}_{g}(s)$. Then the contribution to the gauged theory's ground state degeneracy from the sector $s$ is $\mathcal{N}_{g}^{G}(s) = \mathcal{I}_{g}(s) + \frac{1}{2}[\mathcal{N}_{g}(s) - \mathcal{I}_{g}(s)]$. The trivial defect sector $s=0$ has basis states labeled by charges in $\mathcal{C}_{\bf 0}= \mathcal{B}\boxtimes\mathcal{B}$ and, thus, has ground state degeneracy $\mathcal{N}_{g}(0) = (\mathcal{N}^{\mathcal{B}}_g)^2$, where $\mathcal{N}^\mathcal{B}_g$ is the ground state degeneracy of $\mathcal{B}$ on a genus $g$ surface and invariant basis state degeneracy $\mathcal{I}_{g}(0) = \mathcal{N}^\mathcal{B}_g$ (given by the diagonally labeled states). As discussed in Section~\ref{sec:higher_genus}, when $G = \mathbb{Z}_2$, all nontrivial defect sectors can be mapped onto each other using Dehn twists and, thus, have the same number of ground states and contribution to the gauge theory's ground state degeneracy. As such, it is sufficient to consider the contribution of one nontrivial sector $s\neq 0$, e.g. that with a single defect branch line around a single cycle. This sector has $\mathcal{N}_{g}(s) = \mathcal{N}^{\mathcal{B}}_{2g-1}$ ground state degeneracy, which can be easily seen in the basis with topological flux lines label by charges in $\mathcal{C}_{\bf 0}$, with one cycle's flux line crossing the (layer interchange) $\mathbb{Z}_2$ branch line. The number of invariant basis state is given by $\mathcal{I}_{g}(s) = \mathcal{N}^\mathcal{B}_g$ (given by the diagonally labeled states).
It follows that the gauged theory's ground state degeneracy $\mathcal{N}_g^{G}$ on a genus $g$ surface is given by
\begin{eqnarray}
\mathcal{N}^{G}_g  &=& \sum_{s} \mathcal{N}_{g}^{G}(s) \notag \\
&=& \frac{1}{2} \left[(2^{2g} - 1) \mathcal{N}^{\mathcal{B}}_{2g-1} + (\mathcal{N}^{\mathcal{B}}_g)^{2} + 2^{2g} \mathcal{N}^{\mathcal{B}}_g   \right]
. \quad
\end{eqnarray}
This matches the ground state degeneracy obtained from Eq.~(\ref{eq:genus_g_Verlinde}), using the gauged theory's quantum dimensions.

An interesting example of a layer interchange symmetric double layer system is given by letting $\mathcal{B}=$Fib, the MTC of Fibonnaci anyons. The quantum dimensions of the gauged theory agree exactly with the $\overline{\text{SU}(2)_8}$ theory. One can also check that, starting from $\overline{\text{SU}(2)_8}$ and condensing the highest spin boson ($j=4$) results in $\text{Fib} \boxtimes \text{Fib}$~\cite{bais2009}, so this is, indeed, one of the two gauged theories. The other gauged theory (corresponding to the other defectification class) is the closely related Jones-Kauffman theory at level 8 (JK$_{8}$), see, e.g. \Ref{Levaillant2015} for the basic data.

As another example, consider the case where $\mathcal{B} = \mathrm{D}(H)$, where $\mathcal{B}$ is the quantum double of a discrete group $H$. In this case, the gauged theory is $(\mathcal{B} \boxtimes \mathcal{B})/\mathbb{Z}_2 = \mathrm{D}((H \times H) \rtimes \mathbb{Z}_2)$. For example, when $H = \mathbb{Z}_2$, $\mathcal{B}$ is the $\mathbb{Z}_2$ toric code phase, and gauging the $\mathbb{Z}_2$ symmetry of two layers of toric code gives the quantum double of $(\mathbb{Z}_2 \times \mathbb{Z}_2) \rtimes \mathbb{Z}_2$, which is the dihedral group of order $8$, $\mathbb{D}_8$. This theory has, for example, $22$ states on a torus.

\subsection{$S_3$-Gauge Theory D$(S_3)$ with $\mathbb{Z}_2$ Symmetry}
\label{sec:D(S_3)}

In this section, we consider the $\mathrm{D}(S_{3})$ UMTC, which corresponds to a $S_3$ discrete gauge theory or quantum double. Physically, it can be realized by the $S_3$ generalization of Kitaev's toric code model~\cite{Kitaev03}, or as $S_3$ lattice gauge theory. First, we briefly review the anyon model of D$(S_3)$. The topological charges are labeled by a pair $([a], \pi_a)$ where $[a]$ is the conjugacy class of $a\in S_3$ and $\pi_a$ is an irreducible representation of the centralizer of $a$ in $S_3$. There are three conjugacy classes in $S_3$: $[e]=\{e\}$, $[(12)]=\{(12),(23), (13)\}$, and $[(123)]=\{(123),(132)\}$. For $[e]$, the centralizer of the identity element $e$ is simply $S_3$, which has three irreducible representations: trivial, alternating, and $2$-dimensional. The corresponding anyon labels are denoted by $I, B, C$, where $I$ is the vacuum. For $[(12)]$, we pick the representative $(12)$, whose centralizer is $\mathbb{Z}_2$, so we have two anyon labels $D$ and $E$ corresponding to the trivial and alternating irreps of $\mathbb{Z}_2$, respectively. For $[(123)]$, we pick the representative $(123)$, whose centralizer is $\mathbb{Z}_3$, so we have three anyon labels $F$, $G$, and $H$, corresponding to the trivial and two nontrivial (corresponding to 3rd roots of unity) irreps of $\mathbb{Z}_3$, respectively. Altogether, we have $8$ topological charges. For a complete list of the fusion rules, $F$-symbols, $R$-symbols, and the modular data, we refer the readers to \Ref{Cui2014} (we note that the basic data in \Ref{Cui2014} actually corresponds to the topological charges $D$ and $E$ carrying the alternating and trivial irreps of $\mathbb{Z}_2$, respectively).

The topological symmetry group is $\mathrm{Aut}(\mathrm{D}(S_{3})) = \mathbb{Z}_{2}$, where the nontrivial topological symmetry permutes the topological charges $C$ and $F$. (All other topological charges have distinct fusion, quantum dimensions, and/or twists, and so cannot be permuted.) One can think of this nontrivial symmetry as a kind of ``electric-magnetic duality,'' since $C = ([e],2)$ is a pure charge (irrep) and $F = ([(123)],1)$ is a pure flux (conjugacy class). Since $I$ and $B$ are the only two Abelian anyons, $\mathcal{A}=\mathbb{Z}_{2}$.

We consider a $G=\mathbb{Z}_{2}$ global symmetry. The symmetry action can either act trivially or as electric-magnetic duality. Since the symmetry action cannot permute $I$ or $B$, both possible symmetry actions yield $H^2_{[\rho]}(\mathbb{Z}_2, \mathbb{Z}_2)=\mathbb{Z}_{2}$. Thus, there are two symmetry fractionalization classes, corresponding to $\coho{w}(\mb{1},\mb{1})=I$ or $B$, respectively. Since $H^3(\mathbb{Z}_2, \text{U}(1))=\mathbb{Z}_2$, there are two defectification classes (for each fractionalization class). The $G$-crossed extensions for trivial symmetry action can be obtained from the results of Sec.~\ref{sec:no_perm}, so we only focus on the case of $\mathbb{Z}_2$ electric-magnetic duality symmetry action. In this case, there will be 6 distinct types of defects, since $|\mathcal{C}_{\bf 1}| = |\mathcal{C}_{\bf 0}^{\bf 1}|=6$.

We start by considering the fusion rules of the symmetry defects. Na\"{i}vely, one might guess the following fusion rules for one of the symmetry defects $\sigma \in \mathcal{C}_{\bf 1}$: $\sigma\times\overline{\sigma}=I+G+H$. However, this is incorrect. To see the inconsistency, let us consider the fusion $C\times C\times \sigma\times\overline{\sigma}$. Consistency between associativity and the $G$-crossed action requires that
\begin{equation}
C\times C\times\sigma\times\overline{\sigma}=C\times F\times\sigma\times\overline{\sigma},
\label{eqn:DS3fusion2}
\end{equation}
since $C\times \sigma= \sigma\times ^\mb{1}\!C=\sigma\times F$.
Using the na\"ive guess, the left-hand side would equal
\begin{equation}
(I+B+C)\times(I+G+H) = I+B+C+2F+3G+3H,
\end{equation}
while the right-hand side would equal
\begin{equation}
(G+H)\times(I+G+H) = 2(I+B+C+F+G+H)
.
\end{equation}
This proves that the na\"ive guess does not yield a consistent fusion theory.
Indeed, we can check that $\sigma\times\overline{\sigma}=I+G$ and $\sigma\times\overline{\sigma}=I+H$ both satisfy \eqref{eqn:DS3fusion2}, so we postulate that both of these fusion rules are realized by different types of defects in $\mathcal{C}_\mb{1}$.

Rather than continuing to derive all the data from consistency, we use a short-cut by noticing that D$(S_3)$ is obtained from either $\overline{\text{JK}_4} \boxtimes \text{JK}_4$ and $\text{SU}(2)_4 \boxtimes \ol{\text{SU}(2)_4}$ by condensing the $(4,4)$ boson in these theories (here, we use the integer label convention; see, e.g. \Ref{Levaillant2015} for the basic data of $\text{JK}_4$ and $\text{SU}(2)_4$). Indeed, these are two of the four possible gauged theories for this example. We notice that these two gauged theories correspond to different defectification classes of $G$-crossed theories, as the Frobenius-Schur indicators of $\overline{\text{JK}_4} \boxtimes \text{JK}_4$ are $\varkappa_{(a_{\text{L}},a_{\text{R}})}=1$, while those of $\text{SU}(2)_4 \boxtimes \ol{\text{SU}(2)_4}$ are  $\varkappa_{(a_{\text{L}},a_{\text{R}})}=(-1)^{a_{\text{L}}+a_{\text{R}}}$. These gauged theories correspond to one of the fractionalization classes, which may be associated with $\coho{w}(\mb{1},\mb{1})=I$, and there will be two more corresponding to the other fractionalization class, associated with $\coho{w}(\mb{1},\mb{1})=B$.

Condensing the $(4,4)$ anyon in $\overline{\text{JK}_4} \boxtimes \text{JK}_4$ or $\text{SU}(2)_4 \boxtimes \ol{\text{SU}(2)_4}$, we recognize the mapping of topological charges to the $G$-crossed theory as giving the eight quasiparticle charges
\begin{equation}
\begin{array}{ll}
(0,0)\sim (4,4) \mapsto I, & (4,0) \sim (0,4) \mapsto B, \\
(1,1)\sim (3,3) \mapsto D, & (3,1) \sim (1,3) \mapsto E, \\
(2,0) \sim (2,4) \mapsto G, & (0,2)\sim (4,2) \mapsto H, \\
(2,2) \mapsto C \text{ and } F,  &
\end{array}
\end{equation}
and six defect charges
\begin{equation}
\begin{array}{ll}
(1,0)\sim (3,4) \mapsto \sigma_{\text{L}}, & (3,0)\sim (1,4) \mapsto \sigma'_{\text{L}}, \\
(0,1)\sim (4,3) \mapsto \sigma_{\text{R}}, & (0,3)\sim (4,1) \mapsto \sigma'_{\text{R}}, \\
(1,2)\sim (3,2) \mapsto \tau_{\text{L}}, & (2,1) \sim (2,3) \mapsto \tau_{\text{R}}
.
\end{array}
\end{equation}
It is then straightforward to determine various properties, such as fusion rules, quantum dimensions, and twist factors, for the $G$-crossed theories from the data of gauged theories. For example, we find the fusion rules
\begin{eqnarray}
\sigma_{\text{L}} \times \sigma_{\text{L}} &=&\sigma'_{\text{L}} \times \sigma'_{\text{L}} = I + G ,  \\
\sigma_{\text{R}} \times \sigma_{\text{R}} &=&\sigma'_{\text{R}} \times \sigma'_{\text{R}} = I + H ,  \\
\tau_{\text{L}} \times \tau_{\text{L}} &=& I + B+ C+ F + 2G + H ,  \\
\tau_{\text{R}} \times \tau_{\text{R}} &=& I + B+ C+ F + G + 2H,
\end{eqnarray}
where the first two lines were previously discussed. The quantum dimensions of the defects are $d_{\sigma_{\text{L}}}=d_{\sigma'_{\text{L}}}=d_{\sigma_{\text{R}}}=d_{\sigma'_{\text{R}}}=\sqrt{3}$ and $d_{\tau_{\text{L}}} = d_{\tau_{\text{R}}} = 2\sqrt{3}$.

The fractionalization class associated with $\coho{w}(\mb{1},\mb{1})=B$ will yield $G$-crossed extensions with the same number of defects and quantum dimensions, but whose defects have different fusion rules. For example, we instead find
\begin{eqnarray}
\sigma_{\text{L}} \times \sigma_{\text{L}} &=&\sigma'_{\text{L}} \times \sigma'_{\text{L}} = B + G ,  \\
\sigma_{\text{R}} \times \sigma_{\text{R}} &=&\sigma'_{\text{R}} \times \sigma'_{\text{R}} = B + H ,  \\
\tau_{\text{L}} \times \tau_{\text{L}} &=& I + B+ C+ F + 2G + H ,  \\
\tau_{\text{R}} \times \tau_{\text{R}} &=& I + B+ C+ F + G + 2H,
\end{eqnarray}
and
\begin{eqnarray}
\sigma_{\text{L}} \times \sigma'_{\text{L}} &=& I + G ,  \\
\sigma_{\text{R}} \times \sigma'_{\text{R}} &=& I + H .
\end{eqnarray}
We emphasize that, since $\overline{\sigma_{\text{L}}} = \sigma'_{\text{L}}$ and $\overline{\sigma_{\text{R}}} = \sigma'_{\text{R}}$ for these two $G$-crossed extensions, they are necessarily distinct (even when allowing for relabeling of topological charges) from the previous two extensions, whose topological charges are all self-dual. These $G$-crossed extensions will similarly yield gauged theories that have the same number of quasiparticles and quantum dimensions as the other two, but with different fusion rules. To the best of our knowledge, these rank 25 MTCs were not previously known.

In order to further verify the relation to the gauged theories while providing an example exhibiting nontrivial features of the ground state degeneracy calculations, we determine the gauged theories' ground state degeneracy from the defect theory, for surfaces of genus $g=1$ and $2$. We use similar methods as in Sec.~\ref{sec:genons}, but must also allow for nontrivial phase factors due to the symmetry action. Again, we label the defect sectors by $s$, with $s=0$ corresponding to the sector with no defect branch lines, and we denote the ground state degeneracy for each defect sector by $\mathcal{N}_{g}(s)$ and its contribution to the gauged theory's ground state degeneracy by $\mathcal{N}_{g}^{G}(s)$. The $s\neq 0$ sectors for $G = \mathbb{Z}_2$ all contribute the same ground state degeneracy, since they can all be mapped to each other by Dehn twists, so it suffices to compute the contribution from one nontrivial sector, e.g. one with a single defect branch line around a single nontrivial cycle.

For the torus (genus $g=1$), D$(S_3)$ has ground state degeneracy $\mathcal{N}_{1}(0)=8$, but the subspace that is invariant under the $\mathbb{Z}_2$ symmetry transformation is reduced to $\mathcal{N}_{1}^{G}(0)=7$, by the symmetrization of the basis states labeled by $C$ and $F$. There are three nontrivial defect sectors. Considering a sector $s\neq 0$ with a branch line around the $m$-cycle and quasiparticle flux lines around the $l$-cycle, we see that the ground state degeneracy is equal to the number of $\mathbb{Z}_2$ invariant anyons, and they are all invariant under the $\mathbb{Z}_2$ action, so $\mathcal{N}^{G}_{1}(s)=\mathcal{N}_{1}(s)= |\mathcal{C}_{\bf 0}^{\bf 1} | =6$. Thus, we obtain the ground state degeneracy for the gauged theories on the torus
\begin{equation}
\mathcal{N}_1^{G} = \sum_{s} \mathcal{N}_{1}^{G}(s) = 7 + 3\times 6 = 25.
\end{equation}
This agrees with the torus ground state degeneracy of $\overline{\text{JK}_4} \boxtimes \text{JK}_4$ and $\text{SU}(2)_4 \boxtimes \ol{\text{SU}(2)_4}$, as expected.

For the genus $g=2$ surface, there are 16 defect sectors, corresponding to the possible configurations of nontrivial $\mathbb{Z}_2$ branch lines around nontrivial cycles of the surface. Let us choose a basis for the trivial defect sector ($s=0$) that is specified diagrammatically (as in Sec.~\ref{sec:generalGenus}) by
\begin{equation}
|\Phi_{a_1, a_2; z}^{(0)}\rangle\equiv
\pspicture[shift=-0.6](-1.7,-0.4)(2.3,1.5)
	\psset{linewidth=1.0pt,arrowscale=1.3,arrowinset=0.15,linestyle=solid, border=1.5pt}
	\psarc{<-}(-0.3,1){0.45}{-195}{4}
	\psarc{<-}(1.6,1){0.45}{-195}{4}
	\psset{linewidth=1.0pt,arrowscale=1.3,arrowinset=0.15,linestyle=solid, border=0}
	\psarc(-0.3,1){0.45}{0}{215}
	\psarc(1.6,1){0.45}{0}{215}
	\psline(-0.3,-0.1)(-0.3,0.55)
	\psline{->}(-0.3,0.25)(-0.3,0.3)
	\psline(1.6,-0.3)(1.6,0.55)
	\psline{->}(1.6,0.20)(1.6,0.25)
    \psline(-0.3,-0.1)(1.6,-0.3)
	\rput[c]{0}(-0.3,1){\scalebox{1}{$\otimes$}}
	\rput[c]{0}(1.6,1){\scalebox{1}{$\otimes$}}
	\rput[tr]{0}(-0.85,1.1){\scalebox{1}{$a_{1}$}}
	\rput[tr]{0}(1.05,1.1){\scalebox{1}{$a_{2}$}}
	\rput[tl]{0}(-0.2,0.3){$z$}
	\rput[tl]{0}(1.7,0.2){$\bar{z}$}
\endpspicture
,
\end{equation}
where $a_1, a_2, z \in \mathcal{C}_{\bf 0}$. This state space has ground state degeneracy
\begin{equation}
\mathcal{N}_{2}(0) = \sum_{a_1, a_2, z} N_{a_1 \overline{a_1}}^{z} N_{a_2 \overline{a_{2}}}^{\bar{z}} = 116
.
\end{equation}
The number of basis states that are mapped to themselves by the $\mathbb{Z}_{2}$ action (i.e. the states with no labels equal to $C$ or $F$) is given by
\begin{equation}
\mathcal{I}_{2}(0) = \sum_{a_1, a_2, z \neq C, F} N_{a_1 \overline{a_1}}^{z} N_{a_2 \overline{a_{2}}}^{\bar{z}} = 58
.
\end{equation}
However, the $\mathbb{Z}_{2}$ symmetry acts on these states (with $a_1, a_2, z \neq C, F$) as $R_{\bf 1} |\Phi_{a_1, a_2; z}^{(0)}\rangle = \pm |\Phi_{a_1, a_2; z}^{(0)}\rangle $, so not all of them are necessarily invariant under the symmetry. We denote the number of such basis states that transform with a $-1$ factor by $\mathcal{A}_{2}(0)$. It turns out that $\mathcal{A}_{2}(0) =2$, where the two corresponding basis states are $|\Phi_{G, H; B}^{(0)}\rangle$ and $|\Phi_{H,G; B}^{(0)}\rangle$. To see this, we first notice that the $\mathbb{Z}_{2}$ symmetry action is given in terms of the topological symmetry action by
\begin{eqnarray}
&& R_{\bf 1} |\Phi_{a_1, a_2; z}^{(0)}\rangle  \\
&&\quad = \frac{ U_{\bf 1}(a_1 , \overline{a_{1}} ; z) U_{\bf 1}(a_2 , \overline{a_{2}} ; \bar{z}) U_{\bf 1}(z , \bar{z} ; I) }{ U_{\bf 1}(a_1 , \overline{a_{1}} ; I) U_{\bf 1}(a_2 , \overline{a_{2}} ; I) } |\Phi_{a_1, a_2; z}^{(0)}\rangle
.
\notag
\end{eqnarray}

In order to compute the $U_{\bf 1}$-symbols, we use the $F$-symbols and $R$-symbols given in \Ref{Cui2014}, and the constraints imposed by their invariance, as in Eqs.~(\ref{eq:rho_g_F}) and (\ref{eq:rho_g_R}). Invariance of the $R$-symbols indicates that $U_{\bf 1}(a,b;c) = U_{\bf 1}(b,a;c)$. Invariance of the $F$-symbols further constrains the $U_{\bf 1}$-symbols to yield the claimed properties for the basis states. Let us consider the transformation of $|\Phi_{G, H; B}\rangle$ in more detail; the relevant $F$-symbols are
\begin{equation}
F^{GGH}_H=\frac{1}{\sqrt{2}} \left[
	\begin{array}{rr}
		1 & 1\\
		1 & -1
	\end{array} \right]
,
\end{equation}
where the rows of the matrix are indexed by $I,B$ and the columns are indexed by $C,F$, and
\begin{equation}
F^{BHH}_{I}=F^{BHH}_{B}= 1
.
\end{equation}
Invariance of these $F$-symbol under the $\mathbb{Z}_2$ symmetry action gives
\begin{eqnarray}
\frac{U_{\bf 1}(G,G;I)}{U_{\bf 1}(G,H;C) U_{\bf 1}(G,C;H)} &=& 1 , \\
\frac{ U_{\bf 1}(G,G;I)}{U(G,H;F)U(G,F;H)} &=& 1 , \\
\frac{U_{\bf 1}(G,G;B) U_{\bf 1}(B,H;H)}{ U_{\bf 1}(G,H;C) U_{\bf 1}(G,C;H)}&=& -1, \\
\frac{U_{\bf 1}(G,G;B) U_{\bf 1}(B,H;H) }{U_{\bf 1}(G,H;F) U_{\bf 1}(G,F;H)}&=& -1, \\
\frac{U_{\bf 1}(B,H; H) U_{\bf 1}(H,H;I)}{U_{\bf 1}(H,H;B) U_{\bf 1}(B,B;I)} &=& 1, \\
\frac{U_{\bf 1}(B,H; H) U_{\bf 1}(H,H;B)}{U_{\bf 1}(H,H;I)} &=& 1,
\end{eqnarray}
from which we conclude that
\begin{equation}
\frac{ U_{\bf 1}(G , G ; B) U_{\bf 1}(H ,H ;B) U_{\bf 1}(B , B ; I) }{ U_{\bf 1}(G,G ; I) U_{\bf 1}(H,H ; I) } =-1
,
\end{equation}
and $R_{\bf 1} |\Phi_{G, H; B}^{(0)}\rangle = -|\Phi_{G, H; B}^{(0)}\rangle$. A similar calculation yields $R_{\bf 1} |\Phi_{H,G; B}^{(0)}\rangle = -|\Phi_{H,G; B}^{(0)}\rangle$.

The basis states that include labels equal to $C$ or $F$ are mapped to each other in pairs and can be symmetrized with respect to the $\mathbb{Z}_{2}$ symmetry action. Putting this all together, we find the contribution from the $s=0$ sector to the gauged theories' ground state degeneracy to be
\begin{equation}
\mathcal{N}^{G}_{2}(0) = \frac{1}{2} \left[ \mathcal{N}_{2}(0) - \mathcal{I}_{2}(0) \right]  + \mathcal{I}_{2}(0) - \mathcal{A}_{2}(0) = 85
.
\end{equation}

There are 15 nontrivial defect sectors ($s\neq 0$) for the genus $g=2$ surface. We consider the sector ($s=1$) with a single defect branch line around a single cycle, for which we can write the basis states as
\begin{equation}
|\Phi_{a_1, a_2; z}^{(1)}\rangle\equiv
\pspicture[shift=-0.6](-1.7,-0.4)(2.3,1.5)
    \psset{linewidth=1.3pt,linecolor=blue,arrowscale=1.2,arrowinset=0.15, linestyle=dashed, dash=3pt 1.8pt}
	\psellipse(2.05,1)(0.5,0.15)
    \psset{linewidth=1.0pt,linecolor=black,arrowscale=1.3,arrowinset=0.15,linestyle=solid, border=1.5pt}
	\psarc{<-}(-0.3,1){0.45}{-195}{4}
	\psarc{<-}(1.6,1){0.45}{-195}{4}
	\psset{linewidth=1.0pt,arrowscale=1.3,arrowinset=0.15,linestyle=solid, border=0}
	\psarc(-0.3,1){0.45}{0}{215}
	\psarc(1.6,1){0.45}{32}{215}
	\psline(-0.3,-0.1)(-0.3,0.55)
	\psline{->}(-0.3,0.25)(-0.3,0.3)
	\psline(1.6,-0.3)(1.6,0.55)
	\psline{->}(1.6,0.20)(1.6,0.25)
    \psline(-0.3,-0.1)(1.6,-0.3)
	\rput[c]{0}(-0.3,1){\scalebox{1}{$\otimes$}}
	\rput[c]{0}(1.6,1){\scalebox{1}{$\otimes$}}
	\rput[tr]{0}(-0.85,1.1){\scalebox{1}{$a_{1}$}}
	\rput[tr]{0}(1.05,1.1){\scalebox{1}{$a_{2}$}}
	\rput[tl]{0}(-0.2,0.3){$z$}
	\rput[tl]{0}(1.7,0.2){$\bar{z}$}
	\rput[br]{0}(2.8,1.1){\scalebox{0.8}{$\mathbf{1}$}}
\endpspicture
,
\end{equation}
where $a_1, a_2, z \in \mathcal{C}_{\bf 0}$. Each nontrivial defect sector has ground state degeneracy
\begin{equation}
\mathcal{N}_{2}(s) = \sum_{a_1, a_2, z} N_{a_1 \overline{a_1}}^{z} N_{a_2 \,^{\bf 1}\overline{a_{2}}}^{\bar{z}} = 98
.
\end{equation}
Similar to the trivial sector, the number of the $s=1$ basis states that are mapped to themselves by the $\mathbb{Z}_{2}$ action (i.e. the states with no labels equal to $C$ or $F$) is given by
\begin{equation}
\mathcal{I}_{2}(1) = \sum_{a_1, a_2, z \neq C, F} N_{a_1 \overline{a_1}}^{z} N_{a_2 \overline{a_{2}}}^{\bar{z}} = 58
\end{equation}
and the number of these states that transform as $R_{\bf 1} |\Phi_{a_1, a_2; z}^{(1)}\rangle = - |\Phi_{a_1, a_2; z}^{(1)}\rangle $ is $\mathcal{A}_{2}(1) =2$, where the two corresponding basis states are $|\Phi_{G, H; B}^{(1)}\rangle$ and $|\Phi_{H,G; B}^{(1)}\rangle$. The basis states that include labels equal to $C$ or $F$ are mapped to each other in pairs and can be symmetrized with respect to the $\mathbb{Z}_{2}$ symmetry action. Thus, we find the contribution from each $s\neq 0$ sector to the gauged theories' ground state degeneracy to be
\begin{equation}
\mathcal{N}^{G}_{2}(s) = \frac{1}{2} \left[ \mathcal{N}_{2}(1) - \mathcal{I}_{2}(1) \right]  + \mathcal{I}_{2}(1) - \mathcal{A}_{2}(1) = 76
.
\end{equation}

Finally, summing the contributions from all defect sectors, we find the ground state degeneracy of the gauged theories on genus $g=2$ surfaces to be
\begin{equation}
\mathcal{N}_{2}^{G} = \sum_{s} \mathcal{N}_{2}^{G}(s) = 85 + 15\times 76 = 1225.
\end{equation}
This agrees with the genus $g=2$ ground state degeneracy of $\overline{\text{JK}_4} \boxtimes \text{JK}_4$ and $\text{SU}(2)_4 \boxtimes \ol{\text{SU}(2)_4}$, as expected.

\subsection{$3$-Fermion Model $\text{SO}(8)_1$ with $S_3$ Symmetry}
\label{sec:3-Fermion}

In this section, we consider the $3$-fermion model $\text{SO}(8)_1$ with global symmetry $G=S_3$ that is non-Abelian. The MTC describing the $3$-fermion model is Abelian, with three fermions that have nontrivial braiding with each other, and chiral central charge $c_- = 4 \mod 8$. Recently, this topological phase has been proposed to exist at the surface of a bosonic 3D time-reversal-invariant topological superconductor~\cite{VishwanathPRX2013, burnell2013}. We also notice that this theory can arise in the following physical way: consider three identical layers of semion theories (with the same chiralities) $\mathbb{Z}_{2}^{(\frac{1}{2} )} \boxtimes \mathbb{Z}_{2}^{(\frac{1}{2} )} \boxtimes \mathbb{Z}_{2}^{(\frac{1}{2} )}$, e.g. three layers of $\nu=\frac{1}{2}$ bosonic Laughlin states. Writing the semion charge of the $j$th layer as $s_j$, we identify a subtheory of the three-layer theory that contains the charges $\{I, s_1 s_2, s_2 s_3,s_1 s_3\}$ as the $\text{SO}(8)_1$ theory with $c_- = 4$. In this way, the three-layer theory can be written as $[\mathbb{Z}_{2}^{(\frac{1}{2} )}]^{3} = \text{SO}(8)_1 \boxtimes \mathbb{Z}_{2}^{(-\frac{1}{2} )}$, where the $\mathbb{Z}_{2}^{(-\frac{1}{2} )}$ semion theory (which has $c_- = -1$) is associated with $\{I, s_1s_2s_3\}$. The ${S}_3$ symmetry in this system is just the permutation symmetry of the three layers, which clearly only acts nontrivially on the three-fermion sector. In fact, this type of layer permutation symmetry and the associated defects have been considered in \Ref{barkeshli2013genon}.

We denote the three fermions by $\psi_j$ for $j=1,2,3$. This MTC is closely related to that of the $\mathbb{Z}_2$ toric code in that they both have a $\mathbb{Z}_2\times\mathbb{Z}_2$ fusion algebra and all $F$-symbols are trivial. The nontrivial $R$-symbols of the 3-fermion model (for a certain choice of gauge) are given by
\begin{eqnarray}
R^{\psi_1 \psi_1}_I = R^{\psi_2 \psi_2}_I = R^{\psi_3 \psi_3}_I &=&-1, \\
R^{\psi_1 \psi_2}_{\psi_3} = R^{\psi_2 \psi_3}_{\psi_1} = R^{\psi_3\psi_1}_{\psi_2}&=& -1,
\end{eqnarray}
and all other allowed $R$-symbols equal 1.

Arbitrary permutations of the three fermion labels leaves the MTC invariant (up to gauge transformations), so the topological symmetry group is $\text{Aut}(\text{SO}(8)_1)=S_{3}$. We represent $S_3$ as the permutation group of three objects, whose six elements are denoted by $\{ \openone, (12), (23), (13), (123), (132) \}$.

We let the global symmetry be described by $G=S_{3}$ with the action on topological charge labels corresponding to the matching permutation of the three fermions. For the pairwise permutations, the $U_{\bf g}$-symbols for the symmetry action on the $\mathcal{C}_{\bf 0}$ sector can be chosen to be given by
\begin{eqnarray}
U_{(12)}(\psi_1 , \psi_2 ; \psi_3) = U_{(12)}(\psi_2 , \psi_1 ; \psi_3)^{-1} &=& i, \notag \\
U_{(12)}(\psi_2 , \psi_3 ; \psi_1) = U_{(12)}(\psi_3 , \psi_2 ; \psi_1)^{-1} &=& -i, \notag \\
U_{(12)}(\psi_3 , \psi_1 ; \psi_2) = U_{(12)}(\psi_1 , \psi_3 ; \psi_2)^{-1}&=& -i, \notag \\
U_{(12)}(\psi_1 , \psi_1 ; I) = U_{(12)}(\psi_2 , \psi_2 ; I) &=& -1, \notag \\
U_{(12)}(\psi_3 , \psi_3 ; I) &=& 1,
\end{eqnarray}
\begin{eqnarray}
U_{(13)}(\psi_1 , \psi_2 ; \psi_3) = U_{(13)}(\psi_2 , \psi_1 ; \psi_3)^{-1} &=& -i, \notag \\
U_{(13)}(\psi_2 , \psi_3 ; \psi_1) = U_{(13)}(\psi_3 , \psi_2 ; \psi_1)^{-1} &=& -i, \notag \\
U_{(13)}(\psi_3 , \psi_1 ; \psi_2) = U_{(13)}(\psi_1 , \psi_3 ; \psi_2)^{-1}&=&  i, \notag \\
U_{(13)}(\psi_1 , \psi_1 ; I) = U_{(13)}(\psi_3 , \psi_3 ; I) &=& -1, \notag \\
U_{(13)}(\psi_2 , \psi_2 ; I) &=& 1,
\end{eqnarray}
\begin{eqnarray}
U_{(23)}(\psi_1 , \psi_2 ; \psi_3) = U_{(23)}(\psi_2 , \psi_1 ; \psi_3)^{-1} &=& -i, \notag \\
U_{(23)}(\psi_2 , \psi_3 ; \psi_1) = U_{(23)}(\psi_3 , \psi_2 ; \psi_1)^{-1} &=& i, \notag \\
U_{(23)}(\psi_3 , \psi_1 ; \psi_2) = U_{(23)}(\psi_1 , \psi_3 ; \psi_2)^{-1}&=&  -i, \notag \\
U_{(23)}(\psi_2 , \psi_2 ; I) = U_{(23)}(\psi_3 , \psi_3 ; I) &=& -1, \notag \\
U_{(23)}(\psi_1 , \psi_1 ; I) &=& 1
.
\end{eqnarray}

For the cyclic permutations of all three objects, the symmetry action leaves the $R$-symbols unchanged, so we can choose $U_{(123)}(a,b;a\times b)=U_{(132)}(a,b;a\times b)=1$ for all $a, b \in \mathcal{C}_{\bf 0}$.

One can calculate $H^2_{[\rho]}(S_3, \mathbb{Z}_2\times\mathbb{Z}_2)=\mathbb{Z}_{1}$ using \Ref{Hartl2007}, so there is exactly one fractionalization class. Since $H^4(S_3, \text{U}(1))=\mathbb{Z}_{1}$, there is no obstruction to defectification. $H^3(S_3, \text{U}(1))=\mathbb{Z}_{6}$, so we expect six defectification classes.

Since $S_3=\mathbb{Z}_3\rtimes \mathbb{Z}_2$, we begin our preparatory analysis by considering the two subgroups.

\subsubsection{$\mathbb{Z}_2$ symmetry}

We consider a $\mathbb{Z}_2$ subgroup of the symmetry. Without loss of generality, we consider the $\mathbb{Z}_2$ action $\rho_{(12)}$, which interchanges $\psi_1$ and $\psi_2$. The analysis of the $G$-crossed extensions is very similar to that of the toric code with electric-magnetic duality symmetry. There is exactly one symmetry fractionalization class, since $H^2_{[\rho]}(\mathbb{Z}_{2}, \mathbb{Z}_2\times\mathbb{Z}_2)=\mathbb{Z}_{1}$, and two defectification classes, since $H^3(\mathbb{Z}_2,\text{U}(1))=\mathbb{Z}_2$.

There are two $(12)$-defect types, since there are two quasiparticle types that are invariant under $(12)$. We label these as $\mathcal{C}_{(12)} = \{ z^{+} , z^{-} \}$. The defect fusion rules are given by
\begin{eqnarray}
z^{\mp} &=& \psi_{1} \times z^{\pm} = \psi_{2} \times z^{\pm}, \\
z^{\pm} &=& \psi_{3} \times z^{\pm} , \\
z^{\pm} \times z^{\pm} &=& I+\psi_{3} , \\
z^{\pm} \times z^{\mp} &=& \psi_{1}+ \psi_{2} .
\end{eqnarray}
The $(12)$-symmetry action on the $(12)$-defects is necessarily trivial, i.e. $\rho_{(12)}(z^{\pm})=z^{\pm}$. The fusion rules indicate that the quantum dimensions of the defects are $d_{z^{s}} = \sqrt{2}$.

While considering the $\mathcal{C}_{(12)}$ sector, we write the quasiparticle topological charges as $a_{\openonesub} = (a^{(1)}, a^{(2)})$, so that $I=(0,0)$, $\psi_{1}=(1,0)$, $\psi_{2}=(0,1)$, and $\psi_{3}=(1,1)$. The nontrivial $F$-symbols of the extended category are given (in a choice of gauge) by
\begin{eqnarray}
F^{a_{\openonesub} b_{(12)} c_{\openonesub}}_{d_{(12)}} &=& (-1)^{a^{(1)} c^{(1)} }, \\
F^{a_{(12)} b_{\openonesub} c_{(12)}}_{d_{\openonesub}} &=& (-1)^{b^{(1)} d^{(1)} }, \\
\left[F^{a_{(12)} b_{(12)} c_{(12)}}_{d_{(12)}}\right]_{e_{\openonesub} f_{\openonesub}} &=& \frac{\varkappa_{z}}{\sqrt{2}}(-1)^{e^{(1)} f^{(1)} },
\end{eqnarray}
where $e_{\openonesub}$ and $f_{\openonesub}$ take values in either $\{I,\psi_{3} \}$ or $\{ \psi_{1},\psi_{2} \}$, depending on the values of $a_{(12)}$, $b_{(12)}$, and $c_{(12)}$. The remaining $F$-symbols allowed by fusion are equal to $1$.

For a convenient choice of gauge (and after removing charge relabeling redundancy), the $R$-symbols are given by
\begin{eqnarray}
R^{\psi_{1} z^{\pm}} &=& \pm i, \quad R^{z^{\pm} \psi_{1}} = i , \\
R^{\psi_{2} z^{\pm}} &=& \mp i, \quad R^{z^{\pm} \psi_{2}} = \pm 1, \\
R^{\psi_{3} z^{\pm}} &=& i , \quad R^{z^{\pm} \psi_{3}} = \pm i , \\
R^{z^{\pm} z^{\pm}}_{I} &=& \pm R^{z^{\pm} z^{\mp}}_{\psi_{2}}  = \pm \left( \sqrt{\varkappa_{z}} e^{i \frac{\pi}{8} } \right)^{\pm 1}, \\
R^{z^{\pm} z^{\pm}}_{\psi_{3}} &=& \pm R^{z^{\pm} z^{\mp}}_{\psi_{3}} = \pm \left( \sqrt{\varkappa_{z}} e^{i \frac{\pi}{8} } \right)^{\mp 3},
\end{eqnarray}
where we let $\sqrt{\varkappa_{z}} = 1$ and $i$, for $\varkappa_{z} = 1$ and $-1$, respectively. The topological twist factors of the defects are
\begin{equation}
\theta_{z^{\pm}} = \pm (\sqrt{\varkappa_{z}} e^{-i \frac{\pi}{8} })^{\pm 1}
.
\end{equation}
The $\eta$-symbols are all trivial, i.e.
\begin{equation}
\eta_{a_{\bf k}}({\bf g},{\bf h}) =1,
\end{equation}
and the $U_{\bf k}$-symbols are given by
\begin{eqnarray}
&& U_{(12)}(a , b ; a\times b) = \frac{\sqrt{\theta_{a}} \sqrt{\theta_{b}}}{\sqrt{\theta_{a\times b}}} (-1)^{a^{(2)} b^{(1)}} , \\
&& U_{(12)}(z^{\pm} , \psi_{1} ; z^{\mp} ) = U_{(12)}(\psi_{2} , z^{\mp} ; z^{\pm}) \notag \\
&& \qquad =U_{(12)}(z^{\pm} , z^{\mp} ; \psi_{2}) = i \varkappa_{z} e^{\pm i \frac{\pi}{4}} , \\
&& U_{(12)}(z^{\pm} , \psi_{2} ; z^{\mp} ) = U_{(12)}(\psi_{1} , z^{\mp} ; z^{\pm}) \notag \\
&& \qquad =U_{(12)}(z^{\pm} , z^{\mp} ; \psi_{1}) = -i \varkappa_{z} e^{\mp i \frac{\pi}{4}} ,\\
&& U_{(12)}(z^{\pm} ,\psi_{3} ; z^{\pm} ) = U_{(12)}(\psi_{3}, z^{\pm} ; z^{\pm}) \notag \\
&& \qquad =U_{(12)}(z^{\pm} , z^{\pm} ; \psi_{3}) = \pm 1 , \\
&& U_{(12)}(z^{\pm} , z^{\pm} ; I) = 1 ,
\end{eqnarray}
where $\sqrt{\theta_{\psi_{j}}} = i$.

Gauging the $\mathbb{Z}_2$ symmetry, we obtain the gauged theories $\text{Ising}^{(1)}\boxtimes \text{Ising}^{(7)}$ and $\text{Ising}^{(3)}\boxtimes \text{Ising}^{(5)}$ for $\varkappa_{z} =\pm 1$, respectively.

The structure of the other two $\mathbb{Z}_2$ symmetry subgroup sectors, whose charges we label as $\mathcal{C}_{(23)} = \{ x^{+} , x^{-} \}$ and $\mathcal{C}_{(13)}= \{ y^{+} , y^{-} \}$, are similar and may be obtained from the data given above by cyclically permuting all $j=1,2,3$ labels.

\subsubsection{$\mathbb{Z}_3$ symmetry}
\label{sec:3-Fermion_Z3}

We consider the $\mathbb{Z}_3$ subgroup of the symmetry group, which acts as cyclic permutation of the three fermions. That is $\rho_{(123)}(\psi_j)=\, ^{(123)}\! \psi_j=\psi_{[j+1]_3}$ and $\rho_{(132)}(\psi_j)=\,^{(132)}\! \psi_j=\psi_{[j-1]_3}$. There is exactly one symmetry fractionalization class, since $H^2_{[\rho]}(\mathbb{Z}_{3}, \mathbb{Z}_2\times\mathbb{Z}_2)=\mathbb{Z}_{1}$, and three defectification classes, since $H^3(\mathbb{Z}_3,\text{U}(1))=\mathbb{Z}_3$. The corresponding defect sectors each have one defect type (since $I$ is the only fixed charge under these symmetries) and they are each others' charge conjugates, so we write them as $\mathcal{C}_{(123)} = \{ w \}$ and $\mathcal{C}_{(132)} = \{ \bar{w} \}$. Their fusion rules are given by
\begin{eqnarray}
w\times \psi_j &=& w, \\
\overline{w}\times\psi_j&=& \overline{w}, \\
w \times \overline{w} &=& I+\psi_1+\psi_2+\psi_3, \\
w\times w &=& 2\overline{w}, \\
\overline{w}\times\overline{w} &=& 2w.
\end{eqnarray}
We note that the first two lines follow from the fact that there is only one defect types in each sector, and they imply the defects' quantum dimensions are $d_w=d_{\overline{w}}=2$. This further implies the third line, i.e. that the fusion rules necessarily include fusion multiplicities. Physically, the fusion multiplicities can be understood from the existence of three mutually anti-commuting Wilson net operators around two $w$ defects. Alternatively, the need for degeneracy can be understood in terms of operators that transfer topological charge $\psi_j$ between the two $w$ defects.

The $F$-symbols are obtained by solving the pentagon equations (and making gauge choices)~\cite{teo2013, Titsworth_private}. The $F$-symbols that do not involve fusion multiplicities are given by
\begin{eqnarray}
F^{abw}_w &=& \chi(b,\,^{(132)}a) \\
F^{w a b}_w &=& \chi(a,\,^{(123)}b) \\
F^{aw b}_w &=& \chi(b,\,^{(123)}a) \chi(a,\,^{(132)}b)\\
F^{w\ol{w}a}_b &=& \chi(a, \,^{(132)}b)\\
F^{aw\ol{w}}_b &=& \chi(a \times b ,\,^{(132)}a)\\
F^{w a\ol{w}}_b &=& \chi(b,\,^{(132)}a) \chi(a,\,^{(123)}[a \times b]) \\
\left[F^{w\ol{w}w}_w \right]_{ab} &=& \frac{1}{2} \chi(b, \,^{(123)} [a \times b]) \chi(a, \,^{(132)} b)
,
\end{eqnarray}
where $\chi$ is a bi-character on the $\mathbb{Z}_{2} \times \mathbb{Z}_{2}$ fusion algebra. From the above equations, we can obtain the corresponding $F$-symbols with $w$ and $\ol{w}$ interchanged by interchanging the group elements $(123)$ and $(132)$ on the right hand sides.

\begin{table*}[t]
	\centering
	\begin{tabular}{|c|c|c|c|c|}
		\hline
		$a$ & $I$ & $\psi_1$ & $\psi_2$ & $\psi_3$\\
		\hline
		[${F}_{\ol{w}}^{aww}]_{(w,0,\mu),(\ol{w},\nu,0)}$ & $\openone$ & $-i\sigma_1$ & $-i\sigma_3$ & $-i\sigma_2$\\
		\hline
	$[{F}_{\ol{w}}^{w aw }]_{(w,0,\mu),(\ol{w},0,\nu)}$ & $\openone$ & $i\sigma_2$ & $i\sigma_1$ & $i\sigma_3$\\
		\hline
		$[{F}_{\ol{w}}^{wwa}]_{(\ol{w},\mu,0),(w,0,\nu)}$ & $\openone$ & $i\sigma_3$ & $i\sigma_2$ & $i\sigma_1$\\
		\hline
		$[{F}_w^{a\ol{w}\,\ol{w}}]_{(\ol{w},0,\mu),({w},\nu,0)}$ & $\openone$ & $-i\sigma_2$ & $-i\sigma_1$ & $-i\sigma_3$\\
		\hline
		$[{F}_w^{\ol{w}a\ol{w}}]_{(\ol{w},0,\mu),({w},0,\nu)}$ & $\openone$ & $-i\sigma_1$ & $-i\sigma_3$ & $-i\sigma_2$\\
		\hline
		$[{F}_w^{\ol{w}\, \ol{w}a}]_{({w},\mu,0),(\ol{w},0,\nu)}$ & $\openone$ & $i\sigma_3$ & $i\sigma_2$ & $i\sigma_1$\\
		\hline
		$[{F}^{www}_a]_{(\ol{ w},\mu,0),(\ol{ w},\nu,0)}$ & $-\alpha e^{-i \frac{\pi}{3\sqrt{3}}(1,1,1)\cdot\bm{\sigma}}$ & $\alpha e^{i\frac{\pi}{3\sqrt{3}}(1,-1,1)\cdot\bm{\sigma}}$ & $\alpha e^{i\frac{\pi}{3\sqrt{3}}(-1,1,1)\cdot\bm{\sigma}}$ &$\alpha e^{i \frac{\pi}{3\sqrt{3}}(1,1,-1)\cdot\bm{\sigma}}$\\
		\hline
$[{F}^{\ol{ www}}_a]_{( w,\mu,0),(\ol{ w},\nu,0)}$  & $-\alpha e^{i\frac{\pi}{3\sqrt{3}}(1,1,1)\cdot\bm{\sigma}}$ & $-\alpha e^{i\frac{\pi}{3\sqrt{3}}(1,-1,-1)\cdot\bm{\sigma}}$ & $-\alpha e^{i\frac{\pi}{3\sqrt{3}}(-1,-1,1)\cdot\bm{\sigma}}$ & $-\alpha e^{i\frac{\pi}{3\sqrt{3}}(-1,1,-1)\cdot\bm{\sigma}}$\\
		\hline
$[F^{ ww\ol{ w}}_w]_{(\ol{w},\mu,\nu),(a,0,0)}$& $\frac{1}{\sqrt{2}}e^{i\frac{\pi}{3\sqrt{3}}(-1,-1,1)\cdot\bm{\sigma}}$ & $-\frac{1}{\sqrt{2}}e^{i\frac{\pi}{3\sqrt{3}}(-1,1,-1)\cdot\bm{\sigma}}$ & $-\frac{1}{\sqrt{2}}e^{i\frac{\pi}{3\sqrt{3}}(1,1,1)\cdot\bm{\sigma}}$ & $\frac{1}{\sqrt{2}}e^{i\frac{\pi}{3\sqrt{3}}(1,-1,-1)\cdot\bm{\sigma}}$\\
		\hline
$[{F}^{\ol{w}\, \ol{w}w}_{\ol{w}}]_{(w,\mu,\nu),(a,0,0)}$ & $\frac{1}{\sqrt{2}}e^{i\frac{\pi}{3\sqrt{3}}(-1,1,1)\cdot\bm{\sigma}}$ & $\frac{1}{\sqrt{2}}e^{i\frac{\pi}{3\sqrt{3}}(1,1,-1)\cdot\bm{\sigma}}$ & $-\frac{1}{\sqrt{2}}e^{-i\frac{\pi}{3\sqrt{3}}(1,1,1)\cdot\bm{\sigma}}$ & $-\frac{1}{\sqrt{2}}e^{i\frac{\pi}{3\sqrt{3}}(1,-1,1)\cdot\bm{\sigma}}$\\
		\hline
$[{F}^{\ol{w}ww}_w]_{(a,0,0),(\ol{w},\nu,\mu)}$ & $-\frac{i \alpha^{\ast}}{\sqrt{2}}\sigma_2$ & $-\frac{\alpha^{\ast}}{\sqrt{2}}\openone$ & $\frac{i \alpha^{\ast}}{\sqrt{2}}\sigma_3$  & $-\frac{i\alpha^{\ast}}{\sqrt{2}}\sigma_1$ \\
		\hline
$[{F}^{{w}\ol{w}\,\ol{w}}_{\ol{w}}]_{(a,0,0),(\ol{w},\nu,\mu)}$ & $\frac{i\alpha^{\ast}}{\sqrt{2}}\sigma_2$ & $-\frac{i\alpha^{\ast}}{\sqrt{2}}\sigma_3$ & $\frac{i\alpha^{\ast}}{\sqrt{2}}\sigma_1$  & $\frac{\alpha^{\ast}}{\sqrt{2}}\openone$ \\
		\hline
	\end{tabular}
	\caption{$F$-symbols that involve fusion multiplicities for the $\mathbb{Z}_{3}$ symmetry defects of the 3-fermion model. The indices $\mu$ and $\nu$ label the (two-fold degenerate) fusion vertex basis states. The $\mu,\nu$ matrix element of each entry gives the value of the corresponding $F$-symbol, where $\sigma_j$ for $j=1,2,3$ are the two-dimensional Pauli matrices. The phase $\alpha = e^{i \frac{2\pi}{3}k}$ is a 3rd root of unity that differentiates the three distinct defectification classes.}
	\label{tab:fsym}
\end{table*}

The bi-character $\chi$ is fixed by the $G$-crossed heptagon equations to be
\begin{equation}
\chi(a,b)=R^{ba}.
\end{equation}

We list the rest of the nontrivial $F$-symbols, which involve fusion multiplicities, in Table~\ref{tab:fsym}. The phase
\begin{equation}
\alpha = e^{i \frac{2\pi}{3}k}
\end{equation}
for $k=0,1,2$ is a 3rd root of unity that distinguishes the three defectification classes, and which is determined by a choice in $H^3(\mathbb{Z}_3,\text{U}(1))=\mathbb{Z}_3$. We note that the 3rd order Frobenius-Schur indicators~\cite{Ng2005} of $w$, which are gauge invariant quantities associated with the fusion space $V^{w w w}_{0} \cong V^{w w}_{\ol{w}} \otimes V^{\ol{w} w}_{0}$, are found to be $\nu_{3,1}(w) = -\alpha$ and $\nu_{3,2}(w) = -\alpha^2$.

We choose a gauge in which
\begin{equation}
\eta_{a_{\bf k}}({\bf g},{\bf h}) =1
\end{equation}
for all $a_{\bf k}$ and ${\bf g},{\bf h},{\bf k}\in \mathbb{Z}_{3}$. Solving the $G$-crossed heptagon equations then yields the $R$-symbols
\begin{eqnarray}
R^{w \psi_{j}} &=& R^{\ol{w} \psi_{j}}=-1, \\
R^{ww}_{\overline{w}} &=& \theta_w e^{-i\frac{\pi}{3\sqrt{3}}(1,1,1)\cdot\bm{\sigma}}, \\
R^{\ol{w} \, \ol{w}}_{{w}} &=& \theta_{\ol{w}} e^{i \frac{\pi}{3\sqrt{3}}(1,1,1)\cdot\bm{\sigma}}, \\
R^{w\ol{w}}_I &=& R^{w\ol{w}}_{\psi_j} = \theta_w^{-1}\\
R^{\ol{w}w}_I &=& \theta_{\ol{w}}^{-1}, \\
R^{\ol{w}w}_{\psi_{j}} &=& -\theta_{\ol{w}}^{-1},
\end{eqnarray}
where $\theta_w$ and $\theta_{\ol{w}}$ are the topological twist factors of the $w$ and $\ol{w}$ defects, respectively. These twist factors satisfy $\theta_w^3=\theta_{\ol{w}}^3=\alpha^{\ast}$, but are only determined up to 3rd roots of unity, as expected. Physically, this uncertainty can be attributed to possible $\mathbb{Z}_3$ charges attached to the defects.

Notice that the $G$-crossed heptagon equations cannot completely fix $R^{\psi_{j} w}_w$ and $R^{\psi_{j} \ol{w}}_{\ol{w}}$, but subjects them to the following conditions:
\begin{eqnarray}
R^{\psi_{j} w}_w R^{\psi_{[j-1]_{3}} \ol{w}}_{\ol{w}} &=& 1 ,\\
R^{\psi_{1} w}_w  R^{\psi_{2} w}_w R^{\psi_{3} w}_w &=& 1.
\end{eqnarray}

We also find the $U_{\bf k}$-symbols
\begin{eqnarray}
\label{eqn:Z3U}
U_{(123)}(w,w;\ol{w}) &=&  -\frac{\theta_{\ol{w}}}{\alpha \theta_{w}} e^{- i\frac{\pi}{3\sqrt{3}}(1,1,1)\cdot\bm{\sigma}}, \\
U_{(123)}(w,\ol{w};I) &=& \frac{\theta_w}{\theta_{\ol{w}}}.
\end{eqnarray}
In other words, the $\mathbb{Z}_3$ symmetry action on the $V^{ww}_{\ol{w}}$ space is nontrivial.

We now consider the gauged theory (for the $\mathbb{Z}_{3}$ symmetry). We take a gauge choice for the $\mathbb{Z}_{3}$-crossed theory in which $\theta_w=\theta_{\ol{w}}$. The $\mathbb{Z}_{3}$ orbits of $\mathbb{Z}_{3}$-crossed theory are all singletons, except for $[\psi_j] = \{\psi_1,\psi_2,\psi_3\}$. The singletons $I$, $w$, and $\ol{w}$ each yield three topological charges in the gauged theory, corresponding to pairing them with an irrep of $\mathbb{Z}_{3}$, their stabilizer subgroup. We label these as $I_{n} \equiv(I, n)$, $W_{n}=(w, n)$, and $\ol{W_{n}}=(\ol{w},[-n]_{3})$, where $n=0,1,2$. The quantum dimensions of these topological charges are $d_{I_{n}}=1$ and $d_{W_{n}}=2$. The orbit $[\psi_j]$ has stabilizer subgroup $\mathbb{Z}_{1}$, so it becomes a single topological charge $\Psi$, which has quantum dimension $d_{\Psi}=3$. Thus, the gauged theory has a total of $10$ topological charge types. The fusion rules between non-Abelian topological charges can be determined using \eqref{eqn:fusion} to be
\begin{eqnarray}
I_{n} \times I_{m} &=& I_{[n+m]_{3}} \\
I_{n} \times W_{m} &=& W_{[n+m]_{3}} \\
I_{n} \times \ol{W_{m}} &=& \ol{W_{[m-n]_{3}}} \\
I_{n} \times \Psi &=& \Psi \\
W_{n} \times W_{m} &=& \ol{W_{[k-n-m-1]_{3}}} + \ol{W_{[k-n-m+1]_{3}}} \\
W_{n} \times \ol{W_{m}} &=& I_{[n-m]_{3}} + \Psi\\
W_{n} \times \Psi &=& W_{0} + W_{1} + W_{2} \\
\ol{W_{n}} \times \ol{W_{m}} &=& W_{[k-n-m-1]_{3}} + W_{[k-n-m+1]_{3}} \\
\ol{W_{n}} \times \Psi &=& \ol{W_{0}} + \ol{W_{1}} + \ol{W_{2}} \\
\Psi \times \Psi &=& I_{0} + I_{1} + I_{2} + 2\Psi .
\end{eqnarray}
Notice that the fusion of $W_{n}$ and $W_{m}$ has two channels, which corresponds to the defect fusion space $V^{ww}_{\ol{w}}$ carrying a nontrivial reducible $2$-dimensional representation of $\mathbb{Z}_3$, as specified by \eqref{eqn:Z3U}.

The fusion category obtained here is identified as that of $\text{SU}(3)_3$ for $k=\pm 1$, e.g. for $k=1$ via the mapping of topological charges: $I_0 \mapsto \mb{1}$, $I_{1} \mapsto \mb{10}$, $W_{0} \mapsto \mb{3}$, $W_{1} \mapsto \mb{15}$, $W_{2} \mapsto \ol{\mb{6}}$, $\Psi \mapsto \mb{8}$.

\subsubsection{$S_3$ symmetry}

We now consider the full $G=S_3$ extension, which has six defect sectors $\mathcal{C}_\mathbf{g}$, where $\mathbf{g}\in S_3$. From the analysis in the previous subsections, we can write down all fusion rules between charges within a sector, so we only need to determine the fusion between different sectors.

The fusion rules between the $(12)$, $(23)$, and $(13)$ sectors are easily determined to be
\begin{eqnarray}
x^r \times y^s &=& {w}, \quad y^s \times x^r = \overline{w}, \\
y^r \times z^s &=& {w}, \quad z^s \times y^r = \overline{w}, \\
z^r \times x^s &=& {w}, \quad x^s \times z^r = \overline{w},
\end{eqnarray}
where $r,s = \pm$.

The fusion rules between the $(jk)$-sectors and the $(jkl)$-sectors is straightforward to determine to be
\begin{eqnarray}
x^r \times w &=& \ol{w} \times x^r= y^+ + y^- ,\\
x^r \times \ol{w} &=& w \times x^r = z^+ + z^-, \\
y^r \times w &=& \ol{w} \times y^r= z^+ + z^- ,\\
y^r \times \ol{w} &=& w \times y^r = x^+ + x^-,\\
z^r \times w &=& \ol{w} \times z^r= x^+ + x^- ,\\
z^r \times \ol{w} &=& w \times z^r = y^+ + y^- ,
\end{eqnarray}
for $r = \pm$. We note that na\"ive guesses for fusion rules such as $x^+ \times w = 2y^+$ can be ruled out by fusing both sides of the equation with the fermions, e.g. $\psi_1 \times x^+ \times w = x^+ \times w$ and $\psi_1 \times y^+= y^-$, which would lead to an inconsistency.

Additionally, we need to understand the symmetry actions on the defect sectors. In general, $\mathbf{g}$-action takes $\mathcal{C}_{\mathbf{h}}$ to $\mathcal{C}_{\mathbf{ghg^{-1}}}$. Since $\mathcal{C}_{(123)}$ and $\mathcal{C}_{(132)}$ each contain one defect, the nontrivial $\mathbb{Z}_2$ action is obviously given by
\begin{equation}
\rho_{(jk)} (w) = \ol{w}, \qquad  \rho_{(jk)} (\ol{w}) = w.
\end{equation}
The $\mathbb{Z}_3$ symmetry has nontrivial actions on the $\mathcal{C}_{(jk)}$ sectors. Since each of these sectors contains two defects, we need to determine the specific action of $(123)$. Let us consider the action on $\mathcal{C}_{(12)}$. The two defects $z^{\pm}$ are distinguished by the eigenvalue of a $\psi_3$-Wilson loop around the defect. The action of $(123)$ maps $\mathcal{C}_{(12)}$ to $\mathcal{C}_{(23)}$ and $\psi_3$ is mapped to $\psi_1$. As such, it is natural to associate defects with the same eigenvalues of the corresponding invariant Wilson loops, that is
\begin{equation}
\rho_{(123)}( z^+ ) = x^+ , \qquad \rho_{(123)}( z^- ) = x^- .
\end{equation}
The action of $(123)$ on the other two sectors can be obtained similarly, as can the action of $(132)$ on these sectors.

\subsubsection{Sequentially gauging the $S_3$ symmetry}

We are now ready to gauge the full $S_3$ symmetry. Our strategy is to break the $S_3$ symmetry into the $\mathbb{Z}_3$ normal subgroup and the $\mathbb{Z}_2$ subgroup, and then gauge them sequentially~\cite{Cui_gauging}. In Sec.~\ref{sec:3-Fermion_Z3}, we found that gauging the $\mathbb{Z}_3$ symmetry gives $\text{SU}(3)_3$-type theories, and we just need to gauge the remaining $\mathbb{Z}_2$ symmetry of these theories. The action of the $\mathbb{Z}_2$ symmetry on the quasiparticles of these $\mathbb{Z}_{3}$-gauged theories can be identified as topological charge conjugation, that is
\begin{equation}
^\mathbf{1} I_{n} = I_{[-n]_3}, \quad ^\mathbf{1} W_{n} = \overline{W_{n}}, \quad ^\mathbf{1} \Psi = \Psi.
\end{equation}

Let us consider the $\mathbb{Z}_2$ symmetry defects of the $\mathbb{Z}_3$-gauged theory. Since there are two topological charges, $I$ and $\Psi$, that are fixed under the symmetry action, there are two defect types, which we denote as $\Sigma^{\pm}$. We first state the conjectured fusion rules:
\begin{eqnarray}
I_{n} \times \Sigma^{\pm} &=& \Sigma^{\pm} , \\
W_{n} \times \Sigma^{\pm} &=& \Sigma^{+} + \Sigma^{-} , \\
\Psi \times \Sigma^{+} &=& \Sigma^{+} + 2 \Sigma^{-} , \\
\Psi \times \Sigma^{-} &=& 2 \Sigma^{+} + \Sigma^{-} , \\
\label{eq:SigmaSigma_fusion}
\Sigma^{\pm} \times \Sigma^{\pm} &=& \Psi + \sum_{n=0,1,2} (I_{n} + W_n + \overline{W_n}) , \\
\Sigma^{\pm} \times \Sigma^{\mp} &=& 2\Psi + \sum_{n=0,1,2} (W_n + \overline{W_n}) .
\end{eqnarray}
From these fusion rules, the quantum dimensions of the defects are determined to be $d_{\Sigma^{\pm}}= 3\sqrt{2}$.

In order to justify these fusion rules, it is useful to revert to the $S_3$-crossed extended category. The $\mathbb{Z}_2$-crossed extensions of $\mathbb{Z}_{3}$-gauged theories should be equivalent to the $S_3$-crossed theories with its $\mathbb{Z}_3$ subgroup gauged.  Armed with this perspective, we immediately see that the $\mathbb{Z}_2$ defects of the $\mathbb{Z}_{3}$-gauged theories are the equivariantized orbits of the $\mathbb{Z}_2$-defects in $\mathcal{C}_{(jk)}$. Schematically, we can write
\begin{equation}
\Sigma^{\pm} \simeq x^{\pm} + y^{\pm} + z^{\pm}
.
\end{equation}
To actually use the general formula \eqref{eqn:fusion}, we will have to solve the entire extended category to obtain the $U_{\bf k}$ symbols. However, we will just use this expression for a heuristic derivation of the fusion rules. For example,
\begin{equation}
	\begin{split}
	\Sigma^{+} \times \Sigma^{+} &\simeq
	(x^+ + y^+ + z^+) \times (x^+ + y^+ + z^+) \\
	&=(I + \psi_1) + (I +\psi_2) + (I + \psi_3) \\
	&\quad + (w + \ol{w}) + (w + \ol{w}) + (w + \ol{w})
	\end{split}
\end{equation}
The three occurrences of the vacuum $I$ should be interpreted as $I_0 + I_1 + I_2$ (where the subscript indicates the value of $\mathbb{Z}_3$ charge). Similarly, the three occurrences of $w$ and $\ol{w}$ should be interpreted as $W_0 + W_1 + W_2$ and $\overline{W_0} + \overline{W_1} + \overline{W_2}$, respectively. Clearly, $\psi_1+\psi_2+\psi_3$ should be identified with $\Psi$. This yields Eq.~(\ref{eq:SigmaSigma_fusion}), and the other fusion rules can be obtained in a similar fashion. We have checked that the fusion rules are associative and satisfy all the symmetry properties.

In addition, without solving the $G$-crossed consistency equations for the complicated $\mathbb{Z}_{3}$-gauged theories, we can directly read off the topological twists of the defects $\Sigma^{\pm}$, since their twists are the same as the $\mathbb{Z}_2$-defects in the $\text{SO}(8)_1$ theory, as suggested by \eqref{eqn:twist}.

We are now ready to attack our final goal of describing the $S_3$-gauged theory. First, we determine the topological charges of the gauged theory. The $\mathbb{Z}_{2}$ orbits of the $\mathbb{Z}_{3}$-gauged theories form singletons and doublets. The singletons $I_0$, $\Psi$, $\Sigma^{+}$, and $\Sigma^{-}$ each yield two topological charges, corresponding to pairing them with a trivial ($+$) or nontrivial ($-$) irrep of $\mathbb{Z}_{2}$, their stabilizer subgroup. The doublets $[I_1] = \{I_1 , I_2 \}$, $[W_0] = \{W_0 , \overline{W_0} \}$, $[W_1] = \{W_1 , \overline{W_1} \}$, and $[W_2] = \{W_2 , \overline{W_2} \}$ all have stabilizer subgroup $\mathbb{Z}_{1}$, so each one becomes a topological charge in the gauged theory. Thus, there are $12$ topological charges in the gauged theory. The quantum dimensions of the anyons derived from singletons are unchanged by gauging, while those of the anyons derived from doublets are multiplied by 2. The twist factors are unchanged for the anyons derived from the ${\bf 0}$-sector, while the ${\bf 1}$-defects $\Sigma^{\pm}$ are $\mathbb{Z}_{2}$-fluxes that braid nontrivially with the nontrivial $\mathbb{Z}_{2}$ irrep, yielding twists that differ by a sign. The anyons of the gauged theory, along with their quantum dimensions and twist factors are listed in Table~\ref{tab:3fermions}.

In order to get the fusion rules of the gauged theory, in principle one needs the full data of the $G$-crossed theories, especially the $U_\mb{k}$ symbols. Fortunately, in this case, we find that merely requiring associativity is enough to constrain the fusion rules obtained by equivariantization. With the fusion rules and the topological twist factors, we can compute the $S$-matrix. There are $6$ possibilities for the topological twists, in accordance with the $H^3(S_3, \text{U}(1))=\mathbb{Z}_6$ classification.

Choosing $\alpha = e^{i \frac{4\pi}{3}}$ and $\nu=1$, the resulting $S$-matrix is~\cite{Cui_gauging}
\begin{eqnarray}
&& \mathcal{D} S = \\
&&
\left[\begin{smallmatrix}
		1 & 1 & 2 & 3 & 3 & 4 & 4 & 4 & 3\sqrt{2} & 3\sqrt{2} & 3\sqrt{2} & 3\sqrt{2}\\
1 & 1 & 2 & 3 & 3 & 4 & 4 & 4 & \sminus 3\sqrt{2} & \sminus 3\sqrt{2} & \sminus 3\sqrt{2} & \sminus 3\sqrt{2}\\
2 & 2 & 4 & 6 & 6 & \sminus 4 & \sminus 4 & \sminus 4 & 0 & 0 & 0 & 0\\
3 & 3 & 6 & \sminus 3 & \sminus 3 & 0 & 0 & 0 & \sminus 3\sqrt{2} & \sminus 3\sqrt{2} & 3\sqrt{2} & 3\sqrt{2}\\
3 & 3 & 6 & \sminus 3 & \sminus 3 & 0 & 0 & 0 & 3\sqrt{2} & 3\sqrt{2} & \sminus 3\sqrt{2} & \sminus 3\sqrt{2}\\
4 & 4 & \sminus 4 & 0 & 0 & b & c & a & 0 & 0 & 0 & 0\\
4 & 4 & \sminus 4 & 0 & 0 & c & a & b & 0 & 0 & 0 & 0\\
4 & 4 & \sminus 4 & 0 & 0 & a & b & c & 0 & 0 & 0 & 0\\
3\sqrt{2} & \sminus 3\sqrt{2} & 0 & \sminus 3\sqrt{2} & 3\sqrt{2} & 0 & 0 & 0 & 0 & 0 & 6 & \sminus 6\\
3\sqrt{2} & \sminus 3\sqrt{2} & 0 & \sminus 3\sqrt{2} & 3\sqrt{2} & 0 & 0 & 0 & 0 & 0 & \sminus 6 & 6\\
3\sqrt{2} & \sminus 3\sqrt{2} & 0 & 3\sqrt{2} & \sminus 3\sqrt{2} & 0 & 0 & 0 & 6 & \sminus 6 & 0 & 0\\
3\sqrt{2} & \sminus 3\sqrt{2} & 0 & 3\sqrt{2} & \sminus 3\sqrt{2} & 0 & 0 & 0 & \sminus 6 & 6 & 0 & 0
\end{smallmatrix}\right]	
\notag
\end{eqnarray}
where $a=-8\cos\frac{2\pi}{9}$, $b=-8\sin\frac{\pi}{9}$, and $c=8\cos\frac{\pi}{9}$. The columns and rows of the $S$-matrix are ordered as in Table~\ref{tab:3fermions}. We will not write the fusion rules explicitly, since they can be obtained easily from the $S$-matrix using the Verlinde formula. To the best of our knowledge, this $12$-particle MTC was not previously known.

\begin{table}
	\centering
	\begin{tabular}{|c|c|c|c|c|}
		\hline
		topological charge $a$ & $d_a$ & $\theta_a$  \\ 
		\hline
		$(I_0 ,+)$ & $1$ & $1$  \\ 
		\hline
		$(I_0,-)$ & $1$ & $1$ \\  
		\hline
		$[I_1]$ & $2$ & $1$\\ 
		\hline
		$(\Psi,+)$ & $3$ & $-1$ \\ 
		\hline
		$(\Psi,-)$ & $3$ & $-1$ \\ 
		\hline
		$[W_0]$ & $4$ & $\alpha^{-1/3}$ \\ 
		\hline
		$[W_1]$ & $4$ & $\omega\alpha^{-1/3}$ \\ 
		\hline
		$[W_2]$ & $4$ & $\omega^2 \alpha^{-1/3}$ \\ 
		\hline
		$(\Sigma^+,+)$ & $3\sqrt{2}$ &  $\sqrt{\varkappa_{\Sigma}} e^{-i \frac{\pi}{8} }$ \\ 
		\hline
		$(\Sigma^+,-)$ & $3\sqrt{2}$ & $-\sqrt{\varkappa_{\Sigma}} e^{-i \frac{\pi}{8} }$ \\ 
		\hline
		$(\Sigma^-, +)$ & $3\sqrt{2}$ & $-(\sqrt{\varkappa_{\Sigma}})^{\ast} e^{i \frac{\pi}{8} }$ \\ 
		\hline
		$(\Sigma^-,-)$ & $3\sqrt{2}$ & $(\sqrt{\varkappa_{\Sigma}})^{\ast} e^{i \frac{\pi}{8} }$ \\ 
		\hline
	\end{tabular}
	\caption{Topological charges, quantum dimensions, and twist factors of the gauged theory $\gauged{C}{S_3}$, where $\mathcal{C} = \text{SO}(8)_1$ is the $3$-Fermion model. Here, $\omega = e^{i \frac{2 \pi}{3}}$, $\alpha = \omega^{k}$, and $\sqrt{\varkappa_{\Sigma}} = 1$ or $i$.}
	\label{tab:3fermions}
\end{table}

\subsubsection{Ground state degeneracy of the gauged theory}

In order to further verify the relation to the gauged theories and exhibit nontrivial features of the ground state degeneracy calculations, we determine the full $S_3$ gauged theories' ground state degeneracy from the $\mathbb{Z}_{2}$-crossed defect theory of the $\mathbb{Z}_{3}$-gauged theories, for surfaces of genus $g=1$ and $2$. We use similar methods as in Secs.~\ref{sec:genons} and \ref{sec:D(S_3)}, but must also allow for nontrivial unitary transformations due to the symmetry action. We label the $\mathbb{Z}_{2}$-defect sectors by $s$, with $s=0$ corresponding to the sector with no defect branch lines, and we denote the ground state degeneracy for each defect sector by $\mathcal{N}_{g}(s)$ and its contribution to the gauged theory's ground state degeneracy by $\mathcal{N}_{g}^{G}(s)$. The $s\neq 0$ sectors for $G = \mathbb{Z}_2$ all contribute the same ground state degeneracy, since they can all be mapped to each other by Dehn twists, so it suffices to compute the contribution from one nontrivial sector, e.g. one with a single defect branch line around a single nontrivial cycle.

For the torus (genus $g=1$), the $\mathbb{Z}_{3}$-gauged theories have ground state degeneracy $\mathcal{N}_{1}(0)=10$, but the subspace that is invariant under the $\mathbb{Z}_2$ symmetry transformation is reduced to $\mathcal{N}_{1}^{G}(0)=6$, by the symmetrization of the basis states labeled by the charges in the $\mathbb{Z}_{2}$ doublets. There are three nontrivial defect sectors. Considering a sector $s\neq 0$ with a branch line around the $m$-cycle and quasiparticle flux lines around the $l$-cycle, we see that the ground state degeneracy is equal to the number of $\mathbb{Z}_2$ invariant anyons, and they are all invariant under the $\mathbb{Z}_2$ action, so $\mathcal{N}^{G}_{1}(s)=\mathcal{N}_{1}(s)= |[\gauged{C}{\mathbb{Z}_3}]_{\bf 0}^{\bf 1} | =2$. Thus, we obtain the ground state degeneracy for the gauged theories on the torus
\begin{equation}
\mathcal{N}_1^{G} = \sum_{s} \mathcal{N}_{1}^{G}(s) = 6 + 3\times 2 = 12.
\end{equation}
This matches the expected ground state degeneracy on a torus, i.e. the number of topological charge types in the MTC.

For the genus $g=2$ surface, there are 16 defect sectors, corresponding to the possible configurations of nontrivial $\mathbb{Z}_2$-defect branch lines around nontrivial cycles of the surface. Let us choose a basis for the trivial defect sector ($s=0$) that is specified diagrammatically (as in Sec.~\ref{sec:generalGenus}) by
\begin{equation}
|\Phi_{a_1, a_2; z, \mu_1, \mu_2}^{(0)}\rangle\equiv
\pspicture[shift=-0.6](-1.7,-0.4)(2.3,1.5)
	\psset{linewidth=1.0pt,arrowscale=1.3,arrowinset=0.15,linestyle=solid, border=1.5pt}
	\psarc{<-}(-0.3,1){0.45}{-195}{4}
	\psarc{<-}(1.6,1){0.45}{-195}{4}
	\psset{linewidth=1.0pt,arrowscale=1.3,arrowinset=0.15,linestyle=solid, border=0}
	\psarc(-0.3,1){0.45}{0}{215}
	\psarc(1.6,1){0.45}{0}{215}
	\psline(-0.3,-0.1)(-0.3,0.55)
	\psline{->}(-0.3,0.25)(-0.3,0.3)
	\psline(1.6,-0.3)(1.6,0.55)
	\psline{->}(1.6,0.20)(1.6,0.25)
    \psline(-0.3,-0.1)(1.6,-0.3)
	\rput[c]{0}(-0.3,1){\scalebox{1}{$\otimes$}}
	\rput[c]{0}(1.6,1){\scalebox{1}{$\otimes$}}
	\rput[tr]{0}(-0.85,1.1){\scalebox{1}{$a_{1}$}}
	\rput[tr]{0}(1.05,1.1){\scalebox{1}{$a_{2}$}}
	\rput[tl]{0}(-0.2,0.3){$z$}
	\rput[tl]{0}(1.7,0.2){$\bar{z}$}
 \scriptsize
  \rput[bl]{0}(-0.65,0.3){$\mu_1$}
  \rput[bl]{0}(1.25,0.3){$\mu_2$}
\endpspicture
,
\end{equation}
where $a_1, a_2, z \in \gauged{C}{\mathbb{Z}_3}$, and we include the vertex labels $\mu_j \in \{ 1,\ldots , N_{a_j \overline{a_j}}^{z} \}$ because the $V^{\Psi \Psi}_{\Psi}$ fusion space has multiplicity. This state space has ground state degeneracy
\begin{equation}
\mathcal{N}_{2}(0) = \sum_{a_1, a_2, z} N_{a_1 \overline{a_1}}^{z} N_{a_2 \overline{a_{2}}}^{\bar{z}} = 166
.
\end{equation}
The number of basis states that are mapped to themselves by the $\mathbb{Z}_{2}$ action (i.e. the states with labels only from the set $[\gauged{C}{\mathbb{Z}_3}]_{\bf 0}^{\bf 1} = \{ I_0 , \Psi \}$) is given by
\begin{equation}
\mathcal{I}_{2}(0) = \sum_{a_1, a_2, z \in \{ I_0 , \Psi \}} N_{a_1 \overline{a_1}}^{z} N_{a_2 \overline{a_{2}}}^{\bar{z}} = 8
.
\end{equation}
However, the $\mathbb{Z}_{2}$ symmetry action on these states is not necessarily trivial, so not all of them are necessarily invariant under the symmetry. We will see that some of these state transform trivially under the symmetry action, while others transform with a $-1$ factor. We denote the number of basis states that acquire a $-1$ factor by $\mathcal{A}_{2}(0)$. It turns out that $\mathcal{A}_{2}(0) =2$, where the two corresponding basis states are $|\Phi_{\Psi, \Psi; \Psi, 1,2}^{(0)}\rangle$ and $|\Phi_{\Psi, \Psi; \Psi, 2, 1}^{(0)}\rangle$. To see this, we first notice that the $\mathbb{Z}_{2}$ symmetry action is given in terms of the topological symmetry action by
\begin{widetext}
\begin{equation}
R_{\bf 1} |\Phi_{a_1, a_2; z, \mu_1 , \mu_2}^{(0)}\rangle  = \sum_{\nu_1 , \nu_2}  \frac{ [U_{\bf 1}(a_1 , \overline{a_{1}} ; z)]_{\mu_1 \nu_1} [U_{\bf 1}(a_2 , \overline{a_{2}} ; \bar{z})]_{\mu_2 \nu_2}  U_{\bf 1}(z , \bar{z} ; I) }{ U_{\bf 1}(a_1 , \overline{a_{1}} ; I) U_{\bf 1}(a_2 , \overline{a_{2}} ; I) }  |\Phi_{a_1, a_2; z , \nu_1 , \nu_2}^{(0)}\rangle
.
\end{equation}
The action on the basis states with $z=I_0$ is trivial, so we only need to consider those with $a_1=a_2=z = \Psi$ in more detail. In this case, the action reduces to
\begin{equation}
R_{\bf 1} |\Phi_{\Psi, \Psi; \Psi, \mu_1 , \mu_2}^{(0)}\rangle  = \sum_{\nu_1 , \nu_2}  \frac{ [U_{\bf 1}(\Psi , \Psi ; \Psi)]_{\mu_1 \nu_1} [U_{\bf 1}(\Psi , \Psi ; \Psi)]_{\mu_2 \nu_2} }{ U_{\bf 1}(\Psi , \Psi ; I) }  |\Phi_{\Psi, \Psi ; \Psi , \nu_1 , \nu_2}^{(0)}\rangle
.
\end{equation}
\end{widetext}

In order to compute the necessary $U_{\bf 1}$-symbols, we use the $F$-symbols and $R$-symbols given in \cite{Ardonne2010}, and the constraints imposed by their invariance, as in Eqs.~(\ref{eq:rho_g_F}) and (\ref{eq:rho_g_R}). Invariance of
\begin{equation}
R^{\Psi \Psi}_{\Psi} = \left[
\begin{matrix}
-i & 0\\
0 & i
\end{matrix} \right]
\end{equation}
implies that $U_{\bf 1}(\Psi , \Psi ; \Psi) \sigma_{3} U_{\bf 1}(\Psi , \Psi ; \Psi)^{-1} = \sigma_{3}$, i.e. $U_{\bf 1}(\Psi , \Psi ; \Psi)$ is diagonal. Invariance of the $F$-symbols
\begin{equation}
F^{\Psi \Psi \Psi}_{I_{1}}= \left[
\begin{matrix}
    -\frac{1}{2} & -\frac{\sqrt{3}}{2} \\
    \frac{\sqrt{3}}{2} & -\frac{1}{2}
\end{matrix}
\right]
, \quad
F^{\Psi \Psi \Psi}_{I_{2}}=\left[
\begin{matrix}
-\frac{1}{2} & \frac{\sqrt{3}}{2} \\
-\frac{\sqrt{3}}{2} & -\frac{1}{2}
\end{matrix}
\right] ,
\end{equation}
implies that $U_{\bf 1}(\Psi , \Psi ; \Psi) \sigma_{2} U_{\bf 1}(\Psi , \Psi ; \Psi)^{-1} = -\sigma_{2}$. Combining these gives $U_{\bf 1}(\Psi , \Psi ; \Psi) = e^{i \phi} \sigma_{3}$. We note that this shows the symmetry action
\begin{equation}
\label{eq:SU(3)3_nontrivial_symm_action}
\rho_{\bf 1} \ket{\Psi,\Psi;\Psi,\mu} = \sum_\nu U_{\bf 1}(\Psi,\Psi;\Psi)_{\mu\nu}\ket{\Psi,\Psi;\Psi,\nu},
\end{equation}
on the two-dimensional fusion space $V^{\Psi \Psi}_{\Psi}$ is nontrivial.

Finally, invariance of
\begin{equation}
[F^{\Psi \Psi \Psi}_{\Psi}]_{(\Psi , \mu_1 , \mu_2) I_0} =
\left[
\begin{matrix}
\frac{1}{\sqrt{3}} \\
0 \\
0 \\
\frac{1}{\sqrt{3}}
\end{matrix}
\right]_{(\mu_1 , \mu_2)}
,
\end{equation}
where rows are labeled by $(\mu_1 , \mu_2)$ in the order $(1,1)$, $(1,2)$, $(2,1)$, $(2,2)$, combined with $U_{\bf 1}(\Psi , \Psi ; \Psi) = e^{i \phi} \sigma_{3}$, gives $e^{i 2 \phi} = U_{\bf 1}(\Psi , \Psi ; I_{0})$. Thus, we obtain
\begin{equation}
R_{\bf 1} |\Phi_{\Psi, \Psi; \Psi, \mu_1 , \mu_2}^{(0)}\rangle  = (-1)^{\mu_1 + \mu_2}  |\Phi_{\Psi, \Psi ; \Psi , \mu_1 , \mu_2}^{(0)}\rangle
.
\end{equation}
In other words, the action on these basis states is symmetric for $\mu_1 = \mu_2$ and antisymmetric for $\mu_1 \neq \mu_2$.

The basis states that include labels other than $I_0$ and $\Psi$ are mapped to each other in pairs and can be symmetrized with respect to the $\mathbb{Z}_{2}$ symmetry action. Putting this all together, we find the contribution from the $s=0$ sector to the gauged theories' ground state degeneracy to be
\begin{equation}
\mathcal{N}^{G}_{2}(0) = \frac{1}{2} \left[ \mathcal{N}_{2}(0) - \mathcal{I}_{2}(0) \right]  + \mathcal{I}_{2}(0) - \mathcal{A}_{2}(0) = 85
.
\end{equation}

There are 15 nontrivial defect sectors ($s\neq 0$) for the genus $g=2$ surface. We consider the sector ($s=1$) with a single defect branch line around a single cycle, for which we can write the basis states as
\begin{equation}
|\Phi_{a_1, a_2; z,\mu_1 , \mu_2}^{(1)}\rangle\equiv
\pspicture[shift=-0.6](-1.7,-0.4)(2.3,1.5)
    \psset{linewidth=1.3pt,linecolor=blue,arrowscale=1.2,arrowinset=0.15, linestyle=dashed, dash=3pt 1.8pt}
	\psellipse(2.05,1)(0.5,0.15)
    \psset{linewidth=1.0pt,linecolor=black,arrowscale=1.3,arrowinset=0.15,linestyle=solid, border=1.5pt}
	\psarc{<-}(-0.3,1){0.45}{-195}{4}
	\psarc{<-}(1.6,1){0.45}{-195}{4}
	\psset{linewidth=1.0pt,arrowscale=1.3,arrowinset=0.15,linestyle=solid, border=0}
	\psarc(-0.3,1){0.45}{0}{215}
	\psarc(1.6,1){0.45}{32}{215}
	\psline(-0.3,-0.1)(-0.3,0.55)
	\psline{->}(-0.3,0.25)(-0.3,0.3)
	\psline(1.6,-0.3)(1.6,0.55)
	\psline{->}(1.6,0.20)(1.6,0.25)
    \psline(-0.3,-0.1)(1.6,-0.3)
	\rput[c]{0}(-0.3,1){\scalebox{1}{$\otimes$}}
	\rput[c]{0}(1.6,1){\scalebox{1}{$\otimes$}}
	\rput[tr]{0}(-0.85,1.1){\scalebox{1}{$a_{1}$}}
	\rput[tr]{0}(1.05,1.1){\scalebox{1}{$a_{2}$}}
	\rput[tl]{0}(-0.2,0.3){$z$}
	\rput[tl]{0}(1.7,0.2){$\bar{z}$}
	\rput[br]{0}(2.8,1.1){\scalebox{0.8}{$\mathbf{1}$}}
 \scriptsize
  \rput[bl]{0}(-0.65,0.3){$\mu_1$}
  \rput[bl]{0}(1.25,0.3){$\mu_2$}

\endpspicture
,
\end{equation}
where $a_1, a_2, z \in \gauged{C}{\mathbb{Z}_3}$, and $\mu_j \in \{ 1,\ldots , N_{a_j \overline{a_j}}^{z} \}$. Each nontrivial defect sector has ground state degeneracy
\begin{equation}
\mathcal{N}_{2}(s) = \sum_{a_1, a_2, z} N_{a_1 \overline{a_1}}^{z} N_{a_2 \,^{\bf 1}\overline{a_{2}}}^{\bar{z}} = 40
.
\end{equation}

Similar to the trivial sector, the number of the $s=1$ basis states that are mapped to themselves by the $\mathbb{Z}_{2}$ action (i.e. the states with labels only from the set $\{ I_0 , \Psi \}$) is given by
\begin{equation}
\mathcal{I}_{2}(1) = \sum_{a_1, a_2, z \in \{ I_0 , \Psi \}} N_{a_1 \overline{a_1}}^{z} N_{a_2 \overline{a_{2}}}^{\bar{z}} = 8
\end{equation}
and the number of these states that transform as $R_{\bf 1} |\Phi_{a_1, a_2; z , \mu_1 , \mu_2}^{(1)}\rangle = - |\Phi_{a_1, a_2; z, \mu_1 , \mu_2}^{(1)}\rangle $ is $\mathcal{A}_{2}(1) =2$, where the two corresponding basis states are $|\Phi_{\Psi, \Psi; \Psi, 1,2}^{(0)}\rangle$ and $|\Phi_{\Psi, \Psi; \Psi, 2, 1}^{(0)}\rangle$. The basis states that include labels not equal to $I_0$ or $\Psi$ are mapped to each other in pairs and can be symmetrized with respect to the $\mathbb{Z}_{2}$ symmetry action. Thus, we find the contribution from each $s\neq 0$ sector to the gauged theories' ground state degeneracy to be
\begin{equation}
\mathcal{N}^{G}_{2}(s) = \frac{1}{2} \left[ \mathcal{N}_{2}(1) - \mathcal{I}_{2}(1) \right]  + \mathcal{I}_{2}(1) - \mathcal{A}_{2}(1) = 22
.
\end{equation}

Finally, summing the contributions from all defect sectors, we find the ground state degeneracy of the gauged theories on genus $g=2$ surfaces to be
\begin{equation}
\mathcal{N}_{2}^{G} = \sum_{s} \mathcal{N}_{2}^{G}(s) = 85 + 15\times 22 = 415.
\end{equation}
This matches the genus $g=2$ ground state degeneracy obtained from Eq.~(\ref{eq:genus_g_Verlinde}), using the (final) gauged theory's quantum dimensions.

\subsection{$\text{Rep}(\mathbb{D}_{10})$ with $\mathbb{Z}_2$ Symmetry: An $H^3_{[\rho]}(G,\mathcal{A})$ Obstruction}
\label{ex_sec:H3_obstruction}

We provide an example of the $H^3_{[\rho]}(G,\mathcal{A})$ obstruction (i.e. obstruction to symmetry fractionalization) in a pre-modular category~\cite{Cegarra2001}. Consider the dihedral group $\mathbb{D}_{10}=\mathbb{Z}_{10}\rtimes\mathbb{Z}_2$ generated from two elements $r, s$ with group relations $r^{10}=s^2=1$ and $srs=r^{-1}$. It has $8$ irreducible representations, four of which are $1$-dimensional and four of which are $2$-dimensional. We will consider the BTC $\mathcal{C}=\text{Rep}(\mathbb{D}_{10})$. The fusion rules of $\mathcal{C}$ can be easily deduced from the character table of $\mathbb{D}_{10}$, which we spell out explicitly here:
There are four Abelian topological charges, $I$, $A$, $B$, and $C= A \times B$, which form a $\mathbb{Z}_2\times\mathbb{Z}_2$ fusion subalgebra. There are $4$ non-Abelian topological charges $X_{j}$, where $j=1,2,3,4$, which have quantum dimension $2$, such that
\begin{equation}
	\begin{gathered}
	A\times X_j=X_j\\
	B\times X_j=C\times X_j =X_{5-j}\\
	X_j\times X_k=
	\begin{cases}
		X_{j+k}+ X_{|j-k|} & j+k\leq 5\\
		X_{10-j-k} + X_{|j-k|} & j+k>5
	\end{cases}
	\end{gathered}
\end{equation}
where we define $X_0=I+ A$ and $X_5=B+ C$, to make the expressions more compact. The $F$-symbols (or Wigner $6j$-symbols) of this category can be computed from the Clebsch-Gordon coefficients.

In addition, this category also admits braiding. In fact, the representation category of any finite group can be endowed with symmetric braiding, i.e. all topological charges have twist factors $\theta_a = 1$ and $S_{ab}=\frac{d_ad_b}{\mathcal{D}}$ for all topological charges $a, b$, which shows that the BTC is clearly not modular. For a modular theory that contains this as a subcategory, one can embed it in the quantum double $\text{D}(\mathbb{D}_{10})$ as the charge sector, as can always be done for any representation category $\text{Rep}(G)$ of a finite group $G$. (It is further known that the quantum double $\text{D}(G)$ is the minimal modular extension of $\text{Rep}(G)$~\cite{muger2003}.)

We now define an obstructed $\mathbb{Z}_2$ symmetry on Rep$(\mathbb{D}_{10})$. We first define an automorphism $\rho$ on the group $\mathbb{D}_{10}$ by: $\rho(r)=r^7, \rho(s)=r^5s$. We can easily check that $\rho\circ\rho(r)=r^{-1}=srs$ and $\rho\circ\rho(s)=s=sss$, and thus $\rho\circ\rho$ is conjugation by $s$. Therefore, although $\rho$ is not an exact $\mathbb{Z}_2$ automorphism on the group (only a $\mathbb{Z}_2$ outer automorphism), it still induces a $\mathbb{Z}_2$ action on the representations, since representations are defined up to similarity transformations. The explicit action on the label set is found to be
\begin{equation}
\rho_{\bf g}(B)=C, \quad \rho_{\bf g}(X_1) = X_3, \quad \rho_{\bf g}(X_2)=X_4.
\end{equation}
One can check that the fusion rules and modular data are all invariant under this symmetry.

However, by directly checking the definition of the symmetry action, we find that this $\mathbb{Z}_2$ symmetry is not fractionalizable. In other words, it is impossible to fractionalize the symmetry in a manner as described in Sec.~\ref{sec:symmetryfrac}. Therefore, the symmetry is obstructed. Notice that because the Rep$(\mathbb{D}_{10})$ category is not modular, we can not directly relate the obstruction to an obstruction class in $H^3_{[\rho]}(\mathbb{Z}_{2}, \mathcal{A})$. However, as described in Sec. \ref{sec:PSG}, the group outer automorphism can actually be turned into a topological $\mathbb{Z}_2$ symmetry of the quantum double $\text{D}(\mathbb{D}_{10})$. Restricting this topological symmetry to the charge sector of $\text{D}(\mathbb{D}_{10})$, i.e. the $\text{Rep}(\mathbb{D}_{10})$ category, is precise the obstructed symmetry action described, and thus the topological symmetry action on $\text{D}(\mathbb{D}_{10})$ is also obstructed.

\acknowledgements
We would like to thank M.~Hermele, A.~Kitaev, M.~Metlitski, V.~Ostrik, J.~Slingerland, M.~Titsworth, and K.~Walker for enlightening discussions and especially X.~Cui, C.~Galindo, and J.~Plavnik for sharing unpublished work. We would like to thank A.~Tran and Y.-Z.~You for help on figures. This work was performed in part at the Aspen Center for Physics, which is supported by National Science Foundation grant PHY-1607611.

\emph{Note added}: During the preparation of this manuscript, we learned of related unpublished works~\cite{Fidkowski-unpublished,Teo-unpublished}.

\appendix

\section{Review of Group Cohomology}
\label{app:cohomology}

In this appendix, we provide a brief review of group cohomology. (See \Ref{brown} for more details.)

Given a finite group $G$, let $M$ be an Abelian group equipped with a $G$ action $\rho: G \times M \rightarrow M$, which is compatible with group multiplication. In particular, for any $\mathbf{g},\mathbf{h}\in G$ and $a,b \in M$, we have
\begin{eqnarray}
\rho_\mathbf{g} \left( \rho_\mathbf{h}(a) \right) &=& \rho_\mathbf{gh}(a)
\\
\rho_\mathbf{g}(ab) &=& \rho_\mathbf{g}(a) \rho_\mathbf{g}(b)
.
\end{eqnarray}
(We leave the group multiplication symbols implicit.) Such an Abelian group $M$ with $G$ action $\rho$ is called a $G$-module.

Let $\omega(\mathbf{g}_1, \dots,\mathbf{g}_n)\in M$ be a function of $n$ group elements $\mathbf{g}_j \in G$ for $j=1,\dots,n$. Such a function is called a $n$-cochain and the set of all $n$-cochains is denoted as $C^n(G, M)$. They naturally form a group under multiplication,
\begin{equation}
  (\omega\cdot\omega')(\mb{g}_1, \dots, \mb{g}_n)=\omega(\mb{g}_1, \dots, \mb{g}_n)\omega'(\mb{g}_1, \dots, \mb{g}_n),
\end{equation}
and the identity element is the trivial cochain $\omega(\mb{g}_1,\dots,\mb{g}_n)=1$.

We now define the ``coboundary'' map $\mathrm{d}: C^n(G, M) \rightarrow C^{n+1}(G, M)$ acting on cochains to be
\begin{equation}
\begin{split}
\mathrm{d}\omega &(\mathbf{g}_1,\dots,\mathbf{g}_{n+1})= \rho_{\mathbf{g}_1}[\omega(\mathbf{g}_2,\dots,\mathbf{g}_{n+1})] \\
	&\times \prod_{j=1}^n [ \omega(\mathbf{g}_1,\dots,\mathbf{g}_{j-1},\mathbf{g}_j\mathbf{g}_{j+1},\mathbf{g}_{j+2},\dots,\mathbf{g}_{n+1})]^{(-1)^j} \\
    &\times [\omega(\mathbf{g}_1,\dots,\mathbf{g}_{n})]^{(-1)^{n+1}}
.
\end{split}
\end{equation}
One can directly verify that $\mathrm{d} \mathrm{d}\omega=1$ for any $\omega \in C^n(G, M)$, where $1$ is the trivial cochain in $C^{n+2}(G, M)$. This is why $\mathrm{d}$ is considered a ``boundary operator.''

With the coboundary map, we next define $\omega\in C^n(G, M)$ to be an $n$-cocycle if it satisfies the condition $\mathrm{d}\omega=1$. We denote the set of all $n$-cocycles by
\begin{equation}
\begin{split}
Z^n_{\rho}(G, M) &= \text{ker}[\mathrm{d}: C^n(G, M) \rightarrow C^{n+1}(G, M)]  \\
& = \{ \, \omega\in C^n(G, M) \,\, | \,\, \mathrm{d}\omega=1 \, \}.
\end{split}
\end{equation}
We also define $\omega\in C^n(G, M)$ to be an $n$-coboundary if it satisfies the condition $\omega= \mathrm{d} \mu $ for some $(n-1)$-cochain $\mu \in C^{n-1}(G, M)$. We denote the set of all $n$-coboundaries by
\begin{equation}
\begin{split}
& B^n_{\rho}(G, M)  = \text{im}[ \mathrm{d}: C^{n-1}(G, M) \rightarrow C^{n}(G, M) ] \\
& =\{ \, \omega\in C^n(G, M) \,\, | \,\, \exists \mu \in C^{n-1}(G, M) : \omega = \mathrm{d}\mu \, \}
.
\end{split}
\end{equation}

Clearly, $B^n_{\rho}(G, M) \subset Z^n_{\rho}(G, M) \subset C^n(G, M)$. In fact, $C^n$, $Z^n$, and $B^n$ are all groups and the co-boundary maps are homomorphisms. It is easy to see that $B^n_{\rho}(G, M)$ is a normal subgroup of $Z^n_{\rho}(G, M)$. Since d is a boundary map, we think of the $n$-coboundaries as being trivial $n$-cocycles, and it is natural to consider the quotient group
\begin{equation}
H^n_{\rho}(G, M)=\frac{Z^n_{\rho}(G, M)}{B^n_{\rho}(G, M)}
,
\end{equation}
which is called the $n$-th cohomology group. In other words, $H^n_{\rho}(G, M)$ collects the equivalence classes of $n$-cocycles that only differ by $n$-coboundaries.

It is instructive to look at the lowest several cohomology groups. Let us first consider $H^1_{\rho}(G, M)$:
\begin{equation}
	\begin{split}
Z^1_{\rho}(G, M) &= \{\, \omega \, \,| \, \, \omega(\mathbf{g}_1)\rho_\mathbf{g}[\omega(\mathbf{g}_2)]=\omega(\mathbf{g}_1\mathbf{g}_2) \,\} \\
B^1_{\rho}(G, M) &= \{\, \omega \,\, | \,\, \omega(\mathbf{g})=\rho_\mathbf{g}(\mu)\mu^{-1} \, \}
.
\end{split}
\end{equation}
If the $G$-action on $M$ is trivial, then $B^1_{\rho}(G, M) = \{ 1 \}$ and $Z^1_{\rho}(G, M)$ is the group homomorphisms from $G$ to $M$. In general, $H^1_{\rho}(G, M)$ classifies ``crossed group homomorphisms'' from $G$ to $M$.

For the second cohomology, we have
\begin{equation}
	\begin{split}
		Z^2_{\rho}(G, M)&=\{ \, \omega \,\, | \,\, \rho_{\mathbf{g}_1}[\omega(\mathbf{g}_2,\mathbf{g}_3)]\omega(\mathbf{g}_1,\mathbf{g}_2\mathbf{g}_3)\\
		& \qquad \qquad \qquad  = \omega(\mathbf{g}_1,\mathbf{g}_2)\omega(\mathbf{g_1}\mathbf{g}_2,\mathbf{g}_3) \,\}\\
		B^2_{\rho}(G, M)&=\{\, \omega \,\, | \,\, \omega(\mathbf{g}_1,\mathbf{g}_2) \\
& \qquad \qquad =\rho_{\mathbf{g}_1}[\varepsilon(\mathbf{g}_2)][\varepsilon(\mathbf{g}_1\mathbf{g}_2)]^{-1} \varepsilon(\mathbf{g}_1) \, \}
.
\end{split}
\end{equation}
If $M=\mathrm{U}(1)$, it is well-known that $Z^2(G, \mathrm{U}(1))$ is exactly the factor sets (also known as the Schur multipliers) of projective representations of $G$, with the cocycle condition coming from the requirement of associativity. $H^2(G, \mathrm{U}(1))$ classifies all inequivalent projective representations of $G$.

For the third cohomology, we have
\begin{equation}
\begin{split}
Z^3_{\rho}(G, M)=\{ \, \omega \, \, |\,\, \omega(\mathbf{g}_1\mathbf{g}_2, \mathbf{g}_3, \mathbf{g}_4)\omega(\mathbf{g}_1, \mathbf{g}_2, \mathbf{g}_3\mathbf{g}_4)\\
    	=\rho_{\mathbf{g}_1}[\omega(\mathbf{g}_2, \mathbf{g}_3, \mathbf{g}_4)]\omega(\mathbf{g}_1, \mathbf{g}_2\mathbf{g}_3, \mathbf{g}_4)\omega(\mathbf{g}_1, \mathbf{g}_2, \mathbf{g}_3) \, \}
\end{split}
\end{equation}
For $M=\mathrm{U}(1)$ and trivial $G$ action, $Z^3(G, \mathrm{U}(1))$ is the set of $F$-symbols for the fusion category $\text{Vec}_G$, with the $3$-cocycle condition being the Pentagon identity. $B^3(G, \mathrm{U}(1))$ is identified with all the $F$-symbols that are gauge-equivalent to the trivial one. $H^3(G, \mathrm{U}(1))$ then classifies the gauge-equivalent classes of $F$-symbols on $\text{Vec}_G$.

\section{Projective Representations of Finite Groups}
\label{sec:Projective_reps}

In this appendix, we briefly summarize some basic results of the theory of projective representations of finite groups over the complex numbers $\mathbb{C}$ and discuss the unitary case without loss of generality. For proofs, we refer the readers to \Ref{Karpilovsky}.

Consider a finite group $G$ and a normalized $2$-cocycle $\omega \in Z^{2}(G,\text{U}(1))$.  Suppose $V$ is a non-zero vector space over $\mathbb{C}$. A $\omega$-representation of $G$ over the vector space $V$  is a map $\pi: G\rightarrow \mathrm{GL}(V)$ such that
\begin{equation}
  \begin{gathered}
	\pi(\mb{g})\pi(\mb{h})=\omega(\mb{g},\mb{h})\pi(\mb{gh}), \quad \forall \mb{g},\mb{h}\in G\\
	\pi(\mb{0})=\openone.
  \end{gathered}
\end{equation}

We denote the $\omega$-projective representative by a triple $(\omega, \pi, V)$, or for brevity $(\pi, V)$ or simply $\pi$ below.  Also $n_\pi \equiv \mdim V$.

Two $\omega$-representations $(\pi_1, V_1)$ and $(\pi_2, V_2)$ are $\omega$-isomorphic, denoted as $\pi_1\sim_\omega \pi_2$, if and only if there exits an isomorphism $S$ between $V_1$ and $V_2$ such that $S\pi_1(\mb{g})S^{-1}=\pi_2(\mb{g}), \forall \mb{g}\in G$.

Given two $\omega$-representations $(\pi_1, V_1)$ and $(\pi_2, V_2)$, we can form their direct sum, which is a $\omega$-representation of $G$ over $V_1\oplus V_2$. In matrix form, we have
\begin{equation}
  (\pi_1\oplus \pi_2)(\mb{g})\equiv
  \left[\begin{matrix}
	\pi_1(\mb{g}) & 0\\
	0  & \pi_2(\mb{g})
  \end{matrix}\right].
\end{equation}

Clearly $\pi_1\oplus \pi_2$ also has the same factor set $\omega$. However, there is no natural way of defining a direct sum of a $\omega$-representation and a $\omega'$-representation when $\omega\neq \omega'$.

One can also define a tensor product of two projective representations.  Given two projective representations $(\omega_1, \pi_1, V_1)$ and $(\omega_2, \pi_2, V_2)$, their tensor product $\pi_1\otimes\pi_2$ is defined as $(\pi_1\otimes \pi_2)(\mb{g})=\pi_1(\mb{g})\otimes \pi_2(\mb{g})$ over the vector space $V_1\otimes V_2$.  The factor set of the tensor product $\pi_1\otimes\pi_2$ is $\omega_1\omega_2$.

Similar to linear representations, one can define reducible and irreducible projective representations. A projective representation $(\omega, \pi, V)$ is called irreducible if the vector space $V$ has no invariant subspace under the map $\pi$ other than $0$ or $V$. A projective representation is reducible if it is not irreducible. A reducible projective representation always decomposes into a direct sum of irreducible projective representations with the same factor set.

Given a projective representation $\pi$ of $G$, its character $\chi_\pi:G\rightarrow \mathbb{C}$ is defined to be
\begin{equation}
\chi_\pi(\mb{g})=\tr\big[\pi(\mb{g})\big].
\end{equation}
It follows that
\begin{eqnarray}
  \chi_{\pi}(\mb{0})&=& n_\pi \\
\chi_{\pi}(\mb{g}^{-1}) &=& \omega(\mb{g},\mb{g}^{-1})\chi^{*}_{\pi}(\mb{g})
\label{eq:chi_pi_g_inverse}
\end{eqnarray}
where we use the identity $\omega({\bf g, g}^{-1}) = \omega({\bf g}^{-1}{\bf, g})$.

Another more nontrivial relation is
\begin{equation}
\chi_{\pi}(\mb{h}\mb{g}\mb{h}^{-1}) = \frac{\omega(\mb{h}^{-1},\mb{hgh^{-1}})}{\omega(\mb{g},\mb{h}^{-1})}\chi_{\pi}(\mb{g}),
\end{equation}
which reveals an important difference between projective and regular characters, because regular characters depend only on the conjugacy classes.

Given two $\omega$-representations $\pi_1$ and $\pi_2$,  obviously, one has $\chi_{\pi_1\oplus\pi_2}=\chi_{\pi_1}+\chi_{\pi_2}$ and $\chi_{\pi_1\otimes\chi_2}=\chi_{\pi_1}\chi_{\pi_2}$.

As in the theory of linear representations, characters are important because they distinguish the isomorphism classes of irreducible projective representations:

\textit{Two $\omega$-representations are $\omega$-isomorphic if and only if they have the same character.}

Analogous to the familiar character theory of linear representations, one can show that the projective characters satisfy some orthogonality relations. We give the first orthogonality relation here and discuss the second one later. For two irreducible $\omega$-representations $\pi_1$ and $\pi_2$, we have
\begin{equation}
	\frac{1}{|G|}\sum_{\mb{g}\in G}\chi_{\pi_1}(\mb{g})\chi_{\pi_2}^*(\mb{g})=
	\begin{cases}
		1 & \text{if} \;\;\; \pi_1\sim_\omega \pi_2\\
		0 & \text{otherwise}
	\end{cases}
\end{equation}

One can use the characters to decompose projective representations. Namely, fix a factor set $\omega$, let $\pi$ be a projective representation (not necessarily irreducible) of $G$ and $\pi'$ an irreducible projective representation. The multiplicity of $\pi'$ in $\pi$ can be computed by
\begin{equation}
  m(\pi',\pi)=\frac{1}{|G|}\sum_{\mb{g}\in G}\chi_{\pi'}(\mb{g})\chi_{\pi}^*(\mb{g}).
  \label{eqn:multiplicity}
\end{equation}

In general, given two $\omega$-representations $\pi$ and $\pi'$ (neither of which is necessarily irreducible), we define the multiplicity $m(\pi',\pi)$ as
\begin{equation}
  m(\pi',\pi)=\mathrm{dim}\,\Hom_G(V_{\pi'},V_\pi).
\end{equation}
Here $\Hom_G(V_{\pi'},V_\pi)$ is the space of intertwining operators, i.e. linear maps between $V_{\pi'}$ and $V_\pi$ which commute with the $G$ actions. Note that the $G$ action on $V_\pi$ is given exactly by the representation $\pi$. Schur's lemma implies that if $\pi$ is an irrep, then $\Hom_G(V_\pi, V_\pi)=\mathbb{C}\openone_{V_\pi}$, i.e. all intertwiners are scalar multiplications. If $\pi$ and $\pi'$ are irreducible representations that are not isomorphic, then $\Hom_G(V_\pi, V_{\pi'})=0$. Therefore, given two $\omega$-representations $\pi$ and $\pi'$, we can decompose them into the direct sum of $\omega$-irreps: $\pi=\oplus_{j}N_j \pi_j, \pi'=\oplus_{j}N_j' \pi_j$, where $\pi_j$ is the complete set of $\omega$-irreps, and $N_j,N_j'$ are multiplicities, respectively.  Then a general intertwiner $\Phi \in \Hom_G(V_{\pi'}, V_{\pi})$ is of the form
\begin{equation}
  \Phi=\bigoplus_j (M_j \otimes \openone_{V_{\pi_j}}).
\end{equation}
Here $M_j$ is a linear map between $\mathbb{C}^{N_j}$ and $\mathbb{C}^{N_j'}$, i.e. an $N_j\times N_j'$ complex matrix, which can be thought as a vector in an $N_j N_j'$-dimensional complex vector space. It follows that
\begin{equation}
  \mdim \Hom_G(V_{\pi'}, V_{\pi})=\sum_j N_j N_j'.
\end{equation}
For applications, we can show that \eqref{eqn:multiplicity} applies to the general case, too.

A special projective representation, the $\omega$-regular representation, is defined as $R(\mb{g})e_\mb{h}=\omega(\mb{g},\mb{h})e_{\mb{gh}}$, where $\{e_\mb{g}|\mb{g}\in G\}$ is a basis for a $|G|$-dimensional vector space. Its character $\chi_R(\mb{g})=|G|\delta_{\mb{g0}}$. Using \eqref{eqn:multiplicity}, we see that the $\omega$-regular representation is reducible and
 each irreducible projective representation $\pi$ appears exactly $n_\pi$ times in its decomposition. Consequently, we have the following two relations
\begin{equation}
	\sum_{\pi} n_\pi^2=|G|,\;\;\;
  \sum_{\pi} n_\pi \chi_\pi(\mb{g})=|G|\delta_{\mb{g0}}.
\end{equation}
The sum is over all irreducible $\omega$-projective representations $\pi$.

An element $\mb{g}\in G$  is called an $\omega$-regular element if and only if $\omega(\mb{g},\mb{h})=\omega(\mb{h},\mb{g})$ for all $\mb{h}\in N_\mb{g}$, where $N_\mb{g}$ is the centralizer of $\mb{g}$ in $G$. Moreover, $\mb{g}$ is $\omega$-regular if and only if all elements in its conjugacy class $[\mb{g}]$ are $\omega$-regular. This property follows from the $2$-cocycle condition.

Now consider $\mb{h}\in N_{\mb{g}}$, so
\begin{equation}
  \chi_\pi(\mb{g})=\chi_\pi(\mb{h^{-1}gh})=\frac{\omega(\mb{h},\mb{g})}{\omega(\mb{g},\mb{h})}\chi_\pi(\mb{g})
\end{equation}
Therefore, if $\mb{g}$ is not $\omega$-regular, then $\chi_\pi(\mb{g})=0$.
In fact, one can show that an element $\mb{g}$ is $\omega$-regular if and only if $\chi_\pi(\mb{g})\neq 0$ for some irreducible representation $\pi$.
We thus have the following important result:

\textit{For a given factor set $\omega$, the number of non-isomorphic irreducible projective $\omega$-representations of $G$ is equal to the number of $\omega$-regular conjugacy classes of $G$.}

We can now state the second orthogonality relation: Let $\mb{g}_1, \mb{g}_2, \dots$ be a complete set of representatives for $\omega$-regular classes of $G$. For any two $\omega$-regular elements $\mb{g_j}$ and $\mb{g_k}$,
\begin{equation}
  \sum_{\pi}\chi_\pi(\mb{g}_j)\chi_\pi^*(\mb{g}_k)=|N_{\mb{g}_j}|\delta_{jk}.
  \label{eqn:ortho2}
\end{equation}
The sum is over all irreducible $\omega$-projective representations $\pi$.

If two factor sets $\omega$ and $\omega'$ belong to the same equivalence class in $H^2(G,\text{U}(1))$, then we have
\begin{equation}
	\omega'(\mb{g},\mb{h})=\frac{\mu(\mb{g})\mu(\mb{h})}{\mu(\mb{gh})}\omega(\mb{g},\mb{h})
\end{equation}
for some $\mu(\mb{g}): G\rightarrow \text{U}(1)$ with $\mu(\mb{0})=\openone$.

Given a $\mu$ as above and an irreducible $\omega$-projective representation $\pi$, we can
then construct another $\omega'$-projective representation
$\pi'(\mb{g})=\mu(\mb{g})\pi(\mb{g})$. Clearly, the two procedures above define a one-to-one correspondence. Their characters also differ by $\mu$, that is
$\chi_{\pi'}(\mb{g})=\mu(\mb{g})\chi_{\pi}(\mb{g})$.

\section{Gluing $G$-Crossed Theories}
\label{sec:gluing}

In this appendix, we describe a construction that we call ``gluing'' $G$-crossed theories, which takes two $G$-crossed theories (with the same symmetry group $G$) and forms a new $G$-crossed theory by combing objects from the two theories that have the same group label ${\bf g}\in G$. Mathematically, this construction is the diagonally $G$-graded product of the two theories. We begin by considering the product $\mathcal{C}_{G}^{(1)\times} \boxtimes \mathcal{C}_{G}^{(2)\times}$ of two $G$-crossed BTCs (labeled 1 and 2). This yields a $(G\times G)$-crossed BTC, whose ${\bf 0}$-sector is $\mathcal{C}_{\bf 0}^{(1)} \boxtimes \mathcal{C}_{\bf 0}^{(2)}$. It is clear that the basic data of the product may be expressed simply as the product of the basic data of the two theories. Next we take the restriction of the product to the subcategory in which the group labels of topological charges are in the diagonal of $G \times G$, that is $({\bf g}^{(1)},{\bf g}^{(2)} ) = ({\bf g},{\bf g})$ for ${\bf g} \in G$. The result is the glued theory
\begin{equation}
\mathcal{C}_{G}^{\times} =  \mathcal{C}_{G}^{(1)\times} \underset{G}{\boxtimes} \mathcal{C}_{G}^{(2)\times} = \left. \mathcal{C}_{G}^{(1)\times} \boxtimes \mathcal{C}_{G}^{(2)\times} \right|_{{\bf g}^{(1)}={\bf g}^{(2)}}
.
\end{equation}
Since the diagonal of $G \times G$ is a subgroup that is isomorphic to $G$, it is clear that the glued theory is a (closed) subcategory of the product theory and that it forms a $G$-crossed BTC. Similarly, $\mathcal{C}_{G}^{\times}$ is modular if and only if both $\mathcal{C}_{G}^{(1)\times}$ and $\mathcal{C}_{G}^{(2)\times}$ are modular. Since the glued theory $\mathcal{C}_{G}^{\times}$ can be written as the restriction of the product theory, the basic data of the glued theory can be expressed as the product of the basic data of $\mathcal{C}_{G}^{(1)\times}$ and $\mathcal{C}_{G}^{(2)\times}$ (while respecting the restriction); for example
\begin{equation}
R^{ (a_{\bf g}^{(1)}  , a_{\bf g}^{(2)}  ) (b_{\bf h}^{(1)}  , b_{\bf h}^{(2)}  )}_{(c_{\bf gh}^{(1)}  , c_{\bf gh}^{(2)}  )}
=R^{ a_{\bf g}^{(1)} b_{\bf h}^{(1)}  }_{c_{\bf gh}^{(1)}} R^{ a_{\bf g}^{(2)} b_{\bf h}^{(2)} }_{c_{\bf gh}^{(2)} }
.
\end{equation}

The ${\bf 0}$-sector of the glued theory is $\mathcal{C}_{\bf 0} = \mathcal{C}_{\bf 0}^{(1)} \boxtimes \mathcal{C}_{\bf 0}^{(2)}$, so the gluing construction provides a method of generating $G$-crossed extensions of $\mathcal{C}_{\bf 0}^{(1)} \boxtimes \mathcal{C}_{\bf 0}^{(2)}$ from $G$-crossed extensions of $\mathcal{C}_{\bf 0}^{(1)}$ and $\mathcal{C}_{\bf 0}^{(2)}$. Moreover, when $\mathcal{C}_{\bf 0}^{(1)}$ and $\mathcal{C}_{\bf 0}^{(2)}$ are MTCs, the gluing construction produces \emph{all} possible $G$-crossed extensions of $\mathcal{C}_{\bf 0}^{(1)} \boxtimes \mathcal{C}_{\bf 0}^{(2)}$ in which the symmetry action does not interchange topological charge labels between the two theories, i.e. we can write $\rho = (\rho^{(1)},\rho^{(2)})$, where $\rho^{(j)} : G \rightarrow \text{Aut}(\mathcal{C}_{\bf 0}^{(j)})$. In order to see this, we note that, for such symmetry actions, $H^{2}_{[\rho]}(G,\mathcal{A}) = H^{2}_{[\rho^{(1)}]}(G,\mathcal{A}^{(1)}) \times H^{2}_{[\rho^{(2)}]}(G,\mathcal{A}^{(2)})$ and the associated classification of symmetry fractionalization respects the product structure. Furthermore, the $H^{3}(G,\text{U}(1))$ defectification classification is recovered from the glued theories by observing that the torsorial action of gluing an SPT$_{G}^{[\alpha]}$ state to $\mathcal{C}_{G}^{\times}$ is the same as gluing it to either $\mathcal{C}_{G}^{(1)\times}$ or $\mathcal{C}_{G}^{(2)\times}$ prior to gluing them together.

\section{Categorical Formulation of Symmetry Fractionalization, Defects, and Gauging}
\label{sec:Cat_Formulation}

In this appendix, $G$ will always denote a finite group and $\mcc$ a unitary modular tensor category (UMTC) unless otherwise stated explicitly.  Also ${\mathrm{Aut}(\mathcal{C})}$ below is ${\mathrm{Aut_{0,0}}(\mathcal{C})}$ in the main text.  For a category $\mcc$, $x\in \mcc$ means that $x$ is an object of $\mcc$, and $\overline{\mcc}$ is the complex conjugate category of $\mcc$.  The materials in this appendix are distilled from \Refs{turaev2010,muger2004,kirillov2004,DGNO2009,ENO2009,Cui_gauging}.

\subsection{Categorical Topological and Global Symmetry}

A categorical-group $\mathcal{G}$ is a monoidal category $\mathcal{G}$ whose objects and morphisms are all invertible.  The complete invariant of a categorical-group $\mcg$ is the triple $(\pi_1(\mcg),\pi_2(\mcg), \phi(\mcg))$, where $\pi_1(\mcg)$ is the group of the isomorphism classes of objects of $\mcg$, $\pi_2(\mcg)$ the abelian group of the automorphisms of the tensor unit $\mathbf{1}$ of $\mcg$, and $\phi(\mcg) \in H^3(\pi_1(\mcg);\pi_2(\mcg))$ the group $3$-cocyle that represents the associativity of the tensor product $\otimes$ of $\mcg$  ($\pi_1(\mcg)$ acts on $\pi_2(\mcg)$ and they form a cross module as the notation suggests)~\cite{baez2004higher}.

A group $G$ can be promoted to a categorical-group $\underline{G}$ as follows:  the objects of $\underline{G}$ are the group elements of $G$, and the morphism set $\textrm{Hom}(\mb{g},\mb{h})$ of two objects $\mb{g,h}$ is empty if $\mb{g}\neq \mb{h}$ and contains only the identity if $\mb{g}=\mb{h}$.  We will use $\underline{\mathrm{Aut}(\mathcal{C})}$ to denote the categorical-group of braided tensor autoequivalences of $\mathcal{C}$.  The tensor product of two braided tensor autoequivalences is their composition.  The morphism between two braided tensor autoequivalences are the natural isomorphisms between the two functors.  We will call $\underline{\mathrm{Aut}(\mathcal{C})}$ the categorical topological symmetry group of $\mcc$.

Given a UMTC $\mcc$, $\pi_2(\mcc)$, i.e. $\pi_2(\mcg)$ for $\mcg=\underline{\mathrm{\mathrm{Aut}}(\mathcal{C})}$, is isomorphic to the group of the invertible object classes of $\mcc$ as an abstract finite abelian group, which we denote by $\mathcal{A}$ in the main text, but the finite group $\pi_1(\mcc)$, i.e. $\pi_1(\mcg)$ for $\mcg=\underline{\mathrm{Aut}(\mathcal{C})}$, is difficult to determine in general except for abelian modular categories.

We will also use $\mathrm{Aut}(\mathcal{C})$ to denote $\pi_1(\mcc)$: the group of equivalence classes of braided tensor autoequivalences of $\mcc$.  This ordinary group is the demotion (or decategorification) of the categorical-group $\underline{\mathrm{Aut}(\mathcal{C})}$ and is called the topological symmetry of $\mcc$.

\begin{definition}

Given a group $G$, a monoidal functor $\underline{\rho}: \underline{G}\rightarrow \underline{\mathrm{Aut}(\mathcal{C})}$ is called a categorical global symmetry of $\mcc$.

We will denote the categorical global symmetry as $(\underline{\rho}, G)$ or simply $\underline{\rho}$ and say that $G$ acts categorically on $\mcc$.

A categorical global symmetry can be demoted to a group homomorphism $\rho: G\rightarrow \mathrm{Aut}(\mathcal{C})$, which is called a global symmetry of $\mcc$.

\end{definition}

To understand a categorical-group action $G$ on a UMTC $\mcc$, we will start with a global symmetry $\rho: G \rightarrow \mathrm{Aut}(\mathcal{C})$.  It is not true that we can always lift such a group homomorphism to a categorical-group functor $\underline{\rho}$.  The obstruction for the existence of such a lifting is the pull-back group cohomology class ${\rho}^*(\phi(\mcc)) \in H^3(G;\pi_2(\mcc))$ of $\phi(\mcc) \in H^3(\pi_1(\mcc),\pi_2(\mcc))$ by $\rho$.  If this obstruction class does not vanish, then $G$ cannot act categorically on $\mcc$ so that the decategorified homomorphism is $\rho$.  If this obstruction does vanish, then there are liftings of $\rho$ to categorical-group actions, but such liftings are not necessarily unique.  The equivalence classes of all liftings form a torsor over $H^2_{\rho}(G,\pi_2(\mcc))$.  We will denote the categorical global symmetry $\underline{\rho}$ also by a pair $(\rho, \coho{t})$, where $\rho: G \rightarrow \mathrm{Aut}(\mathcal{C})$ and $\coho{t}\in H^2_{\rho}(G,\pi_2(\mcc))$.

\subsection{Symmetry Defects}

A module category $\mathcal{M}$ over a UMTC $\mcc$ is a categorical representation of $\mcc$.  A left module category $\mcm$ over $\mcc$ is a semi-simple category with a bi-functor $\alpha_{\mcm}: \mcc\times \mcm \rightarrow \mcm$ that satisfies the analogues of pentagons and the unit axiom.  Similarly for a right module category.
A bi-module category is a simultaneously left and right module category such that the left and right actions are compatible.  Bi-module categories can be tensored together just like bi-modules over algebras.  When $\mcc$ is braided, a left module category naturally becomes a bi-module category by using the braiding.  A bi-module category $\mcm$ over $\mcc$ is invertible if there is another bi-module category $\mathcal{N}$ such that $\mcm \boxtimes \mathcal{N}$ and $\mathcal{N} \boxtimes \mcm$ are both equivalent to $\mcc$---the trivial bi-module category over $\mcc$.  The invertible (left) module categories over a modular category $\mcc$ form the Picard categorical-group $\underline{\textrm{Pic}(\mcc)}$ of $\mcc$.  The Picard categorical-group $\underline{\textrm{Pic}(\mcc)}$ of a modular category $\mcc$ is monoidally equivalent to the categorical-group $\underline{\mathrm{Aut}(\mathcal{C})}$~\cite{ENO2009}.  This one-one correspondence between braided auto-equivalences and invertible module categories is an important relation between symmetry and extrinsic topological defects.

Given a categorical global symmetry $(\underline{\rho}, G)$ of a UMTC $\mcc$ and an isomorphism of categorical groups $\Pic$ with $\underline{\mathrm{Aut}(\mathcal{C})}$, then each $\underline{\rho_{\mb{g}}}\in \underline{\mathrm{Aut}(\mathcal{C})}$ corresponds to an invertible bi-module category $\mcc_\mb{g}\in \Pic$.

\begin{definition}

  An extrinsic topological defect of flux $\mb{g}\in G$ is a simple object in the invertible module category $\mathcal{C}_\mb{g}\in \underline{\mathrm{Pic}(\mcc)}$ over $\mcc$ corresponding to the braided tensor autoequivalence $\underline{\rho_\mb{g}}\in \underline{\mathrm{Aut}(\mathcal{C})}$.

\end{definition}

The analogue of the Picard categorical-group of a modular category for a fusion category $\mcc$ is the Brauer-Picard categorical-group of invertible bi-module categories over $\mcc$.  But invertible bi-module categories over a fusion category $\mcc$ is in one-one correspondence with braided auto-equivalences of the Drinfeld center $\mathrm{D}(\mcc)$ of $\mcc$ (also known as the quantum double of $\mcc$ in physics literature)~\cite{ENO2009}, not tensor auto-equivalences of $\mcc$ itself.  When $\mcc$ is modular, then $\mathrm{D}(\mcc)\cong \mcc\otimes \overline{\mcc}$.  Note that $\Pic$ is naturally included in the Brauer-Picard group of $\mcc$ and $\underline{\mathrm{Aut}(\mathcal{C})}$ included naturally in the categorical-group of braided tensor auto-equivalences of $\mathrm{D}(\mcc)$.  The images of the two inclusions intersect trivially.

The topological defects in the $\mb{g}$-flux sector form an invertible bi-module category $\mcc_\mb{g}$ over the UMTC $\mcc$.  Defects can be fused and their fusion corresponds to the tensor product of bi-module categories. Since all defects arise from the same physics, fusions of defects for all flux sectors should be consistent.  Such a consistency is encoded as the collection $\{\mcc_\mb{g}\}, \mb{g}\in G$ of flux sectors gives rise to an extension of $\mcc$ to a unitary $G$-crossed modular category.
Given a categorical global symmetry $(\rho, \coho{t})$, it is not always possible to define defect fusions so that we could obtain such an extension.  Given fluxes $\mb{g,h}$, we need to choose an identification $M_{\mb{gh}}: \mcc_\mb{g}\boxtimes \mcc_\mb{h} \cong \mcc_{\mb{gh}}$.  For four fluxes $\mb{g,h,k,l}\in G$, the two paths of the pentagon using the $\{M_{\mb{gh}}\}$'s to identify $((\mcc_\mb{g}\boxtimes \mcc_\mb{h}) \boxtimes \mcc_\mb{k}) \boxtimes \mcc_\mb{l}$ with $\mcc_\mb{g}\boxtimes (\mcc_\mb{h} \boxtimes (\mcc_\mb{k} \boxtimes \mcc_\mb{l}))$ could differ by a phase.  The collection of those phases forms a cohomology class in $H^4(G;\mathrm{U}(1))$, which is the obstruction class to consistent pentagons for the flux sectors.  If this obstruction class vanishes, then we need to choose a group cohomology class $\alpha\in H^3(G;\mathrm{U}(1))$ to specify the associativity of the flux sectors.   A subtle point here is that the consistency requirement via pentagons for flux sectors $\mcc_\mb{g}$ is strictly stronger than that for all defects separately.

Given a triple $(\rho, \coho{t}, \alpha)$ as above when the obstruction class in $H^4(G,\mathrm{U}(1))$ vanishes, where $(\rho, \coho{t})$ is a categorical global symmetry and $\alpha\in H^3(G,\mathrm{U}(1))$ specifies associtivity of the flux sectors, we can construct a $G$-crossed modular extension of $\mcc$, which describes the extrinsic topological defects of $\mcc$. In the following, we will call such a triple $(\rho, \coho{t}, \alpha)$ {\it a gauging data}.   The extension $\mcc_G^{\times}=\mcc_{(\rho,\cohosub{t}, \alpha)}=\bigoplus_{\mb{g}\in G}\mcc_\mb{g}$ of $\mcc=\mcc_\mb{0}$ is a unitary $G$-crossed modular category---a unitary $G$-crossed fusion category with a compatible non-degenerate $G$-braiding.

A $G$-grading of a fusion category $\mathcal{C}$ is a decomposition of $\mcc$ into $\bigoplus_{\mb{g}\in G}\mathcal{C}_\mb{g}$.  We will consider only faithful $G$-gradings so that none of the components $\mcc_\mb{g}=0$.  The tensor product respects the grading in the sense $\mathcal{C}_\mb{g}\boxtimes \mathcal{C}_\mb{h}\subset \mathcal{C}_{\mb{gh}}$.  Since $\mcc_{\mb{g}^{-1}}$ is the inverse of $\mcc_\mb{g}$, $\mcc_\mb{g}$ is naturally an invertible bi-module category over $\mcc_\mb{0}$, where $\mb{0}\in G$ is the identity element. A categorical action $\rho$ of $G$ on $\mathcal{C}$ is compatible with the grading if $\underline{\rho}(\mb{g}) \mathcal{C}_\mb{h} \subset \mathcal{C}_{\mb{ghg}^{-1}}$.  A $G$-graded fusion category $\mcc$ with a compatible $G$-action is called a $G$-crossed fusion category.

Suppose $\mcc_G^{\times}=\bigoplus_{\mb{g}\in G}\mcc_\mb{g}$ is an extension of a unitary fusion category $\mcc_\mb{0}$, i.e. $\mcc_G^{\times}$ is a unitary $G$-crossed fusion category.  Let $I_\mb{g}, \mb{g}\in G$ be the set of isomorphism classes of simple objects in $\mcc_\mb{g}$ and $\mathrm{Irr}(\mcc_\mb{g})=\{X_i\}_{i\in I_\mb{g}}$ be a set of representatives of simple objects of $\mcc_\mb{g}$.  The cardinality of $I_\mb{g}$ is called the rank of the component $\mcc_\mb{g}$, and $\mathcal{D}_\mb{g}=\sqrt{\sum_{i\in I_\mb{g}} d^2_i}$ is the total quantum dimension of component $\mcc_\mb{g}$, where $d_i$ is the quantum dimension of $X_i \in \mathrm{Irr}(\mcc_\mb{g})$.

\begin{theorem}[{[\onlinecite{turaev2010,kirillov2004}]}]
  Let $\mcc=\bigoplus_{\mb{g}\in G}\mcc_\mb{g}$ be an extension of a unitary fusion category $\mcc_\mb{0}$.  Then
\begin{enumerate}

  \item The rank of $\mcc_\mb{g}$ is the number of fixed points of the action of $\mb{g}$ on $I_\mb{0}$.

  \item $\mathcal{D}^2_{\mb{g}}=\mathcal{D}^2_{\mb{h}}$ for all $\mb{g,h}\in G$.

\end{enumerate}
\end{theorem}

The extension $\mcc_G^{\times}=\bigoplus_{\mb{g}\in G}\mcc_\mb{g}$ of a UMTC $\mcc_\mb{0}$ for the symmetry $(\rho, \coho{t})$, while not braided in general, has a $G$-crossed braiding.  Given a $G$-crossed fusion category $\mcc$ with categorical $G$-action $\underline{\rho}$, we will denote $\underline{\rho_\mb{g}}(Y)$ for an object $Y$ of $\mcc$ by ${}^\mb{g}Y$.  A $G$-braiding is a collection of natural isomorphisms $c_{X,Y}: X\otimes Y \rightarrow {}^\mb{g}Y\otimes X$ for all $X\in \mcc_\mb{g}, Y\in \mcc$, which satisfies a generalization of the Hexagon equations.

A UMTC is a unitary fusion category with a non-degenerate braiding.  A unitary $G$-crossed modular category is a unitary $G$-crossed fusion category with a non-degenerate $G$-braiding.  An easy way to define non-degeneracy of braiding is through the non-degeneracy of the modular $S$-matrix.  To define the non-degeneracy of the $G$-crossed braiding, we will introduce the extended $G$-crossed $\mathcal{S}$ and $\mathcal{T}$ operators on an extended Verlinde algebra.  Likewise, the extended $\mathcal{S}$ and $\mathcal{T}$ operators will give rise to a projective representation of $\mathrm{SL}(2,\mathbf{Z})$.  We believe that the $\mathcal{S}$ and $\mathcal{T}$ operators will determine the unitary $G$-crossed modular category $\mcc_G^{\times}$.

\begin{theorem}[{[\onlinecite{ENO2009}]}]

The unitary $G$-crossed fusion category extension $\mcc_G^{\times}$ of a UMTC $\mcc$ has a canonical $G$-braiding and categorical $G$-action that make $\mcc_G^{\times}$ into a unitary $G$-crossed modular category.

\end{theorem}

Given a categorical global symmetry $\underline{\rho}: \underline{G}\rightarrow \underline{\mathrm{Aut}(\mathcal{C})}$ of a UMTC $\mcc$, an extension of $\mcc$ to a non-degenerate braided fusion category corresponds to a lifting of $\underline{\rho}$ to a categorical $2$-group functor $\underline{\underline{\rho}}: \underline{\underline{G}} \rightarrow \underline{\underline{\mathrm{Aut}(\mathcal{C})}}$.  The existence of such liftings has an obstruction in $H^4(G;\mathrm{U}(1))$, which is the same as the obstruction for solving pentagons of flux sectors.   When the obstruction class vanishes, the choices correspond to cohomology classes in $H^3(G;\mathrm{U}(1))$.  If we choose a cohomology class $\alpha\in H^3(G;\mathrm{U}(1))$, then we have a lifting to a categorical $2$-group morphism.  Since all other higher obstruction classes vanish, the categorical global symmetry can be lifted to a morphism of any higher categorical number.  As extended $G$-action and $G$-braiding are higher categorical-number morphisms, so they can always be lifted.  Furthermore, since all higher obstruction classes vanish, the liftings are unique.

To see the $G$-action and $G$-crossed braiding concretely, consider the functor category $\textrm{Fun}(\mcc_\mb{g}, \mcc_{\mb{gh}})$.  On one hand, this category can be identified as $\mcc_\mb{h}$ by $\mcc_\mb{g}\boxtimes \mcc_\mb{h} \cong \mcc_{\mb{gh}}$, and on the other hand, as $\mcc_{\mb{ghg}^{-1}}$ by $\mcc_{\mb{ghg}^{-1}}\boxtimes \mcc_\mb{g}\cong \mcc_{\mb{gh}}$.  Therefore, we have an isomorphism $\mcc_\mb{g}\cong \mcc_{\mb{ghg^{-1}}}$.  This defines an extended action of $G$ on $\mcc_G^{\times}$.  By the same consideration, we have $\mcc_\mb{g}\boxtimes \mcc_\mb{h} \cong \mcc_{\mb{ghg^{-1}}} \boxtimes \mcc_\mb{g}$.  This defines the $G$-crossed braiding of $\mcc_G^{\times}$.

To define the extended $\mathcal{S}, \mathcal{T}$-operators, we first define an extended Verlinde algebra.  For each pair $\mb{g,h}$ of commuting elements of $G$, we define the following extended Verlinde algebra component:
$\mathcal{V}_{\mb{g,h}}(\mcc)=\bigoplus_{i\in I_\mb{h}} \textrm{Hom}(X_i, {}^\mb{g}X_i).$

Then the extended Verline algebra is
$$\mathcal{V}({\mcc})=\bigoplus_{\{(\mb{g,h})|\mb{gh}=\mb{hg}\}} \mathcal{V}_{\mb{g},\mb{h}}(\mcc).$$

Note that $\mathcal{V}_{\mb{0},\mb{0}}$ is the Verlinde algebra of $\mcc$, which has a canonical basis given by the identity morphisms of $\textrm{Hom}(X_i,X_i), i\in I_\mb{0}$.  Unlike the usual Verlinde algebra of $\mcc$, the extended Verlinde algebra does not have such canonical basis.  One choice of basis is $\underline{\rho_\mb{g}}: X_i\rightarrow {}^\mb{g}X_i$, and they will give rise to extended $G$-crossed $\mathcal{S}$ and $\mathcal{T}$ transformations.  However, this depends on the choice of cocycle representative of $\alpha$.  Therefore, the extended $\mathcal{S}$ and $\mathcal{T}$ operators are not canonically matrices.  We call a $G$-crossed braided spherical fusion category $G$-crossed modular if the extended $\mathcal{S}$ operator is invertible.

\subsection{Gauging Categorical Global Symmetry}

Let $\underline{\underline{G}}$ be the promotion of a group $G$ to a categorical $2$-group, and $\underline{\underline{\mathrm{Aut}(\mathcal{C})}}$ be the categorical $2$-group of braided tensor auto-equivalences.

\begin{definition}

A categorical global symmetry $\underline{\rho}: \underline{G}\rightarrow \underline{\mathrm{Aut}(\mathcal{C})}$ can be gauged if $\underline{\rho}$ can be lifted to a categorical $2$-group functor $\underline{\underline{\rho}}: \underline{\underline{G}} \rightarrow \underline{\underline{\mathrm{Aut}(\mathcal{C})}}$.

\end{definition}

Given a categorical global symmetry $(\underline{\rho},G)$ of a UMTC $\mcc$, gauging $G$ is possible only when the obstruction as above in $H^4(G;\mathrm{U}(1))$ vanishes.  Then the gauging result in general depends on a gauging data $(\rho, \coho{t}, \alpha)$.  Given a gauging data $(\rho, \coho{t}, \alpha)$,  gauging is defined as the following two-step process: first extend $\mcc$ to a unitary $G$-crossed modular category $\mcc_G^{\times}$ with a categorical $G$ action; Then perform the equivariantization of the categorical $G$ action on $\mcc_G^{\times}$, which results in a UMTC ${(\mcc_G^{\times})}^G$, also simply denoted as $\mcc/G$.  The \lq\lq bosonic" symmetric category $\mathrm{Rep}(G)$ is always contained in $\mcc/G$ as a Tannakian subcategory.  Therefore, gauging actually leads to a pair $\mathrm{Rep}(G)\subset \mcc/G$.

Suppose $\mcc$ is a fusion category with a $G$ action.  The equivariantization of $\mcc$, denoted as $\mcc^G$, is also called orbifolding.  The result of equivariantization of a $G$-action on a fusion category $\mcc$ is a fusion category whose objects are $(X, \{\phi_\mb{g}\}_{\mb{g}\in G})$, where $X$ is an object of $\mcc$ and $\phi_\mb{g}: {}^\mb{g}X\rightarrow X$ an isomorphism such that $\phi_\mb{0}=\mathrm{id}$ and $\phi_\mb{g}\cdot \underline{\rho_\mb{g}}(\phi_\mb{h})=\phi_{\mb{gh}}\cdot \kappa_{\mb{g},\mb{h}}$, where $\kappa_{\mb{g},\mb{h}}$ identifies $\underline{\rho_\mb{h}}\cdot\underline{\rho_\mb{g}}$ with $\underline{\rho_\mb{gh}}$.  Morphisms between two objects $(X, \{\phi_\mb{g}\}_{\mb{g}\in G})$ and $(Y, \{\psi_\mb{g}\}_{\mb{g}\in G})$ are morphisms $f: X\rightarrow Y$ such that $f\cdot \phi_\mb{g}=\phi_\mb{h}\cdot \underline{\rho_\mb{g}}(f)$.

The simple objects of $\mcc^G$ are parameterized by pairs $([X],\pi_X)$, where $[X]$ is an orbit of the $G$-action on simple objects of $\mcc$, and $\pi_X$ is an irreducible projective representation of $G_X$---the stabilizer group of $X$.  The quantum dimension of $([X],\pi_X)$ is $\textrm{dim}(\pi_X)\cdot N_{[X]}\cdot d_X$, where $N_{[X]}$ is the size of the orbit $[X]$.  Fusion rules can be similarly described using algebraic data~\cite{Burciu2013}.

In general, gauging is difficult to perform explicitly.  The first extension step is very difficult.  The second equivariantization step is easier if the $6j$-symbols of the gauged UMTC $\mcc/G$ are not required explicitly. Different triples of gauging data might lead to the same gauged UMTC.

Gauging has an inverse process, which is the condensation of anyons in the Tannakian subcategory $\mathrm{Rep}(G)$.  This condensation process is mathematically called taking the core of the pair $\mathrm{Rep}(G)\subset \mcc/G$~\cite{DGNO2009}.  Taking a core is a powerful method to verify a guess for gauging because anyon condensation is sometimes easier to carry out than gauging.

When $\mcc$ is a $G$-crossed modular category with faithful grading, then its equivariantization is also modular and vise versa~\cite{kirillov2004}.  There is the forgetful functor $F: \mcc^G\rightarrow \mcc$ by $F(X, \{\phi_\mb{g}\}_{\mb{g}\in G})=X$ and its adjoint $G(X)=\bigoplus_{\mb{g}\in G} ({}^\mb{g}X, \{(\mu_X)_\mb{g}\})$, where $(\mu_X)_\mb{g}=\kappa_{\mb{g},\mb{h}}$.  They intertwine the extended $\mathcal{S}, \mathcal{T}$ operators.

Our equivariantization in gauging is applied to a $G$-crossed extension $\mcc_G^{\times}$ of a modular category $\mcc$.  When $\mcc_G^{\times}$ has a faithful grading, then the non-degeneracy of the braiding of $\mcc$ is equivalent to the non-degeneracy of the braiding of $\mcc_G^{\times}$~\cite{DGNO2009}.

\subsection{General Properties of Gauging}

Gauging and its inverse -- condensation of anyons -- are interesting constructions of new modular categories from old ones.  The resulted new modular categories have many interesting relations with the old ones.

\begin{theorem}[{[\onlinecite{DGNO2009}]}]

Let $\mcc$ be a UMTC with a categorical global symmetry $(\rho,G)$.  Then
$\mcc \otimes \overline{\mcc/G}\cong \mathrm{D}(\mcc_G^{\times})$.

It follows that

\begin{enumerate}

\item Chiral topological central charge is invariant under gauging (mod $8$).

\item The total quantum dimension $\mathcal{D}_{\mcc/G}=|G| \mathcal{D}_{\mcc}$.

\end{enumerate}
\end{theorem}

The following theorem says that gauging a quantum double results in a quantum double.

\begin{theorem}
  Suppose $G$ acts categorically on $\mathrm{D}(\mcc)$. Then $\mathrm{D}(\mcc)/G=\mathrm{D}(\mcc_G^{\times})$.
  \label{theorem:gaugedouble}
\end{theorem}

When the symmetry group $G$ has a normal subgroup $N$, then we can first gauge $N$, and then gauge their quotient $H=G/N$.   This sequentially gauging is useful for gauging non-abelian groups $G$ such as $S_3$.

\begin{theorem}
  \label{theorem:seqgauge}
  Let $\rho : \underline{\underline{G}} \longrightarrow \underline{\Pic}$, then there exist $\rho_1: \underline{\underline{N}} \longrightarrow \underline{\Pic}, \, \rho_2 : \underline{\underline{H}} \longrightarrow \underline{\underline{\mathrm{Pic}({\mcc/N})}}$, such that $(\mcc/N)/H$ is braided equivalent to $\mcc/G$.
\end{theorem}

Proofs of Theorems \ref{theorem:gaugedouble} and \ref{theorem:seqgauge} will appear in \Ref{Cui_gauging}.

The construction from a modular category $\mcc$ with a $G$-action to a modular category $\mcc/G$ with a Tannakian subcategory $\mathrm{Rep}(G)$ by gauging can be regarded as a new way to construct interesting modular categories in the same Witt class.  When $\mcc$ is weakly integral, then the gauged category $\mcc/G$ is also weakly integral.  The inverse process of condensation implies that pairs $(\mcc, \underline{\rho})$ and $(\mcc, \mathrm{Rep}(G))$ are in one-one correspondence.

\subsection{New Mathematical Results}

In higher category theory, it is common practice to strictify categories as much as possible by turning natural isomorphisms into identities.  This is desirable because strictification simplifies many computations and does not lose any generality when we are interested in gauge invariant quantities in classification problems.  The drawback is that we have to work with many objects.  In this paper, our preference is the opposite, in the sense that we would like to work with as few objects as possible.  Hence, our goal is to have a skeletal formulation with full computational power, so that we can calculate numerical quantities, such as amplitudes of quantum processes, which are not necessarily gauge invariant.  A category is skeletal if there is only one object in each isomorphism class, and in general strictness and skeletalness cannot be obtained simultaneously, as may be demonstrated, for example, by the semion theory $\mathbb{Z}_{2}^{(1/2)}$. Therefore, we need to skeletonize the existing mathematical theories.  The situation is analogous to the one of a connection or gauge field: mathematically it is good to define a connection as a horizontal distribution, while, in practice, it is better to work with Christoffel symbols, especially in physics.

Our first mathematical result is a skeletonization of $G$-crossed braided fusion category in Sec.~\ref{sec:Algebraic_Theory}. We provide a definition of a $G$-crossed braided category using a collection of quantities organized into a basic data set: $N^c_{ab}, F^{abc}_d, R^{ab}_c, {U}_{\bf k}\left( a ,b ;c \right)$, and ${\eta}_{x}({\bf g,h})$ that satisfy certain consistency polynomial equations.  The fusion coefficients $N^c_{ab}$ and associativity $F$-symbols $F^{abc}_d$ are as usual, but the $R$-symbols $R^{ab}_c$ are extended to incorporate the $G$-crossed structure. The new data ${U}_{\bf k}\left( a ,b ;c \right)$ and ${\eta}_{x}({\bf g,h})$, respectively encode the categorical symmetry: monoidal functors and natural identifications $\rho_{\bf gh}$ with $\rho_{\bf g}\rho_{\bf h}$.  A good example of new consistency equations are our $G$-crossed Heptagon Eqs.~(\ref{eq:heptagon+}) and (\ref{eq:heptagon-}), which generalize the usual Hexagon equations.

Our numericalization of a $G$-crossed braided fusion category provides the full computational power for any theory using diagrammatical recouplings, though care has to be taken when strands pass over/under local extremals. This computational tool is especially useful for dealing with gauge-dependent quantities, which, in the $G$-crossed theory, include the important extended $G$-crossed modular $\mathcal{S}$ and $\mathcal{T}$ transformations.  As an application, we generalize the Verlinde formulas to the $G$-crossed Verlinde formulas Eqs. (\ref{eq:G-crossed_Verlinde_1}) and (\ref{eq:G-crossed_Verlinde_2}).  The diagrammatical recouplings also allow us to prove the theorems mentioned above in an elementary way.  In particular, we prove that the extended $G$-crossed modular $\mathcal{S}$ and $\mathcal{T}$ transformations indeed give rise to projective representations of $\textrm{SL}(2,\mathbb{Z})$.  We also conjecture the topological twists for the gauged (equivariantized) theory and derive the modular $S$-matrix of the gauged theory.

In Sec.~\ref{exampleSec}, we catalog many examples.  Those examples illustrate our theory and also potentially lead to new modular categories.  An interesting example is the gauging of the $S_3$-symmetry of the three-fermion theory $\textrm{SO}(8)_1$.  The resulting rank $12$ weakly integral modular tensor category has not previously appeared in the literature. It would be interesting to see if the triality of the Dynkin diagram $D_4$ would provide insight into the construction of this new modular category from $\textrm{SO}(8)\times S_3$.

\end{document}